\documentclass[fleqn,usenatbib]{mnras}
\usepackage{newtxtext,newtxmath}
\usepackage[T1]{fontenc}
\usepackage{ae,aecompl}
\usepackage{graphicx}	
\usepackage{amsmath}	
\usepackage{amssymb}	

\title[ETV analysis of {\em OGLE-IV} EBs]{Eclipse timing variation analysis of {\em OGLE-IV} eclipsing binaries toward the Galactic Bulge.\\ I.
Hierarchical triple system candidates}

\author[T. Hajdu et al.]{
T. Hajdu$^{1,2,3,4}$\thanks{E-mail: t.hajdu@astro.elte.hu},
T. Borkovits$^{5,3}$,
E. Forg\'{a}cs-Dajka$^{1}$,
J. Sztakovics$^{6,1}$,
G. Marschalk\'{o}$^{5,3}$, \newauthor
G. Kutrov\'{a}tz $^{1}$
\\\\
$^{1}$E\"{o}tv\"{o}s Lor\'{a}nd University, Department of Astronomy, H-1118 P\'{a}zm\'{a}ny P\'{e}ter stny. 1/A, Budapest, Hungary  \\
$^{2}$Konkoly Observatory, Research Centre for Astronomy and Earth Sciences, Hungarian Academy of Sciences, H-1121 Budapest,\\ Konkoly Thege Mikl\'{o}s \'{u}t 15-17, Hungary\\
$^{3}$Wigner Research Centre for Physics of HAS, PO Box 49, H-1525, Budapest, Hungary \\
$^{4}$MTA CSFK Lend\"ulet Near-Field Cosmology Research Group \\
$^{5}$Baja Astronomical Observatory of Szeged University, H-6500 Baja, Szegedi \'{u}t, Kt. 766, Hungary \\
$^{6}$ Eszterh\'{a}zy K\'{a}roly University, Department of Physics, H-3300 Eszterh\'{a}zy t\'{e}r 1 Eger, Hungary 
}

\date{Accepted XXX. Received YYY; in original form ZZZ}

\pubyear{2019}

\begin{document}
\label{firstpage}
\pagerange{\pageref{firstpage}--\pageref{lastpage}}
\maketitle

\begin{abstract}
We report a study of the eclipse timing variation (ETV) of short period  ($P_1\le 6^d$) eclipsing binaries (EB) monitored during the photometric survey {\em OGLE-IV}. From the 425\,193 EBs we selected approximately 80\,000 binaries that we found suitable for further examination. Among them  we identified 992 potential hierarchical triple (or multiple ) system candidates exhibiting light-travel-time effect (LTTE).
Besides, we obtained the orbital parameters of these systems and carried out statistical analyses on the properties of these candidates. 
We found that 
(i) there is a significant lack of triple systems where the outer period is less than 500 days; 
(ii) the distribution of the outer eccentricities has a maximum around $e_2\approx0.3$; 
(iii) the outer mass ratio calculated from an estimated minimum mass of the third component is lower than $q_2\sim 0.5$ for the majority of the sample. 
 We also present some systems that deserve special attention. (i) There are four candidates that show double periodic ETV, which we explain by the presence of a fourth companion. (ii) For two systems  the perturbations of the third component are also found to be significant therefore we give a combined dynamical and LTTE ETV solution. (iii) For one system the third component is found to be probably in the substellar mass domain. 

\end{abstract}


\section{Introduction}

The study of variable stars has a long history going back to antiquity \citep{2013ApJ...773....1J}.  The recent boom of the discovery of thousands of new variable stars is a natural byproduct of the large stellar surveys such as e.\,g. NASA's Kepler Mission \citep{2011AJ....141...83P} and OGLE \citep{OGLE}.

The investigation of triple stellar systems plays a significant role not only  in the understanding of the orbital evolution of close binaries, but also in the study of their whole life (from their formation to the death of the binary components) \citep{2016ComAC...3....6T}.  Furthermore, the various formation theories of close binary systems, e.\,g. the so called Kozai Cycles with Tidal Friction (KCTF) mechanism \citep[see, e.\,g.][]{Kiselevaetal98,fabryckytremaine07,naozfabrycky14}, as well as the recently proposed different disk and core fragmentation procedures \citep{tokovinin18,moekratter18}, require the presence of an additional, third stellar component for the explanation of the large number of the shortest (less than a few days) period, non-evolved binary stars.

One of the best known methods for the identification of a third companion around an eclipsing binary (EB) is based on the detection and analysis of the eclipse timing variations (ETV) of the binary.  If an EB has a distant, third companion, its distance from the observer varies periodically due to the EB's revolution around the common center of mass of the triple (or multiple) system. As a natural consequence, the light-travel time effect (LTTE) occurs, which manifests in periodic fluctuations in the observed times of the eclipses. Such kind of periodic ETVs have been found in hundreds of EBs in the last 60-70 years.

Several surveys have provided excellent photometry for ETV analysis. Besides the investigations of ultraprecise space photometry like {\em Kepler} {\citep{rappaportetal13,conroyetal14,Borkovits2015,Borkovits2016}} and {\em CoRoT} {\citep{hajduetal17} there are several studies that used ground-based survey data for searching multiple stellar systems, see e.g. \citet{2018arXiv180411272L}.

{ The Optical Gravitational Lensing Experiment ({\em OGLE}) was designed for  discovering dark matter using the microlensing technique in 1992 {\citep{udalskietal92}}. Observations of the currently running project {\em OGLE-IV} are made at Las Campanas Observatory, Chile with the 1.3-m Warsaw Telescope, which is currently equipped with a mosaic CCD camera. The majority of the observations were carried out in Cousins $I$ photometric band with an exposure time of 100 s, while a much smaller part of them were made in Johnson  $V$-band with the exposure time of 150 s. Recent and past OGLE surveys were found to be useful e.g. for exoplanet exploration \citep{2004A&A...421L..13B} and for the investigation of variable stars \citep{OGLE}.

The Galactic Bulge part of the {\em OGLE-IV} survey with its approximately half million EBs, which were identified by \citet{OGLE} gives us a good chance to increase the number of the candidates of hierarchical triple stellar systems. The authors also mentioned some potential triple systems in their paper that were also found by our algorithm.

Recently \citet{ZascheLMC, ZascheSMC} investigated LCs of the  {\em OGLE}  Large and the Small Magellanic Cloud eclipsing binaries and found some  additional components and determined their orbit.

In this paper we are searching for hierarchical triple star candidates towards the Galactic Bulge with the analysis of ETVs of EBs observed during the {\em OGLE-IV} survey. For  this study we use publicly available {\em OGLE- IV} photometric data\footnote{\url{ftp://ftp.astrouw.edu.pl/ogle/ogle4/OCVS/blg/ecl/phot\_ogle4/}}.

 In section \ref{effects} we shortly formulate the mathematical background of the third-body effected ETV analysis.

 In Section \ref{data processing} we outline the steps of our investigation, starting with the methods used for data acquisition and automatic $O-C$ curve generation, then continuing with the system selection and, finally, closing with a short description of some details of the applied ETV and the auxiliary light curve analyses as well.

The results of the analysis of the ETVs of the new hierarchical triple candidates, as well as some other interesting by-products of our research, are discussed in Section \ref{results}.

Finally, a short summary is given in Section \ref{summary}.

\section{Effects of a third body on the ETV}\label{effects}

In this paper we define ETV  as the difference between the observed and calculated times of minima (which is also called as $O-C$):
\begin{equation}
\Delta=T(E)-T_0-P_\mathrm{s}E,
\end{equation}
where $T(E)$ denotes the time of the $E$th eclipse, $T_0=T(0)$ is the time of the 'zeroth' eclipse and $P_\mathrm{s}$ denotes the orbital period of the binary. Our basic model is given by
\begin{equation}
\Delta =\sum_{i=0}^2c_iE^i+\left[\Delta_\mathrm{LTTE}+\Delta_\mathrm{dyn}\right]_0^E
\label{Eq:jobboldal}
\end{equation}
The constant and linear terms of the polynomial in $E$ give corrections to $T_0$ and $P_\mathrm{s}$. The second order coefficient 
provides the half rate of the linear variation in period. Note that this term had significant value only in a few cases. These systems are marked in the corresponding tables.
The second and third components in the right side of the equation, $\Delta_\mathrm{LTTE}$ and $\Delta_\mathrm{dyn}$, refer to the contributions of LTTE and short-timescale dynamical third-body perturbations. 
Note that an ETV and, therefore, Eq.\,(\ref{Eq:jobboldal}) may contain additional components, for example the effect of the apsidal motion in an eccentric EB. However the wast majority of the short period EBs in our sample revolve on circular orbit and, therefore, this component does not play any role.

In the following two subsections we describe shortly the LTTE and dynamical terms.

\subsection{The light-travel time effect}

According to our knowledge, \citet{Chandler1888} was the first to mention LTTE as a possible origin of the observed ETVs of Algol. Later the widely used mathematical description of an LTTE forced ETV was published by \citet{Irwin1952}. He also gave a graphical fitting procedure for determining the elements of the light-time orbit from the ETVs that had been investigated by the use of the eclipse timing diagrams. Traditionally these diagrams were called {\em O--C} diagrams (see e.g. \citealp{Sterken2005} for a short review on the advantages of {\em O--C} diagrams in the analyses of period variations). 

There are several other mechanisms capable of producing ETVs in EBs, and some of them may even strongly mimic LTTE-like behavior. 
Therefore the detection of the third component with this method is not an easy matter.
In this regard \citet{Frieboes-Conde1973} listed four criteria that an ETV curve should fulfill for an LTTE solution.
These criteria can be summarized as follows. (1) The shape of the ETV curve must follow the analytical form of an LTTE solution. (2) The ETVs of the primary and secondary minima must be consistent in both amplitude and phase with each other. (3) The estimated mass or the minimum mass of the third component, derived from the amplitude of the LTTE solution, must be in accord with photometric measurements or limits on third light in the system. (4) Variation of the systemic radial velocity (if it is available) should be consistent with the LTTE solution. 

According to \citet{Irwin1952} the LTTE contribution takes the following form 
\begin{equation}
\label{LTTEfunction}
\Delta_\mathrm{LTTE}=-\frac{a_\mathrm{AB}\sin i_2}{c}\frac{\left(1-e_2^2\right)\sin\left(v_2+\omega_2\right)}{1+e_2\cos v_2},
\end{equation}
where $a_\mathrm{AB}$ denotes the semi-major axis of the EB's center of mass around the center of mass of the triple system, while $i_2$, $e_2$, $\omega_2$ stand for the inclination, eccentricity, and argument of periastron of the relative outer orbit, respectively. Furthermore, $c$ is the speed of light and $v_2$ is the true anomaly of the third component. Note the negative sign on the right hand side, which arises from the use of the {\em companion's} argument of periastron, instead of the argument of periastron of the light time orbit of the EB ($\omega_\mathrm{AB}=\omega_2+\upi$).

The amplitude of the LTTE takes the form

\begin{equation}
\label{LTTEamplitude}
\mathcal{A}_\mathrm{LTTE}=\frac{a_\mathrm{AB}\sin i_2}{c}\sqrt{1-e_2^2\cos^2\omega_2},
\end{equation}

while the mass function $f(m_\mathrm{C})$, analogous to its spectroscopic counterpart for single-line spectroscopic binaries, is usually defined as
\begin{equation}
\label{Mass_function_eq}
f(m_\mathrm{C})=\frac{m_C^3\sin^3i_2}{m^2_\mathrm{ABC}}=\frac{4\pi^2a^3_\mathrm{AB}\sin^3i_2}{GP_2^2},
\end{equation}
and can be calculated from the parameters of the LTTE solution.  Using the mass function, the amplitude of the LTTE can be approximated as
\begin{equation}
\label{LTTEamplitudeapprox}
\mathcal{A}_\mathrm{LTTE}\approx 1.1 \times 10^{-4} f(m_C)^{1/3}P_2^{2/3}\sqrt{1-e_2^2\cos^2\omega_2},
\end{equation}
where $f(m_\mathrm{C})$ should be expressed in solar masses, $P_2$ in days, and $\mathcal{A}_\mathrm{LTTE}$ is also resulted in days.

Note that if the mass of the eclipsing binary is known, the minimum mass ($i_2=90^\circ$) of the third component can be determined based on the mass function ($f(m_C)$).

\subsection{Dynamical perturbation of the third component on the ETV}
In tight hierarchical triple stellar systems  the short-timescale three-body perturbations on the Keplerian two-body motion of the inner binary  may also alter the ETVs significantly. This dynamical ETV contribution was analytically described in a series of papers by \citet{Borkovits2003,Borkovits2011,Borkovits2015}. 

The dynamical ETV component ($\Delta_\mathrm{dyn}$) has a complex dependence on the orbital elements of the inner and outer orbits, and their relative configurations as well. Furthermore, for eccentric inner orbits even the orbits' relative orientation to the observer becomes an additional important factor. A comprehensive description of these effects can be found in \cite{Borkovits2015}.  In our sample, however, we calculate dynamical effects only for circular EBs. Therefore, for our purposes it is satisfactory to use the substantially simpler formula of \citet{Borkovits2003} that takes the following form:

\begin{equation}
\Delta_\mathrm{dyn}=\mathcal{A}_\mathrm{dyn}\left[\left(1-\frac{3}{2}\sin^2i_\mathrm{m}\right)\mathcal{M}+\frac{3}{4}\sin^2i_\mathrm{m}\mathcal{S}\right],
\end{equation}
where
\begin{eqnarray}
\mathcal{M}&=&v_2-l_2+e_2\sin v_2, \\
\mathcal{S}&=&\sin(2v_2+2g_2)+e_2 \nonumber \\ 
&&\times\left[\sin(v_2+2g_2)+\frac{1}{3}\sin(3v_2+2g_2)\right].
\end{eqnarray}
 stand for the time-dependent functions of the true anomaly ($v_2$) as well as the mean anomaly ($l_2$) of the outer body on its relative orbit around the EB's center of mass, while $g_2$ is the tertiary's argument of periastron measured from the intersection of the inner and outer orbital planes. Furthermore, $i_\mathrm{m}$ denotes the mutual (relative) inclination of the inner and outer orbits, while the mass and period ratios of the inner and outer subsystems occur in the amplitude-like quantity
\begin{equation}
\mathcal{A}_\mathrm{dyn}=\frac{1}{2\upi}\frac{m_\mathrm{C}}{m_\mathrm{ABC}}\frac{P_1^2}{P_2}\left(1-e_2^2\right)^{-3/2}.
\label{Eq:Adyn}
\end{equation}
Note that despite the fact that the true magnitude of the dynamical ETV can be significantly altered by the two eccentricities and the mutual inclination as well \citep{Borkovits2011}, it was found by \citet{Borkovits2016} that in most cases $\mathcal{A}_\mathrm{dyn}$ gives a reasonable estimation at least for the magnitude of the short-term dynamical ETV contribution.

In this paper we present two short periodic hierarchical triple stellar system candidates with significant dynamical contribution (see Sec. \ref{subsecdyn}). 

\section{Basic steps of the analysis}\label{data processing}
\subsection{System selection}
The {\em{OGLE-IV}} survey provided around 425\,193 light curves of EBs that were observed in \textit{I} and \textit{V} bands. Since the \textit{I} band light curves typically contain much more points than \textit{V} band ones, we relied only on the former photometric band data for our analysis.  
Our main goal was to find hierarchical triple stellar candidates whose outer period ($P_2$) is shorter than, or comparable to, the length of the observations.
Unfortunately most of the data trains do not contain enough data points (more than the half of the cases contain less than 1000 points) for a detailed examination.  Therefore we investigated only those systems whose light curves contain more than 4000 points. Applying  this criterion we reduced our sample from the original $425\,193$ EBs to $\sim80\,000$ systems.

\subsection{Determination of times of minima}
To determine the times of minima we used a slightly modified version of the algorithm used by \citet{hajduetal17}. Note that a similar algorithm was used by \citet{2018arXiv180602247B}. 

For the determination of the minima times we used phase-folded and binned light curves of the systems. The initial periods for the phase-folding processes were taken from the original OGLE site. In some cases these  periods were updated and then the phase-folding was reiterated by the use of the period corrections obtained from the ETV analyses.

The folded light curves were binned into 1000 equally spaced phase-cells, according to the orbital phases of each measured points. Then the weighted arithmetic mean magnitudes were calculated cell by cell, and associated to the phase of the cell midpoints.  (For the weighting the uncertainties of each individual data points were used). In the next step, we formed 12-th order polynomial template functions for the primary and secondary eclipses of the folded and binned light curves. We intended to use these templates to determine the mid-times of the individual eclipse events.

 Unfortunately, in most cases the individual cycle to cycle light curves are badly covered and, therefore, we were not able to determine individual times of minima with the necessary accuracy. Thus, we decided to use normal minima. After some attempts we came to the conclusion that calculating one normal minimum for every 17 consecutive binary cycles would provide the best compromise between the requested phase coverage of any individual eclipsing minima, and the decreasing time-resolution, and smoothing of the ETVs.

We found that by using the templates of the two types of eclipses at once we can  reduce the uncertainties of the ETV data. In our work we used these ETVs for analyses but we also used the primary and the secondary ETVs for confirmations. Note that this dual fit process is applicable only if the eccentricity of the inner orbit is zero so there is no apsidal motion.

\subsection{ETV analysis}

To select those systems  that exhibit periodic ETVs we  applyed a Levenberg-Marquardt (LM) based process fitting a sine function together with a second order polynomial in the following form:

\begin{equation}
\label{sin}
f(x)=a_0+a_1\cdot x+a_2\cdot x^2+a_3\cdot\sin\left(a_4+\frac{2\pi}{a_5}\cdot x\right).	
\end{equation}

The initial period  ($a_5$)} was varied between $80\times P_1$ and 4000 days with a step size of $10\times P_1$, and the best fit was chosen via $\chi^2$ search. The former value was found to be an appropriate lower limit for reasonable outer periods, while the latter one is twice of the maximum length of the data series.

Then  with the use of a combined grid-search and LM method our code searches for an LTTE solution (Eq. \ref{LTTEfunction})  together with the parabolic
term.  For this process the initial values of some of the parameters (third-body period, polynomial coefficients) are taken from the previous, sine-fit.

To get the best fitted  ETV solution the initial values of the eccentricity ($e_2$), argument of the periastron ($\omega_2$) and periastron passage time ($\tau_2$) were  set to 4-6 different, evenly spaced values within their physically realistic range.

Then, in the last stage, the goodness of the solutions was tested and, therefore, the selection of the triple-candidate systems was carried out also in an automatic manner. The first criterion was that the amplitude ($\mathcal{A}_\mathrm{LTTE}$) of the LTTE solution (Eq. \ref{LTTEamplitude}) has to be higher than one and the half times the average absolute difference between successive $O-C$ points.  The other criterion was based on the normalized $\chi^2$ value which was counted in the following form

\begin{equation}
\chi^2 = \frac{1}{N}\sum_{i=1}^N\frac{\left(y_i-f_i\right)^2}{\sigma_i^2},
\end{equation}
 where N is the number of the $O-C$ points, $y_i$ is the value of the $i$th $O-C$ point, $f_i$ represents the $i$th $O-C$ value derived from the ETV solution and finally $\sigma_i$ is the uncertainty of the $i$th $O-C$ point. 
  Finally, this list was corrected (basically reduced) through a manual inspection.

\section{Results}
\label{results}

In conclusion, we have found 992 potential hierarchical multiple stellar system candidates in the photometric data of the {\em OGLE-IV} survey. We divided these candidates into two groups. Into the first group of the more probable triples we put basically those systems for which 
 the outer period is less than 1500 days and the amplitude is at least 3 times higher than the variance of the residual, or else the period is lower than 1000 days, while the other group contains the remaining, less confident cases.

The results obtained for the two sets of our candidates are listed in Tables\,\ref{Triplestable260} and \ref{Triples}, respectively. These tables provide the OGLE ID, the epoch ($T_0$), the inner and outer periods ($P_1$, $P_2$), the eccentricity ($e_2$), argument of the periastron ($\omega_2$) and periastron passage time ($\tau_2$) of the third companion,  the projected semi-major axis of the light-time orbit ($a_\mathrm{AB}\sin i_2$), the mass function ($f(m_\mathrm{C})$) and the parabolic term ($\Delta P_1$) where it is significant. Furthermore, we plot the ETVs together with the LTTE solutions and the raw and folded light curves in Fig.\,\ref{Triplesplot260} as well.

 In what follows, after enumerating some individual systems with special interests (Sects.\,\ref{subsecdyn},\,\ref{doubleperiodic},\,\ref{subsecsub}) we carry out detailed statistical analyses of the properties of our candidates in Sect.\,\ref{stat}.

\begin{table*}
\caption{LTTE solutions for the 258 most certain hierarchical triple star candidates of the OGLE\,IV sample. (The full table can be obtained in machine readable form in the electronic edition of the paper.)}
\label{Triplestable260}
\centering
\resizebox{\textwidth}{!}{%
\begin{tabular}{c c c c @{$\pm$} c c @{$\pm$} c c @{$\pm$} c c @{$\pm$} c c @{$\pm$} c c @{$\pm$} c c @{$\pm$} c}
\hline
ID & $T_0$ & $P_1$ & \multicolumn{2}{c}{$P_2$} & \multicolumn{2}{c}{ $a_\mathrm{AB}\cdot\sin(i_2)$}& \multicolumn{2}{c}{$e_2$} & \multicolumn{2}{c}{$\omega_2$} & \multicolumn{2}{c}{$\tau_2$} & \multicolumn{2}{c}{$f(m_\mathrm{C})$} & \multicolumn{2}{c}{$\Delta P_1$}\\ 
 & [HJD-2450000 days] & [days] & \multicolumn{2}{c}{[days]}  & \multicolumn{2}{c}{[$R_{\odot}$]} & \multicolumn{2}{c}{ }
  & \multicolumn{2}{c}{[deg]} & \multicolumn{2}{c}{[days]} & \multicolumn{2}{c}{} & \multicolumn{2}{c}{[$\times10^{-10}\frac{d}{c}$]}    \\ \hline
28238 & 5265.711315 & 0.336880 & 424.5 & 9.0 & 123.9 & 10.7 & 0.61 & 0.10 & 31.7 & 0.9 & 5606.1 & 9.3 & 0.1413 & 0.03339 \\
32148 & 5265.437046 & 0.311293 & 762.5 & 425.4 & 143.8 & 100.3 & 0.41 & 0.40 & 239.9 & 4.2 & 5484.1 & 329.2 & 0.0685 & 0.21020 \\
35547 & 5265.221608 & 1.745663 & 648.3 & 36.0 & 151.8 & 8.1 & 0.43 & 0.13 & 75.7 & 1.6 & 5965.1 & 21.4 & 0.1116 & 0.03050 \\
47614 & 5265.569008 & 0.290397 & 1406.0 & 7.7 & 142.3 & 5.3 & 0.53 & 0.04 & 10.3 & 0.3 & 6026.3 & 10.3 & 0.0196 & 0.00178 \\
117862 & 5260.079711 & 1.048052 & 438.1 & 13.7 & 101.9 & 5.3 & 0.05 & 0.11 & 157.0 & 11.6 & 5458.2 & 105.9 & 0.0740 & 0.01464 \\
118172 & 5260.619707 & 0.289513 & 859.7 & 211.1 & 28.5 & 4.1 & 0.20 & 0.42 & 93.3 & 7.0 & 5743.9 & 164.1 & 0.0004 & 0.00041 \\
118178 & 5260.599311 & 0.330831 & 772.0 & 55.6 & 70.1 & 3.9 & 0.24 & 0.18 & 49.7 & 4.6 & 5578.7 & 71.7 & 0.0077 & 0.00252 & 32.4 & 5.7\\
...\\
\hline
\end{tabular}%
}
\end{table*}

\begin{table*}
\caption{ LTTE solutions for the remaining, less certain hierarchical triple star candidates of the OGLE\,IV sample. (The full table can be obtained in machine readable form in the electronic edition of the paper.)}
\label{Triples}
\centering
\resizebox{\textwidth}{!}{%
\begin{tabular}{c c c c @{$\pm$} c c @{$\pm$} c c @{$\pm$} c c @{$\pm$} c c @{$\pm$} c c @{$\pm$} c}
\hline
ID & $T_0$ & $P_1$ & \multicolumn{2}{c}{$P_2$} & \multicolumn{2}{c}{ $a_\mathrm{AB}\cdot\sin(i_2)$}& \multicolumn{2}{c}{$e_2$} & \multicolumn{2}{c}{$\omega_2$} & \multicolumn{2}{c}{$\tau_2$} & \multicolumn{2}{c}{$f(m_\mathrm{C})$} \\ 
 & [HJD-2450000 days] & [days] & \multicolumn{2}{c}{[days]}  & \multicolumn{2}{c}{[$R_{\odot}$]} & \multicolumn{2}{c}{ }
  & \multicolumn{2}{c}{[deg]} & \multicolumn{2}{c}{[days]}    \\ \hline
65 & 5260.516383 & 0.212209 & 2050.5 & 197.6 & 31.6 & 4.5 & 0.56 & 0.25 & 254.4 & 4.9 & 5811.5 & 184.5 & 0.0001 & 0.00006 \\
72 & 5260.717277 & 0.213604 & 2240.5 & 637.2 & 72.5 & 56.2 & 0.59 & 0.94 & 44.9 & 10.2 & 6000.6 & 311.4 & 0.0010 & 0.00240 \\
74 & 5261.737899 & 0.198408 & 1207.2 & 70.3 & 64.1 & 25.2 & 0.77 & 0.32 & 42.9 & 3.0 & 5741.3 & 51.1 & 0.0024 & 0.00231 \\
106 & 5260.691963 & 0.218233 & 2154.3 & 886.4 & 37.0 & 62.8 & 0.08 & 1.57 & 327.5 & 79.5 & 6396.4 & 3534.0 & 0.0001 & 0.00046 \\
28702 & 5265.512011 & 0.446362 & 1794.3 & 73.9 & 100.1 & 37.2 & 0.86 & 0.25 & 308.2 & 4.6 & 5403.9 & 38.9 & 0.0042 & 0.00364 \\
28743 & 5265.461422 & 0.303213 & 1642.9 & 32.5 & 186.5 & 62.3 & 0.74 & 0.14 & 8.4 & 1.0 & 5240.9 & 29.6 & 0.0322 & 0.02342 \\
30609 & 5265.235221 & 0.562353 & 2044.0 & 341.3 & 241.0 & 309.0 & 0.31 & 0.09 & 278.3 & 1.5 & 5333.8 & 178.3 & 0.0449 & 0.13763 \\
...\\
\hline
\end{tabular}
}
\end{table*}

\begin{figure*}
\caption{ The raw $I$-band (upper left panel) and the folded (right panel) light curve and the ETV data together with the LTTE solution (lower left panel) for the 255 most certain candidate systems. (The full figure for all the triples can be obtained in the electronic edition of the paper.)}

\includegraphics[width=0.95\columnwidth]{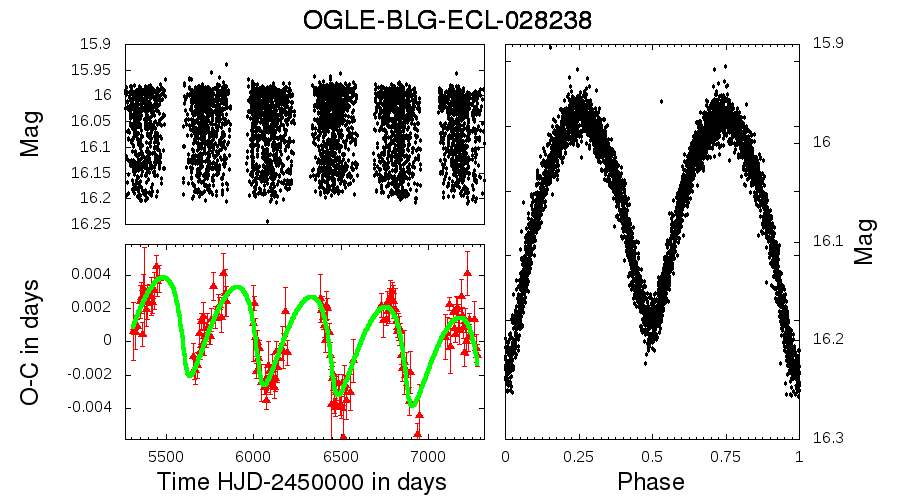}
\includegraphics[width=0.95\columnwidth]{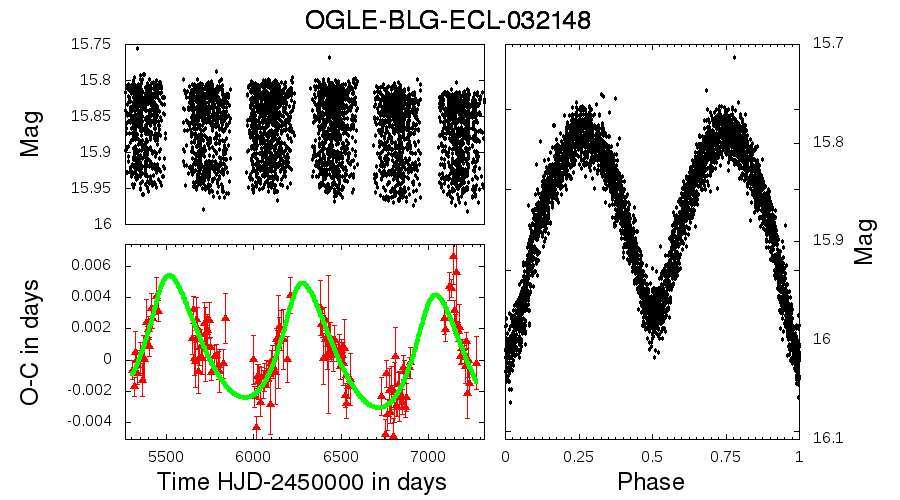}

\includegraphics[width=0.95\columnwidth]{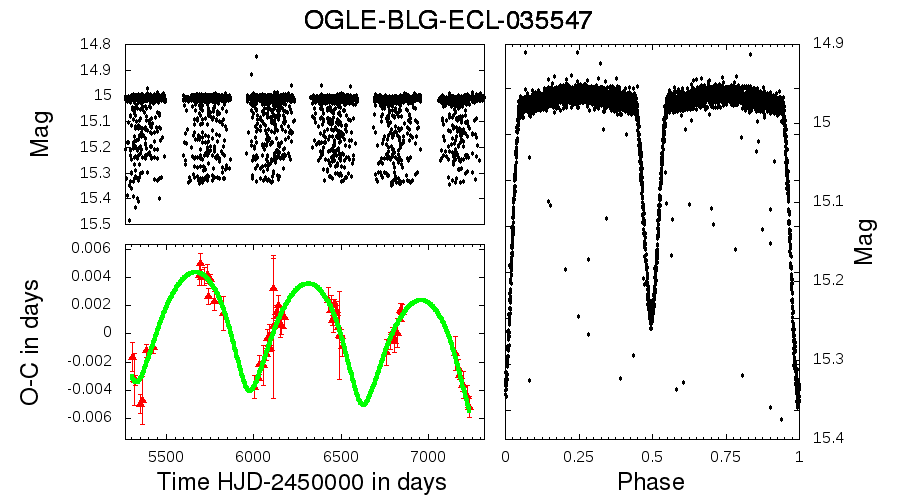}
\includegraphics[width=0.95\columnwidth]{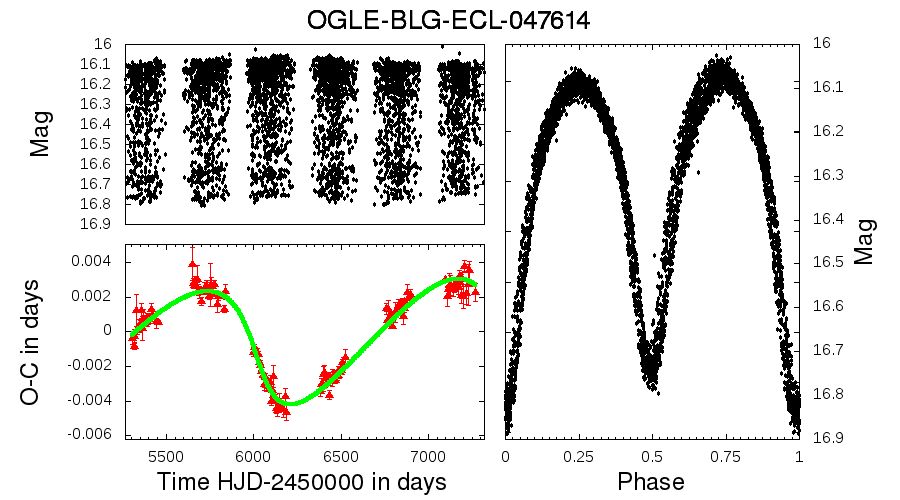}

\includegraphics[width=0.95\columnwidth]{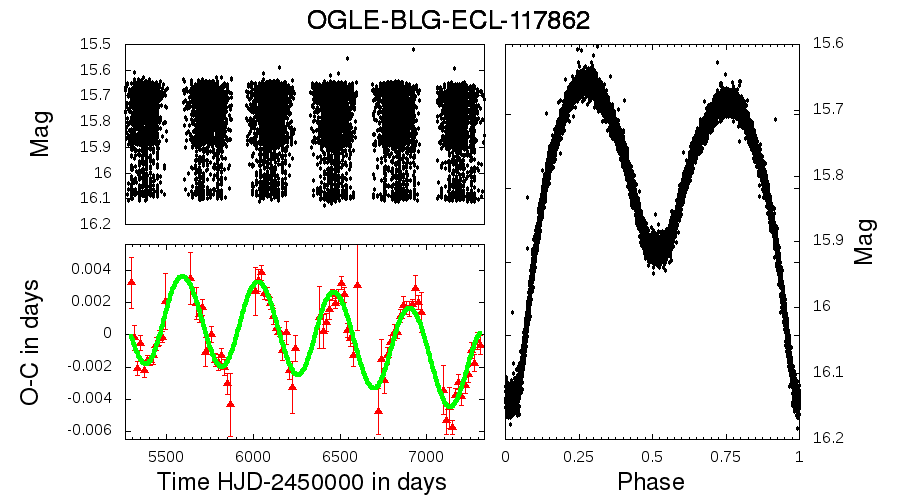}
\includegraphics[width=0.95\columnwidth]{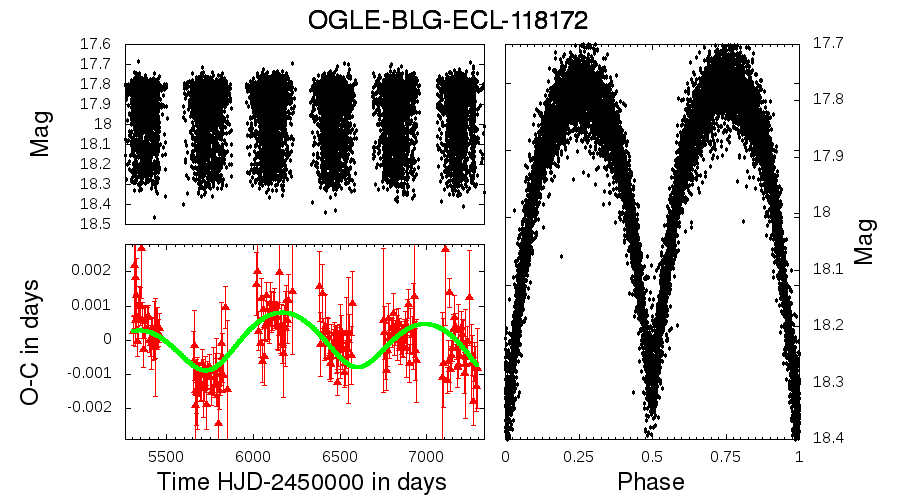}

\includegraphics[width=0.95\columnwidth]{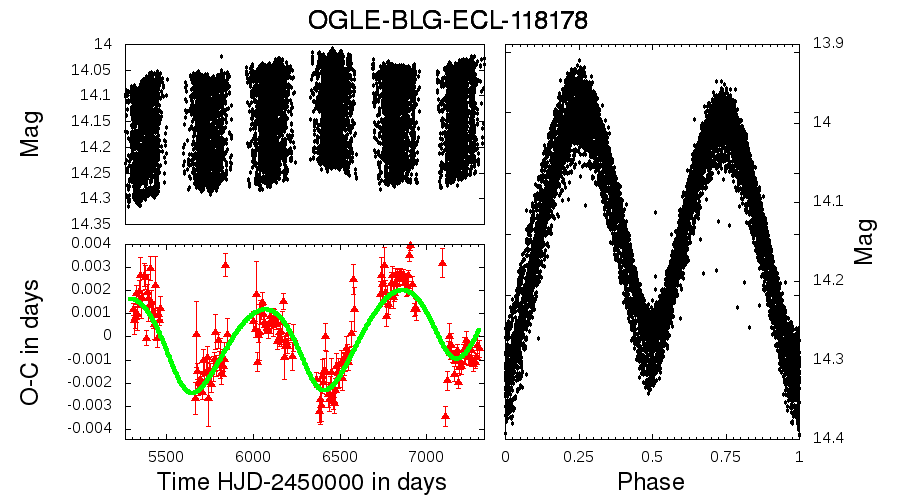}
\includegraphics[width=0.95\columnwidth]{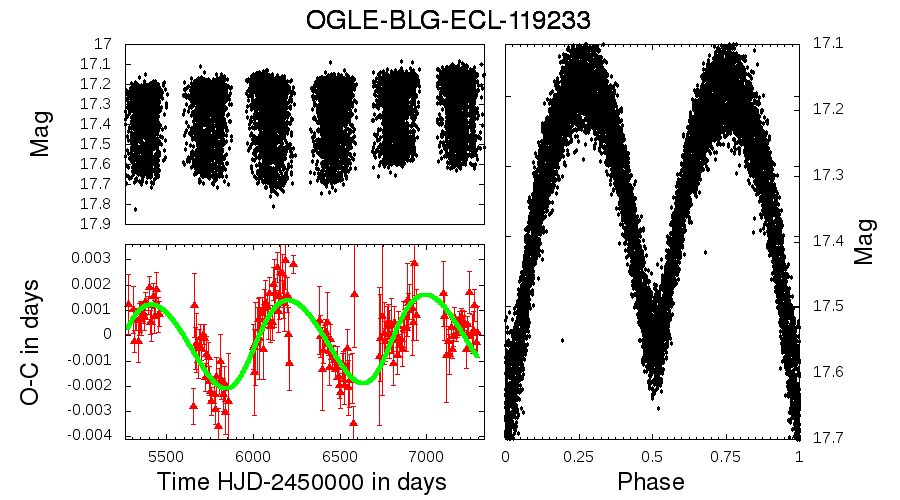}

\includegraphics[width=0.95\columnwidth]{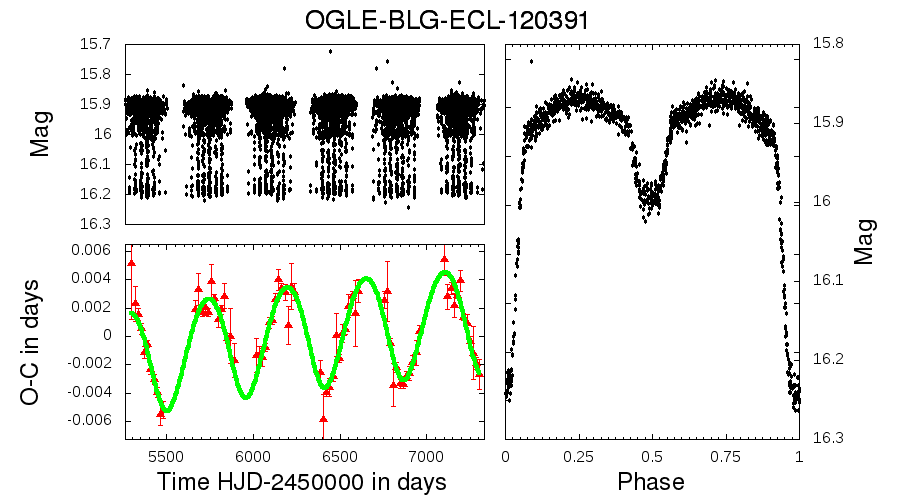}
\includegraphics[width=0.95\columnwidth]{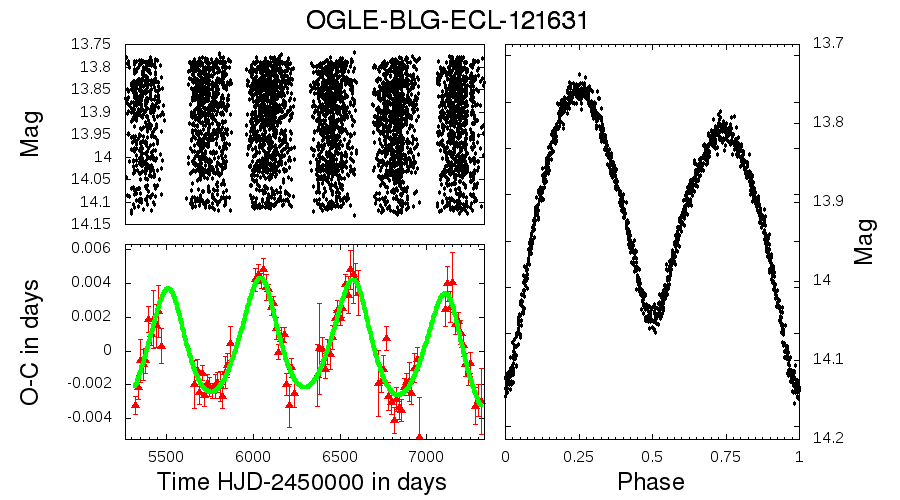}
\label{Triplesplot260}
\end{figure*}

\subsection{Systems with significant dynamical effect}
\label{subsecdyn}
 For the vast majority of the investigated systems, in comparison to the LTTE term, the dynamical contribution to the ETV can safely be ignored. This fact was far not unexpected, as the amplitude of the dynamical ETV is scaled with the EB's period, hence our sample systems with their typical period of $P_1<1$-d are strongly unfavourable for the detection of such effect. Despite this, we detected two systems where the results of the preliminary LTTE analysis have predicted significant dynamical ETV contribution.  

For these two systems, in theory, we were able to determine system masses, though only with large uncertainties. The parameters of these systems are tabulated in Table\,\ref{dyntable} and the LTTE+dynamical ETV solutions are presented in Figs.\,\ref{fig1dyn} and \ref{fig2dyn}.

\begin{table}
\caption{Orbital elements from combined dynamical and LTTE solutions. } 
\label{dyntable}
\centering 
\begin{tabular}{l c c c}
\hline
ID &  & 143356 & 169255\\
$T_0$ & [days & 5258.631022 & 5258.554303 \\
$P_1$ & [days] & 2.442595 & 2.804854\\
$P_2$ & [days] & 202.43 $\pm$ 0.20 &  339.28 $\pm$ 1.03\\
$a_2$ & [R$_{\sun}$] & 254.9 $\pm$ 15.32 & 330.19 $\pm$ 27.49\\
$e_2$ &  & 0.32 $\pm$ 0.02 & 0.3 $\pm$ 0.04\\
$\omega_2$ & [$\degr$]  & 239.19 $\pm$ 8.12  & 288.05 $\pm$ 11.31\\
$\tau_2$  & [days& 5218.02 $\pm$ 5.5 & 5245.44 $\pm$ 13.21\\
$\mathcal{A}_\mathrm{LTTE}$ & [days] & 0.0021 $\pm$ 0.0002  & 0.0031 $\pm$ 0.0005\\
$\mathcal{A}_\mathrm{dyn}$ & [days] & 0.0016 & 0.0013 	\\
$\frac{\mathcal{A}_\mathrm{dyn}}{\mathcal{A}_\mathrm{LTTE}}$ & & 0.78 & 0.44 \\
$f(m_\mathrm{C})$ &  & 0.16 $\pm$ 0.05 & 0.18 $\pm$ 0.09\\
$\frac{m_\mathrm{C}}{m_\mathrm{ABC}}$ &  & 0.31 $\pm$ 0.03  &  0.35 $\pm$ 0.05\\
$m_\mathrm{AB}$ & [M$_{\sun}$] & 3.74 $\pm$ 0.86 & 2.74 $\pm$ 0.87\\
$m_\mathrm{C}$ & [M$_{\sun}$] & 1.69 $\pm$ 0.45 & 1.46 $\pm$ 0.58\\
\hline
\end{tabular}
\end{table}

\begin{figure}
\includegraphics[width=\columnwidth]{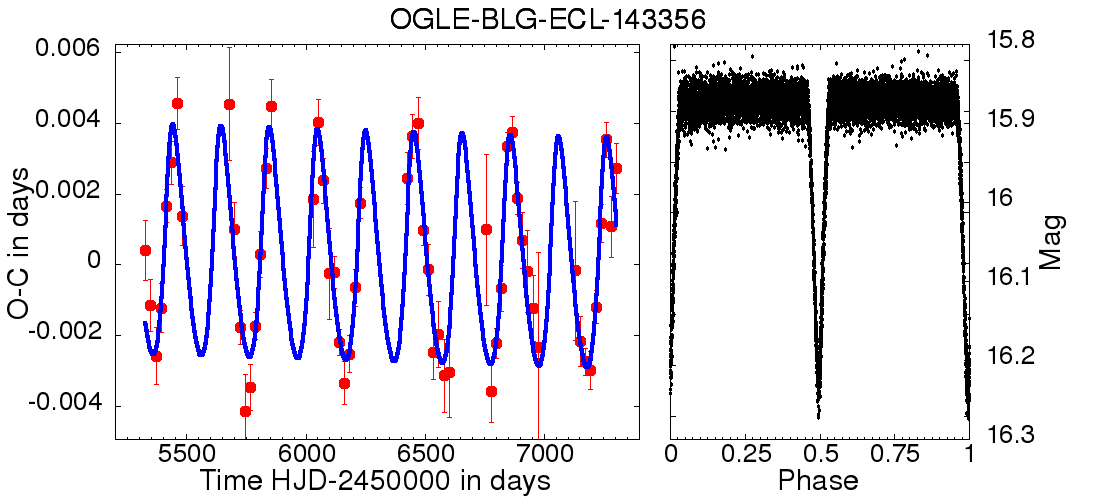}
\caption{ETV of  OGLE-BLG-ECL-143356 (red) together with the LTTE+dynamical ETV solution (blue) in the left panel. The folded light curve in the right panel}
\label{fig1dyn}
\end{figure}

\begin{figure}
\includegraphics[width=\columnwidth]{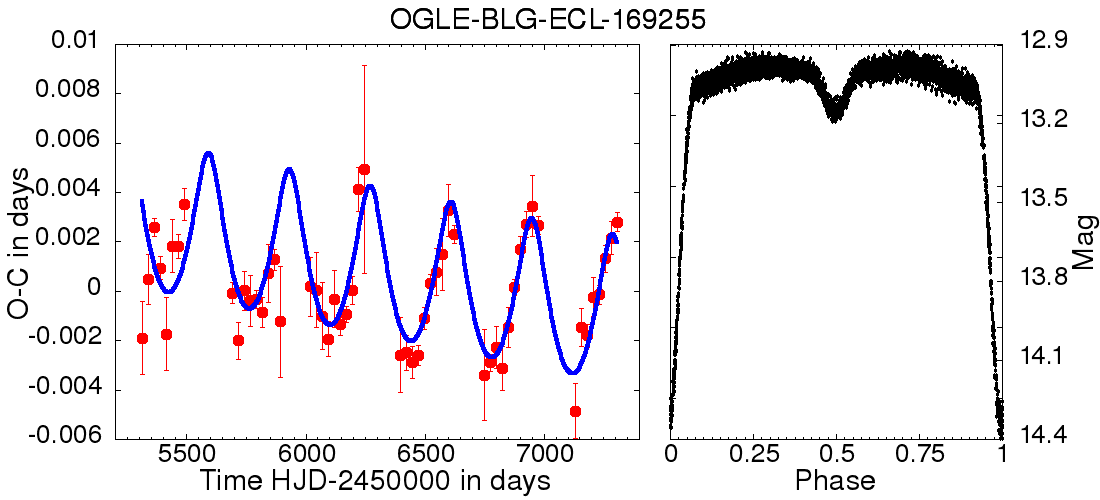}
\caption{ETV of the system OGLE-BLG-ECL-169255 (red) together with the LTTE+dynamical ETV solution (blue) in the left panel. The folded light curve in the right panel}
\label{fig2dyn}
\end{figure}

{\it OGLE-BLG-ECL-143356} is an Algol-type EB with a period of $P_1\sim2\fd4$ and moderately different primary and secondary eclipse depths (see Fig.\,\ref{fig1dyn}, right panel). The amplitude of the dynamical effect is $\sim80\%$ of LTTE amplitude. Our results imply that the system is formed by three stars more massive than our Sun. The total mass of the inner binary was found to be $\sim3.7\pm0.9\,M_{\sun}$, while the mass of the third component was found to be $\sim1.7\pm0.5\,M_{\sun}$. Furthermore, the folded light curve suggests that the inner binary is formed by two similar stars and, therefore, the whole triple might be made up of components of almost equal masses.

{\it OGLE-BLG-ECL-169255} is an another Algol-type EB with  unequally bright components (see Fig.\,\ref{fig2dyn} right panel). According to our solution the third component has a mass of $\sim1.5\pm0.6\,M_{\sun}$. Note that the amplitude of the dynamical delay is less than the half of the amplitude of the LTTE.

\subsection{Systems with double periodic ETVs }
\label{doubleperiodic}

\begin{table*}
\centering
\caption{Results of double LTTE solution fit}
\label{quadrupoltable}
\resizebox{\textwidth}{!}{%
\centering
\begin{tabular}{c c c c @{$\pm$} c c @{$\pm$} c c @{$\pm$} c c @{$\pm$} c c @{$\pm$} c c @{$\pm$} c}
\hline
ID & $T_0$ & $P_1$ & \multicolumn{2}{c}{$P_2$} & \multicolumn{2}{c}{ $a\cdot\sin(i_2)$}& \multicolumn{2}{c}{$e_2$} & \multicolumn{2}{c}{$\omega_2$} & \multicolumn{2}{c}{$\tau_2$} &  \multicolumn{2}{c}{$f(m_\mathrm{C})$ }\\

 & [days] & [HJD-2450000 days] & \multicolumn{2}{c}{[days]}  & \multicolumn{2}{c}{[$R_{\odot}$]} & \multicolumn{2}{c}{ }
  & \multicolumn{2}{c}{[deg]} & \multicolumn{2}{c}{[days]}   \\\hline
$136469_c$ & 5260.409845 & 0.681115 & 149.23 & 0.16 & 69.97 & 1.87 & 0.08 & 0.05 & 112.46 & 37.97 & 
5327.52 & 15.85 & 1.3128 & 0.075\\
$153291_c$ & 5260.916183 & 0.283085 & 120.06 & 0.15 & 31.42 & 1.32 & 0.28 & 0.07 & 178.47 & 14.21 & 
5357.23 & 5.04 & 0.5764 & 0.051\\
$165849_c$ & 5260.553794 & 0.276678 & 367.34 & 0.54 & 182.57 & 22.72 & 0.46 & 0.05 & 178.46 & 34.77 & 
5469.94 & 9.06 & 2.2016 & 0.558\\
$259162_c$ & 5376.300087 & 0.355920 & 192.53 & 0.54 & 83.40 & 3.71 & 0.15 & 0.08 & 3.80 & 32.75  & 
5296.71 & 18.00	& 1.3231 & 0.129\\
\hline
ID & & & \multicolumn{2}{c}{$P_3$} & \multicolumn{2}{c}{ $a\cdot\sin(i_3)$}& \multicolumn{2}{c}{$e_3$} & \multicolumn{2}{c}{$\omega_3$} & \multicolumn{2}{c}{$\tau_3$} & \multicolumn{2}{c}{$f(m_\mathrm{D})$}\\

 &  &  & \multicolumn{2}{c}{[days]}  & \multicolumn{2}{c}{[$R_{\odot}$]} & \multicolumn{2}{c}{ }
  & \multicolumn{2}{c}{[deg]} & \multicolumn{2}{c}{[days]}   \\\hline
$136469_d$ & & & 1633.45 & 115.30 & 45.26 & 3.25 & 0.23 & 0.12 & 265.78 & 33.36 & 5853.13 & 226.57 & 	0.1283 & 0.046\\
$153291_d$ & & & 2077.10 & 339.21 & 46.27 & 7.29 & 0.16 & 0.07 & 101.71 & 61.23 & 
6645.60 & 947.48 & 0.1111 & 0.089\\
$165849_d$ & & & 2026.46 & 298.60 & 203.77 & 37.09 & 0.3 & 0.05 & 45.25 & 12.70 & 
6601.04 & 636.86 & 0.5666 & 0.457\\
$259162_d$ & & & 2036.21 & 224.98 & 188.41 & 17.37 & 0.30 & 0.09 & 23.88 & 21.56 & 
5103.95 & 485.25 & 0.5152 & 0.266\\
\hline
\end{tabular}
}
\end{table*}

\begin{figure*}
\includegraphics[width=\columnwidth]{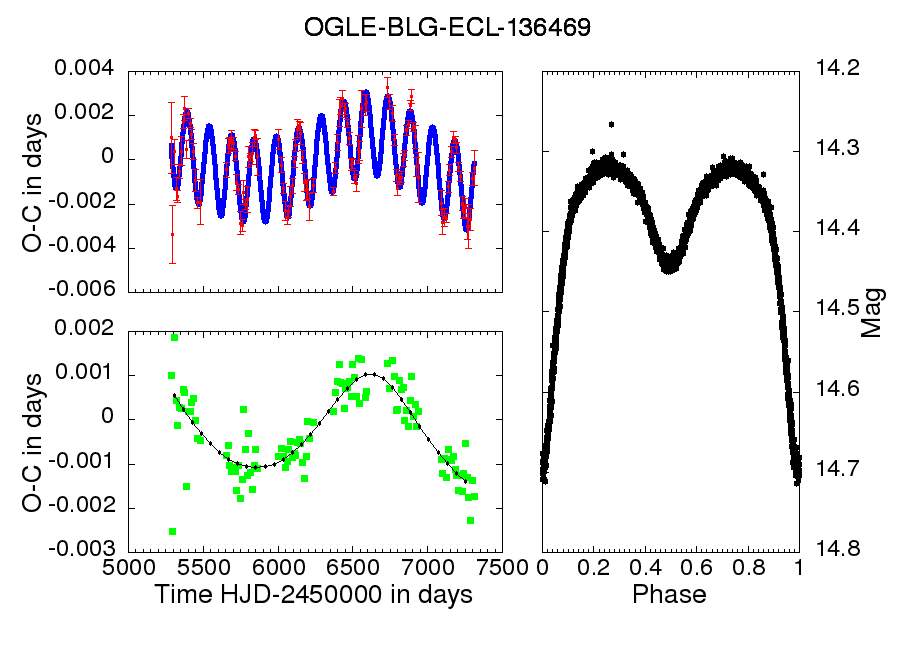}
\includegraphics[width=\columnwidth]{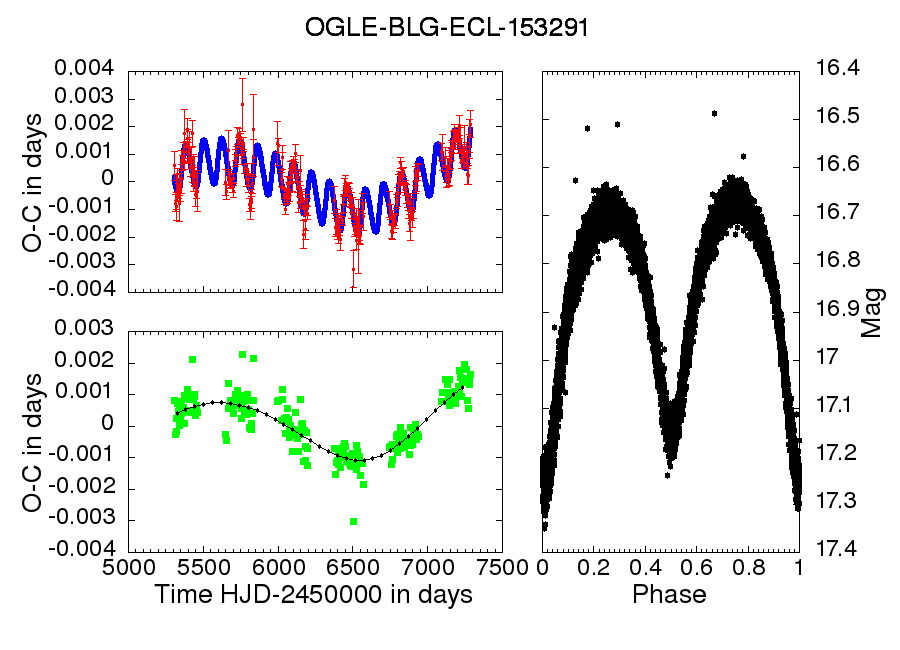}

\includegraphics[width=\columnwidth]{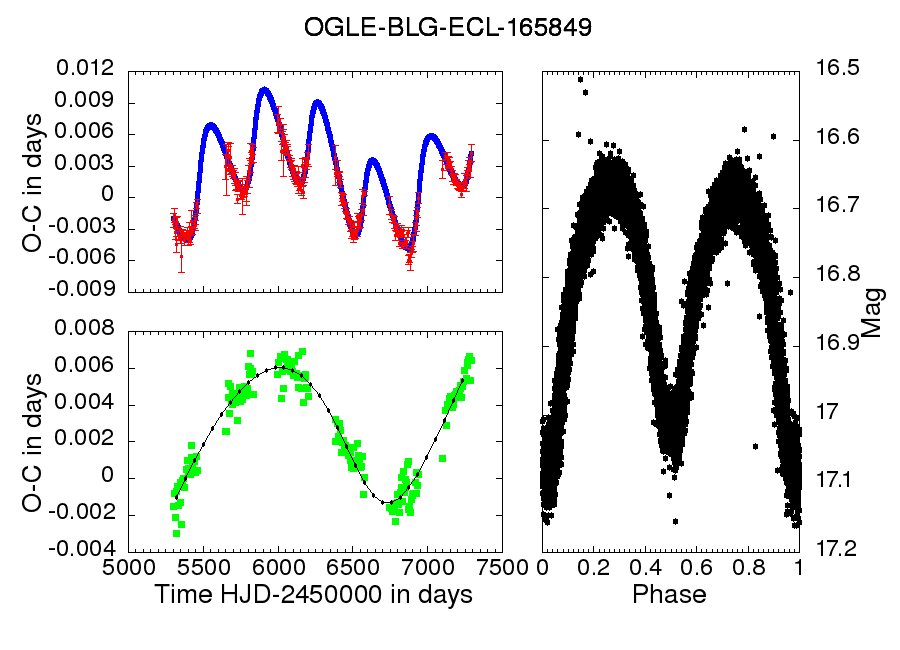}
\includegraphics[width=\columnwidth]{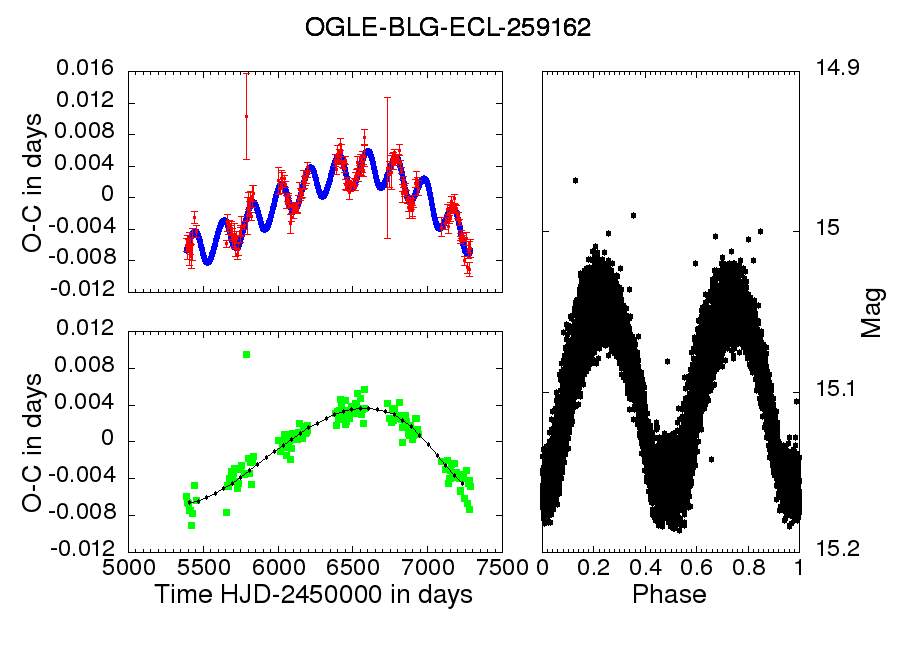}
\caption{Hierarchical four body candidates with double LTTE solution (blue line in the upper left panel) and the residual of the short periodic solution (green squares in the bottom left panel). Here the black dotted lines show the LTTE solution of the third orbit. The folded light curve of the system is in the right panel. }
\label{quadfig}
\end{figure*}

We found four systems where the ETV analyses suggest double periodic solutions. Similarly to the investigation of \citet{ZascheSMC} about OGLE-SMC-ECL-4024 we interpret these ETVs as manifestations of two independent LTTEs occurring in (dynamically) non-interacting (2+1)+1 hierarchicity type quadruple stellar systems.

Therefore, we fit a double LTTE solution (via LM) with some appropriate initial guesses.  The results of our process are plotted in Fig. \ref{quadfig} and the orbital parameters are shown in Table \ref{quadrupoltable}. The parameters for the  middle orbit are in the upper part of the table, while the bottom part contains the parameters of the outer LTTE solution. 
 
Note that our model neglects the dynamical effects, therefore the computed orbital parameters are only indicative. This is especially true in the case of {\em OGLE-BLG-ECL-165849} where the ratio of the outer periods is small ($P_3/P_2<6$), which might indicate strong dynamical effects or instability particularly with regards to these orbital parameters ($e_2\sim0.46$).  Further complex examinations are required to understand the true nature of these systems. 

While {\it OGLE-BLG-ECL-136469} shows a $\beta$ Lyrae-type light variation, the other three short period ($P_1<0.4$\,d) EBs seem to be typical W UMa-type stars with small differences between the depths of the primary and secondary eclipses. There is a possibility that the fourth components have also significant effect on the evolution of the short periodic binaries and triples as well.

\subsection{System with possible substellar companion}
\label{subsecsub}
{\em OGLE-BLG-ECL-200302} is a short-periodic ($P_1=0.24^d$) W UMa type eclipsing binary for which  our ETV solution (Fig \ref{substellar_im}) gives a very low value of the mass function ($f(m_C)=0.00002 \pm 0.00003 $). In order to estimate the minimum mass of the third component we calculated the total mass of the binary by applying the empirical period-mass relation of short periodic binaries \citep{2015MNRAS.448.2890D}. Such a way we got $m_{AB}= 1.29M_\odot$.  In this way the minimum mass of the third companion is found to be $m_{Cmin}=0.034\pm0.044M_\odot$, which is in the substellar domain. The mass of the tertiary will remain in the substellar remain if $28^\circ \le i_2 \le 152^\circ$, therefore we can conclude that this is most likely a brown dwarf.

\begin{figure}
\includegraphics[width=\columnwidth]{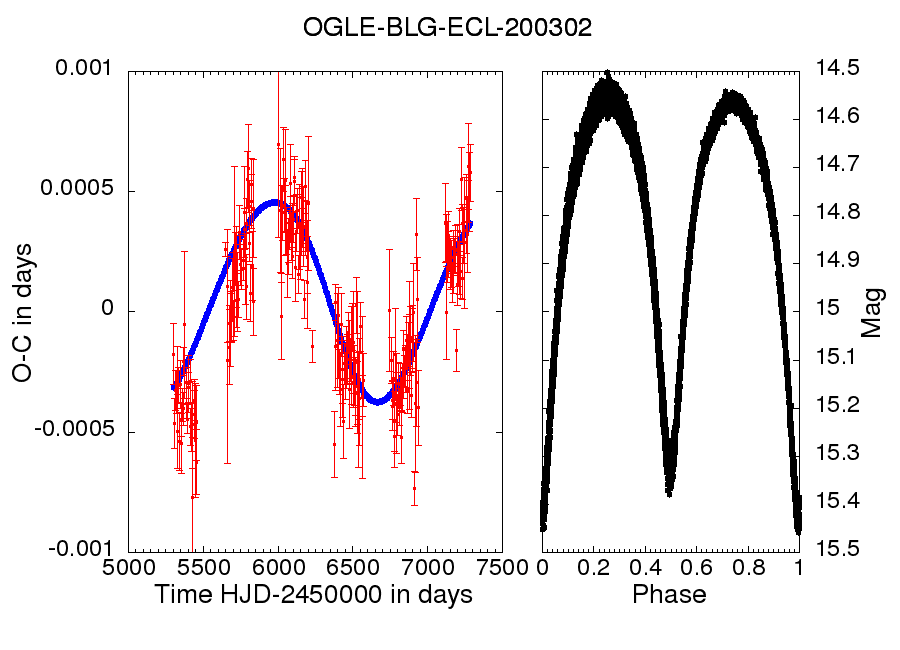}
\caption{ETV of {\em OGLE-BLG-ECL-200302} and the LTTE solution (blue line) which suggests the presence of a substellar third body companion (left panel). Folded and binned light curve of the system is in the right panel.}
\label{substellar_im}
\end{figure}

\subsection{Statistical analysis}
\label{stat}
Due to the large number of triple system candidates it is worthwhile to examine distributions of several parameters that can be determined using only the LTTE delays. These pa\-ra\-me\-ters are the periods ($P_1$ and $P_2$), the outer eccentricity $e_2$, and the mass function $f(m_C)$. These parameters are available for all systems (see Tables \ref{Triplestable260} for the 258 chosen systems  and \ref{Triples} for all the rest).

In spite of the huge amount of potential hierarchical candidates we found, the relatively high uncertainty of parameters suggested that we focus our statistical investigation on those systems where the determinable parameters have lower uncertainty. For the sake of completeness we also present the distributions of the same parameters for the extended sample of all the candidate systems.

\subsubsection{Outer eccentricity}
\begin{figure}
\includegraphics[width=\columnwidth]{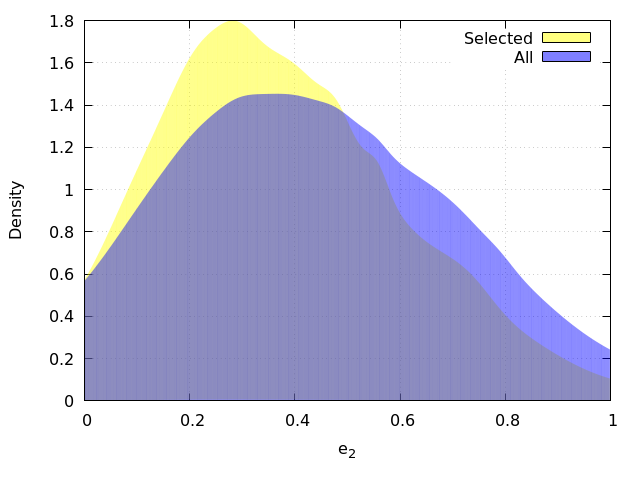}
\caption{ Distribution of outer eccentricity ($e_2$) of the selected systems (yellow) have a peak around $e_2\sim0.3$. For the extended sample of all the candidate systems (blue) the peak value is shifted toward slightly higher eccentricities.}
\label{edistfig}
\end{figure}

 Due to the relatively high uncertainty of our results, similarly to \citet{2018MNRASMurphy}, we use the Kernel Density Estimation (KDE) method to determine its dispersion. This takes the functional form

\begin{equation}
f(e) = \frac{1}{N}\sum_{i=1}^NK(e,e_i,\sigma_i),
\end{equation} 
where
\begin{equation}
K(e,e_i,\sigma_i) = \frac{1}{\sigma_i\sqrt{2\pi}}\exp\left(-\frac{(e-e_i)^2}{2\sigma_i^2}\right)
\end{equation}
is the kernel function, while $e_i$ and $\sigma_i$ are the $i$-th measured eccentricity and its uncertainty.

Fig.\,\ref{edistfig} shows that the distribution has a significant peak around $e_2\approx0.3$, which is consistent whith the results of \citet{Borkovits2016}. Including all systems we got a slightly higher $e_2\approx0.4$ value. This slight increase in the eccentricity as a function of the period can also be observed in the case of wide binary systems (see e.g. \citet{2016MNRAS.456.2070T_e_distrib_wide_bin}).

\begin{figure}
\includegraphics[width=\columnwidth]{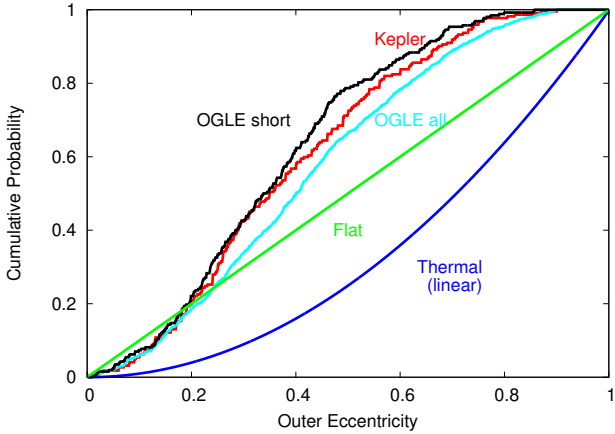}
\caption{Cumulative distribution of the outer eccentricities ($e_2$) of our selected 258 triple candidates (black), all of our triple candidates (cyan), and 222 {\em Kepler} triple candidates of \citet{Borkovits2016} (red), respectively. The green curve, shown for comparison, represents the cumulative distribution expected for an uniformly distributed set of eccentricities between zero and 1. The blue curve is for an eccentricity distribution that increases linearly with $e_2$. None of the comparison curves give a good match with the observed distribution, which results from the distribution of the eccentricity having a peak between $e_2=0.2$ and $e_2=0.4$. For comparison to the eccentricities of unperturbed wide field binaries in the same period regime, see \citep{2013ARA&A..51..269D}.}
\label{e2cumulative}
\end{figure}

As one can see in Fig.\,\ref{e2cumulative} the cumulative distribution is inconsistent both with the 'thermal' distribution that would be linearly rising with $e_2$ (originally posited by \citealt{1919MNRAS..79..408J}) and also with the uniform (flat) distribution. Similarly to binary systems reviewed in \citet{2013ARA&A..51..269D}, none of the `thermal' and `flat' curves represent the true outer eccentricity distribution of our sampled triples.

\subsubsection{Tertiary period}
In Fig. \ref{P2distribution} we present the distribution of the outer orbital periods of the 258 selected systems (yellow) and those systems from the full list where $P_2$ is lower than $2000^d$ (blue). 
This histogram shows a flat maximum between $P_2\approx500-800^d$ which lower limit may be explained  by observational selection effects. 
Furthermore, in general, the shorter the outer period the lower the LTTE amplitude (Eq.\,\ref{LTTEamplitudeapprox}), which acts against the detection of the lowest outer period third companions. The number of candidates is rising with the outer period. The significant peak around $P_2\sim1500^d$ may come from the fitting procedure as it is more likely to converge to this value if the period ($P_2$) is longer than the observation.

\begin{figure}
\includegraphics[width=\columnwidth]{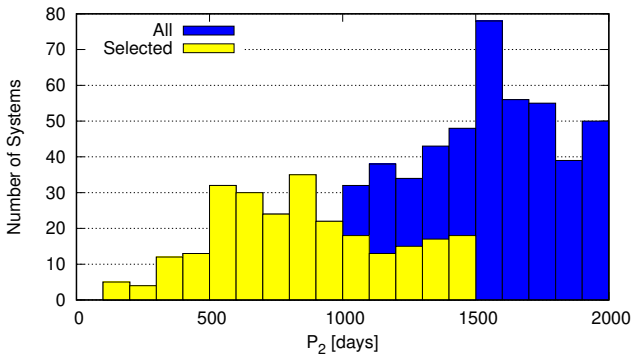}
\caption{ Distribution of the outer orbital periods ($P_2$) from LTTE solution for the sample of the 258 most certain (yellow) as well as for all (blue)  candidate triple systems whose period is lower than 2000 days.}
\label{P2distribution}
\end{figure}

Fig \ref{Fig:P1vsP2KepCor} shows the correlation plot of the outer vs inner periods.  Besides our triple candidates, for comparison we plot also the locations of hundreds of other hierarchical triples, most of those discovered in the prime {\em Kepler} field. For better clarity blue lines denote the limits of the regions where the amplitudes of the LTTE ($\mathit{A_{LTTE}}$ and dynamical $\mathit{A_{dyn}}$) effects are likely to exceed 50 s, a value which roughly approximates the threshold of the probable detection of an ETV. In this figure  the vast majority of our candidate systems are located in a well-defined area that is mainly dominated by LTTE.  The center of the group is really close to the observation length ($\sim 2000^d$), perhaps because the reliability of our LTTE searching algorithm decreases if the outer orbital period becomes longer than time span of the data. Another possibility is that if the real period is significantly longer than our data length then the LM-fit more likely converges to a lower period value that is closer to the duration of the time span.

\begin{figure}
\includegraphics[width=\columnwidth]{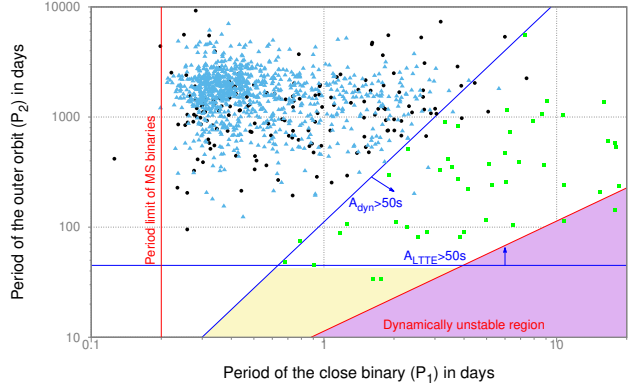}
  \caption{The location of the 992 triple star candidates (blue triangles) in the $P_1$ vs $P_2$ plane. For comparison we plotted those short-period {\em Kepler}, {\em K2} and {\em CoRoT}-triple system candidates for which the inner and outer periods are $P_1\leq 20\,$d and $P_2\leq10000\,$d. Following the work of \citet{Borkovits2016}, the pure LTTE systems are marked with black circles, while triples with combined LTTE+dynamical ETV solution are plotted with green squares. 
 The blue lines show the borders of the domains where the amplitudes of the LTTE and dynamical terms may exceed $\sim50\,$sec, which can be regarded as a limit for an unambiguous detection. These limits were calculated for a hypothetical triple system of three, equally solar mass stars, with a typical outer eccentricity of $e_2=0.35$, and quite arbitrarily, $i_2=60\degr$ and $\omega_2=\pm90\degr$. 
The shaded yellow area means that no LTTE can be detected, though dynamical effects may be significant and, therefore, certainly detectable.
The purple region is a dynamically unstable region, in the sense of the stability criteria of \citet{mardlingaarseth01}.}
  \label{Fig:P1vsP2KepCor}
\end{figure}

\subsubsection{Frequency of triple systems}
We compare the period distributions of the investigated 78\,912 EBs and the detected triple system candidates in the left panel of Fig.\,\ref{EB-triple}. A significant peak occurs around $P_1=0.4^d$ in both cases.  The lack of $P_1<0.2$\,d-period systems is consistent with the theoretical lower limit of the period of contact binaries \citep{Rucinski}.

The right panel represents the percentage of triples in relation to the investigated EBs. 	
It is clearly visible that at lower periods the probability of the triplicity is significantly higher.  This supports the
idea that close binary systems need a third component for their formation, although the forming mechanisms might be various as noted in the introduction.
}

\begin{figure*}
\includegraphics[width=\columnwidth]{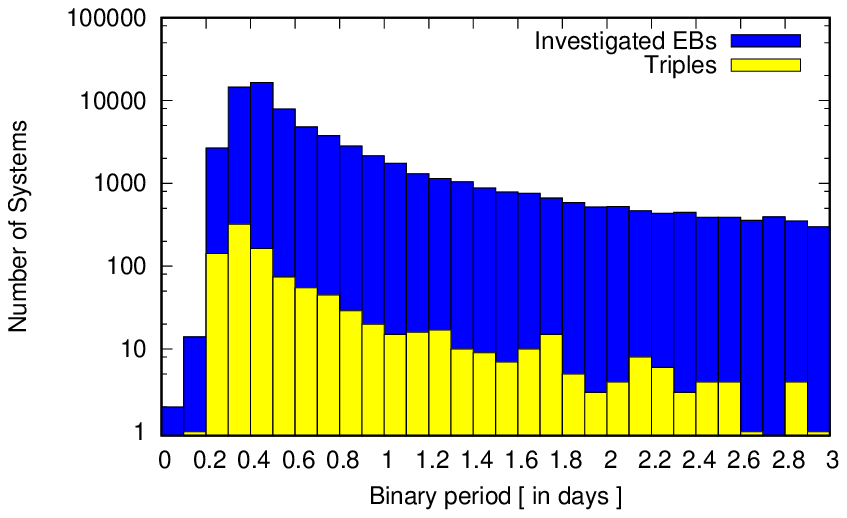}
\includegraphics[width=\columnwidth]{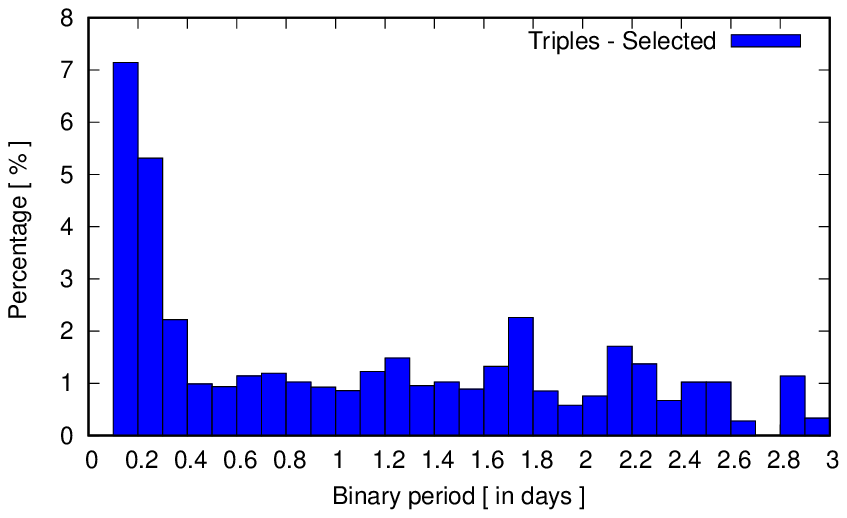}
\caption{Number of all EBs observed by {\em OGLE-IV},  the investigated EB systems, and found triple stellar systems as a function of the binaries' orbital period ($P_1$) in the left panel. The right hand panel shows the percentage of the triples in relation to the investigated EBs and the percentage of the investigated EBs in relation to all observed systems. }
\label{EB-triple}
\end{figure*} 

\subsubsection{Minimum mass}

\begin{figure}
\includegraphics[width=\columnwidth]{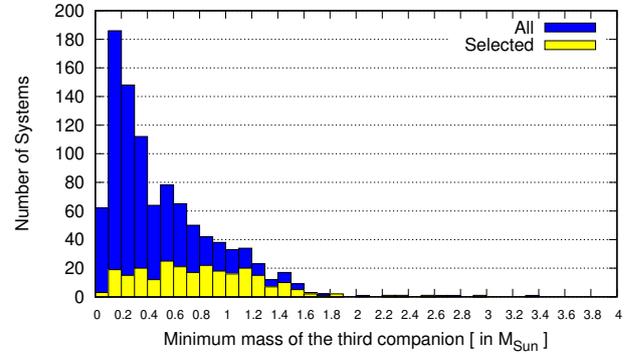}
\caption{ Distribution of the minimum tertiary masses, $m_{Cmin}$ for triple systems found in {\em OGLE-IV}. The tertiary masses are
calculated from the LTTE solutions with the assumption of $m _{AB}\simeq 2 M_{\odot}$.}
\label{Mass_of_C}
\end{figure}

 In the absence of the true binary masses in our sample, the
minimum masses were estimated from the mass function $f(m_\mathrm{C})$ with the assumption that $m_{AB}\simeq 2M_{\odot}$. We plot the distribution of the predicted minimum masses  of the third bodies in Fig.\,\ref{Mass_of_C}. As far as we consider only the narrower sample of the most certain triples we find a mostly flat distribution.  There are 7 candidate systems where the mass of the third component is higher than the mass of the eclipsing binary. These systems are listed in Table \ref{mC_max}.

\begin{table}
\caption{The 7 systems where the third component has higher mass than the eclipsing binary.}
\label{mC_max}
\begin{tabular}{l c c@{$\pm$}c c@{$\pm$}c c@{$\pm$}c}
\hline
ID & $P_1$ & \multicolumn{2}{c}{$P_2$} & \multicolumn{2}{c}{$f(m_C)$} & \multicolumn{2}{c}{$m_C$}\\
 & [days] & \multicolumn{2}{c}{[days]} & \multicolumn{2}{c}{ } & \multicolumn{2}{c}{$M_\odot$}\\
\hline
133733 & 2.173444 & 746.1 & 43.3   & 0.79 & 0.81  & 2.53 & 1.38\\
136328 & 0.363304 & 1016.4 & 10.8  & 0.70 & 0.05  & 2.38 & 0.09\\
150450 & 5.646726 & 1807.2 & 145.1 & 0.53 & 0.28 & 2.07 & 0.55\\
172418 & 1.107474 & 1488.7 & 39.4 & 0.90 & 0.34 & 2.71 & 0.56\\
209134 & 0.436740 & 1298.1 & 39.1 & 1.32 & 0.96 & 3.36 & 1.40 \\
270588 & 0.420899 & 834.4 &  20.6 & 1.02 & 0.17 & 2.90 & 0.26\\
301085 & 0.641771 & 4072.7 & 33.8 & 0.86 & 0.66 & 2.64 & 1.10\\
\end{tabular}
\end{table}

Regarding the total sample (blue) one can find that the vast majority of the candidate systems have minimum outer masses less than $1M_\odot$, i.e. an outer mass ratio of $q_{2min}<0.5$, which suggests that the third component in most cases is a lower mass object.

 As shown in Fig.\,\ref{Mass_of_C} there is a lack of systems whose $m_{Cmin}$ is lower than $0.1 M_\odot$. This may be either because the amplitudes of these systems are too low to detect with our method, or because they are actually uncommon. To decide the question we examined the amplitude distribution of the candidates (see Fig.\,\ref{Amplitudefig}). This suggests that we are able to find systems with amplitudes less than $1^m$. Using   Eq.\,(\ref{LTTEamplitudeapprox}) with the following parameters: $m_{AB}=2M_\odot$, $m_C=0.1$, $\mathcal{A}_\mathrm{LTTE}=1^m $ we were able to estimate a minimum period necessary to detect such a small third component. It resulted in $P_2=1050^d$, which is notably shorter than the used observation series. Based on this we can conclude that substellar components in such EB systems are fairly rare.

\subsubsection{Amplitude of the LTTE}
\label{Amplitude_sec}
\begin{figure}
\includegraphics[width=\columnwidth]{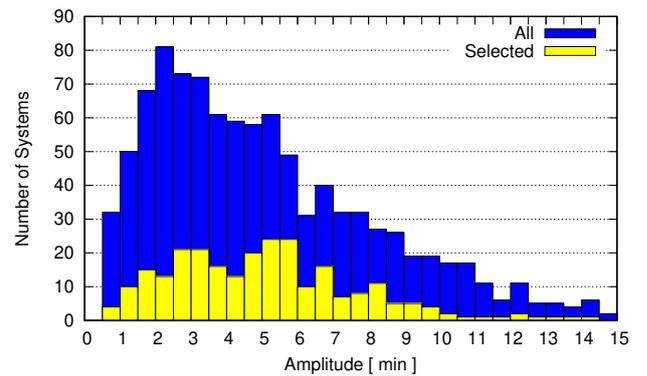}
\caption{Distribution of the amplitudes from the LTTE solution for all systems (blue) and for the systems from the shorter list (yellow).}
\label{Amplitudefig}
\end{figure}

Despite that we used ground based photometry for our research we identified a significant number of hierarchical triple stellar candidates with relatively low  LTTE amplitudes ($0^m.5\le \mathcal{A}_\mathrm{LTTE} \le 1^m.0$). This is due to the fact that most of these systems have the deepest eclipses depths among our candidates, which increases the precision of the fitting method. Nevertheless, we haven't identified any potential candidate with amplitude lower than half a minute.  The distribution of the amplitudes of the LTTE solutions (see Fig. \ref{Amplitudefig}) shows gamma-distribution with a maximum around $\mathcal{A}_\mathrm{LTTE}\approx 2^m$.

\section{Summary and conclusions}\label{summary}

In this paper we reported the results of our search for close, third stellar companions of eclipsing binaries towards the galactic bulge derived from the photometric survey {\em OGLE-IV} via ETV. Owing to the long term observations we were able to find 992 third body candidates. 

For four of them (OGLE-BLG-ECL-136469, OGLE-BLG-ECL-153291,  OGLE-BLG-ECL-165849 and OGLE-BLG-ECL-259162) we came to that conclusion that their ETVs can be well modeled with a double LTTE solution rather than a simple hierarchical stellar system solution. 
 However, since our model neglects dynamic effects, the resulting parameters are only indicative.

We also found two systems with significant dynamical amplitudes (OGLE-BLG-ECL-143356 and OGLE-BLG-ECL-169255). 

Furthermore, a potential substellar third component was also identified in system {\em OGLE-BLG-ECL-200302.}

We investigated the orbital parameter distribution of our systems. For the more reliable results we selected 258 systems where the period and the eccentricity were estimated with lower uncertainties. Besides, we also worked with the full list for comparison. Though we found a very strong peak in the distribution of the eccentricities near $e_2\approx0.3$, and for the full list a bit higher. The number of systems shows a strong increase with the rise of the outer period ($P_2$).
Through our investigations we found potential third components with relative high ($\sim 1.8M_{\odot}$) and low ($0.6M_{\odot}$) minimum masses even in the short period case. We also determined our sensitivity limit for the selection that is around half a minute ($\mathcal{A}_\mathrm{LTTE}\approx 0.5^m$).

There is a great deal of follow-up work that can be carried out in the future, such as the search for systems with apsidal motion. We also plan to investigate the interesting systems through light curve modeling.

\section*{Acknowledgments}
This project has partly been supported by the HAS Wigner RCP-GPU-Lab, the Hungarian National Research, Development and Innovation Office, NKFIH-OTKA grants K-113117, K-115709 and KH-130372, and the Lend\"ulet Program of the Hungarian Academy of Sciences, project No. LP2018-7/2018. The authors are grateful to K. Perger and S. Pint\'er for their valuable comments and suggestions. We are also grateful for L. Dobos for his help in the numerical alignment especially with python fitting algorithms.

\bibliographystyle{mnras}
\bibliography{Hajdu_etal_rev1}


\appendix
\section{Hierarchical triple star candidates}\label{Appendix_A}

\begin{table*}                           
\caption{The selected 258 hierarchical triple star candidates with pure LTTE solution and their fitted orbital parameters.}           
\centering
\resizebox{\textwidth}{!}{%
\begin{tabular}{c c c c @{$\pm$} c c @{$\pm$} c c @{$\pm$} c c @{$\pm$} c c @{$\pm$} c c @{$\pm$} c c @{$\pm$} c}
\hline
ID & $T_0$ & $P_1$ & \multicolumn{2}{c}{$P_2$} & \multicolumn{2}{c}{ $a_\mathrm{AB}\cdot\sin(i_2)$}& \multicolumn{2}{c}{$e_2$} & \multicolumn{2}{c}{$\omega_2$} & \multicolumn{2}{c}{$\tau_2$} & \multicolumn{2}{c}{$f(m_\mathrm{C})$} & \multicolumn{2}{c}{$\Delta P_1$}\\ 
 & [HJD-2450000 days] & [days] & \multicolumn{2}{c}{[days]}  & \multicolumn{2}{c}{[$R_{\odot}$]} & \multicolumn{2}{c}{ }
  & \multicolumn{2}{c}{[deg]} & \multicolumn{2}{c}{[days]} & \multicolumn{2}{c}{} & \multicolumn{2}{c}{[$\times10^{-10}\frac{d}{c}$]}    \\ \hline
28238 & 5265.711315 & 0.336880 & 424.5 & 9.0 & 123.9 & 10.7 & 0.61 & 0.10 & 31.7 & 0.9 & 5606.1 & 9.3 & 0.1413 & 0.03339 \\
32148 & 5265.437046 & 0.311293 & 762.5 & 425.4 & 143.8 & 100.3 & 0.41 & 0.40 & 239.9 & 4.2 & 5484.1 & 329.2 & 0.0685 & 0.21021 \\
35547 & 5265.221608 & 1.745663 & 648.3 & 36.0 & 151.8 & 8.1 & 0.43 & 0.13 & 75.7 & 1.6 & 5965.1 & 21.4 & 0.1116 & 0.03050 \\
47614 & 5265.569008 & 0.290397 & 1406.0 & 7.7 & 142.3 & 5.3 & 0.53 & 0.04 & 10.3 & 0.3 & 6026.3 & 10.3 & 0.0196 & 0.00178 \\
117862 & 5260.079711 & 1.048052 & 438.1 & 13.7 & 101.9 & 5.3 & 0.05 & 0.11 & 157.0 & 11.6 & 5458.2 & 105.9 & 0.0740 & 0.01464 \\
118172 & 5260.619707 & 0.289513 & 859.7 & 211.1 & 28.5 & 4.1 & 0.20 & 0.42 & 93.3 & 7.0 & 5743.9 & 164.1 & 0.0004 & 0.00041 \\
118178 & 5260.599311 & 0.330831 & 772.0 & 55.6 & 70.1 & 3.9 & 0.24 & 0.18 & 49.7 & 4.6 & 5578.7 & 71.7 & 0.0077 & 0.00252 \\
119233 & 5260.695072 & 0.380490 & 790.7 & 86.2 & 64.3 & 6.9 & 0.18 & 0.20 & 157.7 & 7.5 & 5183.2 & 111.6 & 0.0057 & 0.00309 \\
120391 & 5259.602258 & 0.970351 & 455.4 & 16.4 & 138.2 & 4.2 & 0.14 & 0.08 & 63.3 & 3.0 & 5475.2 & 33.3 & 0.1706 & 0.02880 \\
121631 & 5262.191075 & 0.776057 & 534.2 & 33.7 & 121.8 & 6.0 & 0.29 & 0.13 & 286.3 & 2.1 & 5526.2 & 21.3 & 0.0848 & 0.02440 \\
123182 & 5262.264267 & 0.476579 & 1019.5 & 14.3 & 210.5 & 12.3 & 0.46 & 0.09 & 177.2 & 0.8 & 5841.9 & 19.0 & 0.1202 & 0.01911 \\
123244 & 5260.616248 & 0.266444 & 1436.4 & 81.3 & 133.3 & 7.3 & 0.08 & 0.14 & 182.4 & 9.7 & 6592.9 & 311.8 & 0.0154 & 0.00430 \\
123404 & 5260.585383 & 0.875111 & 691.4 & 108.0 & 99.3 & 19.8 & 0.68 & 0.34 & 76.0 & 2.7 & 5366.2 & 65.6 & 0.0275 & 0.02385 \\
123475 & 5260.030734 & 0.516229 & 1016.1 & 72.4 & 147.0 & 10.0 & 0.32 & 0.17 & 230.4 & 2.4 & 5270.1 & 79.3 & 0.0412 & 0.01441 \\
123874 & 5260.366187 & 1.663551 & 1084.1 & 32.7 & 145.4 & 38.5 & 0.69 & 0.19 & 199.8 & 1.7 & 6122.0 & 29.1 & 0.0350 & 0.02170 \\
124024 & 5260.490812 & 0.460257 & 755.5 & 8.5 & 134.7 & 4.2 & 0.15 & 0.07 & 181.9 & 2.4 & 5783.1 & 40.4 & 0.0574 & 0.00552 \\
124514 & 5260.811957 & 0.309294 & 1385.5 & 16.3 & 145.0 & 4.8 & 0.42 & 0.05 & 187.3 & 0.7 & 5524.8 & 22.6 & 0.0213 & 0.00216 \\
126461 & 5260.637967 & 0.371450 & 154.8 & 4.4 & 66.8 & 2.1 & 0.07 & 0.09 & 222.5 & 6.8 & 5417.9 & 22.6 & 0.1670 & 0.02474 \\
126789 & 5282.840812 & 0.227151 & 1282.7 & 24.2 & 92.7 & 5.4 & 0.24 & 0.09 & 168.8 & 2.3 & 5237.5 & 60.4 & 0.0065 & 0.00113 \\
127373 & 5259.669323 & 1.079916 & 850.8 & 24.9 & 178.7 & 15.8 & 0.28 & 0.16 & 343.1 & 3.1 & 5660.2 & 55.2 & 0.1057 & 0.02797 \\
127744 & 5260.239383 & 0.892071 & 1464.3 & 62.9 & 283.4 & 9.5 & 0.60 & 0.09 & 114.0 & 0.7 & 5886.5 & 41.2 & 0.1422 & 0.02786 \\
128093 & 5260.786625 & 0.274143 & 1242.0 & 13.6 & 76.6 & 2.0 & 0.33 & 0.07 & 21.2 & 1.2 & 5956.2 & 33.9 & 0.0039 & 0.00033 \\
128118 & 5259.298154 & 1.604095 & 530.8 & 84.6 & 149.2 & 17.6 & 0.25 & 0.23 & 84.4 & 2.9 & 5498.4 & 59.8 & 0.1579 & 0.11275 \\
128204 & 5273.644229 & 0.422809 & 515.6 & 16.1 & 26.0 & 3.0 & 0.50 & 0.16 & 12.5 & 1.3 & 5680.9 & 16.4 & 0.0009 & 0.00029 \\
128770 & 5259.905570 & 2.304310 & 1056.4 & 29.2 & 104.9 & 9.2 & 0.41 & 0.15 & 141.1 & 2.1 & 5415.9 & 40.9 & 0.0139 & 0.00359 \\
129718 & 5260.471151 & 0.449365 & 123.6 & 15.7 & 49.4 & 4.6 & 0.24 & 0.22 & 92.9 & 3.0 & 5312.1 & 10.0 & 0.1057 & 0.05996 \\
130223 & 5260.590311 & 0.312704 & 589.3 & 36.5 & 172.6 & 12.7 & 0.29 & 0.17 & 223.8 & 3.4 & 5540.8 & 53.2 & 0.1986 & 0.06613 \\
131110 & 5260.459002 & 0.731550 & 504.6 & 6.9 & 160.1 & 5.7 & 0.39 & 0.07 & 331.6 & 0.8 & 5830.3 & 8.0 & 0.2160 & 0.02424 \\
131281 & 5259.443695 & 2.082515 & 938.1 & 24.5 & 253.9 & 18.4 & 0.36 & 0.14 & 30.8 & 1.9 & 5846.0 & 35.0 & 0.2492 & 0.05564 \\
131326 & 5320.818580 & 2.294011 & 637.5 & 38.5 & 120.2 & 10.6 & 0.55 & 0.20 & 61.5 & 1.6 & 5704.5 & 31.7 & 0.0573 & 0.02049 \\
131392 & 5260.543929 & 0.344992 & 895.1 & 38.7 & 194.4 & 19.4 & 0.39 & 0.17 & 159.6 & 1.6 & 6363.3 & 36.2 & 0.1228 & 0.04044 \\
132637 & 5261.871266 & 0.917014 & 193.8 & 27.0 & 64.2 & 5.4 & 0.05 & 0.23 & 82.8 & 16.5 & 5321.4 & 71.0 & 0.0945 & 0.05539 & 42.4 & 3.9 \\
133128 & 5260.532248 & 0.357497 & 537.9 & 16.2 & 47.4 & 3.5 & 0.27 & 0.14 & 192.1 & 2.8 & 5781.0 & 33.1 & 0.0049 & 0.00117 \\
133733 & 5258.660799 & 2.173444 & 746.1 & 43.3 & 320.4 & 97.2 & 0.67 & 0.20 & 191.9 & 1.3 & 5859.2 & 48.0 & 0.7918 & 0.61827 \\
133862 & 5260.427950 & 0.587533 & 812.4 & 2.6 & 172.2 & 2.2 & 0.56 & 0.02 & 322.3 & 0.2 & 6123.2 & 2.5 & 0.1036 & 0.00364 \\
133954 & 5260.447546 & 0.516212 & 1365.3 & 19.1 & 153.1 & 8.3 & 0.56 & 0.07 & 218.2 & 0.6 & 6483.1 & 24.3 & 0.0258 & 0.00388 \\
134206 & 5260.561049 & 0.420053 & 601.5 & 15.9 & 94.8 & 6.6 & 0.37 & 0.13 & 21.5 & 1.6 & 5825.2 & 22.9 & 0.0316 & 0.00691 \\
135472 & 5260.633002 & 0.549421 & 762.7 & 11.6 & 76.6 & 3.9 & 0.34 & 0.08 & 339.2 & 1.6 & 5320.4 & 24.0 & 0.0104 & 0.00153 \\
136113 & 5260.113353 & 1.247158 & 1365.7 & 18.8 & 290.6 & 16.2 & 0.40 & 0.10 & 188.8 & 1.2 & 5519.1 & 31.5 & 0.1764 & 0.02695 \\
136328 & 5261.569072 & 0.363304 & 1016.4 & 10.8 & 377.9 & 6.6 & 0.40 & 0.05 & 310.5 & 0.6 & 5144.9 & 12.0 & 0.7001 & 0.04677 \\
136387 & 5260.472852 & 0.297563 & 671.7 & 5.1 & 131.2 & 2.1 & 0.19 & 0.04 & 220.2 & 1.5 & 5425.8 & 22.1 & 0.0672 & 0.00368 \\
136950 & 5260.365906 & 0.328786 & 1337.2 & 414.2 & 282.5 & 323.2 & 0.77 & 0.33 & 115.6 & 2.6 & 6346.1 & 123.7 & 0.1689 & 0.54342 \\
137482 & 5261.707495 & 0.434935 & 695.8 & 183.1 & 37.6 & 6.3 & 0.32 & 0.44 & 92.3 & 6.9 & 6017.5 & 80.3 & 0.0015 & 0.00169 & 30.9	& 3.8 \\
137746 & 5260.592591 & 0.245151 & 1211.8 & 185.5 & 62.6 & 6.6 & 0.14 & 0.20 & 92.9 & 5.4 & 6707.2 & 131.7 & 0.0022 & 0.00147 \\
138215 & 5259.781065 & 0.738575 & 1449.9 & 42.3 & 236.5 & 11.7 & 0.39 & 0.13 & 303.9 & 1.6 & 6080.8 & 42.9 & 0.0843 & 0.01572 \\
138843 & 5260.113999 & 0.646781 & 763.3 & 76.0 & 105.5 & 10.5 & 0.59 & 0.28 & 282.2 & 2.9 & 5378.1 & 41.8 & 0.0270 & 0.01344 \\
139873 & 5260.447730 & 1.418994 & 313.3 & 22.3 & 122.5 & 5.7 & 0.15 & 0.11 & 94.9 & 2.5 & 5531.1 & 16.7 & 0.2511 & 0.07699 \\
141034 & 5260.455211 & 0.461049 & 1214.7 & 20.4 & 240.7 & 6.3 & 0.37 & 0.07 & 151.6 & 0.6 & 5482.5 & 16.0 & 0.1267 & 0.01302 \\
141410 & 5260.765407 & 0.619069 & 558.8 & 44.8 & 86.0 & 6.3 & 0.13 & 0.21 & 26.5 & 7.6 & 5605.2 & 101.1 & 0.0273 & 0.01057 \\
141564 & 5260.843911 & 0.305721 & 1102.7 & 37.9 & 107.0 & 4.0 & 0.64 & 0.10 & 267.1 & 0.5 & 5706.6 & 21.5 & 0.0135 & 0.00240 & -156.9 & 4.3 \\
141967 & 5259.408264 & 2.293642 & 643.4 & 79.7 & 158.1 & 25.0 & 0.64 & 0.17 & 73.8 & 1.0 & 5905.3 & 52.6 & 0.1279 & 0.08798 \\
142450 & 5260.849632 & 0.917783 & 199.3 & 5.4 & 84.1 & 3.5 & 0.13 & 0.10 & 197.5 & 4.0 & 5283.2 & 17.1 & 0.2010 & 0.03307 & -17.5 & 1.8\\
143356 & 5258.631022 & 2.442595 & 201.3 & 8.2 & 123.3 & 5.3 & 0.30 & 0.12 & 232.1 & 2.2 & 5436.5 & 11.9 & 0.6210 & 0.1307\\
144092 & 5260.663331 & 0.370236 & 461.6 & 7.7 & 152.8 & 4.2 & 0.24 & 0.05 & 320.1 & 1.2 & 5691.7 & 10.9 & 0.2246 & 0.02359 & 48.4 & 6.2\\
145736 & 5260.451467 & 0.294348 & 1315.5 & 33.0 & 29.7 & 3.5 & 0.53 & 0.16 & 147.2 & 1.5 & 6170.3 & 34.6 & 0.0002 & 0.00006 \\
146569 & 5260.389788 & 0.286587 & 1329.7 & 155.8 & 47.4 & 3.1 & 0.31 & 0.24 & 239.9 & 2.5 & 5978.4 & 127.4 & 0.0008 & 0.00039 \\
146682 & 5260.889201 & 0.379488 & 1325.0 & 79.7 & 139.1 & 6.0 & 0.22 & 0.13 & 106.8 & 4.5 & 6769.9 & 110.1 & 0.0206 & 0.00549 \\
146753 & 5260.430657 & 0.289919 & 887.4 & 56.4 & 167.9 & 8.8 & 0.41 & 0.13 & 81.0 & 1.4 & 5959.6 & 41.7 & 0.0806 & 0.02382 \\
148498 & 5260.423952 & 0.388181 & 1125.8 & 17.9 & 131.6 & 7.2 & 0.55 & 0.09 & 139.3 & 0.9 & 5738.4 & 18.0 & 0.0241 & 0.00379 \\
148919 & 5275.432549 & 0.384067 & 800.4 & 15.5 & 127.9 & 14.5 & 0.59 & 0.12 & 6.0 & 0.9 & 5728.2 & 18.5 & 0.0437 & 0.01245 \\
149325 & 5260.208015 & 0.792015 & 1094.2 & 19.5 & 90.1 & 9.6 & 0.49 & 0.13 & 0.9 & 1.3 & 5453.3 & 31.1 & 0.0082 & 0.00219 \\
\end{tabular}                           
}
\end{table*}                           
                           
\begin{table*}                           
\centering
\resizebox{\textwidth}{!}{%
\begin{tabular}{c c c c @{$\pm$} c c @{$\pm$} c c @{$\pm$} c c @{$\pm$} c c @{$\pm$} c c @{$\pm$} c c @{$\pm$} c}
\hline
ID & $T_0$ & $P_1$ & \multicolumn{2}{c}{$P_2$} & \multicolumn{2}{c}{ $a_\mathrm{AB}\cdot\sin(i_2)$}& \multicolumn{2}{c}{$e_2$} & \multicolumn{2}{c}{$\omega_2$} & \multicolumn{2}{c}{$\tau_2$} & \multicolumn{2}{c}{$f(m_\mathrm{C})$} & \multicolumn{2}{c}{$\Delta P_1$}\\ 
 & [HJD-2450000 days] & [days] & \multicolumn{2}{c}{[days]}  & \multicolumn{2}{c}{[$R_{\odot}$]} & \multicolumn{2}{c}{ }
  & \multicolumn{2}{c}{[deg]} & \multicolumn{2}{c}{[days]} & \multicolumn{2}{c}{} & \multicolumn{2}{c}{[$\times10^{-10}\frac{d}{c}$]}    \\ \hline
149911 & 5259.815577 & 2.505057 & 1450.8 & 657.4 & 167.0 & 127.2 & 0.65 & 0.21 & 313.3 & 3.0 & 5326.0 & 120.0 & 0.0297 & 0.08562 \\
150704 & 5260.574775 & 0.386493 & 1427.1 & 8.9 & 210.7 & 11.9 & 0.69 & 0.04 & 158.0 & 0.3 & 6089.8 & 7.0 & 0.0615 & 0.00810 \\
151941 & 5262.527650 & 0.371683 & 863.8 & 12.5 & 119.7 & 3.8 & 0.14 & 0.08 & 349.2 & 3.9 & 5433.2 & 71.4 & 0.0308 & 0.00329 \\
152811 & 5260.727204 & 0.679217 & 1092.0 & 36.8 & 184.0 & 11.2 & 0.68 & 0.15 & 116.8 & 1.0 & 5768.5 & 18.9 & 0.0700 & 0.01560 \\
154020 & 5264.760589 & 0.376478 & 821.6 & 35.0 & 208.3 & 7.3 & 0.43 & 0.08 & 92.4 & 0.9 & 5660.3 & 20.3 & 0.1794 & 0.03550 \\
154199 & 5260.548047 & 0.357652 & 324.2 & 17.5 & 136.9 & 6.4 & 0.18 & 0.11 & 293.3 & 2.6 & 5302.3 & 17.1 & 0.3274 & 0.08363 \\
155453 & 5260.797617 & 0.285681 & 1269.6 & 103.4 & 50.4 & 4.4 & 0.29 & 0.16 & 297.1 & 4.1 & 5873.2 & 103.2 & 0.0011 & 0.00046 \\
156385 & 5260.413943 & 0.310578 & 1200.9 & 55.6 & 108.8 & 14.3 & 0.65 & 0.14 & 157.9 & 0.8 & 5935.2 & 19.1 & 0.0120 & 0.00482 \\
157195 & 5260.730841 & 0.301976 & 1073.3 & 75.3 & 283.1 & 12.6 & 0.17 & 0.13 & 273.2 & 2.3 & 5306.4 & 51.6 & 0.2639 & 0.07903 \\
158062 & 5260.585937 & 0.383252 & 937.8 & 79.2 & 228.1 & 18.0 & 0.49 & 0.16 & 262.0 & 1.4 & 6011.2 & 53.5 & 0.1809 & 0.07438 \\
158089 & 5260.367593 & 1.267915 & 480.4 & 27.1 & 119.6 & 11.2 & 0.23 & 0.21 & 24.2 & 5.5 & 5436.7 & 63.8 & 0.0992 & 0.03537 \\
159527 & 5261.587903 & 0.482246 & 845.8 & 86.8 & 111.4 & 16.5 & 0.55 & 0.28 & 52.1 & 2.8 & 5617.0 & 85.7 & 0.0259 & 0.01565 \\
159989 & 5260.339124 & 1.350938 & 1273.7 & 145.4 & 214.9 & 20.6 & 0.29 & 0.24 & 95.9 & 1.8 & 5712.0 & 65.7 & 0.0820 & 0.04380 \\
160193 & 5260.531237 & 0.537802 & 1169.1 & 4.3 & 348.6 & 7.5 & 0.48 & 0.02 & 194.1 & 0.4 & 5380.1 & 8.7 & 0.4152 & 0.02245 \\
160721 & 5260.240182 & 0.607679 & 212.9 & 8.6 & 80.2 & 3.1 & 0.13 & 0.11 & 305.3 & 4.3 & 5323.8 & 18.0 & 0.1523 & 0.03023 \\
160935 & 5259.573454 & 0.965671 & 929.9 & 76.3 & 51.5 & 9.6 & 0.37 & 0.38 & 329.4 & 3.9 & 5267.0 & 71.9 & 0.0021 & 0.00130 \\
161329 & 5260.780781 & 0.266650 & 1446.9 & 14.8 & 76.3 & 2.5 & 0.43 & 0.07 & 139.8 & 0.9 & 5385.7 & 22.7 & 0.0028 & 0.00027 \\
162442 & 5260.416547 & 0.613347 & 926.7 & 24.0 & 86.5 & 3.4 & 0.22 & 0.10 & 205.8 & 2.2 & 6039.7 & 47.9 & 0.0101 & 0.00158 \\
162463 & 5260.506494 & 0.948193 & 527.2 & 40.5 & 68.4 & 8.3 & 0.01 & 0.30 & 169.3 & 185.1 & 5554.2 & 2044.3 & 0.0155 & 0.00733 \\
163296 & 5260.776504 & 0.318124 & 1020.6 & 50.8 & 31.6 & 3.9 & 0.61 & 0.18 & 235.1 & 2.1 & 5164.7 & 58.9 & 0.0004 & 0.00016 \\
164225 & 5277.565106 & 0.312660 & 1414.5 & 19.3 & 82.2 & 3.1 & 0.02 & 0.14 & 358.4 & 20.8 & 5763.0 & 617.5 & 0.0037 & 0.00043 \\
166506 & 5259.887877 & 1.253322 & 922.9 & 45.3 & 178.3 & 23.6 & 0.28 & 0.23 & 192.1 & 4.9 & 5851.4 & 96.2 & 0.0892 & 0.03675 \\
166978 & 5260.874850 & 0.410614 & 688.4 & 93.2 & 213.4 & 13.5 & 0.06 & 0.43 & 102.0 & 27.6 & 5441.9 & 388.6 & 0.2748 & 0.14638 \\
167675 & 5260.269221 & 1.096171 & 1003.1 & 156.7 & 157.7 & 17.8 & 0.36 & 0.28 & 267.1 & 2.9 & 6219.2 & 114.1 & 0.0522 & 0.03625 \\
168649 & 5260.394722 & 0.360857 & 351.6 & 10.1 & 71.8 & 10.4 & 0.11 & 0.12 & 357.4 & 10.4 & 5337.2 & 75.2 & 0.0401 & 0.01507 & -37.0 & 3.5 \\
168949 & 5261.733049 & 0.320427 & 1172.4 & 121.7 & 192.6 & 12.0 & 0.17 & 0.17 & 247.7 & 2.9 & 5140.8 & 113.9 & 0.0696 & 0.03035 \\
169126 & 5260.940368 & 1.059594 & 530.8 & 38.9 & 114.5 & 12.0 & 0.21 & 0.26 & 221.8 & 7.1 & 5573.7 & 84.7 & 0.0713 & 0.03062 \\
169255 & 5258.554303 & 2.804854 & 346.2 & 26.7 & 95.5 & 7.7 & 0.20 & 0.26 & 235.2 & 5.9 & 5541.2 & 48.9 & 0.0974  & 0.0386 & 6.7 & 1.3\\
170133 & 5260.497840 & 0.708668 & 866.6 & 25.8 & 211.8 & 16.3 & 0.31 & 0.16 & 218.2 & 2.8 & 5425.2 & 56.4 & 0.1697 & 0.04128 \\
172176 & 5260.752122 & 0.261559 & 871.5 & 30.3 & 72.6 & 7.7 & 0.40 & 0.17 & 159.7 & 2.3 & 5577.7 & 41.9 & 0.0067 & 0.00212 \\
172213 & 5260.892138 & 0.467339 & 1085.3 & 8.8 & 138.3 & 27.8 & 0.86 & 0.06 & 172.6 & 0.3 & 5805.0 & 6.7 & 0.0301 & 0.01283 \\
172555 & 5260.020425 & 0.527723 & 988.3 & 30.0 & 37.8 & 3.9 & 0.42 & 0.14 & 28.0 & 2.2 & 5507.3 & 50.2 & 0.0007 & 0.00021 \\
173335 & 5260.744317 & 0.301014 & 764.5 & 40.8 & 242.5 & 39.6 & 0.43 & 0.28 & 328.9 & 2.4 & 5798.6 & 36.9 & 0.3268 & 0.15905 \\
174906 & 5260.834200 & 0.410644 & 494.2 & 21.3 & 129.9 & 10.2 & 0.25 & 0.17 & 329.3 & 3.5 & 5413.5 & 34.8 & 0.1203 & 0.03445 \\
175309 & 5260.292121 & 0.873500 & 1048.5 & 122.1 & 215.6 & 15.0 & 0.15 & 0.21 & 262.5 & 5.9 & 6197.0 & 174.6 & 0.1223 & 0.05974 \\
175712 & 5313.477775 & 0.409733 & 797.4 & 72.3 & 127.0 & 13.5 & 0.20 & 0.28 & 20.0 & 7.4 & 5960.5 & 112.3 & 0.0432 & 0.02094 \\
176186 & 5259.698016 & 0.851832 & 1016.8 & 72.8 & 160.2 & 9.7 & 0.06 & 0.20 & 316.3 & 16.2 & 6248.8 & 359.8 & 0.0533 & 0.01790 \\
176472 & 5260.602275 & 0.291087 & 1448.1 & 46.7 & 162.2 & 3.8 & 0.34 & 0.07 & 253.9 & 1.2 & 6103.6 & 56.6 & 0.0273 & 0.00392 \\
180562 & 5311.712354 & 0.362373 & 693.2 & 17.9 & 121.1 & 4.7 & 0.26 & 0.09 & 61.1 & 2.6 & 5789.7 & 41.7 & 0.0495 & 0.00768 \\
182381 & 5260.448051 & 0.439597 & 689.4 & 150.6 & 80.6 & 11.3 & 0.37 & 0.42 & 107.8 & 4.3 & 5563.9 & 113.5 & 0.0148 & 0.01385 & 58.0 & 5.6\\
182412 & 5260.819061 & 0.374112 & 995.9 & 34.8 & 20.2 & 2.9 & 0.56 & 0.17 & 209.7 & 1.9 & 6833.5 & 48.5 & 0.0001 & 0.00004 \\
182426 & 5260.523379 & 0.367003 & 794.3 & 18.8 & 200.8 & 16.8 & 0.23 & 0.14 & 178.8 & 4.3 & 5783.7 & 72.8 & 0.1721 & 0.04102 \\
182476 & 5260.268382 & 0.707814 & 912.5 & 114.0 & 87.8 & 8.2 & 0.23 & 0.22 & 262.9 & 2.3 & 5829.5 & 82.9 & 0.0109 & 0.00612 \\
182609 & 5260.667301 & 0.878974 & 581.6 & 30.4 & 154.8 & 7.9 & 0.20 & 0.13 & 136.5 & 4.0 & 5245.3 & 44.1 & 0.1469 & 0.03803 & 27.2 & 4.8\\
183737 & 5260.520484 & 0.348501 & 761.0 & 50.8 & 178.7 & 10.5 & 0.29 & 0.14 & 111.3 & 3.7 & 5637.8 & 45.3 & 0.1321 & 0.04198 & -111.4 & 11.8\\
184129 & 5260.489465 & 0.554548 & 533.9 & 9.0 & 204.5 & 10.1 & 0.46 & 0.07 & 12.5 & 0.8 & 5308.9 & 10.2 & 0.4019 & 0.06002 \\
184203 & 5260.350652 & 0.688049 & 565.2 & 30.4 & 94.9 & 8.9 & 0.46 & 0.19 & 134.0 & 2.3 & 5259.4 & 25.7 & 0.0359 & 0.01253 \\
184596 & 5260.742124 & 0.521784 & 552.0 & 4.4 & 99.8 & 1.7 & 0.22 & 0.03 & 155.5 & 0.8 & 5160.5 & 9.2 & 0.0437 & 0.00253 \\
184764 & 5260.883297 & 2.381377 & 858.8 & 64.3 & 160.2 & 14.5 & 0.69 & 0.20 & 75.3 & 1.6 & 6005.2 & 43.8 & 0.0748 & 0.03034 \\
185892 & 5260.378912 & 0.727322 & 1383.5 & 94.6 & 56.3 & 2.4 & 0.18 & 0.12 & 81.9 & 3.0 & 5358.2 & 90.7 & 0.0012 & 0.00035 \\
186015 & 5258.690271 & 1.683499 & 667.2 & 40.7 & 83.2 & 9.5 & 0.32 & 0.27 & 49.5 & 4.9 & 5830.9 & 77.6 & 0.0174 & 0.00716 \\
186990 & 5260.613727 & 0.788449 & 529.4 & 14.6 & 57.2 & 6.3 & 0.46 & 0.15 & 165.9 & 1.6 & 5601.1 & 18.4 & 0.0089 & 0.00270 \\
188182 & 5260.568592 & 0.290304 & 615.8 & 32.7 & 111.1 & 14.4 & 0.43 & 0.20 & 139.0 & 4.1 & 5615.0 & 48.3 & 0.0485 & 0.02030 \\
188183 & 5260.474973 & 1.312443 & 776.5 & 25.8 & 64.9 & 6.3 & 0.26 & 0.16 & 182.2 & 3.2 & 5261.3 & 49.4 & 0.0061 & 0.00179 \\
188296 & 5262.744336 & 0.260866 & 1379.9 & 54.7 & 44.5 & 1.7 & 0.35 & 0.10 & 109.9 & 1.0 & 6031.9 & 28.7 & 0.0006 & 0.00012 \\
188383 & 5260.620449 & 0.359649 & 721.3 & 46.3 & 53.8 & 10.0 & 0.40 & 0.24 & 155.3 & 2.8 & 5288.9 & 36.5 & 0.0040 & 0.00226 \\
188476 & 5262.402364 & 1.598581 & 610.6 & 45.5 & 117.6 & 8.2 & 0.17 & 0.22 & 319.6 & 7.2 & 5563.3 & 93.7 & 0.0585 & 0.02124 \\
189570 & 5259.941270 & 1.404715 & 481.3 & 37.4 & 160.7 & 8.8 & 0.28 & 0.12 & 251.2 & 2.0 & 5452.5 & 36.5 & 0.2402 & 0.08230 \\
190863 & 5260.725258 & 0.383568 & 1284.5 & 139.0 & 333.4 & 30.7 & 0.28 & 0.21 & 94.4 & 1.5 & 5648.8 & 73.2 & 0.3010 & 0.15315 \\
191271 & 5259.889478 & 0.931319 & 1124.2 & 25.1 & 162.7 & 21.2 & 0.45 & 0.15 & 344.4 & 2.3 & 5378.2 & 49.9 & 0.0457 & 0.01497 \\
191509 & 5259.808539 & 0.823592 & 517.1 & 14.4 & 76.7 & 3.8 & 0.20 & 0.12 & 4.5 & 3.2 & 5288.3 & 34.1 & 0.0226 & 0.00413 \\
\end{tabular}                           
}
\end{table*}                           
                           
\begin{table*}                           
\centering
\resizebox{\textwidth}{!}{%
\begin{tabular}{c c c c @{$\pm$} c c @{$\pm$} c c @{$\pm$} c c @{$\pm$} c c @{$\pm$} c c @{$\pm$} c c @{$\pm$} c}
\hline
ID & $T_0$ & $P_1$ & \multicolumn{2}{c}{$P_2$} & \multicolumn{2}{c}{ $a_\mathrm{AB}\cdot\sin(i_2)$}& \multicolumn{2}{c}{$e_2$} & \multicolumn{2}{c}{$\omega_2$} & \multicolumn{2}{c}{$\tau_2$} & \multicolumn{2}{c}{$f(m_\mathrm{C})$} & \multicolumn{2}{c}{$\Delta P_1$}\\ 
 & [HJD-2450000 days] & [days] & \multicolumn{2}{c}{[days]}  & \multicolumn{2}{c}{[$R_{\odot}$]} & \multicolumn{2}{c}{ }
  & \multicolumn{2}{c}{[deg]} & \multicolumn{2}{c}{[days]} & \multicolumn{2}{c}{} & \multicolumn{2}{c}{[$\times10^{-10}\frac{d}{c}$]}    \\ \hline
192016 & 5260.434915 & 0.639250 & 1163.4 & 35.2 & 242.8 & 7.1 & 0.58 & 0.07 & 118.0 & 0.5 & 6246.1 & 23.8 & 0.1417 & 0.02115 \\
192503 & 5260.313621 & 0.801608 & 626.6 & 7.5 & 181.8 & 6.6 & 0.31 & 0.06 & 7.6 & 1.1 & 5256.0 & 14.5 & 0.2051 & 0.02226 \\
194779 & 5260.093152 & 1.202517 & 1185.8 & 24.2 & 210.7 & 9.9 & 0.24 & 0.08 & 337.7 & 2.6 & 6006.0 & 62.2 & 0.0891 & 0.01383 \\
195031 & 5260.769046 & 0.285789 & 975.6 & 16.1 & 229.9 & 9.5 & 0.17 & 0.08 & 203.7 & 3.4 & 5813.0 & 68.2 & 0.1711 & 0.02261 \\
195283 & 5260.211979 & 0.605712 & 651.5 & 28.9 & 173.9 & 16.3 & 0.38 & 0.20 & 39.0 & 3.0 & 5450.3 & 42.4 & 0.1662 & 0.05327 \\
195325 & 5260.750597 & 0.686939 & 720.2 & 15.2 & 171.5 & 3.0 & 0.10 & 0.08 & 294.5 & 3.4 & 5402.0 & 48.9 & 0.1304 & 0.01282 \\
195990 & 5261.149103 & 1.922126 & 1026.0 & 119.6 & 117.5 & 10.6 & 0.07 & 0.15 & 291.9 & 10.5 & 5703.1 & 205.4 & 0.0207 & 0.01097 \\
196649 & 5262.077414 & 1.662246 & 564.1 & 69.3 & 106.7 & 10.1 & 0.02 & 0.28 & 311.8 & 97.9 & 5264.6 & 1149.3 & 0.0511 & 0.02851 \\
198111 & 5260.557431 & 0.300113 & 691.2 & 61.5 & 108.6 & 37.7 & 0.67 & 0.34 & 145.3 & 2.3 & 5536.8 & 37.1 & 0.0360 & 0.03460 \\
199573 & 5260.681845 & 0.348137 & 1228.9 & 16.9 & 151.0 & 8.6 & 0.26 & 0.09 & 8.7 & 1.6 & 6106.6 & 42.3 & 0.0306 & 0.00475 \\
199904 & 5260.759128 & 0.361163 & 860.6 & 59.3 & 26.6 & 9.1 & 0.39 & 0.54 & 185.3 & 6.8 & 5253.0 & 128.0 & 0.0003 & 0.00027 \\
199986 & 5260.323611 & 0.749582 & 676.0 & 57.8 & 60.0 & 21.2 & 0.64 & 0.30 & 161.3 & 2.1 & 5591.7 & 56.3 & 0.0063 & 0.00607 \\
200302 & 5260.548679 & 0.240949 & 1477.0 & 433.5 & 15.5 & 3.4 & 0.07 & 0.31 & 243.5 & 15.9 & 5855.4 & 732.9 & 0.00002 & 0.00003 \\
202708 & 5260.568356 & 0.427509 & 475.6 & 31.5 & 96.9 & 7.9 & 0.44 & 0.18 & 53.9 & 1.8 & 5624.8 & 29.6 & 0.0539 & 0.01950 \\
203087 & 5260.563646 & 0.362337 & 458.1 & 15.0 & 147.8 & 10.0 & 0.30 & 0.13 & 150.2 & 2.2 & 5567.1 & 20.0 & 0.2061 & 0.04813 \\
203169 & 5260.610530 & 0.322835 & 810.5 & 24.1 & 211.4 & 9.9 & 0.15 & 0.11 & 226.5 & 5.2 & 5759.3 & 96.7 & 0.1928 & 0.03526 \\
203853 & 5260.752376 & 0.301235 & 1207.8 & 92.4 & 88.4 & 5.6 & 0.47 & 0.20 & 253.9 & 1.8 & 5584.4 & 64.7 & 0.0063 & 0.00224 \\
203961 & 5260.862537 & 0.265592 & 517.9 & 10.9 & 167.6 & 6.2 & 0.35 & 0.08 & 140.4 & 1.1 & 5446.8 & 12.2 & 0.2353 & 0.03227 \\
204640 & 5282.576788 & 0.426572 & 676.7 & 32.9 & 140.8 & 10.8 & 0.26 & 0.23 & 224.1 & 4.7 & 5763.4 & 69.7 & 0.0817 & 0.02445 \\
204860 & 5260.605158 & 0.321399 & 722.1 & 438.9 & 185.9 & 94.3 & 0.61 & 0.63 & 178.0 & 6.3 & 5148.0 & 429.6 & 0.1652 & 0.46883 \\
205208 & 5260.661049 & 0.629577 & 305.0 & 54.7 & 74.2 & 10.2 & 0.25 & 0.26 & 71.4 & 4.8 & 5525.3 & 54.6 & 0.0590 & 0.04796 \\
206206 & 5260.466174 & 0.531936 & 808.5 & 27.9 & 195.5 & 16.0 & 0.36 & 0.17 & 143.5 & 2.7 & 5694.0 & 43.3 & 0.1533 & 0.04096 \\
207016 & 5260.673787 & 0.364235 & 527.8 & 8.6 & 130.1 & 4.5 & 0.32 & 0.07 & 340.1 & 1.0 & 5657.0 & 11.5 & 0.1058 & 0.01249 & -70.5 & 4.7\\
208021 & 5260.153461 & 0.657449 & 253.4 & 7.0 & 90.4 & 2.3 & 0.38 & 0.07 & 120.0 & 1.0 & 5460.4 & 4.9 & 0.1543 & 0.02064 \\
208059 & 5260.410831 & 0.413251 & 1055.9 & 22.8 & 142.4 & 3.6 & 0.26 & 0.06 & 62.0 & 1.5 & 5481.4 & 37.6 & 0.0347 & 0.00400 \\
208470 & 5260.704571 & 0.333492 & 760.4 & 183.5 & 153.8 & 28.3 & 0.29 & 0.32 & 301.1 & 2.8 & 5132.2 & 63.5 & 0.0844 & 0.09216 \\
209684 & 5260.176546 & 1.349187 & 552.6 & 40.0 & 148.0 & 22.3 & 0.59 & 0.27 & 127.1 & 2.7 & 5312.3 & 25.9 & 0.1422 & 0.07373 \\
210234 & 5260.468438 & 0.717307 & 920.6 & 48.6 & 161.8 & 13.7 & 0.38 & 0.18 & 68.6 & 3.4 & 5407.4 & 75.2 & 0.0669 & 0.02192 \\
210312 & 5260.551587 & 0.607592 & 854.0 & 27.4 & 170.6 & 32.2 & 0.58 & 0.20 & 195.2 & 1.4 & 5189.1 & 27.6 & 0.0912 & 0.04321 \\
210447 & 5260.038040 & 0.898156 & 928.6 & 25.7 & 272.6 & 16.4 & 0.32 & 0.13 & 152.2 & 1.7 & 6189.5 & 30.8 & 0.3150 & 0.06406 & -65.3 & 10.0 \\
212096 & 5260.527778 & 0.344485 & 1461.2 & 58.7 & 342.9 & 9.7 & 0.39 & 0.09 & 233.8 & 0.9 & 5592.4 & 47.6 & 0.2531 & 0.04482 \\
212462 & 5259.520132 & 1.709583 & 581.7 & 19.2 & 92.6 & 3.5 & 0.13 & 0.11 & 207.1 & 4.5 & 5932.6 & 55.1 & 0.0314 & 0.00548 \\
213521 & 5260.606305 & 1.755518 & 1256.5 & 96.6 & 82.5 & 4.2 & 0.45 & 0.16 & 253.6 & 1.8 & 6437.9 & 48.5 & 0.0048 & 0.00160 \\
217270 & 5258.793587 & 1.622832 & 776.9 & 77.4 & 131.3 & 28.6 & 0.58 & 0.26 & 21.9 & 1.5 & 5535.3 & 66.8 & 0.0503 & 0.03695 \\
218203 & 5260.407767 & 0.310802 & 923.6 & 40.2 & 31.8 & 9.8 & 0.74 & 0.24 & 38.3 & 2.0 & 5539.0 & 36.8 & 0.0005 & 0.00037 \\
219110 & 5260.017996 & 0.740710 & 1136.3 & 14.5 & 111.3 & 2.7 & 0.25 & 0.05 & 214.6 & 1.4 & 6170.0 & 37.2 & 0.0143 & 0.00124 \\
219338 & 5258.906724 & 1.890380 & 952.5 & 75.9 & 197.5 & 21.9 & 0.32 & 0.26 & 309.6 & 4.2 & 6423.8 & 72.2 & 0.1138 & 0.05244 \\
219531 & 5260.518930 & 0.331798 & 360.0 & 8.6 & 147.3 & 19.2 & 0.20 & 0.17 & 231.8 & 3.3 & 5369.6 & 25.5 & 0.7864 & 0.26137 \\
219947 & 5260.681227 & 0.272917 & 1289.2 & 154.0 & 45.5 & 10.5 & 0.79 & 0.31 & 282.0 & 2.2 & 6148.1 & 72.8 & 0.0008 & 0.00066 \\
220827 & 5260.533302 & 0.285405 & 695.1 & 5.8 & 136.6 & 10.2 & 0.72 & 0.06 & 9.8 & 0.3 & 5733.3 & 6.6 & 0.0707 & 0.01233 \\
220951 & 5260.497427 & 0.307168 & 455.1 & 9.6 & 89.0 & 6.3 & 0.45 & 0.10 & 37.6 & 1.2 & 5803.1 & 12.3 & 0.0456 & 0.00934 \\
223244 & 5260.049048 & 0.804559 & 862.4 & 26.0 & 209.1 & 17.2 & 0.48 & 0.11 & 162.6 & 0.9 & 5855.2 & 22.2 & 0.1648 & 0.04202 \\
223955 & 5260.426551 & 0.367006 & 1451.8 & 121.2 & 40.8 & 1.9 & 0.04 & 0.14 & 273.4 & 19.0 & 5754.6 & 566.7 & 0.0004 & 0.00014 \\
224377 & 5259.885007 & 1.145178 & 571.2 & 17.9 & 154.3 & 18.5 & 0.45 & 0.18 & 343.8 & 1.7 & 5428.6 & 19.8 & 0.1510 & 0.05040 & 25.4 & 4.9 \\
224836 & 5260.504332 & 0.367318 & 1387.8 & 216.3 & 152.2 & 19.7 & 0.42 & 0.38 & 67.1 & 3.4 & 6655.8 & 96.8 & 0.0245 & 0.01780 \\
227716 & 5260.669645 & 0.229971 & 516.0 & 10.4 & 123.0 & 4.1 & 0.23 & 0.08 & 155.1 & 1.9 & 5868.1 & 20.4 & 0.0937 & 0.01191 & 194.6 & 8.9 \\
228342 & 5260.619788 & 0.253410 & 577.0 & 66.2 & 166.5 & 18.4 & 0.41 & 0.19 & 103.5 & 1.5 & 5270.4 & 25.1 & 0.1858 & 0.10502 \\
228421 & 5259.719199 & 0.887966 & 963.8 & 60.4 & 53.7 & 5.3 & 0.43 & 0.24 & 40.7 & 2.0 & 6050.5 & 52.8 & 0.0022 & 0.00085 \\
229158 & 5260.338081 & 0.620128 & 727.2 & 41.1 & 71.1 & 14.7 & 0.36 & 0.41 & 216.3 & 5.2 & 5706.3 & 68.7 & 0.0091 & 0.00531 \\
231054 & 5260.560905 & 0.441326 & 611.0 & 13.2 & 75.4 & 1.6 & 0.27 & 0.06 & 49.2 & 1.4 & 5878.7 & 23.1 & 0.0154 & 0.00165 \\
231108 & 5260.505286 & 0.309563 & 603.3 & 6.4 & 179.1 & 3.2 & 0.08 & 0.05 & 24.1 & 2.8 & 5318.1 & 35.7 & 0.2115 & 0.01429 \\
231134 & 5259.394302 & 1.542671 & 551.3 & 110.6 & 98.9 & 14.9 & 0.31 & 0.43 & 75.3 & 5.6 & 5487.5 & 91.4 & 0.0426 & 0.03847 \\
231174 & 5260.852317 & 0.306145 & 725.3 & 18.3 & 165.3 & 5.1 & 0.66 & 0.07 & 260.7 & 0.7 & 5406.3 & 13.4 & 0.1152 & 0.01583 & -89.0 & 4.8\\
231675 & 5260.680243 & 0.275610 & 310.7 & 10.4 & 95.1 & 4.7 & 0.22 & 0.11 & 52.9 & 2.6 & 5418.0 & 18.4 & 0.1193 & 0.02377 \\
231868 & 5260.839336 & 0.338490 & 1256.3 & 30.5 & 65.9 & 5.7 & 0.65 & 0.13 & 133.1 & 1.2 & 5568.5 & 24.1 & 0.0024 & 0.00059 \\
233210 & 5258.777482 & 1.909852 & 607.7 & 30.0 & 166.5 & 30.1 & 0.44 & 0.27 & 178.6 & 2.5 & 5764.9 & 33.3 & 0.1676 & 0.08542 \\
234407 & 5260.030964 & 0.805788 & 1312.6 & 20.5 & 149.9 & 5.2 & 0.47 & 0.06 & 210.6 & 0.7 & 5644.2 & 21.5 & 0.0262 & 0.00305 \\
234703 & 5260.588167 & 0.390871 & 623.0 & 45.2 & 152.0 & 8.9 & 0.46 & 0.14 & 282.8 & 1.4 & 5603.9 & 24.2 & 0.1212 & 0.04057 & 49.9 & 7.8\\
236519 & 5260.532099 & 0.554674 & 674.9 & 74.5 & 44.0 & 4.3 & 0.22 & 0.18 & 317.8 & 4.7 & 5296.2 & 53.5 & 0.0025 & 0.00132 \\
236848 & 5260.421097 & 0.429959 & 825.6 & 48.7 & 35.7 & 7.5 & 0.52 & 0.23 & 39.1 & 2.5 & 5301.8 & 54.2 & 0.0009 & 0.00054 \\
\end{tabular}                           
}
\end{table*}                           
                           
\begin{table*}                           
\centering
\resizebox{\textwidth}{!}{%
\begin{tabular}{c c c c @{$\pm$} c c @{$\pm$} c c @{$\pm$} c c @{$\pm$} c c @{$\pm$} c c @{$\pm$} c c @{$\pm$} c}
\hline
ID & $T_0$ & $P_1$ & \multicolumn{2}{c}{$P_2$} & \multicolumn{2}{c}{ $a_\mathrm{AB}\cdot\sin(i_2)$}& \multicolumn{2}{c}{$e_2$} & \multicolumn{2}{c}{$\omega_2$} & \multicolumn{2}{c}{$\tau_2$} & \multicolumn{2}{c}{$f(m_\mathrm{C})$} & \multicolumn{2}{c}{$\Delta P_1$}\\ 
 & [HJD-2450000 days] & [days] & \multicolumn{2}{c}{[days]}  & \multicolumn{2}{c}{[$R_{\odot}$]} & \multicolumn{2}{c}{ }
  & \multicolumn{2}{c}{[deg]} & \multicolumn{2}{c}{[days]} & \multicolumn{2}{c}{} & \multicolumn{2}{c}{[$\times10^{-10}\frac{d}{c}$]}    \\ \hline
237136 & 5260.744234 & 0.515309 & 800.2 & 82.5 & 176.9 & 15.1 & 0.02 & 0.24 & 109.9 & 69.8 & 5232.9 & 1158.3 & 0.1158 & 0.05559 \\
238112 & 5273.450359 & 0.398630 & 836.8 & 104.2 & 76.9 & 9.5 & 0.46 & 0.31 & 249.1 & 3.3 & 5709.1 & 93.1 & 0.0087 & 0.00540 \\
238744 & 5258.650020 & 1.823186 & 391.5 & 8.4 & 70.9 & 2.7 & 0.21 & 0.09 & 335.1 & 2.9 & 5306.0 & 20.8 & 0.0312 & 0.00438 \\
239017 & 5321.335819 & 0.571381 & 798.0 & 17.7 & 160.5 & 11.0 & 0.35 & 0.13 & 157.0 & 1.8 & 5347.8 & 29.5 & 0.0871 & 0.01773 \\
239476 & 5260.334651 & 0.410181 & 756.5 & 363.8 & 203.7 & 113.5 & 0.32 & 0.43 & 85.5 & 2.1 & 5888.7 & 198.1 & 0.1981 & 0.50656 \\
239993 & 5260.495771 & 0.312929 & 467.6 & 21.4 & 62.2 & 9.2 & 0.41 & 0.23 & 164.7 & 2.7 & 5323.8 & 26.2 & 0.0148 & 0.00641 \\
241936 & 5260.652576 & 0.358846 & 1167.6 & 62.3 & 64.0 & 2.5 & 0.18 & 0.12 & 82.4 & 2.5 & 5652.9 & 74.0 & 0.0026 & 0.00062 \\
242159 & 5260.508397 & 0.471755 & 1153.3 & 58.4 & 243.7 & 10.2 & 0.09 & 0.17 & 226.7 & 7.5 & 5975.2 & 179.8 & 0.1459 & 0.03438 \\
243302 & 5260.899190 & 0.338515 & 830.5 & 13.7 & 62.3 & 2.7 & 0.38 & 0.09 & 139.4 & 1.9 & 5703.0 & 27.5 & 0.0047 & 0.00064 \\
243365 & 5260.273009 & 0.730657 & 680.6 & 25.0 & 134.1 & 4.5 & 0.36 & 0.08 & 117.5 & 1.2 & 6051.3 & 14.2 & 0.0697 & 0.01236 & 17.6 & 2.7\\
243618 & 5260.555652 & 0.296662 & 1331.1 & 222.9 & 47.3 & 5.3 & 0.10 & 0.30 & 64.8 & 15.2 & 5320.6 & 471.1 & 0.0008 & 0.00058 \\
243892 & 5260.790249 & 0.270311 & 1451.1 & 82.0 & 35.0 & 2.0 & 0.24 & 0.11 & 74.3 & 2.6 & 6177.6 & 106.7 & 0.0003 & 0.00009 \\
245557 & 5285.565609 & 0.299509 & 544.4 & 41.8 & 212.2 & 13.3 & 0.17 & 0.08 & 231.3 & 2.7 & 5605.2 & 55.4 & 0.4322 & 0.15373 \\
245692 & 5260.275588 & 0.671522 & 964.0 & 87.1 & 224.2 & 14.6 & 0.44 & 0.26 & 286.5 & 2.2 & 5176.6 & 40.0 & 0.1625 & 0.06521 \\
247165 & 5260.653814 & 0.340779 & 1349.3 & 24.0 & 257.7 & 3.1 & 0.27 & 0.03 & 275.4 & 0.6 & 6463.3 & 15.0 & 0.1260 & 0.00975 \\
247495 & 5260.163774 & 0.828757 & 903.8 & 97.9 & 135.4 & 11.1 & 0.26 & 0.21 & 296.3 & 4.0 & 5666.4 & 65.9 & 0.0408 & 0.01995 \\
248494 & 5260.640682 & 0.607061 & 570.5 & 55.1 & 141.8 & 10.1 & 0.38 & 0.18 & 272.9 & 1.5 & 5329.8 & 29.3 & 0.1174 & 0.05074 & 59.0 & 5.2 \\
248738 & 5259.801309 & 1.219331 & 801.4 & 35.1 & 66.8 & 6.0 & 0.20 & 0.21 & 30.6 & 5.2 & 5753.0 & 97.0 & 0.0062 & 0.00193 \\
249116 & 5260.350023 & 0.392285 & 859.1 & 37.7 & 104.5 & 17.4 & 0.44 & 0.22 & 177.4 & 2.6 & 5325.6 & 49.7 & 0.0207 & 0.00962 \\
250046 & 5260.480221 & 0.595688 & 629.2 & 26.9 & 100.3 & 10.2 & 0.38 & 0.22 & 156.8 & 2.9 & 5490.8 & 34.8 & 0.0341 & 0.01131 \\
252308 & 5262.430400 & 0.453661 & 1108.1 & 185.9 & 211.6 & 41.2 & 0.57 & 0.17 & 120.3 & 2.7 & 5651.5 & 43.8 & 0.1034 & 0.09231 \\
252603 & 5276.731131 & 0.295704 & 855.4 & 86.9 & 256.5 & 20.0 & 0.17 & 0.15 & 264.1 & 2.7 & 5869.5 & 65.1 & 0.3089 & 0.14231 \\
252887 & 5259.297440 & 1.768218 & 503.9 & 37.5 & 43.1 & 5.4 & 0.24 & 0.30 & 207.7 & 6.0 & 5314.3 & 69.5 & 0.0042 & 0.00199 \\
253962 & 5260.689536 & 0.438764 & 1000.1 & 32.2 & 142.6 & 4.3 & 0.15 & 0.10 & 6.3 & 2.9 & 4950.3 & 72.4 & 0.0389 & 0.00610 \\
255449 & 5260.254936 & 0.428242 & 951.8 & 23.0 & 215.8 & 27.6 & 0.55 & 0.15 & 182.1 & 1.6 & 5127.9 & 31.3 & 0.1486 & 0.04878 \\
256377 & 5260.152851 & 0.432889 & 899.9 & 56.7 & 73.3 & 3.1 & 0.37 & 0.14 & 255.7 & 1.3 & 5858.6 & 36.7 & 0.0065 & 0.00178 \\
256437 & 5259.985381 & 0.789466 & 834.5 & 149.4 & 134.1 & 17.3 & 0.20 & 0.28 & 264.0 & 5.2 & 5925.7 & 115.0 & 0.0464 & 0.03689 \\
256668 & 5260.548636 & 0.328324 & 833.4 & 40.5 & 119.3 & 17.3 & 0.74 & 0.21 & 122.8 & 1.9 & 5737.4 & 24.7 & 0.0327 & 0.01425 \\
257317 & 5259.762393 & 0.860510 & 788.1 & 135.2 & 170.0 & 17.2 & 0.09 & 0.43 & 110.5 & 20.8 & 5738.2 & 330.3 & 0.1060 & 0.07600 \\
259530 & 5260.601448 & 0.428596 & 900.4 & 55.5 & 179.1 & 13.9 & 0.30 & 0.17 & 242.8 & 4.3 & 5429.6 & 97.7 & 0.0950 & 0.03231 \\
259591 & 5260.390071 & 0.549346 & 871.8 & 14.9 & 128.3 & 7.9 & 0.45 & 0.08 & 345.4 & 0.8 & 5871.7 & 14.6 & 0.0372 & 0.00649 & 39.0 & 3.3 \\
262145 & 5260.362326 & 0.373863 & 817.4 & 66.0 & 239.1 & 74.4 & 0.76 & 0.30 & 47.9 & 3.1 & 5940.7 & 34.8 & 0.2740 & 0.23689 \\
263423 & 5260.848837 & 0.289787 & 1347.8 & 26.2 & 176.1 & 30.5 & 0.86 & 0.11 & 228.1 & 1.2 & 5853.4 & 24.1 & 0.0403 & 0.01631 \\
264186 & 5260.920392 & 0.299389 & 1331.5 & 100.4 & 51.0 & 2.8 & 0.39 & 0.22 & 115.5 & 2.0 & 5607.4 & 66.8 & 0.0010 & 0.00034 \\
264351 & 5260.331052 & 0.353627 & 1217.3 & 36.7 & 298.7 & 9.5 & 0.49 & 0.05 & 64.5 & 0.6 & 5782.7 & 32.1 & 0.2411 & 0.03714 \\
264468 & 5260.192168 & 0.602293 & 510.2 & 32.7 & 92.5 & 12.5 & 0.12 & 0.29 & 177.6 & 11.7 & 5596.2 & 121.8 & 0.0408 & 0.01887 \\
268481 & 5260.702339 & 0.368992 & 820.7 & 16.5 & 235.5 & 15.9 & 0.47 & 0.11 & 149.4 & 1.3 & 5680.8 & 19.1 & 0.2597 & 0.05073 \\
269340 & 5260.670487 & 0.430312 & 438.4 & 14.9 & 91.0 & 9.0 & 0.34 & 0.16 & 359.9 & 2.2 & 5564.8 & 20.1 & 0.0526 & 0.01577 & -104.4 & 7.4\\
269555 & 5260.463734 & 0.561163 & 890.8 & 17.3 & 177.1 & 12.7 & 0.36 & 0.13 & 356.5 & 1.7 & 5365.2 & 31.4 & 0.0938 & 0.01892 \\
270588 & 5260.687806 & 0.420899 & 834.4 & 20.6 & 375.3 & 14.2 & 0.20 & 0.09 & 220.6 & 2.6 & 5762.6 & 50.9 & 1.0174 & 0.15234 \\
271708 & 5260.453740 & 0.307049 & 949.4 & 40.9 & 69.7 & 2.2 & 0.13 & 0.10 & 54.5 & 5.4 & 5436.3 & 113.4 & 0.0050 & 0.00096 \\
272056 & 5260.320582 & 0.622995 & 1462.3 & 8.9 & 343.1 & 16.5 & 0.74 & 0.04 & 149.5 & 0.3 & 5475.0 & 8.3 & 0.2532 & 0.02898 \\
272885 & 5259.562150 & 1.088641 & 755.7 & 44.7 & 153.5 & 6.9 & 0.33 & 0.14 & 123.6 & 2.3 & 5860.7 & 51.0 & 0.0849 & 0.02270 & 32.8 & 2.4\\
273764 & 5260.850759 & 0.957377 & 459.9 & 27.5 & 101.2 & 5.1 & 0.05 & 0.13 & 71.0 & 12.5 & 5485.4 & 123.6 & 0.0657 & 0.01841 \\
273875 & 5260.666875 & 0.324494 & 1111.3 & 18.1 & 88.3 & 4.4 & 0.42 & 0.08 & 36.5 & 1.3 & 5363.3 & 33.6 & 0.0075 & 0.00111 \\
276746 & 5258.700553 & 1.609378 & 649.8 & 23.3 & 214.1 & 4.9 & 0.19 & 0.06 & 90.0 & 1.2 & 5804.9 & 17.6 & 0.3115 & 0.04777 \\
278769 & 5260.739823 & 0.356270 & 1104.4 & 68.9 & 163.0 & 17.6 & 0.24 & 0.21 & 227.2 & 6.6 & 5434.3 & 163.4 & 0.0476 & 0.01919 \\
279293 & 5262.653125 & 0.285679 & 1427.7 & 61.8 & 45.8 & 3.0 & 0.23 & 0.14 & 63.6 & 5.1 & 5739.3 & 168.8 & 0.0006 & 0.00016 \\
280174 & 5260.501885 & 0.570827 & 921.6 & 75.7 & 171.7 & 21.3 & 0.51 & 0.25 & 52.9 & 2.1 & 5745.2 & 70.6 & 0.0798 & 0.03946 \\
280942 & 5260.621143 & 0.638285 & 951.5 & 42.6 & 108.2 & 24.1 & 0.65 & 0.24 & 156.4 & 1.2 & 5006.4 & 24.6 & 0.0187 & 0.01084 \\
281727 & 5260.266608 & 0.628767 & 479.7 & 9.6 & 135.9 & 31.3 & 0.76 & 0.12 & 344.2 & 0.9 & 6000.1 & 9.3 & 0.1461 & 0.07607 \\
281873 & 5259.784457 & 1.342625 & 539.2 & 11.8 & 72.3 & 2.6 & 0.22 & 0.09 & 40.6 & 2.4 & 5768.7 & 27.1 & 0.0174 & 0.00239 \\
283931 & 5258.837819 & 2.045536 & 1429.5 & 39.8 & 153.4 & 40.0 & 0.80 & 0.14 & 151.6 & 1.2 & 5373.7 & 23.4 & 0.0237 & 0.01434 \\
284300 & 5260.353102 & 0.582255 & 593.9 & 30.5 & 119.9 & 5.6 & 0.46 & 0.12 & 67.0 & 1.2 & 5510.9 & 27.6 & 0.0655 & 0.01621 & 58.7 & 4.6\\
285648 & 5260.781443 & 0.387505 & 558.1 & 15.1 & 161.2 & 17.7 & 0.46 & 0.14 & 357.0 & 1.4 & 5341.2 & 16.6 & 0.1802 & 0.05420 & -52.5 & 10.5\\
285958 & 5260.794247 & 0.436324 & 1279.2 & 55.3 & 186.1 & 15.9 & 0.47 & 0.16 & 320.1 & 2.4 & 5495.1 & 51.8 & 0.0528 & 0.01587 \\
287844 & 5260.389590 & 0.403481 & 678.1 & 65.9 & 171.8 & 24.6 & 0.68 & 0.32 & 248.2 & 2.7 & 5700.8 & 37.9 & 0.1478 & 0.08542 \\
288000 & 5260.524531 & 0.332212 & 1470.4 & 29.8 & 196.8 & 23.9 & 0.68 & 0.11 & 145.6 & 1.0 & 6073.5 & 20.2 & 0.0472 & 0.01433 \\
288075 & 5260.347409 & 0.420139 & 779.4 & 34.4 & 190.8 & 16.5 & 0.26 & 0.18 & 316.9 & 4.3 & 5613.3 & 64.8 & 0.1533 & 0.04681 \\
289401 & 5260.364468 & 0.485770 & 580.8 & 28.3 & 57.6 & 3.6 & 0.26 & 0.16 & 312.8 & 3.7 & 5407.8 & 41.0 & 0.0076 & 0.00206 \\
\end{tabular}                           
}
\end{table*}                           
                           
\begin{table*}                           
\centering
\resizebox{\textwidth}{!}{%
\begin{tabular}{c c c c @{$\pm$} c c @{$\pm$} c c @{$\pm$} c c @{$\pm$} c c @{$\pm$} c c @{$\pm$} c c @{$\pm$} c}
\hline
ID & $T_0$ & $P_1$ & \multicolumn{2}{c}{$P_2$} & \multicolumn{2}{c}{ $a_\mathrm{AB}\cdot\sin(i_2)$}& \multicolumn{2}{c}{$e_2$} & \multicolumn{2}{c}{$\omega_2$} & \multicolumn{2}{c}{$\tau_2$} & \multicolumn{2}{c}{$f(m_\mathrm{C})$} & \multicolumn{2}{c}{$\Delta P_1$}\\ 
 & [HJD-2450000 days] & [days] & \multicolumn{2}{c}{[days]}  & \multicolumn{2}{c}{[$R_{\odot}$]} & \multicolumn{2}{c}{ }
  & \multicolumn{2}{c}{[deg]} & \multicolumn{2}{c}{[days]} & \multicolumn{2}{c}{} & \multicolumn{2}{c}{[$\times10^{-10}\frac{d}{c}$]}    \\ \hline
289420 & 5260.391476 & 0.595642 & 948.2 & 20.0 & 126.4 & 5.1 & 0.77 & 0.05 & 283.8 & 0.4 & 5330.4 & 9.1 & 0.0301 & 0.00433 \\
289678 & 5260.771054 & 0.275781 & 601.2 & 18.1 & 88.4 & 7.0 & 0.49 & 0.14 & 316.2 & 1.3 & 5855.3 & 15.3 & 0.0256 & 0.00637 \\
292115 & 5259.793223 & 1.136677 & 1074.6 & 59.9 & 194.5 & 12.9 & 0.68 & 0.14 & 74.0 & 0.7 & 5745.5 & 40.4 & 0.0854 & 0.02561 \\
292332 & 5260.524127 & 0.593263 & 1001.6 & 58.8 & 125.1 & 5.7 & 0.14 & 0.12 & 114.2 & 5.5 & 5757.8 & 106.0 & 0.0261 & 0.00698 \\
292333 & 5260.813455 & 0.373477 & 330.8 & 3.1 & 111.5 & 2.2 & 0.30 & 0.03 & 197.0 & 0.5 & 5585.1 & 5.2 & 0.1700 & 0.01149 \\
292418 & 5259.951559 & 0.636538 & 380.6 & 17.3 & 132.8 & 15.5 & 0.20 & 0.18 & 161.9 & 3.7 & 5481.0 & 33.0 & 0.2170 & 0.08025 \\
292479 & 5260.721290 & 0.333332 & 359.0 & 62.3 & 46.7 & 5.1 & 0.36 & 0.36 & 271.4 & 4.4 & 5671.1 & 20.5 & 0.0106 & 0.00783 \\
292696 & 5259.647462 & 0.859660 & 647.6 & 148.2 & 115.1 & 19.7 & 0.33 & 0.41 & 104.3 & 5.1 & 5431.9 & 61.7 & 0.0487 & 0.05010 \\
293135 & 5260.434981 & 0.576170 & 1379.7 & 114.2 & 216.7 & 10.9 & 0.16 & 0.14 & 267.5 & 4.2 & 5727.8 & 127.7 & 0.0717 & 0.02502 \\
293299 & 5260.577088 & 0.236086 & 307.2 & 18.7 & 63.0 & 3.5 & 0.27 & 0.11 & 290.8 & 1.7 & 5373.1 & 9.1 & 0.0355 & 0.01043 \\
295315 & 5260.685686 & 0.364502 & 895.8 & 49.5 & 177.2 & 15.3 & 0.14 & 0.22 & 211.4 & 11.8 & 5216.3 & 234.8 & 0.0930 & 0.03148 \\
297206 & 5260.037306 & 1.783266 & 852.8 & 209.3 & 157.0 & 51.7 & 0.62 & 0.39 & 82.2 & 2.2 & 5470.3 & 124.4 & 0.0714 & 0.09959 \\
297211 & 5260.652052 & 0.362440 & 1490.6 & 25.0 & 256.1 & 8.1 & 0.22 & 0.09 & 315.6 & 3.0 & 6856.7 & 87.1 & 0.1013 & 0.01150 \\
297736 & 5260.577283 & 0.951446 & 877.6 & 92.6 & 39.1 & 3.5 & 0.24 & 0.21 & 116.9 & 4.7 & 5889.7 & 77.6 & 0.0010 & 0.00050 \\
303179 & 5260.461239 & 0.287059 & 1351.4 & 76.8 & 93.3 & 10.5 & 0.45 & 0.16 & 192.4 & 1.7 & 5972.4 & 75.3 & 0.0060 & 0.00237 \\
303780 & 5260.709385 & 0.829257 & 324.3 & 17.0 & 46.3 & 5.7 & 0.05 & 0.23 & 196.0 & 24.9 & 5461.0 & 167.5 & 0.0126 & 0.00508 \\
\end{tabular}                         
}  
\end{table*}

\clearpage                           
                           
\begin{figure*}                           
\caption{The light curve, the ETV-s with third body solution and the folded light curve of the first group (258 systems).}       
                           
\includegraphics[width=\columnwidth]{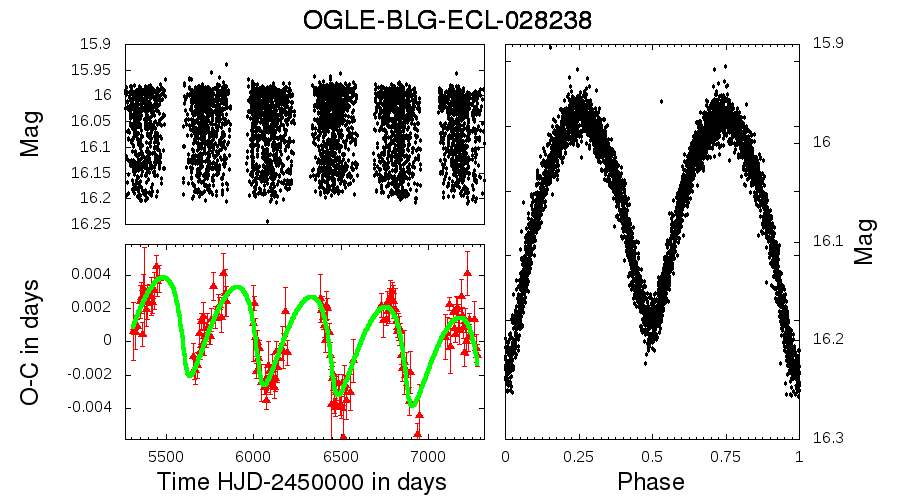}                           
\includegraphics[width=\columnwidth]{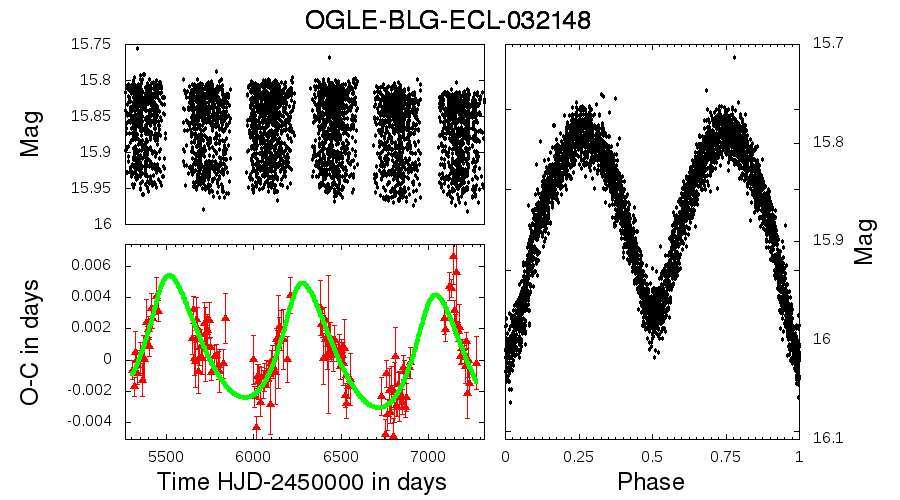}                           
                           
\includegraphics[width=\columnwidth]{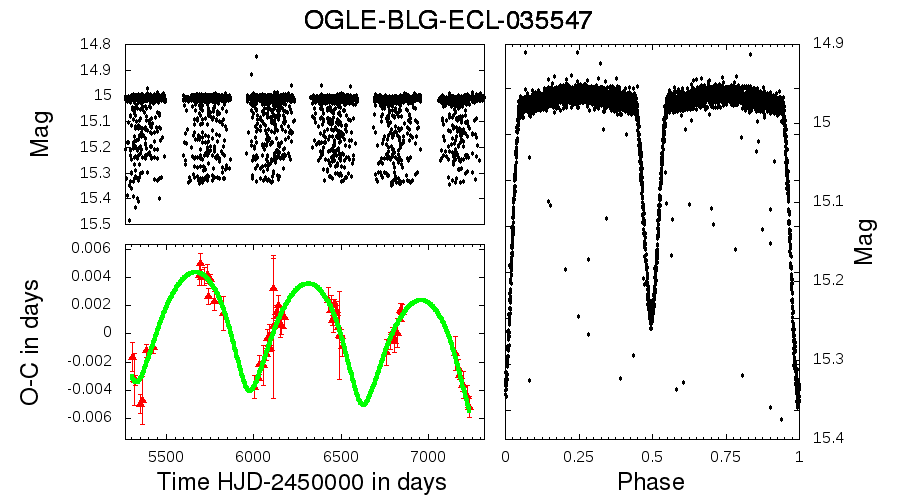}                           
\includegraphics[width=\columnwidth]{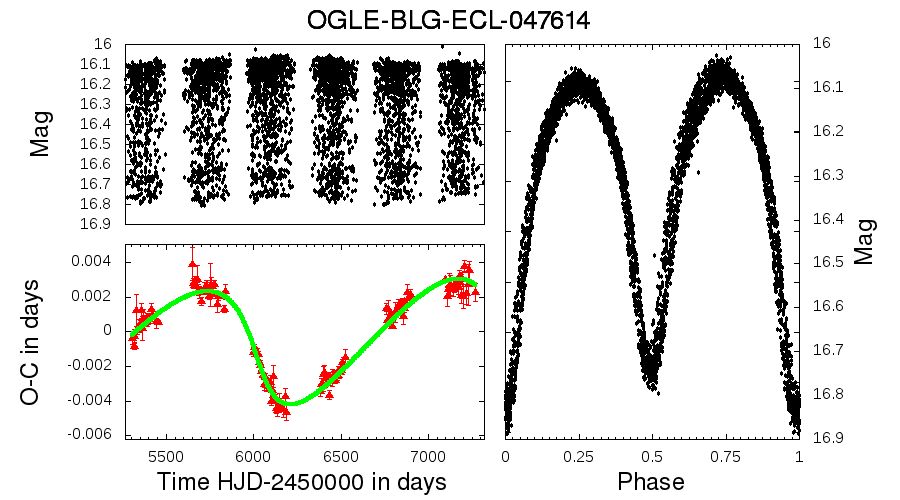}                           
                           
\includegraphics[width=\columnwidth]{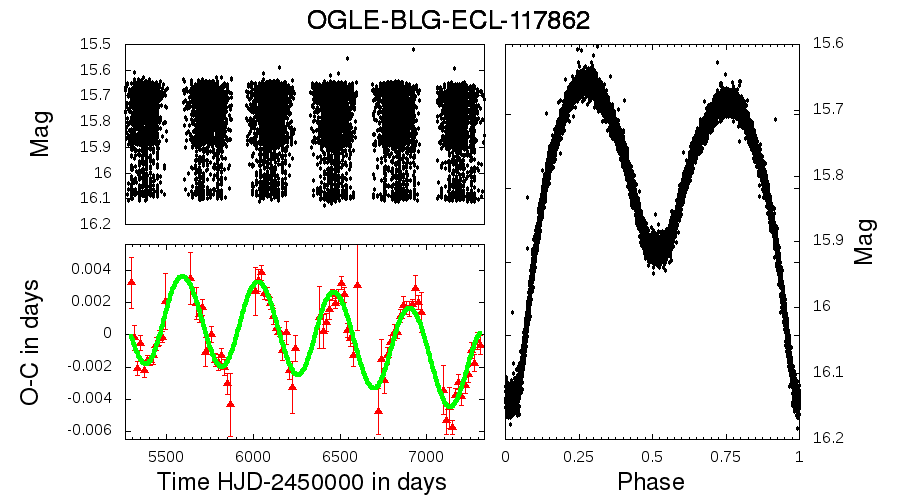}                           
\includegraphics[width=\columnwidth]{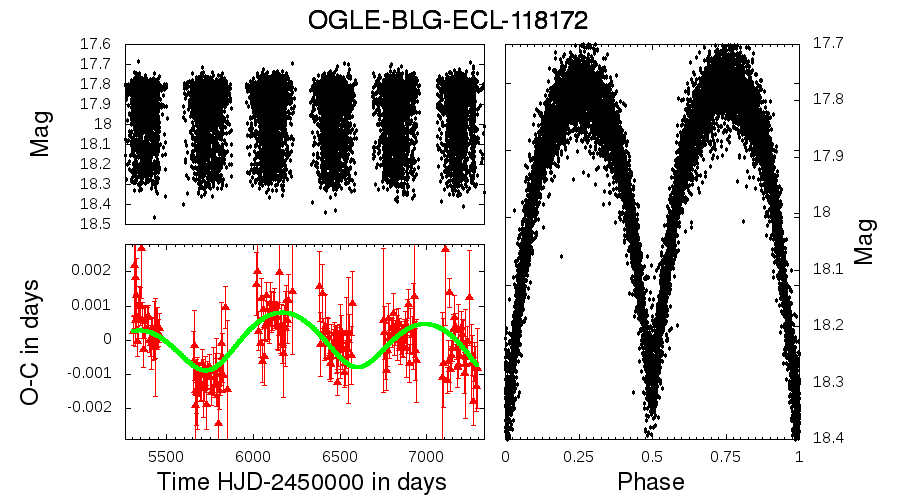}                           
                           
\includegraphics[width=\columnwidth]{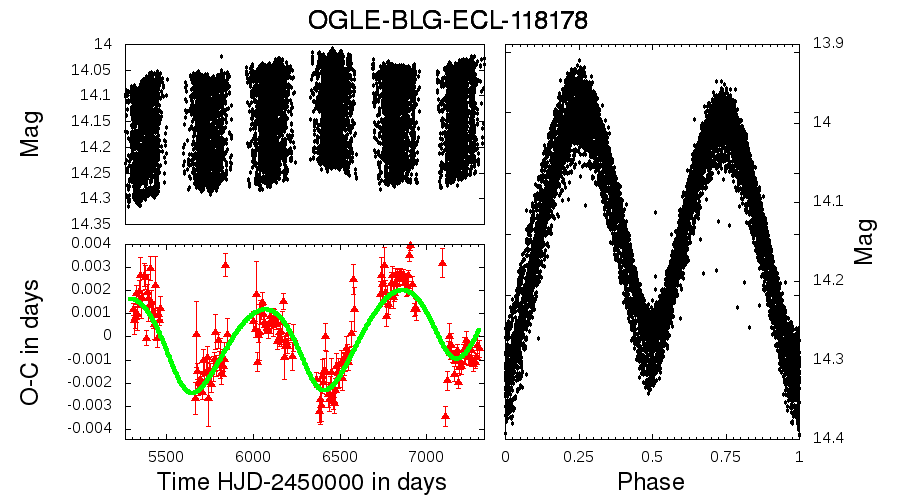}                           
\includegraphics[width=\columnwidth]{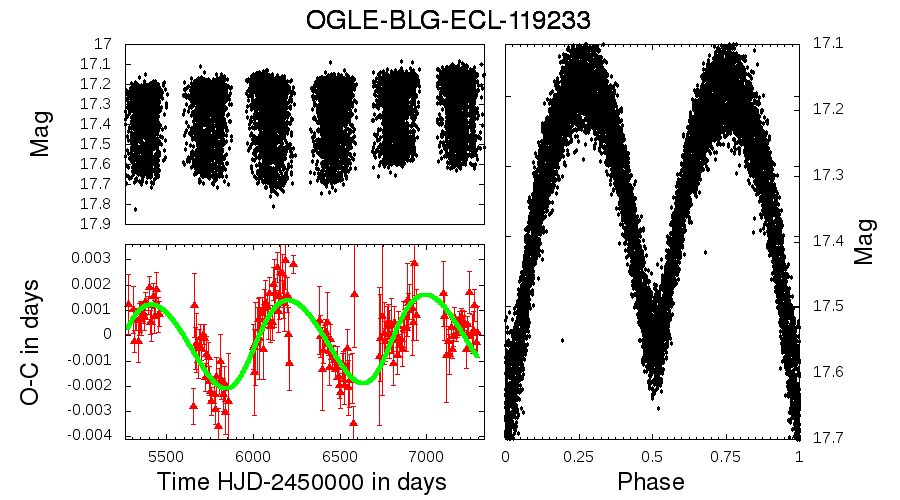}                           
                           
\includegraphics[width=\columnwidth]{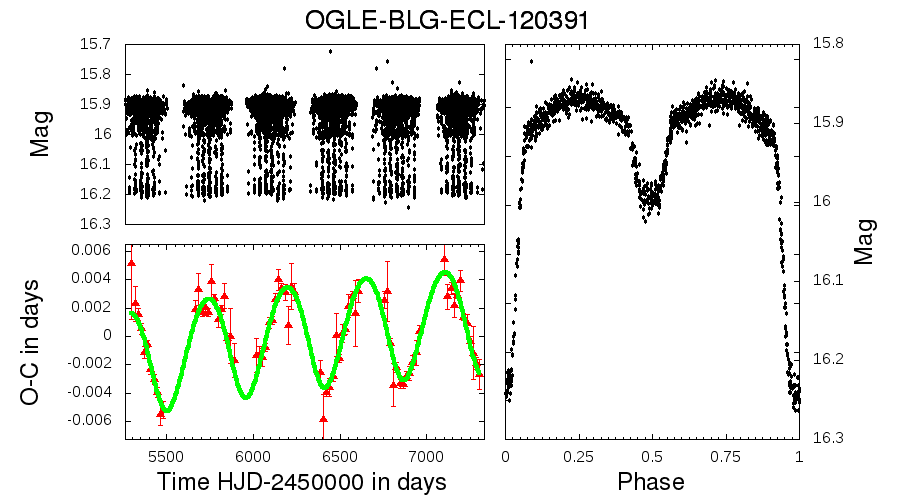}                           
\includegraphics[width=\columnwidth]{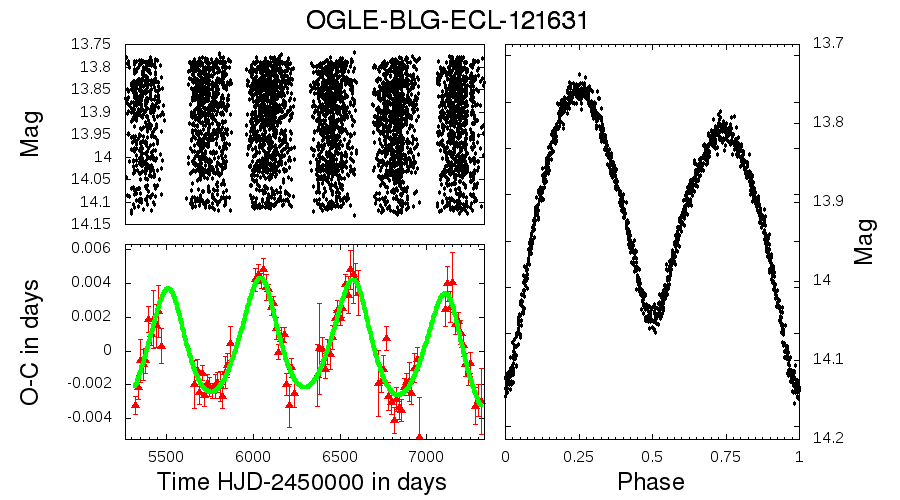}                           
\end{figure*}                           
\clearpage                           
                           
\begin{figure*}                           
                           
\includegraphics[width=\columnwidth]{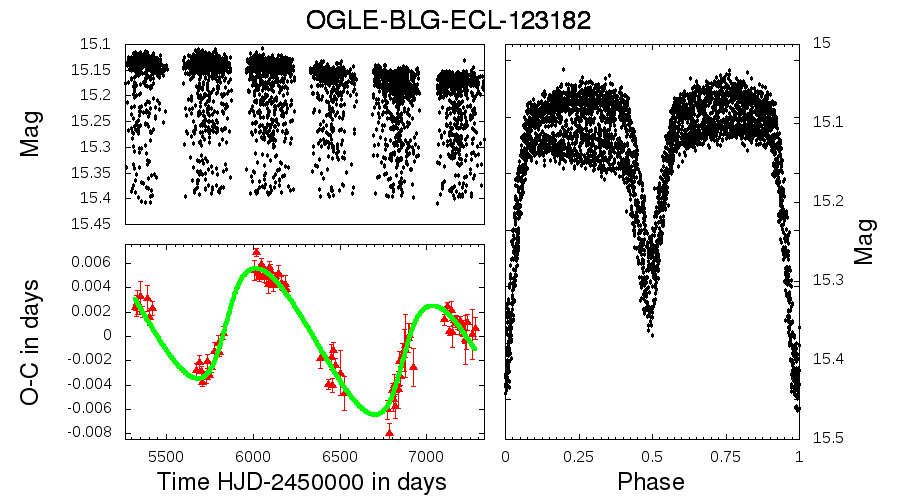}                           
\includegraphics[width=\columnwidth]{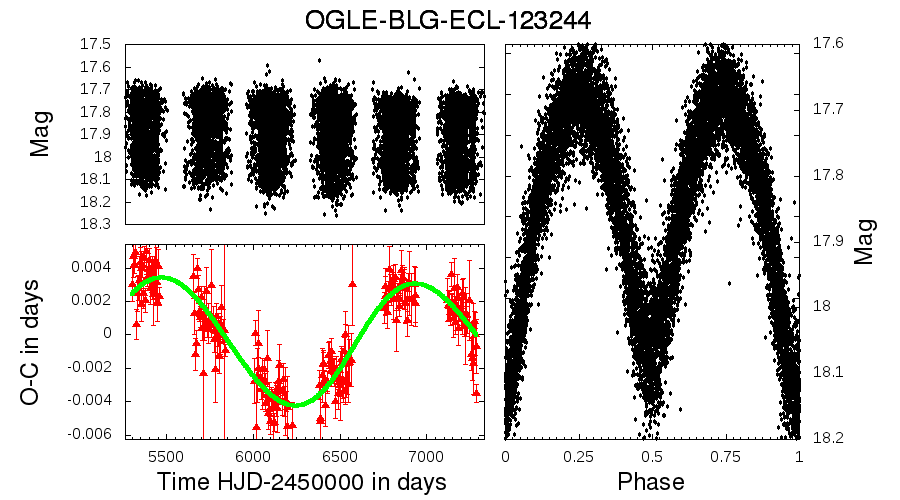}                           
                           
\includegraphics[width=\columnwidth]{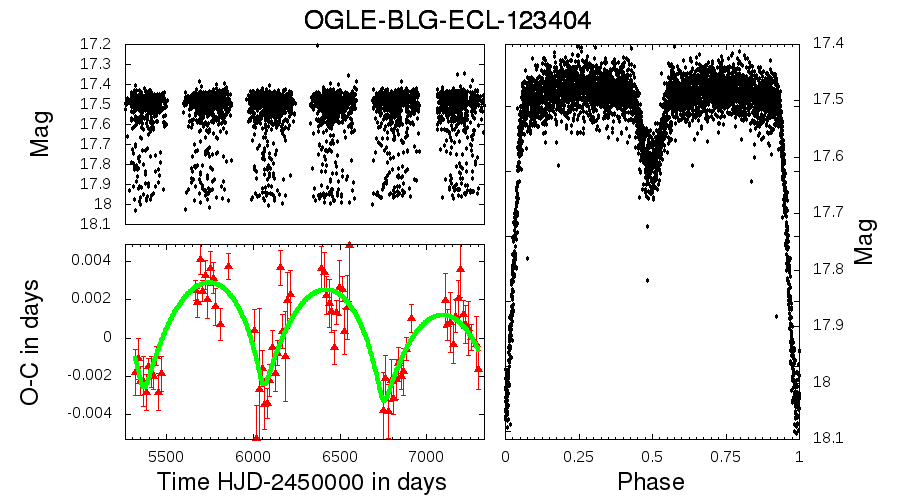}                           
\includegraphics[width=\columnwidth]{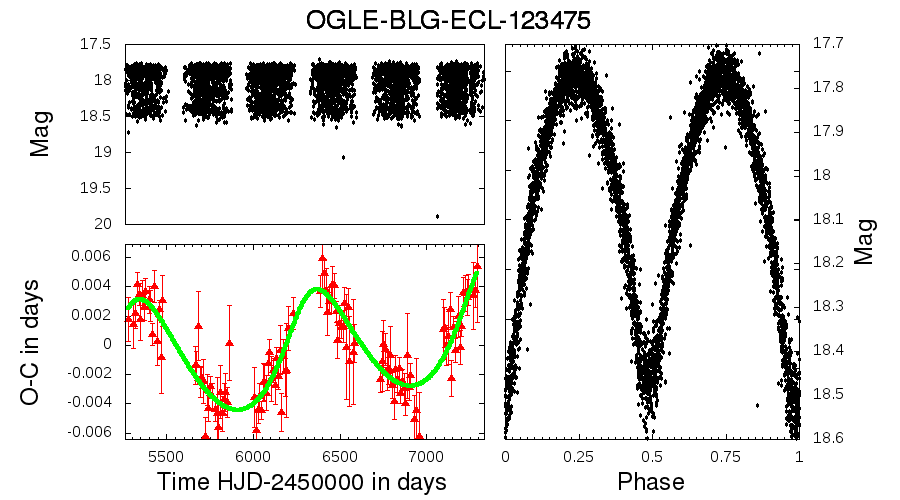}                           
                           
\includegraphics[width=\columnwidth]{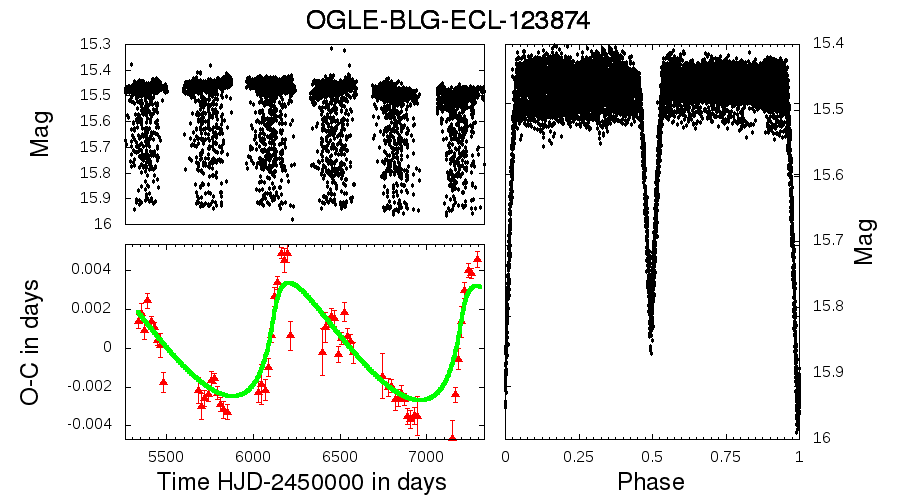}                           
\includegraphics[width=\columnwidth]{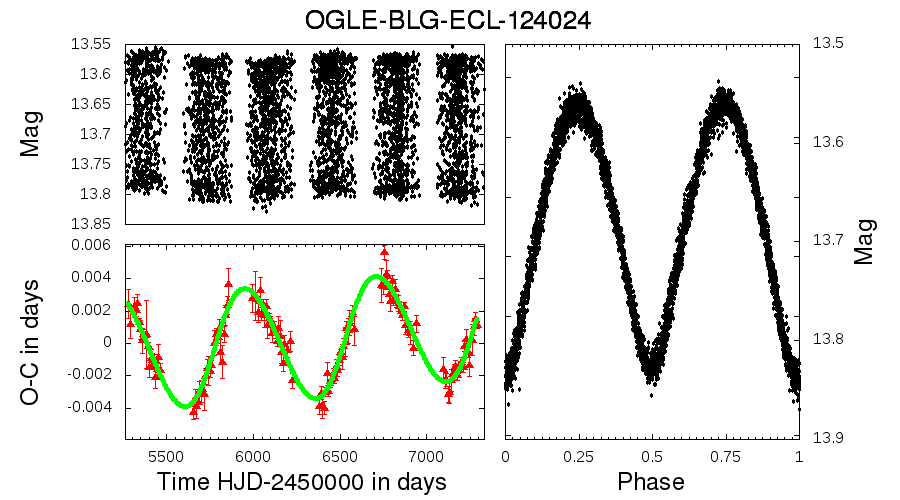}                           
                           
\includegraphics[width=\columnwidth]{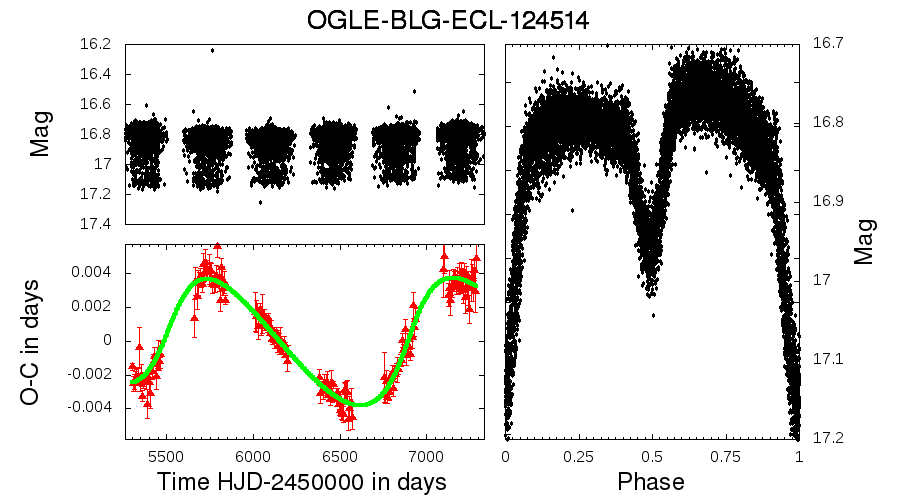}                           
\includegraphics[width=\columnwidth]{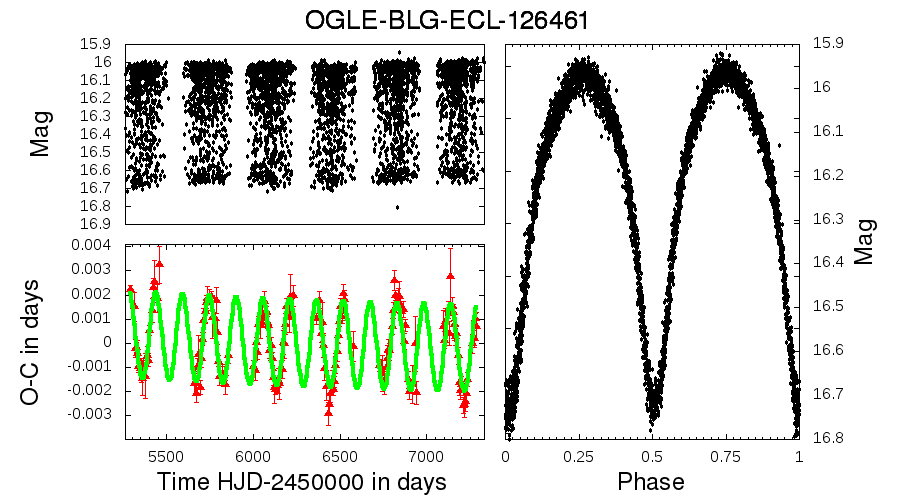}                           
                           
\includegraphics[width=\columnwidth]{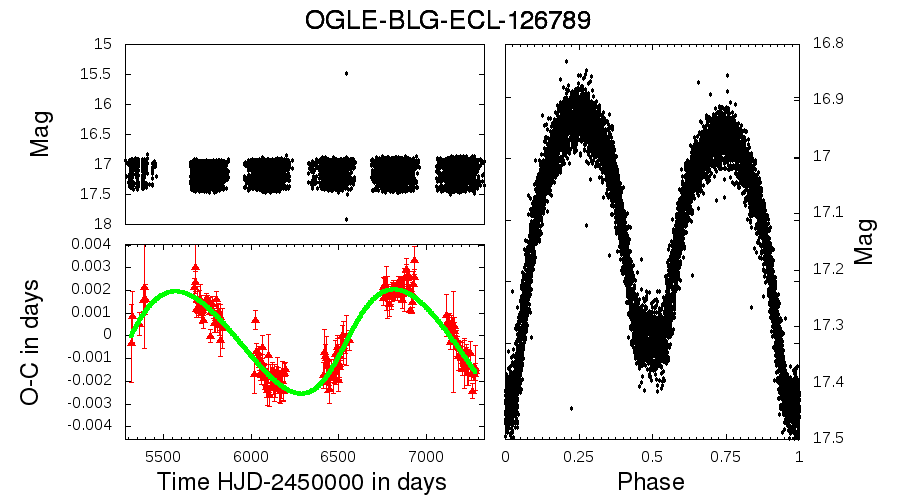}                           
\includegraphics[width=\columnwidth]{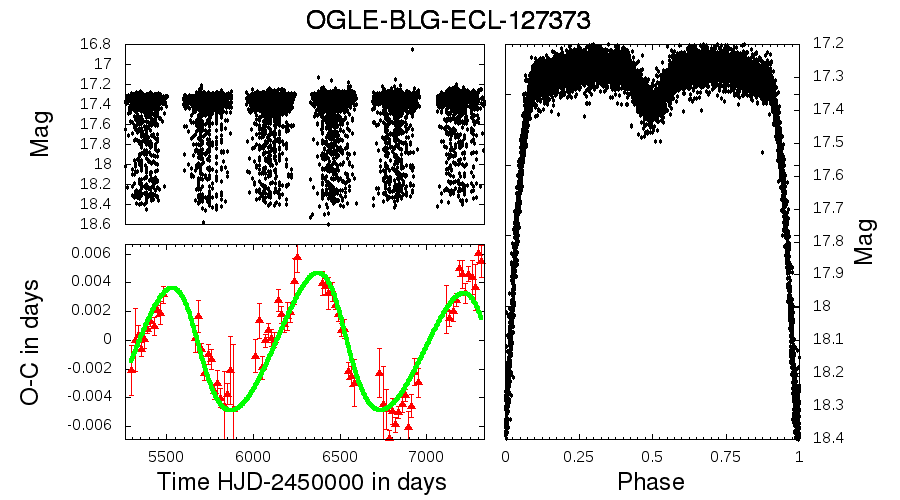}                           
\end{figure*}                           
\clearpage                           
                           
\begin{figure*}                           
                           
\includegraphics[width=\columnwidth]{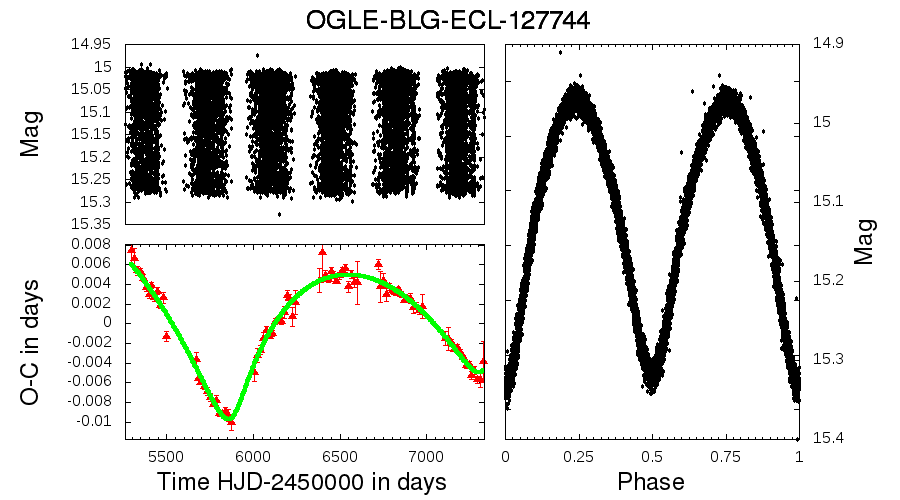}                           
\includegraphics[width=\columnwidth]{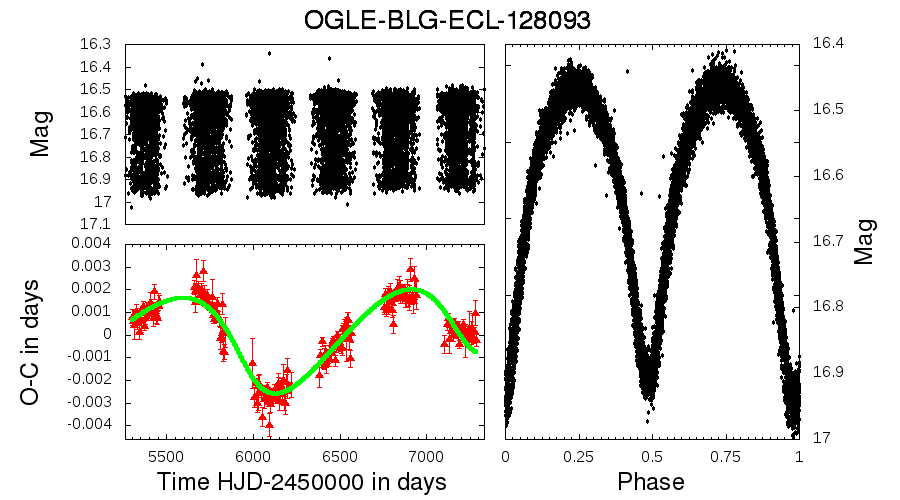}                           
                           
\includegraphics[width=\columnwidth]{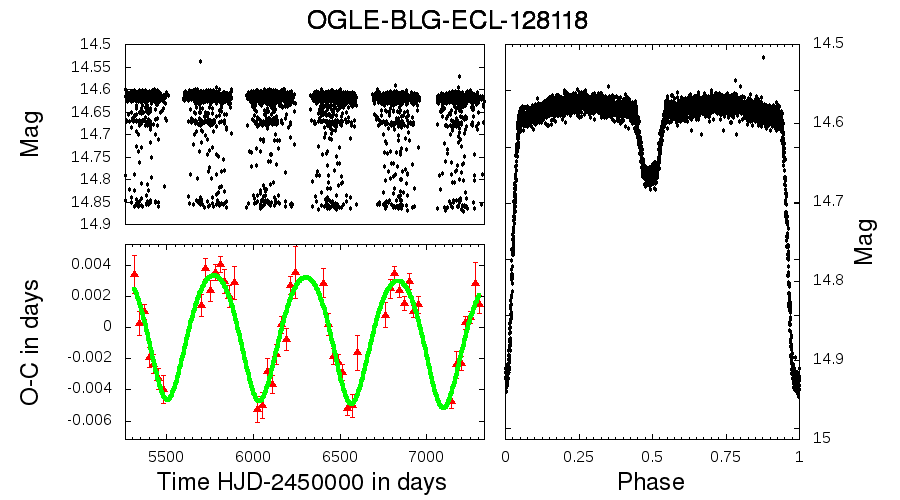}                           
\includegraphics[width=\columnwidth]{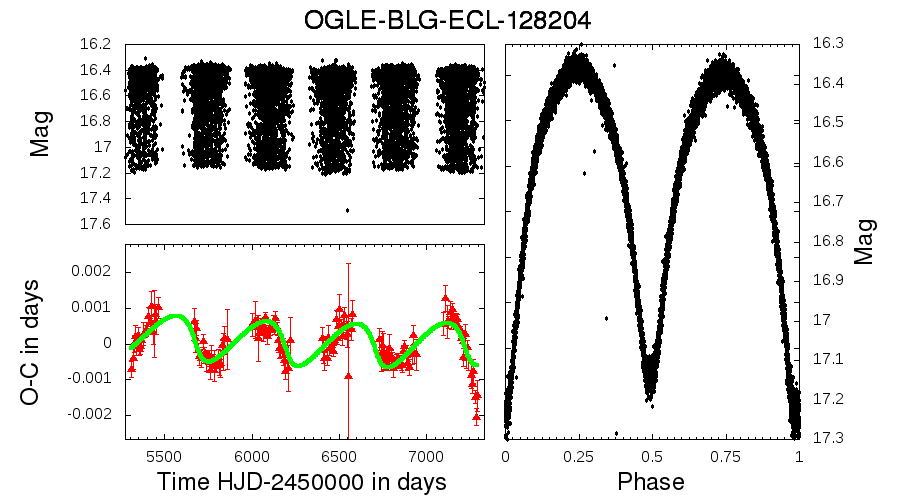}                           
                           
\includegraphics[width=\columnwidth]{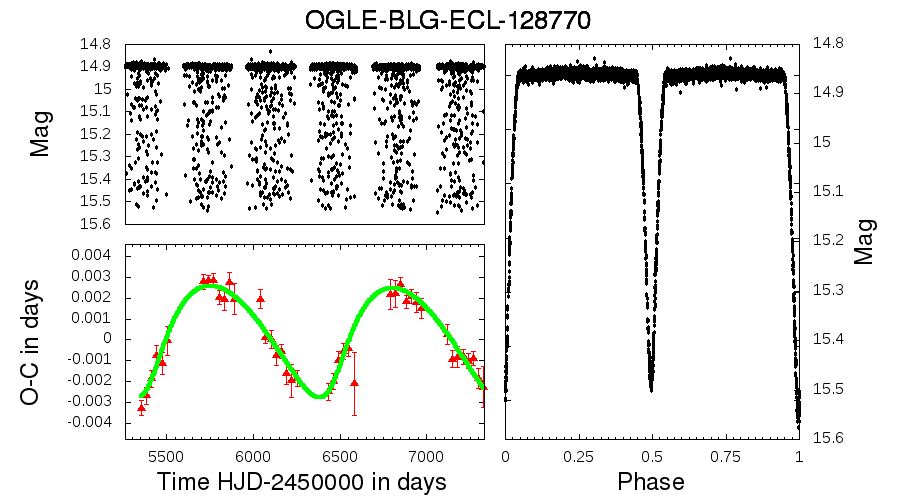}                           
\includegraphics[width=\columnwidth]{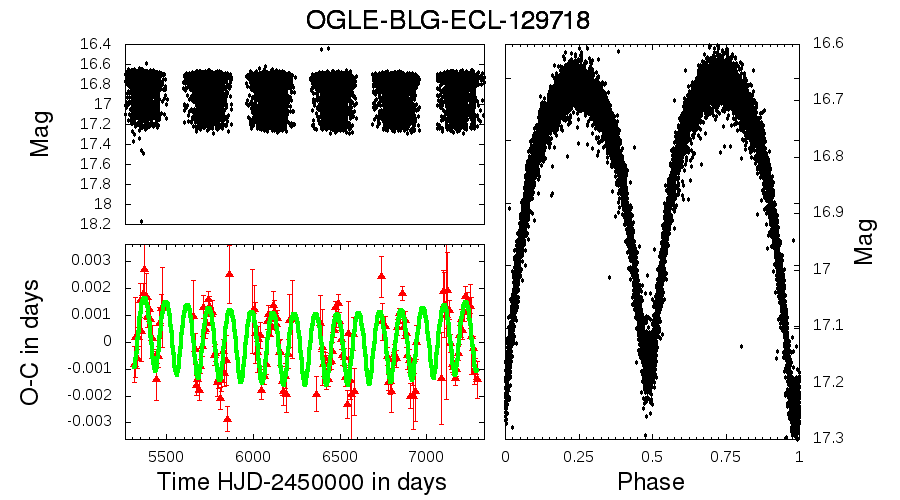}                           
                           
\includegraphics[width=\columnwidth]{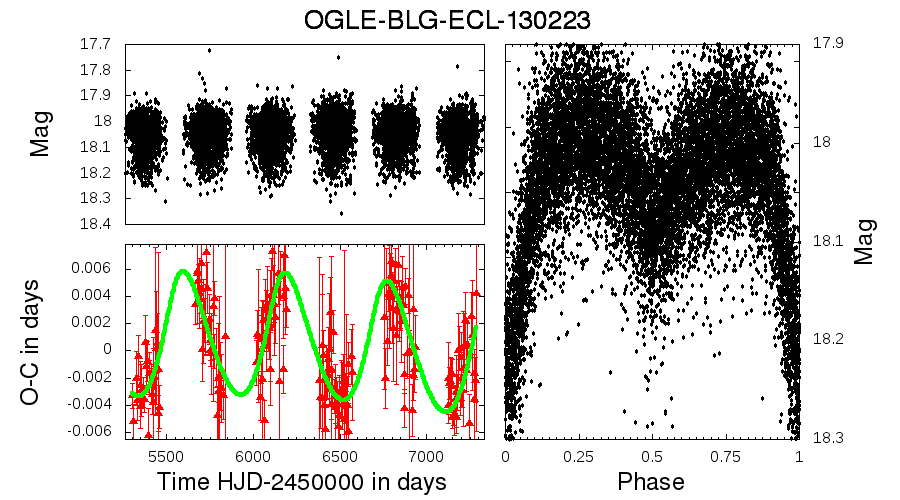}                           
\includegraphics[width=\columnwidth]{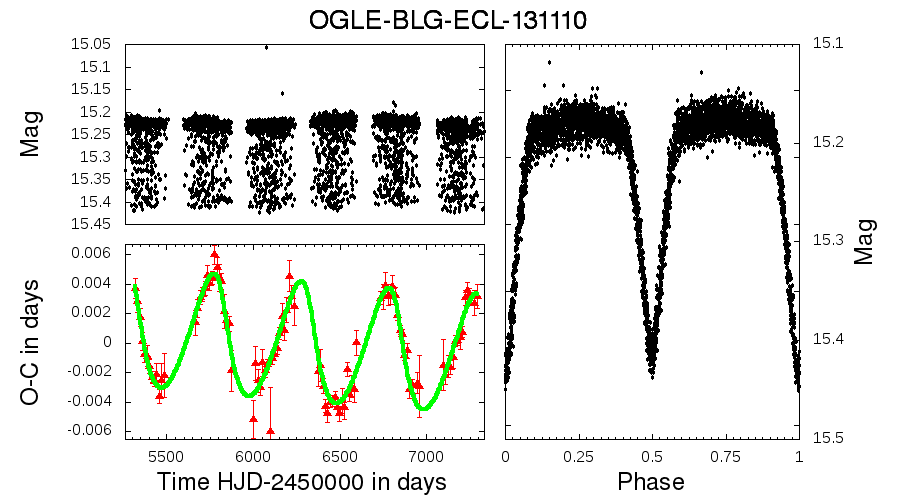}                           
                           
\includegraphics[width=\columnwidth]{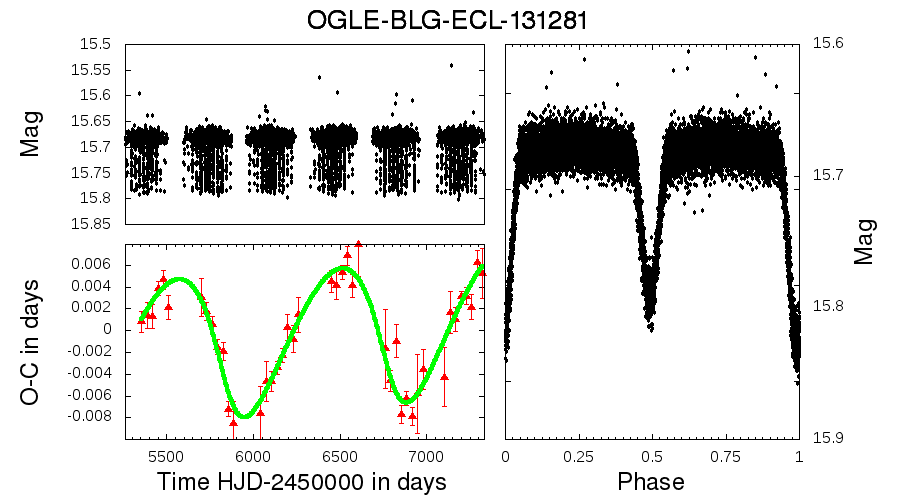}                           
\includegraphics[width=\columnwidth]{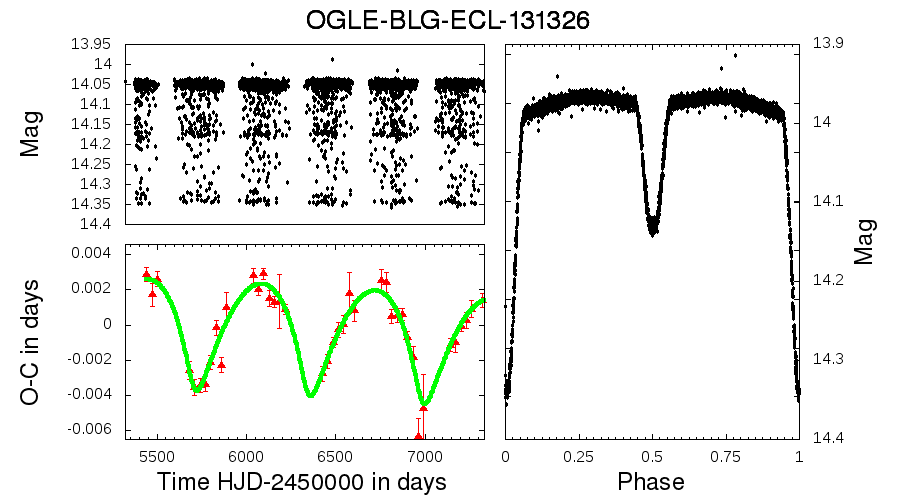}                           
\end{figure*}                           
\clearpage                           
                           
\begin{figure*}                           
                           
\includegraphics[width=\columnwidth]{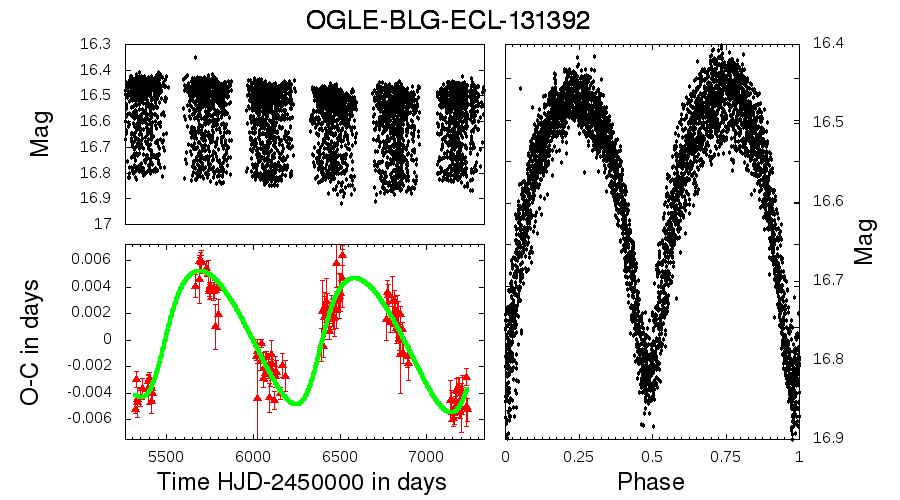}                           
\includegraphics[width=\columnwidth]{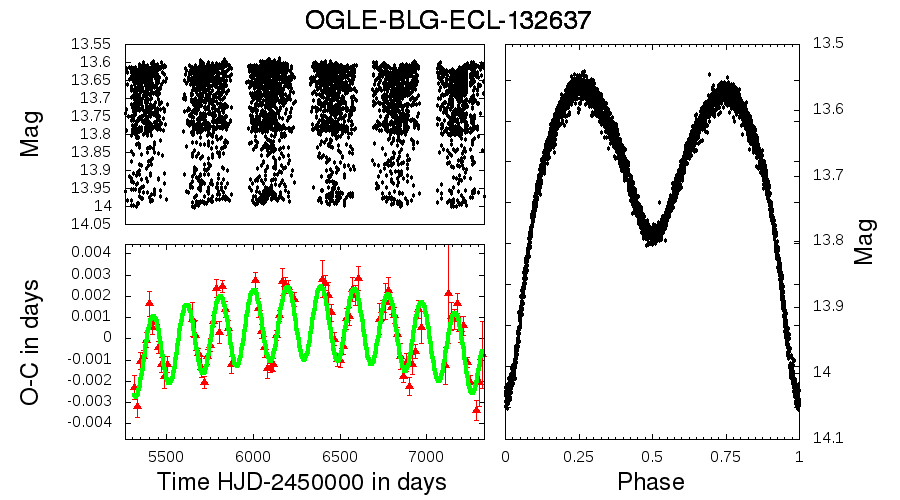}                           
                           
\includegraphics[width=\columnwidth]{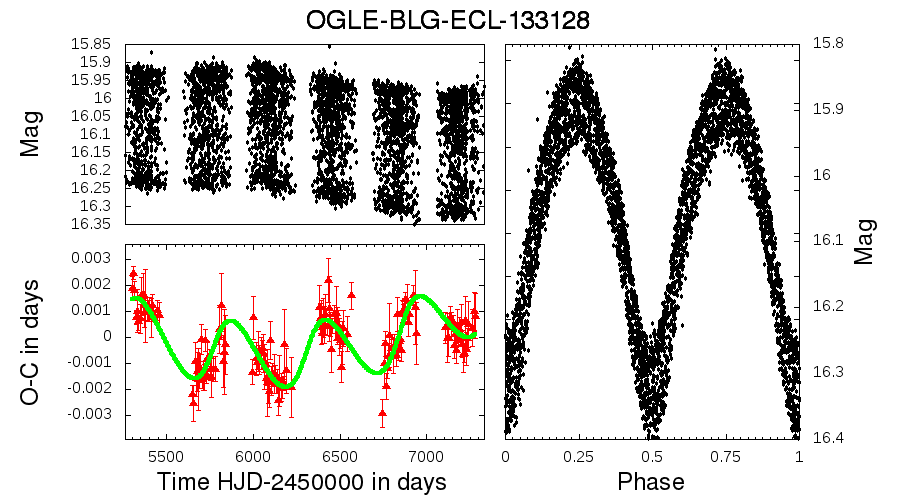}                           
\includegraphics[width=\columnwidth]{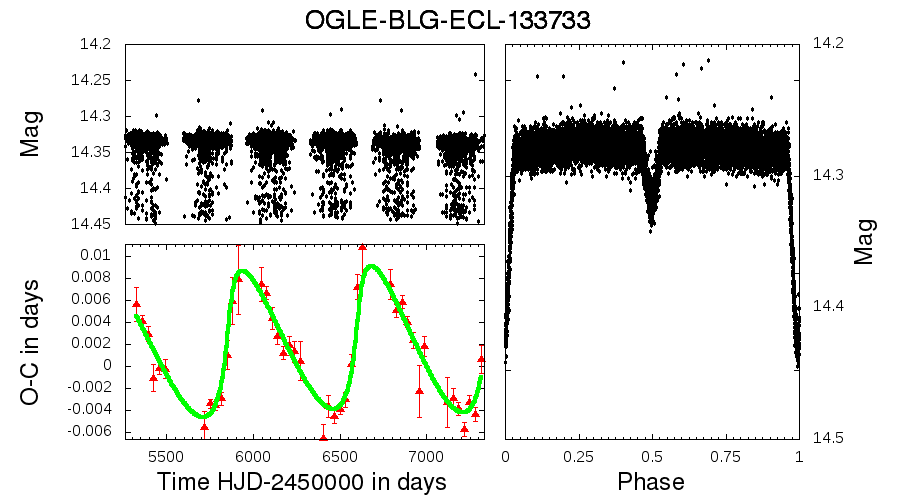}                           
                           
\includegraphics[width=\columnwidth]{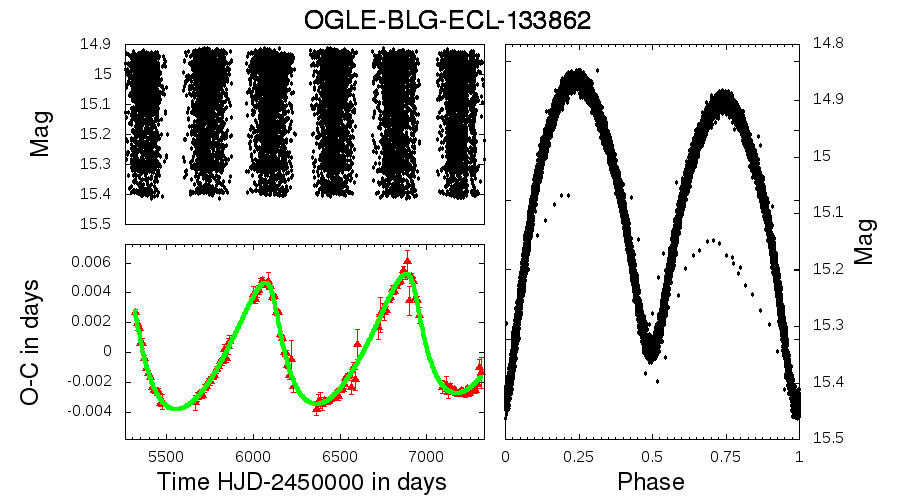}                           
\includegraphics[width=\columnwidth]{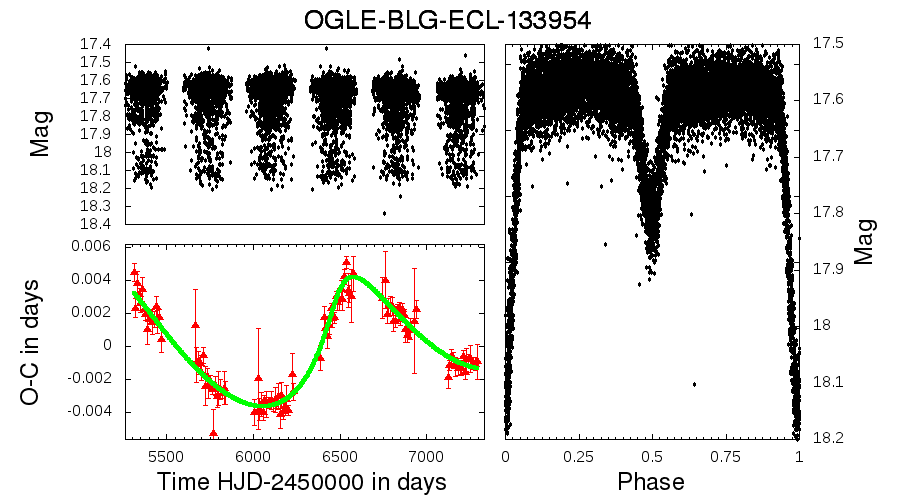}                           
                           
\includegraphics[width=\columnwidth]{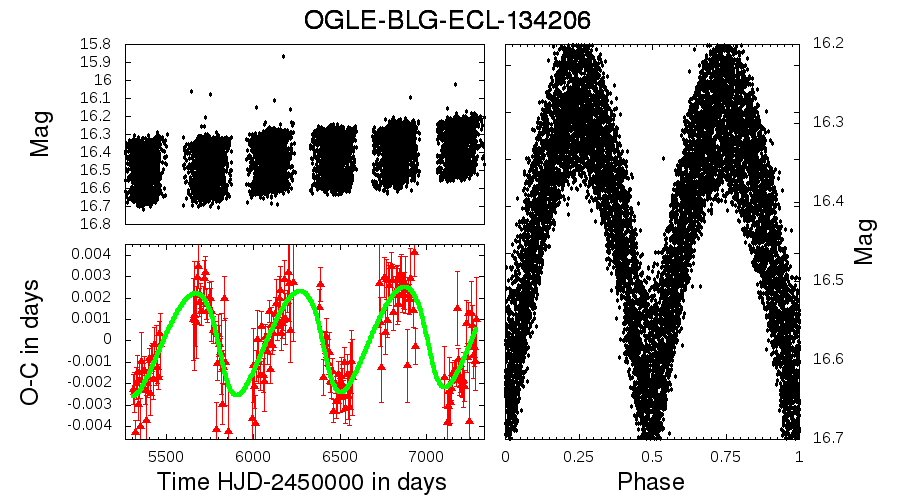}                           
\includegraphics[width=\columnwidth]{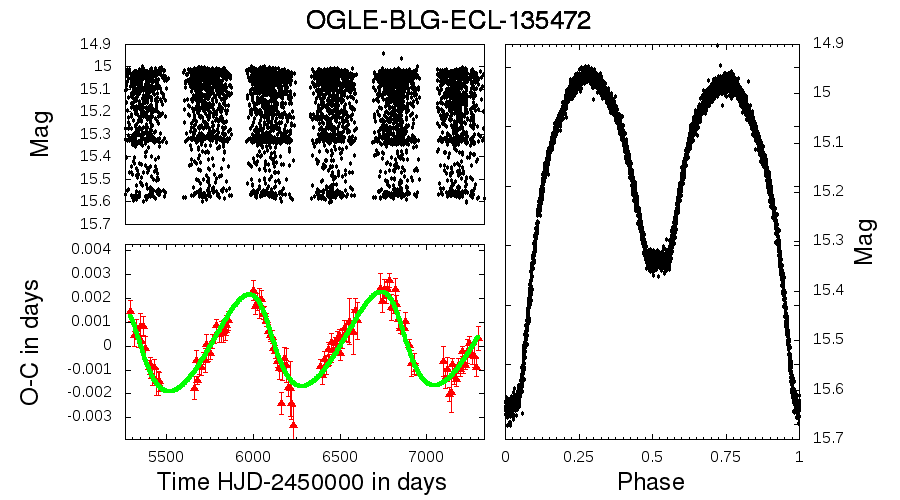}                           
                           
\includegraphics[width=\columnwidth]{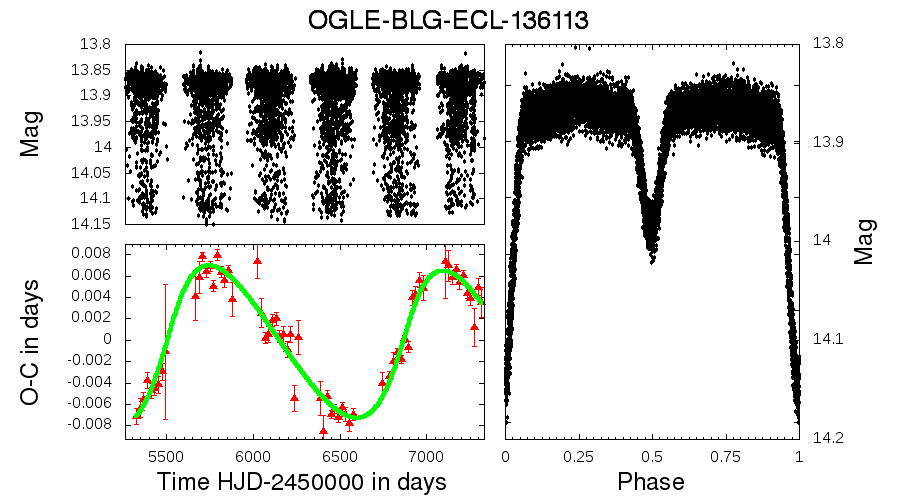}                           
\includegraphics[width=\columnwidth]{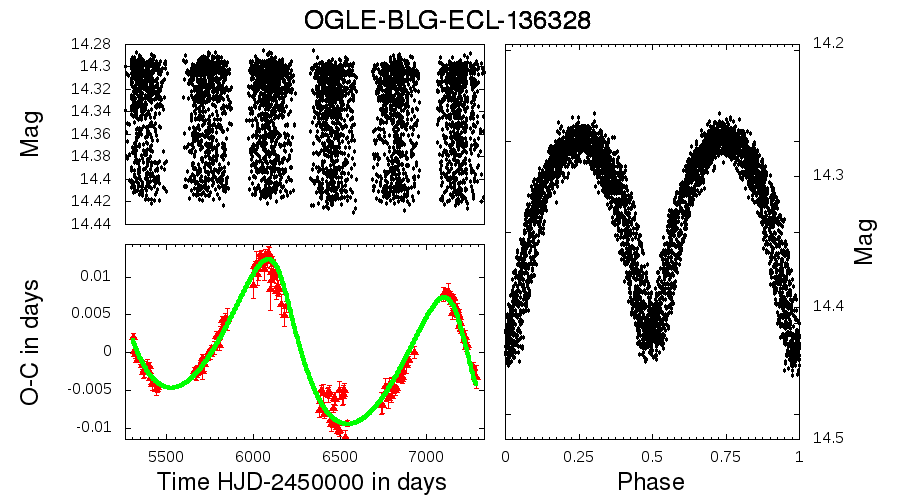}                           
\end{figure*}                           
\clearpage                           
                           
\begin{figure*}                           
                           
\includegraphics[width=\columnwidth]{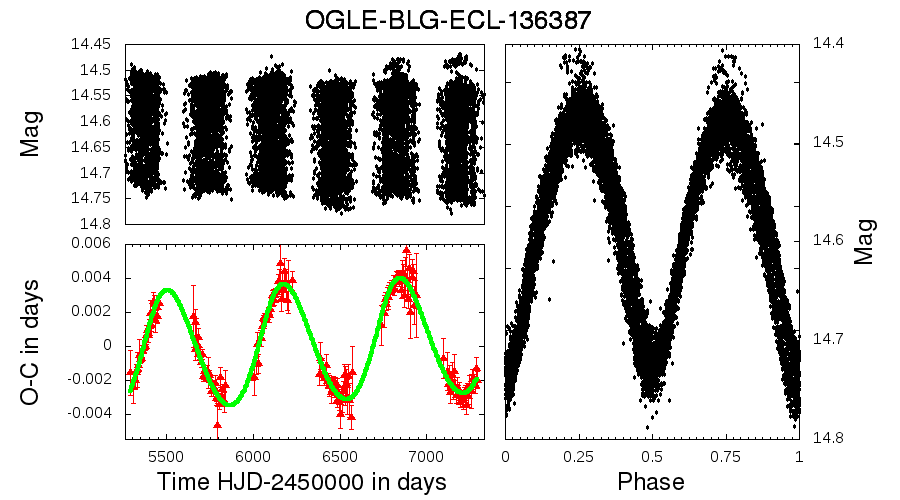}                           
\includegraphics[width=\columnwidth]{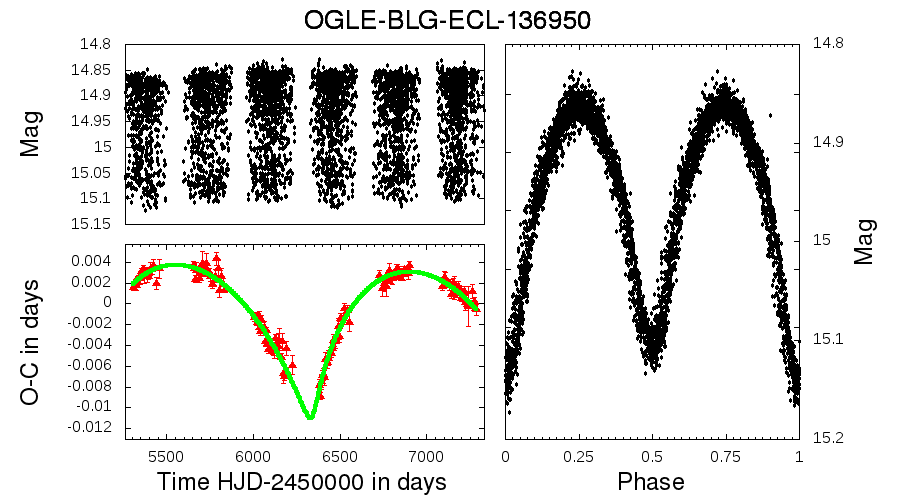}                           
                           
\includegraphics[width=\columnwidth]{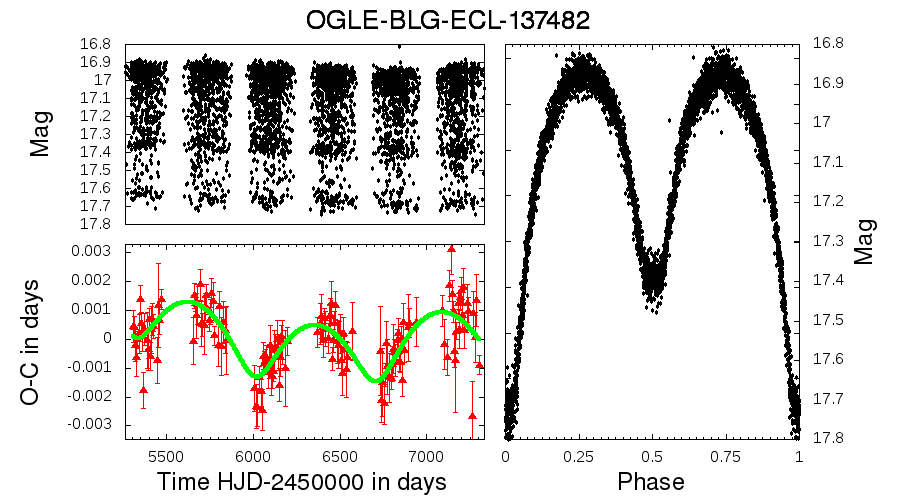}                           
\includegraphics[width=\columnwidth]{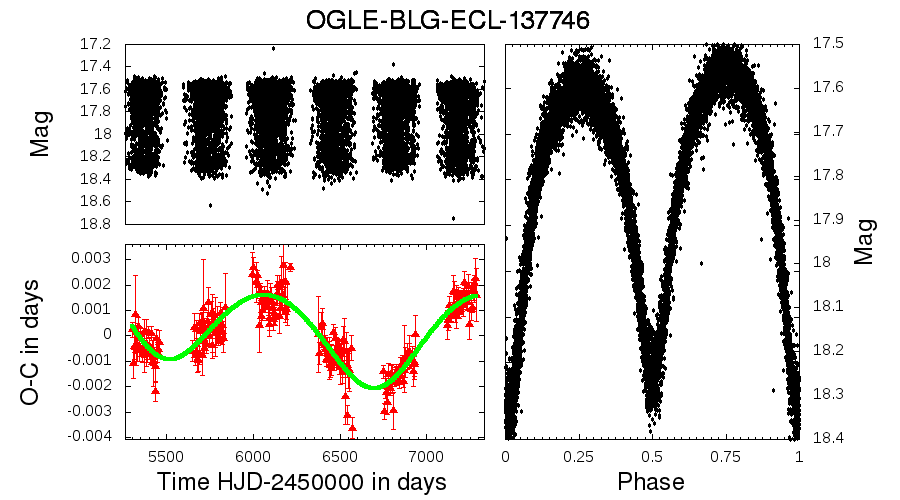}                           
                           
\includegraphics[width=\columnwidth]{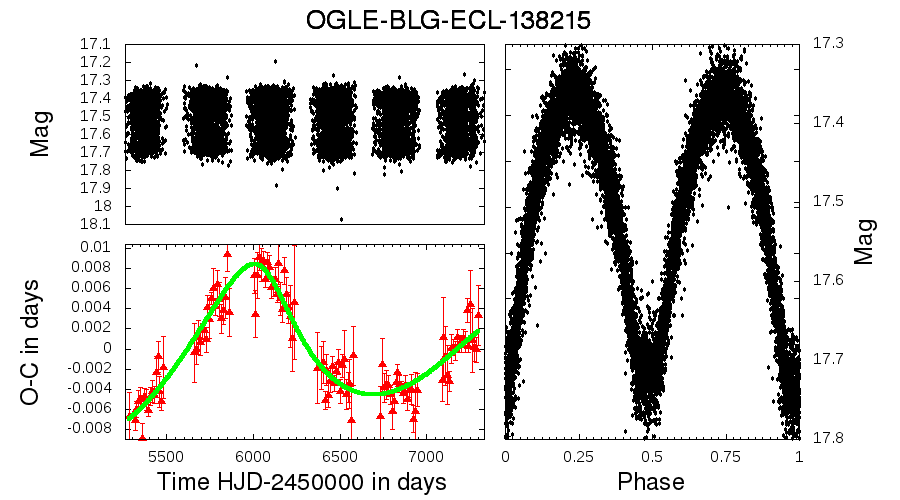}                           
\includegraphics[width=\columnwidth]{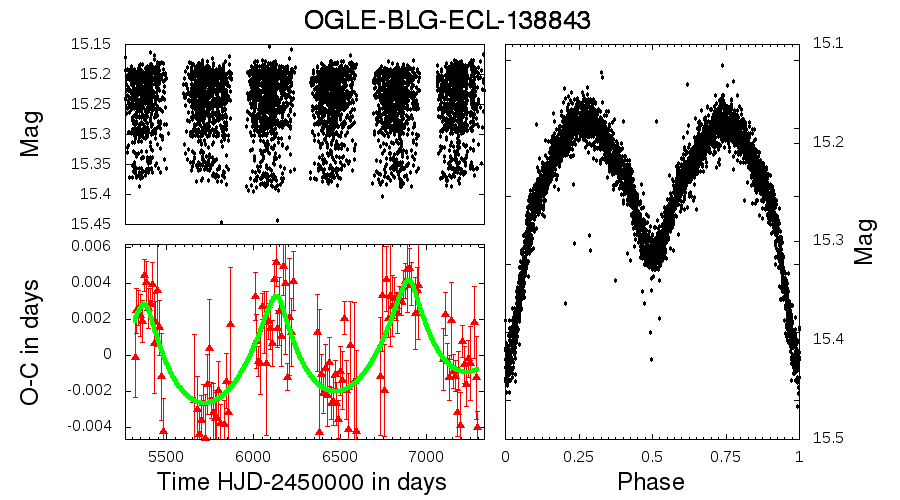}                           
                           
\includegraphics[width=\columnwidth]{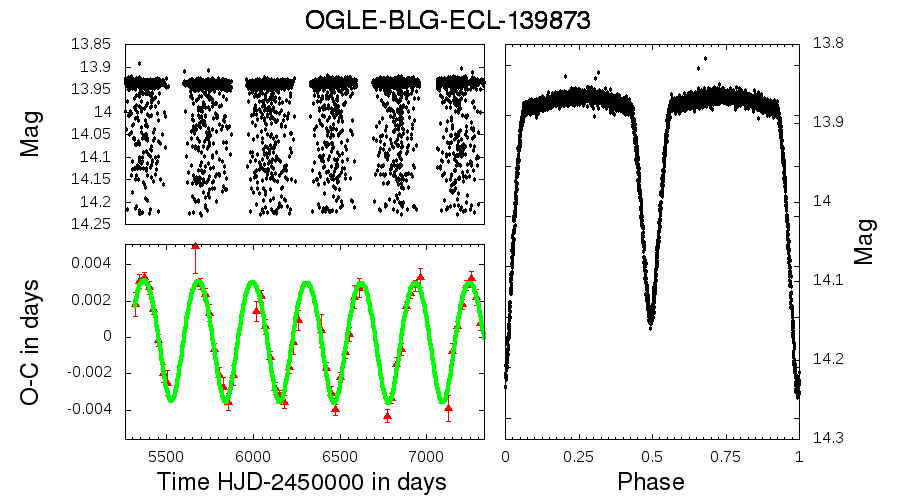}                           
\includegraphics[width=\columnwidth]{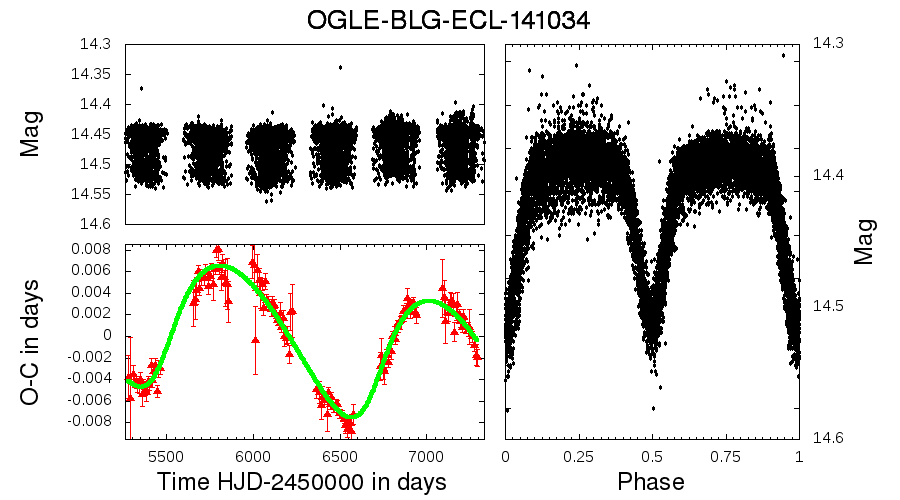}                           
                           
\includegraphics[width=\columnwidth]{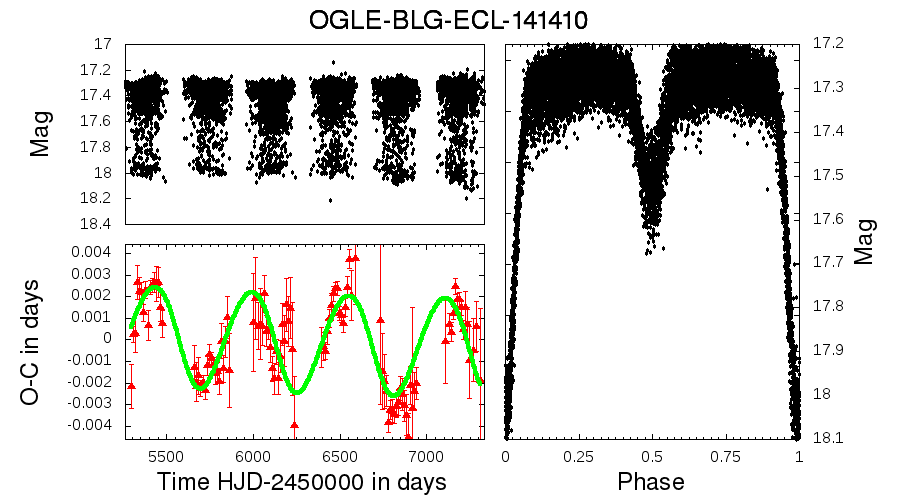}                           
\includegraphics[width=\columnwidth]{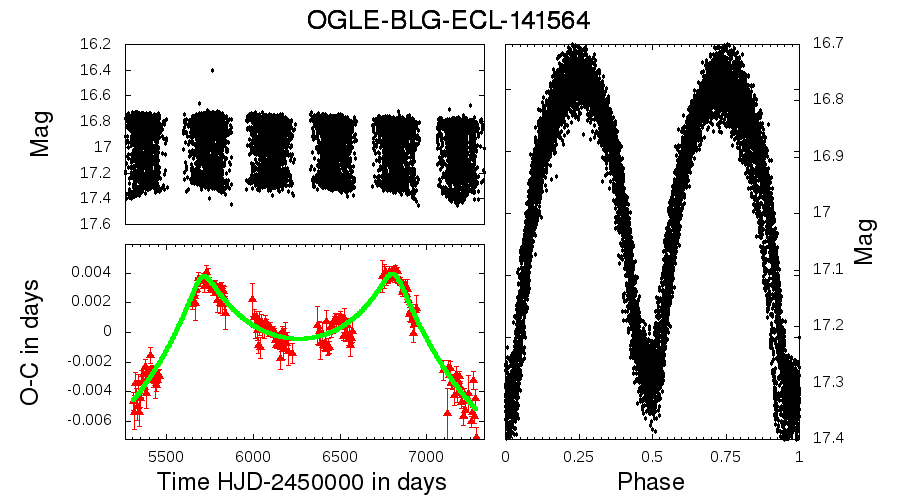}                           
\end{figure*}                           
\clearpage                           
                           
\begin{figure*}                           
                           
\includegraphics[width=\columnwidth]{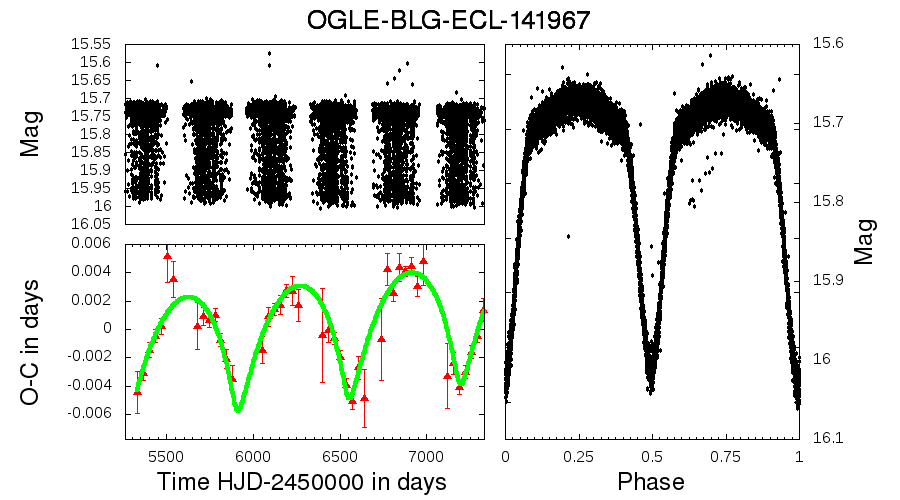}                           
\includegraphics[width=\columnwidth]{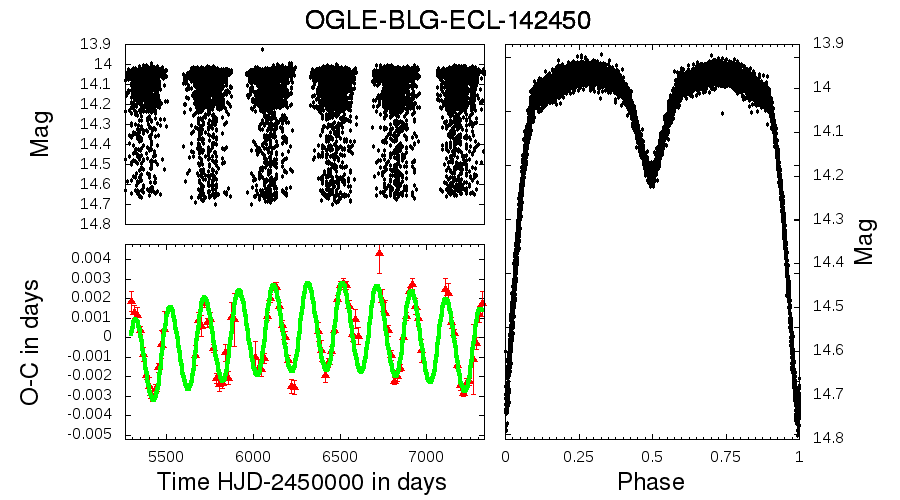}                           
                           
\includegraphics[width=\columnwidth]{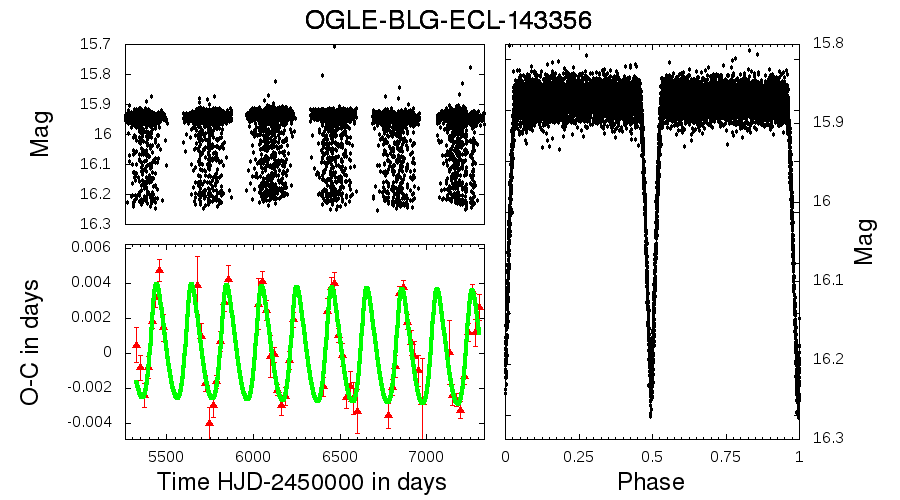}                           
\includegraphics[width=\columnwidth]{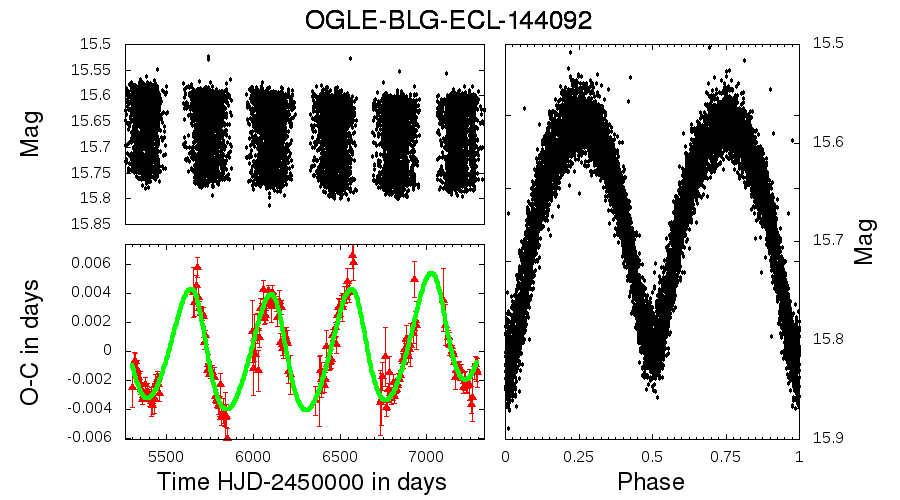}                           
                           
\includegraphics[width=\columnwidth]{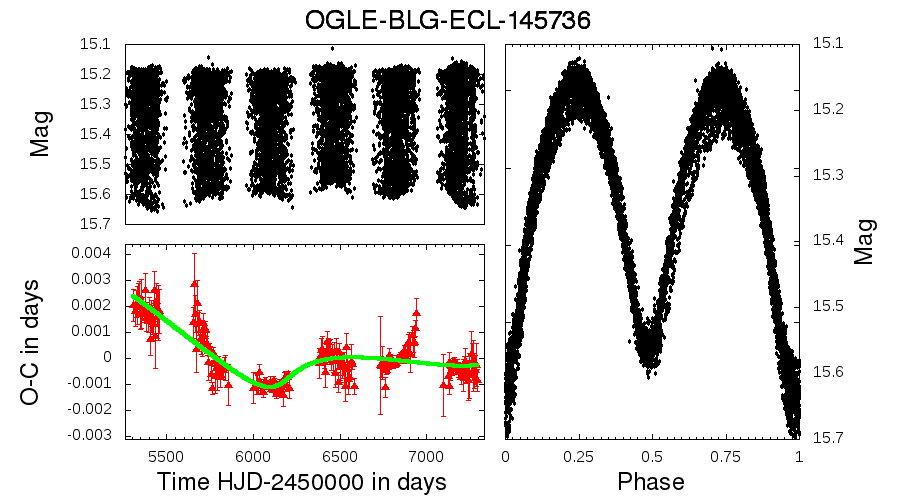}                           
\includegraphics[width=\columnwidth]{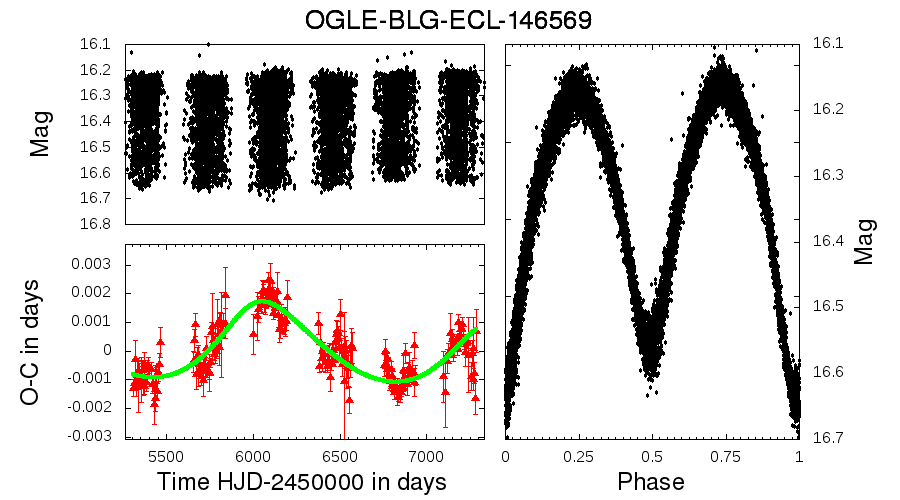}                           
                           
\includegraphics[width=\columnwidth]{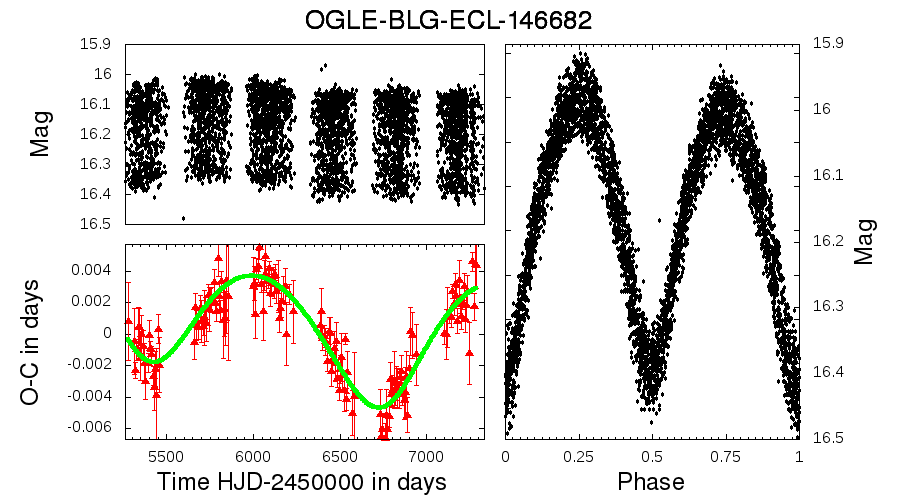}                           
\includegraphics[width=\columnwidth]{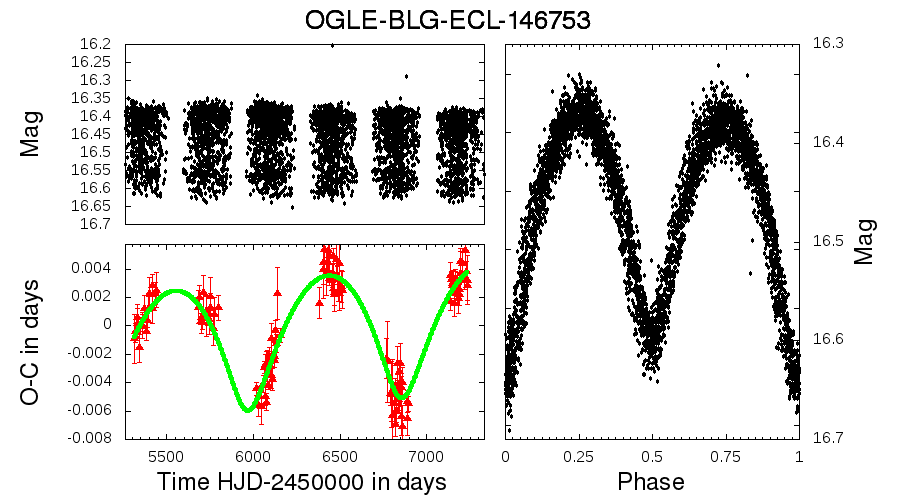}                           
                           
\includegraphics[width=\columnwidth]{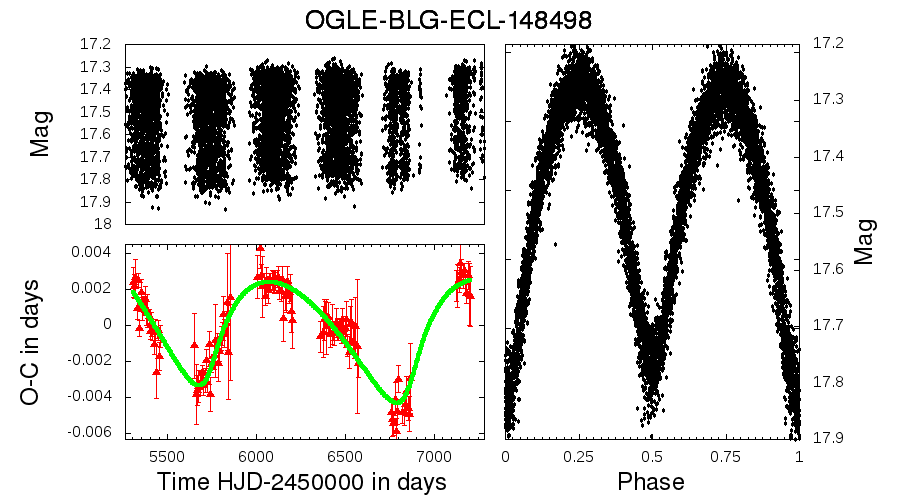}                           
\includegraphics[width=\columnwidth]{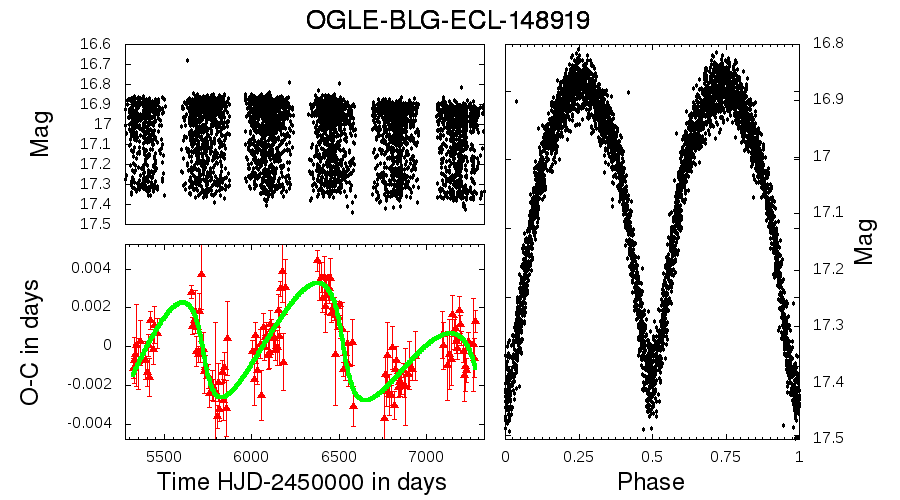}                           
\end{figure*}                           
\clearpage                           
                           
\begin{figure*}                           
                           
\includegraphics[width=\columnwidth]{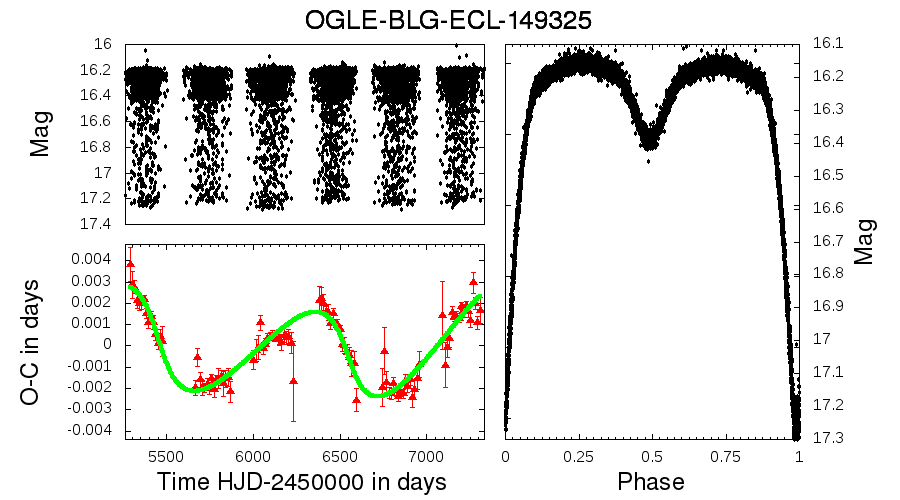}                           
\includegraphics[width=\columnwidth]{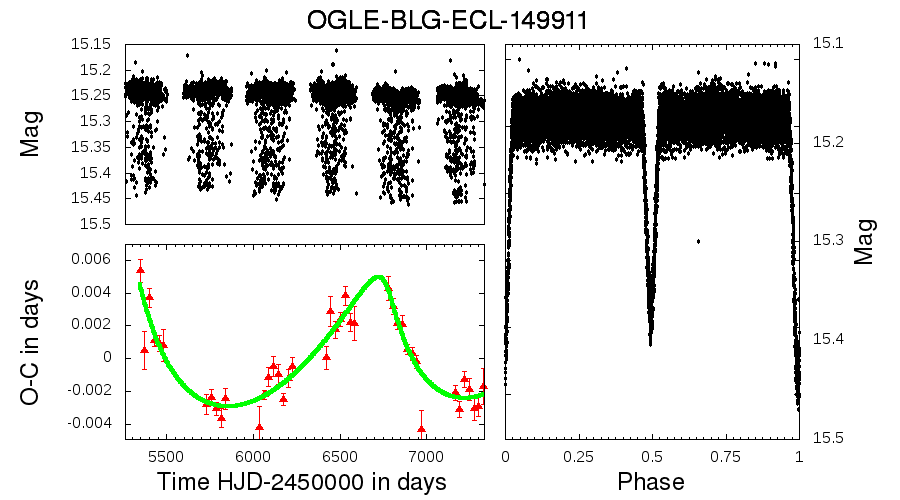}                           
                           
\includegraphics[width=\columnwidth]{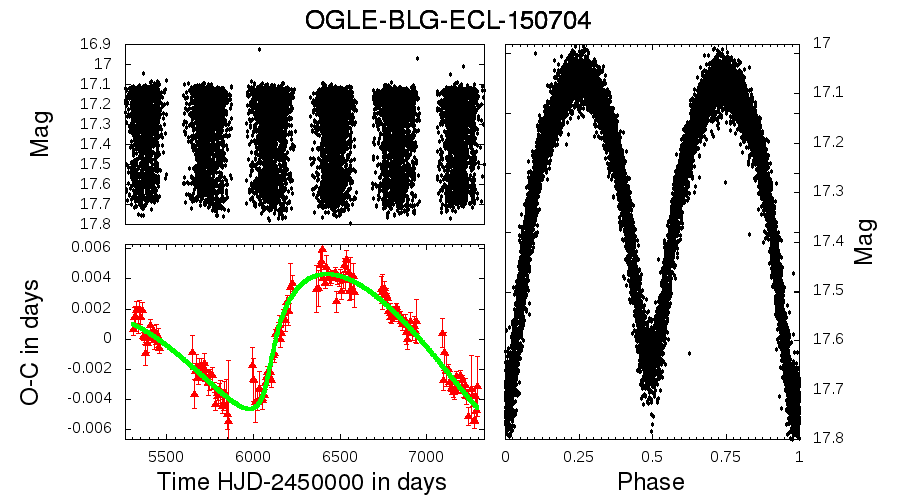}                           
\includegraphics[width=\columnwidth]{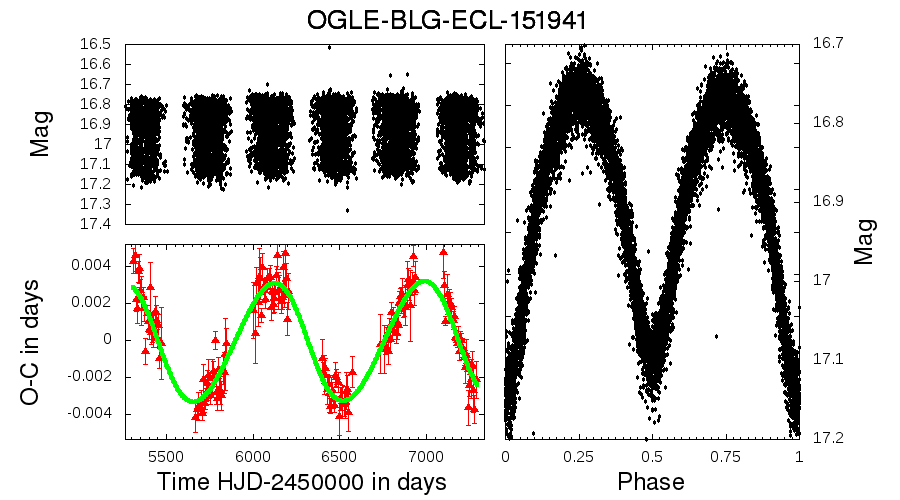}                           
                           
\includegraphics[width=\columnwidth]{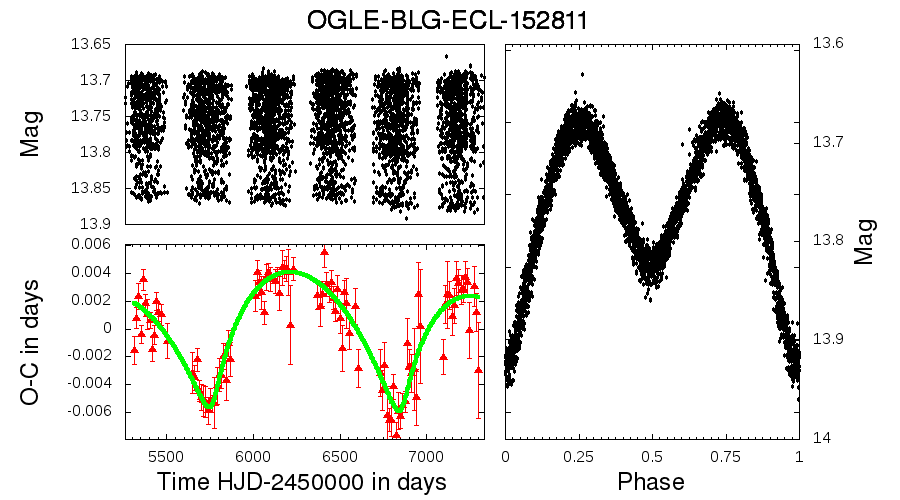}                           
\includegraphics[width=\columnwidth]{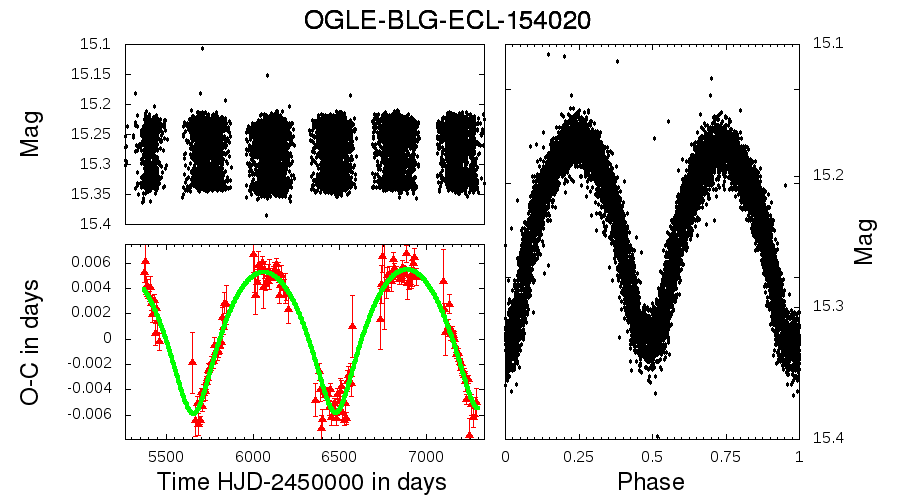}                           
                           
\includegraphics[width=\columnwidth]{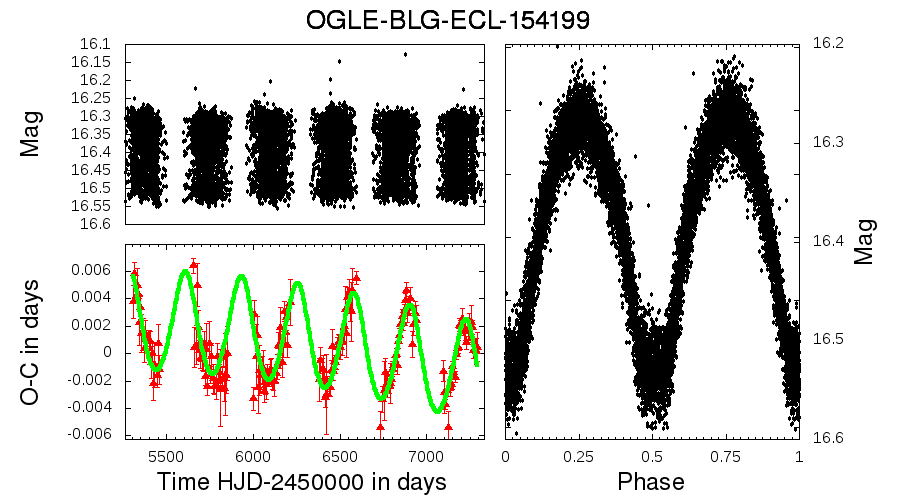}                           
\includegraphics[width=\columnwidth]{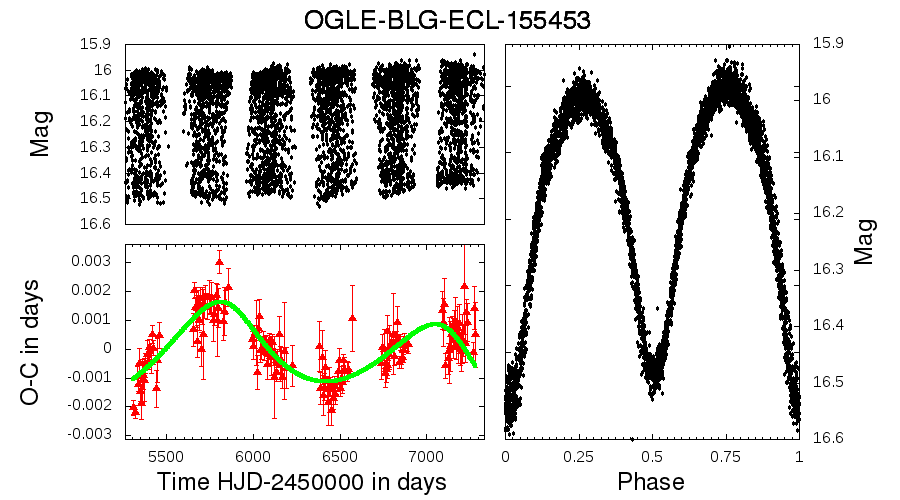}                           
                           
\includegraphics[width=\columnwidth]{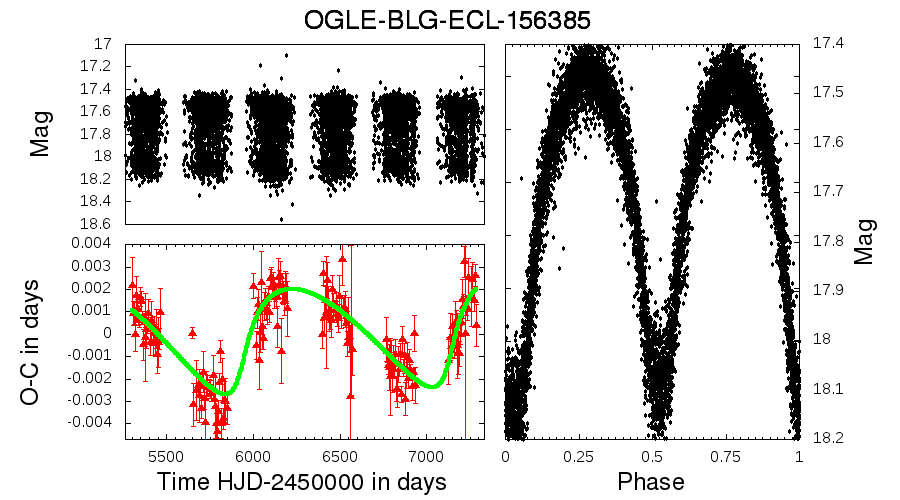}                           
\includegraphics[width=\columnwidth]{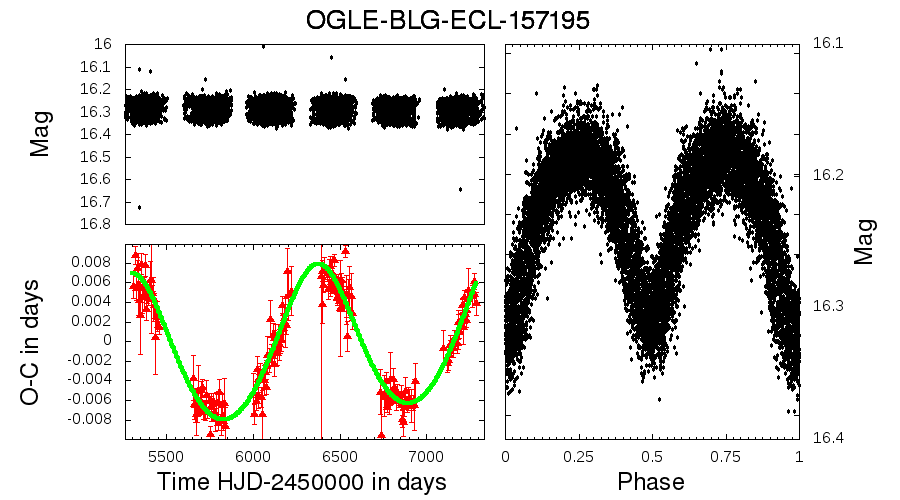}                           
\end{figure*}                           
\clearpage                           
                           
\begin{figure*}                           
                           
\includegraphics[width=\columnwidth]{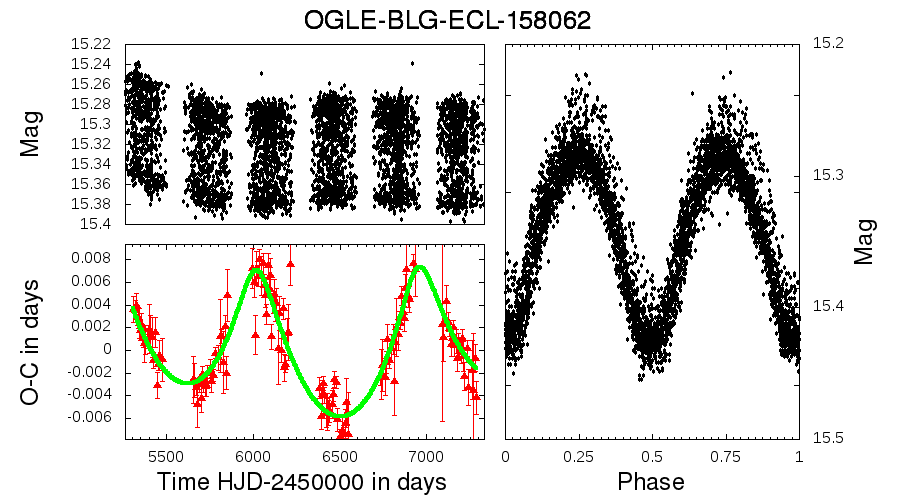}                           
\includegraphics[width=\columnwidth]{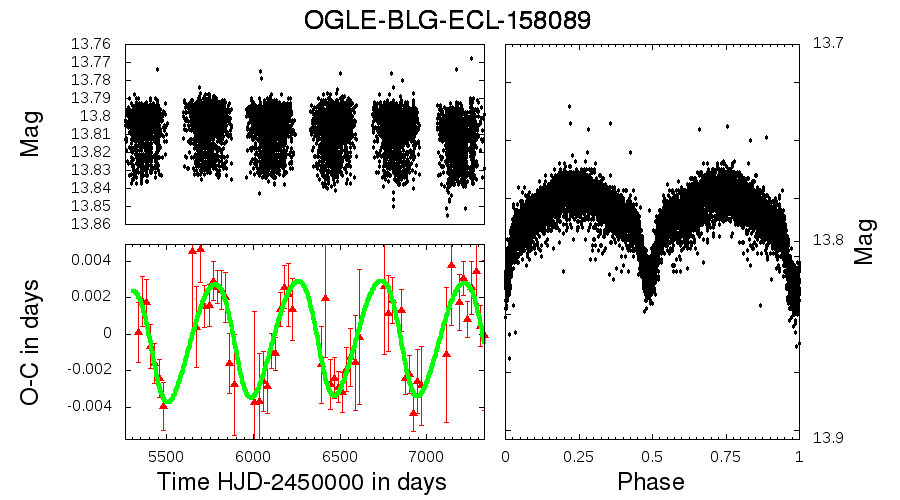}                           
                           
\includegraphics[width=\columnwidth]{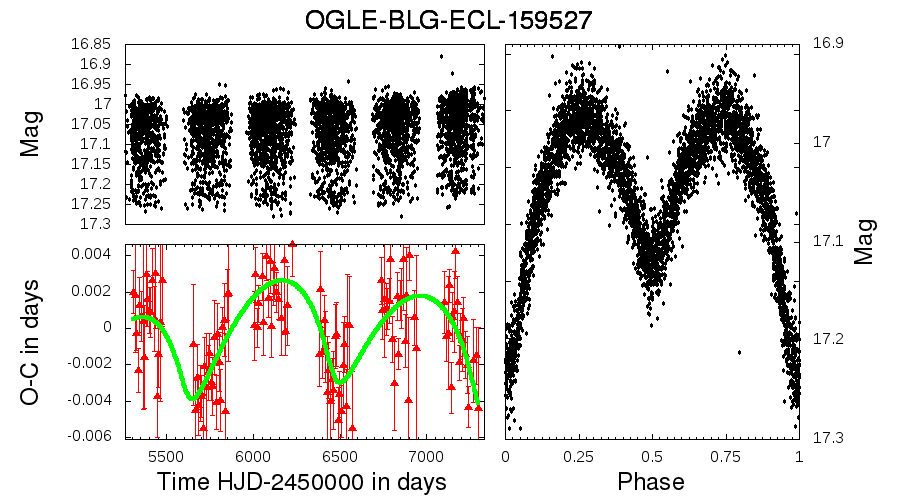}                           
\includegraphics[width=\columnwidth]{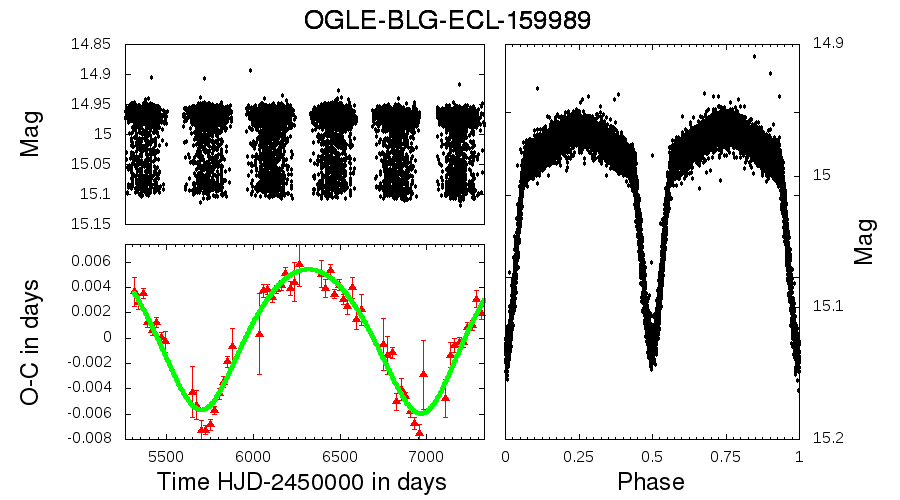}                           
                           
\includegraphics[width=\columnwidth]{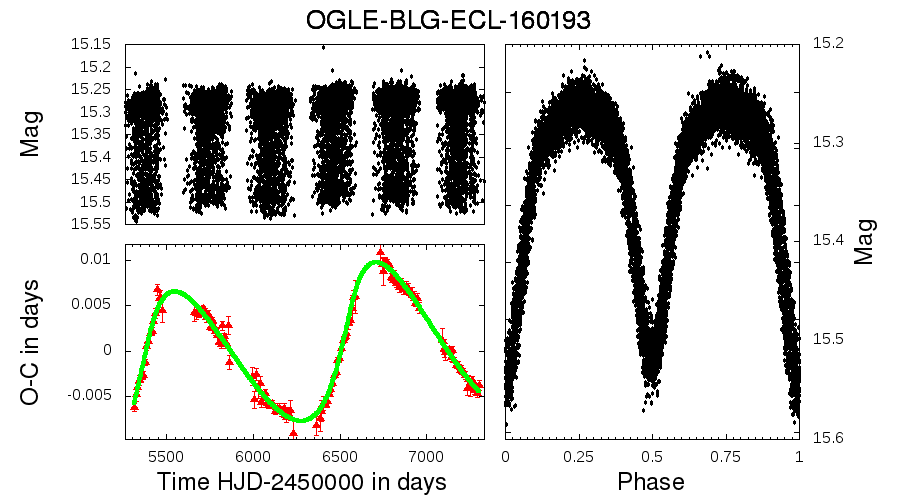}                           
\includegraphics[width=\columnwidth]{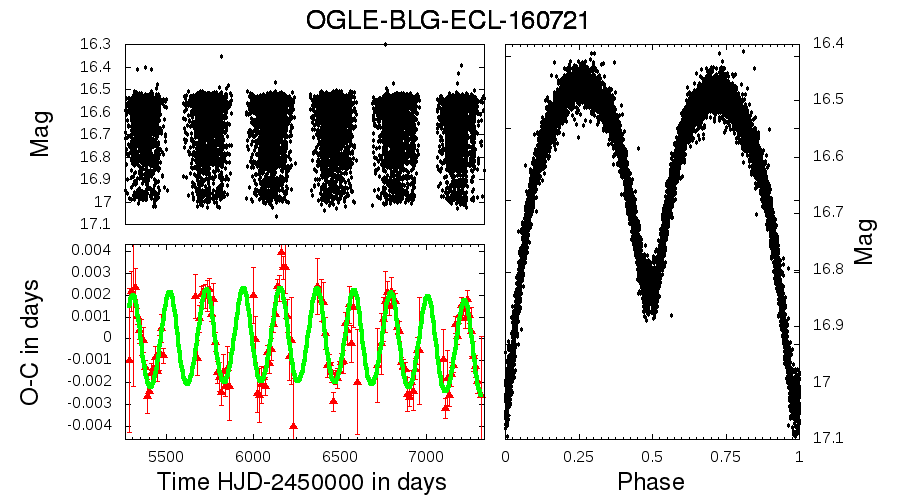}                           
                           
\includegraphics[width=\columnwidth]{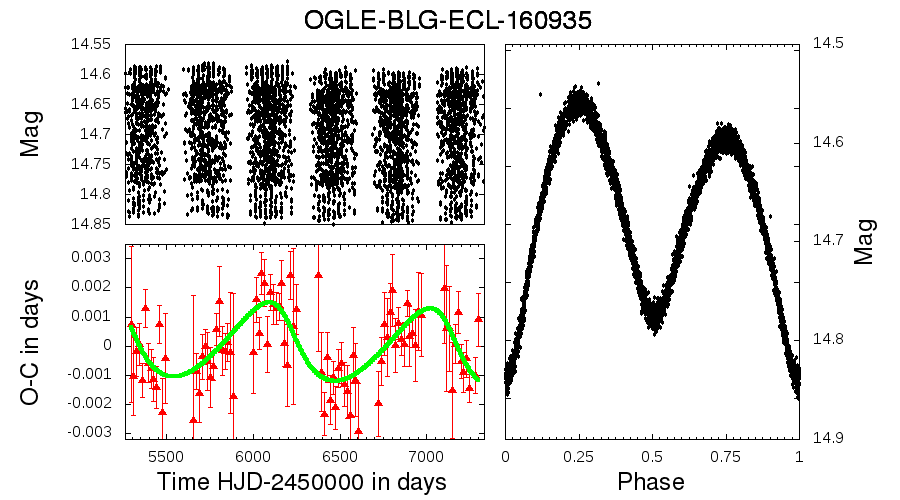}                           
\includegraphics[width=\columnwidth]{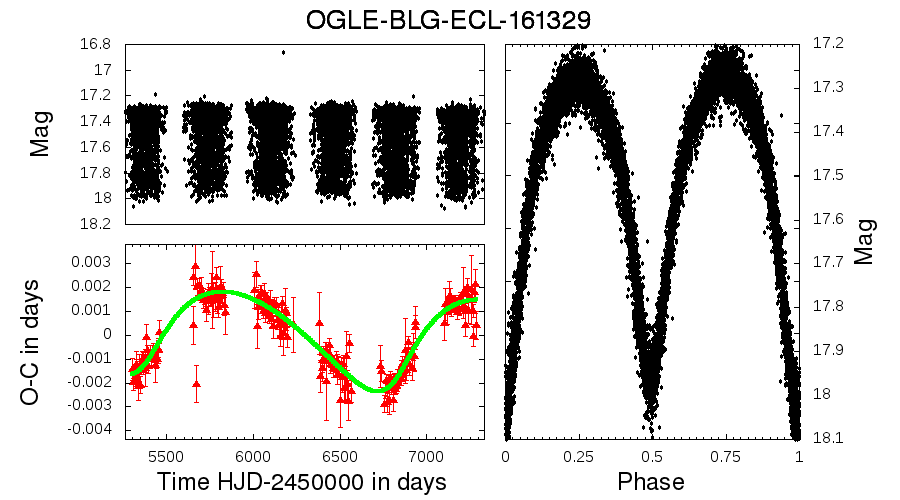}                           
                           
\includegraphics[width=\columnwidth]{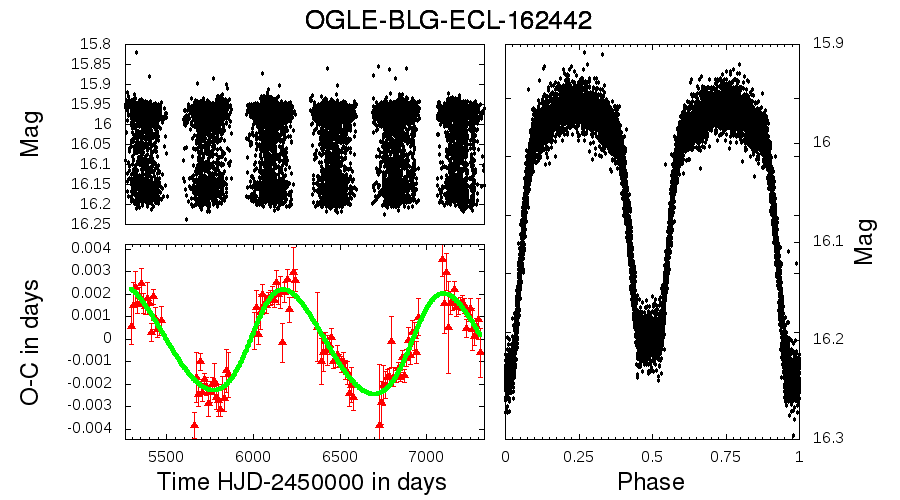}                           
\includegraphics[width=\columnwidth]{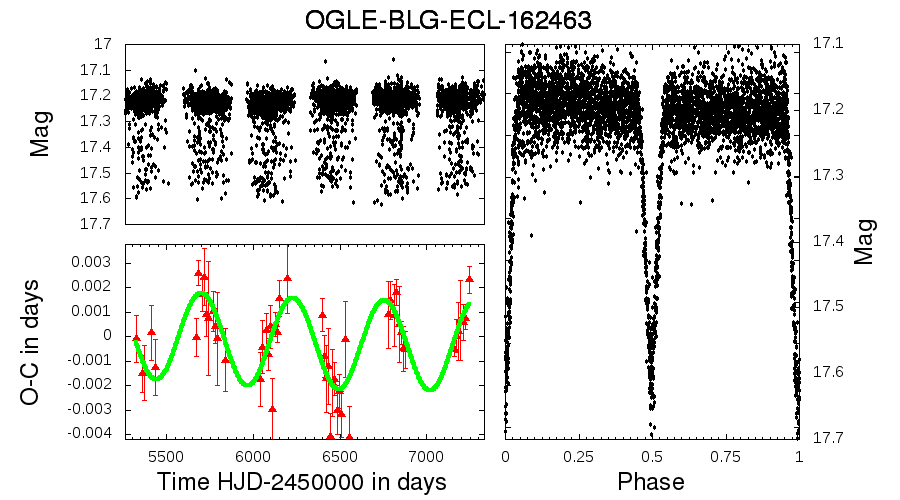}                           
\end{figure*}                           
\clearpage                           
                           
\begin{figure*}                           
                           
\includegraphics[width=\columnwidth]{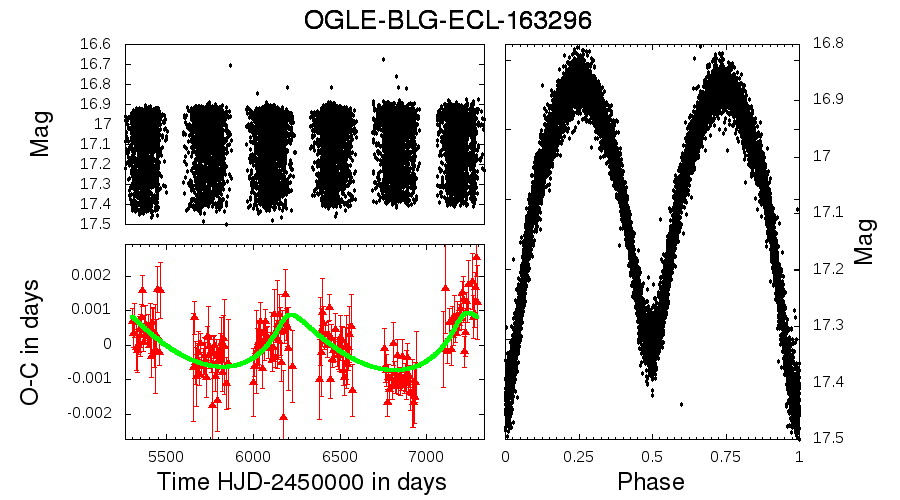}                           
\includegraphics[width=\columnwidth]{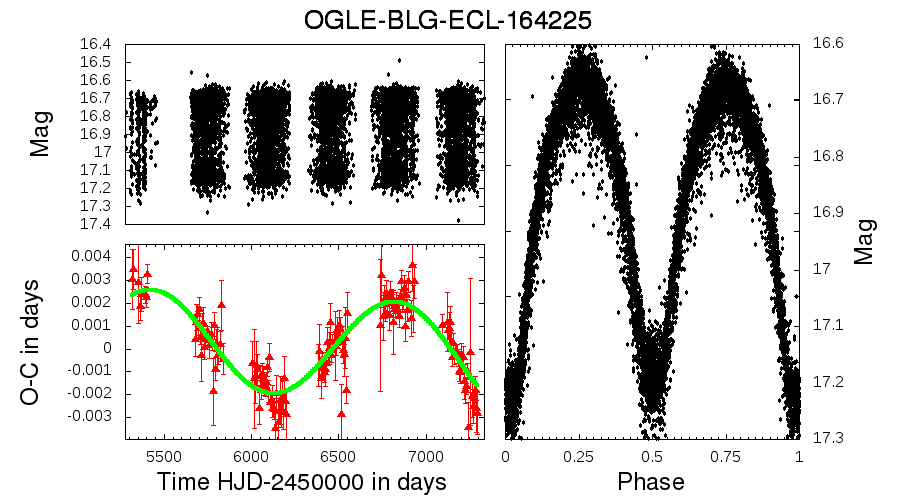}                           
                           
\includegraphics[width=\columnwidth]{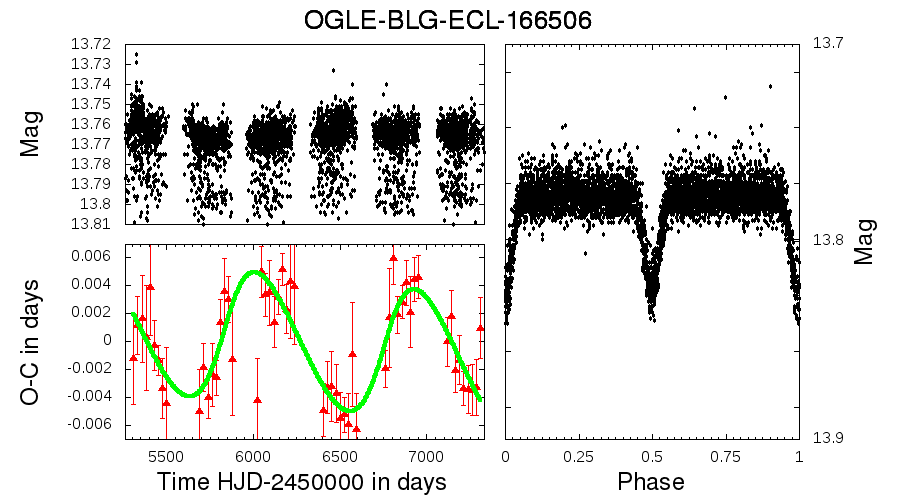}                           
\includegraphics[width=\columnwidth]{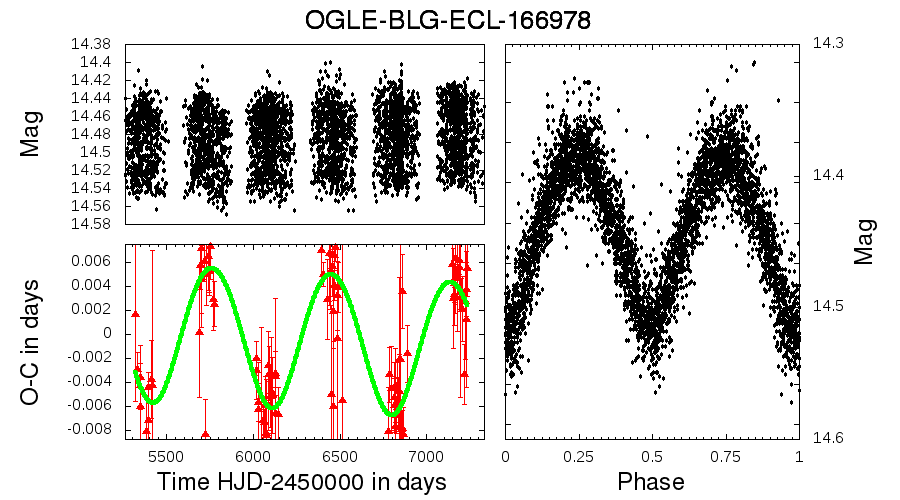}                           
                           
\includegraphics[width=\columnwidth]{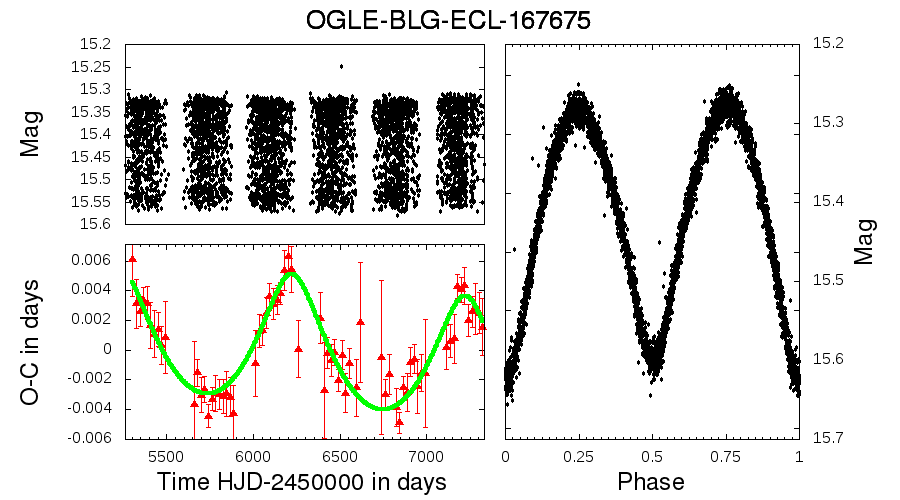}                           
\includegraphics[width=\columnwidth]{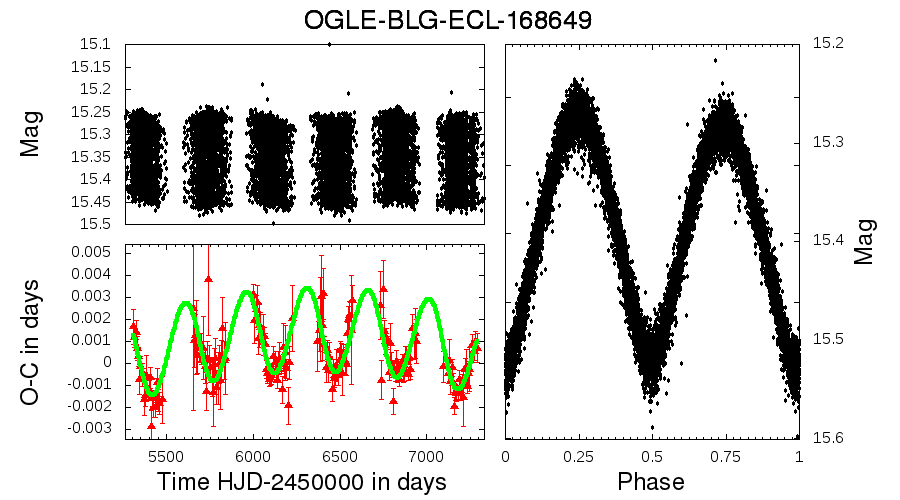}                           
                           
\includegraphics[width=\columnwidth]{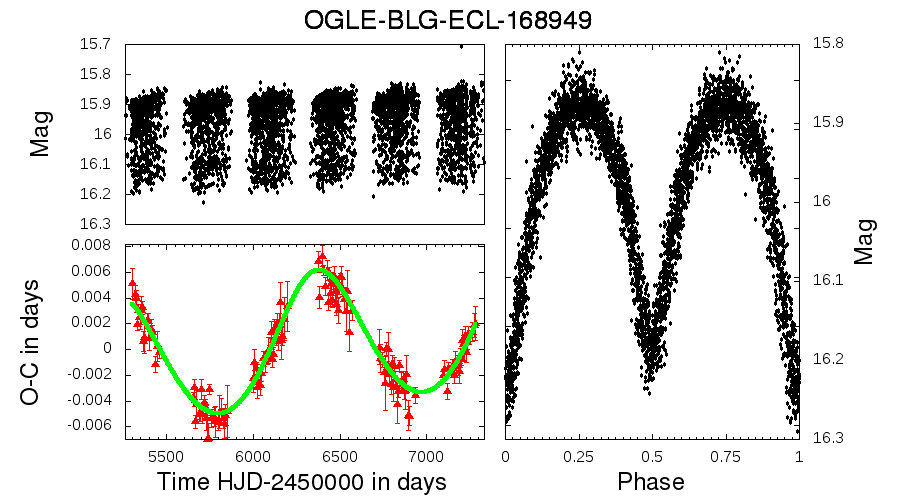}                           
\includegraphics[width=\columnwidth]{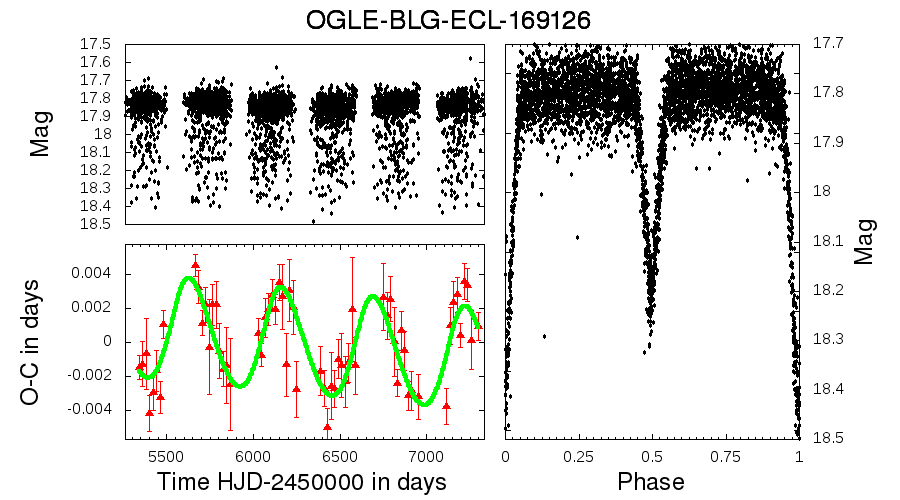}                           
                           
\includegraphics[width=\columnwidth]{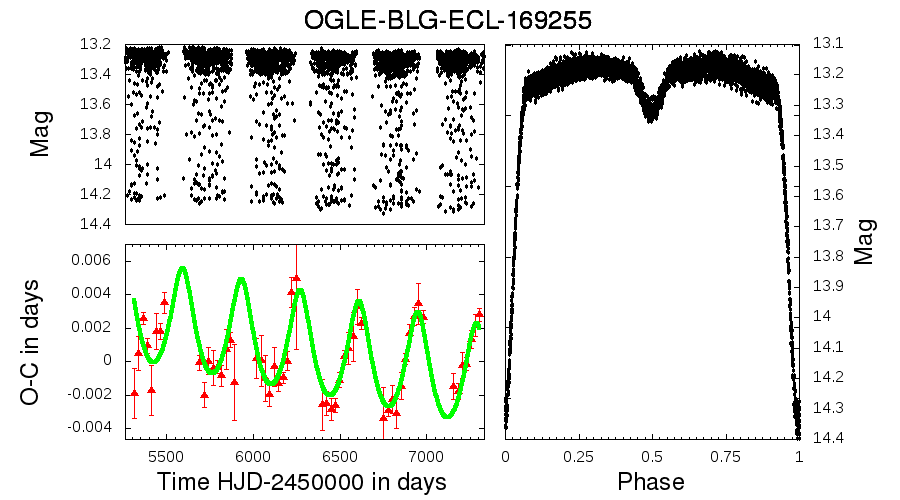}                           
\includegraphics[width=\columnwidth]{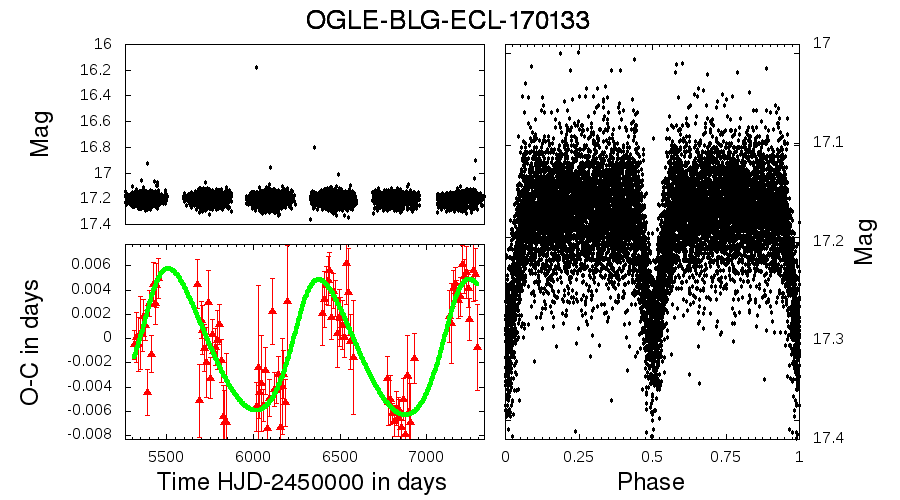}                           
\end{figure*}                           
\clearpage                           
                           
\begin{figure*}                           
                           
\includegraphics[width=\columnwidth]{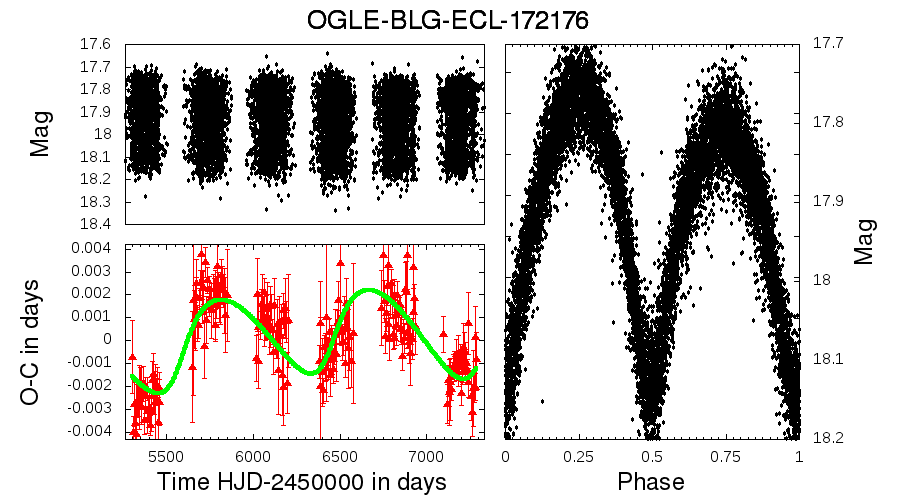}                           
\includegraphics[width=\columnwidth]{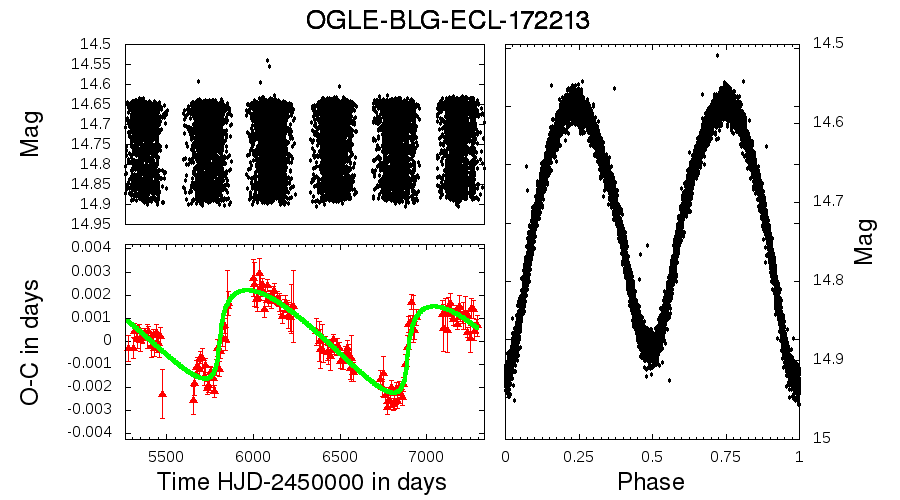}                           
                           
\includegraphics[width=\columnwidth]{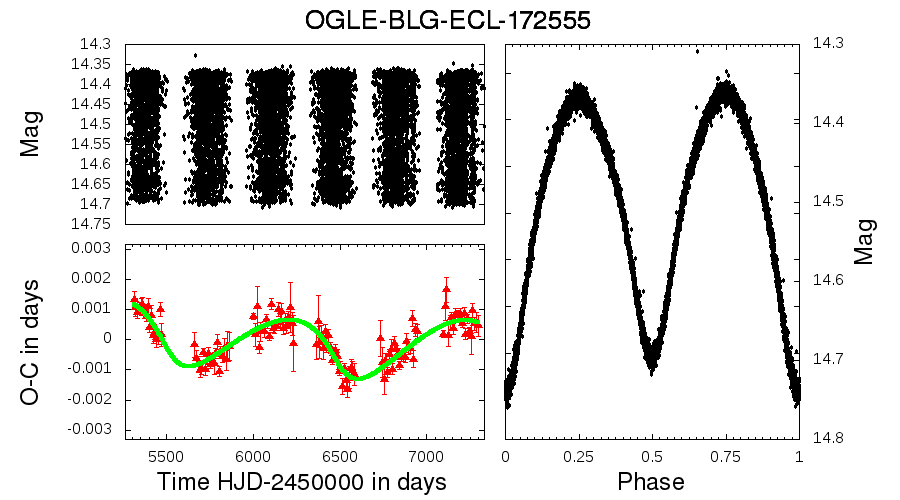}                           
\includegraphics[width=\columnwidth]{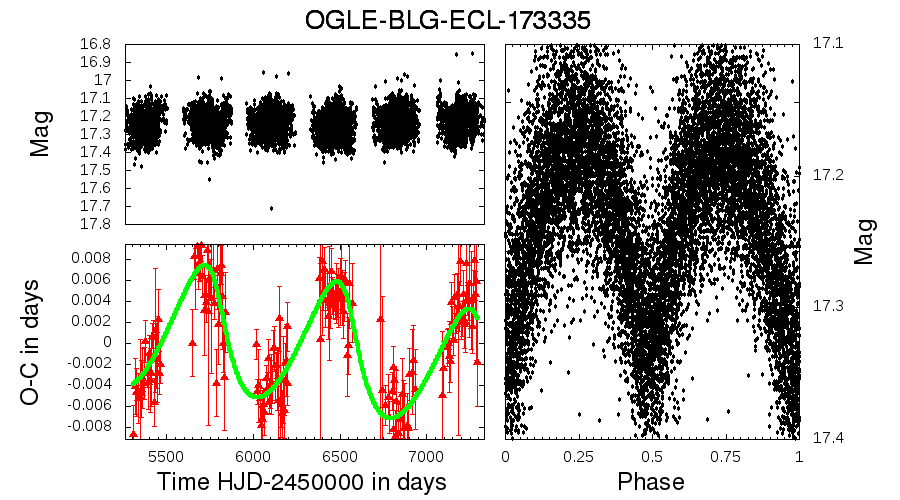}                           
                           
\includegraphics[width=\columnwidth]{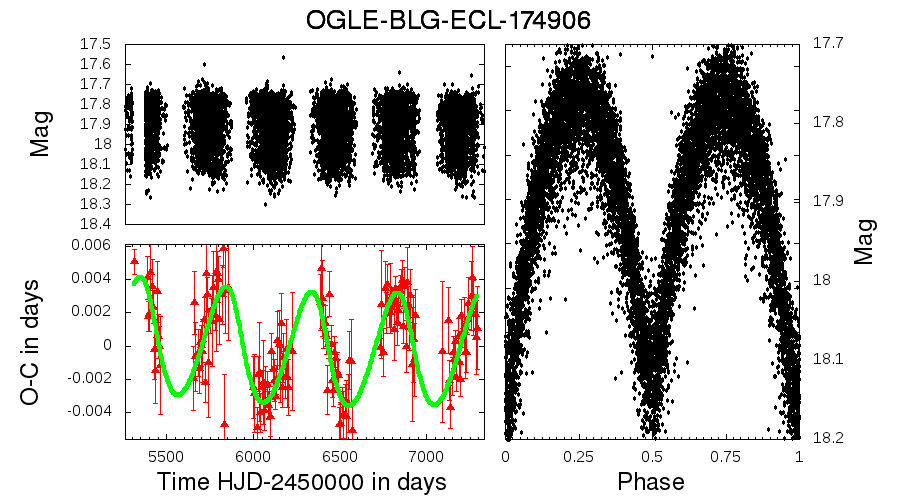}                           
\includegraphics[width=\columnwidth]{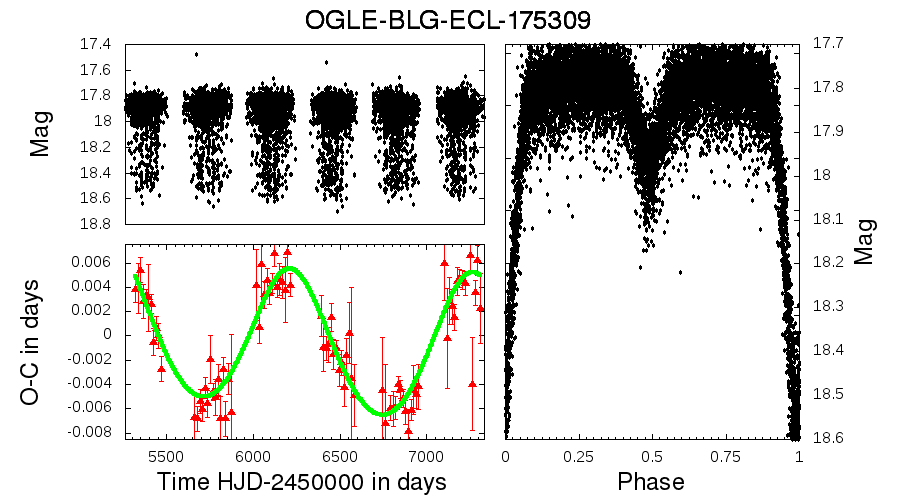}                           
                           
\includegraphics[width=\columnwidth]{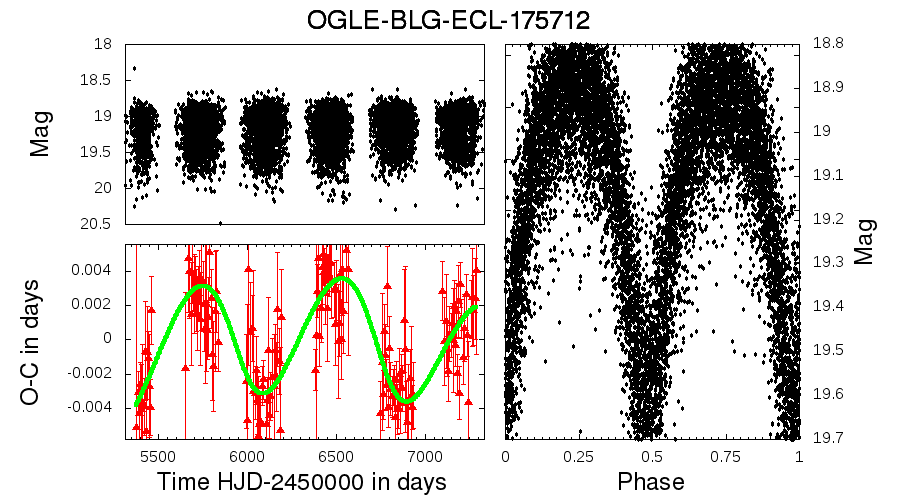}                           
\includegraphics[width=\columnwidth]{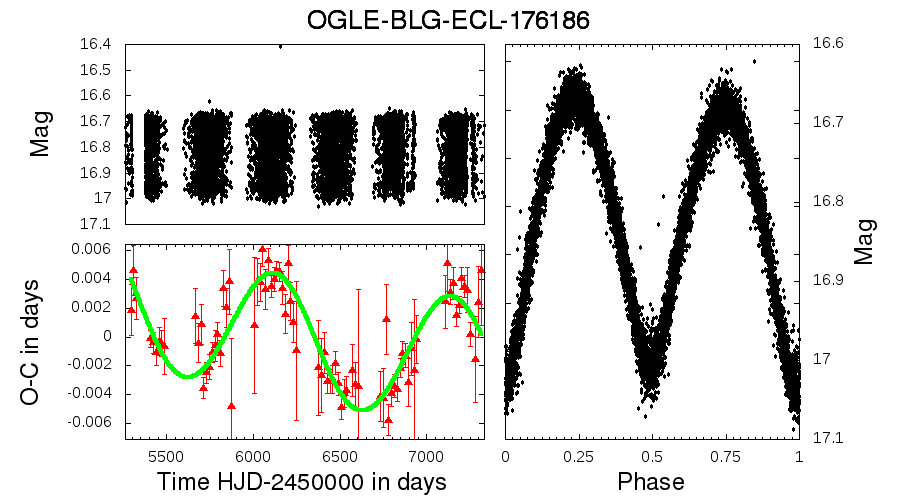}                           
                           
\includegraphics[width=\columnwidth]{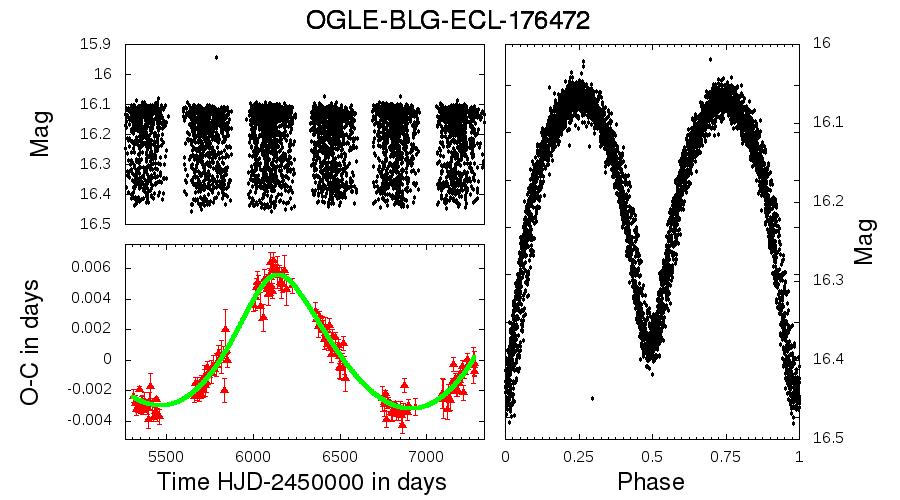}                           
\includegraphics[width=\columnwidth]{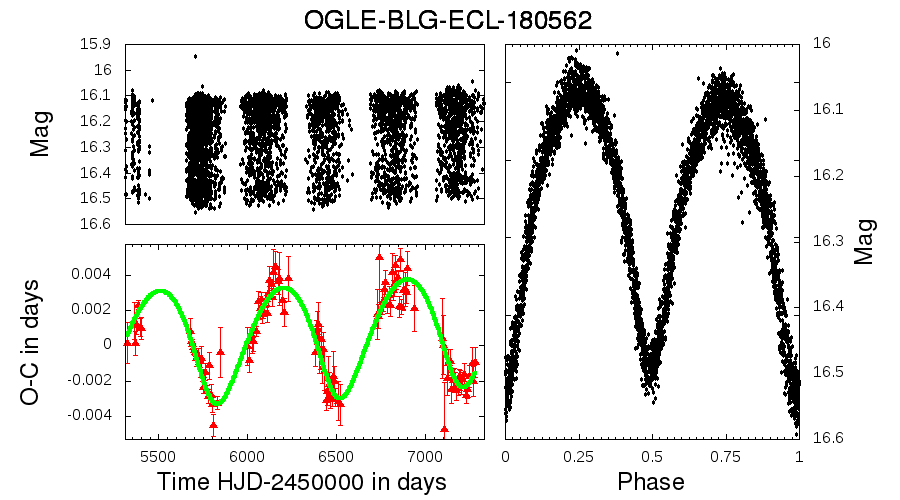}                           
\end{figure*}                           
\clearpage                           
                           
\begin{figure*}                           
                           
\includegraphics[width=\columnwidth]{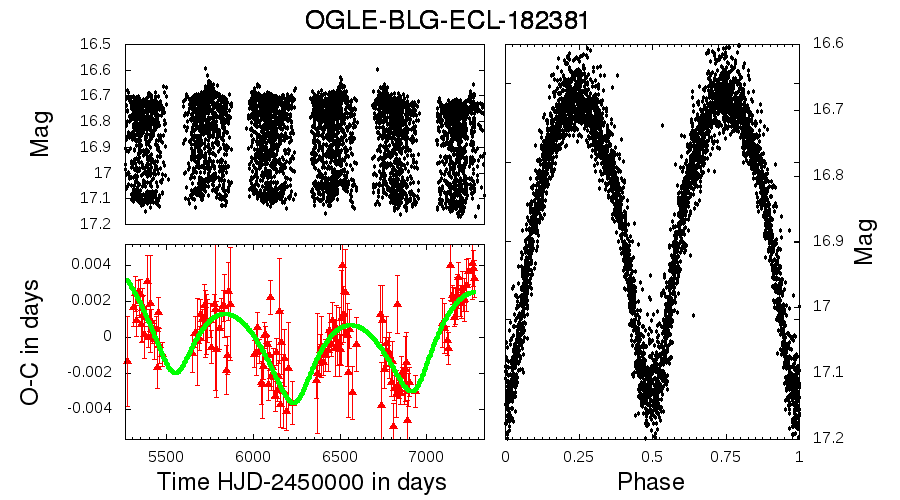}                           
\includegraphics[width=\columnwidth]{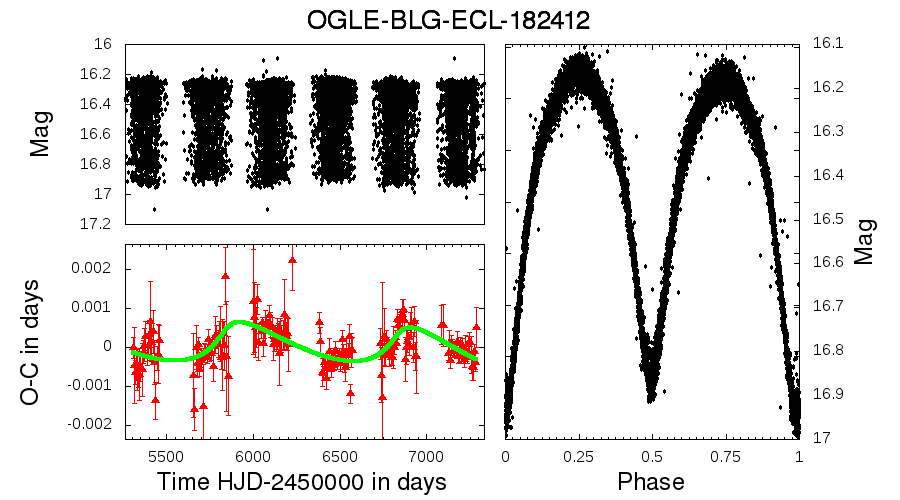}                           
                           
\includegraphics[width=\columnwidth]{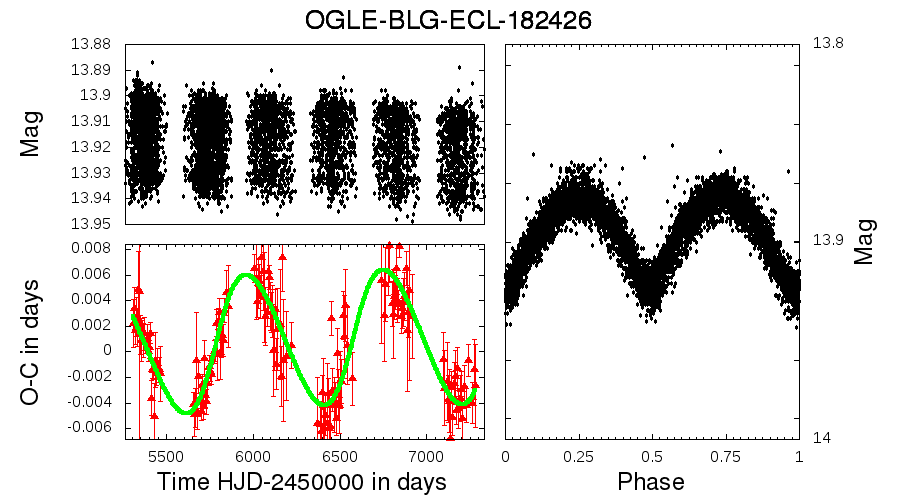}                           
\includegraphics[width=\columnwidth]{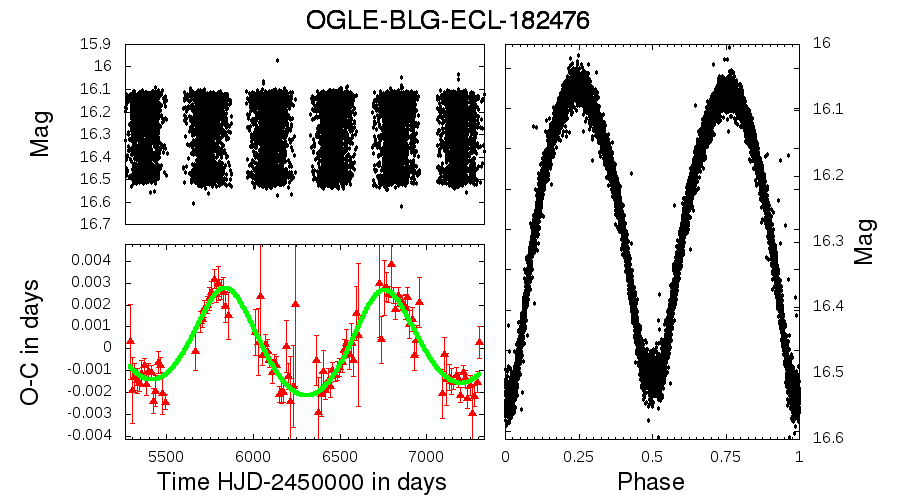}                           
                           
\includegraphics[width=\columnwidth]{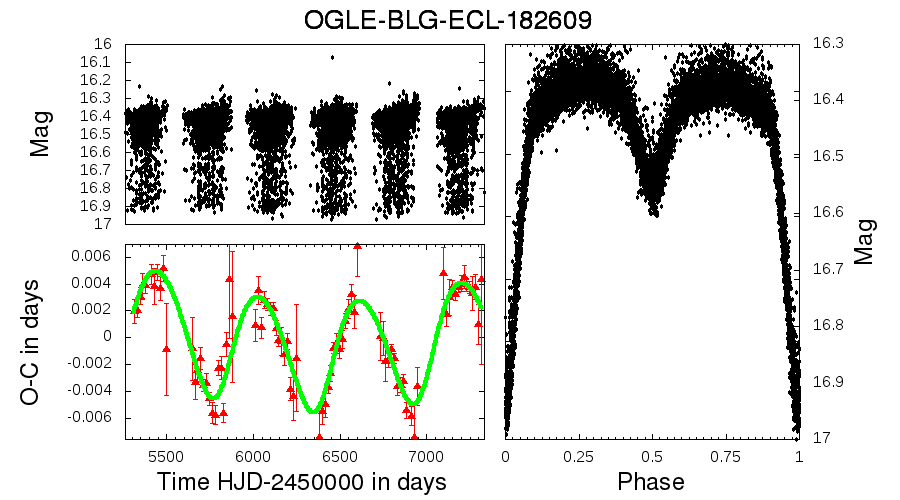}                           
\includegraphics[width=\columnwidth]{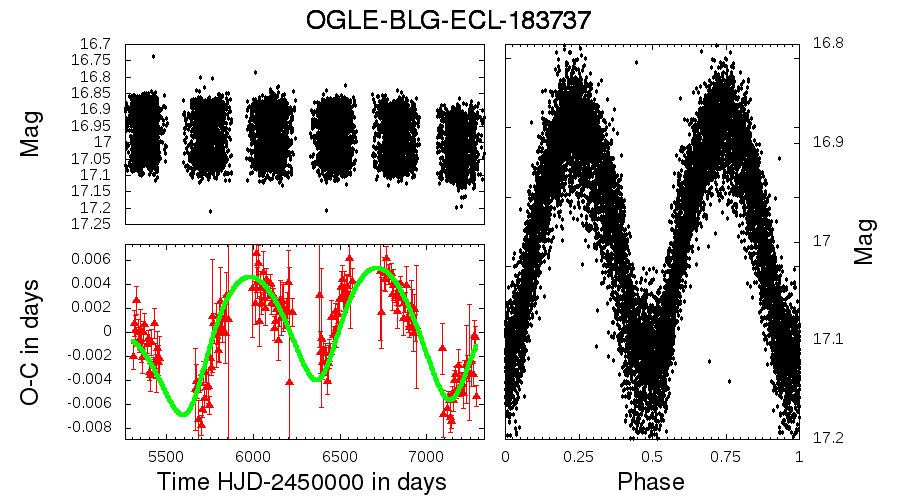}                           
                           
\includegraphics[width=\columnwidth]{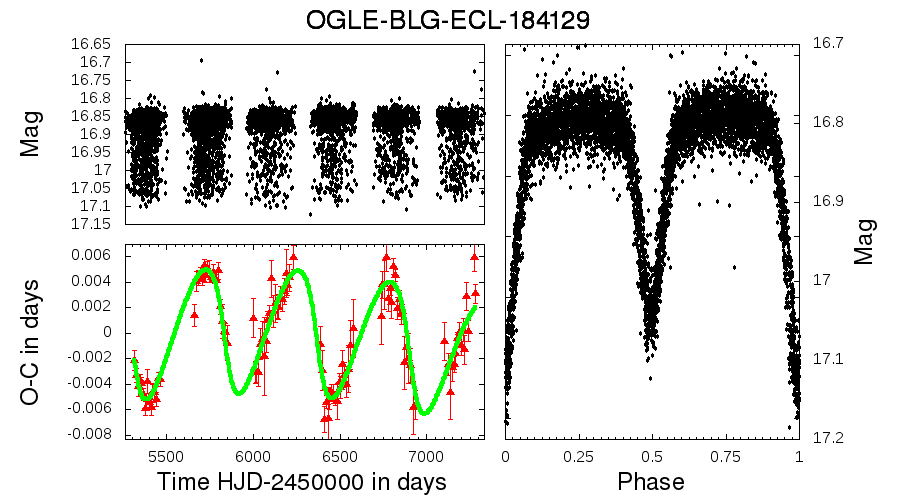}                           
\includegraphics[width=\columnwidth]{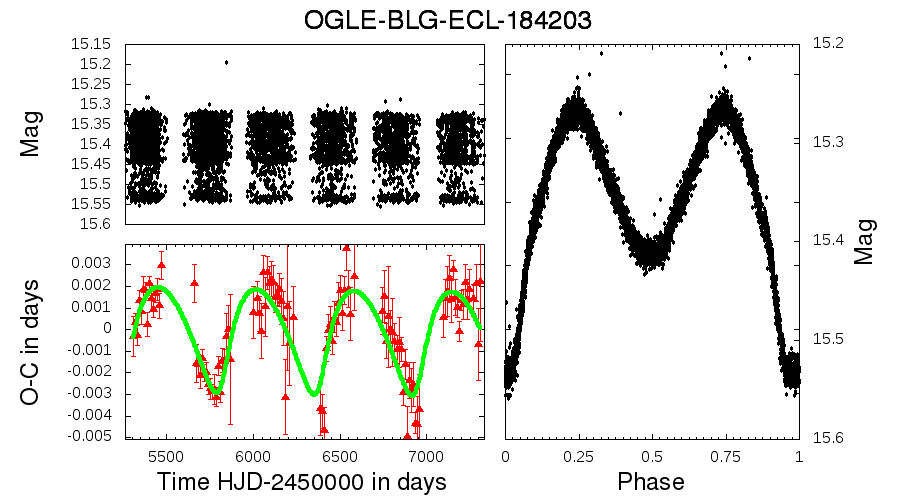}                           
                           
\includegraphics[width=\columnwidth]{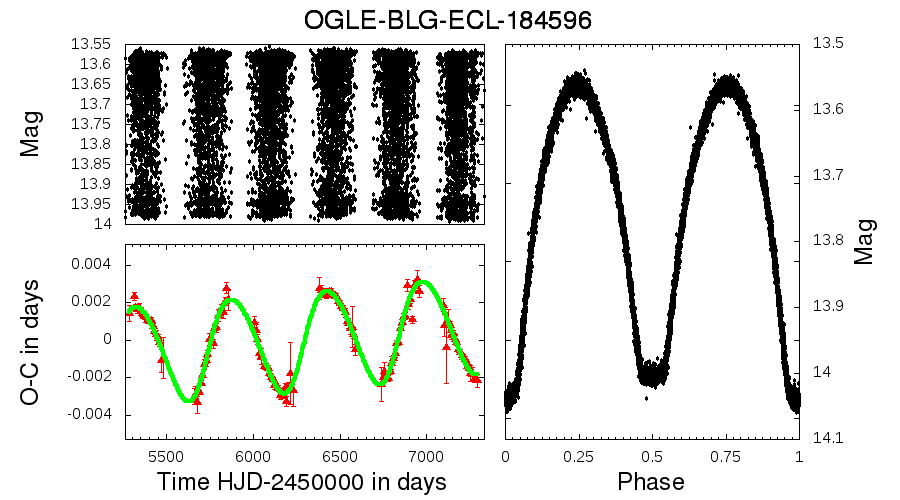}                           
\includegraphics[width=\columnwidth]{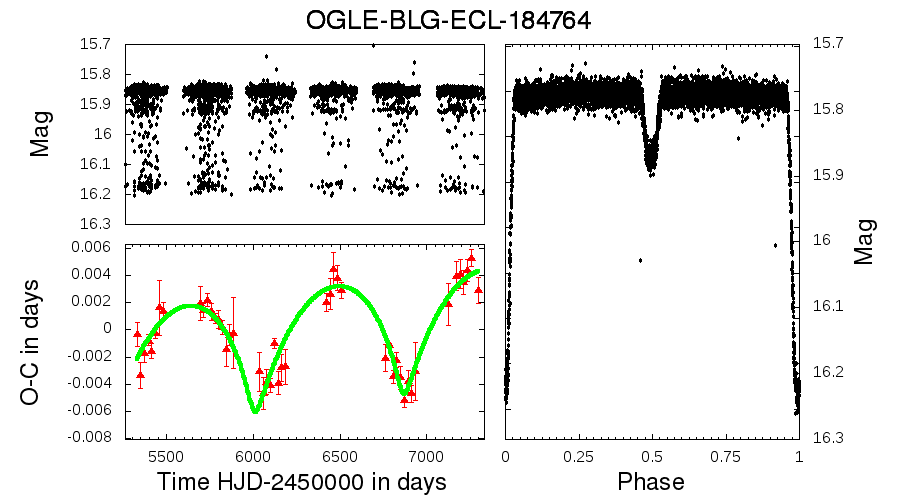}                           
\end{figure*}                           
\clearpage                           
                           
\begin{figure*}                           
                           
\includegraphics[width=\columnwidth]{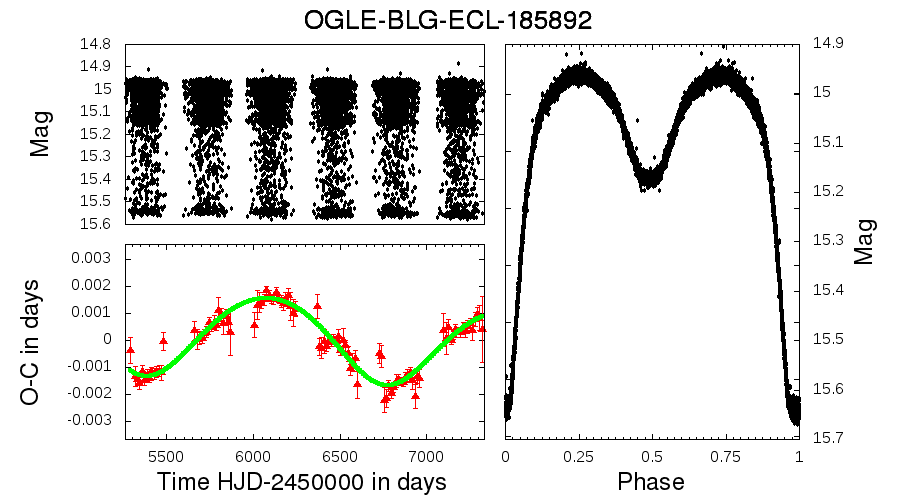}                           
\includegraphics[width=\columnwidth]{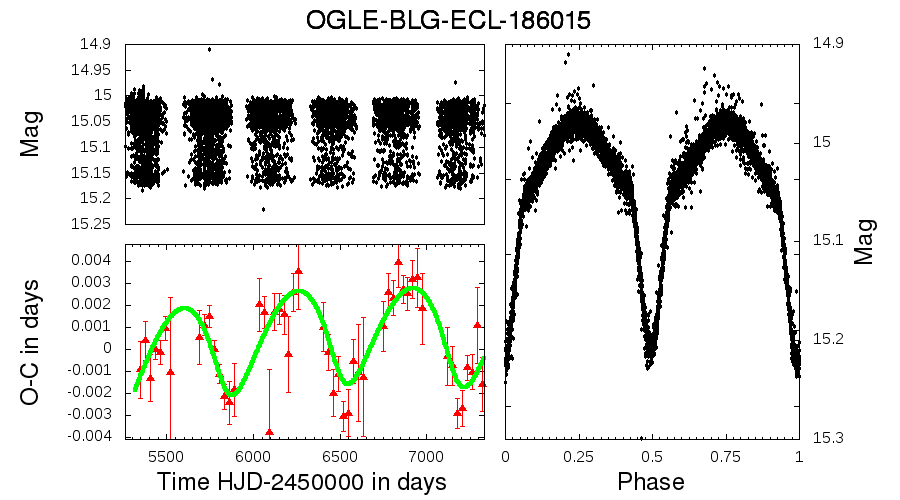}                           
                           
\includegraphics[width=\columnwidth]{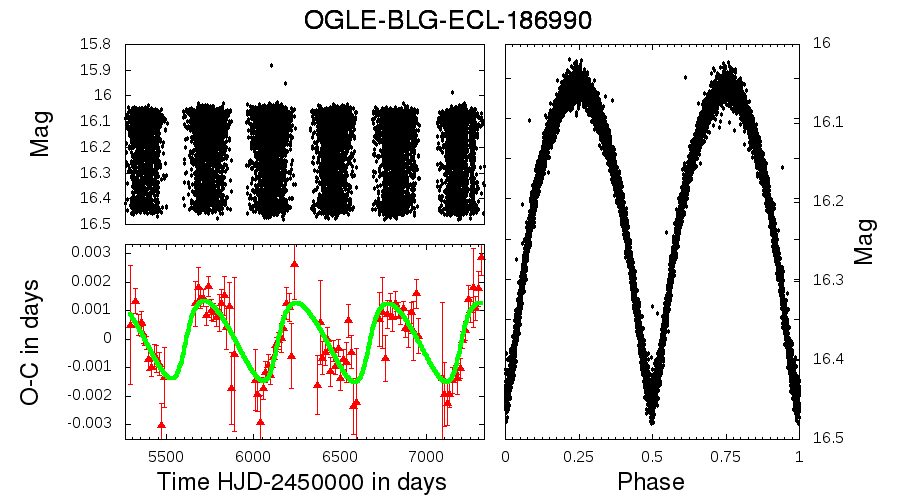}                           
\includegraphics[width=\columnwidth]{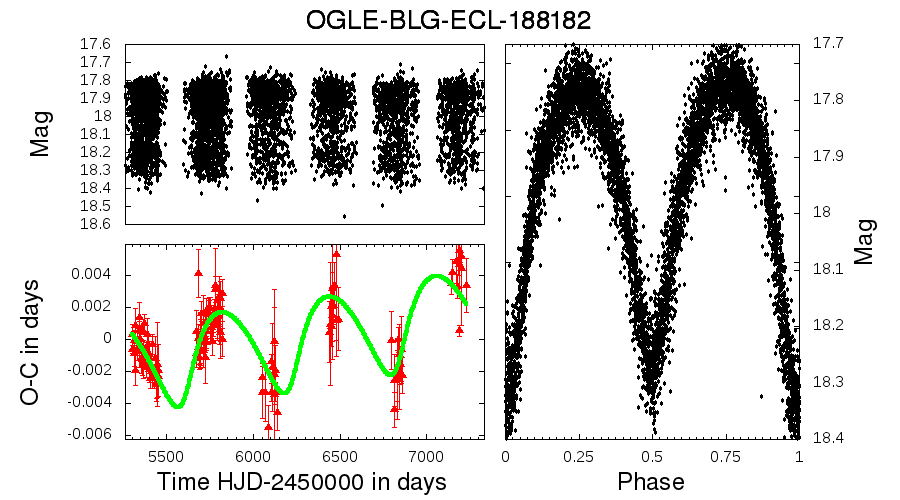}                           
                           
\includegraphics[width=\columnwidth]{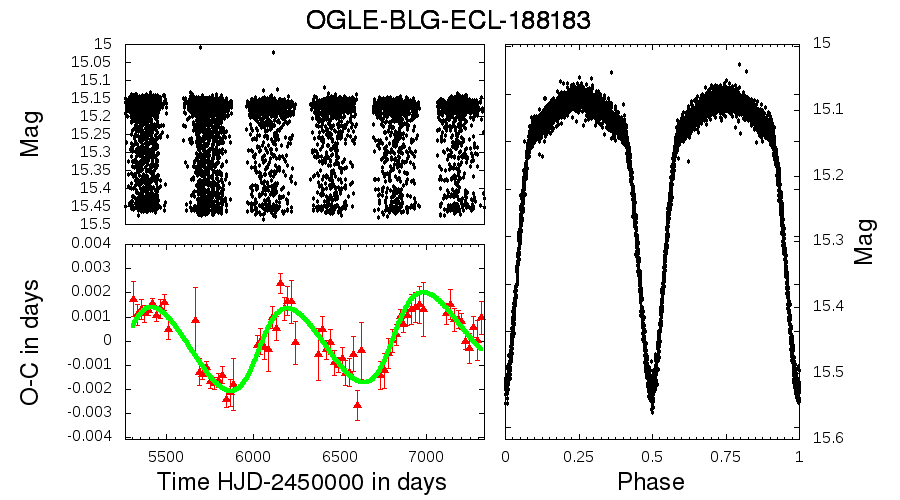}                           
\includegraphics[width=\columnwidth]{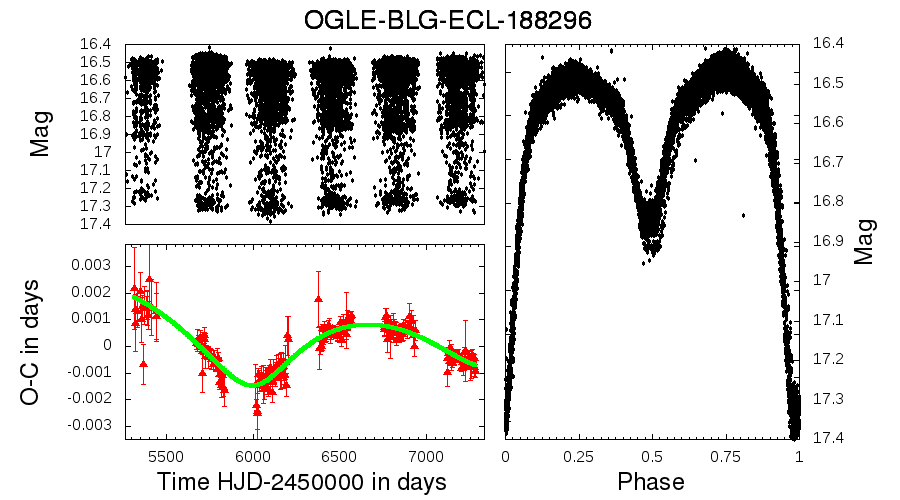}                           
                           
\includegraphics[width=\columnwidth]{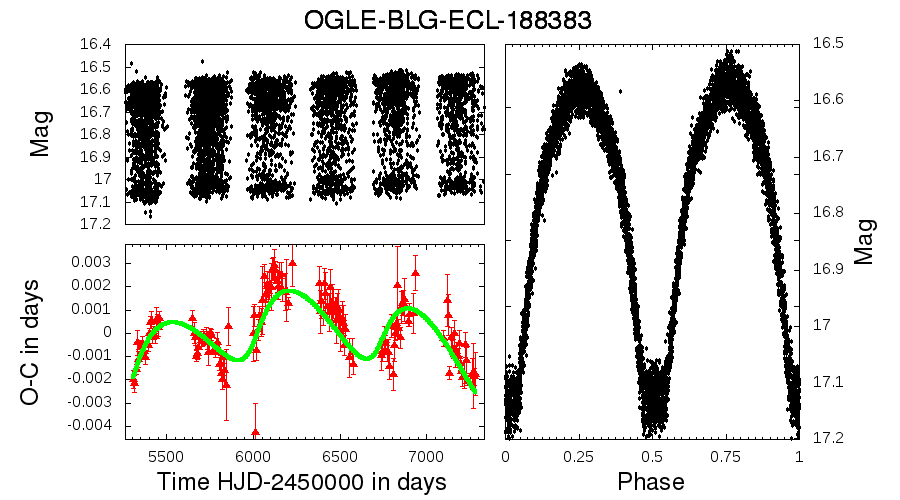}                           
\includegraphics[width=\columnwidth]{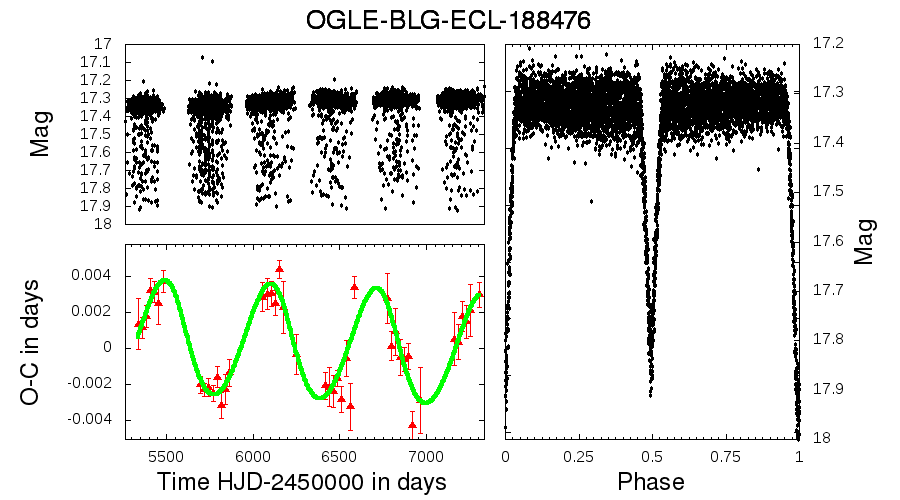}                           
                           
\includegraphics[width=\columnwidth]{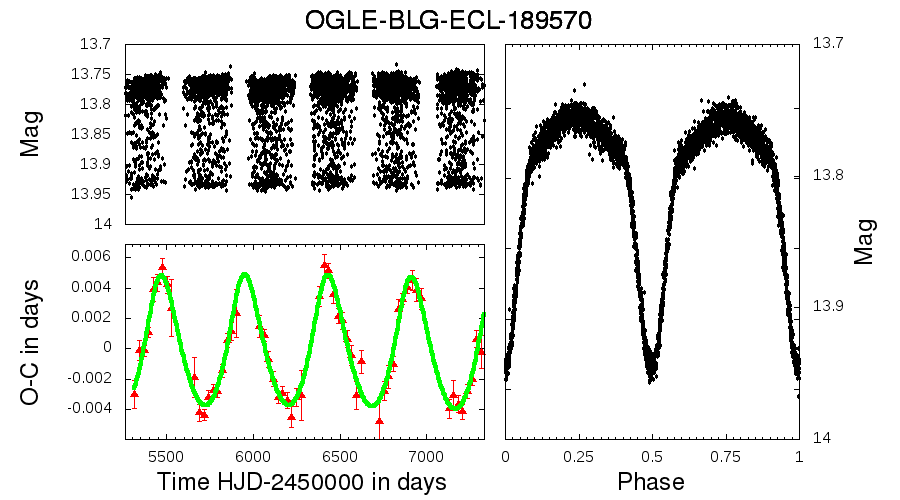}                           
\includegraphics[width=\columnwidth]{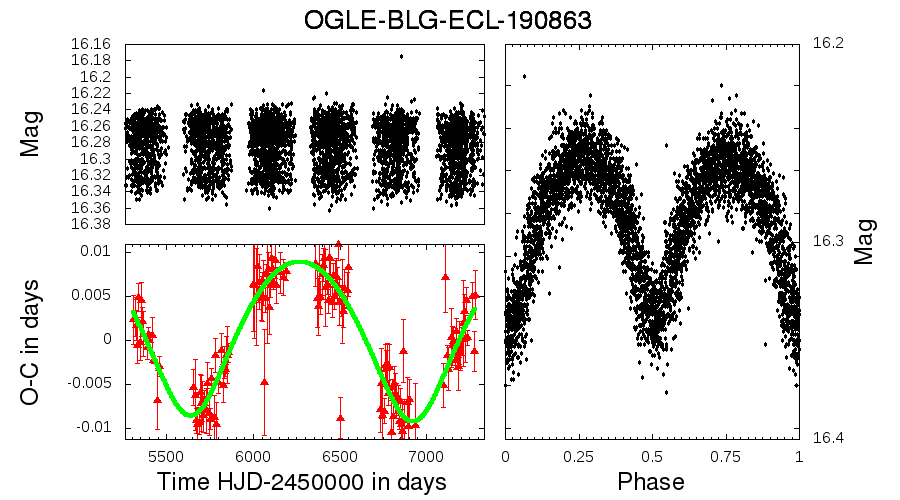}                           
\end{figure*}                           
\clearpage                           
                           
\begin{figure*}                           
                           
\includegraphics[width=\columnwidth]{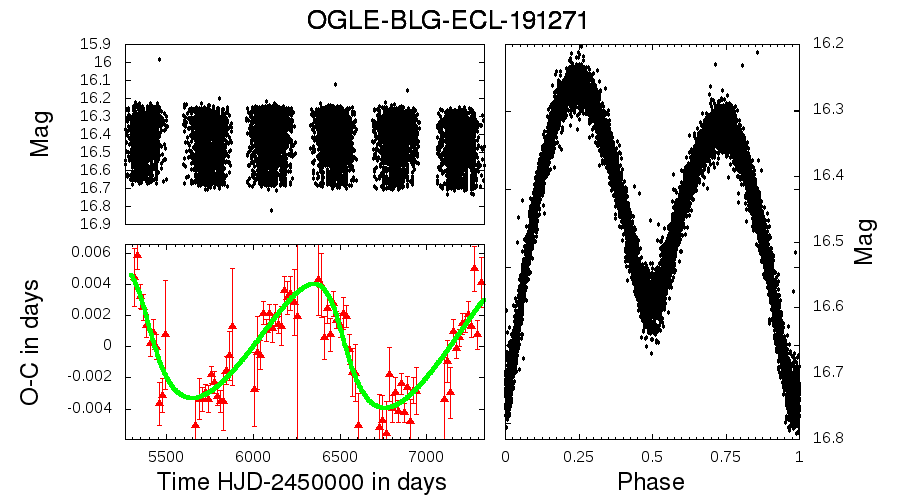}                           
\includegraphics[width=\columnwidth]{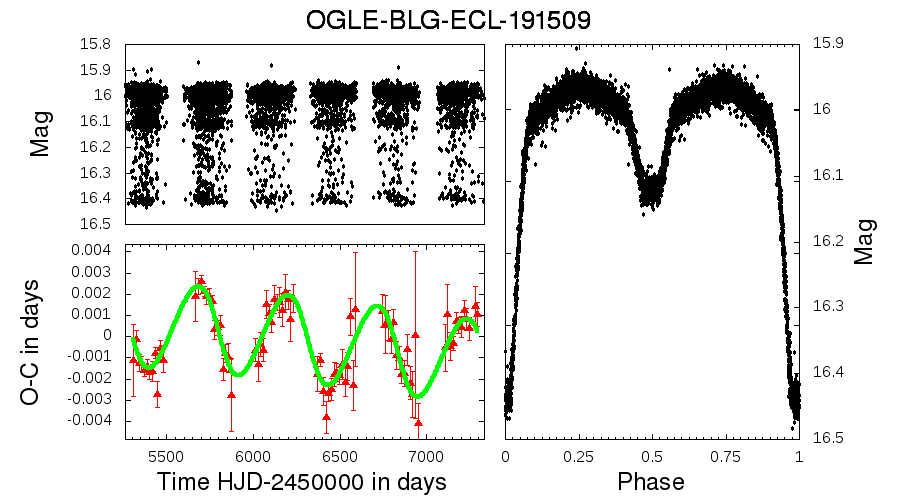}                           
                           
\includegraphics[width=\columnwidth]{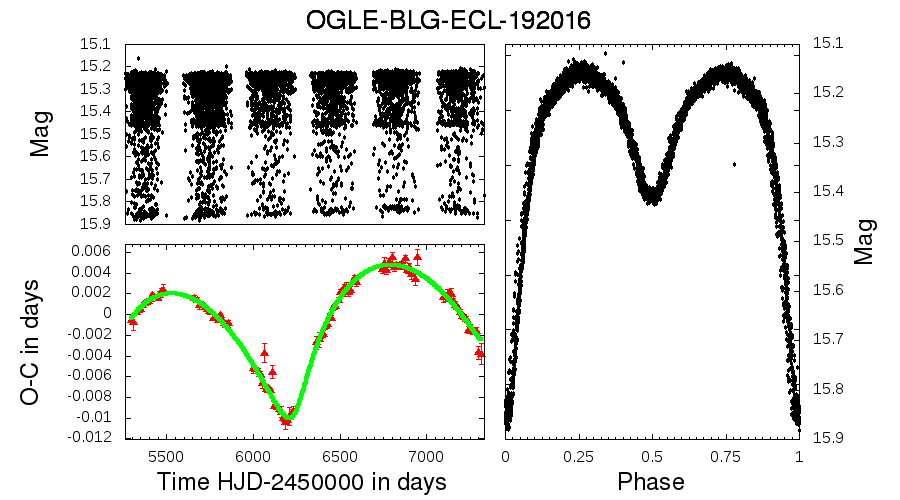}                           
\includegraphics[width=\columnwidth]{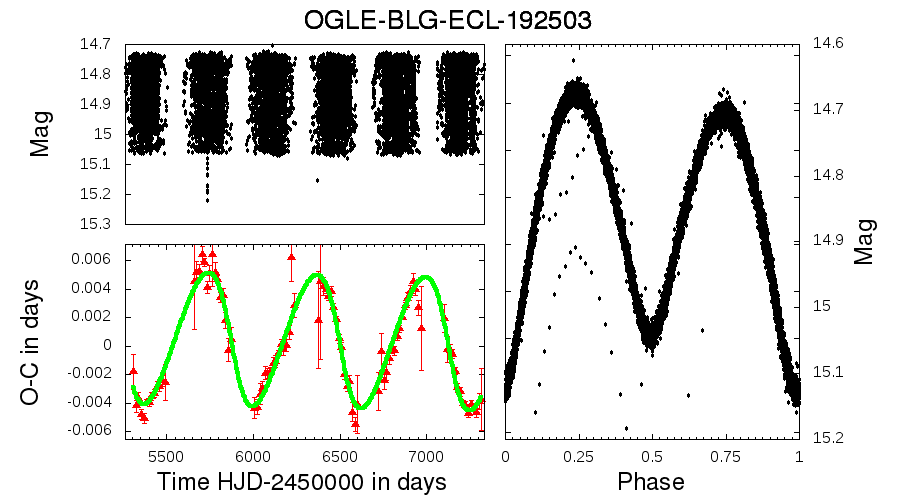}                           
                           
\includegraphics[width=\columnwidth]{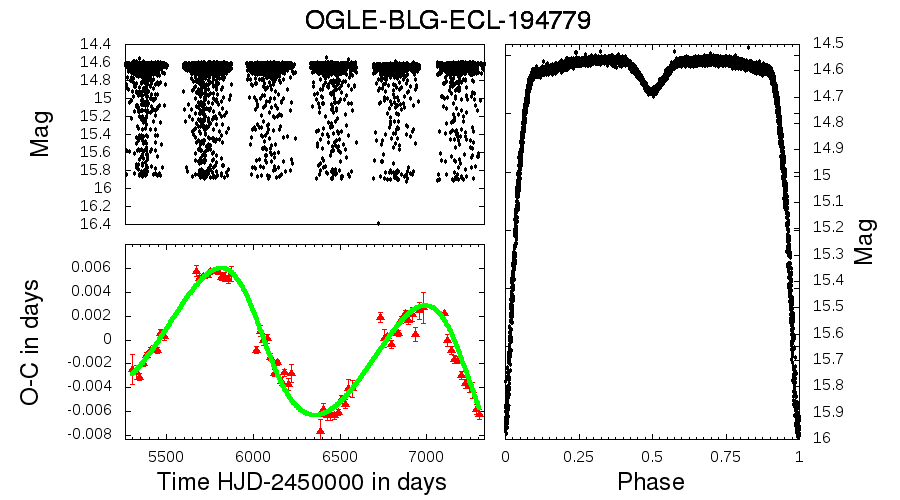}                           
\includegraphics[width=\columnwidth]{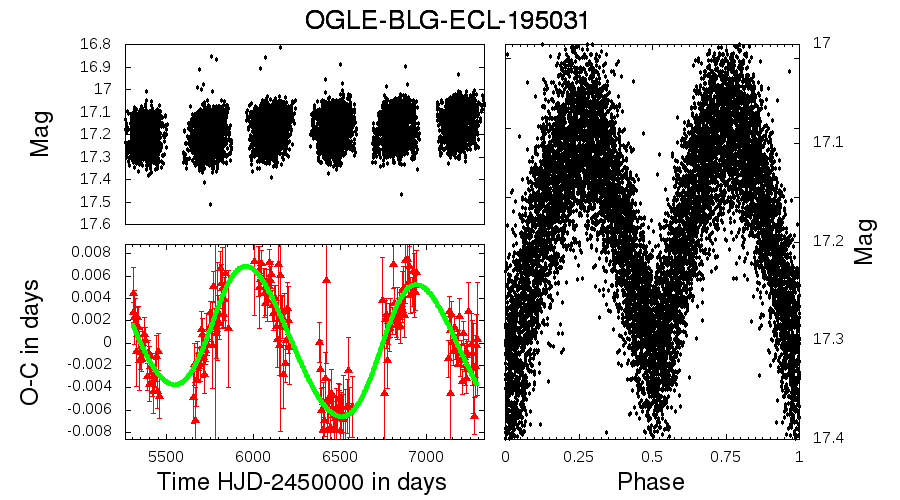}                           
                           
\includegraphics[width=\columnwidth]{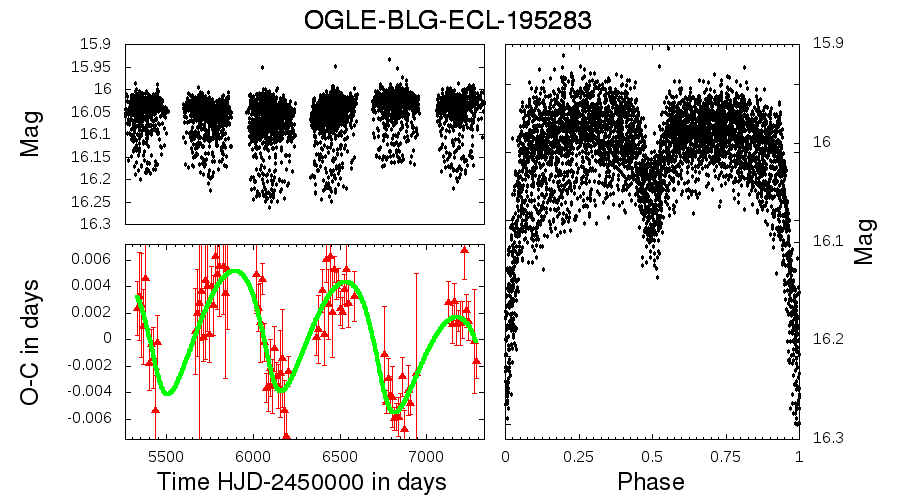}                           
\includegraphics[width=\columnwidth]{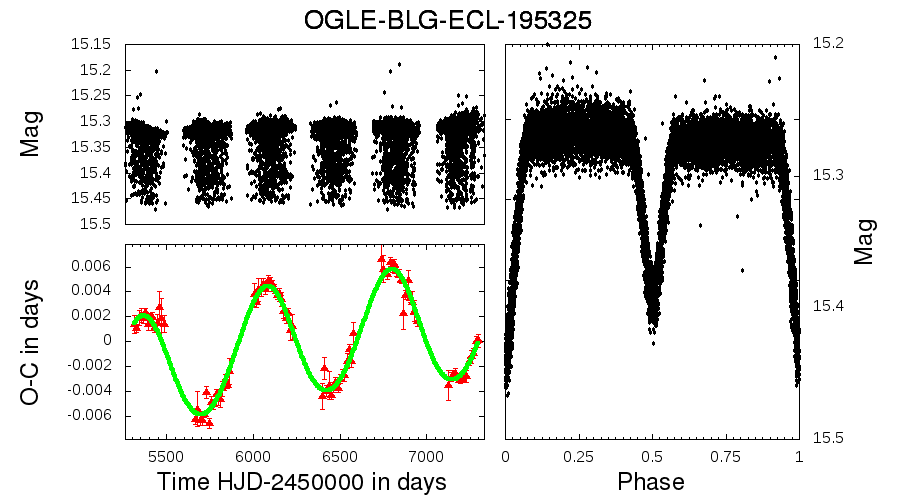}                           
                           
\includegraphics[width=\columnwidth]{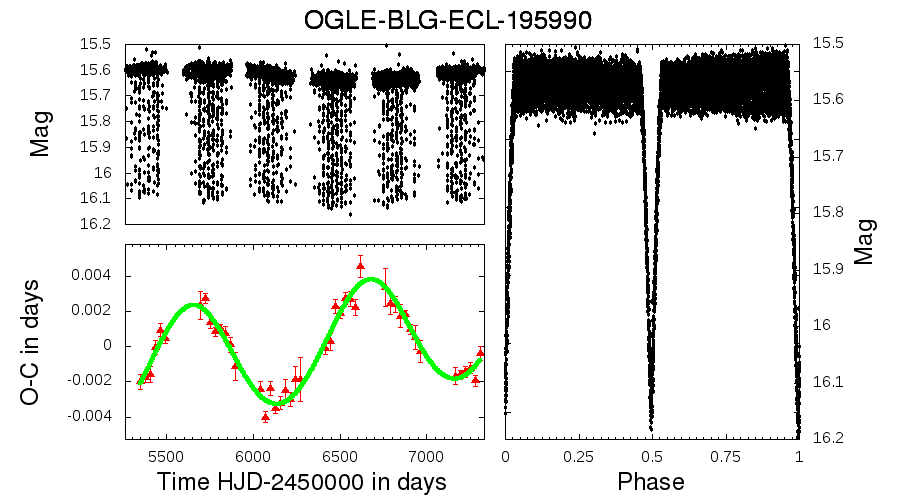}                           
\includegraphics[width=\columnwidth]{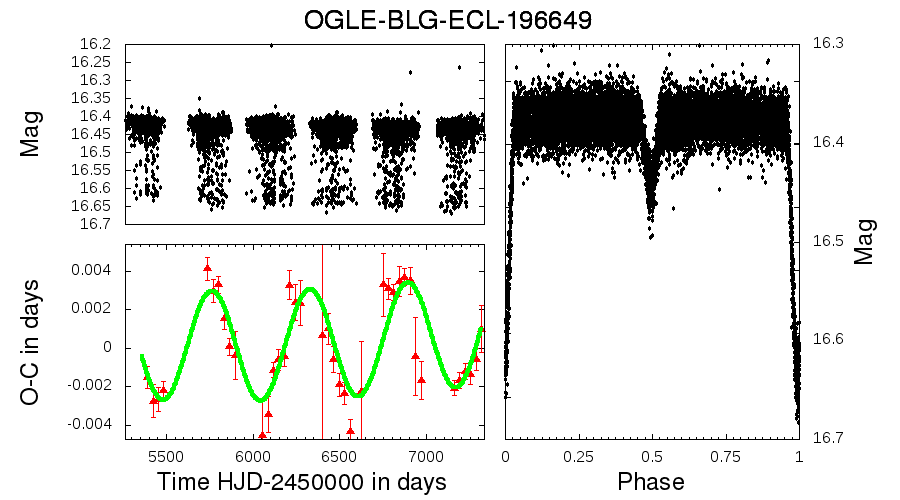}                           
\end{figure*}                           
\clearpage                           
                           
\begin{figure*}                           
                           
\includegraphics[width=\columnwidth]{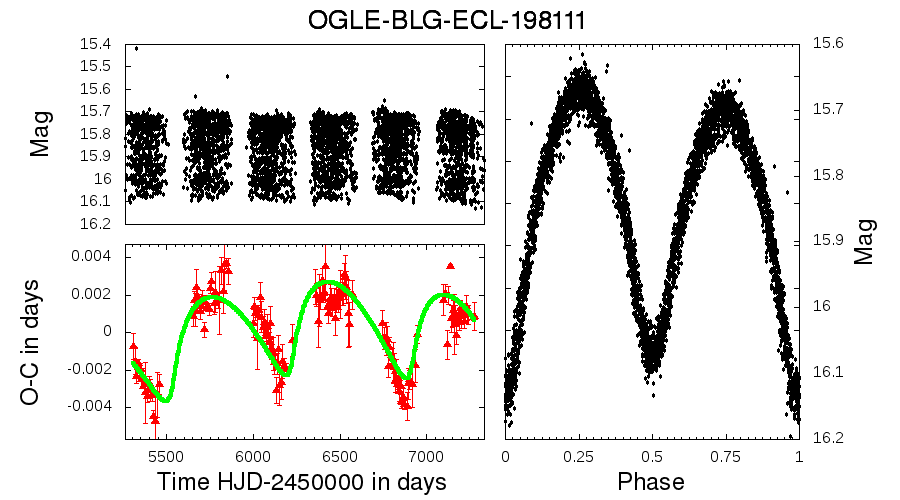}                           
\includegraphics[width=\columnwidth]{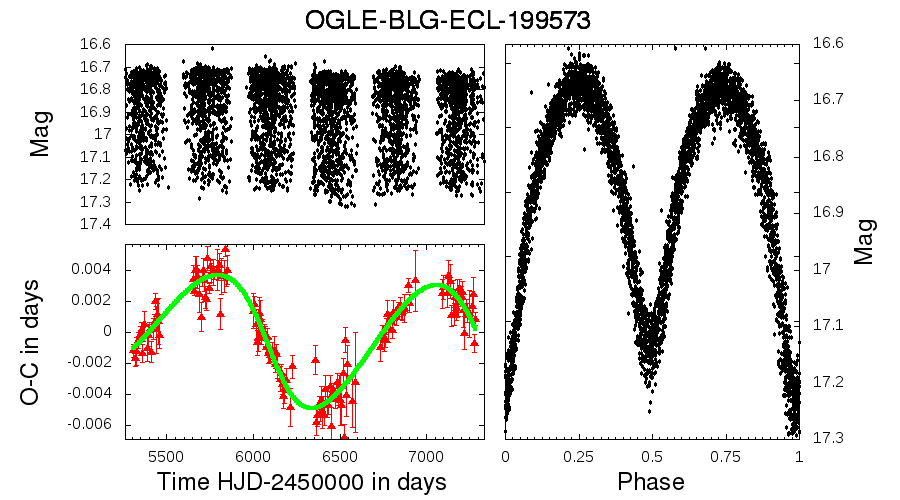}                           
                           
\includegraphics[width=\columnwidth]{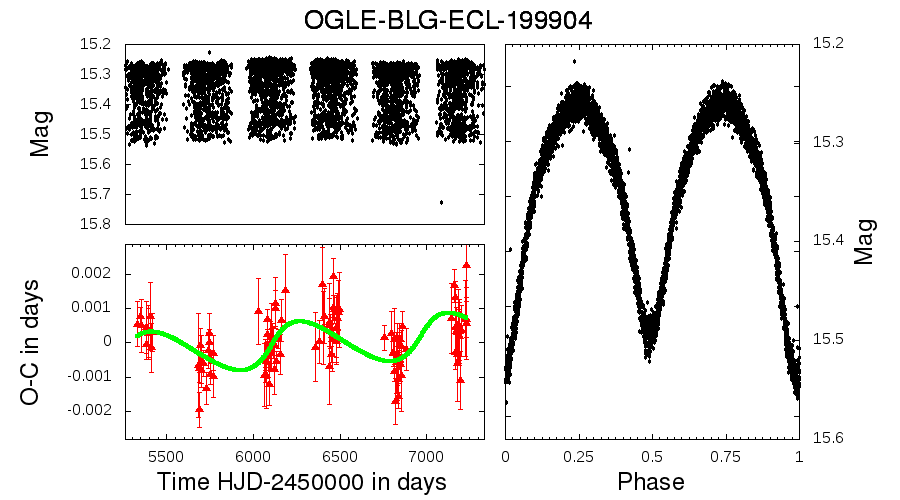}                           
\includegraphics[width=\columnwidth]{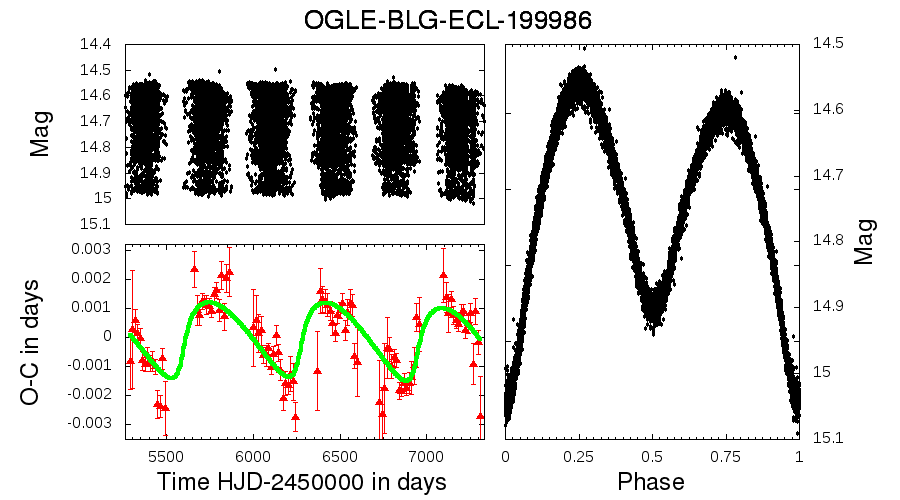}                           
                           
\includegraphics[width=\columnwidth]{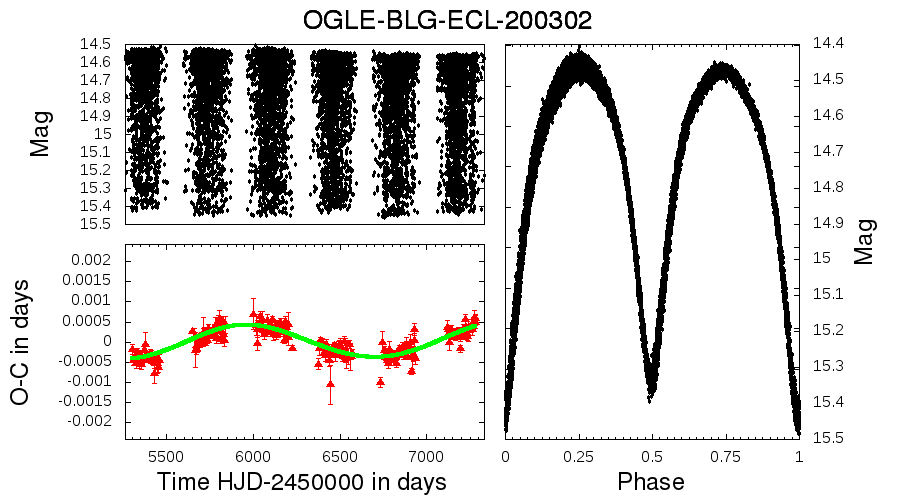}                           
\includegraphics[width=\columnwidth]{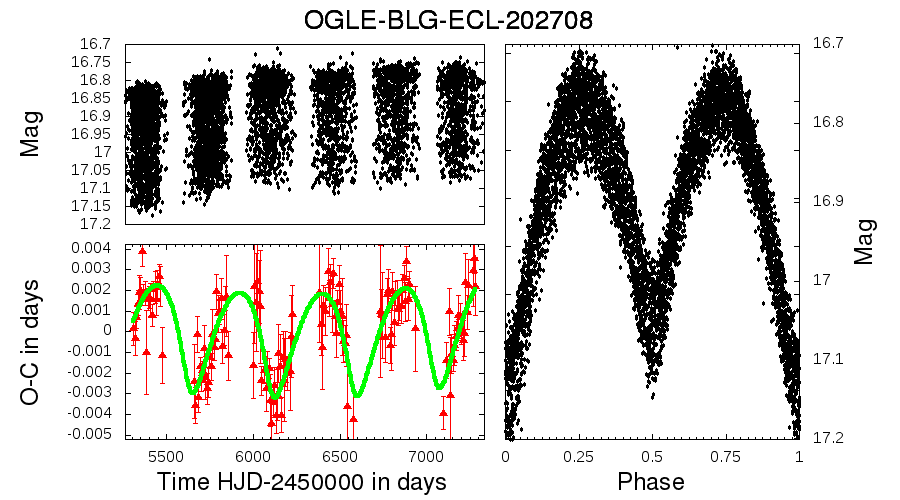}                           
                           
\includegraphics[width=\columnwidth]{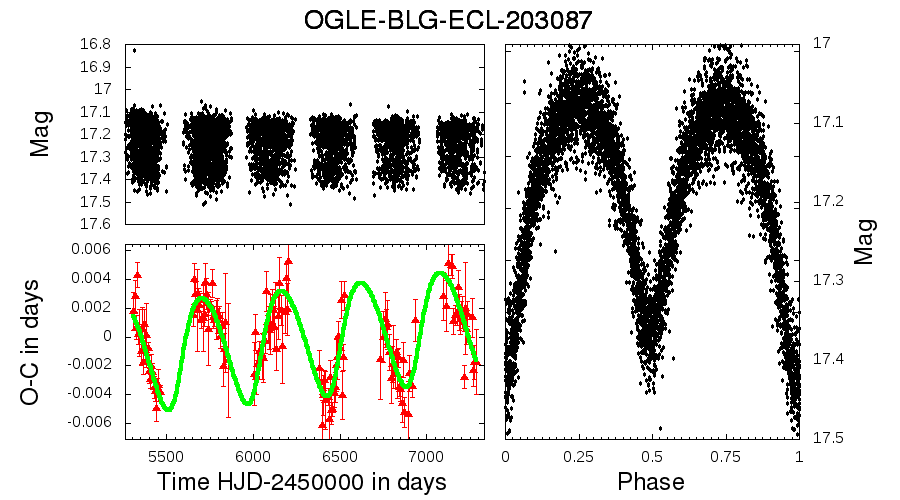}                           
\includegraphics[width=\columnwidth]{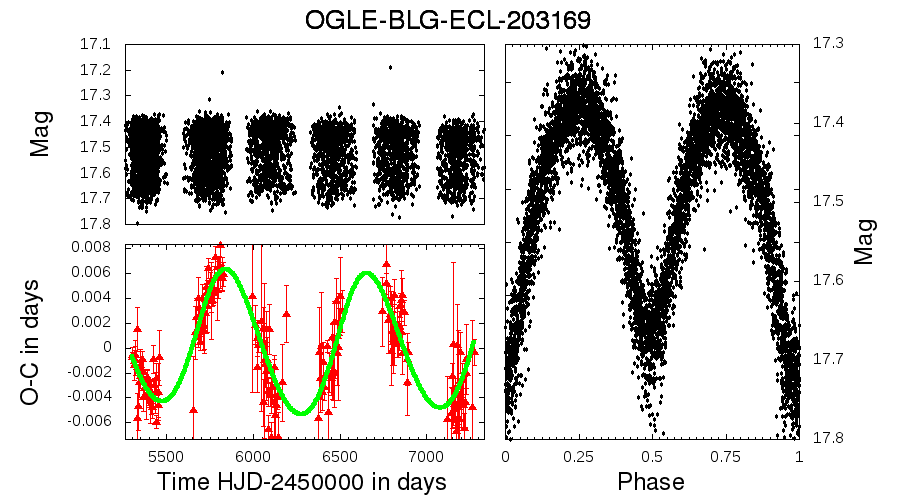}                           
                           
\includegraphics[width=\columnwidth]{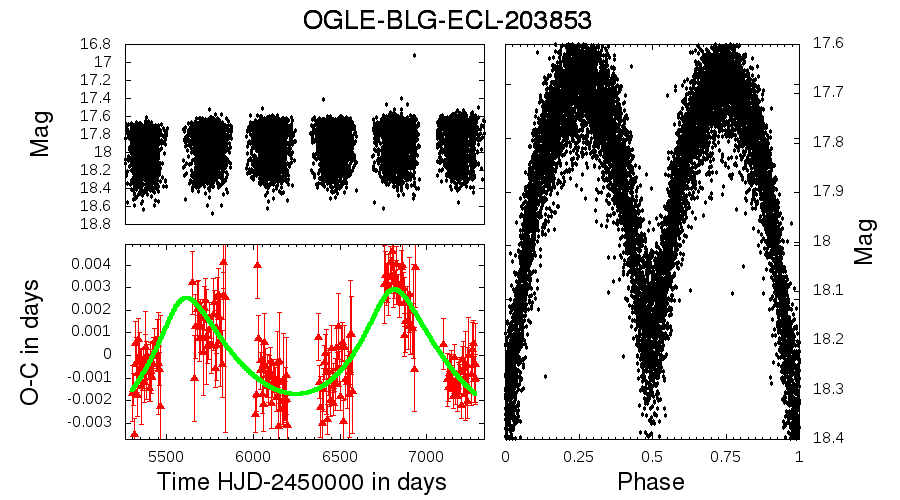}                           
\includegraphics[width=\columnwidth]{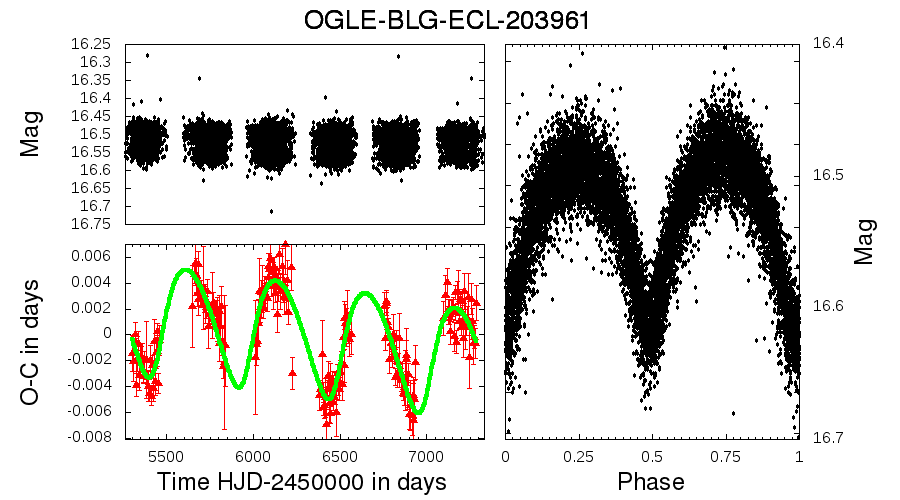}                           
\end{figure*}                           
\clearpage                           
                           
\begin{figure*}                           
                           
\includegraphics[width=\columnwidth]{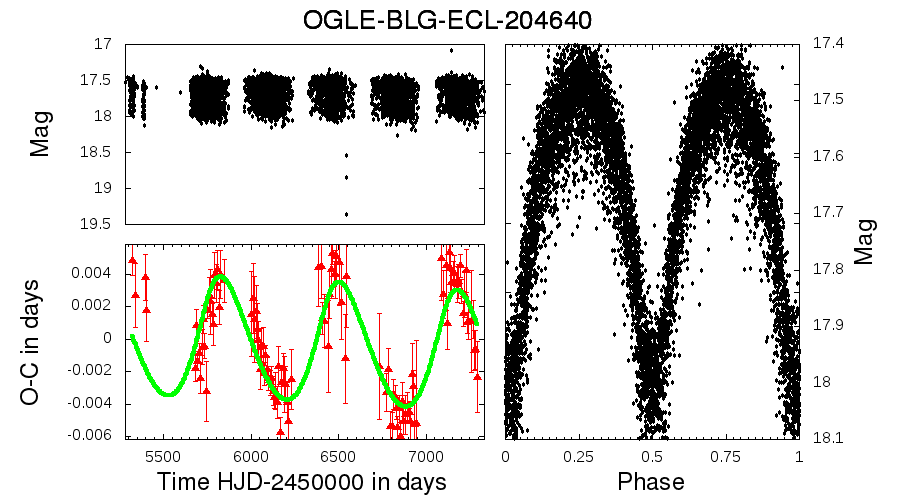}                           
\includegraphics[width=\columnwidth]{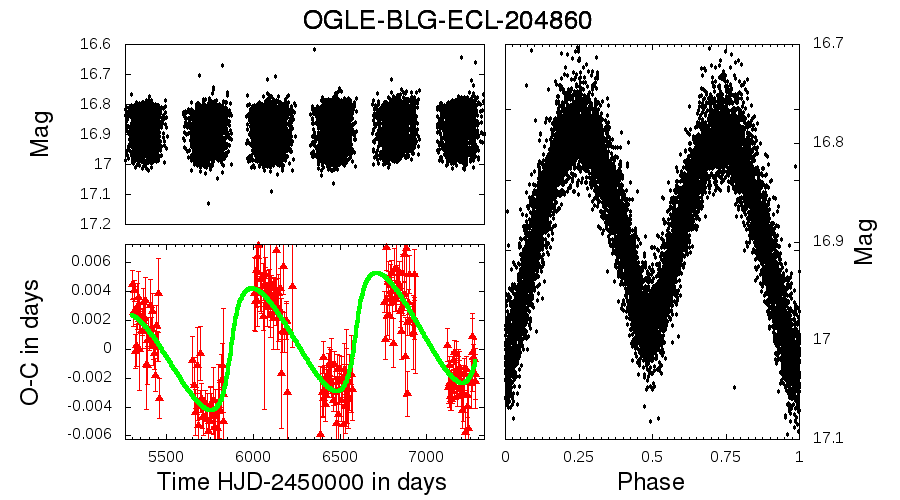}                           
                           
\includegraphics[width=\columnwidth]{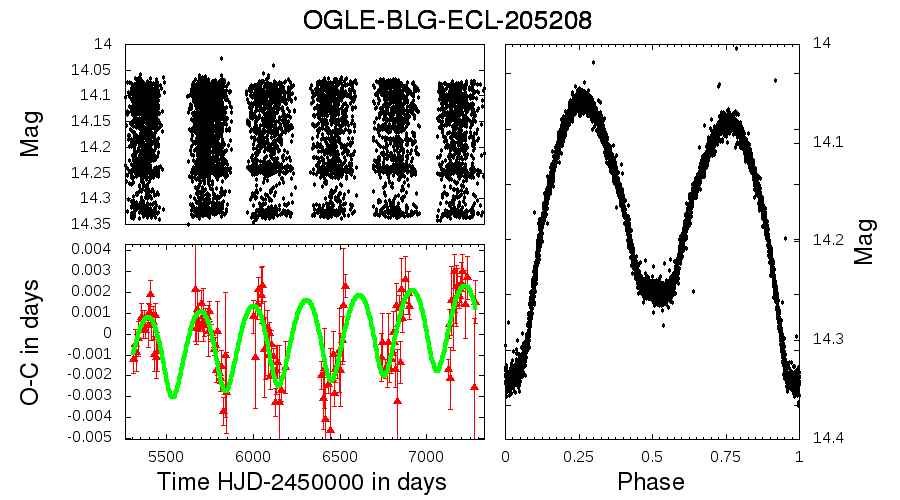}                           
\includegraphics[width=\columnwidth]{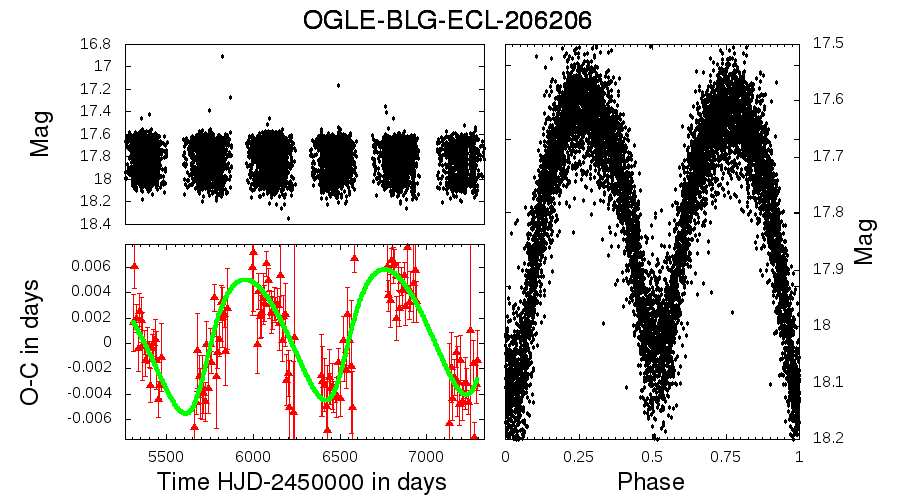}                           
                           
\includegraphics[width=\columnwidth]{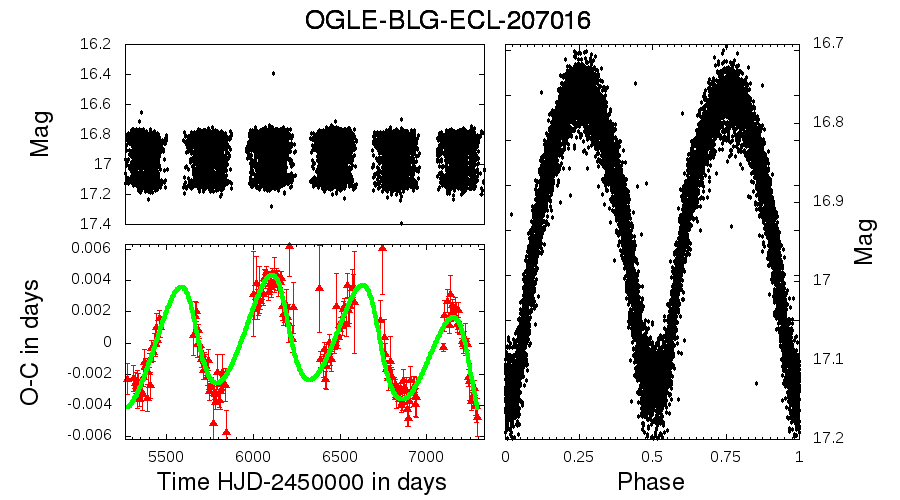}                           
\includegraphics[width=\columnwidth]{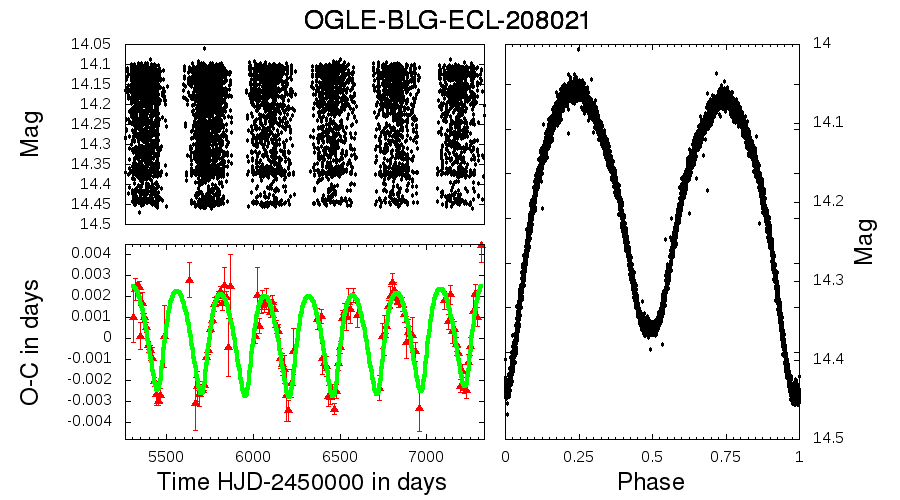}                           
                           
\includegraphics[width=\columnwidth]{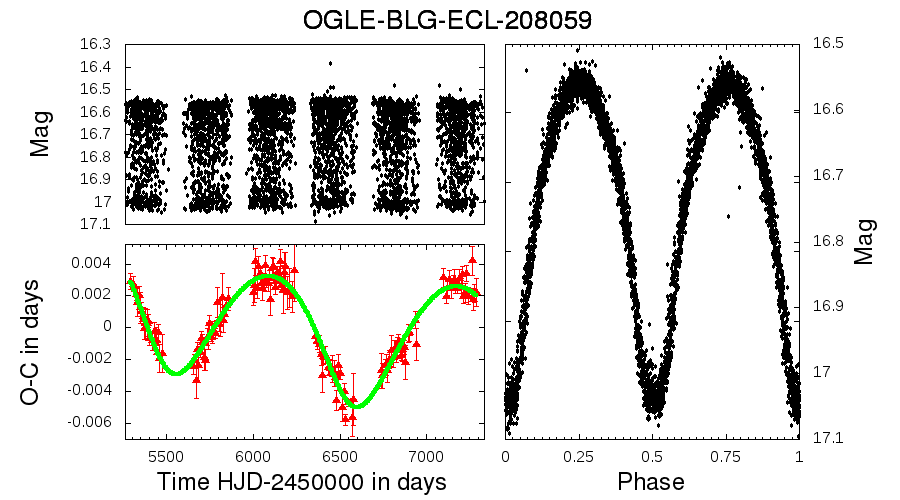}                           
\includegraphics[width=\columnwidth]{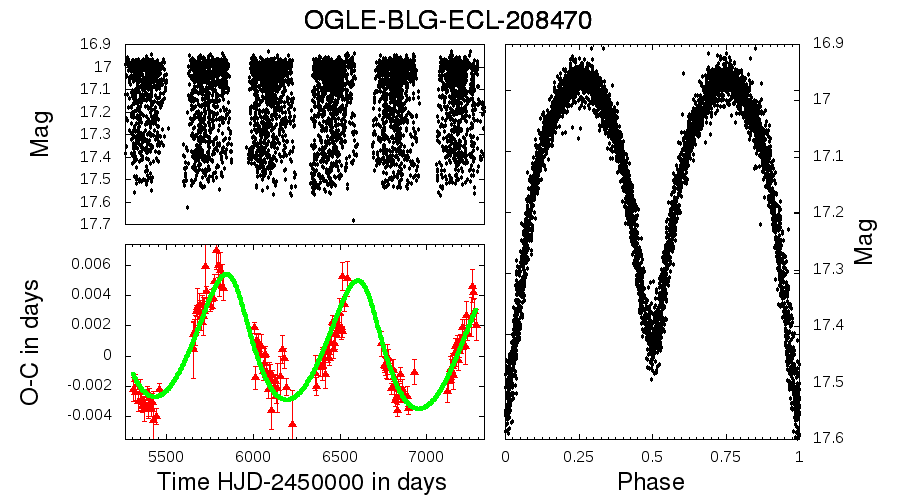}                           
                           
\includegraphics[width=\columnwidth]{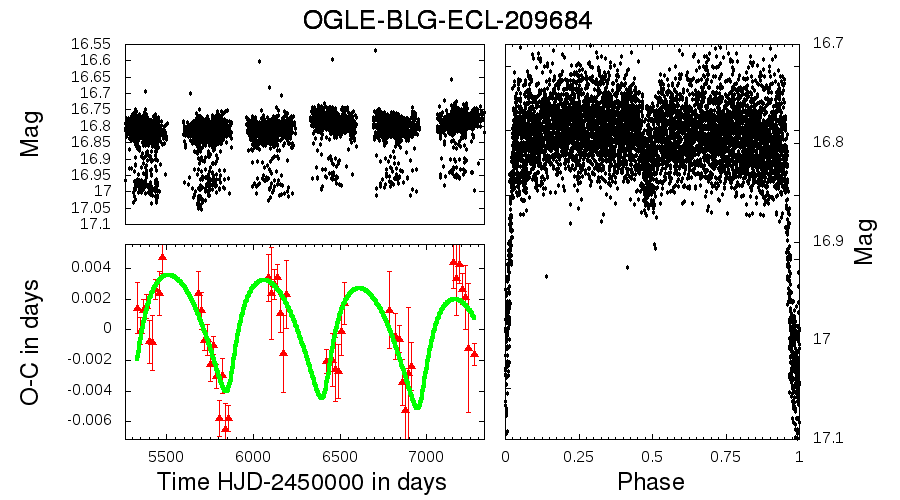}                           
\includegraphics[width=\columnwidth]{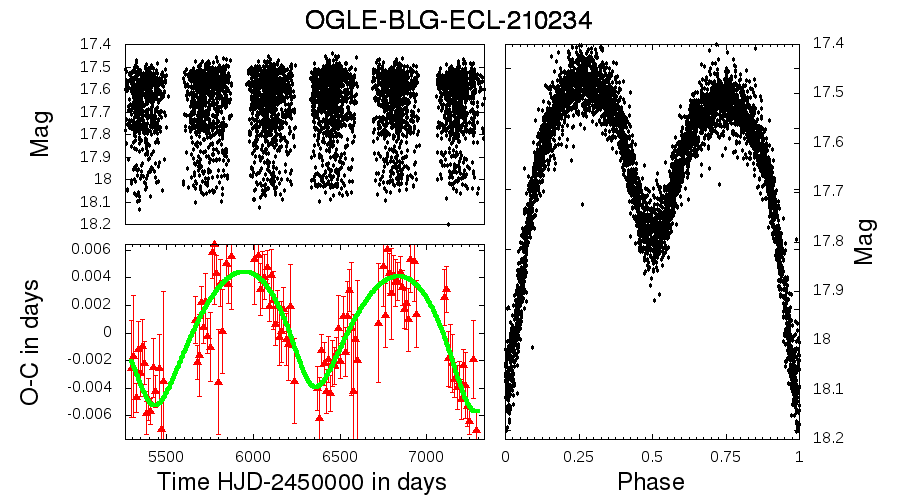}                           
\end{figure*}                           
\clearpage                           
                           
\begin{figure*}                           
                           
\includegraphics[width=\columnwidth]{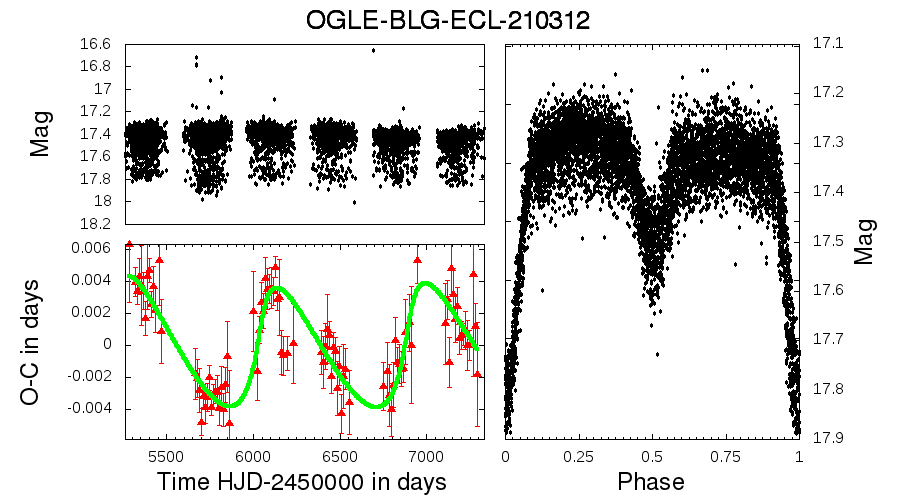}                           
\includegraphics[width=\columnwidth]{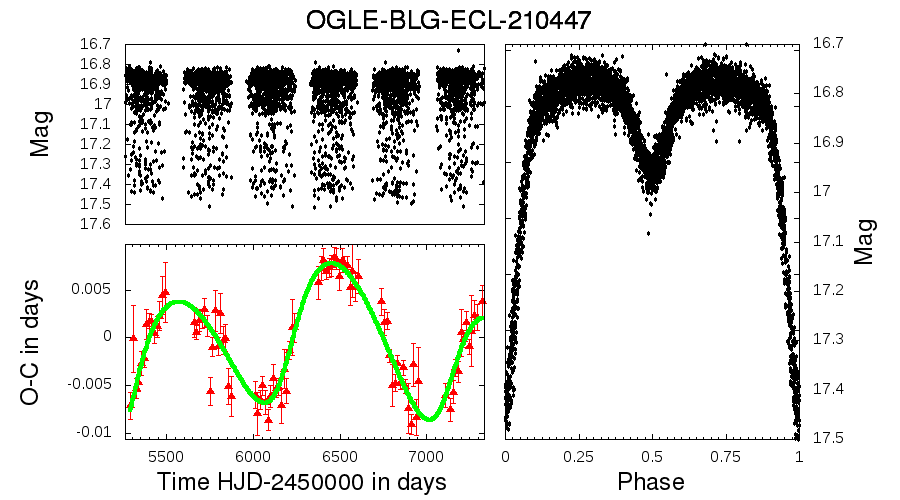}                           
                           
\includegraphics[width=\columnwidth]{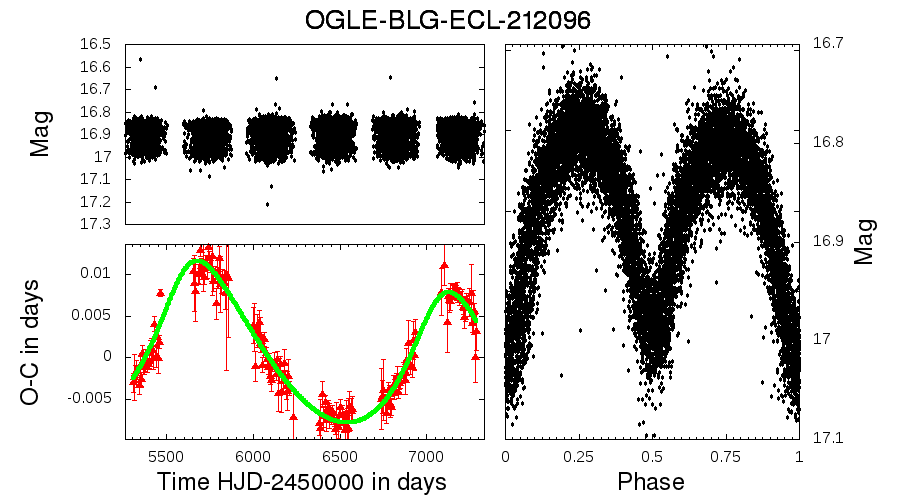}                           
\includegraphics[width=\columnwidth]{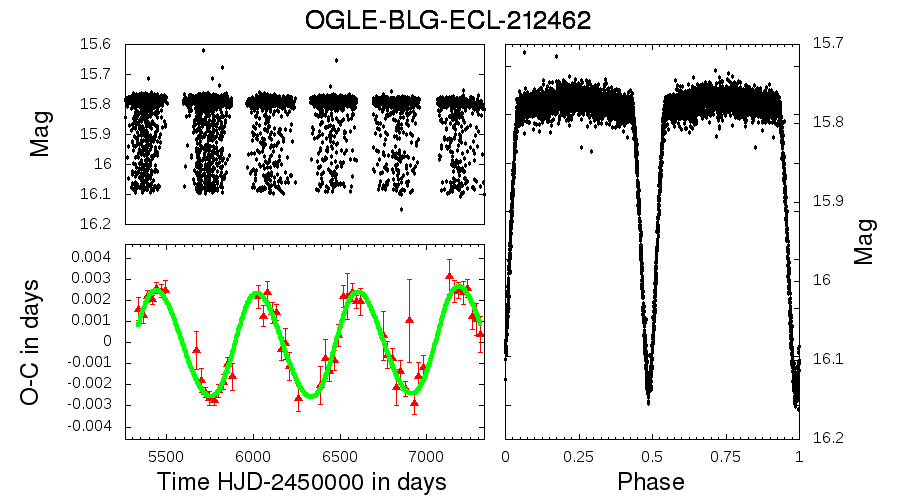}                           
                           
\includegraphics[width=\columnwidth]{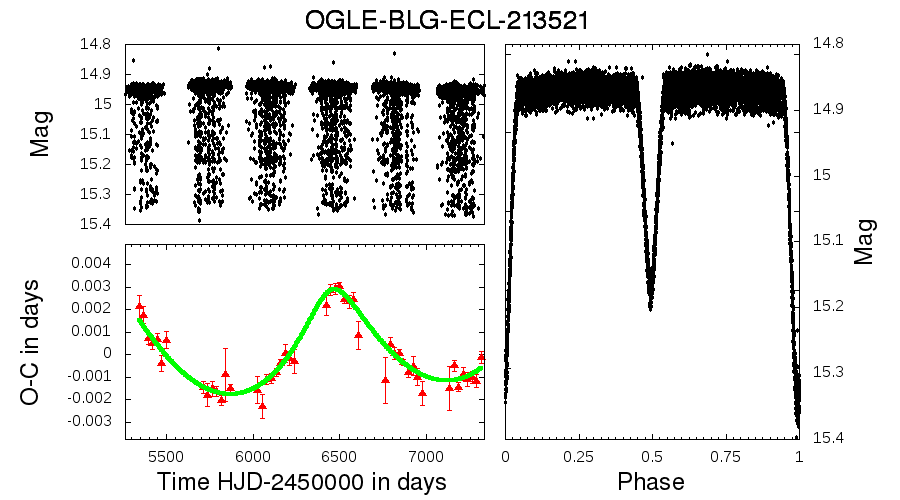}                           
\includegraphics[width=\columnwidth]{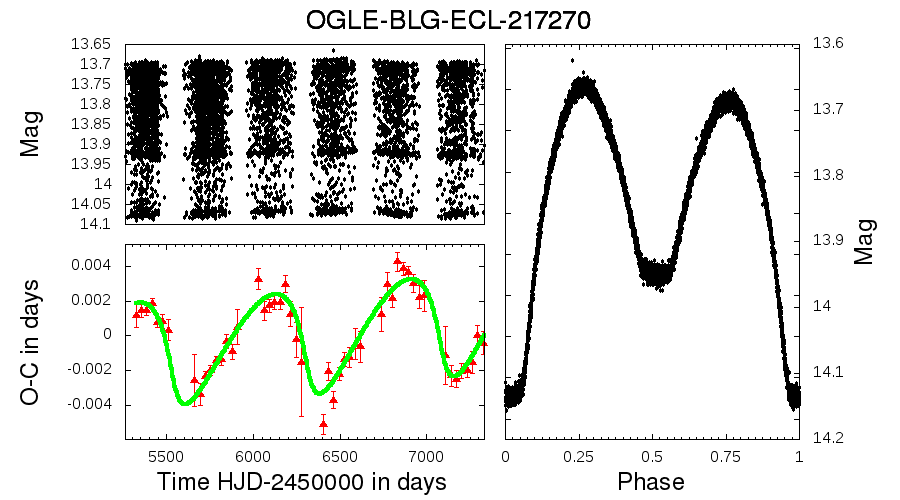}                           
                           
\includegraphics[width=\columnwidth]{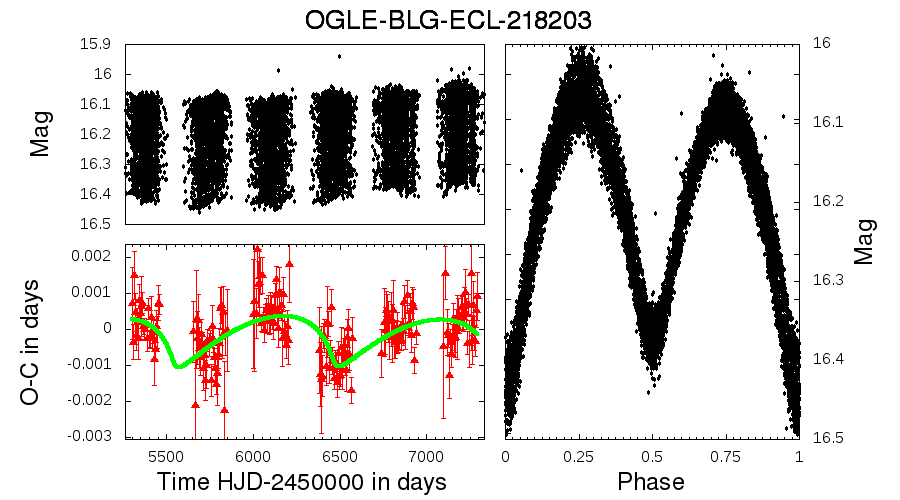}                           
\includegraphics[width=\columnwidth]{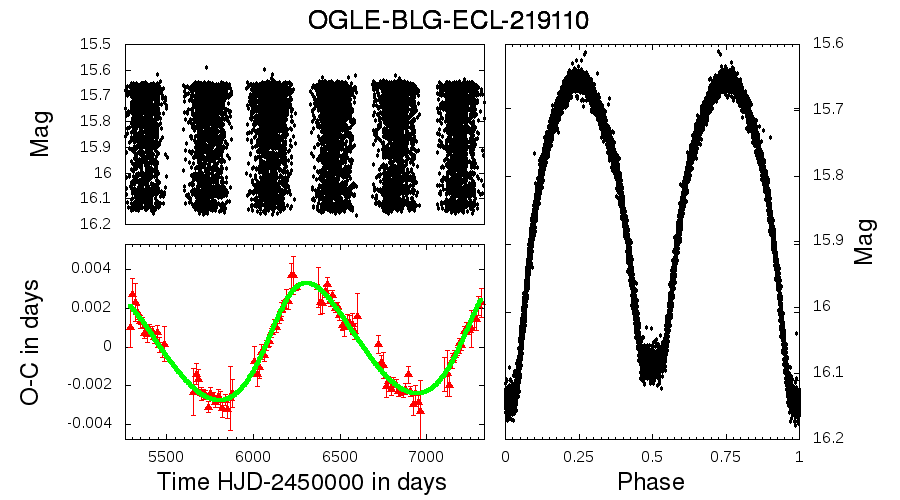}                           
                           
\includegraphics[width=\columnwidth]{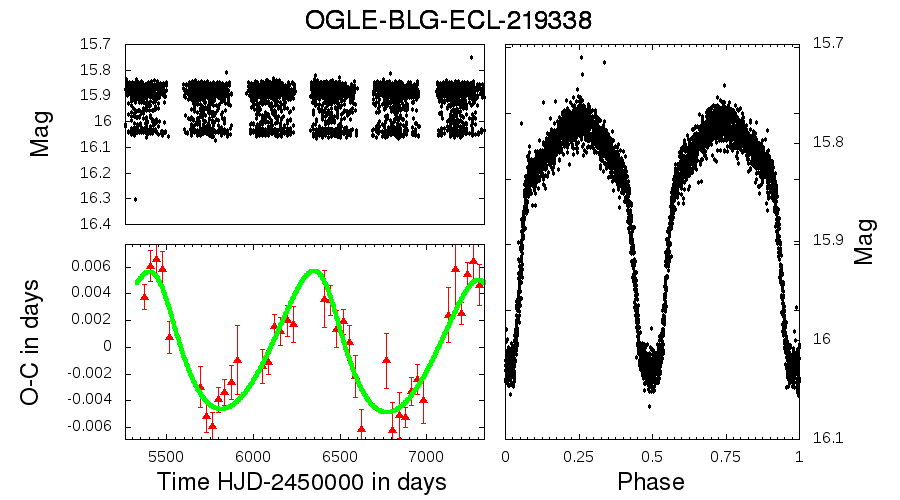}                           
\includegraphics[width=\columnwidth]{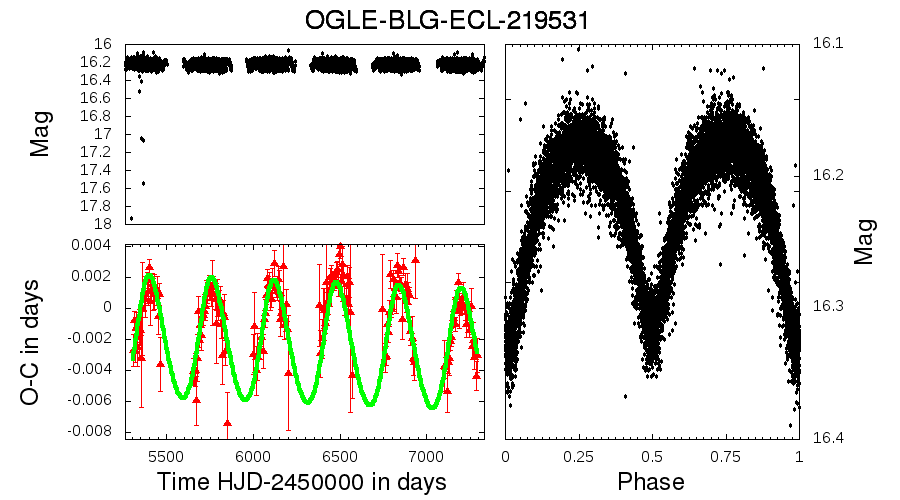}                           
\end{figure*}                           
\clearpage                           
                           
\begin{figure*}                           
                           
\includegraphics[width=\columnwidth]{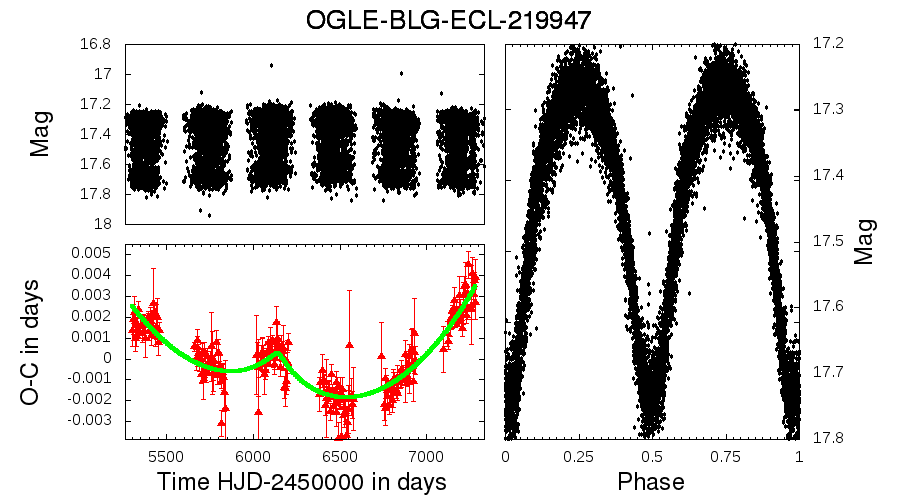}                           
\includegraphics[width=\columnwidth]{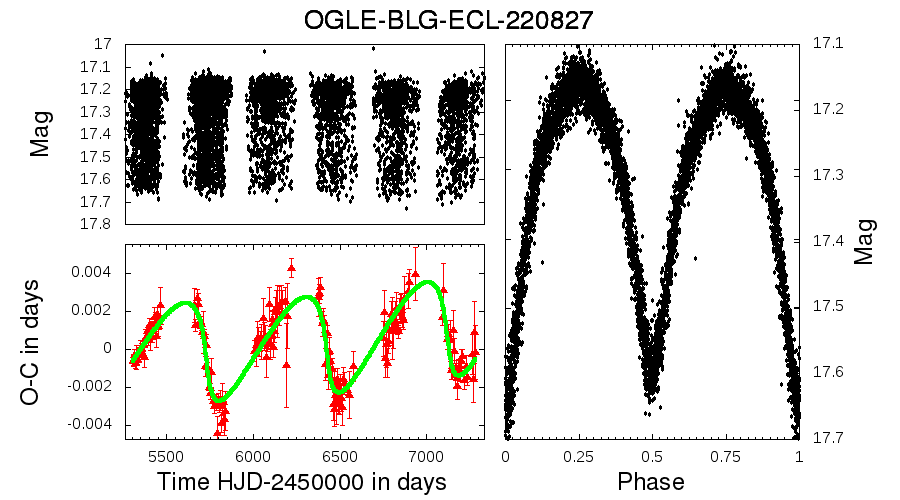}                           
                           
\includegraphics[width=\columnwidth]{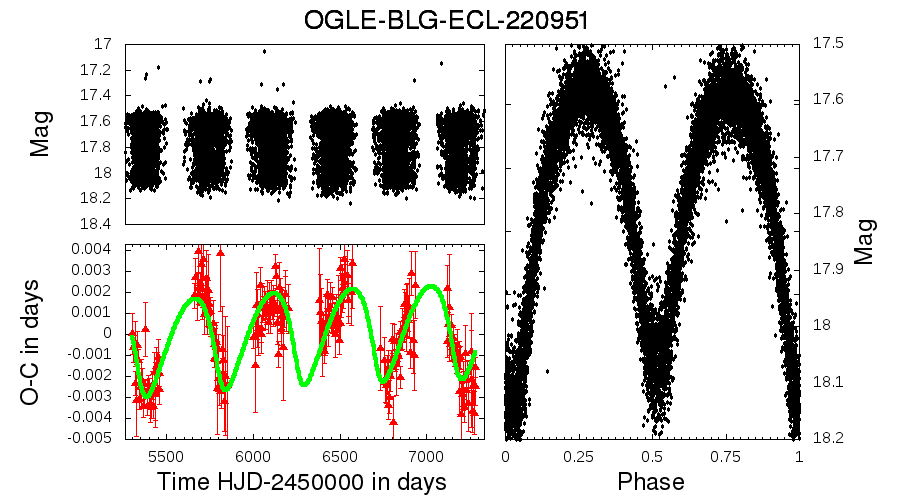}                           
\includegraphics[width=\columnwidth]{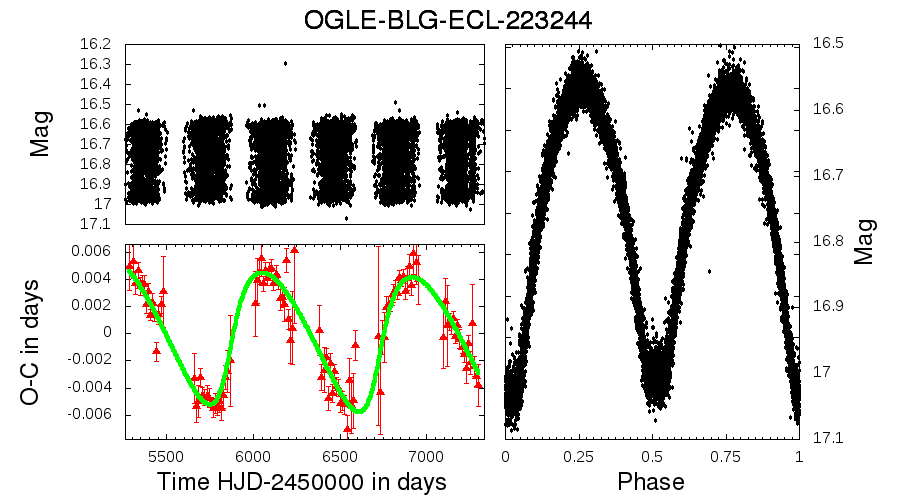}                           
                           
\includegraphics[width=\columnwidth]{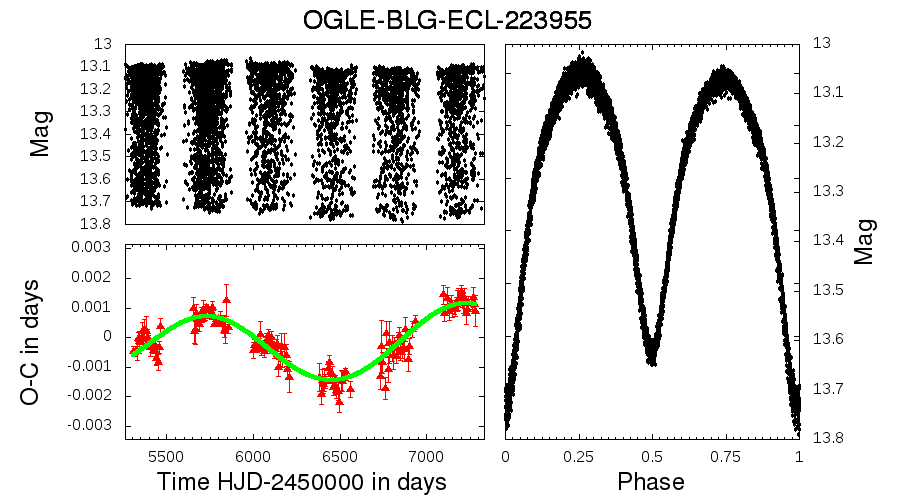}                           
\includegraphics[width=\columnwidth]{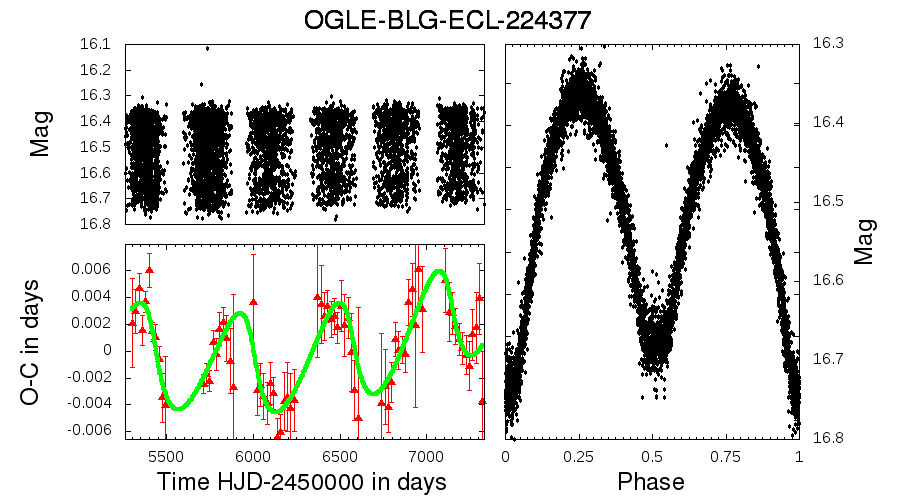}                           
                           
\includegraphics[width=\columnwidth]{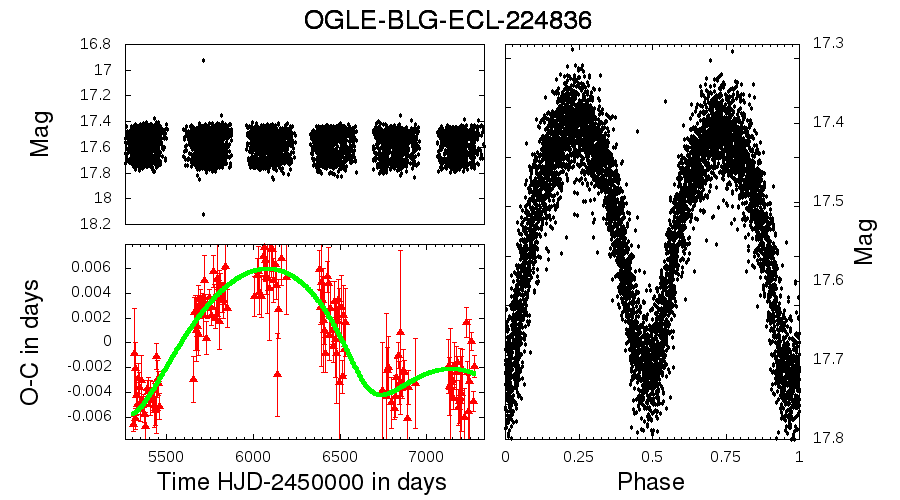}                           
\includegraphics[width=\columnwidth]{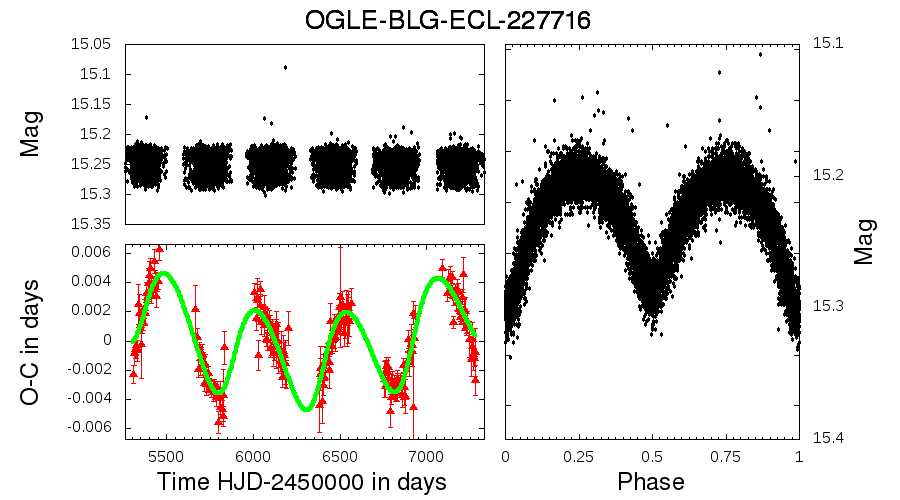}                           
                           
\includegraphics[width=\columnwidth]{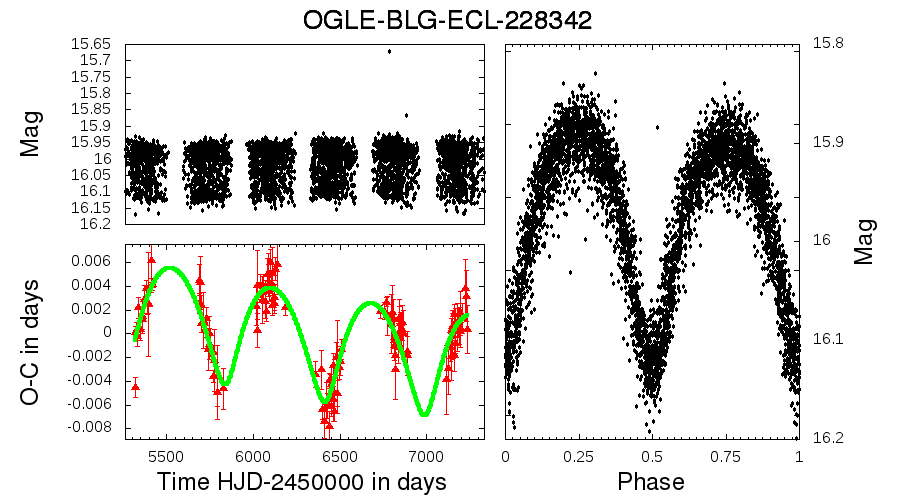}                           
\includegraphics[width=\columnwidth]{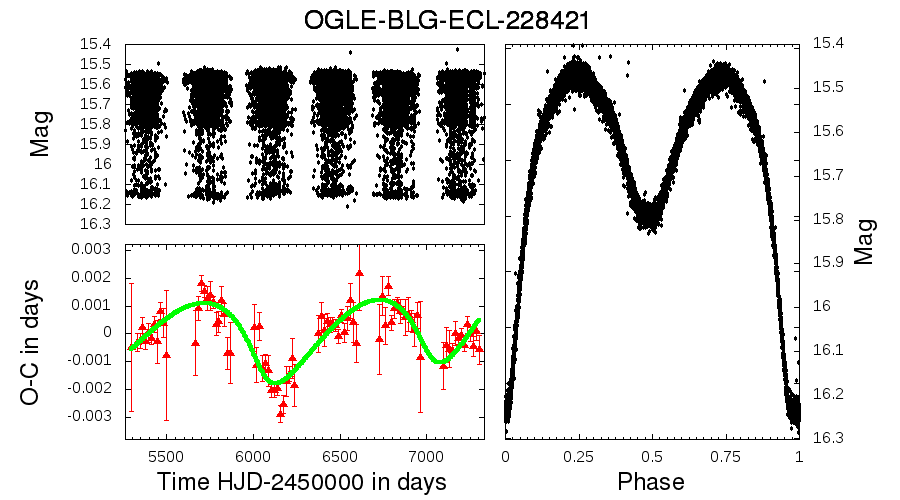}                           
\end{figure*}                           
\clearpage                           
                           
\begin{figure*}                           
                           
\includegraphics[width=\columnwidth]{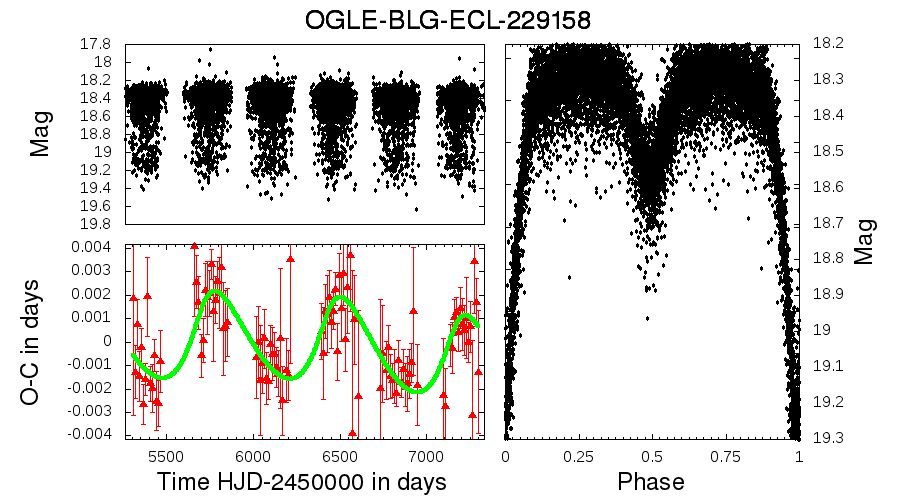}                           
\includegraphics[width=\columnwidth]{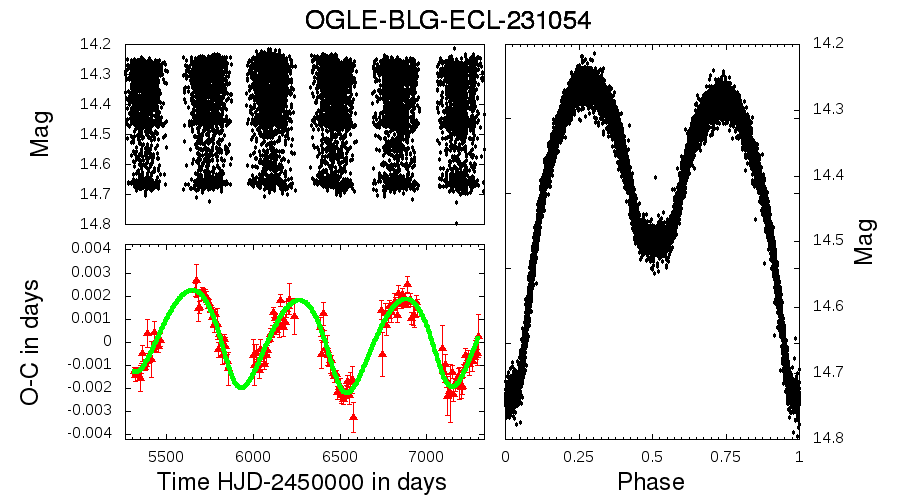}                           
                           
\includegraphics[width=\columnwidth]{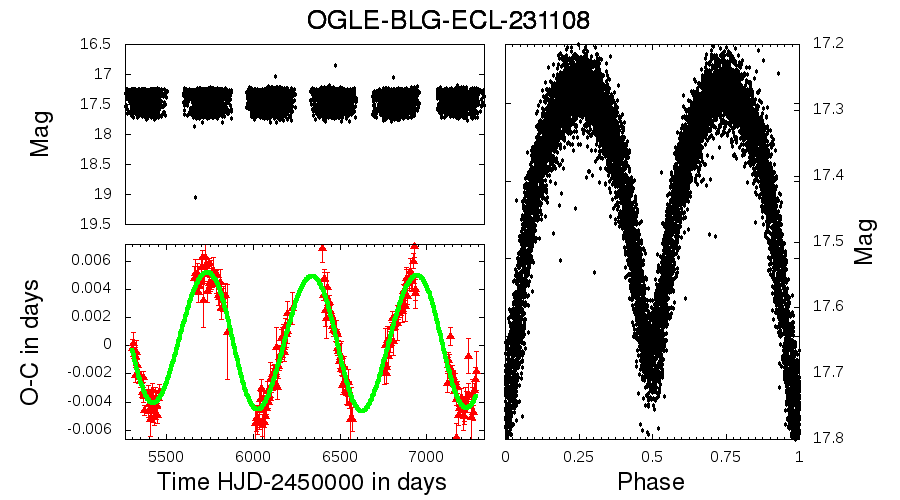}                           
\includegraphics[width=\columnwidth]{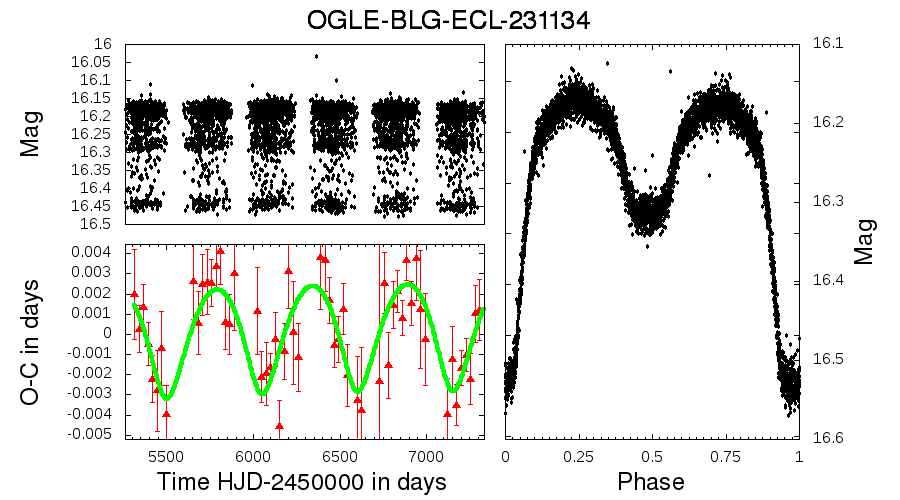}                           
                           
\includegraphics[width=\columnwidth]{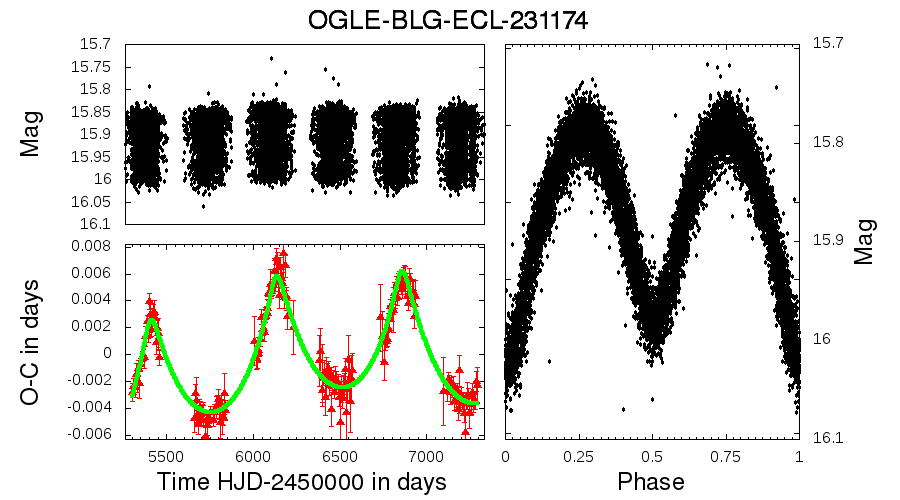}                           
\includegraphics[width=\columnwidth]{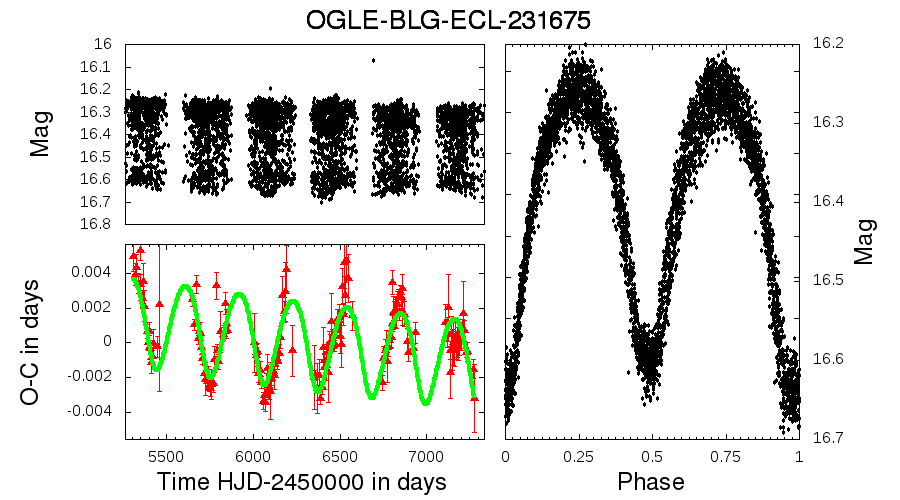}                           
                           
\includegraphics[width=\columnwidth]{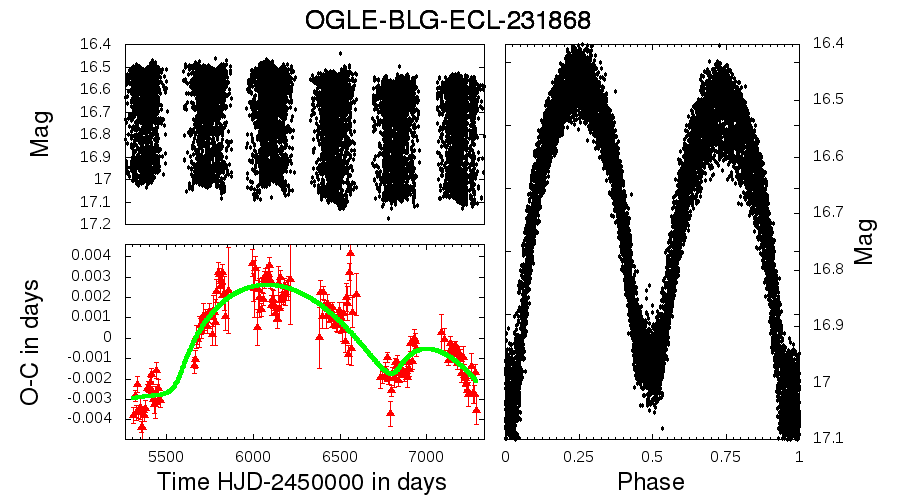}                           
\includegraphics[width=\columnwidth]{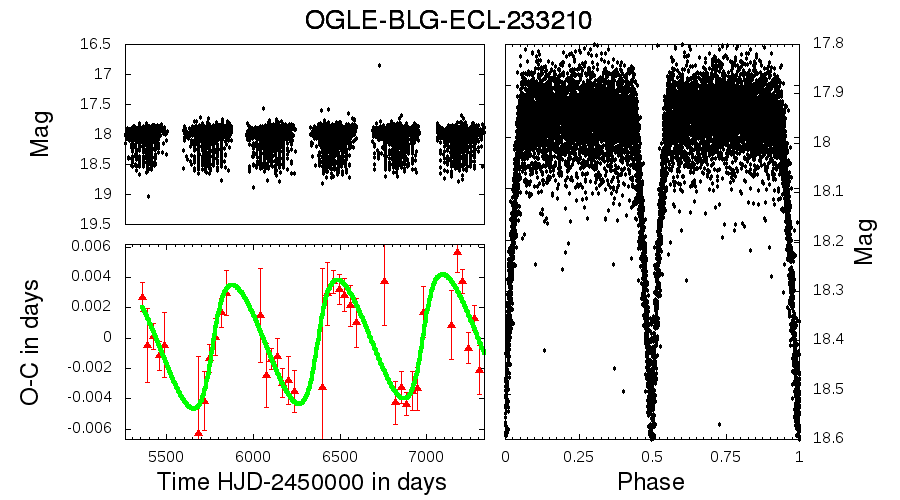}                           
                           
\includegraphics[width=\columnwidth]{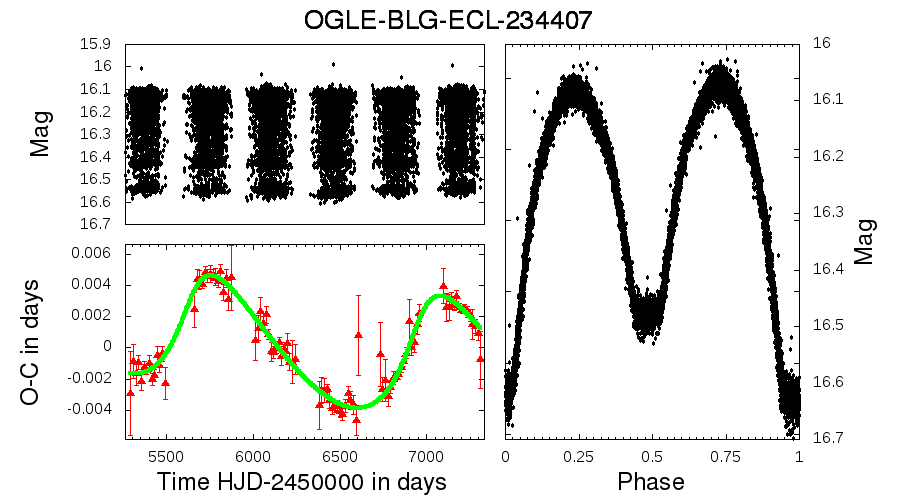}                           
\includegraphics[width=\columnwidth]{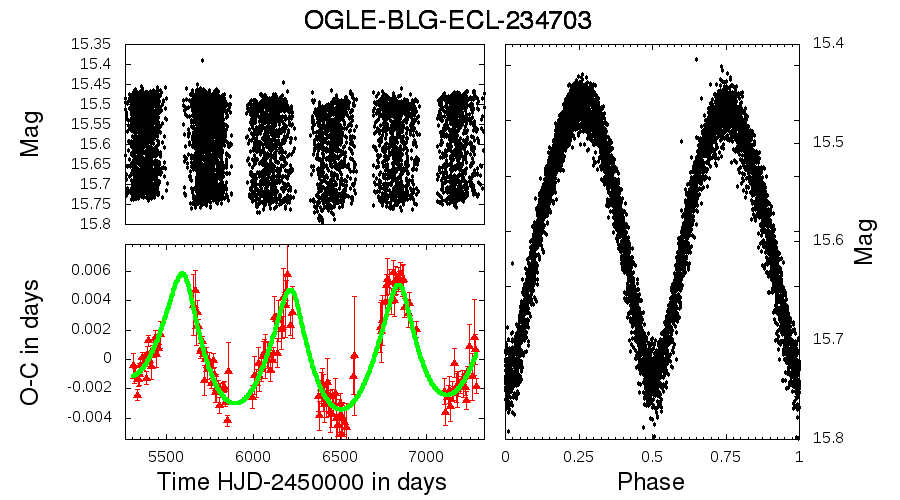}                           
\end{figure*}                           
\clearpage                           
                           
\begin{figure*}                           
                           
\includegraphics[width=\columnwidth]{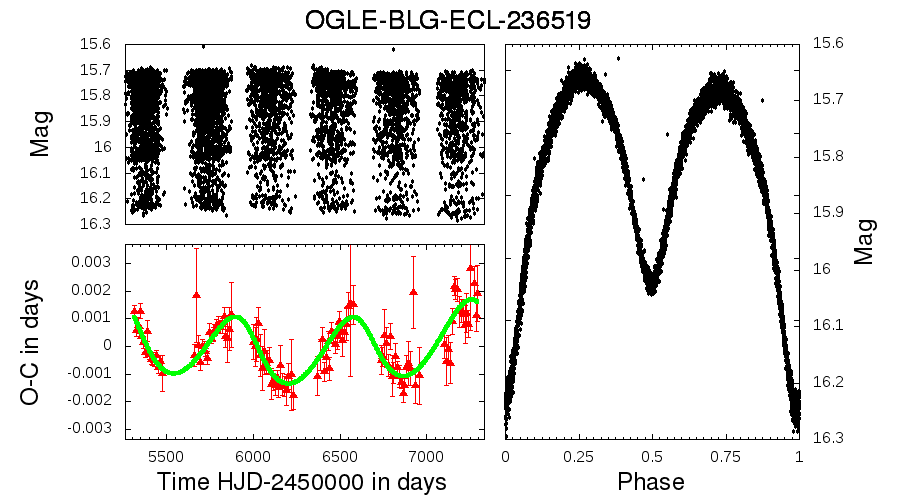}                           
\includegraphics[width=\columnwidth]{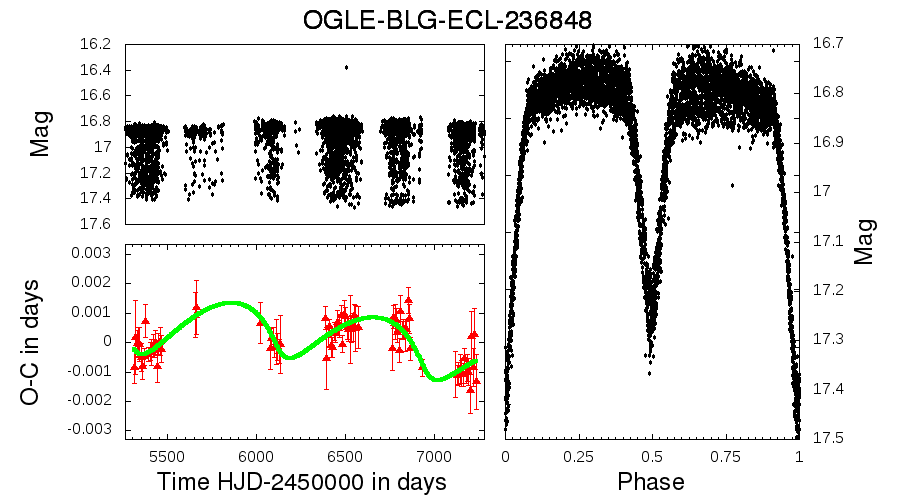}                           
                           
\includegraphics[width=\columnwidth]{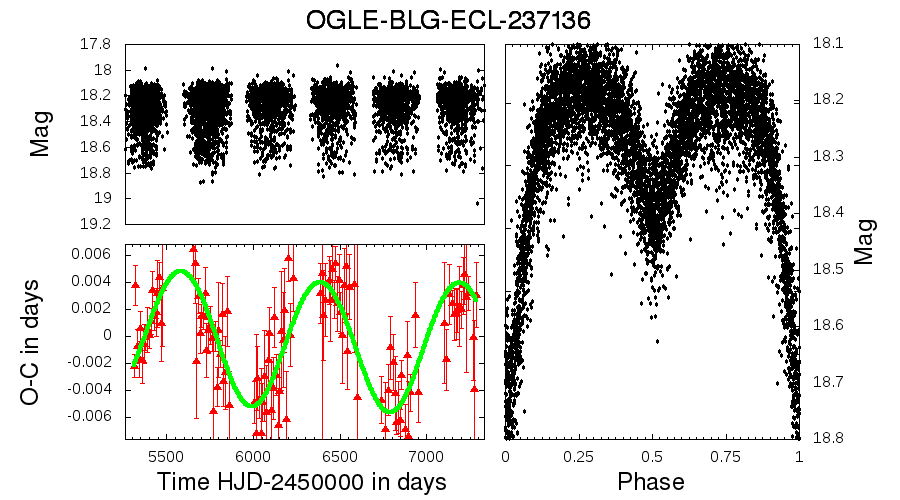}                           
\includegraphics[width=\columnwidth]{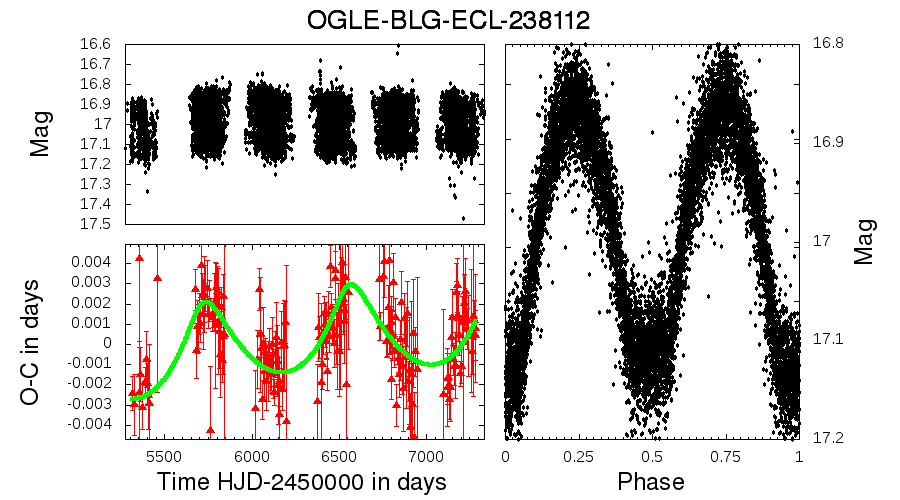}                           
                           
\includegraphics[width=\columnwidth]{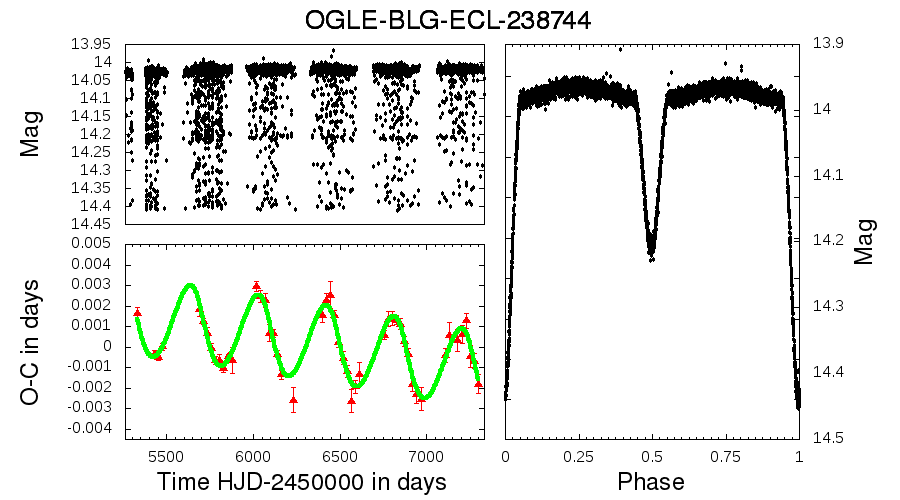}                           
\includegraphics[width=\columnwidth]{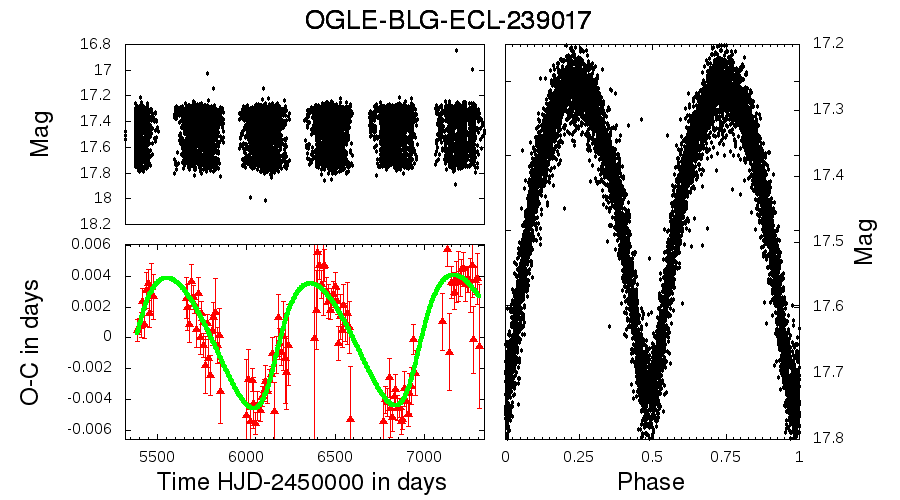}                           
                           
\includegraphics[width=\columnwidth]{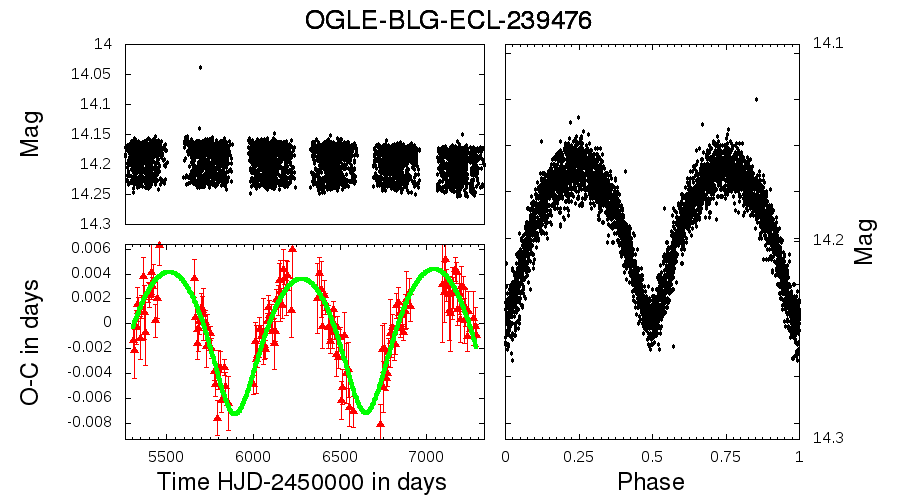}                           
\includegraphics[width=\columnwidth]{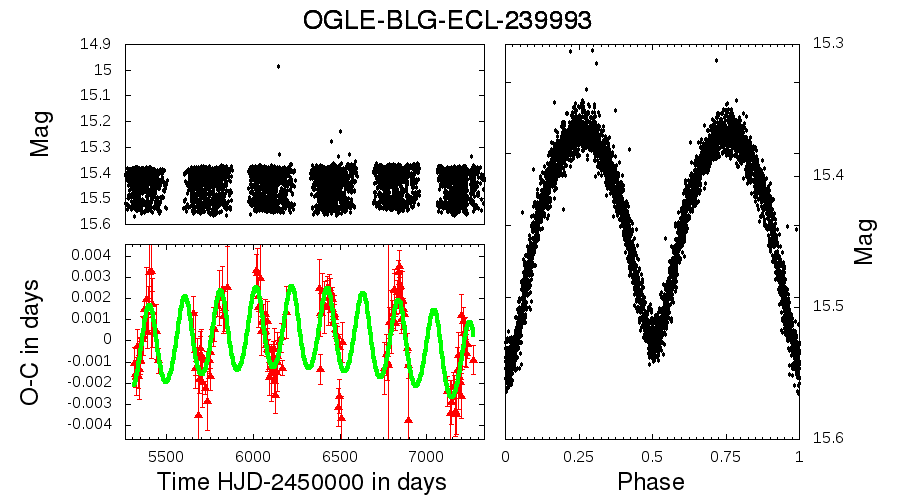}                           
                           
\includegraphics[width=\columnwidth]{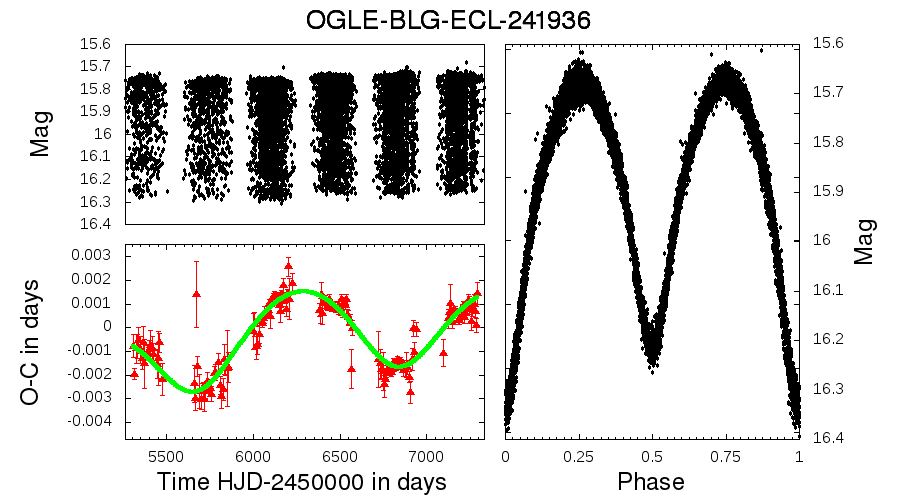}                           
\includegraphics[width=\columnwidth]{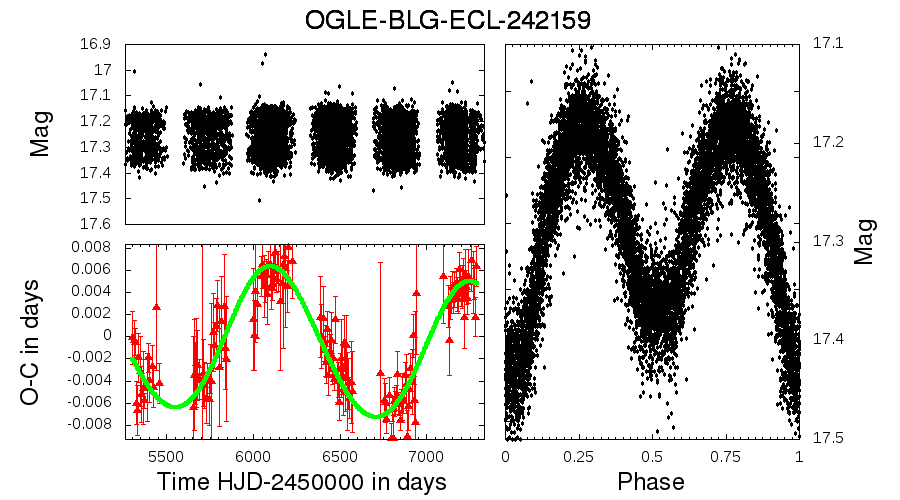}                           
\end{figure*}                           
\clearpage                           
                           
\begin{figure*}                           
                           
\includegraphics[width=\columnwidth]{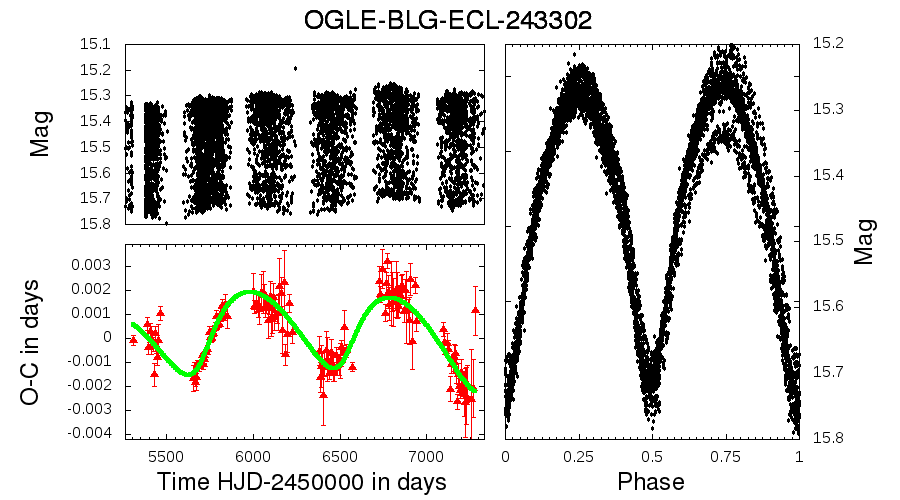}                           
\includegraphics[width=\columnwidth]{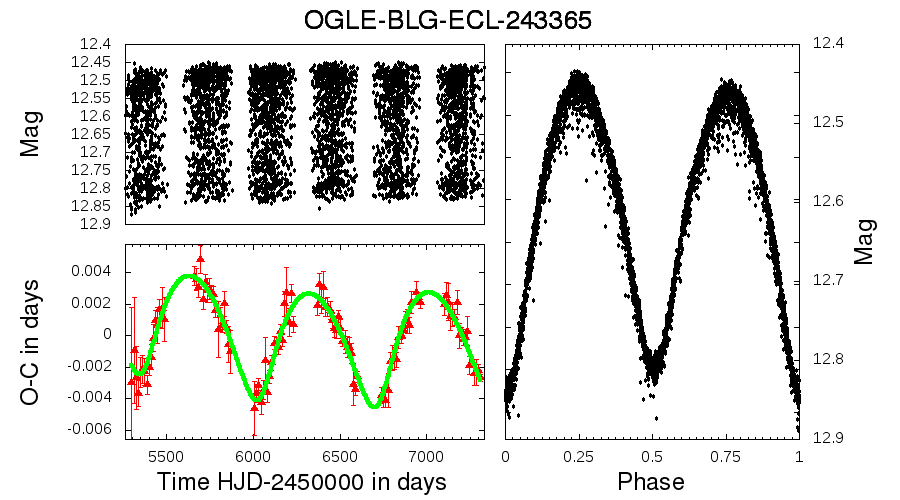}                           
                           
\includegraphics[width=\columnwidth]{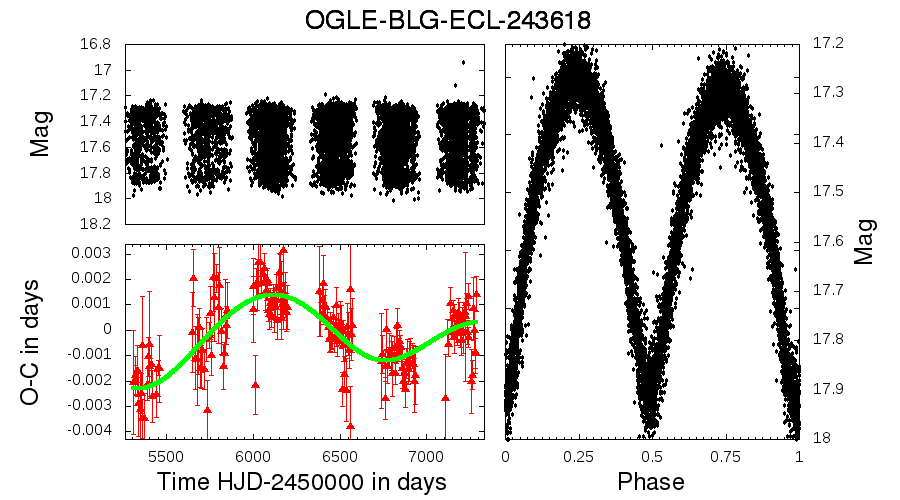}                           
\includegraphics[width=\columnwidth]{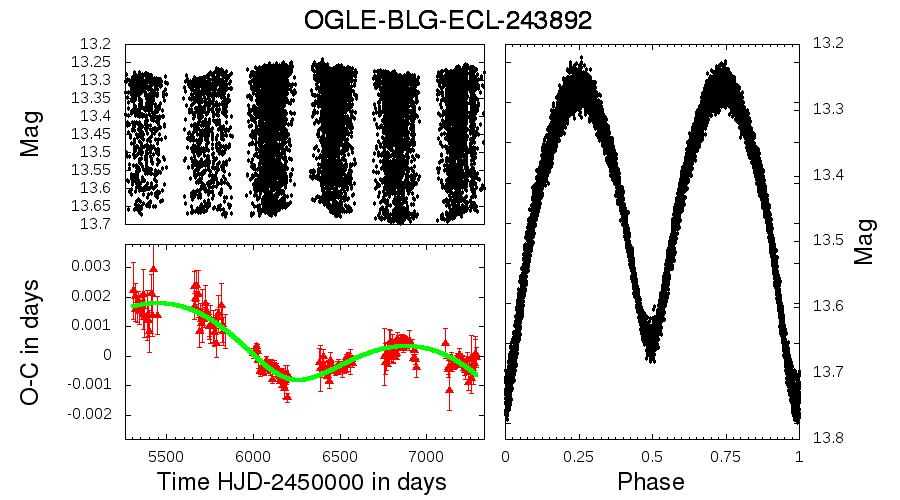}                           
                           
\includegraphics[width=\columnwidth]{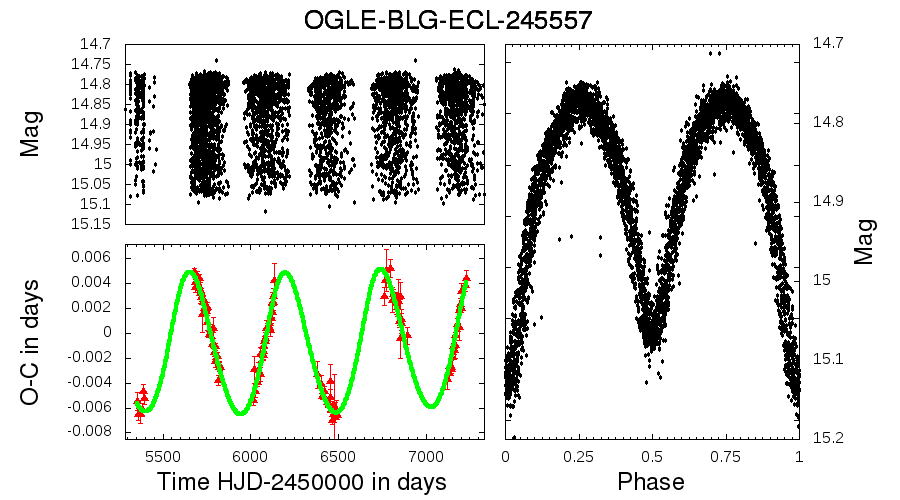}                           
\includegraphics[width=\columnwidth]{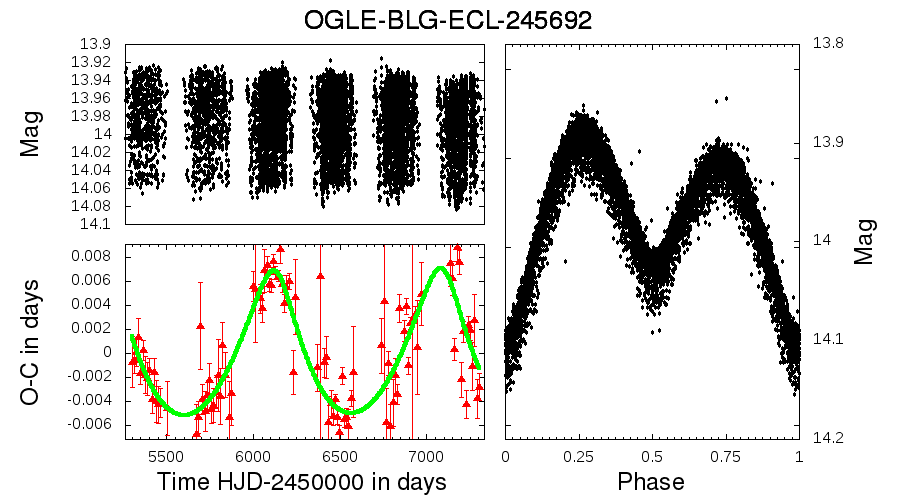}                           
                           
\includegraphics[width=\columnwidth]{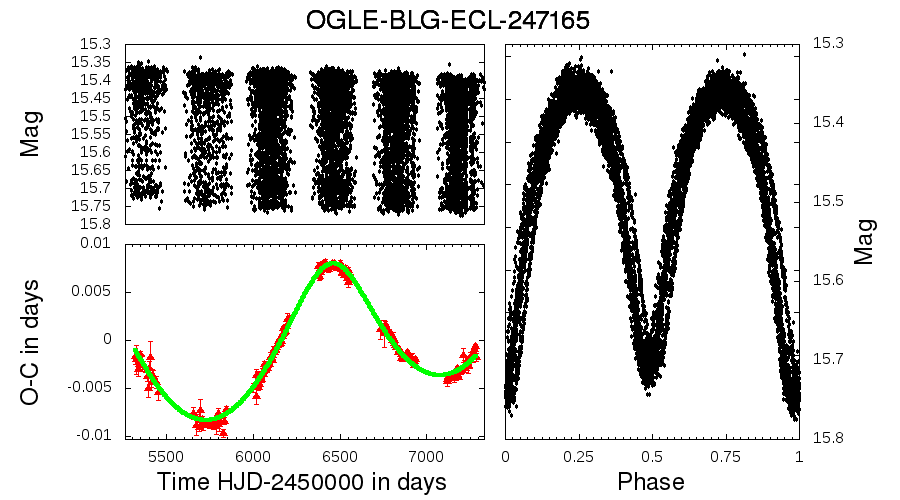}                           
\includegraphics[width=\columnwidth]{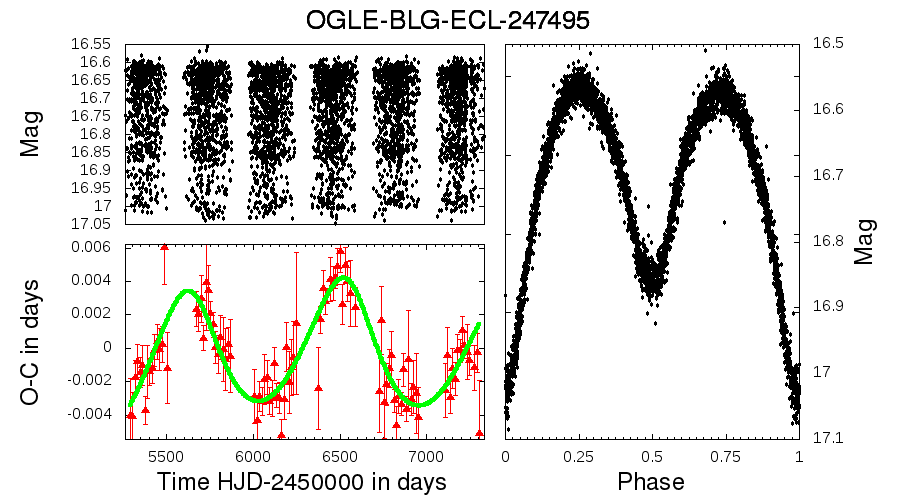}                           
                           
\includegraphics[width=\columnwidth]{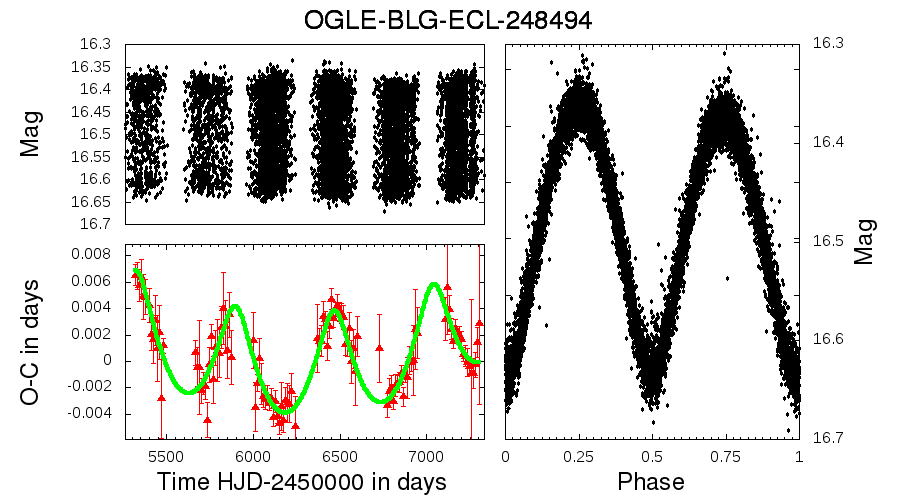}                           
\includegraphics[width=\columnwidth]{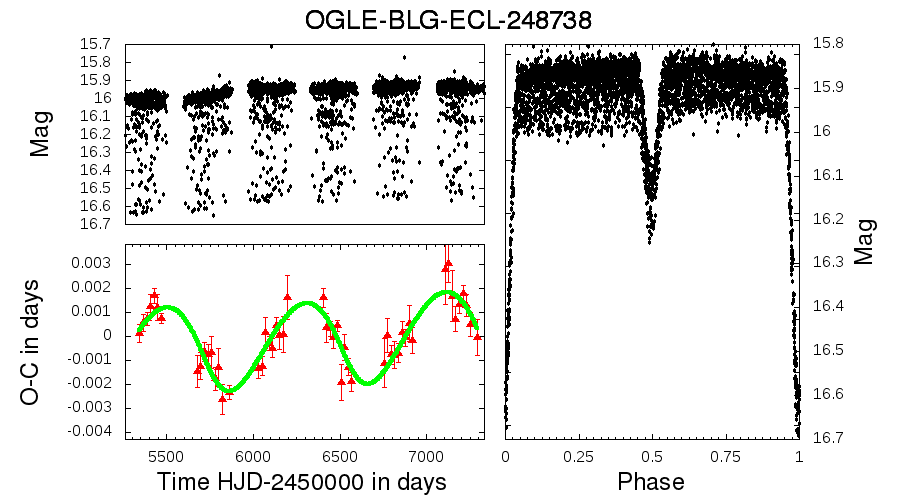}                           
\end{figure*}                           
\clearpage                           
                           
\begin{figure*}                           
                           
\includegraphics[width=\columnwidth]{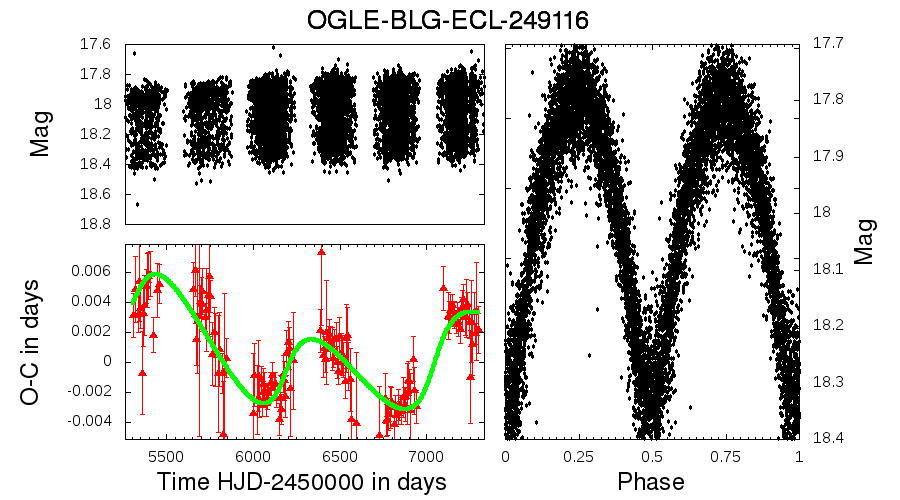}                           
\includegraphics[width=\columnwidth]{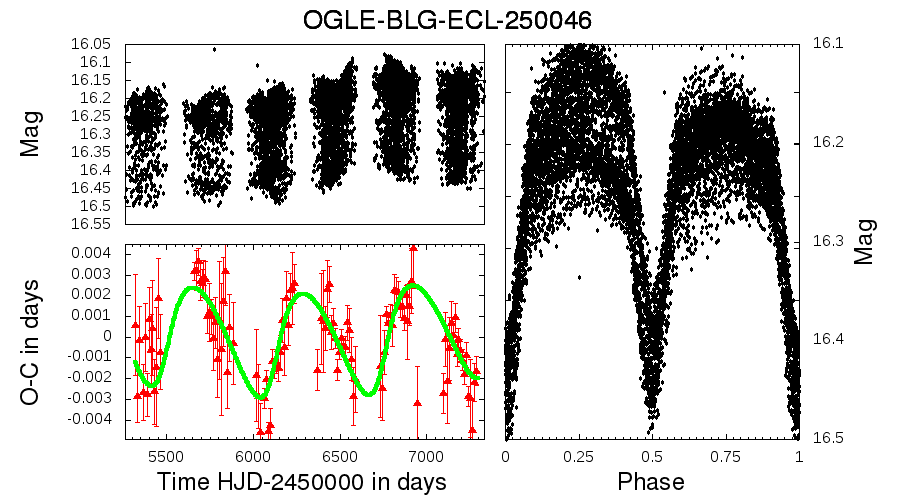}                           
                           
\includegraphics[width=\columnwidth]{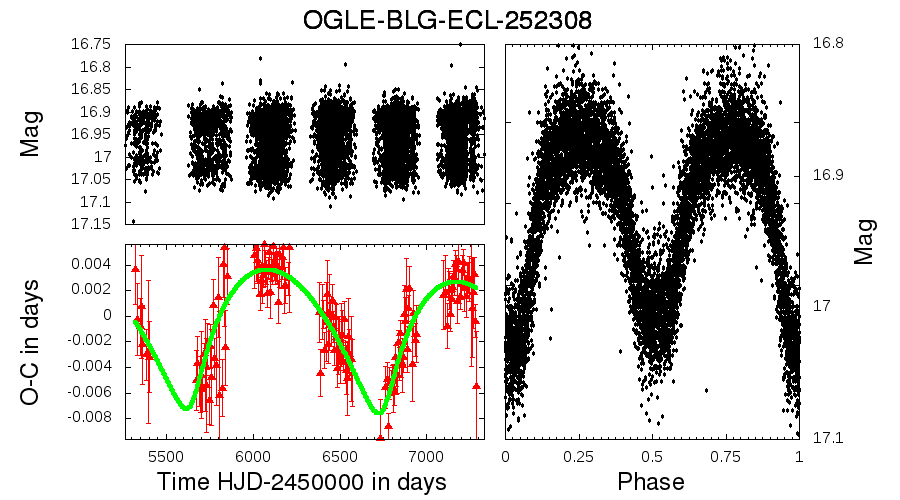}                           
\includegraphics[width=\columnwidth]{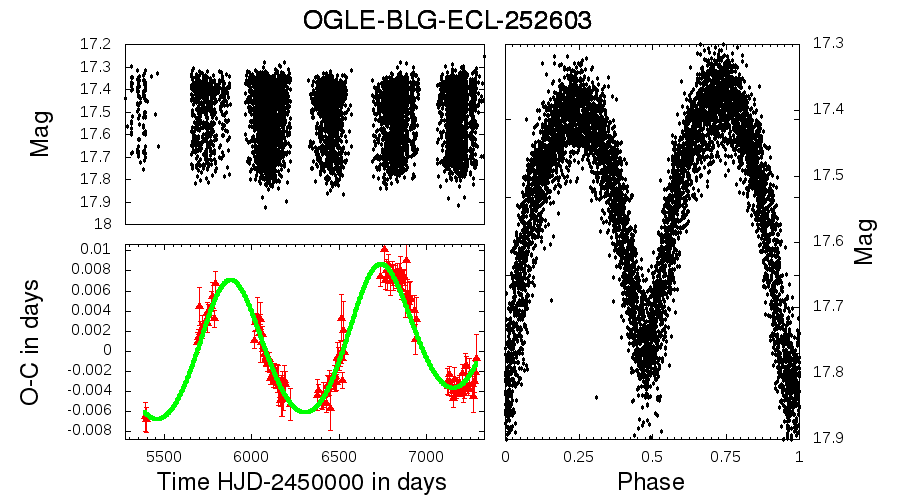}                           
                           
\includegraphics[width=\columnwidth]{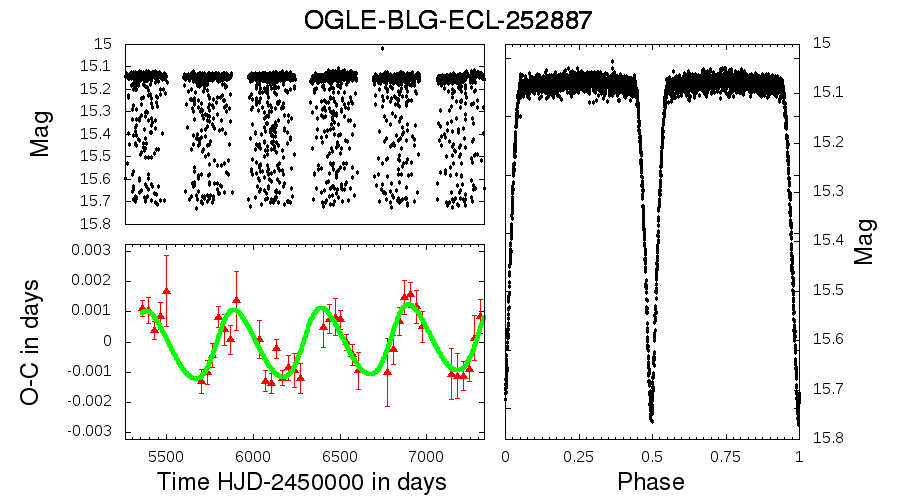}                           
\includegraphics[width=\columnwidth]{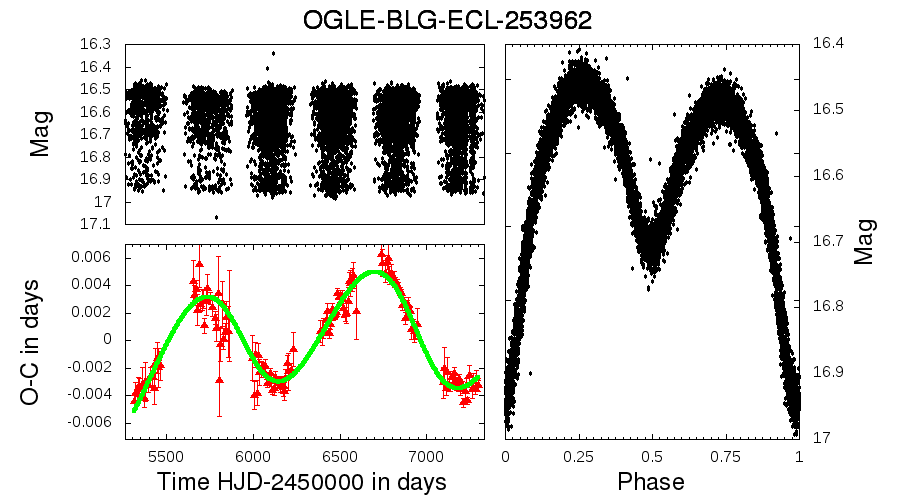}                           
                           
\includegraphics[width=\columnwidth]{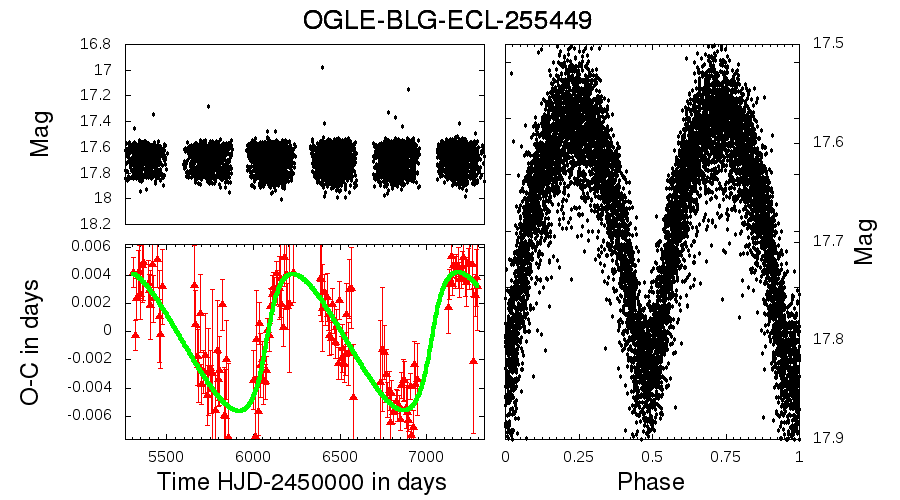}                           
\includegraphics[width=\columnwidth]{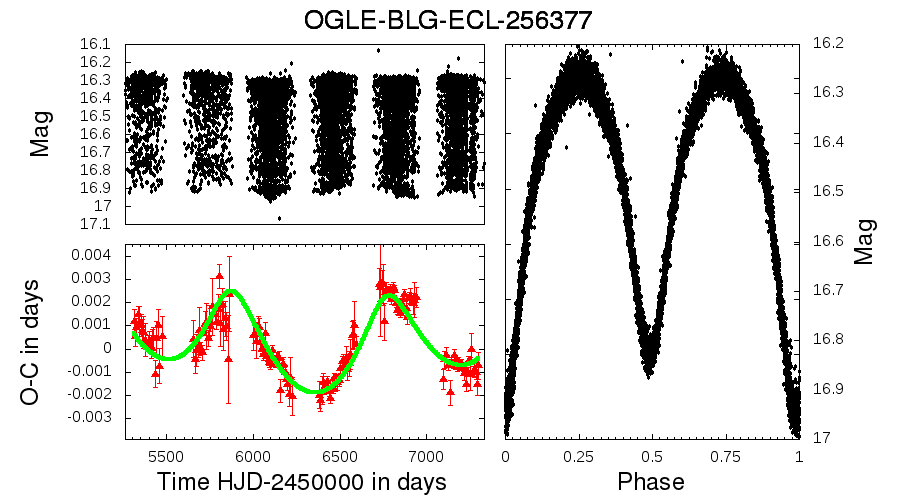}                           
                           
\includegraphics[width=\columnwidth]{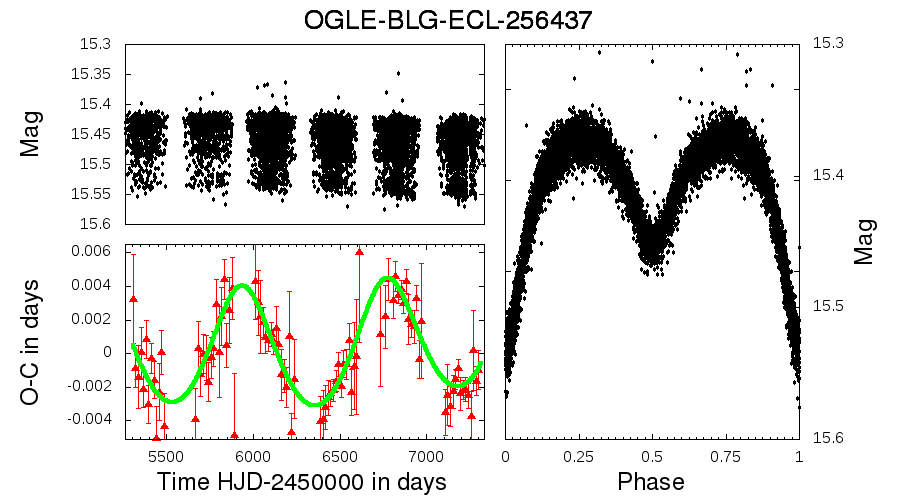}                           
\includegraphics[width=\columnwidth]{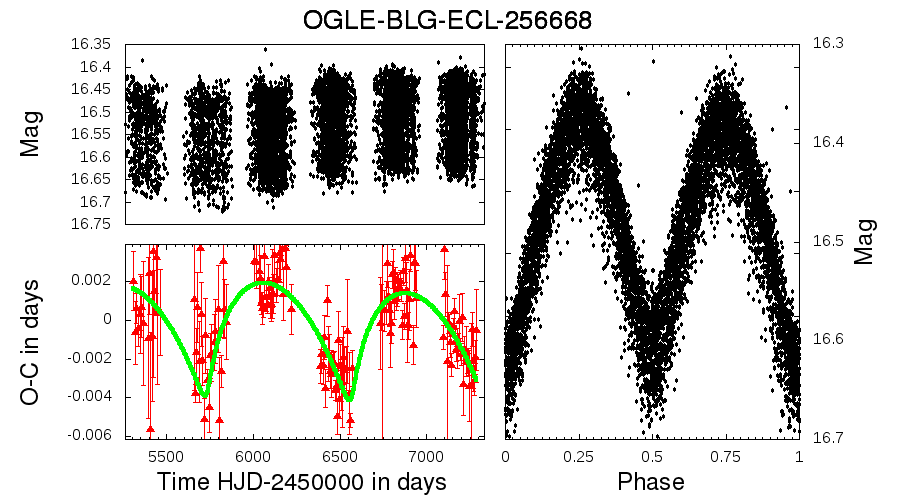}                           
\end{figure*}                           
\clearpage                           
                           
\begin{figure*}                           
                           
\includegraphics[width=\columnwidth]{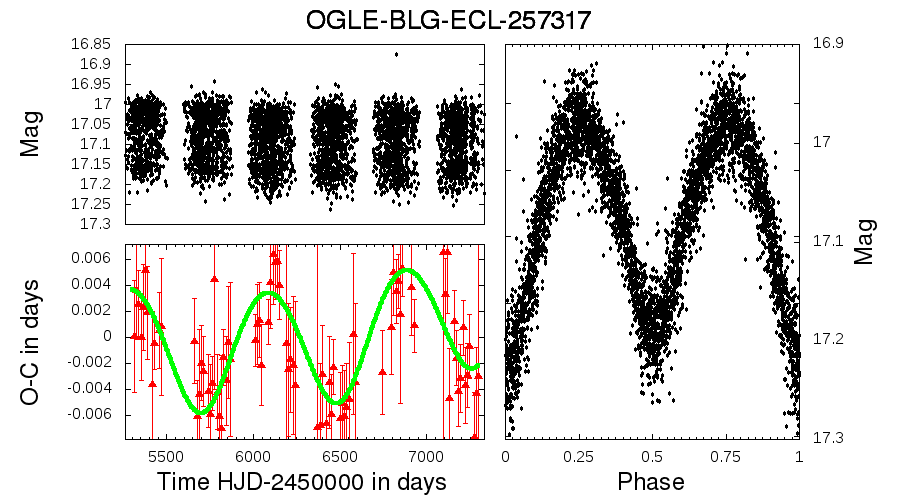}                           
\includegraphics[width=\columnwidth]{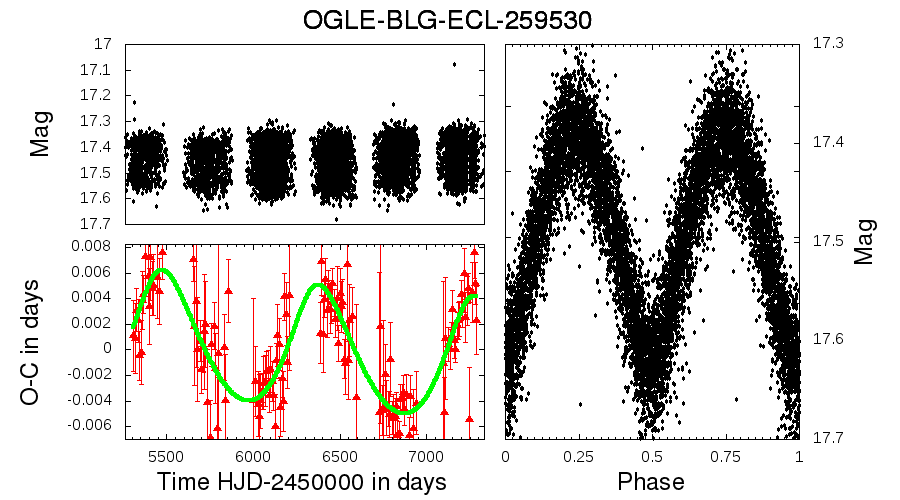}                           
                           
\includegraphics[width=\columnwidth]{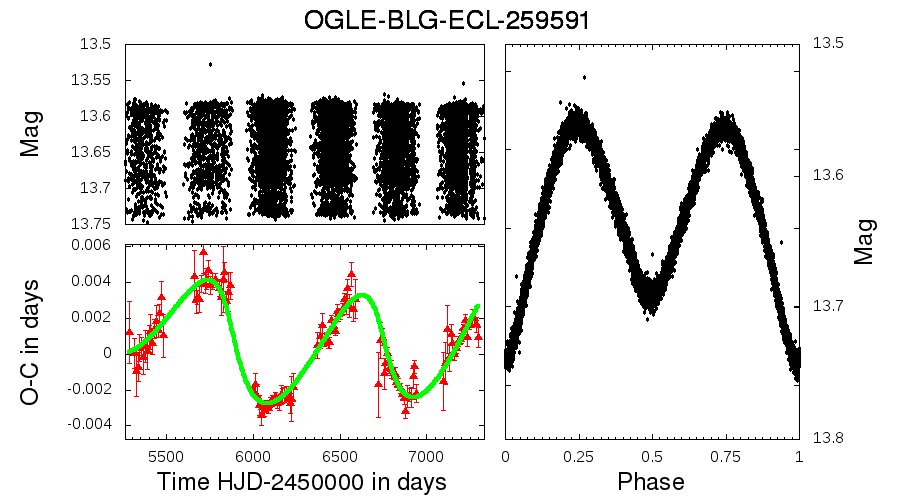}                           
\includegraphics[width=\columnwidth]{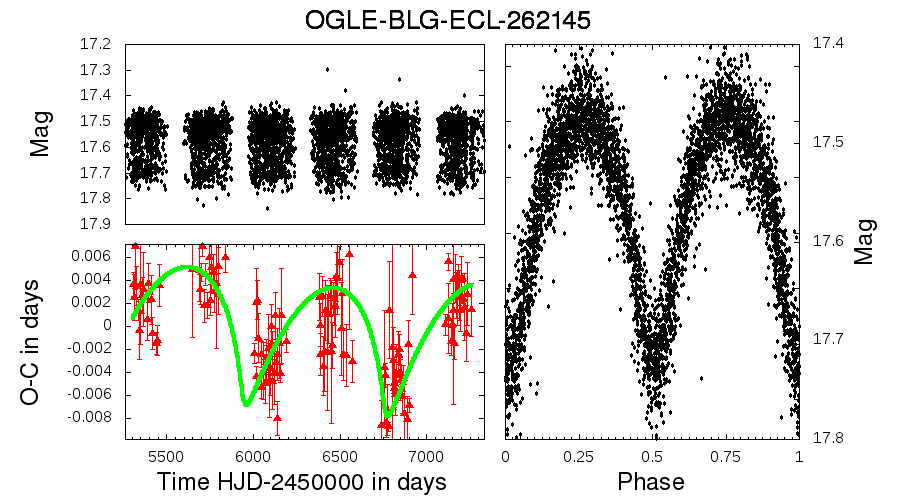}                           
                           
\includegraphics[width=\columnwidth]{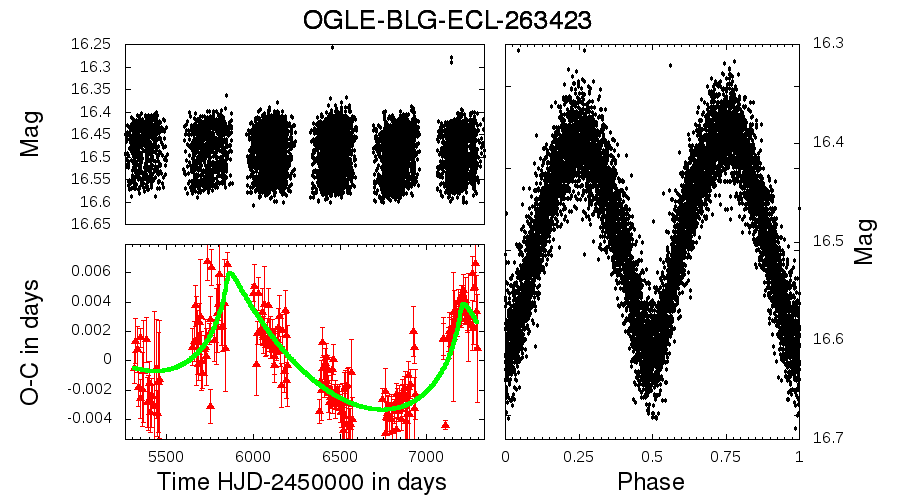}                           
\includegraphics[width=\columnwidth]{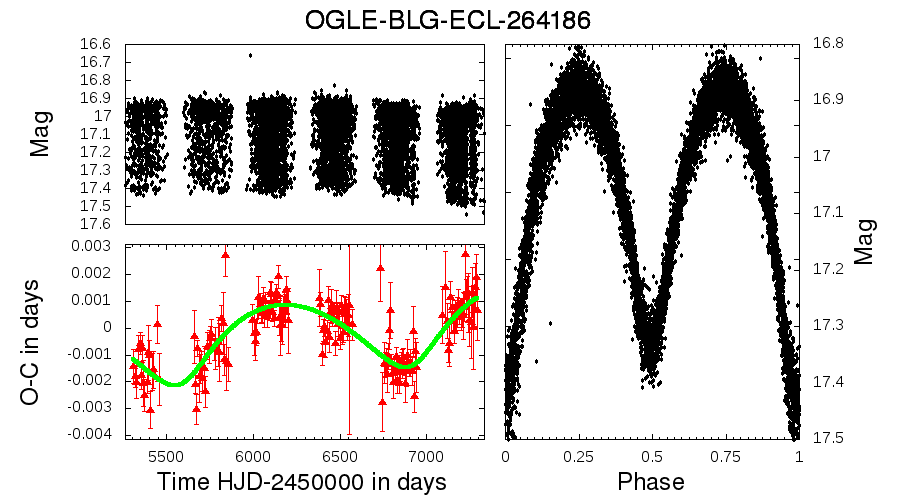}                           
                           
\includegraphics[width=\columnwidth]{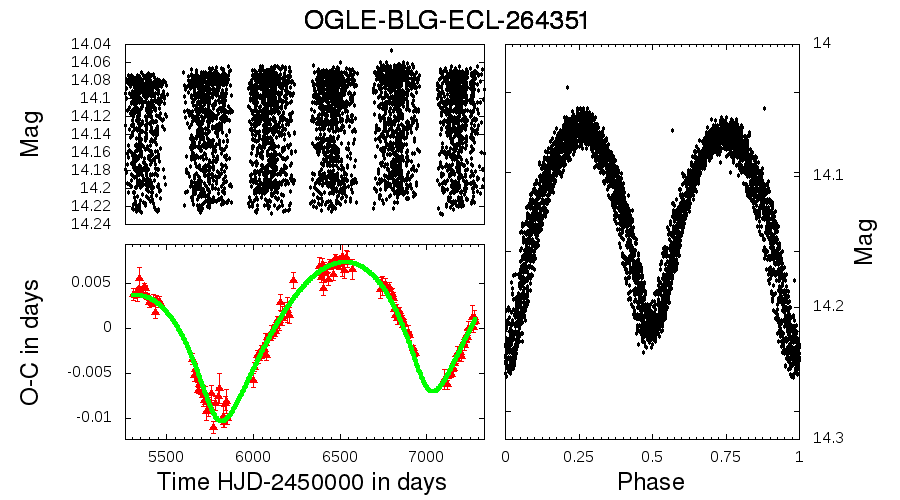}                           
\includegraphics[width=\columnwidth]{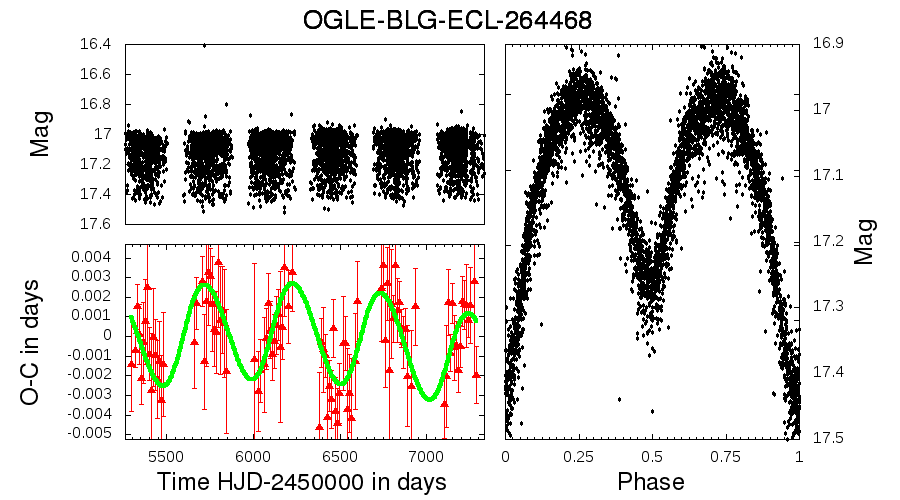}                           
                           
\includegraphics[width=\columnwidth]{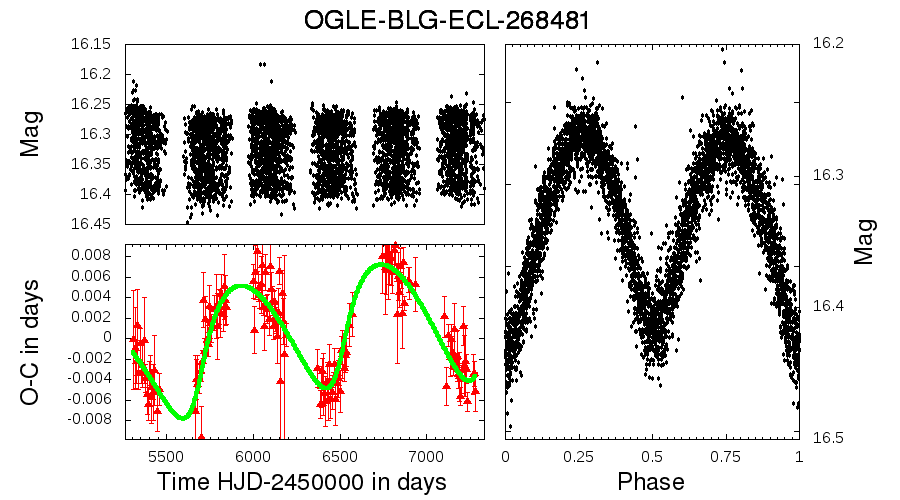}                           
\includegraphics[width=\columnwidth]{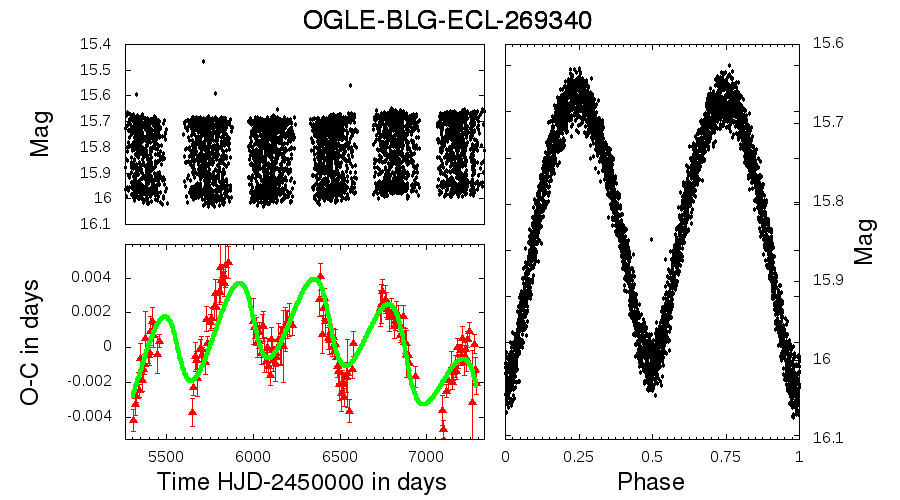}                           
\end{figure*}                           
\clearpage                           
                           
\begin{figure*}                           
                           
\includegraphics[width=\columnwidth]{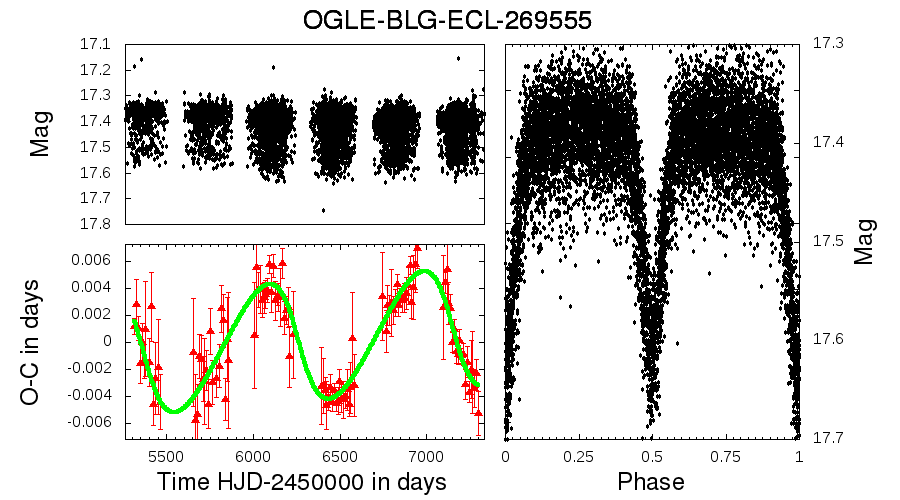}                           
\includegraphics[width=\columnwidth]{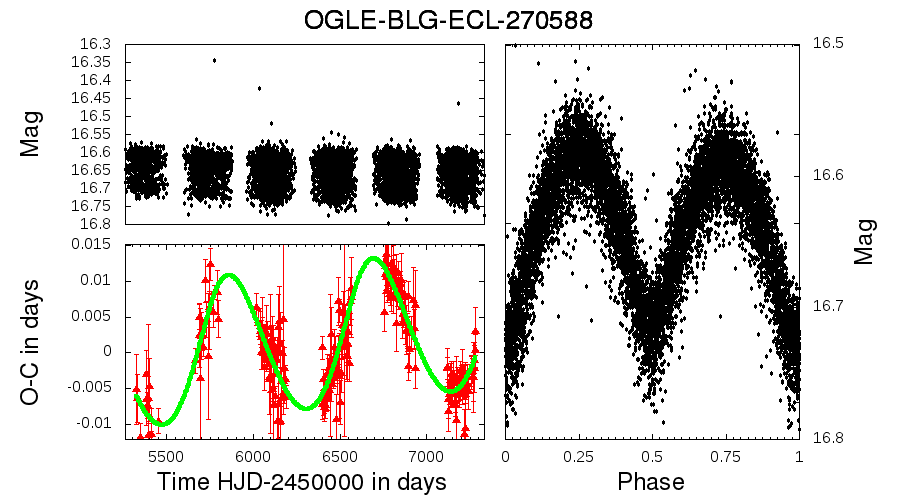}                           
                           
\includegraphics[width=\columnwidth]{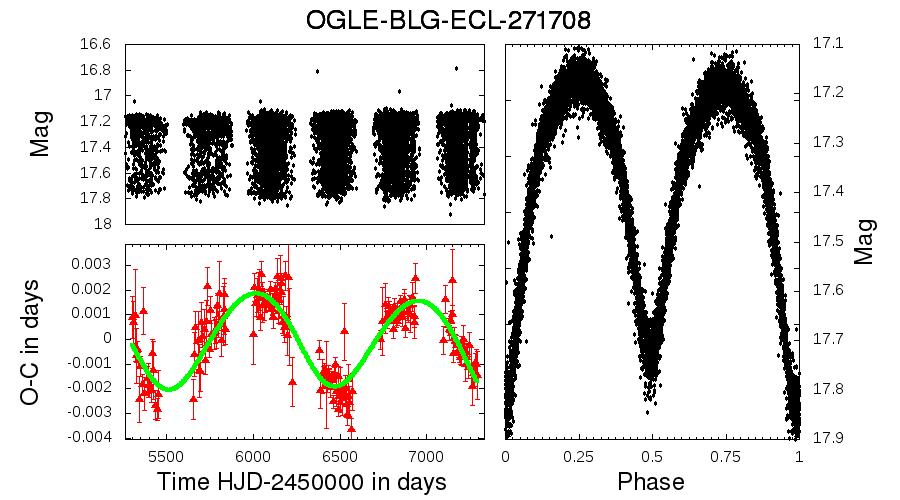}                           
\includegraphics[width=\columnwidth]{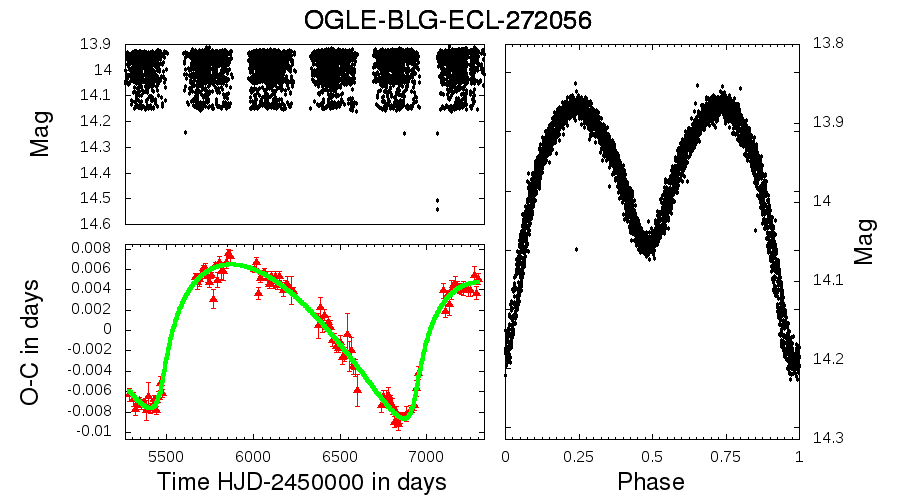}                           
                           
\includegraphics[width=\columnwidth]{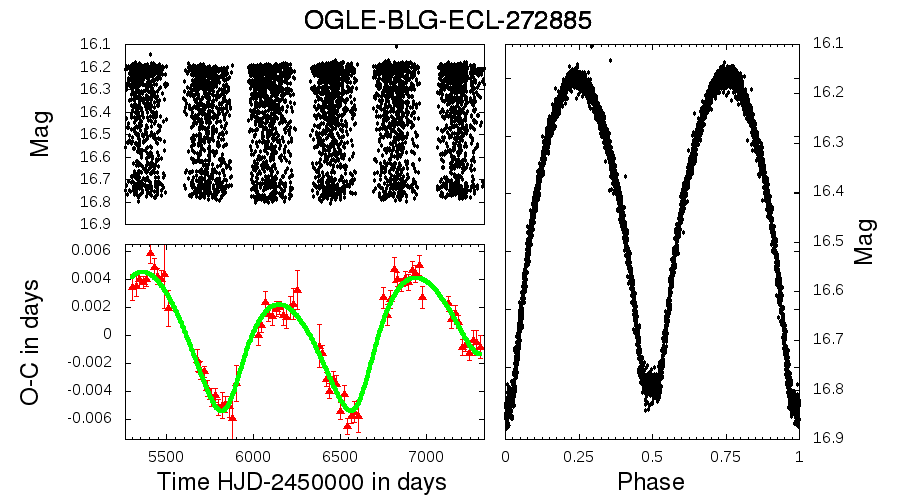}                           
\includegraphics[width=\columnwidth]{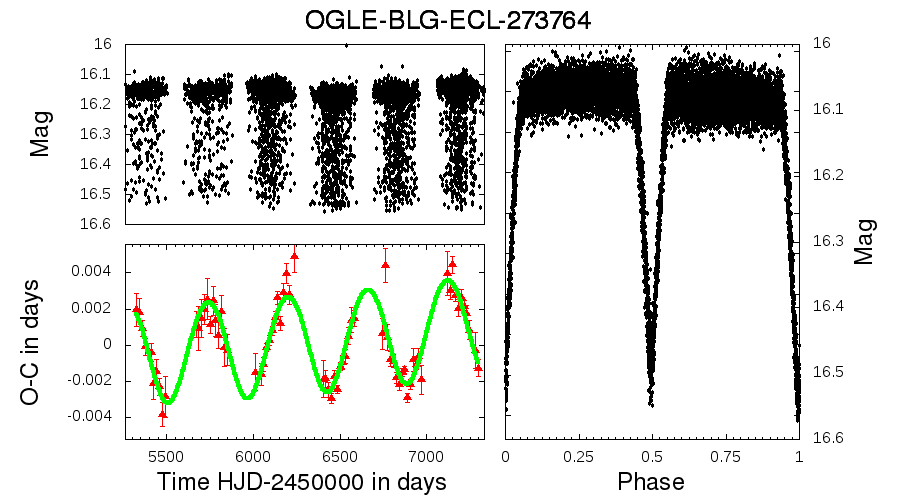}                           
                           
\includegraphics[width=\columnwidth]{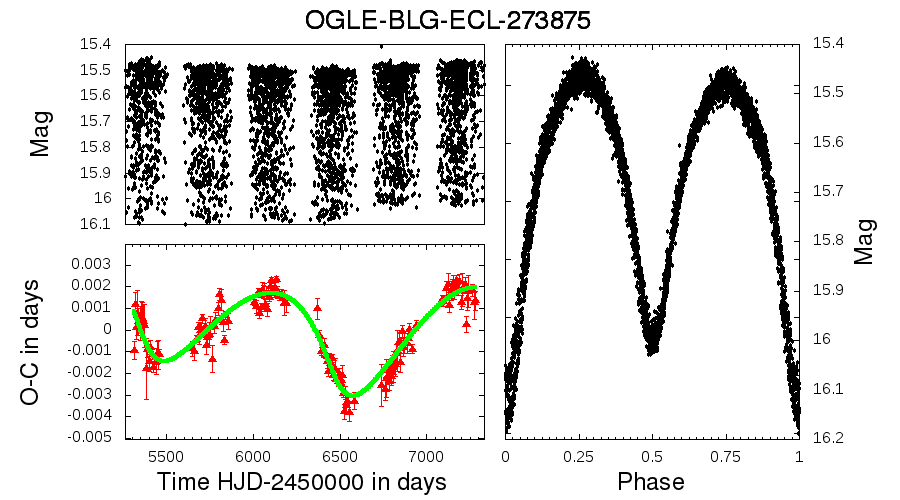}                           
\includegraphics[width=\columnwidth]{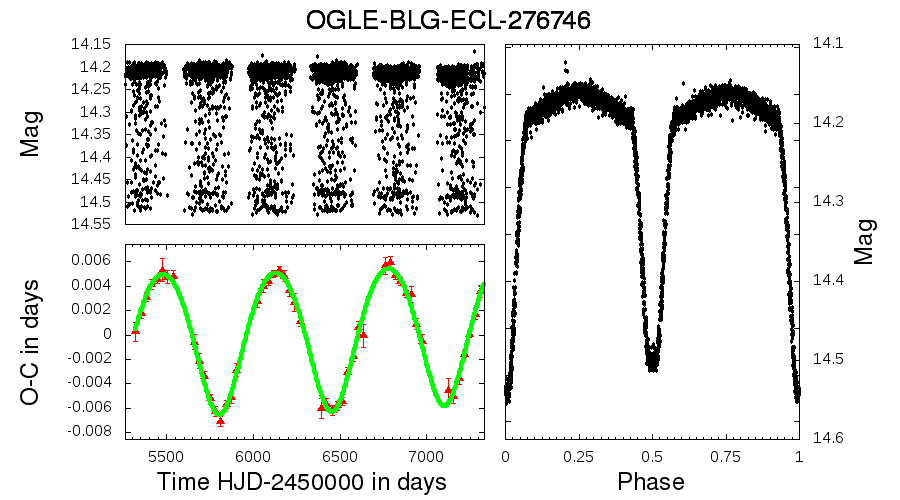}                           
                           
\includegraphics[width=\columnwidth]{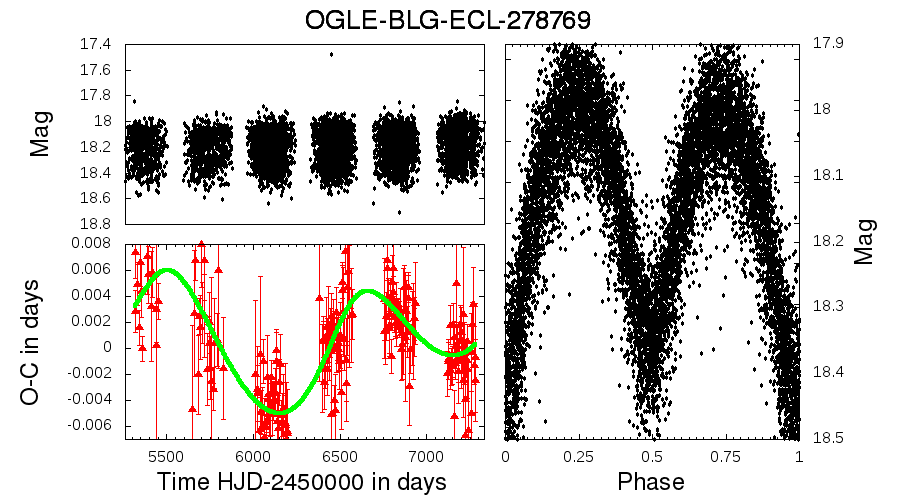}                           
\includegraphics[width=\columnwidth]{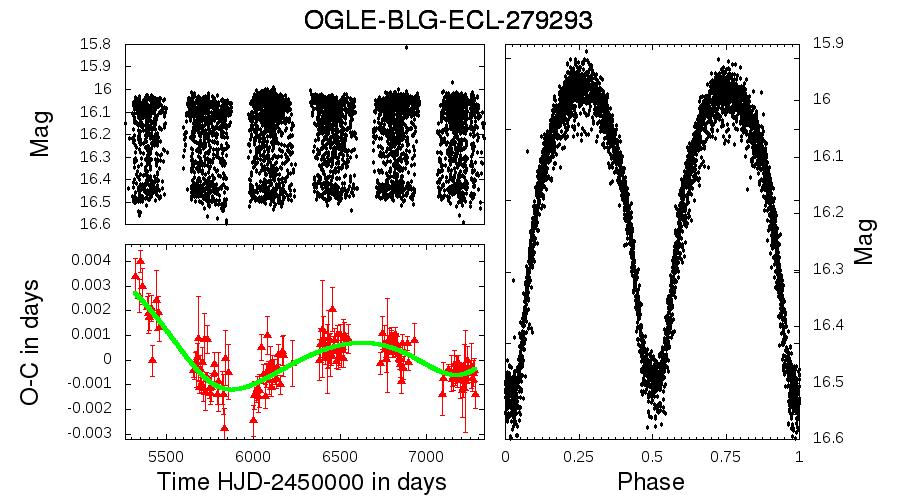}                           
\end{figure*}                           
\clearpage                           
                           
\begin{figure*}                           
                           
\includegraphics[width=\columnwidth]{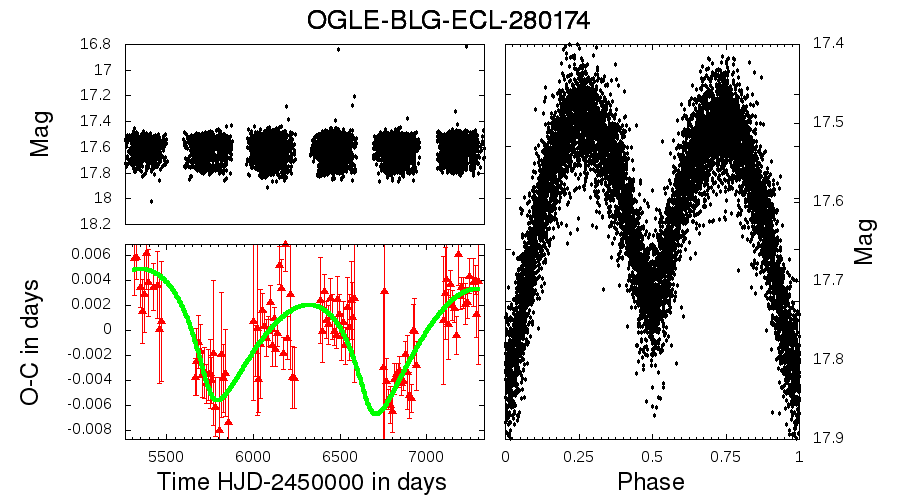}                           
\includegraphics[width=\columnwidth]{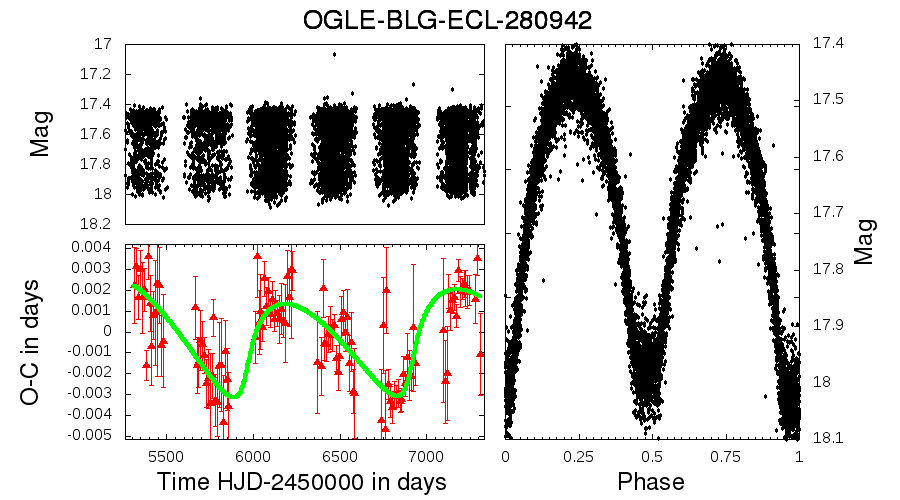}                           
                           
\includegraphics[width=\columnwidth]{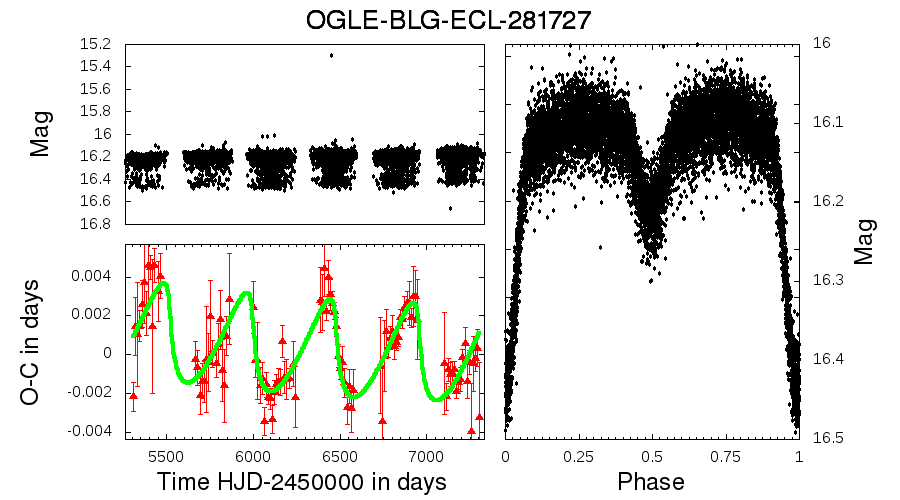}                           
\includegraphics[width=\columnwidth]{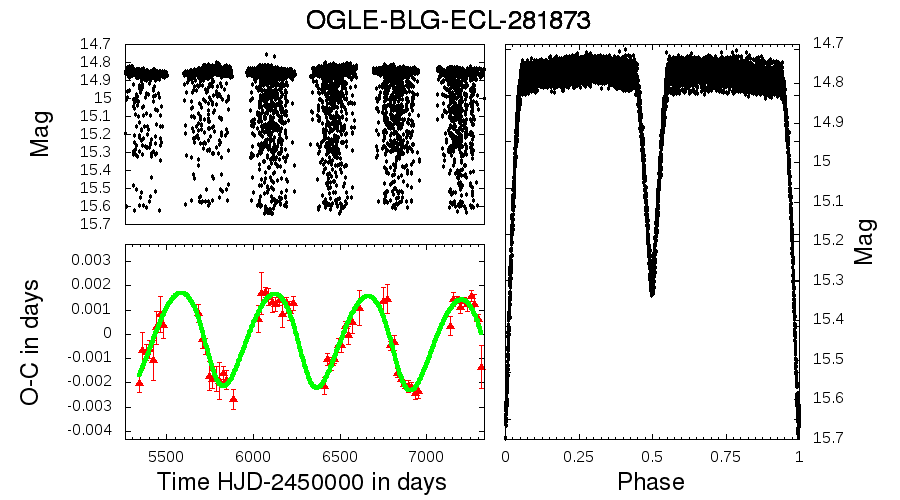}                           
                           
\includegraphics[width=\columnwidth]{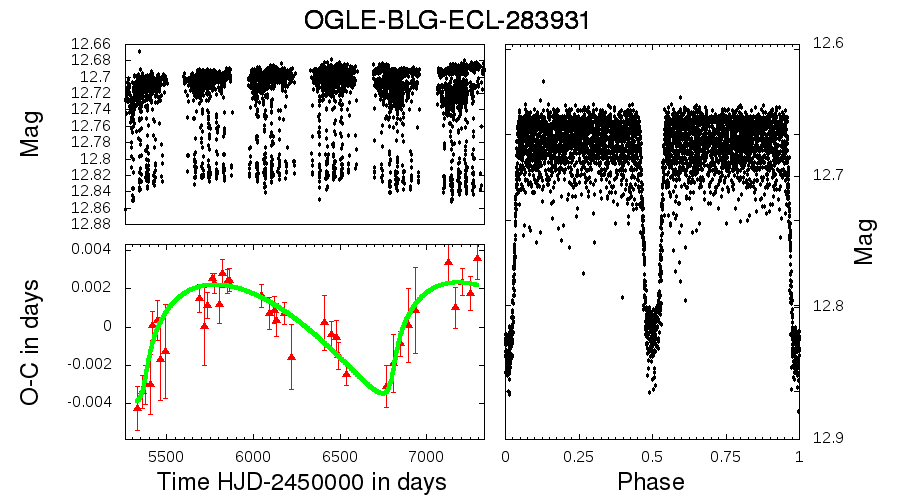}                           
\includegraphics[width=\columnwidth]{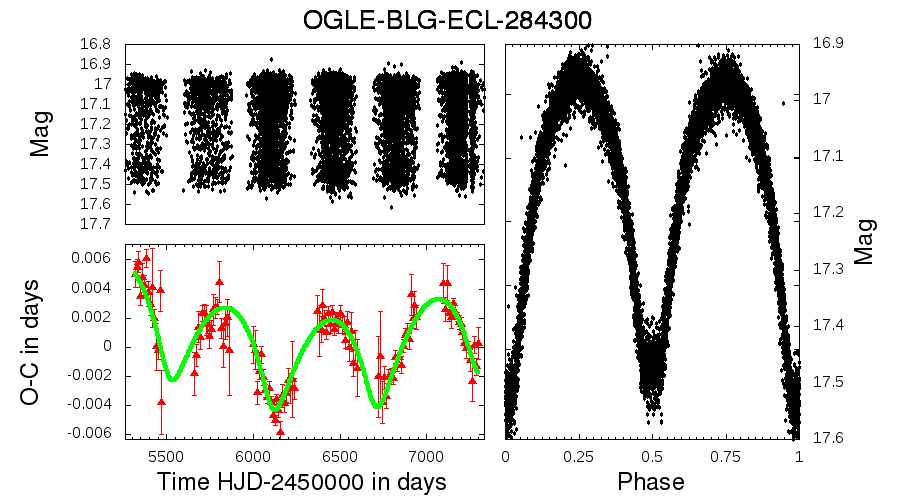}                           
                           
\includegraphics[width=\columnwidth]{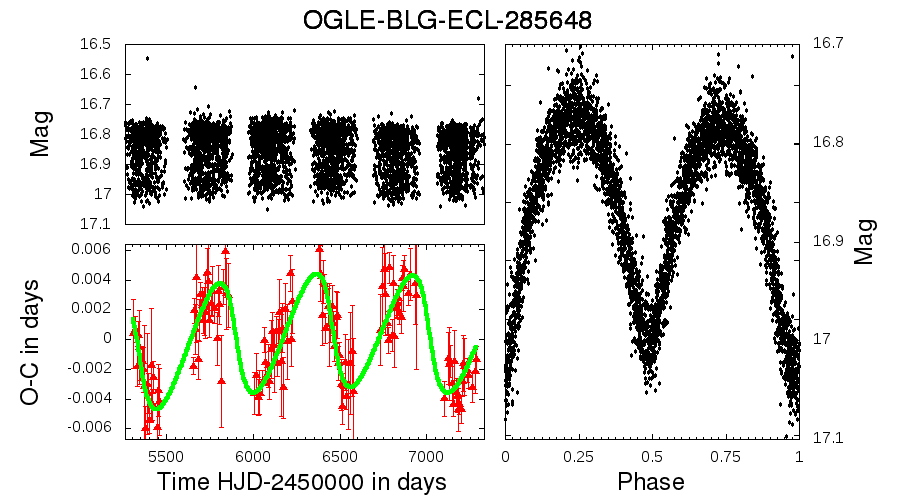}                           
\includegraphics[width=\columnwidth]{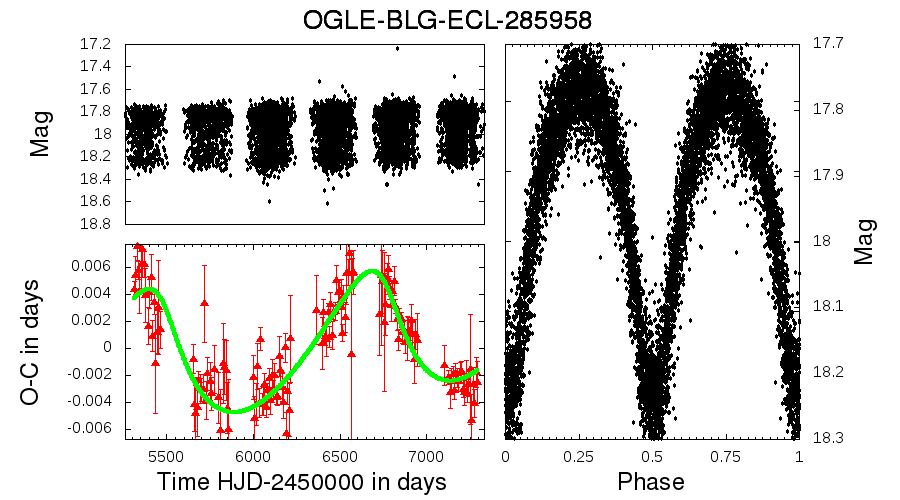}                           
                           
\includegraphics[width=\columnwidth]{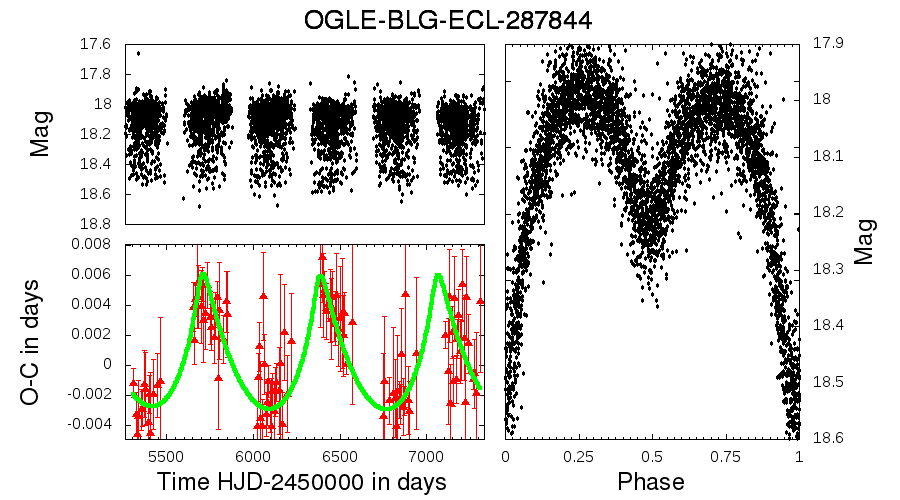}                           
\includegraphics[width=\columnwidth]{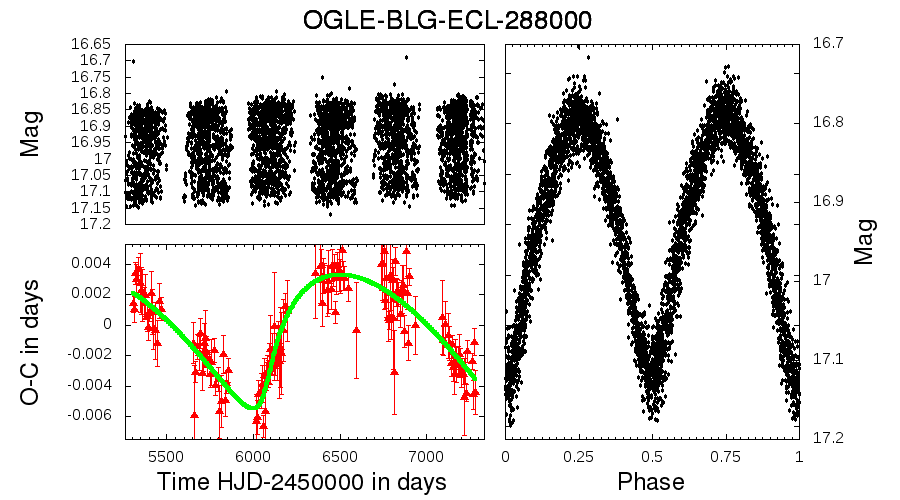}                           
\end{figure*}                           
\clearpage                           
                           
\begin{figure*}                           
                           
\includegraphics[width=\columnwidth]{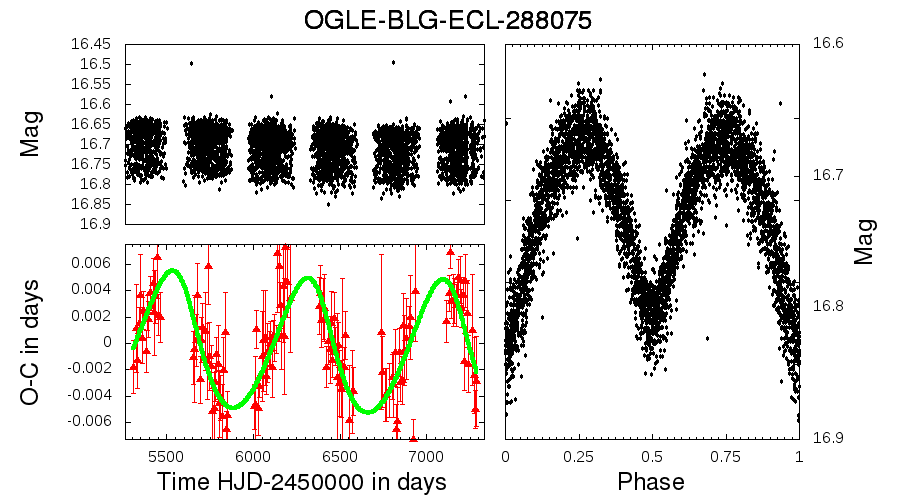}                           
\includegraphics[width=\columnwidth]{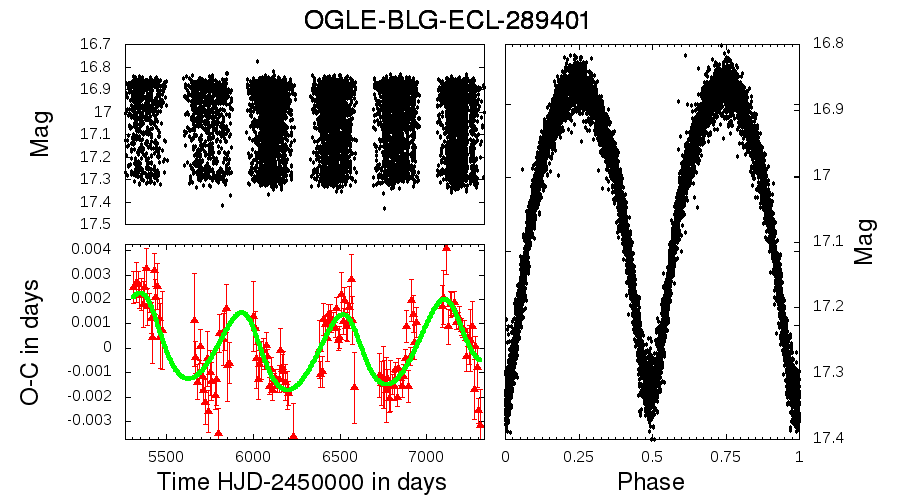}                           
                           
\includegraphics[width=\columnwidth]{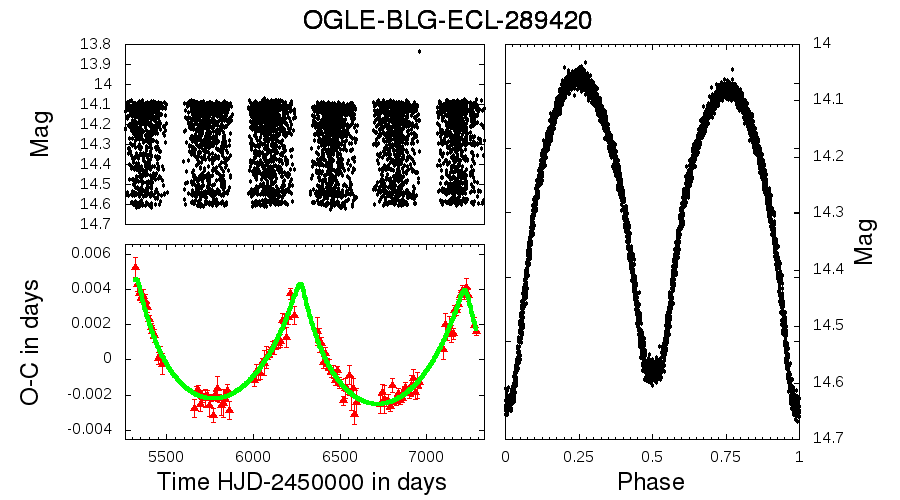}                           
\includegraphics[width=\columnwidth]{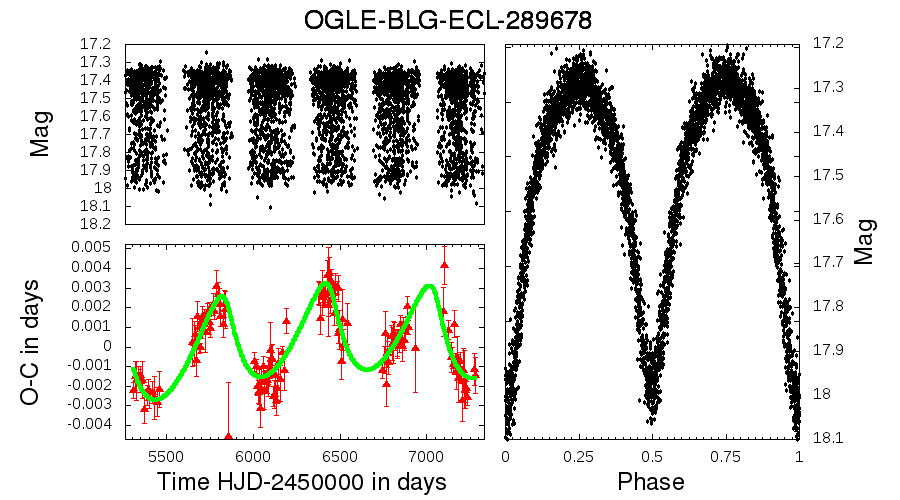}                           
                           
\includegraphics[width=\columnwidth]{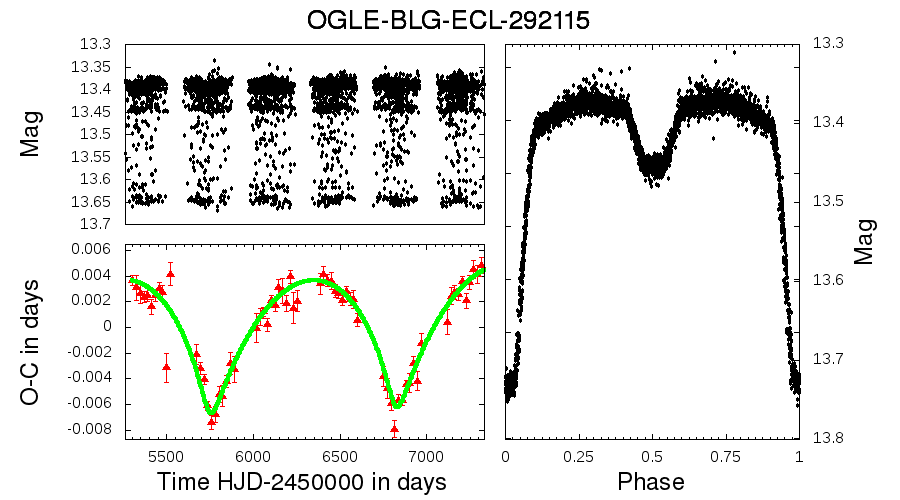}                           
\includegraphics[width=\columnwidth]{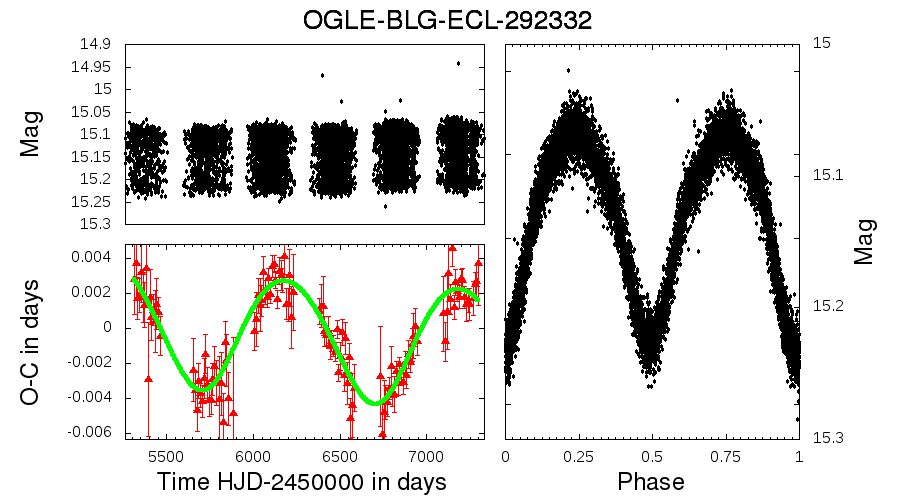}                           
                           
\includegraphics[width=\columnwidth]{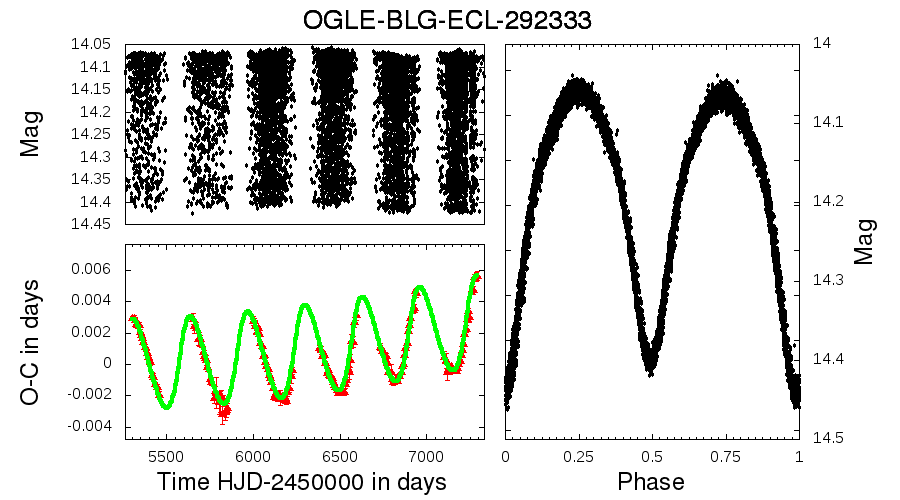}                           
\includegraphics[width=\columnwidth]{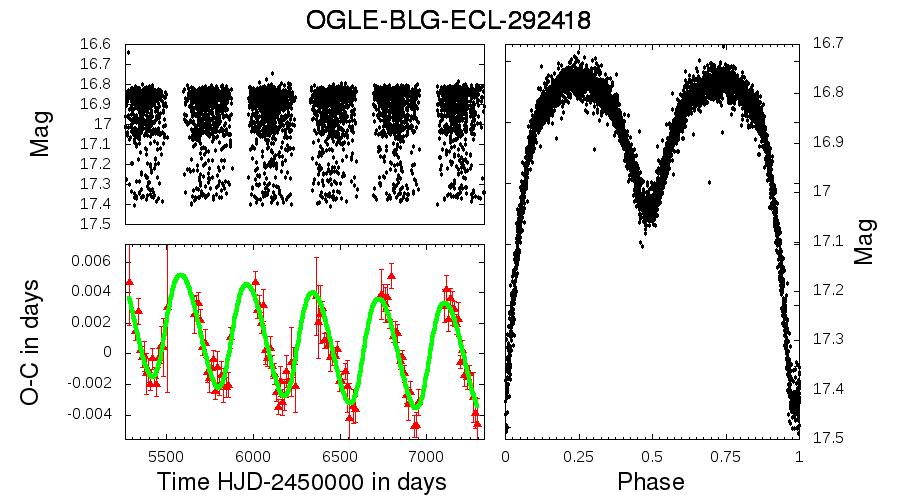}                           
                           
\includegraphics[width=\columnwidth]{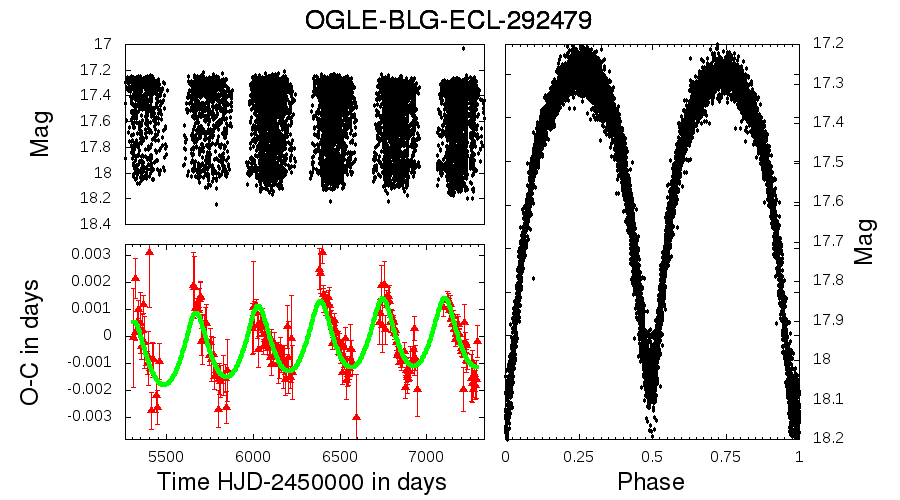}                           
\includegraphics[width=\columnwidth]{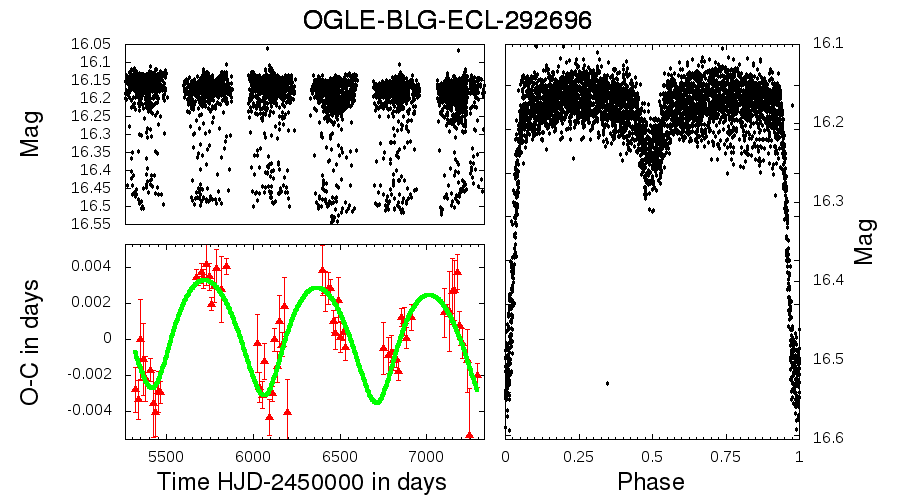}                           
\end{figure*}                           
\clearpage                           
                           
\begin{figure*}                           
                           
\includegraphics[width=\columnwidth]{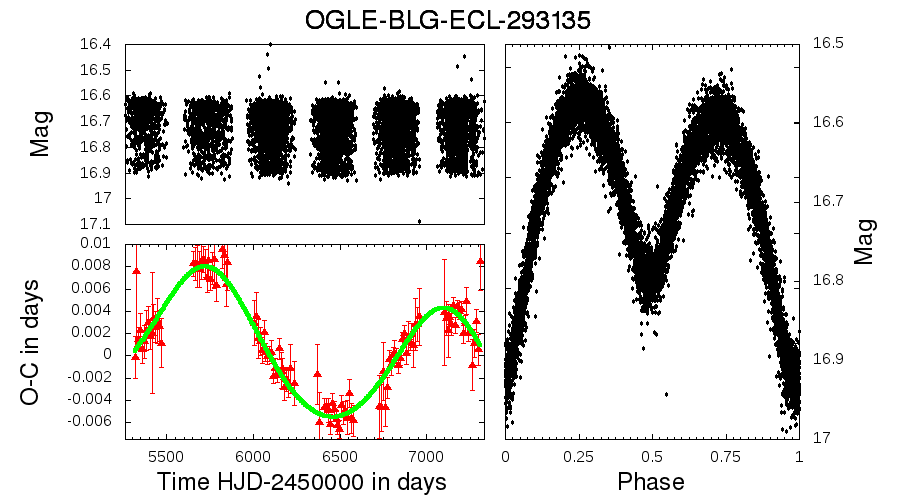}                           
\includegraphics[width=\columnwidth]{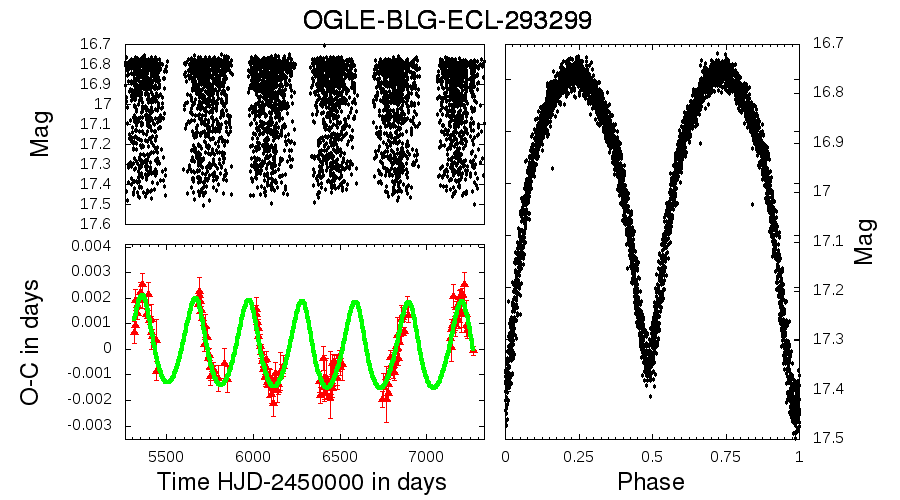}                           
                           
\includegraphics[width=\columnwidth]{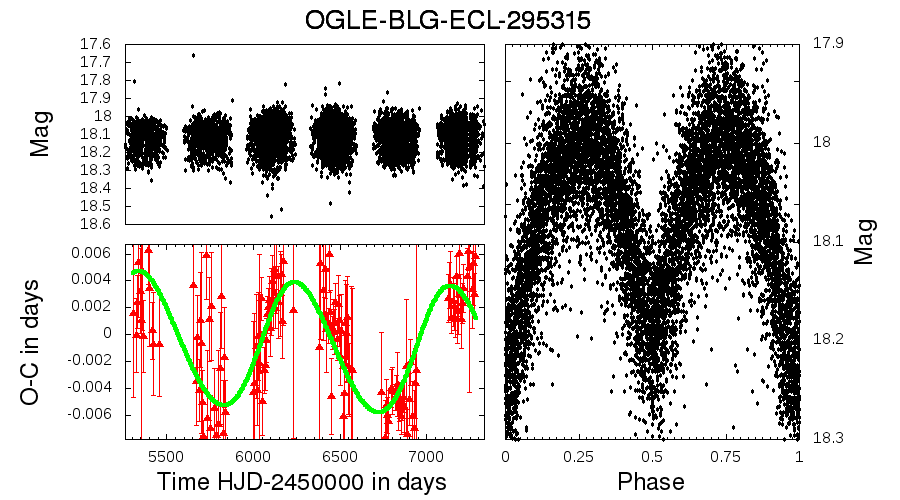}                           
\includegraphics[width=\columnwidth]{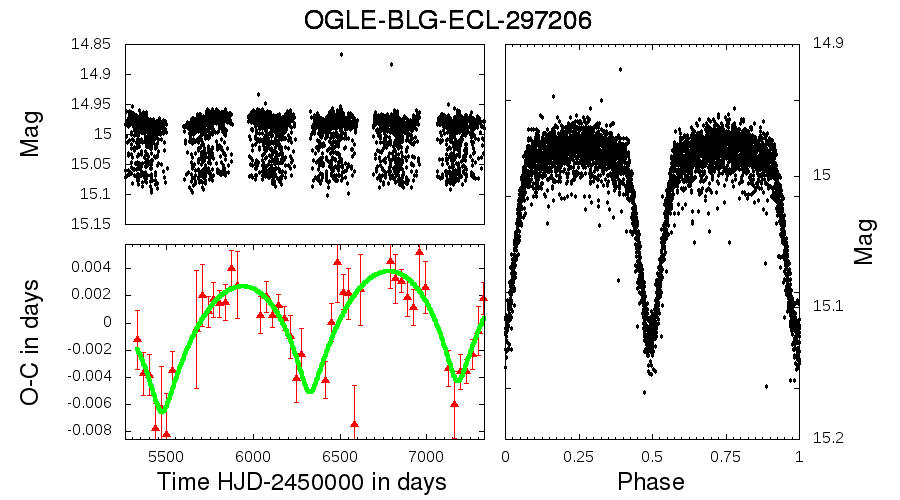}                           
                           
\includegraphics[width=\columnwidth]{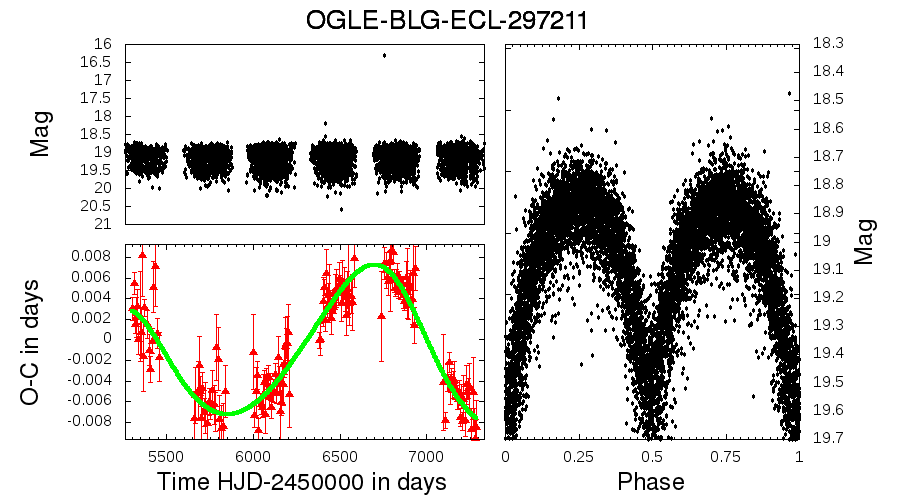}                           
\includegraphics[width=\columnwidth]{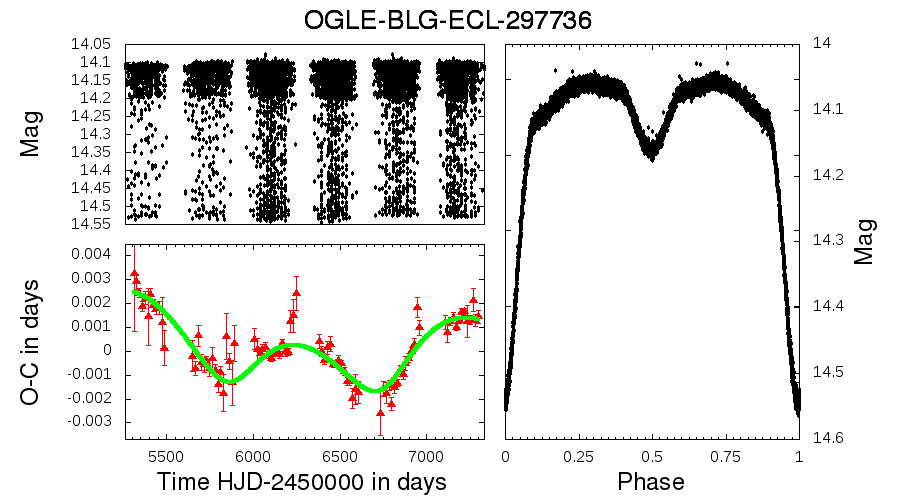}                           
                           
\includegraphics[width=\columnwidth]{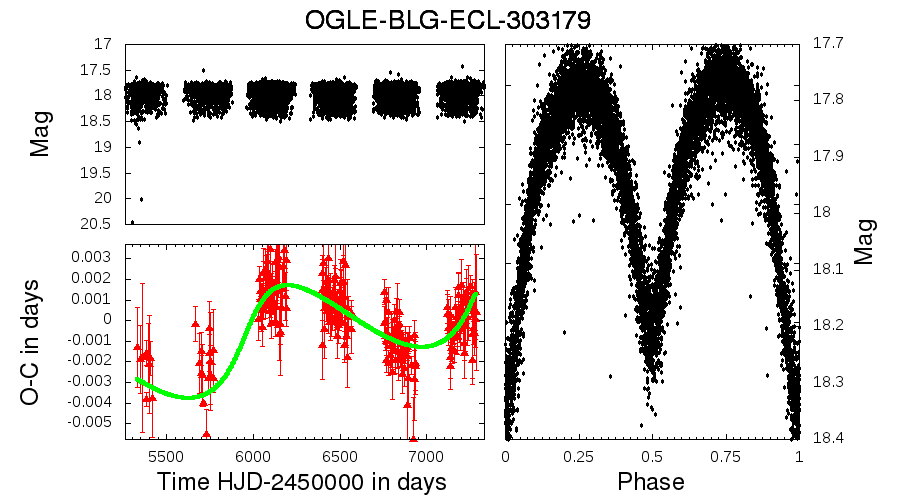}                           
\includegraphics[width=\columnwidth]{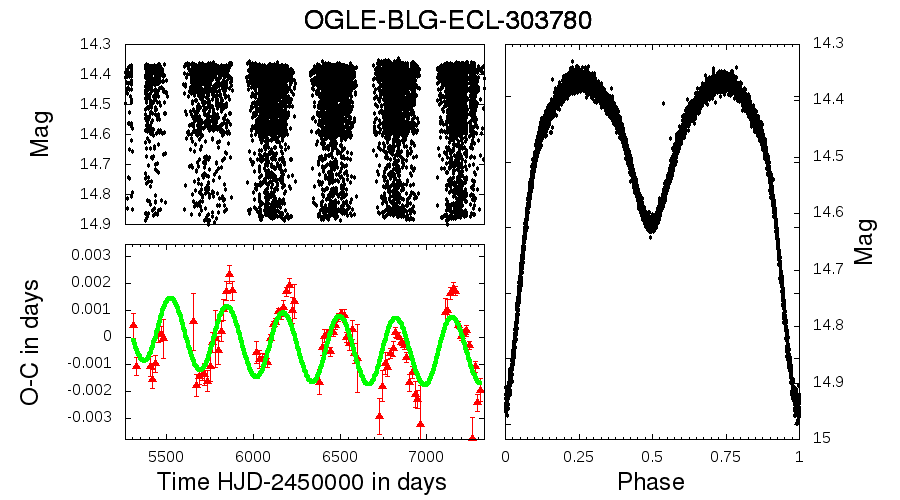}                           
\end{figure*}

\begin{table*}                           
\caption{The remaining hierarchical triple star candidates and their fitted orbital parameters.}
\label{Triples}                           
\centering                           
\begin{tabular}{c c c c @{$\pm$} c c @{$\pm$} c c @{$\pm$} c c @{$\pm$} c c @{$\pm$} c c @{$\pm$} c}         
\hline                           
ID & $T_0$ & $P_1$ & \multicolumn{2}{c}{$P_2$} & \multicolumn{2}{c}{ $a\cdot\sin(i_2)$}& \multicolumn{2}{c}{$e_2$} & \multicolumn{2}{c}{$\omega_2$} & \multicolumn{2}{c}{$\tau_2$} & \multicolumn{2}{c}{$f(m_c)$} \\          
& [HJD-2450000 days] & [days] & \multicolumn{2}{c}{[days]} & \multicolumn{2}{c}{[$R_{\odot}$]} & \multicolumn{2}{c}{ }                
& \multicolumn{2}{c}{[deg]} & \multicolumn{2}{c}{[days]}  \\ \hline                     
65 & 5260.516383 & 0.212209 & 2050.5 & 197.6 & 31.6 & 4.5 & 0.56 & 0.25 & 254.4 & 4.9 & 5811.5 & 184.5 & 0.0001 & 0.00006 \\
72 & 5260.717277 & 0.213604 & 2240.5 & 637.2 & 72.5 & 56.2 & 0.59 & 0.94 & 44.9 & 10.2 & 6000.6 & 311.4 & 0.0010 & 0.00240 \\
74 & 5261.737899 & 0.198408 & 1207.2 & 70.3 & 64.1 & 25.2 & 0.77 & 0.32 & 42.9 & 3.0 & 5741.3 & 51.1 & 0.0024 & 0.00231 \\
106 & 5260.691963 & 0.218233 & 2154.3 & 886.4 & 37.0 & 62.8 & 0.08 & 1.57 & 327.5 & 79.5 & 6396.4 & 3534.0 & 0.0001 & 0.00046 \\
28702 & 5265.512011 & 0.446362 & 1794.3 & 73.9 & 100.1 & 37.2 & 0.86 & 0.25 & 308.2 & 4.6 & 5403.9 & 38.9 & 0.0042 & 0.00364 \\
28743 & 5265.461422 & 0.303213 & 1642.9 & 32.5 & 186.5 & 62.3 & 0.74 & 0.14 & 8.4 & 1.0 & 5240.9 & 29.6 & 0.0322 & 0.02342 \\
\end{tabular}                           
\end{table*}                           
                           
\begin{table*}                           
\centering                           
\begin{tabular}{c c c c @{$\pm$} c c @{$\pm$} c c @{$\pm$} c c @{$\pm$} c c @{$\pm$} c c @{$\pm$} c}         
\hline                           
ID & $T_0$ & $P_1$ & \multicolumn{2}{c}{$P_2$} & \multicolumn{2}{c}{ $a\cdot\sin(i_2)$}& \multicolumn{2}{c}{$e_2$} & \multicolumn{2}{c}{$\omega_2$} & \multicolumn{2}{c}{$\tau_2$} & \multicolumn{2}{c}{$f(m_c)$} \\          
& [HJD-2450000 days] & [days] & \multicolumn{2}{c}{[days]} & \multicolumn{2}{c}{[$R_{\odot}$]} & \multicolumn{2}{c}{ }                
& \multicolumn{2}{c}{[deg]} & \multicolumn{2}{c}{[days]}  \\ \hline                     
125403 & 5261.511565 & 0.410899 & 2500.3 & 542.1 & 67.6 & 48.0 & 0.18 & 0.50 & 77.5 & 23.8 & 5884.9 & 941.7 & 0.0007 & 0.00145 \\
125481 & 5260.641995 & 0.358563 & 2302.7 & 42.5 & 73.6 & 82.3 & 0.63 & 0.12 & 245.8 & 3.8 & 7209.3 & 140.4 & 0.0010 & 0.00229 \\
125770 & 5259.405696 & 2.971372 & 2110.5 & 1311.8 & 236.8 & 253.7 & 0.83 & 0.52 & 163.3 & 3.8 & 6031.1 & 56.2 & 0.0400 & 0.16030 \\
125830 & 5260.721097 & 0.313929 & 1183.3 & 31.1 & 86.6 & 16.3 & 0.76 & 0.15 & 141.4 & 1.3 & 5707.8 & 19.2 & 0.0062 & 0.00282 \\
126315 & 5260.759303 & 0.350349 & 3185.4 & 231.0 & 144.6 & 146.4 & 0.32 & 0.66 & 131.7 & 7.2 & 6503.3 & 379.9 & 0.0040 & 0.00897 \\
126583 & 5273.194085 & 3.026972 & 2752.9 & 28.9 & 390.6 & 108.9 & 0.80 & 0.14 & 175.9 & 1.2 & 6467.0 & 31.3 & 0.1054 & 0.06209 \\
\end{tabular}                           
\end{table*}                           
                           
\begin{table*}                           
\centering                           
\begin{tabular}{c c c c @{$\pm$} c c @{$\pm$} c c @{$\pm$} c c @{$\pm$} c c @{$\pm$} c c @{$\pm$} c}         
\hline                           
ID & $T_0$ & $P_1$ & \multicolumn{2}{c}{$P_2$} & \multicolumn{2}{c}{ $a\cdot\sin(i_2)$}& \multicolumn{2}{c}{$e_2$} & \multicolumn{2}{c}{$\omega_2$} & \multicolumn{2}{c}{$\tau_2$} & \multicolumn{2}{c}{$f(m_c)$} \\          
& [HJD-2450000 days] & [days] & \multicolumn{2}{c}{[days]} & \multicolumn{2}{c}{[$R_{\odot}$]} & \multicolumn{2}{c}{ }                
& \multicolumn{2}{c}{[deg]} & \multicolumn{2}{c}{[days]}  \\ \hline                     
141571 & 5260.602668 & 0.296791 & 2057.3 & 145.9 & 68.4 & 27.0 & 0.18 & 0.16 & 315.8 & 7.9 & 5560.7 & 318.6 & 0.0010 & 0.00100 \\
141640 & 5260.315960 & 1.467467 & 1898.9 & 155.6 & 184.5 & 77.2 & 0.58 & 0.22 & 4.6 & 4.9 & 5214.5 & 96.0 & 0.0233 & 0.02523 \\
142758 & 5260.620701 & 0.357872 & 1540.0 & 47.4 & 51.6 & 8.9 & 0.41 & 0.20 & 155.8 & 2.7 & 5681.9 & 71.0 & 0.0008 & 0.00035 \\
142770 & 5260.614676 & 0.294321 & 1701.7 & 156.7 & 127.3 & 13.0 & 0.23 & 0.16 & 50.7 & 4.1 & 5885.2 & 204.7 & 0.0095 & 0.00456 \\
142919 & 5258.962314 & 2.152708 & 1500.2 & 65.3 & 118.4 & 9.5 & 0.11 & 0.16 & 185.0 & 8.5 & 5856.3 & 280.5 & 0.0099 & 0.00288 \\
142943 & 5260.789703 & 0.334867 & 2076.1 & 264.6 & 115.3 & 67.2 & 0.37 & 0.19 & 152.6 & 6.9 & 3544.0 & 226.0 & 0.0048 & 0.00743 \\
\end{tabular}                           
\end{table*}                           
                           
\begin{table*}                           
\centering                           
\begin{tabular}{c c c c @{$\pm$} c c @{$\pm$} c c @{$\pm$} c c @{$\pm$} c c @{$\pm$} c c @{$\pm$} c}         
\hline                           
ID & $T_0$ & $P_1$ & \multicolumn{2}{c}{$P_2$} & \multicolumn{2}{c}{ $a\cdot\sin(i_2)$}& \multicolumn{2}{c}{$e_2$} & \multicolumn{2}{c}{$\omega_2$} & \multicolumn{2}{c}{$\tau_2$} & \multicolumn{2}{c}{$f(m_c)$} \\          
& [HJD-2450000 days] & [days] & \multicolumn{2}{c}{[days]} & \multicolumn{2}{c}{[$R_{\odot}$]} & \multicolumn{2}{c}{ }                
& \multicolumn{2}{c}{[deg]} & \multicolumn{2}{c}{[days]}  \\ \hline                     
157869 & 5259.494583 & 1.016233 & 3010.6 & 100.8 & 382.9 & 422.6 & 0.27 & 0.29 & 257.4 & 3.7 & 7191.5 & 450.5 & 0.0830 & 0.19155 \\
158162 & 5260.513192 & 0.290645 & 1130.6 & 35.8 & 54.0 & 6.2 & 0.44 & 0.21 & 35.8 & 2.9 & 6074.2 & 67.6 & 0.0017 & 0.00055 \\
158259 & 5261.712132 & 0.309479 & 1505.2 & 39.6 & 141.1 & 27.7 & 0.68 & 0.14 & 355.3 & 0.8 & 5795.1 & 32.3 & 0.0166 & 0.00783 \\
158555 & 5261.536942 & 0.231007 & 1690.4 & 61.1 & 99.2 & 12.2 & 0.59 & 0.19 & 228.5 & 2.9 & 6482.0 & 104.6 & 0.0046 & 0.00163 \\
159473 & 5260.558498 & 0.290938 & 2314.3 & 37.8 & 68.8 & 3.6 & 0.82 & 0.09 & 114.9 & 1.3 & 6106.8 & 19.1 & 0.0008 & 0.00012 \\
159706 & 5260.743536 & 1.171151 & 1592.9 & 25.3 & 167.9 & 28.1 & 0.89 & 0.08 & 132.7 & 1.3 & 5368.4 & 11.0 & 0.0250 & 0.00956 \\
\end{tabular}                           
\end{table*}                           
                           
\begin{table*}                           
\centering                           
\begin{tabular}{c c c c @{$\pm$} c c @{$\pm$} c c @{$\pm$} c c @{$\pm$} c c @{$\pm$} c c @{$\pm$} c}         
\hline                           
ID & $T_0$ & $P_1$ & \multicolumn{2}{c}{$P_2$} & \multicolumn{2}{c}{ $a\cdot\sin(i_2)$}& \multicolumn{2}{c}{$e_2$} & \multicolumn{2}{c}{$\omega_2$} & \multicolumn{2}{c}{$\tau_2$} & \multicolumn{2}{c}{$f(m_c)$} \\          
& [HJD-2450000 days] & [days] & \multicolumn{2}{c}{[days]} & \multicolumn{2}{c}{[$R_{\odot}$]} & \multicolumn{2}{c}{ }                
& \multicolumn{2}{c}{[deg]} & \multicolumn{2}{c}{[days]}  \\ \hline                     
175945 & 5260.495525 & 0.291299 & 1904.6 & 78.9 & 63.3 & 5.2 & 0.28 & 0.08 & 309.7 & 2.1 & 6677.7 & 62.2 & 0.0009 & 0.00026 \\
176931 & 5260.603845 & 0.339735 & 1106.3 & 44.1 & 155.8 & 13.3 & 0.52 & 0.25 & 232.1 & 1.5 & 6421.2 & 41.9 & 0.0414 & 0.01202 \\
176988 & 5306.323717 & 0.788437 & 2304.8 & 67.9 & 286.6 & 972.7 & 0.53 & 0.75 & 291.4 & 19.1 & 4963.4 & 316.4 & 0.0594 & 0.40845 \\
177207 & 5261.591931 & 0.305253 & 2381.5 & 616.2 & 54.4 & 36.6 & 0.30 & 0.59 & 71.2 & 5.0 & 6248.3 & 562.2 & 0.0004 & 0.00085 \\
177583 & 5260.451889 & 0.602832 & 1028.5 & 83.6 & 32.9 & 6.2 & 0.52 & 0.43 & 53.9 & 4.1 & 5444.1 & 102.2 & 0.0005 & 0.00031 \\
177645 & 5260.542598 & 0.318826 & 1457.3 & 78.9 & 81.7 & 42.7 & 0.68 & 0.36 & 12.2 & 2.9 & 6827.0 & 69.1 & 0.0034 & 0.00411 \\
\end{tabular}                           
\end{table*}                           
                           
\begin{table*}                           
\centering                           
\begin{tabular}{c c c c @{$\pm$} c c @{$\pm$} c c @{$\pm$} c c @{$\pm$} c c @{$\pm$} c c @{$\pm$} c}         
\hline                           
ID & $T_0$ & $P_1$ & \multicolumn{2}{c}{$P_2$} & \multicolumn{2}{c}{ $a\cdot\sin(i_2)$}& \multicolumn{2}{c}{$e_2$} & \multicolumn{2}{c}{$\omega_2$} & \multicolumn{2}{c}{$\tau_2$} & \multicolumn{2}{c}{$f(m_c)$} \\          
& [HJD-2450000 days] & [days] & \multicolumn{2}{c}{[days]} & \multicolumn{2}{c}{[$R_{\odot}$]} & \multicolumn{2}{c}{ }                
& \multicolumn{2}{c}{[deg]} & \multicolumn{2}{c}{[days]}  \\ \hline                     190105 & 5260.665725 & 0.307720 & 3543.2 & 92.8 & 181.5 & 256.8 & 0.35 & 0.17 & 207.6 & 17.5 & 9175.9 & 663.3 & 0.0064 & 0.01861 \\
190163 & 5260.408664 & 0.305157 & 1432.8 & 105.0 & 35.8 & 6.0 & 0.52 & 0.30 & 34.7 & 3.5 & 5672.7 & 95.4 & 0.0003 & 0.00017 \\
190462 & 5260.461767 & 0.363808 & 1735.0 & 156.5 & 37.9 & 8.5 & 0.03 & 0.27 & 235.3 & 31.6 & 7044.4 & 1188.0 & 0.0002 & 0.00014 \\
190852 & 5321.353213 & 0.344180 & 1178.0 & 101.9 & 34.9 & 4.8 & 0.42 & 0.28 & 173.9 & 2.4 & 5913.8 & 50.6 & 0.0004 & 0.00021 \\
191088 & 5260.269095 & 0.370798 & 1576.0 & 68.5 & 52.3 & 10.3 & 0.27 & 0.23 & 185.6 & 4.3 & 5278.7 & 132.7 & 0.0008 & 0.00042 \\
191106 & 5260.658524 & 0.404080 & 1677.8 & 131.3 & 204.1 & 41.1 & 0.52 & 0.11 & 89.2 & 0.9 & 5322.7 & 54.7 & 0.0405 & 0.02582 \\
\end{tabular}                           
\end{table*}                           
                           
\begin{table*}                           
\centering                           
\begin{tabular}{c c c c @{$\pm$} c c @{$\pm$} c c @{$\pm$} c c @{$\pm$} c c @{$\pm$} c c @{$\pm$} c}         
\hline                           
ID & $T_0$ & $P_1$ & \multicolumn{2}{c}{$P_2$} & \multicolumn{2}{c}{ $a\cdot\sin(i_2)$}& \multicolumn{2}{c}{$e_2$} & \multicolumn{2}{c}{$\omega_2$} & \multicolumn{2}{c}{$\tau_2$} & \multicolumn{2}{c}{$f(m_c)$} \\          
& [HJD-2450000 days] & [days] & \multicolumn{2}{c}{[days]} & \multicolumn{2}{c}{[$R_{\odot}$]} & \multicolumn{2}{c}{ }                
& \multicolumn{2}{c}{[deg]} & \multicolumn{2}{c}{[days]}  \\ \hline                     
203797 & 5260.802843 & 0.238025 & 2379.2 & 179.6 & 41.7 & 57.6 & 0.24 & 0.38 & 27.7 & 27.7 & 5619.6 & 916.3 & 0.0002 & 0.00060 \\
203804 & 5260.683901 & 0.316744 & 2064.0 & 757.3 & 99.1 & 51.2 & 0.01 & 0.62 & 86.9 & 237.2 & 6515.9 & 10250.1 & 0.0031 & 0.00662 \\
204340 & 5260.911108 & 0.283696 & 1388.8 & 559.5 & 74.8 & 17.5 & 0.33 & 0.63 & 265.6 & 8.3 & 5047.5 & 410.5 & 0.0029 & 0.00486 \\
204388 & 5260.292218 & 0.823053 & 1538.6 & 77.5 & 22.0 & 17.3 & 0.80 & 0.37 & 210.5 & 3.6 & 5364.3 & 62.4 & 0.0001 & 0.00017 \\
204870 & 5260.570736 & 0.299349 & 2288.5 & 77.2 & 66.2 & 19.5 & 0.18 & 0.24 & 205.6 & 7.9 & 6162.1 & 317.8 & 0.0007 & 0.00048 \\
205050 & 5260.589843 & 0.370494 & 1953.3 & 36.5 & 47.5 & 19.3 & 0.41 & 0.15 & 201.1 & 3.9 & 7190.3 & 78.3 & 0.0004 & 0.00035 \\
\end{tabular}                           
\end{table*}                           
                           
\begin{table*}                           
\centering                           
\begin{tabular}{c c c c @{$\pm$} c c @{$\pm$} c c @{$\pm$} c c @{$\pm$} c c @{$\pm$} c c @{$\pm$} c}         
\hline                           
ID & $T_0$ & $P_1$ & \multicolumn{2}{c}{$P_2$} & \multicolumn{2}{c}{ $a\cdot\sin(i_2)$}& \multicolumn{2}{c}{$e_2$} & \multicolumn{2}{c}{$\omega_2$} & \multicolumn{2}{c}{$\tau_2$} & \multicolumn{2}{c}{$f(m_c)$} \\          
& [HJD-2450000 days] & [days] & \multicolumn{2}{c}{[days]} & \multicolumn{2}{c}{[$R_{\odot}$]} & \multicolumn{2}{c}{ }                
& \multicolumn{2}{c}{[deg]} & \multicolumn{2}{c}{[days]}  \\ \hline                     
216572 & 5260.513146 & 0.329990 & 1591.7 & 40.6 & 187.6 & 13.1 & 0.07 & 0.18 & 179.1 & 9.2 & 5336.6 & 310.7 & 0.0350 & 0.00757 \\
216693 & 5264.037358 & 0.537942 & 2361.5 & 123.0 & 39.7 & 248.7 & 0.56 & 0.34 & 242.8 & 22.5 & 7231.1 & 810.3 & 0.0001 & 0.00127 \\
217230 & 5260.731865 & 0.413478 & 1635.2 & 27.3 & 95.3 & 14.9 & 0.48 & 0.11 & 177.4 & 1.5 & 5703.9 & 42.9 & 0.0043 & 0.00156 \\
217430 & 5260.600043 & 0.367850 & 3165.5 & 147.6 & 93.7 & 178.7 & 0.34 & 0.31 & 339.5 & 22.4 & 6932.6 & 968.6 & 0.0011 & 0.00435 \\
217493 & 5258.689296 & 1.728418 & 1372.3 & 13.3 & 203.6 & 12.0 & 0.56 & 0.06 & 188.0 & 0.5 & 5882.6 & 18.2 & 0.0601 & 0.00883 \\
218008 & 5260.418370 & 0.295571 & 2222.0 & 28.7 & 100.5 & 175.4 & 0.54 & 0.37 & 10.2 & 21.6 & 7203.7 & 268.5 & 0.0028 & 0.00988 \\
\end{tabular}                           
\end{table*}                           
                           
\begin{table*}                           
\centering                           
\begin{tabular}{c c c c @{$\pm$} c c @{$\pm$} c c @{$\pm$} c c @{$\pm$} c c @{$\pm$} c c @{$\pm$} c}         
\hline                           
ID & $T_0$ & $P_1$ & \multicolumn{2}{c}{$P_2$} & \multicolumn{2}{c}{ $a\cdot\sin(i_2)$}& \multicolumn{2}{c}{$e_2$} & \multicolumn{2}{c}{$\omega_2$} & \multicolumn{2}{c}{$\tau_2$} & \multicolumn{2}{c}{$f(m_c)$} \\          
& [HJD-2450000 days] & [days] & \multicolumn{2}{c}{[days]} & \multicolumn{2}{c}{[$R_{\odot}$]} & \multicolumn{2}{c}{ }                
& \multicolumn{2}{c}{[deg]} & \multicolumn{2}{c}{[days]}  \\ \hline                     
227184 & 5260.641228 & 0.336874 & 4342.2 & 213.8 & 71.0 & 38.2 & 0.63 & 0.30 & 69.5 & 4.2 & 6054.4 & 230.0 & 0.0003 & 0.00037 \\
227185 & 5260.745360 & 0.397512 & 1598.0 & 48.8 & 226.1 & 4.6 & 0.15 & 0.08 & 302.5 & 1.9 & 5850.3 & 72.3 & 0.0607 & 0.00803 \\
227308 & 5260.032053 & 0.919303 & 4307.4 & 65.1 & 359.2 & 59.3 & 0.82 & 0.18 & 144.1 & 2.1 & 6488.4 & 50.3 & 0.0335 & 0.01258 \\
227323 & 5260.179875 & 0.667560 & 2695.4 & 225.6 & 350.3 & 86.6 & 0.52 & 0.19 & 60.6 & 3.4 & 5899.1 & 233.0 & 0.0793 & 0.05912 \\
227504 & 5260.816465 & 0.360278 & 1237.3 & 64.5 & 21.6 & 6.3 & 0.62 & 0.27 & 24.9 & 2.2 & 5261.4 & 68.9 & 0.0001 & 0.00007 \\
227683 & 5260.649151 & 0.285376 & 1839.1 & 149.1 & 185.0 & 17.1 & 0.40 & 0.20 & 84.2 & 1.7 & 6368.5 & 64.9 & 0.0251 & 0.01074 \\
\end{tabular}                           
\end{table*}                           
                           
\begin{table*}                           
\centering                           
\begin{tabular}{c c c c @{$\pm$} c c @{$\pm$} c c @{$\pm$} c c @{$\pm$} c c @{$\pm$} c c @{$\pm$} c}         
\hline                           
ID & $T_0$ & $P_1$ & \multicolumn{2}{c}{$P_2$} & \multicolumn{2}{c}{ $a\cdot\sin(i_2)$}& \multicolumn{2}{c}{$e_2$} & \multicolumn{2}{c}{$\omega_2$} & \multicolumn{2}{c}{$\tau_2$} & \multicolumn{2}{c}{$f(m_c)$} \\          
& [HJD-2450000 days] & [days] & \multicolumn{2}{c}{[days]} & \multicolumn{2}{c}{[$R_{\odot}$]} & \multicolumn{2}{c}{ }                
& \multicolumn{2}{c}{[deg]} & \multicolumn{2}{c}{[days]} \\ \hline                     
237882 & 5260.289649 & 1.358112 & 1697.9 & 255.9 & 107.5 & 22.0 & 0.73 & 0.28 & 259.1 & 3.0 & 5794.8 & 162.5 & 0.0058 & 0.00500 \\
238059 & 5267.551374 & 0.473851 & 2044.7 & 48.6 & 45.6 & 52.9 & 0.67 & 0.25 & 194.2 & 12.4 & 7171.9 & 190.0 & 0.0003 & 0.00072 \\
238302 & 5259.468237 & 1.634872 & 1528.0 & 238.5 & 365.0 & 62.6 & 0.52 & 0.19 & 40.9 & 2.0 & 5915.8 & 218.1 & 0.2792 & 0.22651 \\
238832 & 5260.825001 & 0.291600 & 1794.9 & 470.4 & 352.4 & 48.0 & 0.16 & 0.34 & 70.4 & 9.0 & 7369.7 & 548.8 & 0.1820 & 0.19267 \\
239500 & 5260.826127 & 0.340064 & 10652.1 & 126.0 & 461.8 & 513.1 & 0.62 & 0.21 & 277.6 & 6.3 & 6503.9 & 254.7 & 0.0116 & 0.02619 \\
239550 & 5260.307036 & 0.424612 & 2072.8 & 33.3 & 53.1 & 12.7 & 0.08 & 0.09 & 222.3 & 4.4 & 5126.8 & 212.8 & 0.0005 & 0.00026 \\
\end{tabular}                           
\end{table*}                           
                           
\begin{table*}                           
\centering                           
\begin{tabular}{c c c c @{$\pm$} c c @{$\pm$} c c @{$\pm$} c c @{$\pm$} c c @{$\pm$} c c @{$\pm$} c}         
\hline                           
ID & $T_0$ & $P_1$ & \multicolumn{2}{c}{$P_2$} & \multicolumn{2}{c}{ $a\cdot\sin(i_2)$}& \multicolumn{2}{c}{$e_2$} & \multicolumn{2}{c}{$\omega_2$} & \multicolumn{2}{c}{$\tau_2$} & \multicolumn{2}{c}{$f(m_c)$} \\          
& [HJD-2450000 days] & [days] & \multicolumn{2}{c}{[days]} & \multicolumn{2}{c}{[$R_{\odot}$]} & \multicolumn{2}{c}{ }                
& \multicolumn{2}{c}{[deg]} & \multicolumn{2}{c}{[days]}  \\ \hline                     
258262 & 5260.588340 & 0.413903 & 2771.4 & 188.1 & 61.2 & 196.1 & 0.13 & 0.30 & 175.0 & 56.9 & 6942.3 & 3248.3 & 0.0004 & 0.00264 \\
258346 & 5260.591066 & 1.058304 & 1976.2 & 139.4 & 152.6 & 82.6 & 0.59 & 0.12 & 94.1 & 0.8 & 5367.3 & 95.4 & 0.0122 & 0.01579 \\
258794 & 5260.740010 & 0.311162 & 1287.4 & 35.9 & 153.9 & 8.7 & 0.41 & 0.12 & 313.3 & 1.6 & 6499.2 & 42.8 & 0.0295 & 0.00580 \\
259022 & 5260.297187 & 1.054545 & 2958.5 & 273.8 & 379.5 & 56.9 & 0.58 & 0.37 & 20.0 & 6.3 & 6618.7 & 378.4 & 0.0837 & 0.04834 \\
259410 & 5262.671664 & 0.373900 & 1677.4 & 105.1 & 56.7 & 6.8 & 0.37 & 0.20 & 220.7 & 2.7 & 5666.3 & 100.4 & 0.0009 & 0.00039 \\
259513 & 5260.570939 & 0.548771 & 1193.3 & 25.2 & 69.8 & 9.8 & 0.57 & 0.15 & 183.7 & 1.1 & 5962.6 & 23.6 & 0.0032 & 0.00110 \\
\end{tabular}                           
\end{table*}                           
                           
\begin{table*}                           
\centering                           
\begin{tabular}{c c c c @{$\pm$} c c @{$\pm$} c c @{$\pm$} c c @{$\pm$} c c @{$\pm$} c c @{$\pm$} c}         
\hline                           
ID & $T_0$ & $P_1$ & \multicolumn{2}{c}{$P_2$} & \multicolumn{2}{c}{ $a\cdot\sin(i_2)$}& \multicolumn{2}{c}{$e_2$} & \multicolumn{2}{c}{$\omega_2$} & \multicolumn{2}{c}{$\tau_2$} & \multicolumn{2}{c}{$f(m_c)$} \\          
& [HJD-2450000 days] & [days] & \multicolumn{2}{c}{[days]} & \multicolumn{2}{c}{[$R_{\odot}$]} & \multicolumn{2}{c}{ }                
& \multicolumn{2}{c}{[deg]} & \multicolumn{2}{c}{[days]} \\ \hline                     
277128 & 5260.590260 & 0.351749 & 2786.3 & 198.7 & 556.3 & 387.8 & 0.42 & 0.47 & 74.1 & 10.7 & 5449.1 & 125.4 & 0.2972 & 0.47794 \\
277772 & 5260.490015 & 0.443780 & 3642.0 & 41.1 & 144.1 & 26.4 & 0.82 & 0.12 & 29.8 & 2.2 & 6460.2 & 57.9 & 0.0030 & 0.00120 \\
277911 & 5260.545544 & 0.350532 & 2286.4 & 29.4 & 82.1 & 101.8 & 0.64 & 0.10 & 221.0 & 8.9 & 7202.7 & 177.8 & 0.0014 & 0.00353 \\
278024 & 5260.910213 & 0.251620 & 1893.2 & 329.4 & 49.7 & 17.5 & 0.40 & 0.29 & 83.5 & 2.5 & 5453.4 & 219.3 & 0.0005 & 0.00061 \\
278125 & 5260.319015 & 0.469390 & 1716.3 & 167.4 & 235.7 & 23.2 & 0.55 & 0.23 & 72.6 & 3.4 & 5783.3 & 139.8 & 0.0596 & 0.02917 \\
278132 & 5260.333000 & 0.582547 & 3200.7 & 157.3 & 208.6 & 279.5 & 0.24 & 0.71 & 2.5 & 19.2 & 5988.3 & 674.6 & 0.0119 & 0.03364 \\
\end{tabular}                           
\end{table*}                           
                           
\begin{table*}                           
\centering                           
\begin{tabular}{c c c c @{$\pm$} c c @{$\pm$} c c @{$\pm$} c c @{$\pm$} c c @{$\pm$} c c @{$\pm$} c}         
\hline                           
ID & $T_0$ & $P_1$ & \multicolumn{2}{c}{$P_2$} & \multicolumn{2}{c}{ $a\cdot\sin(i_2)$}& \multicolumn{2}{c}{$e_2$} & \multicolumn{2}{c}{$\omega_2$} & \multicolumn{2}{c}{$\tau_2$} & \multicolumn{2}{c}{$f(m_c)$} \\          
& [HJD-2450000 days] & [days] & \multicolumn{2}{c}{[days]} & \multicolumn{2}{c}{[$R_{\odot}$]} & \multicolumn{2}{c}{ }                
& \multicolumn{2}{c}{[deg]} & \multicolumn{2}{c}{[days]}  \\ \hline                     
297270 & 5259.846413 & 1.157091 & 1189.1 & 47.9 & 337.9 & 24.8 & 0.13 & 0.16 & 44.9 & 8.0 & 6076.8 & 217.9 & 0.3658 & 0.09790 \\
297569 & 5260.479928 & 0.434196 & 1284.3 & 36.5 & 349.9 & 128.3 & 0.80 & 0.16 & 358.7 & 0.5 & 6007.7 & 28.2 & 0.3482 & 0.28504 \\
298564 & 5267.951063 & 0.952836 & 1705.0 & 36.8 & 277.0 & 13.7 & 0.31 & 0.06 & 33.0 & 0.9 & 5698.9 & 36.2 & 0.0980 & 0.01604 \\
300095 & 5260.799990 & 0.394840 & 1910.9 & 76.2 & 89.5 & 55.9 & 0.53 & 0.19 & 48.9 & 3.7 & 5335.3 & 109.8 & 0.0026 & 0.00356 \\
300670 & 5321.054322 & 0.481134 & 1413.0 & 510.7 & 113.3 & 34.0 & 0.45 & 0.67 & 92.9 & 5.9 & 5822.6 & 358.5 & 0.0098 & 0.01651 \\
300884 & 5260.759065 & 0.269191 & 2217.4 & 78.6 & 107.5 & 10.4 & 0.56 & 0.17 & 139.4 & 2.3 & 6397.3 & 55.5 & 0.0034 & 0.00102 \\
\end{tabular}                           
\end{table*}                           
\clearpage                           
\newpage


\begin{figure*}
\caption{ETV-s with third body solution of the second group.}       

\includegraphics[width=0.64\columnwidth]{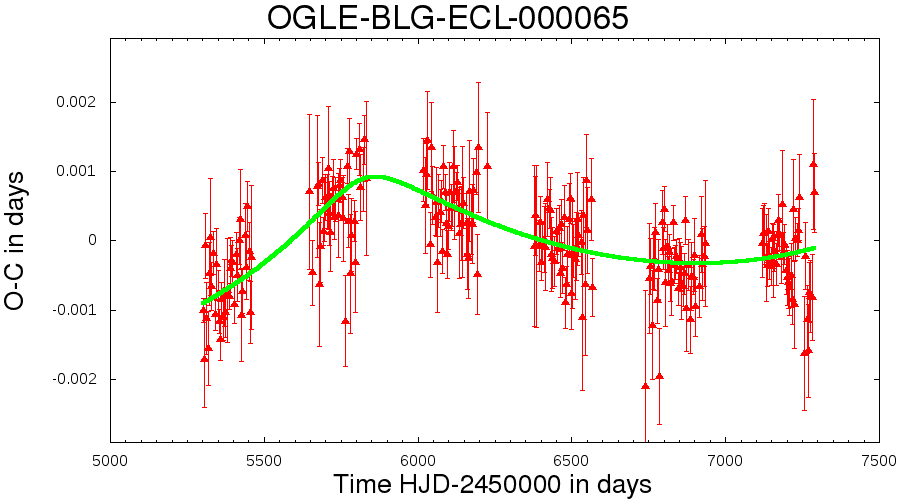}
\includegraphics[width=0.64\columnwidth]{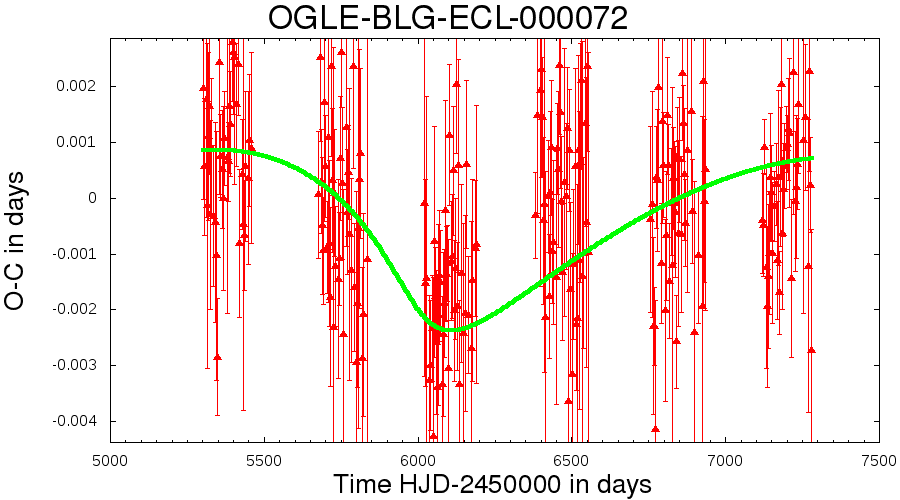}
\includegraphics[width=0.64\columnwidth]{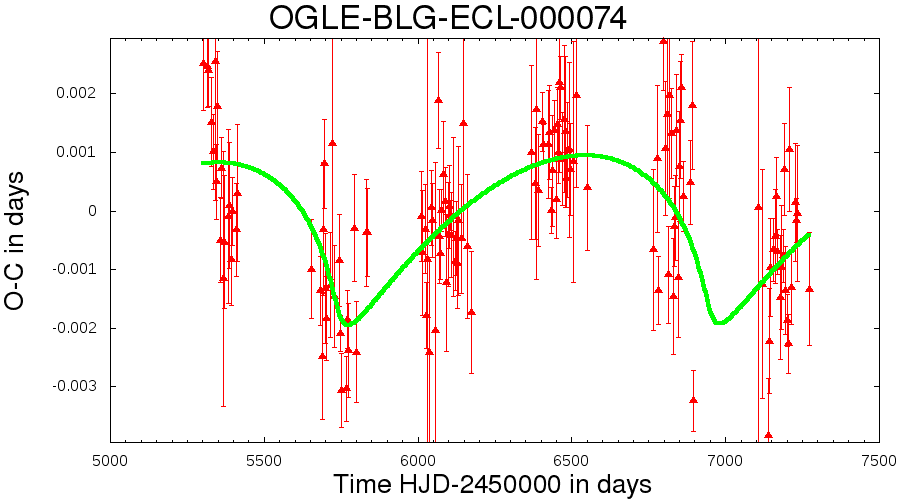}

\includegraphics[width=0.64\columnwidth]{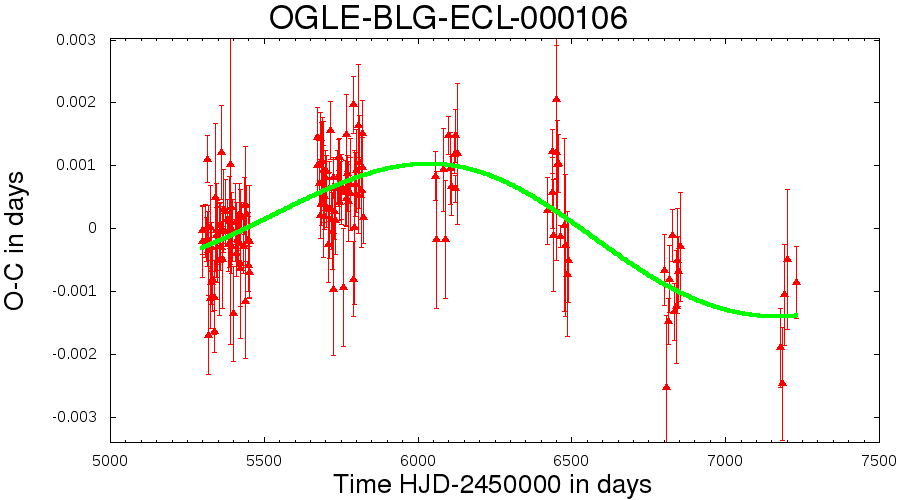}
\includegraphics[width=0.64\columnwidth]{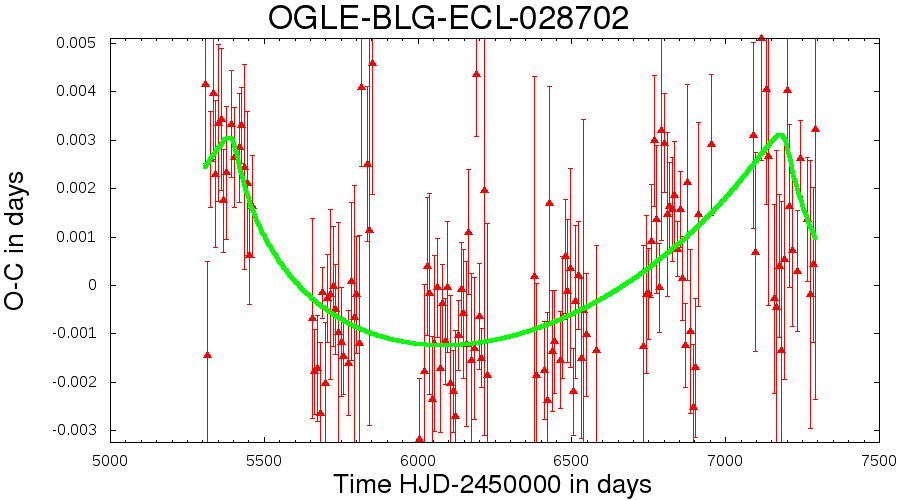}
\includegraphics[width=0.64\columnwidth]{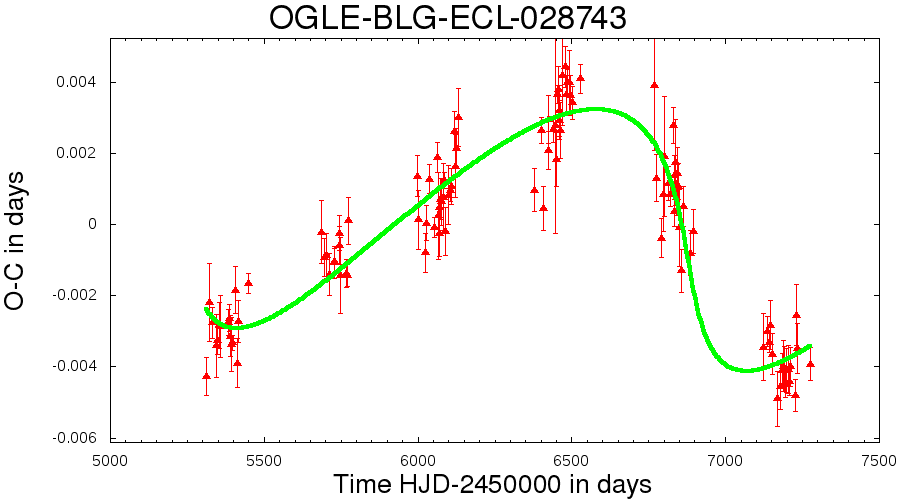}

\includegraphics[width=0.64\columnwidth]{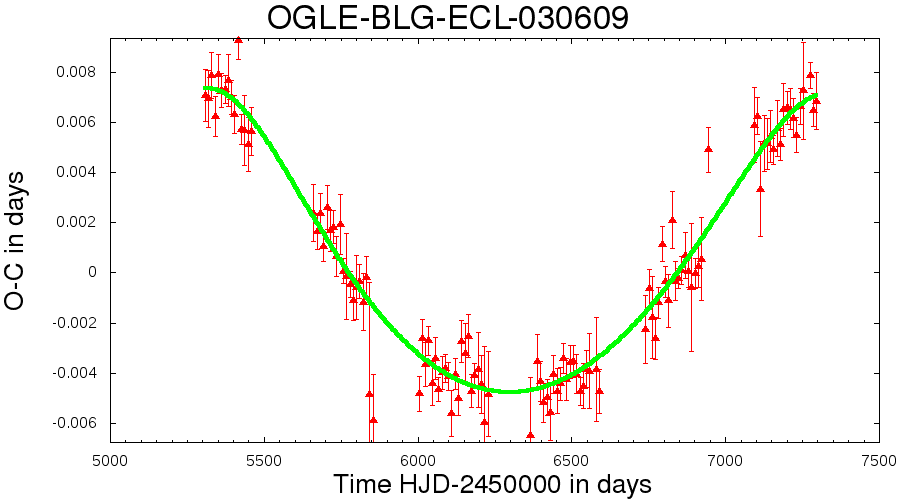}
\includegraphics[width=0.64\columnwidth]{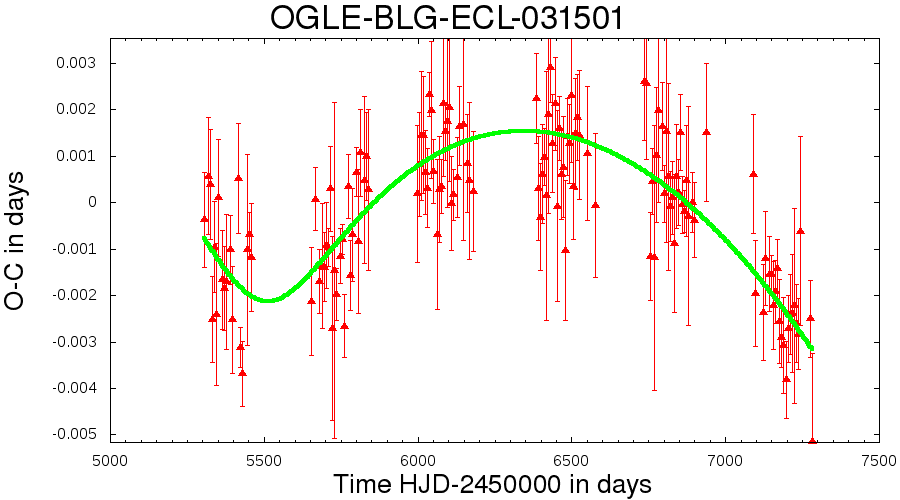}
\includegraphics[width=0.64\columnwidth]{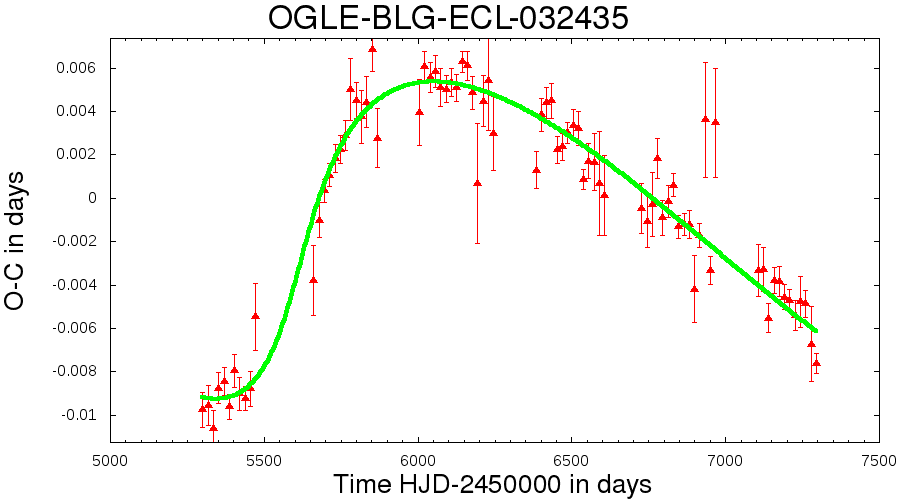}

\includegraphics[width=0.64\columnwidth]{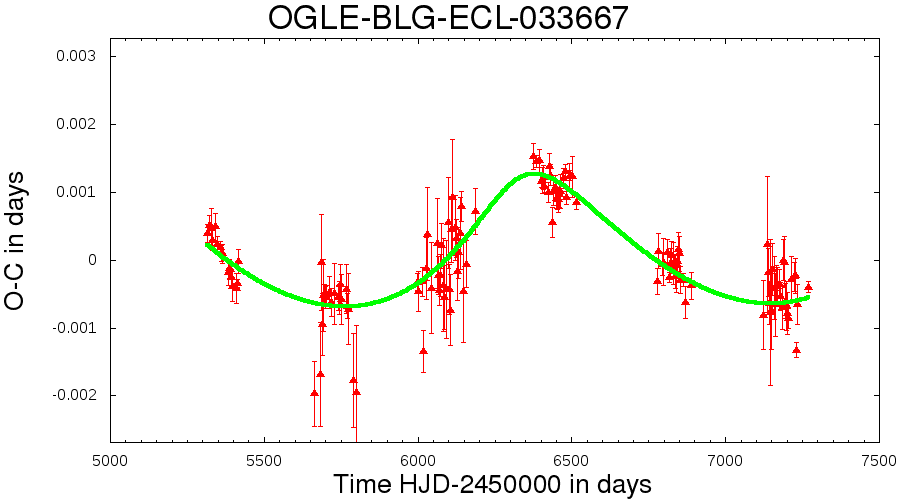}
\includegraphics[width=0.64\columnwidth]{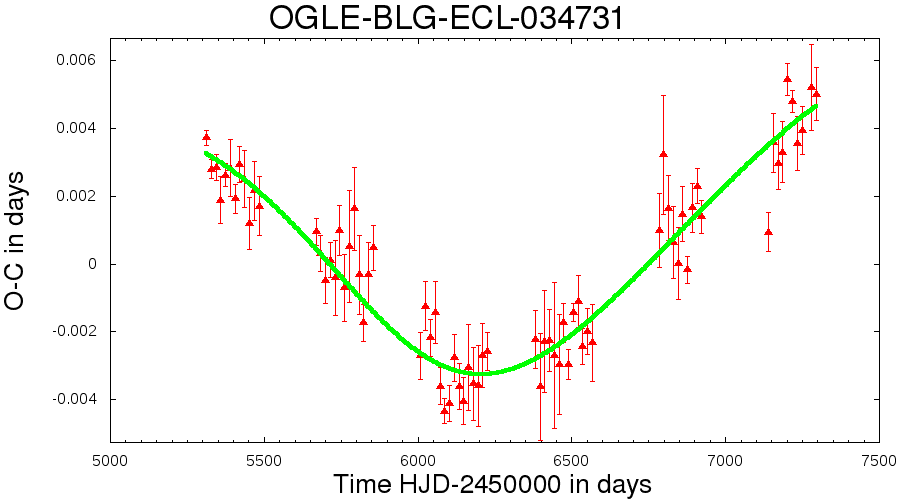}
\includegraphics[width=0.64\columnwidth]{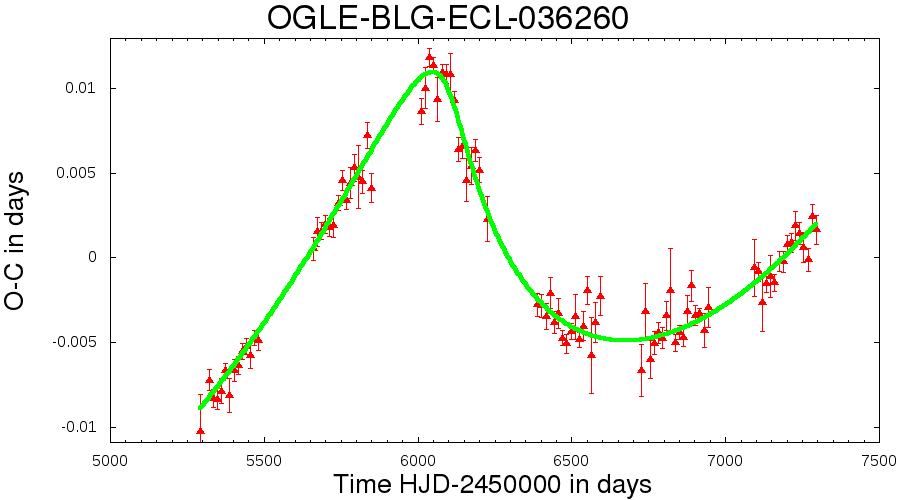}

\includegraphics[width=0.64\columnwidth]{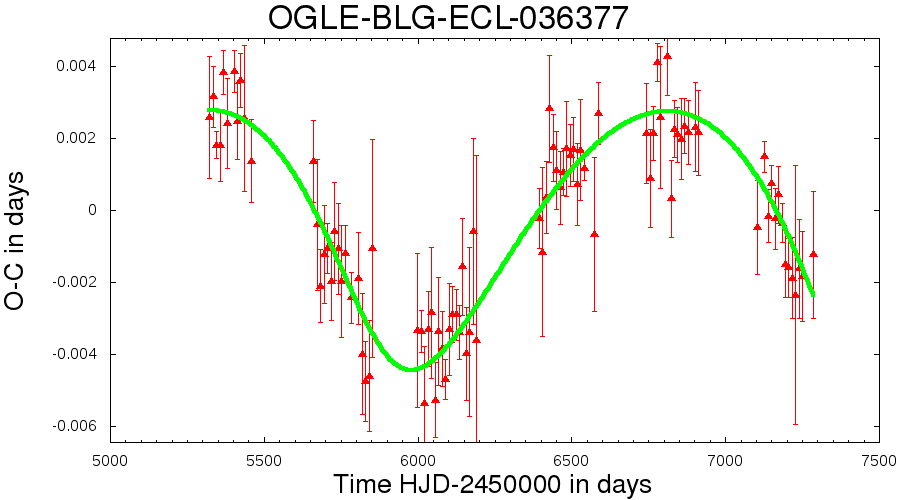}
\includegraphics[width=0.64\columnwidth]{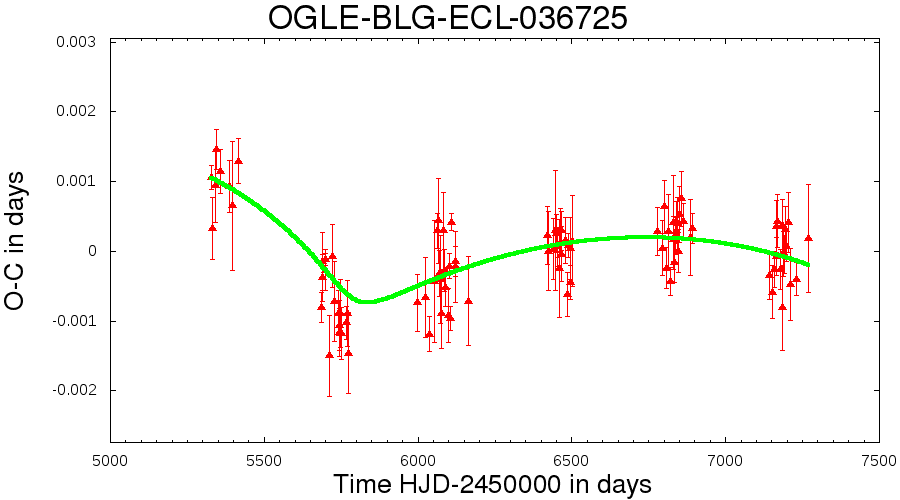}
\includegraphics[width=0.64\columnwidth]{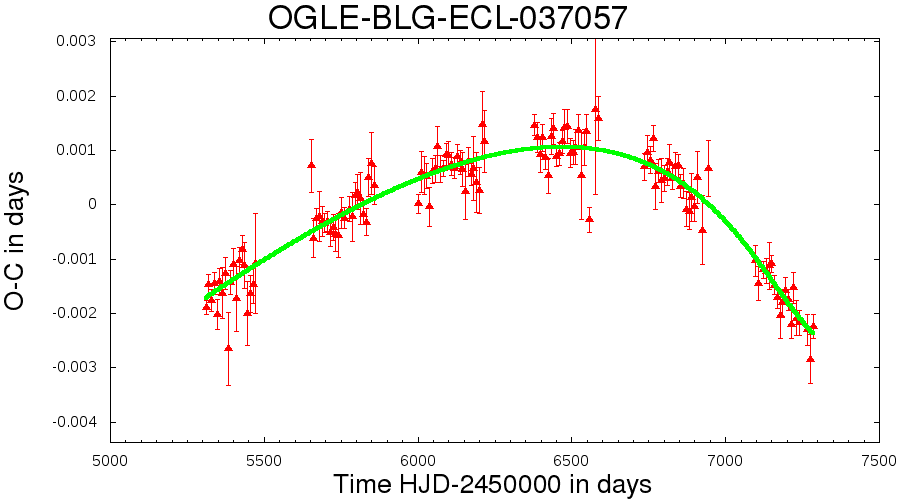}

\includegraphics[width=0.64\columnwidth]{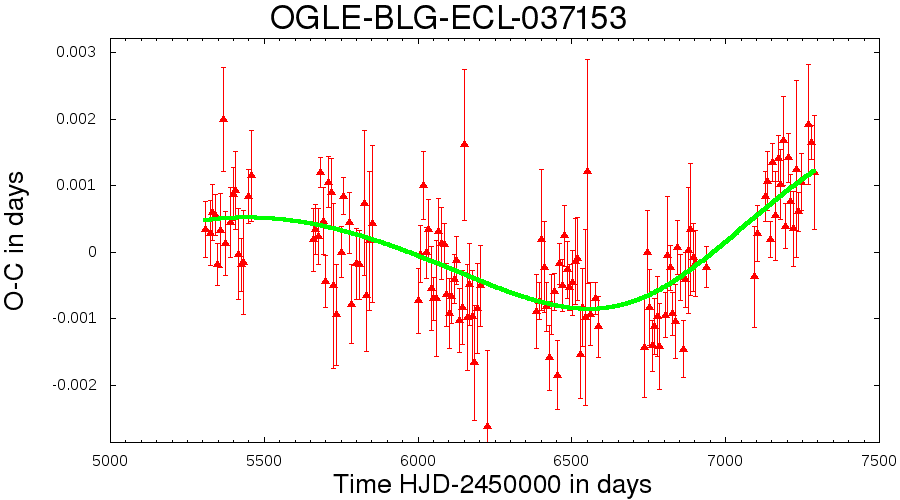}
\includegraphics[width=0.64\columnwidth]{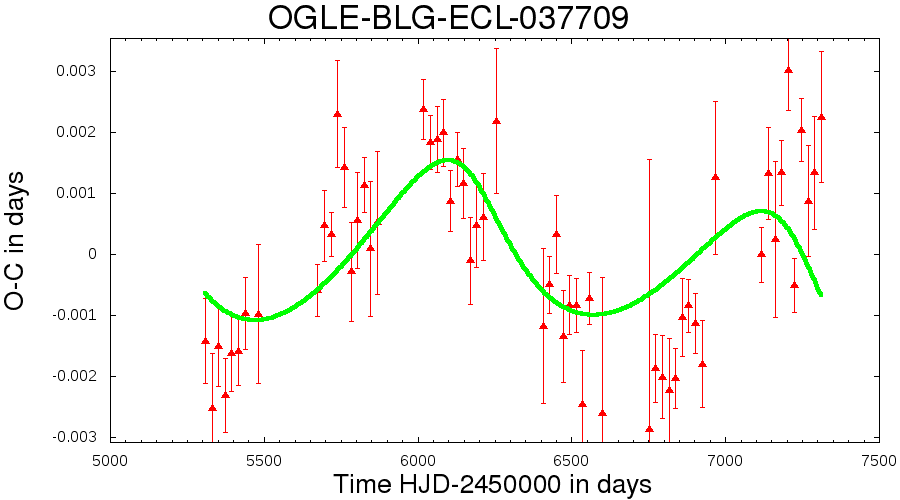}
\includegraphics[width=0.64\columnwidth]{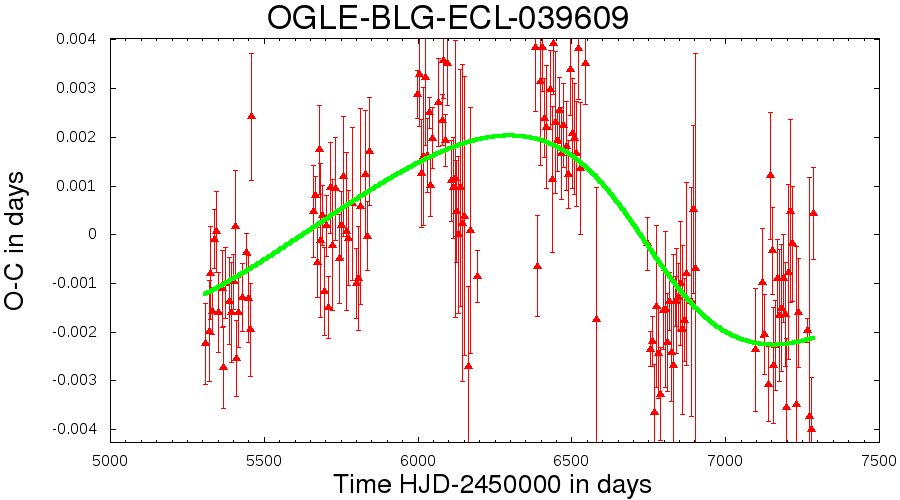}

\includegraphics[width=0.64\columnwidth]{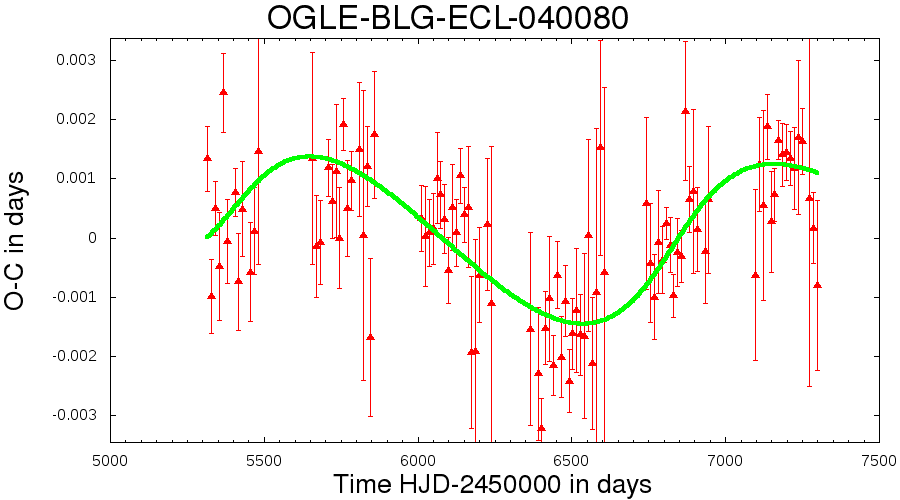}
\includegraphics[width=0.64\columnwidth]{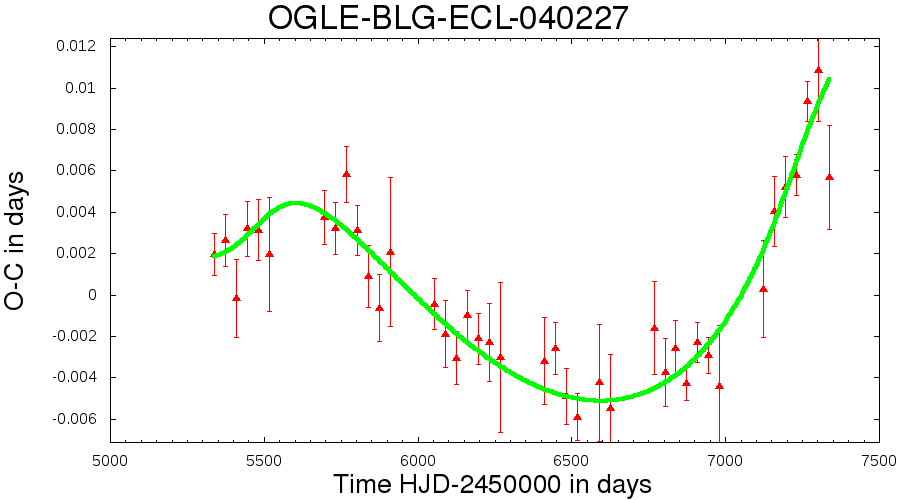}
\includegraphics[width=0.64\columnwidth]{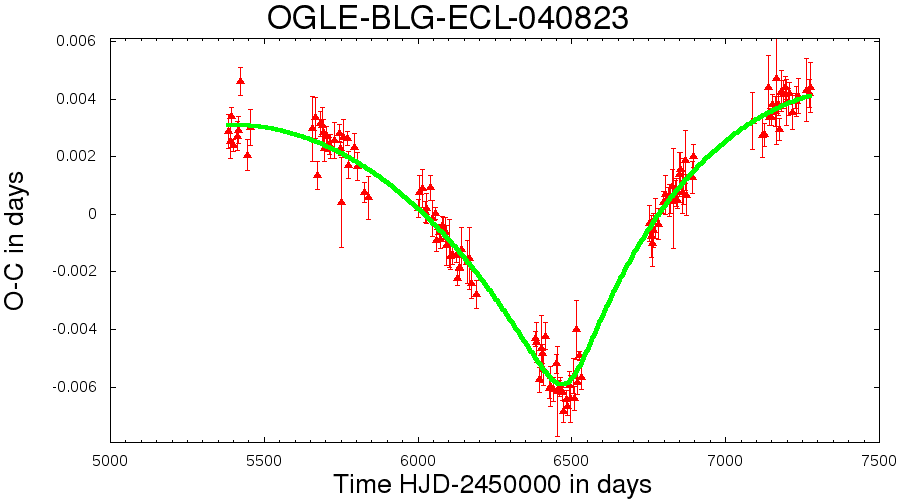}

\end{figure*}
\clearpage

\begin{figure*}
\includegraphics[width=0.64\columnwidth]{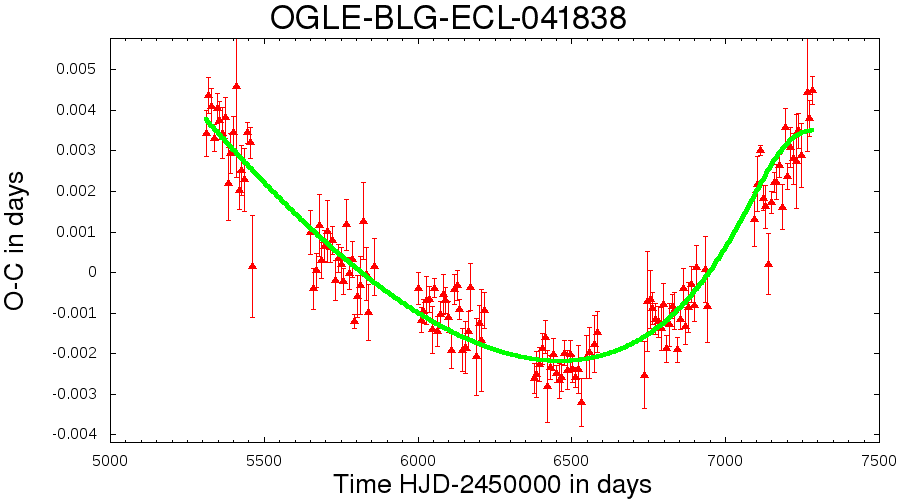}
\includegraphics[width=0.64\columnwidth]{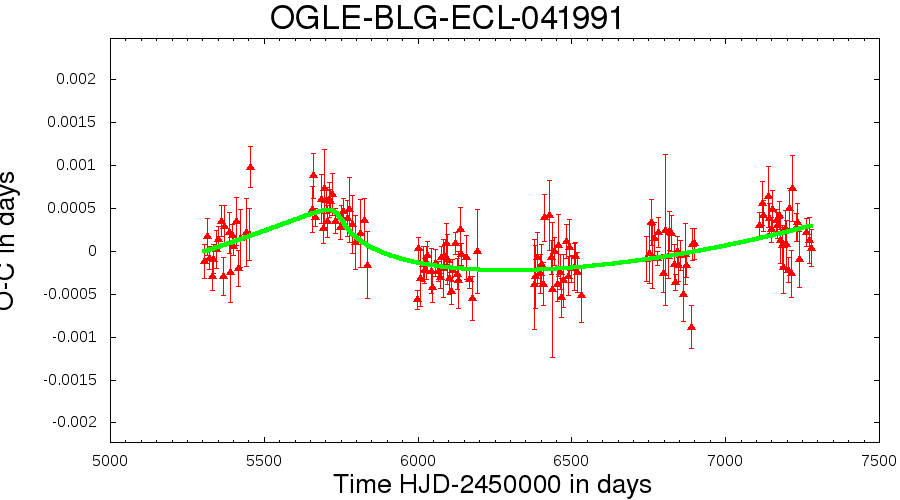}
\includegraphics[width=0.64\columnwidth]{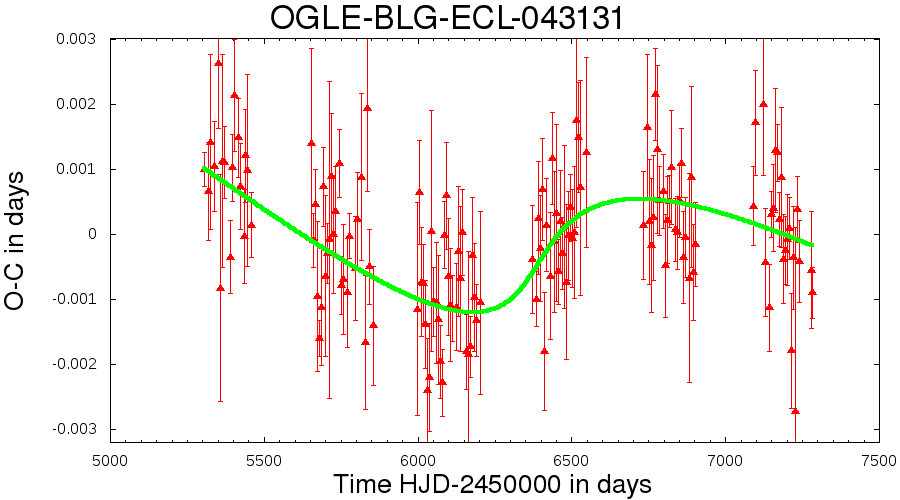}

\includegraphics[width=0.64\columnwidth]{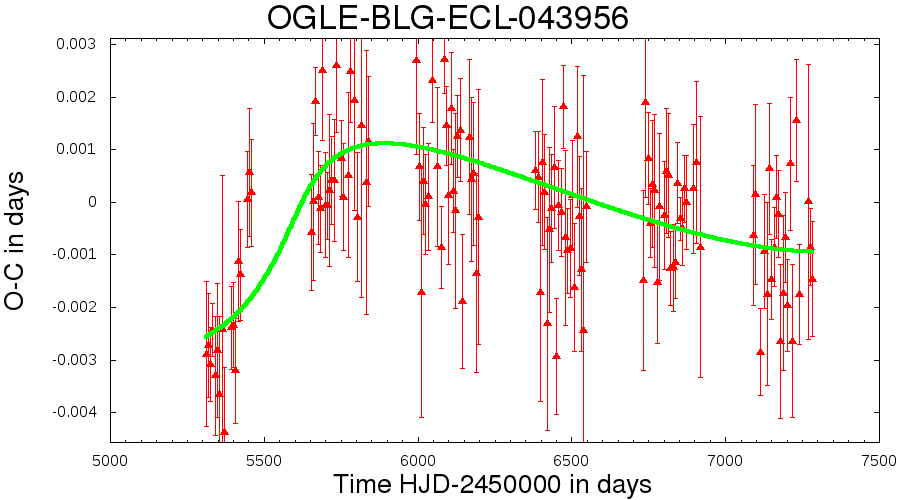}
\includegraphics[width=0.64\columnwidth]{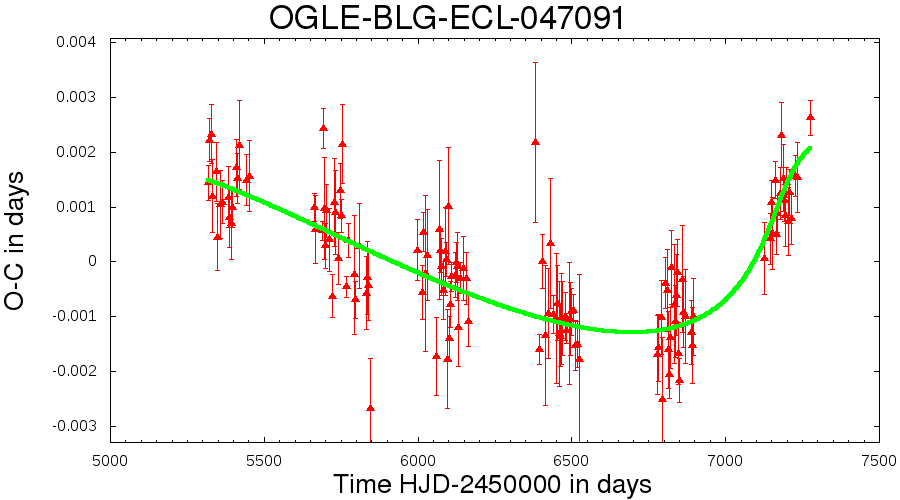}
\includegraphics[width=0.64\columnwidth]{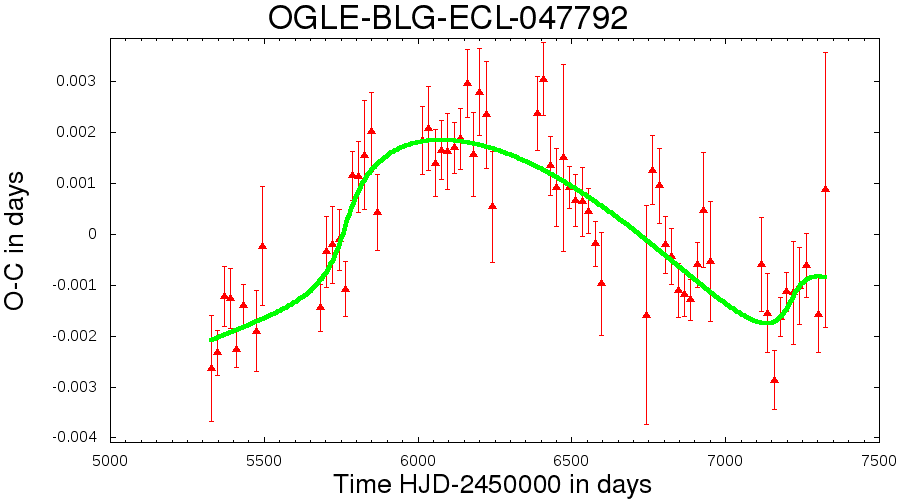}

\includegraphics[width=0.64\columnwidth]{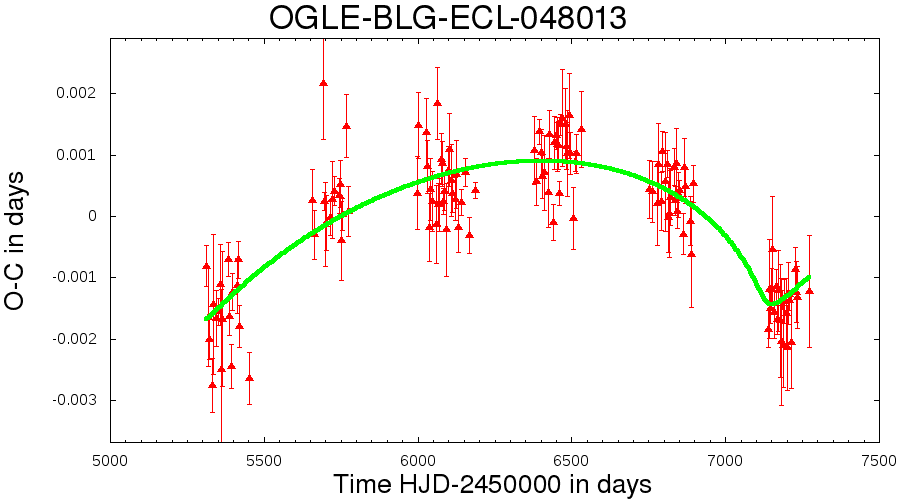}
\includegraphics[width=0.64\columnwidth]{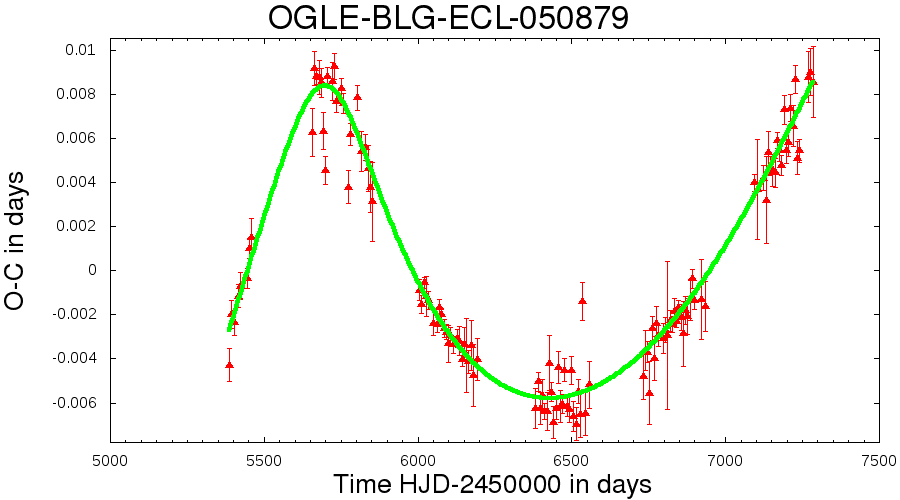}
\includegraphics[width=0.64\columnwidth]{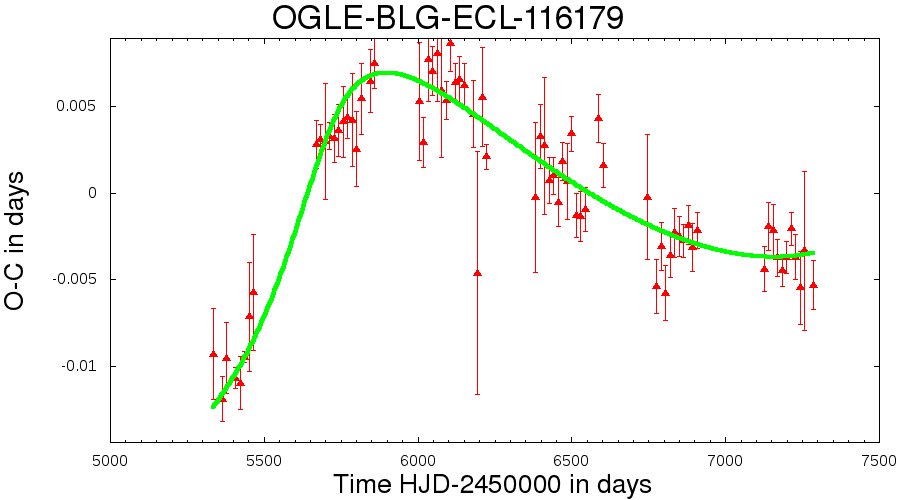}

\includegraphics[width=0.64\columnwidth]{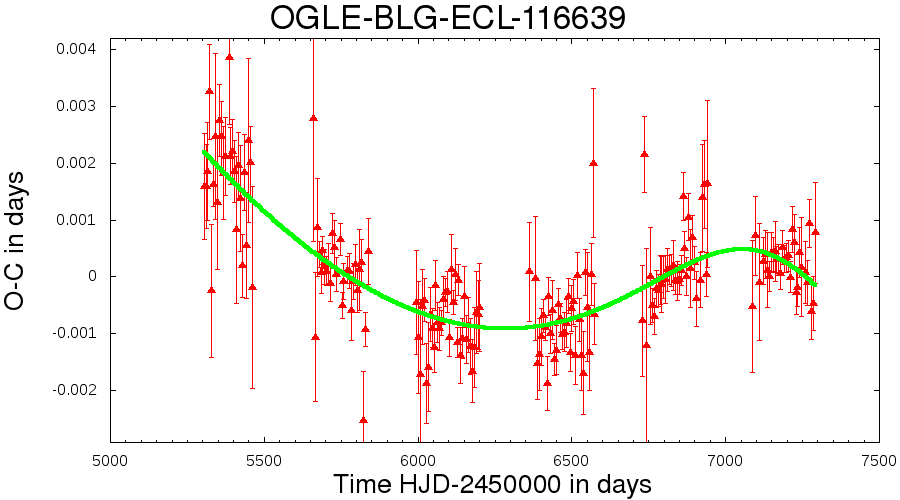}
\includegraphics[width=0.64\columnwidth]{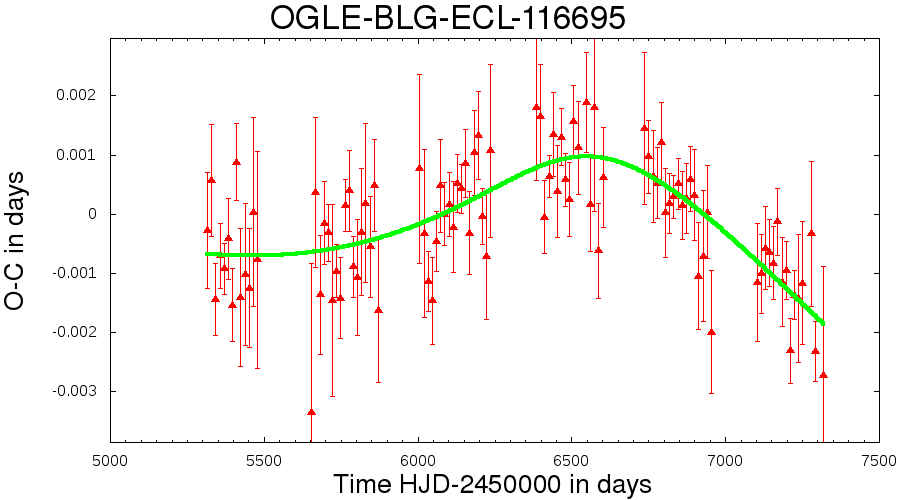}
\includegraphics[width=0.64\columnwidth]{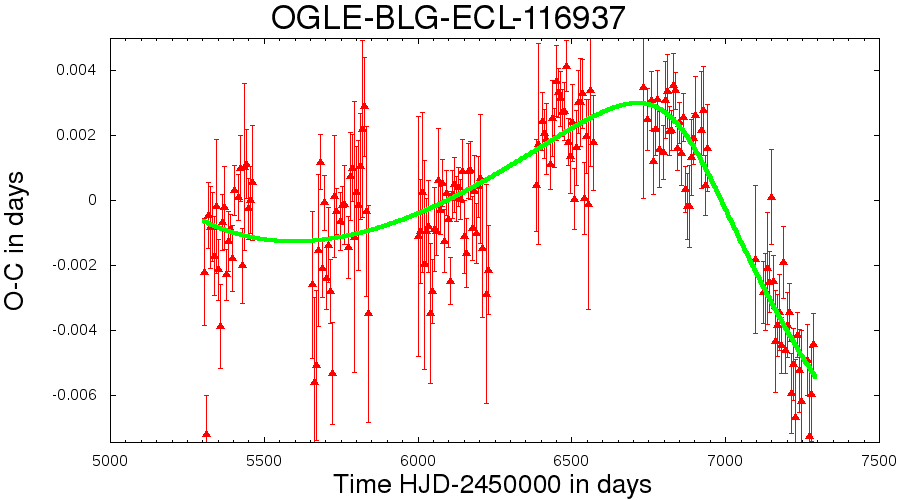}

\includegraphics[width=0.64\columnwidth]{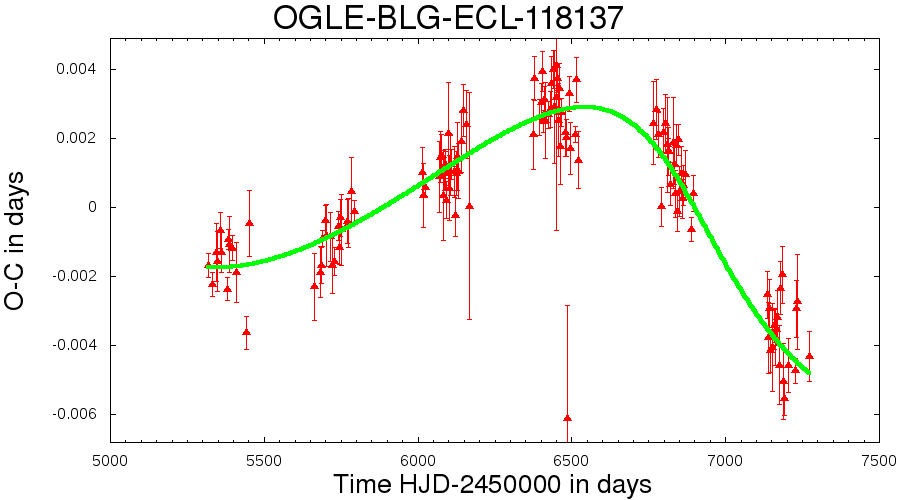}
\includegraphics[width=0.64\columnwidth]{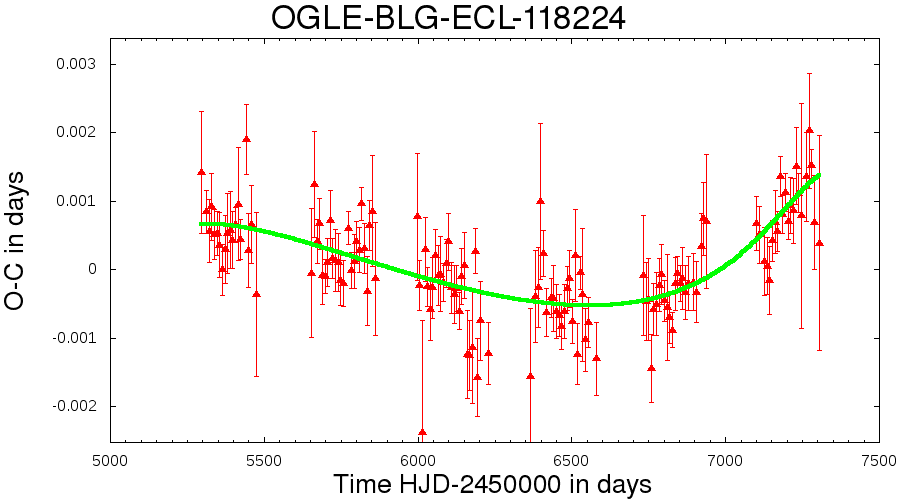}
\includegraphics[width=0.64\columnwidth]{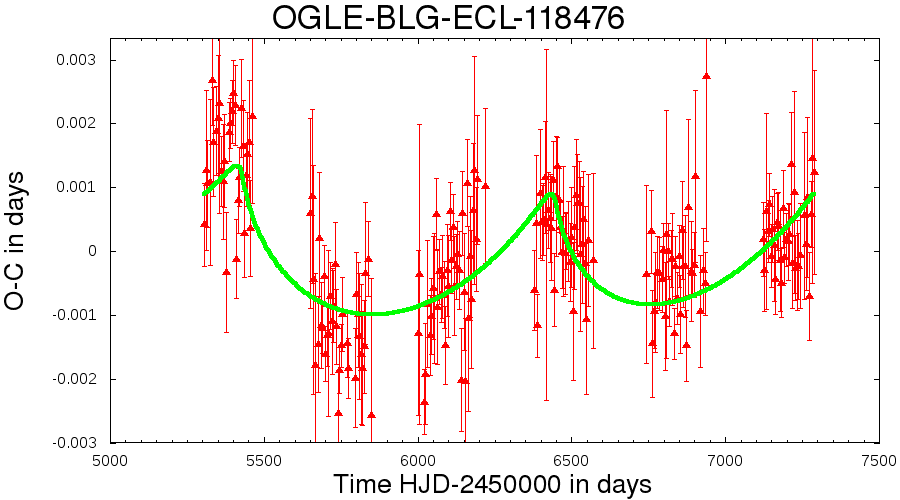}

\includegraphics[width=0.64\columnwidth]{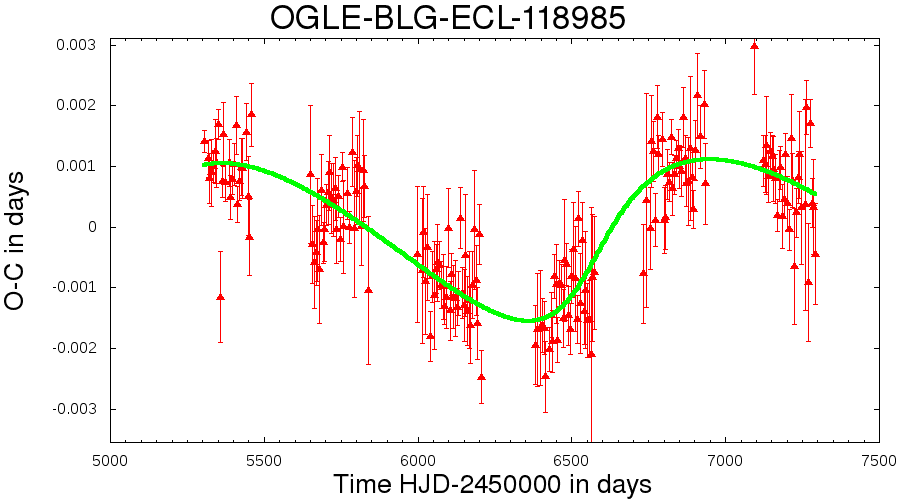}
\includegraphics[width=0.64\columnwidth]{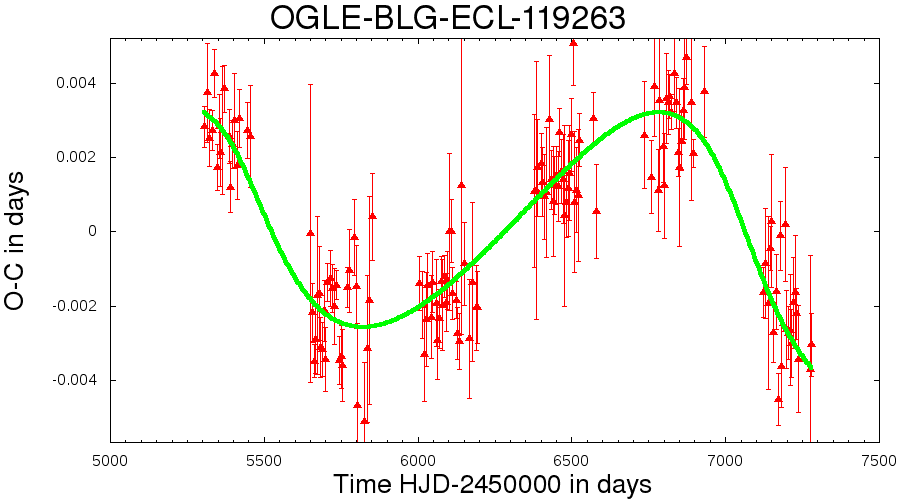}
\includegraphics[width=0.64\columnwidth]{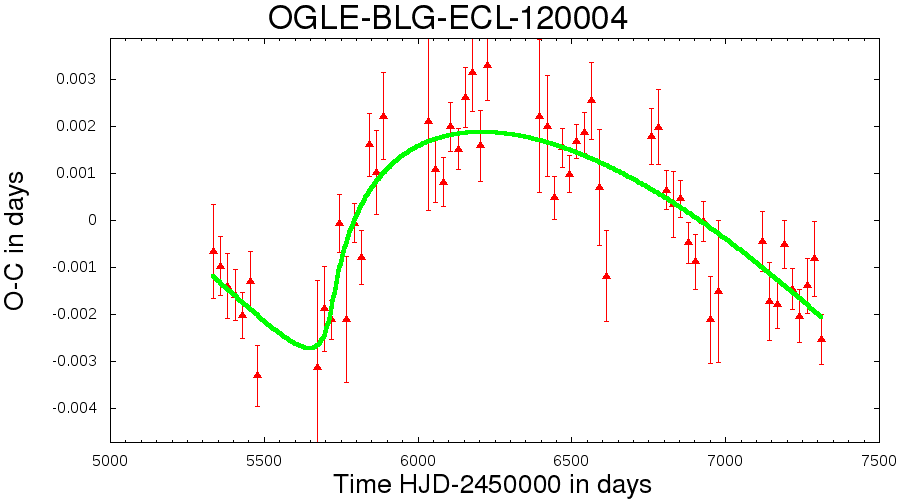}

\includegraphics[width=0.64\columnwidth]{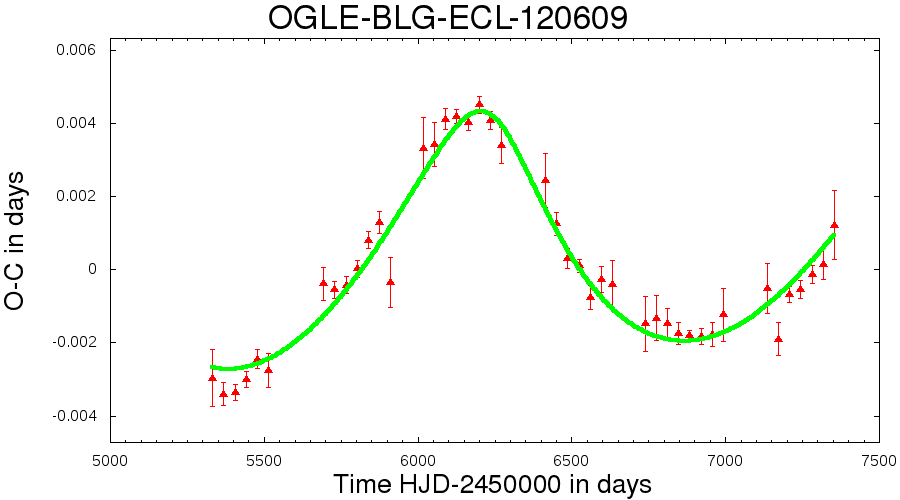}
\includegraphics[width=0.64\columnwidth]{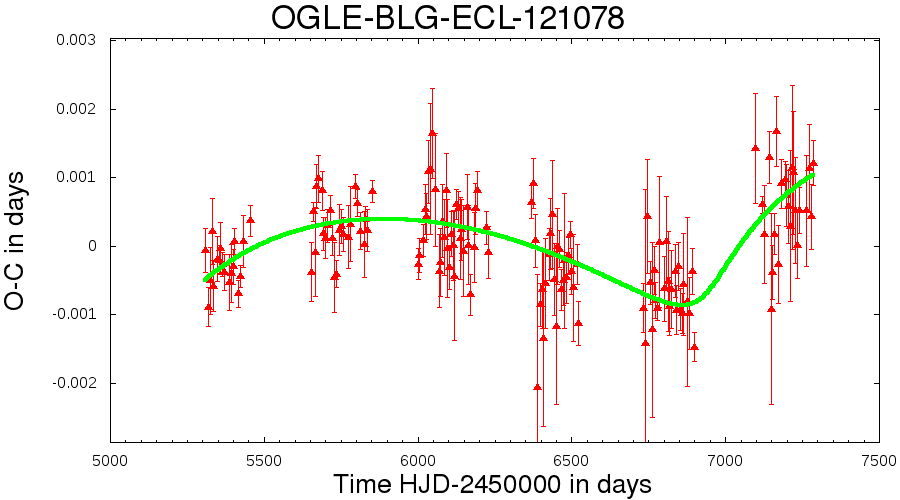}
\includegraphics[width=0.64\columnwidth]{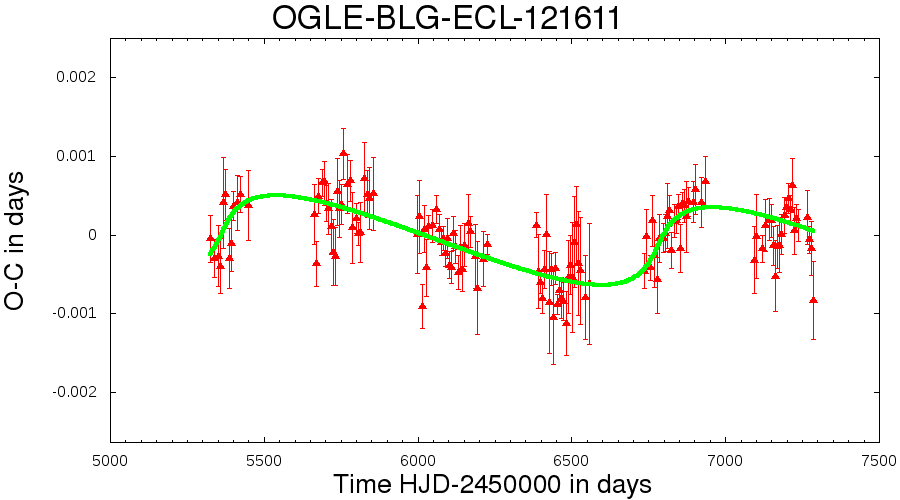}

\includegraphics[width=0.64\columnwidth]{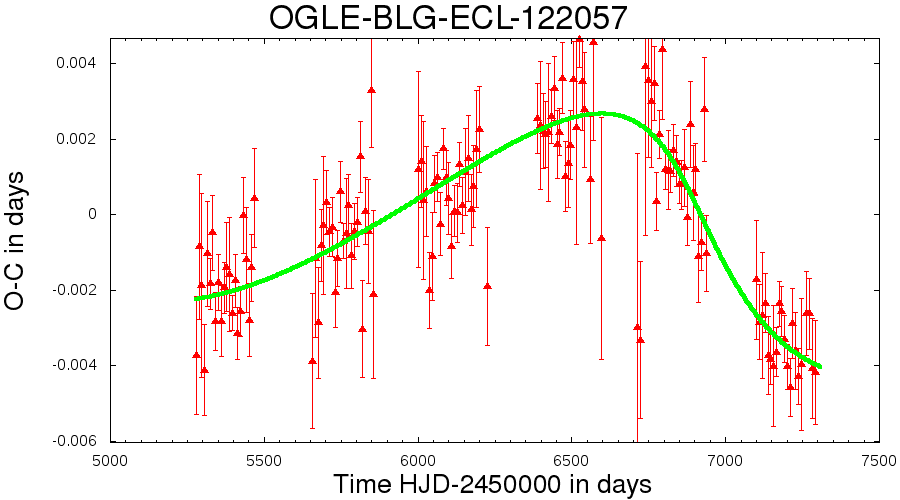}
\includegraphics[width=0.64\columnwidth]{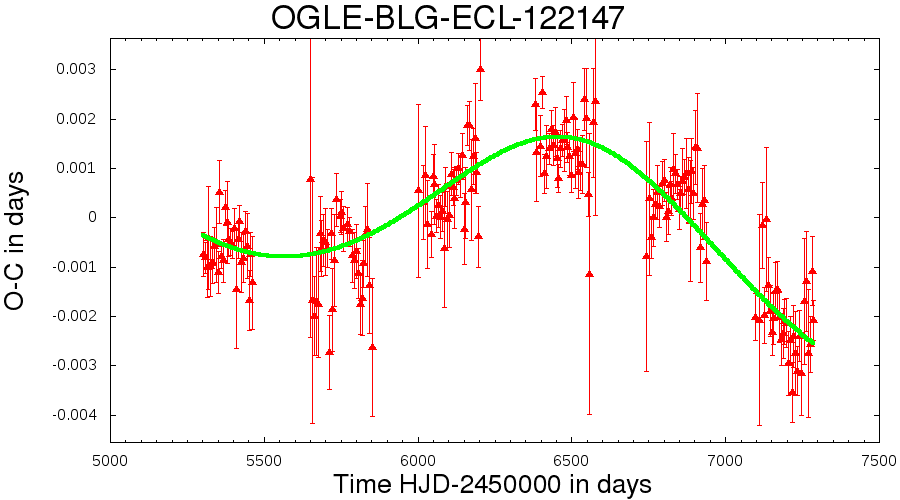}
\includegraphics[width=0.64\columnwidth]{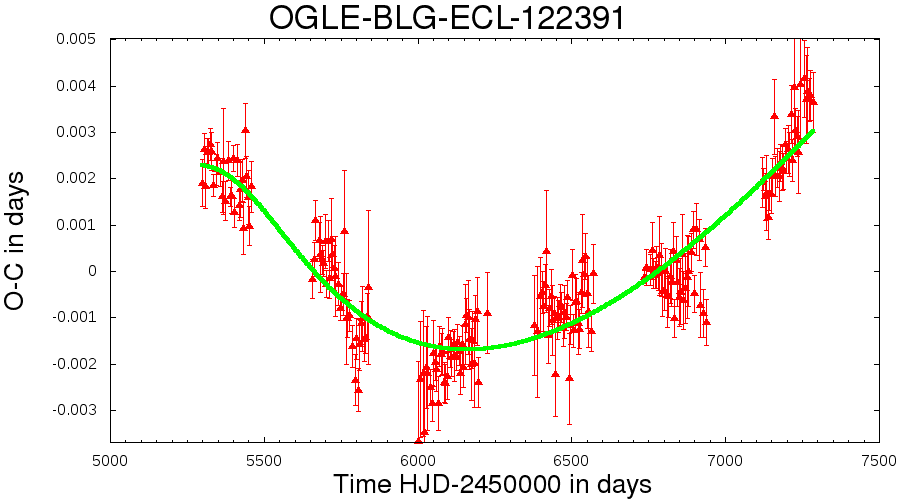}

\end{figure*}
\clearpage

\begin{figure*}
\includegraphics[width=0.64\columnwidth]{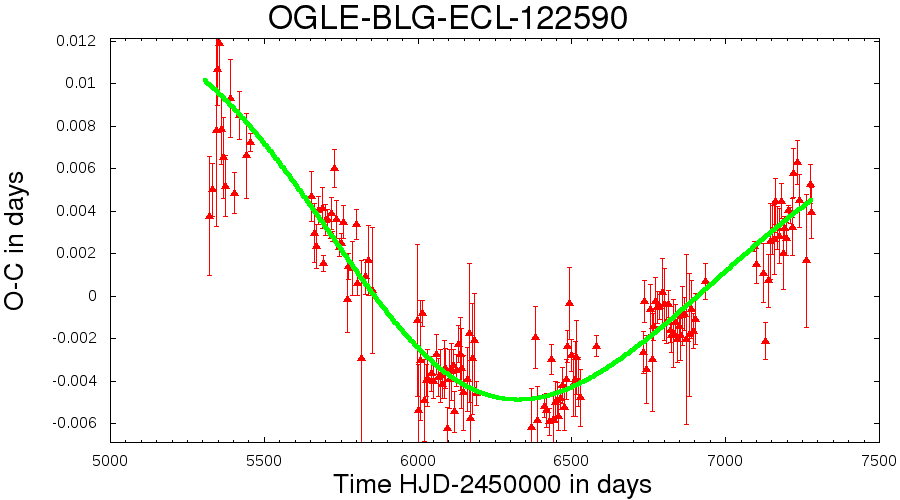}
\includegraphics[width=0.64\columnwidth]{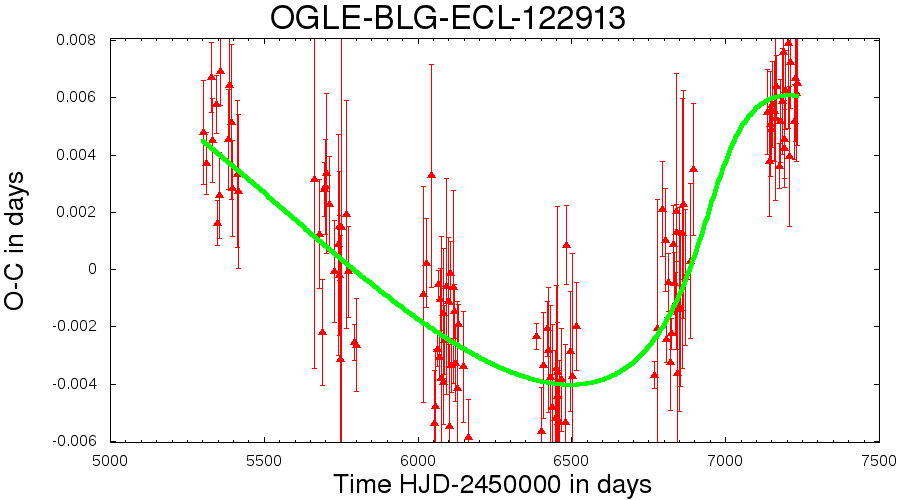}
\includegraphics[width=0.64\columnwidth]{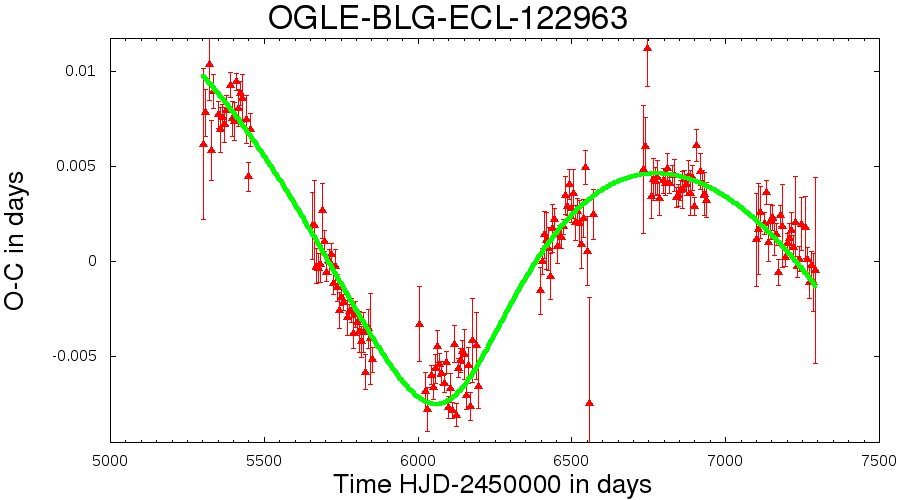}

\includegraphics[width=0.64\columnwidth]{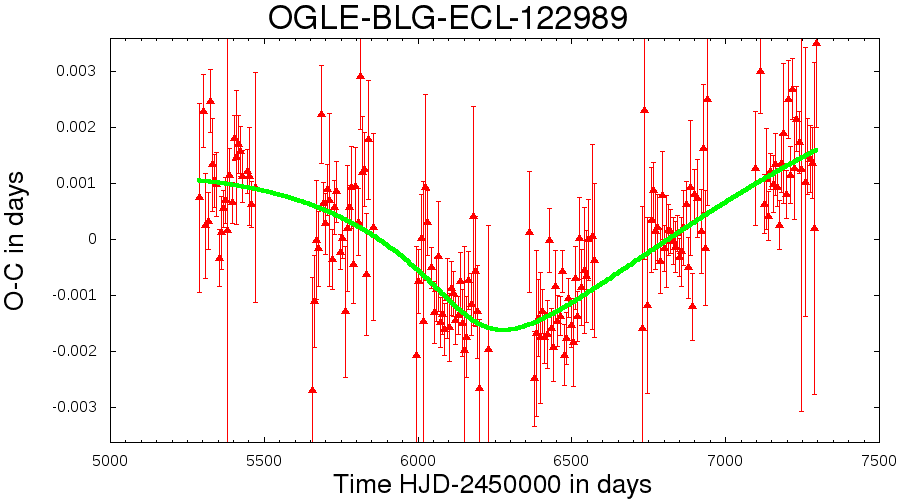}
\includegraphics[width=0.64\columnwidth]{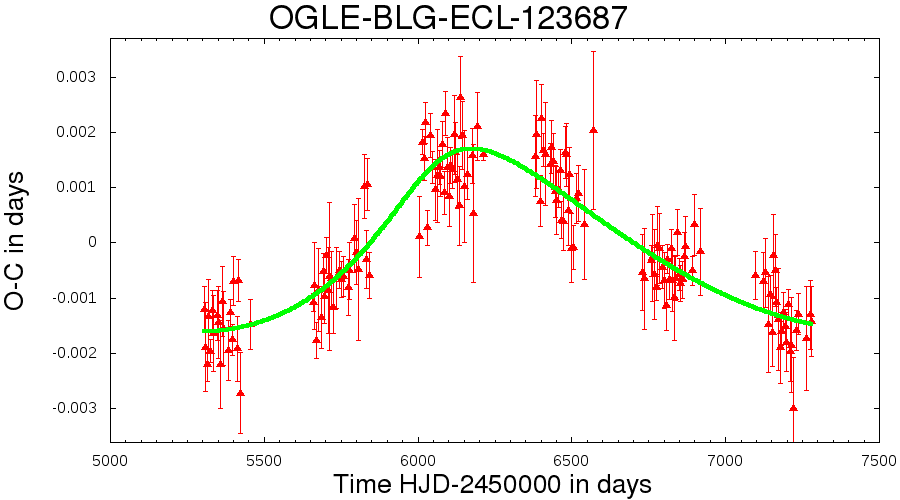}
\includegraphics[width=0.64\columnwidth]{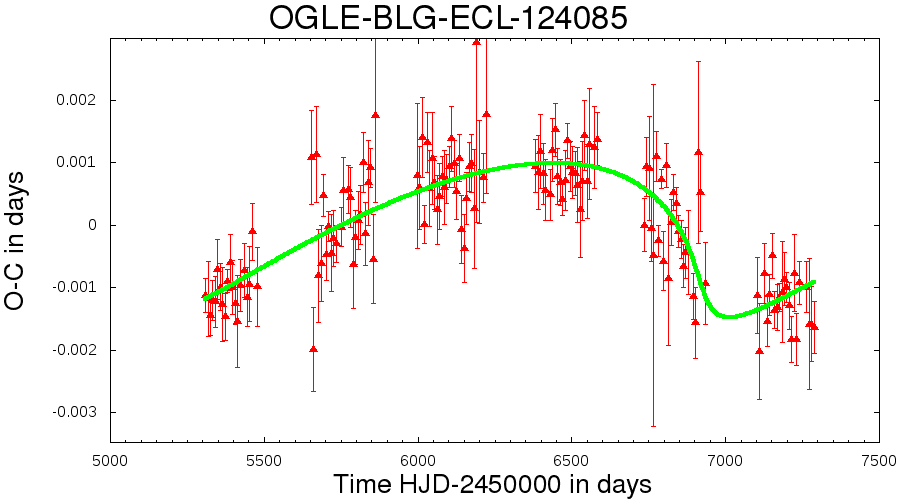}

\includegraphics[width=0.64\columnwidth]{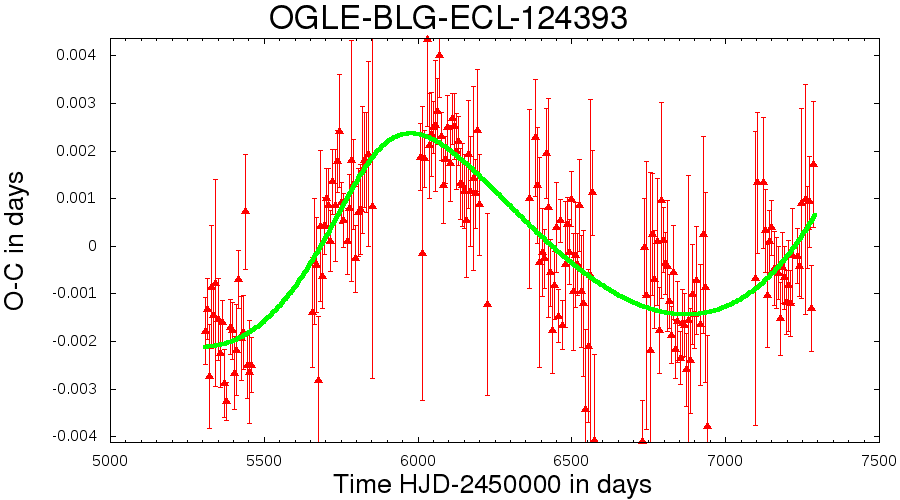}
\includegraphics[width=0.64\columnwidth]{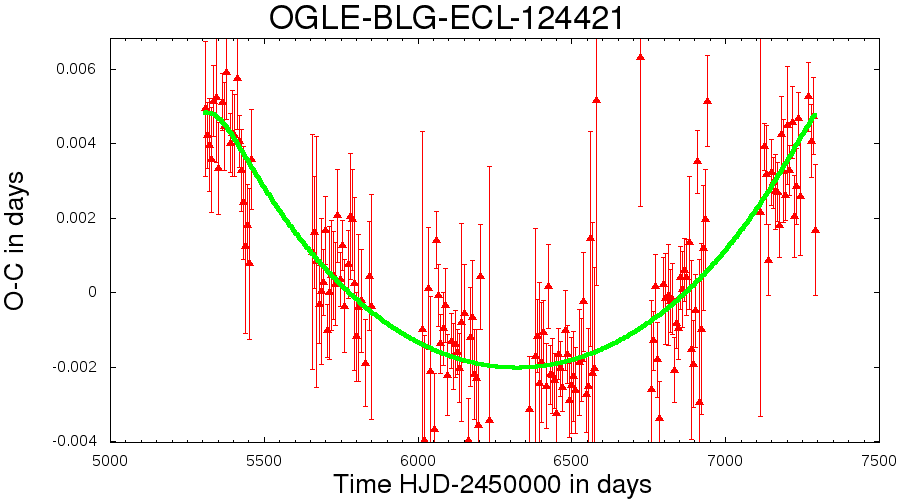}
\includegraphics[width=0.64\columnwidth]{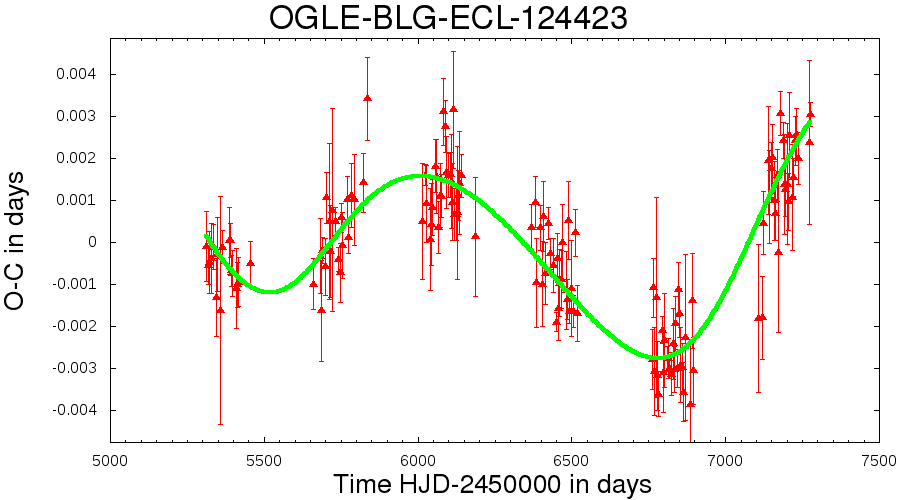}

\includegraphics[width=0.64\columnwidth]{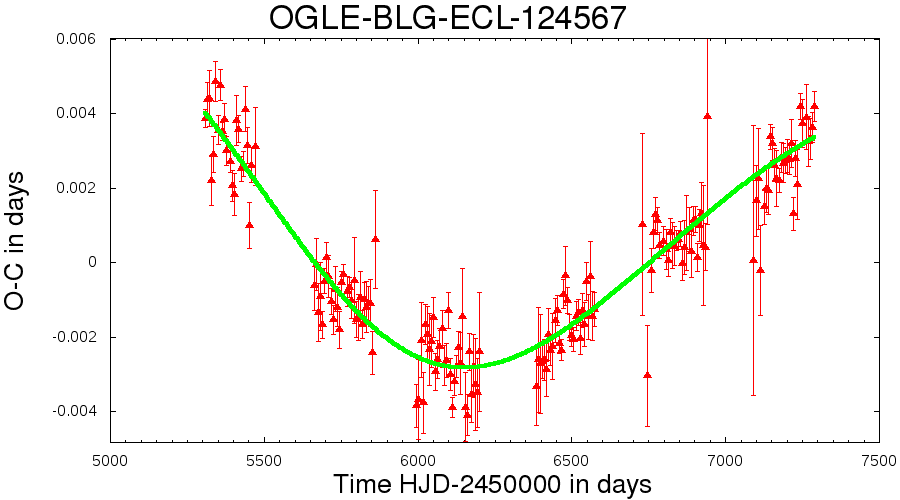}
\includegraphics[width=0.64\columnwidth]{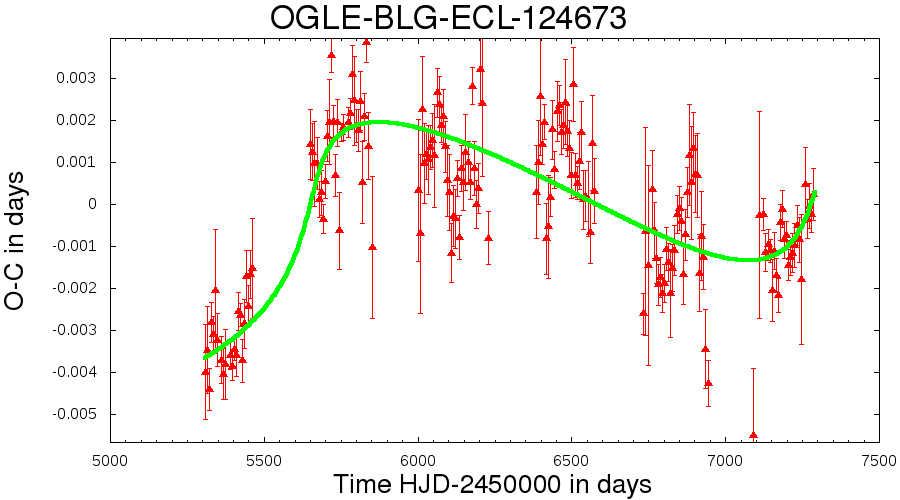}
\includegraphics[width=0.64\columnwidth]{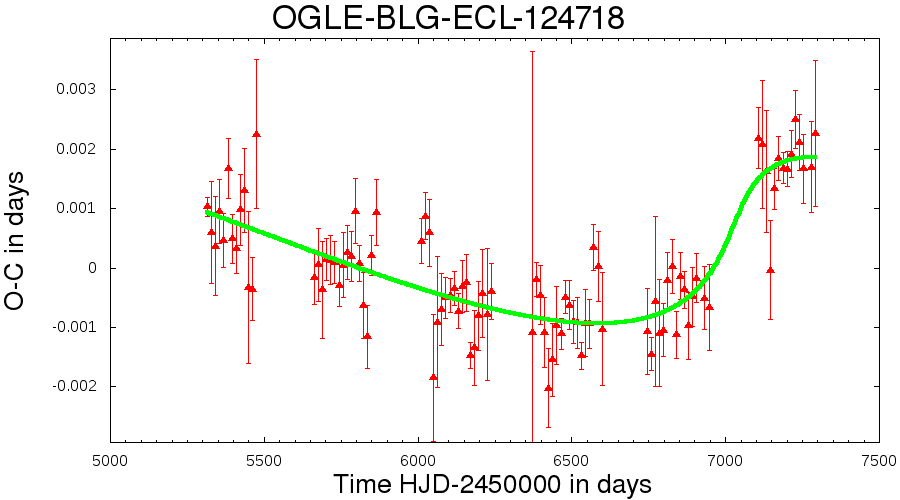}

\includegraphics[width=0.64\columnwidth]{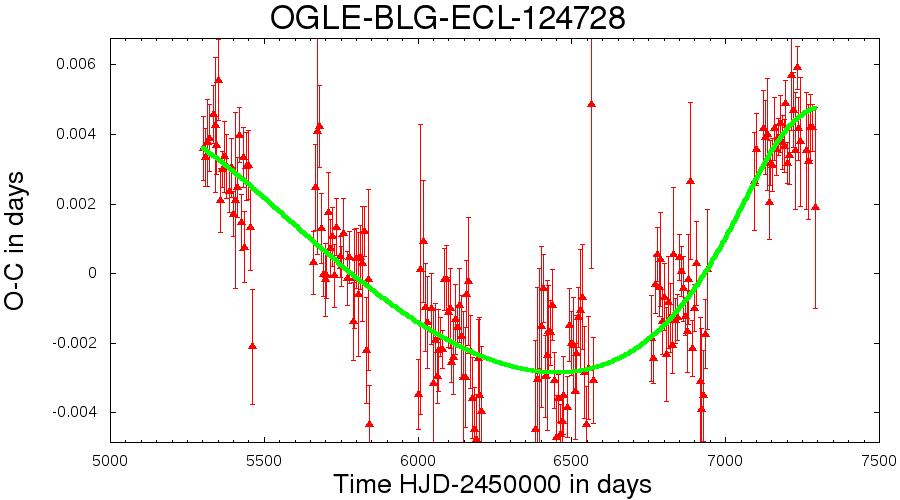}
\includegraphics[width=0.64\columnwidth]{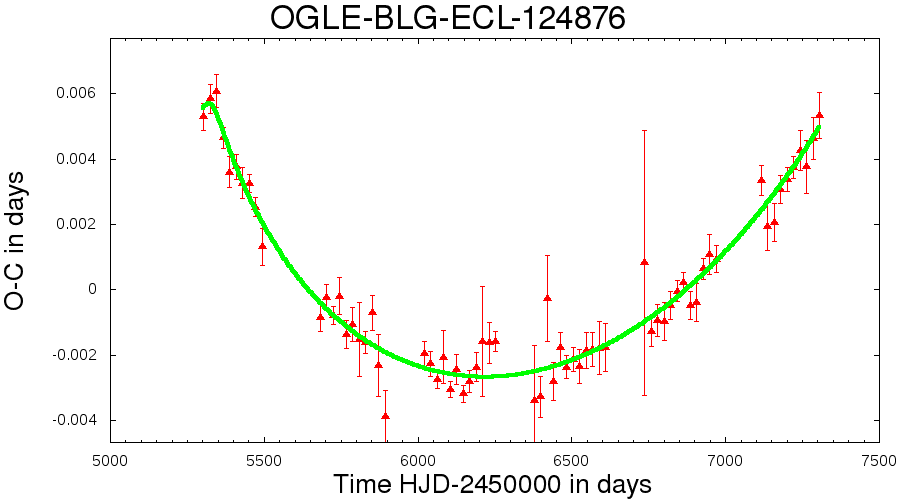}
\includegraphics[width=0.64\columnwidth]{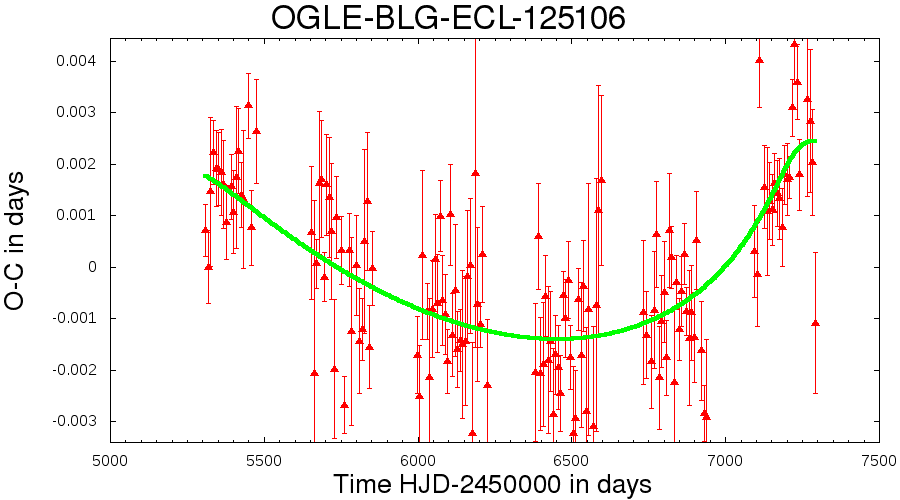}

\includegraphics[width=0.64\columnwidth]{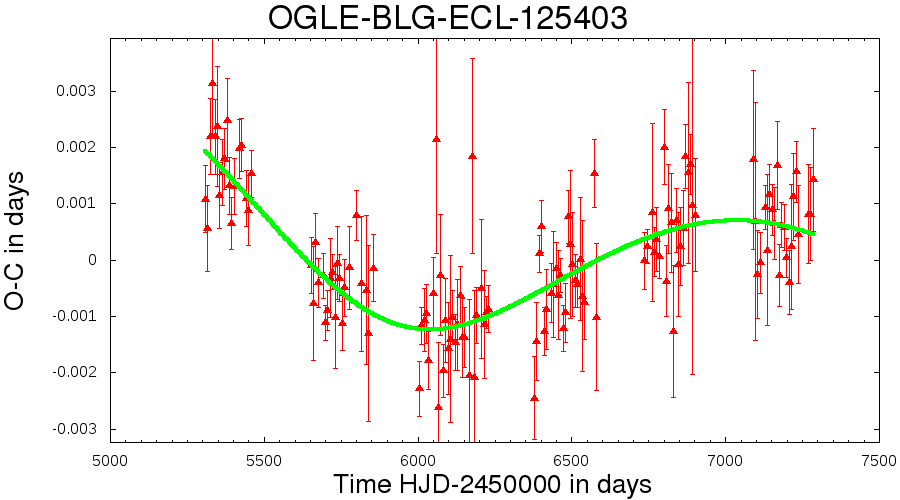}
\includegraphics[width=0.64\columnwidth]{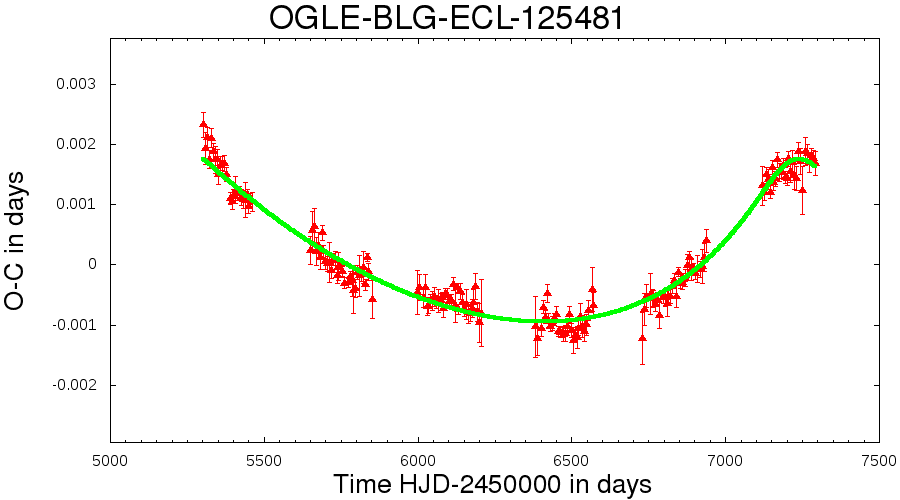}
\includegraphics[width=0.64\columnwidth]{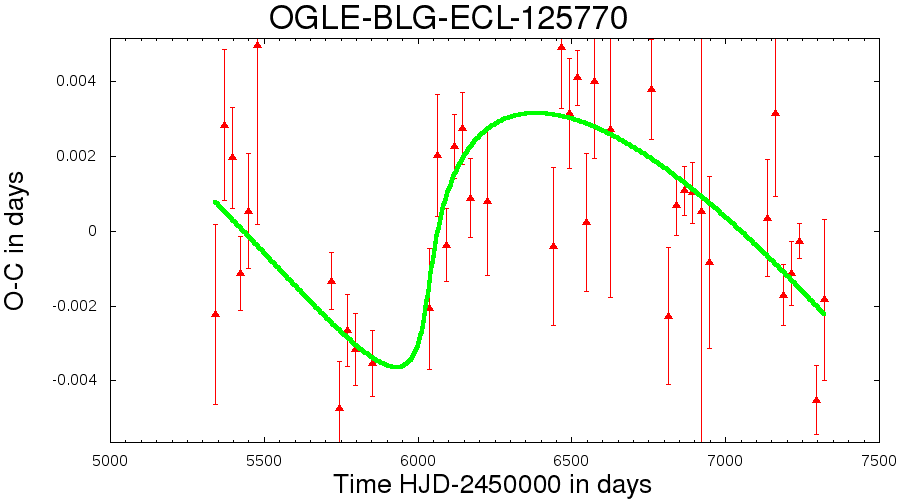}

\includegraphics[width=0.64\columnwidth]{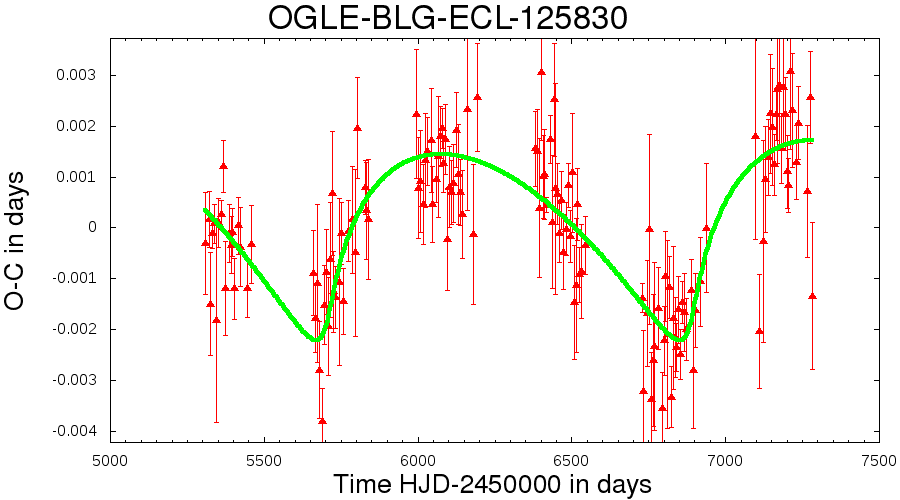}
\includegraphics[width=0.64\columnwidth]{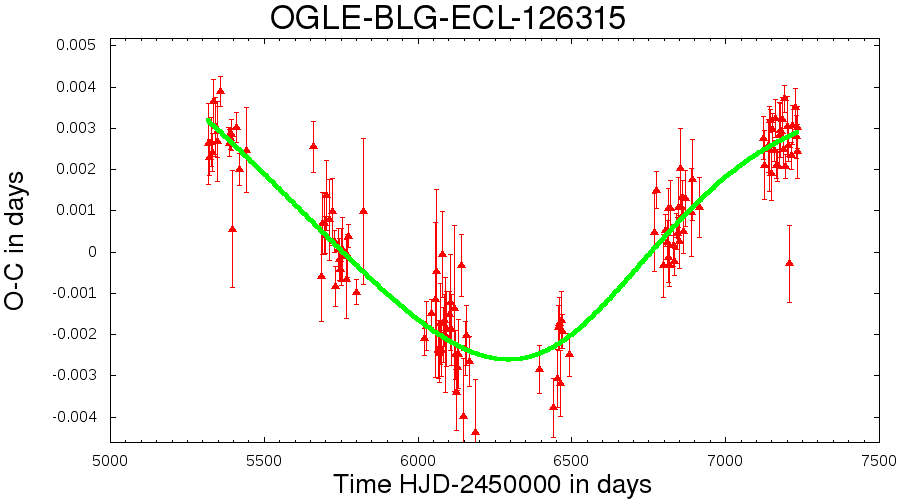}
\includegraphics[width=0.64\columnwidth]{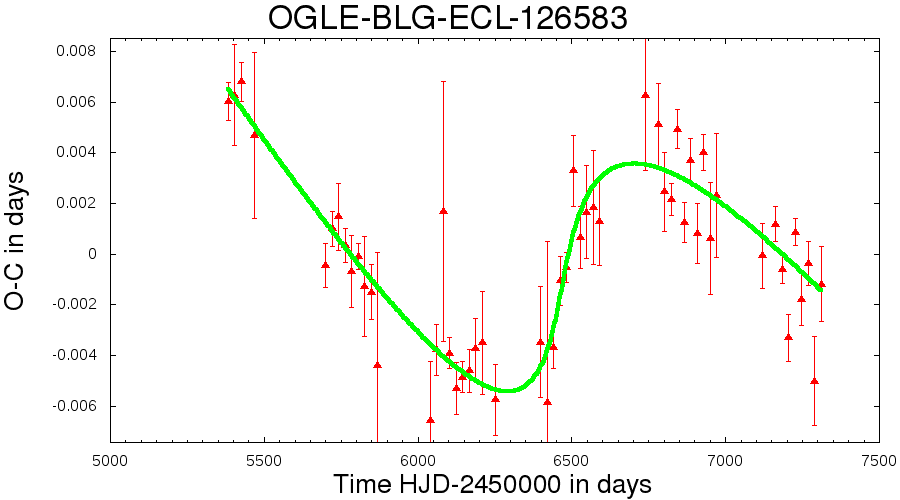}

\includegraphics[width=0.64\columnwidth]{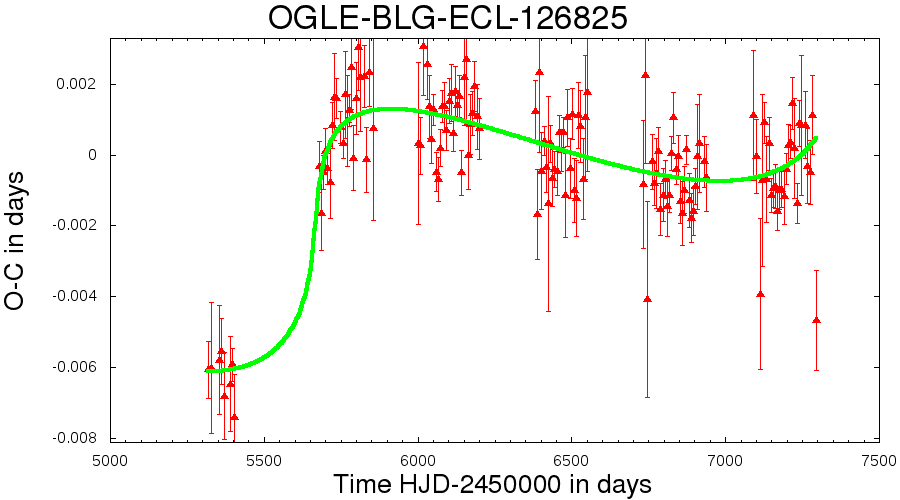}
\includegraphics[width=0.64\columnwidth]{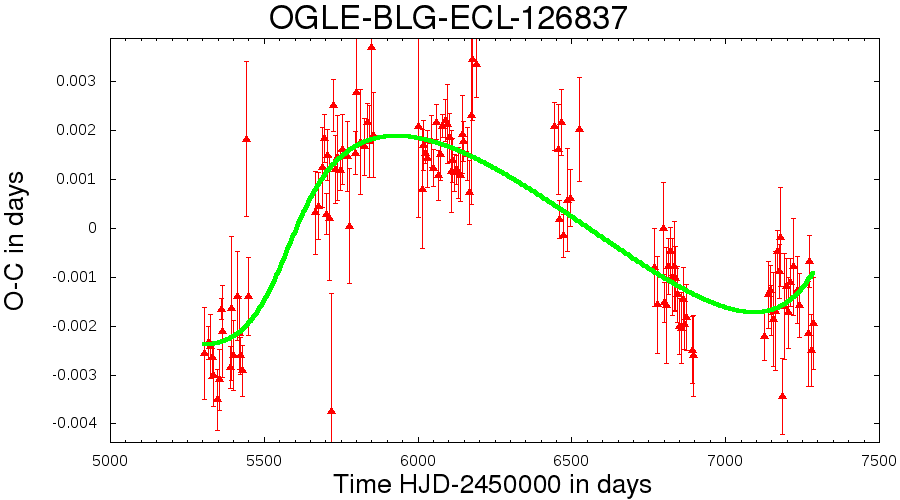}
\includegraphics[width=0.64\columnwidth]{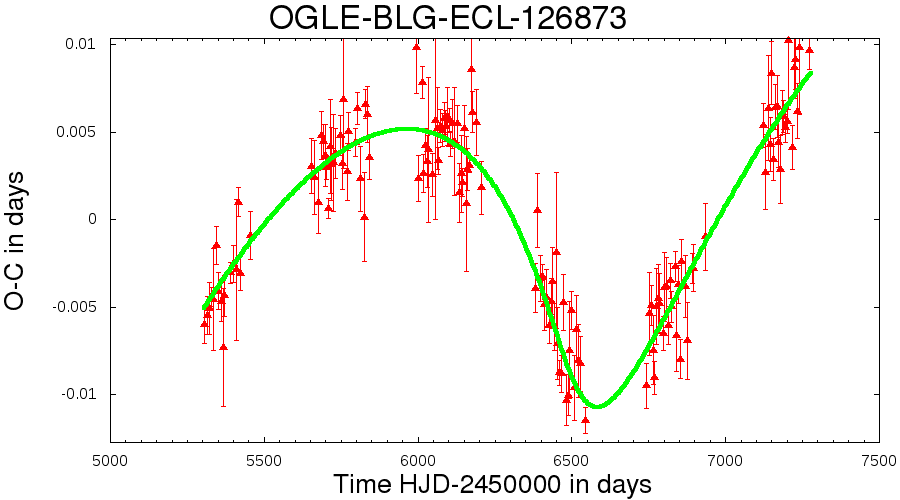}

\end{figure*}
\clearpage

\begin{figure*}
\includegraphics[width=0.64\columnwidth]{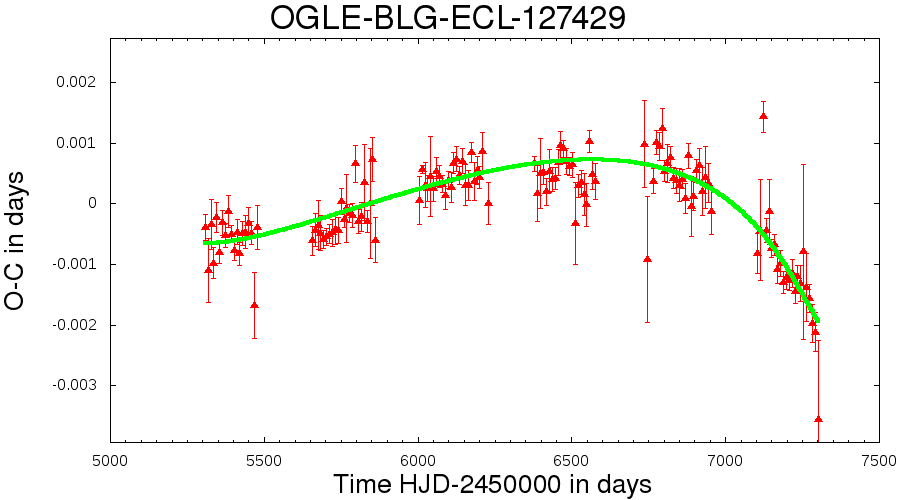}
\includegraphics[width=0.64\columnwidth]{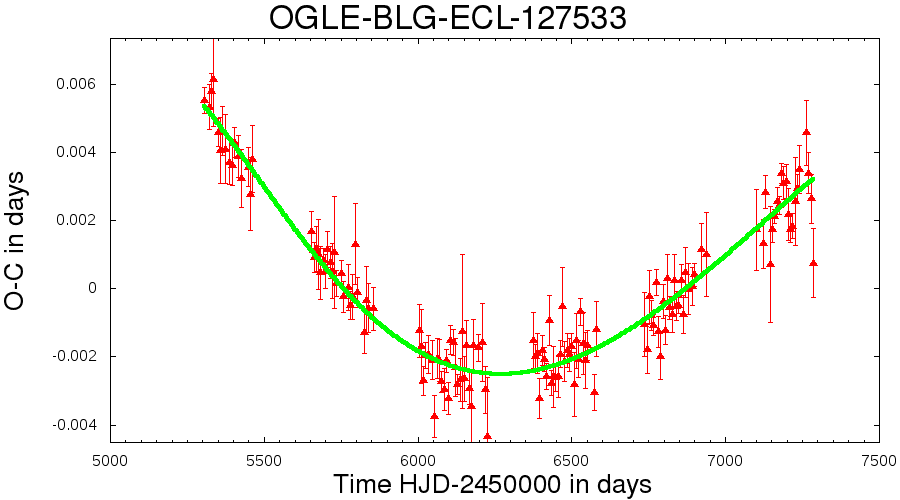}
\includegraphics[width=0.64\columnwidth]{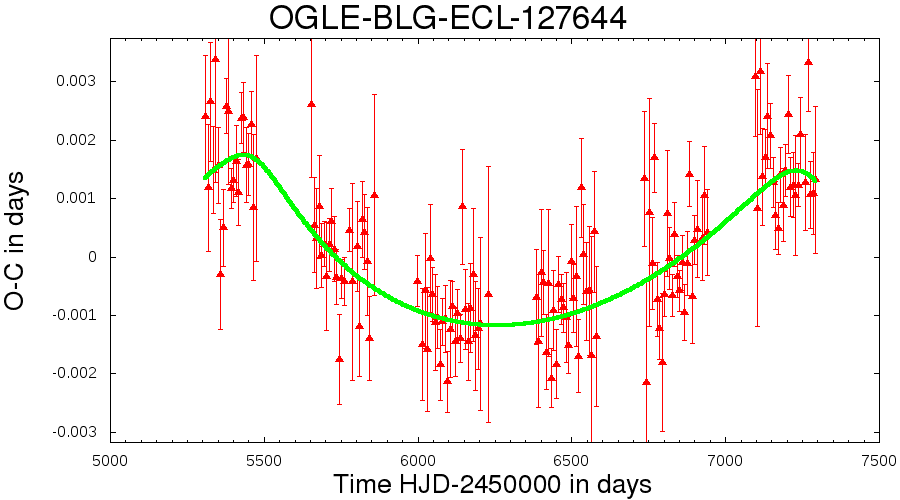}

\includegraphics[width=0.64\columnwidth]{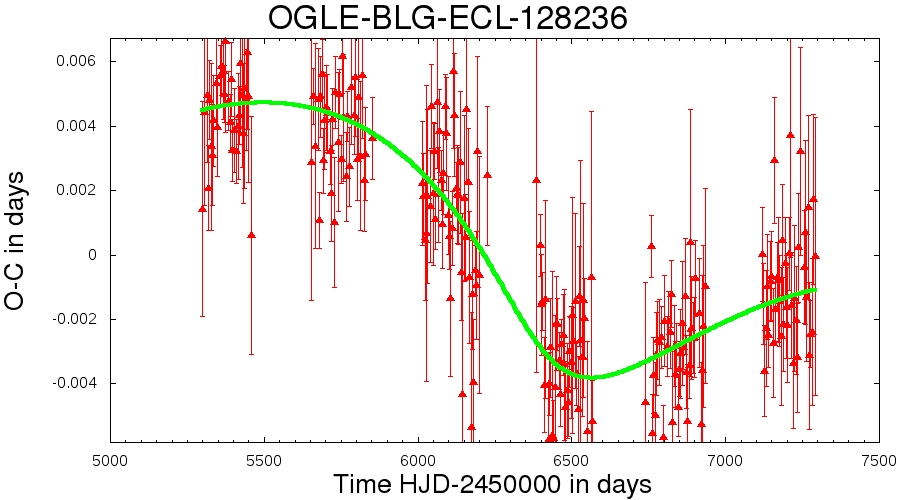}
\includegraphics[width=0.64\columnwidth]{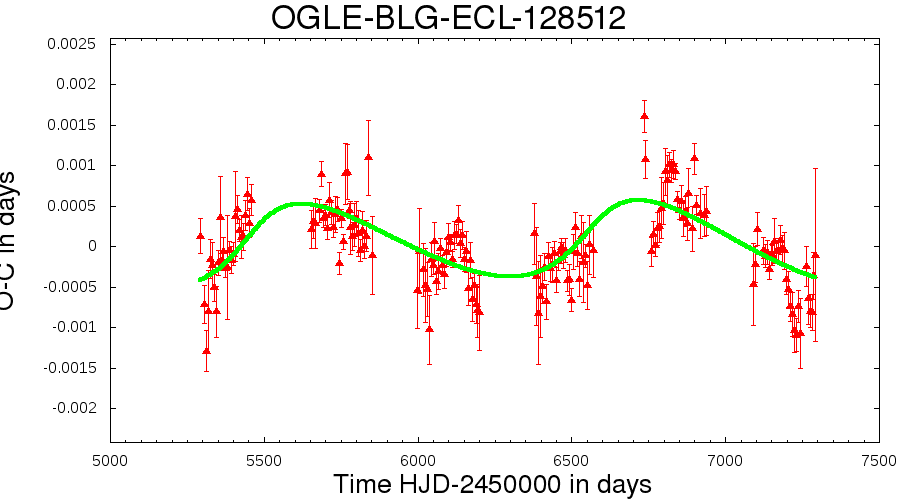}
\includegraphics[width=0.64\columnwidth]{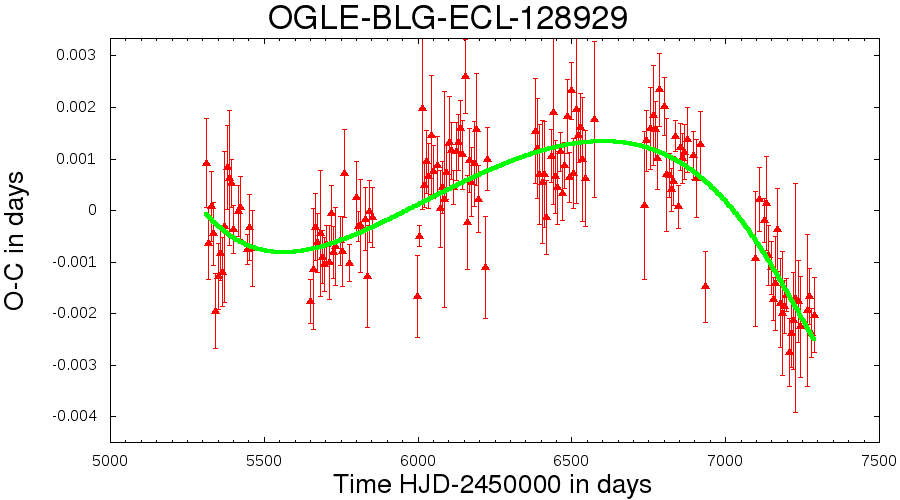}

\includegraphics[width=0.64\columnwidth]{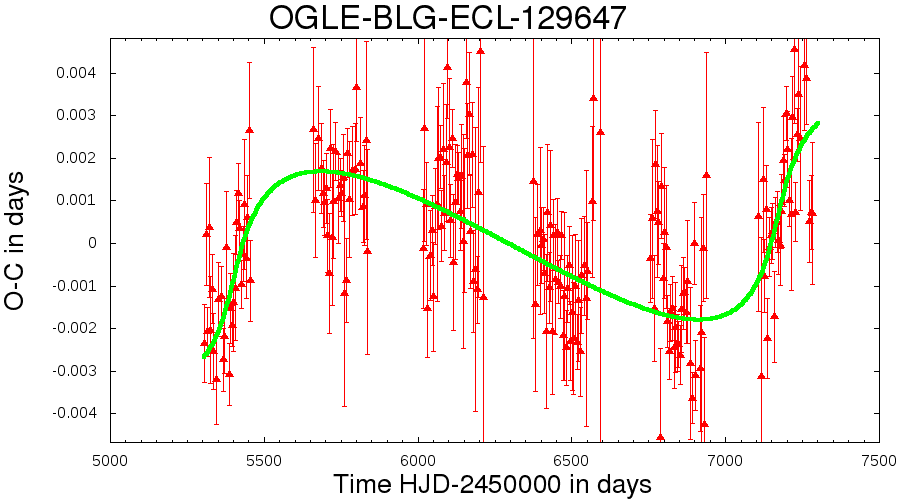}
\includegraphics[width=0.64\columnwidth]{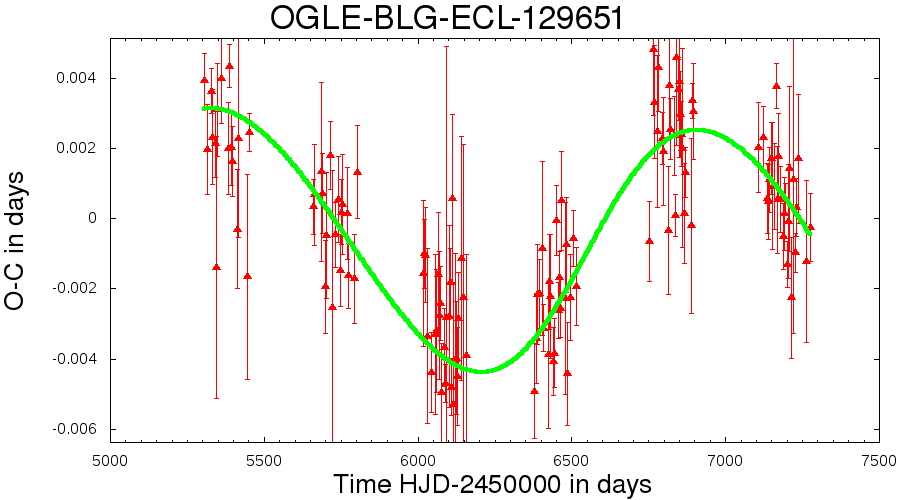}
\includegraphics[width=0.64\columnwidth]{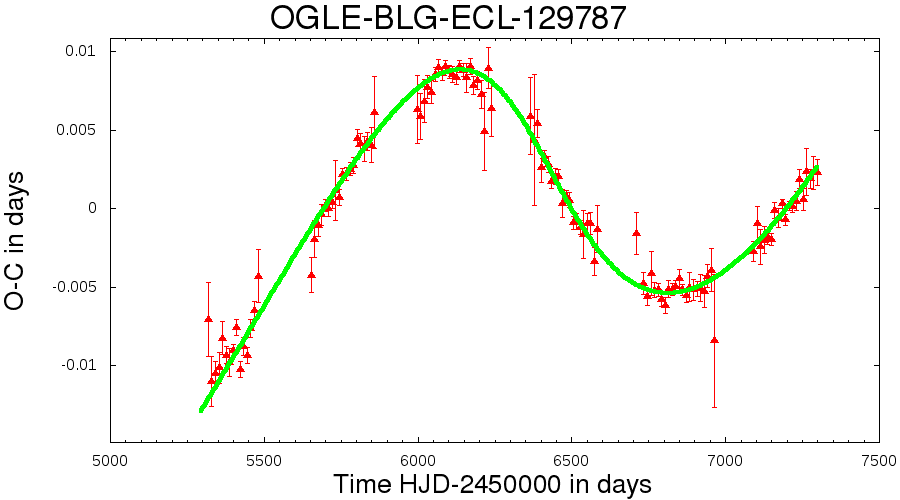}

\includegraphics[width=0.64\columnwidth]{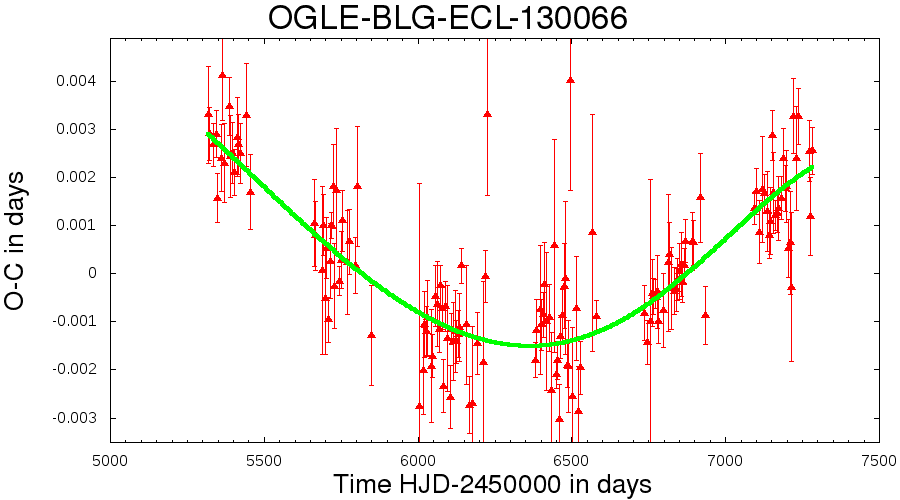}
\includegraphics[width=0.64\columnwidth]{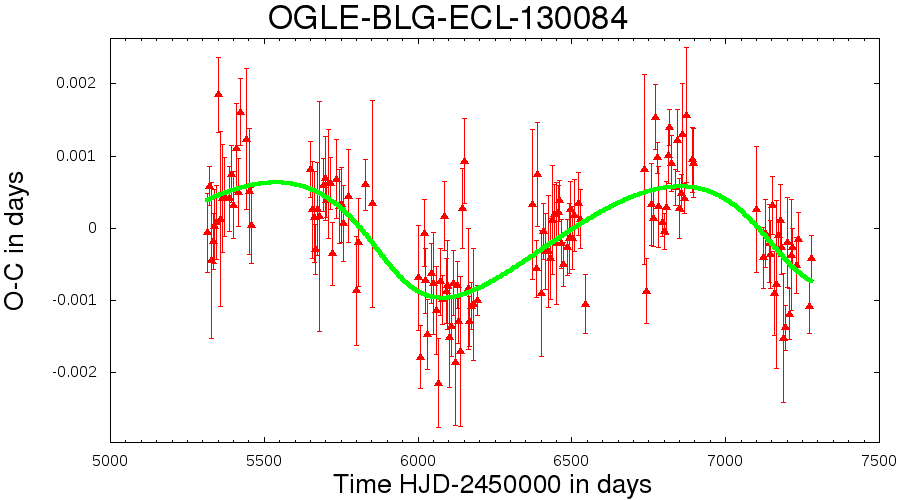}
\includegraphics[width=0.64\columnwidth]{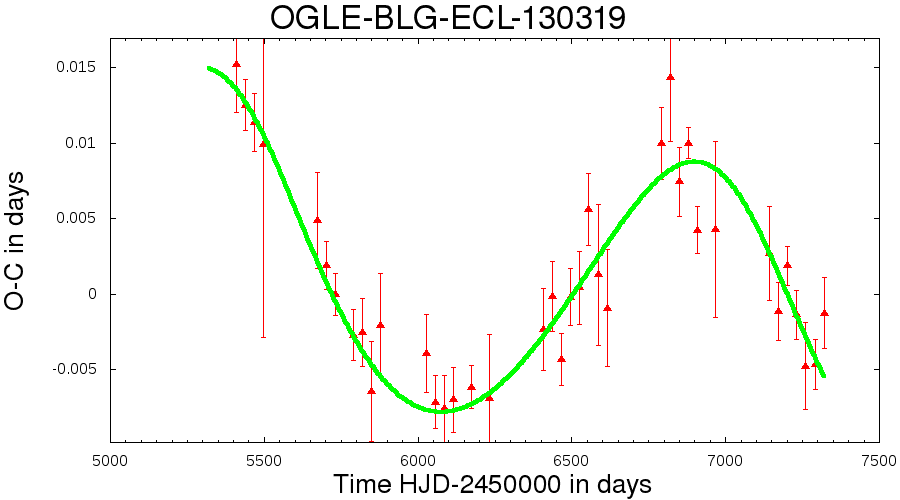}

\includegraphics[width=0.64\columnwidth]{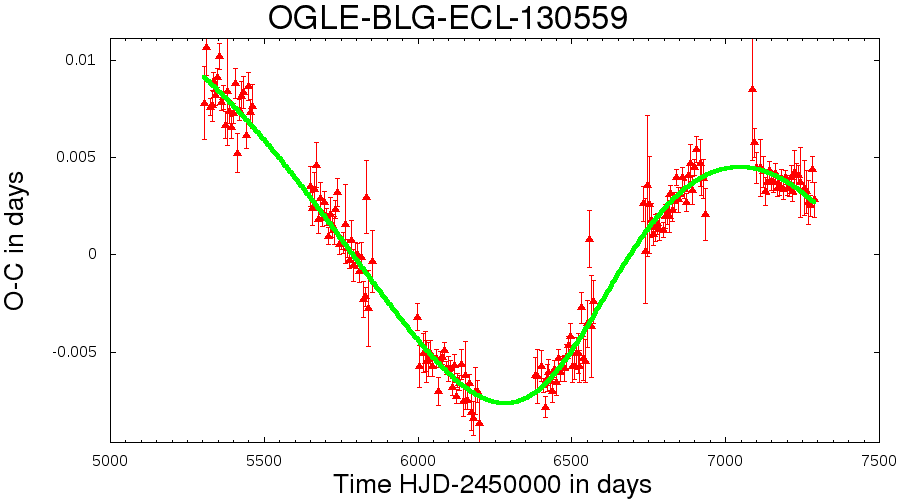}
\includegraphics[width=0.64\columnwidth]{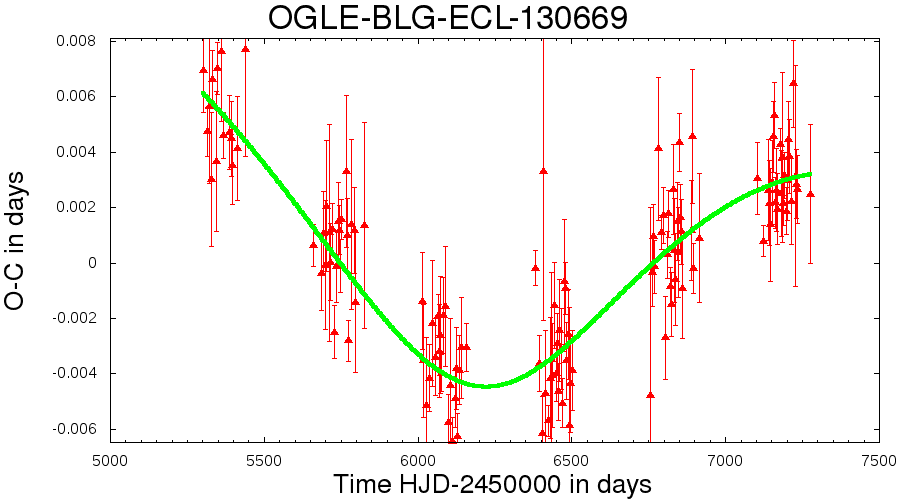}
\includegraphics[width=0.64\columnwidth]{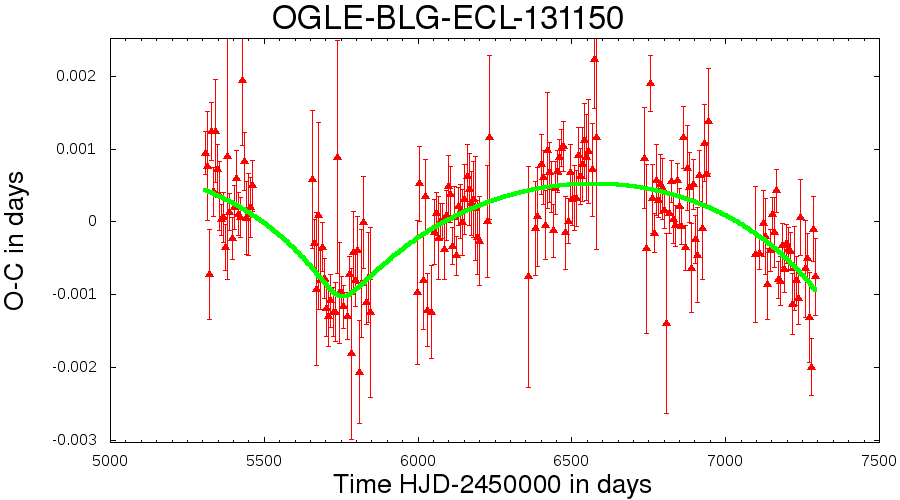}

\includegraphics[width=0.64\columnwidth]{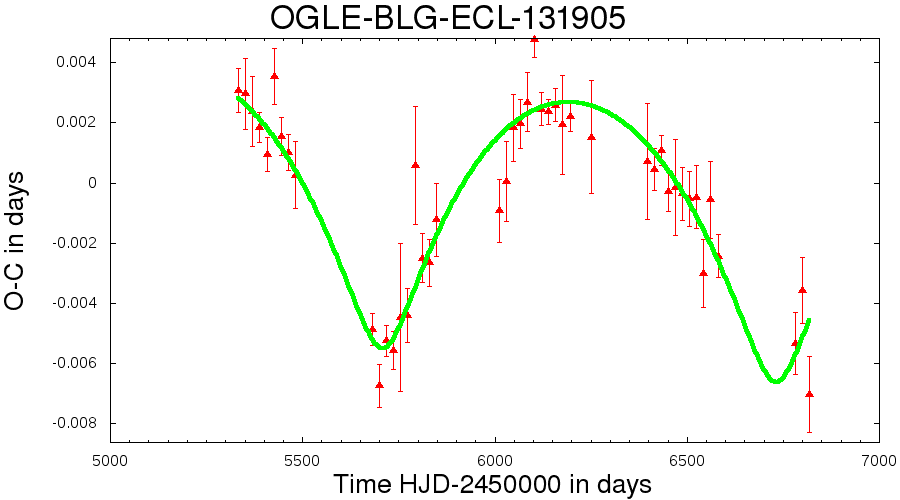}
\includegraphics[width=0.64\columnwidth]{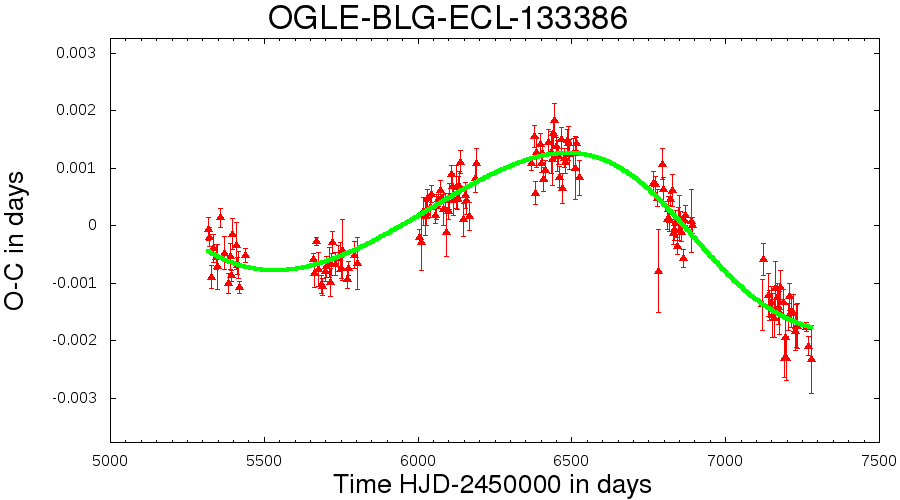}
\includegraphics[width=0.64\columnwidth]{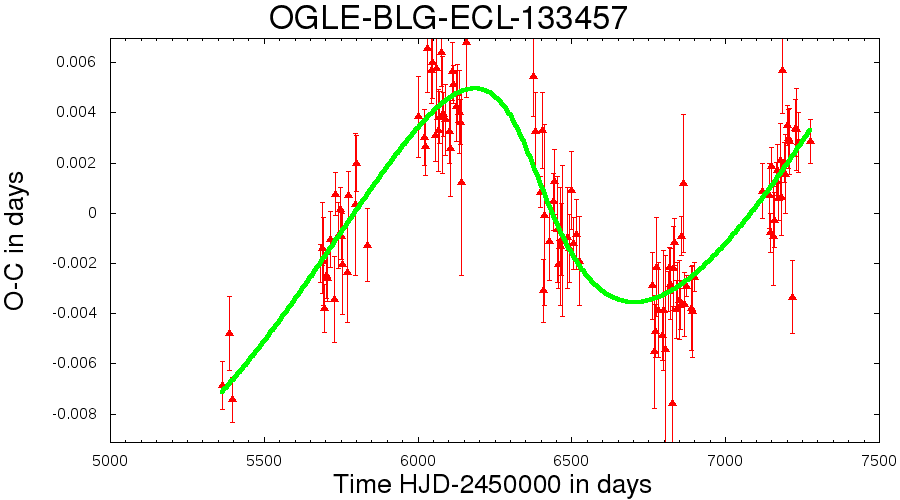}

\includegraphics[width=0.64\columnwidth]{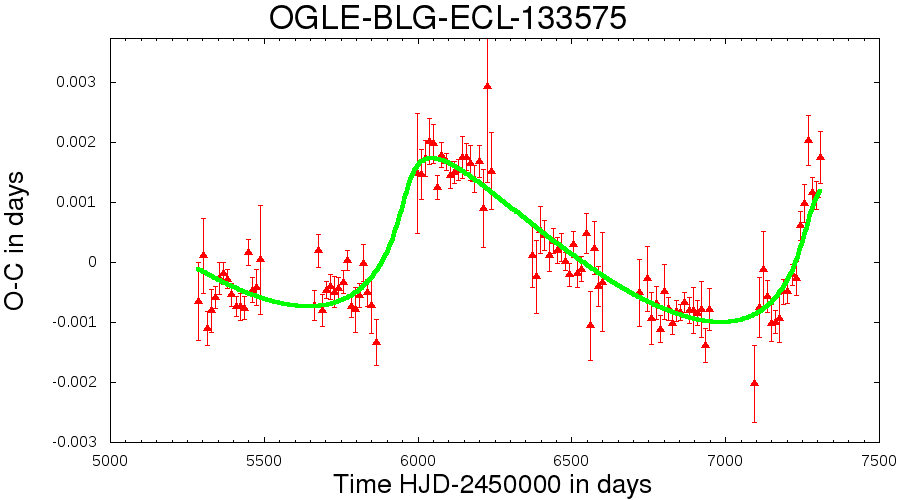}
\includegraphics[width=0.64\columnwidth]{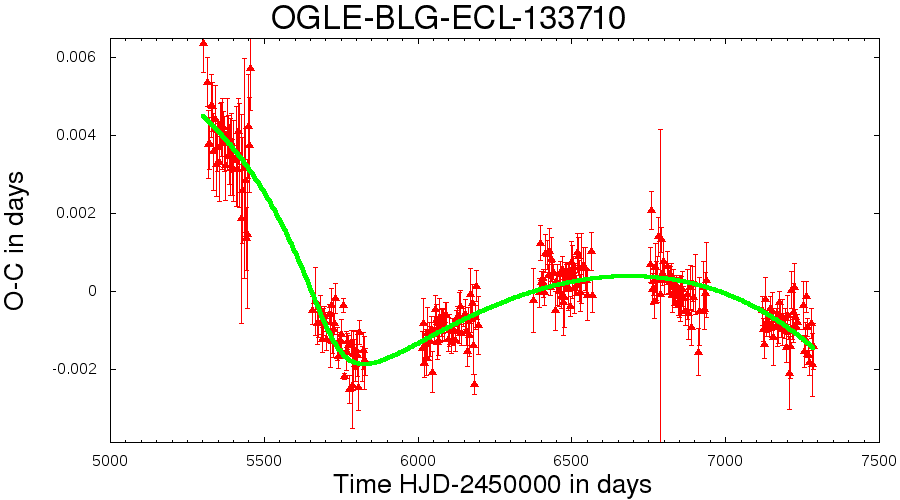}
\includegraphics[width=0.64\columnwidth]{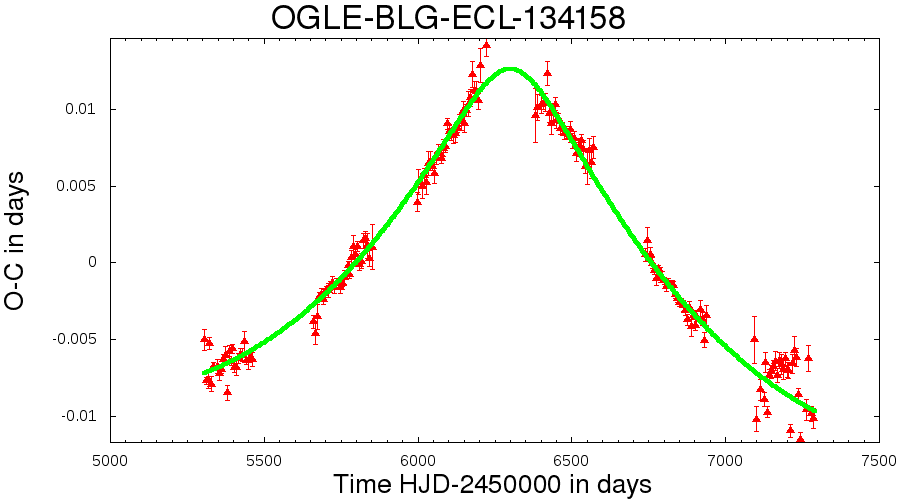}

\includegraphics[width=0.64\columnwidth]{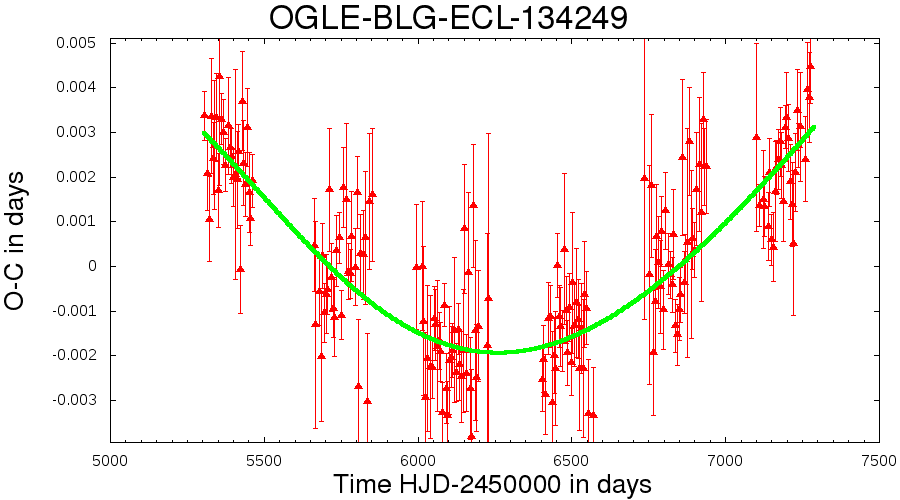}
\includegraphics[width=0.64\columnwidth]{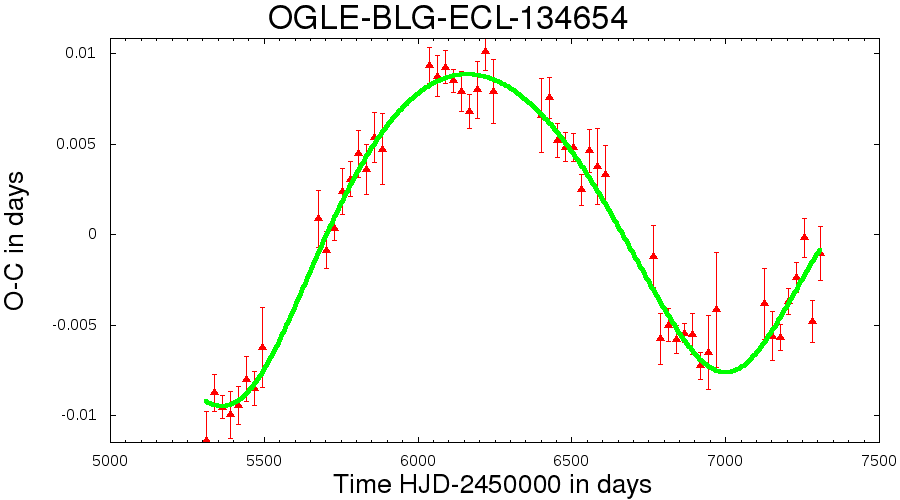}
\includegraphics[width=0.64\columnwidth]{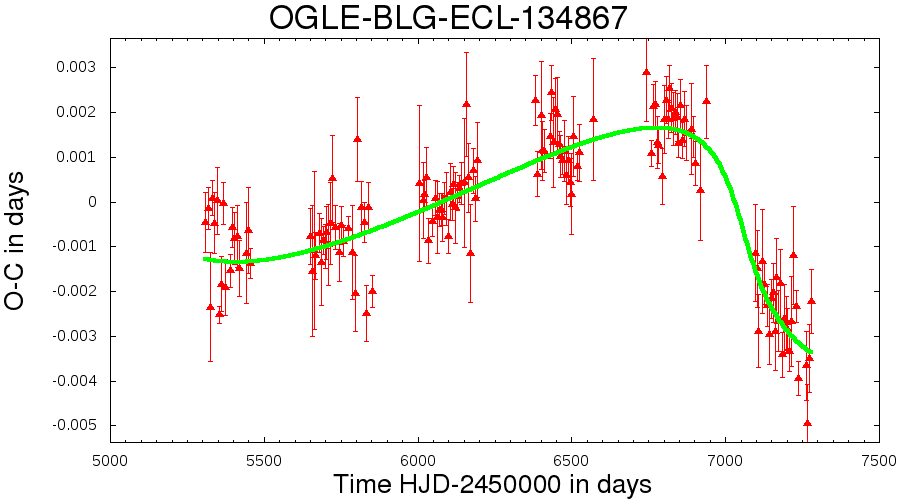}

\end{figure*}
\clearpage

\begin{figure*}
\includegraphics[width=0.64\columnwidth]{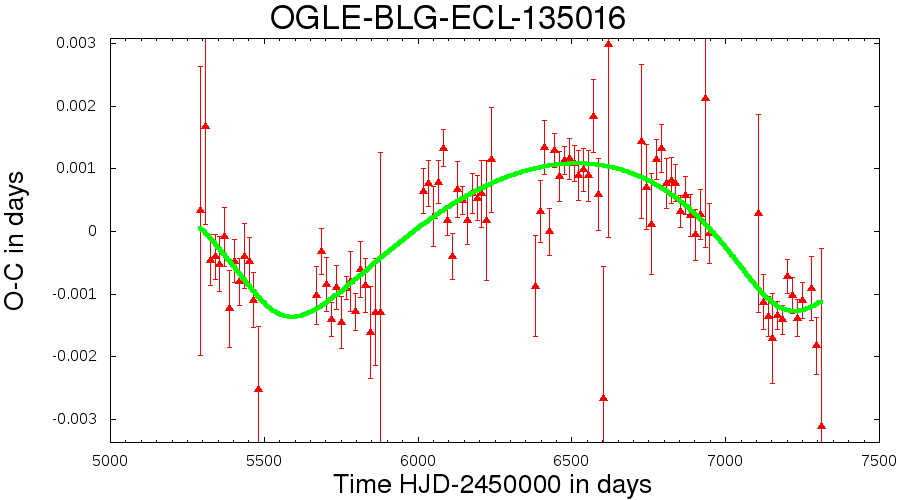}
\includegraphics[width=0.64\columnwidth]{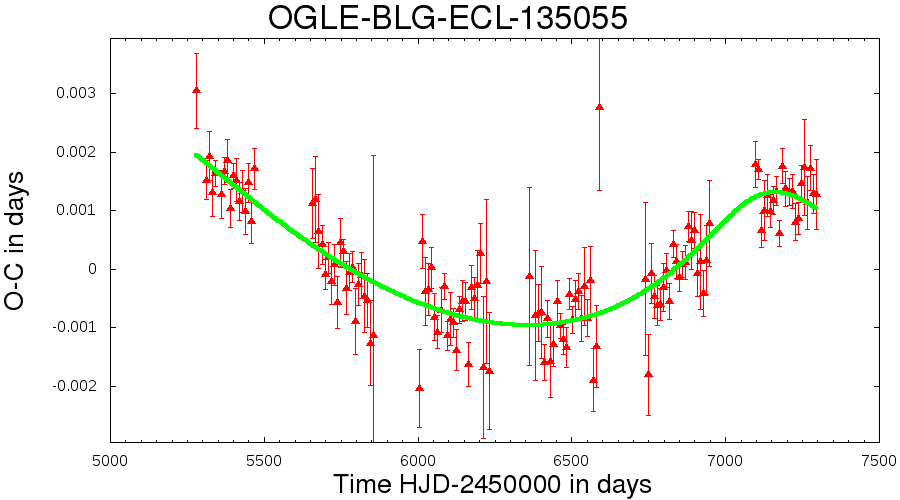}
\includegraphics[width=0.64\columnwidth]{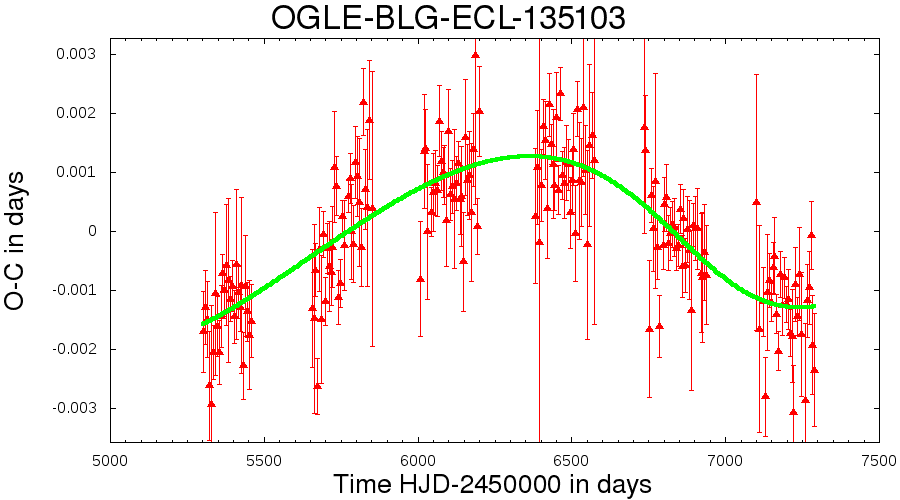}

\includegraphics[width=0.64\columnwidth]{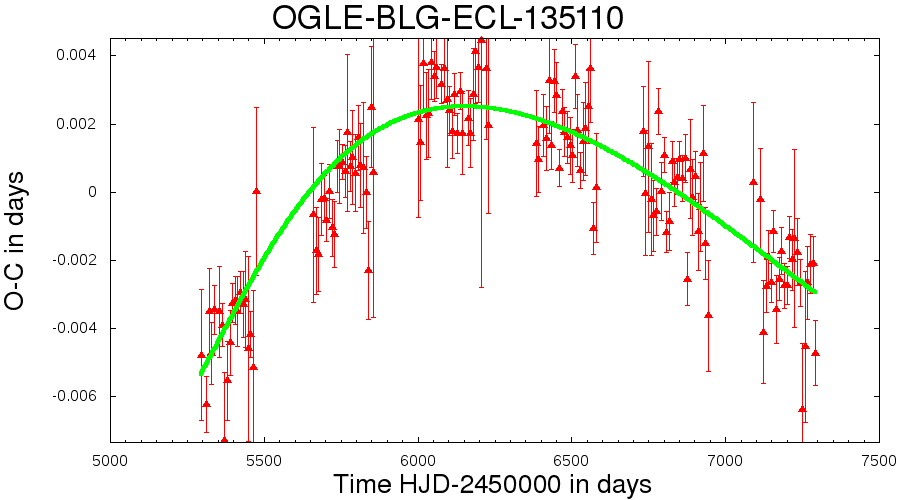}
\includegraphics[width=0.64\columnwidth]{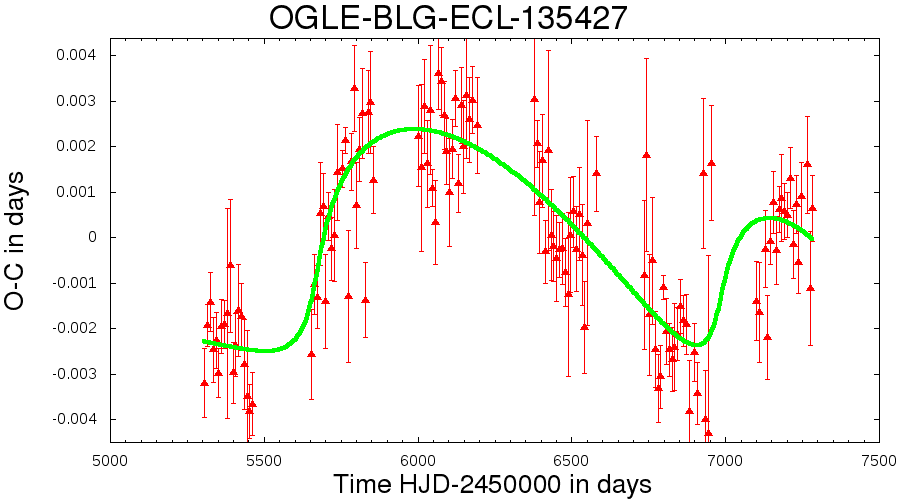}
\includegraphics[width=0.64\columnwidth]{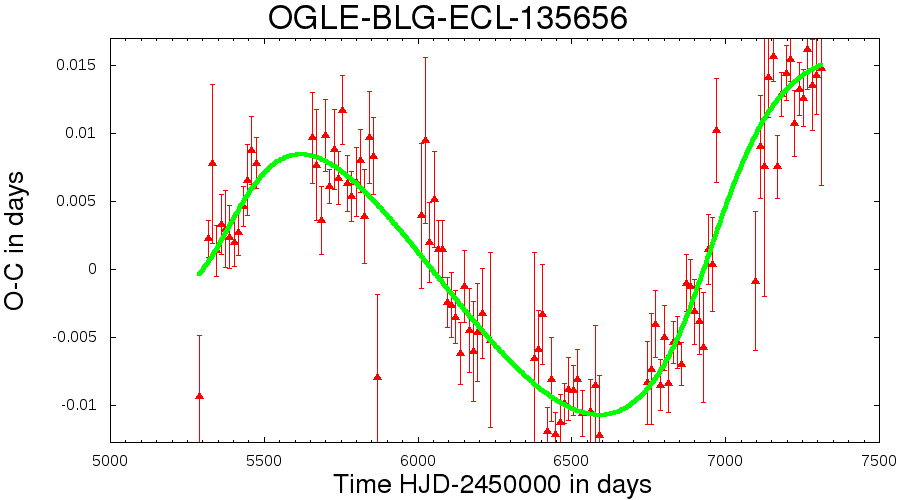}

\includegraphics[width=0.64\columnwidth]{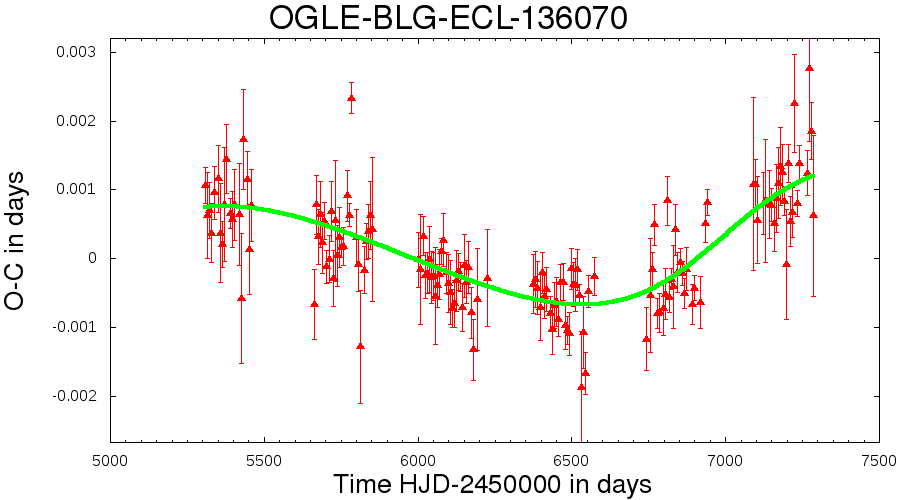}
\includegraphics[width=0.64\columnwidth]{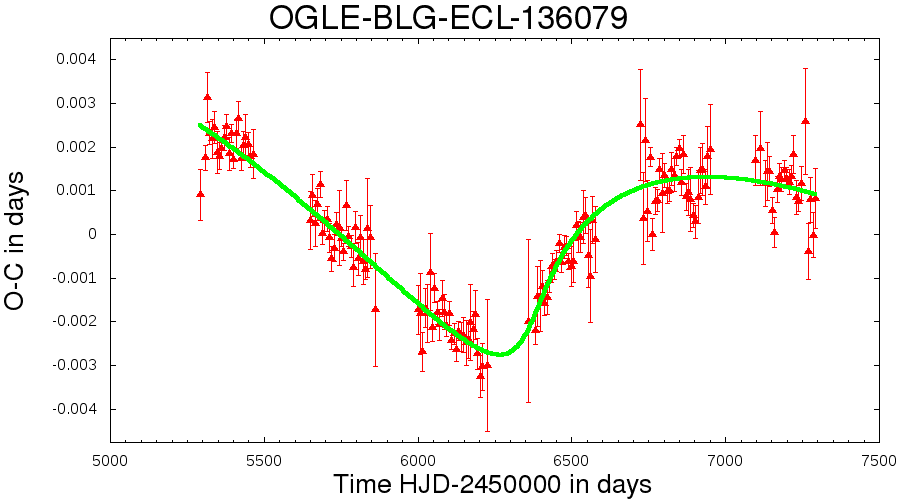}
\includegraphics[width=0.64\columnwidth]{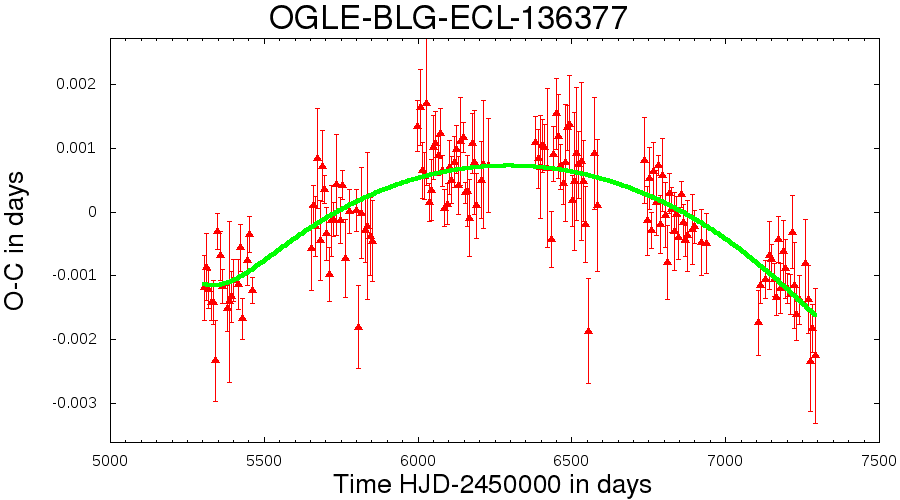}

\includegraphics[width=0.64\columnwidth]{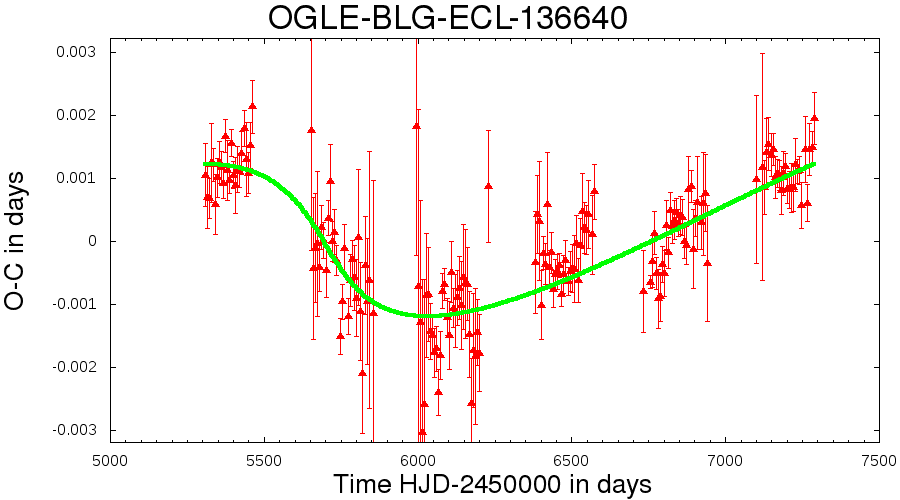}
\includegraphics[width=0.64\columnwidth]{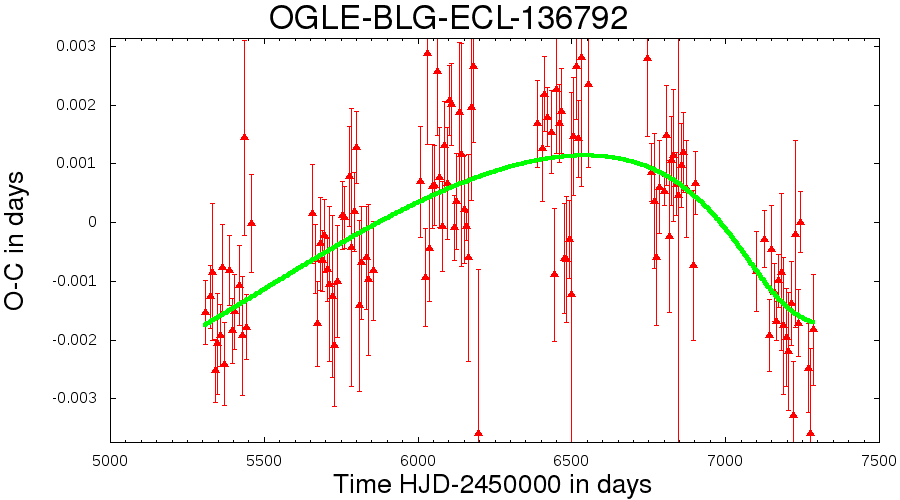}
\includegraphics[width=0.64\columnwidth]{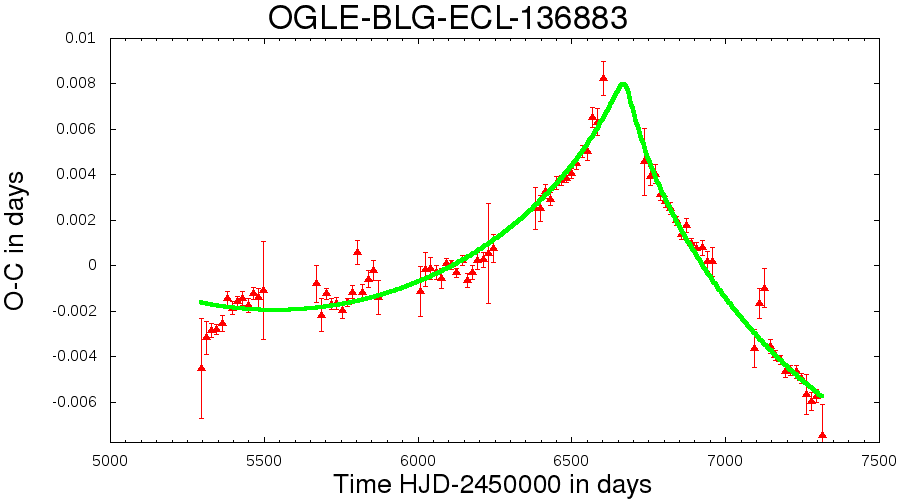}

\includegraphics[width=0.64\columnwidth]{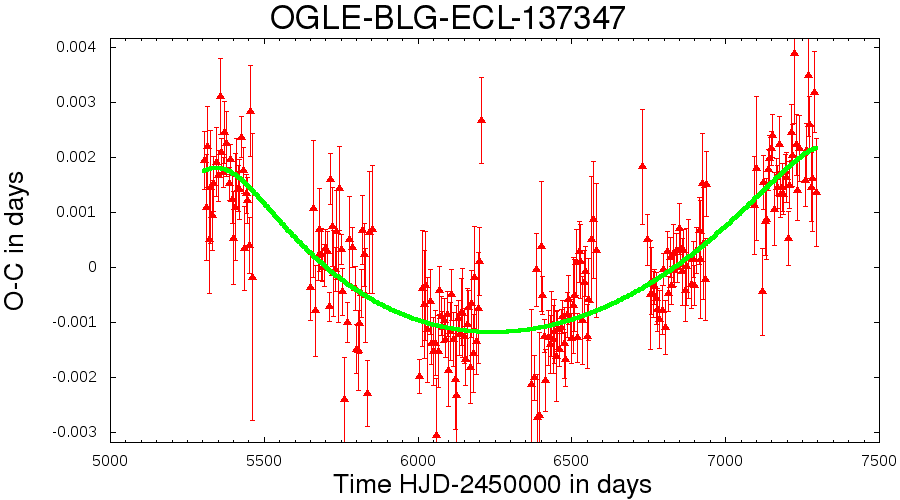}
\includegraphics[width=0.64\columnwidth]{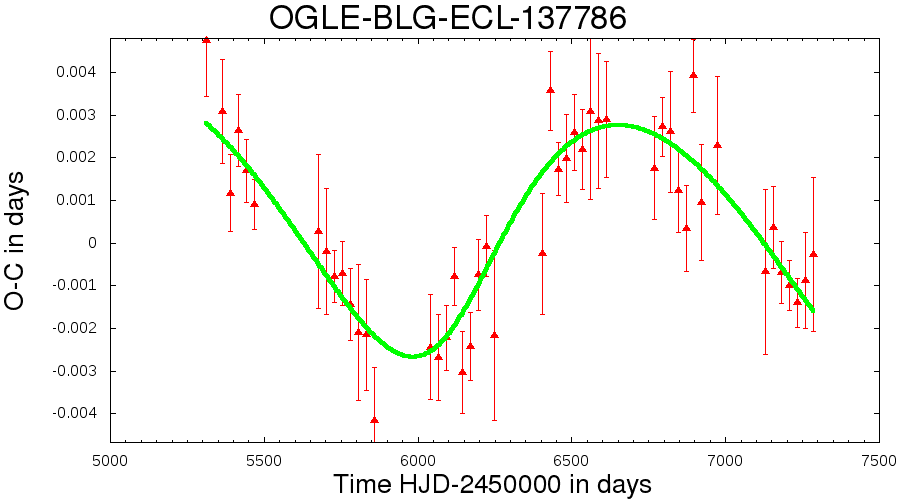}
\includegraphics[width=0.64\columnwidth]{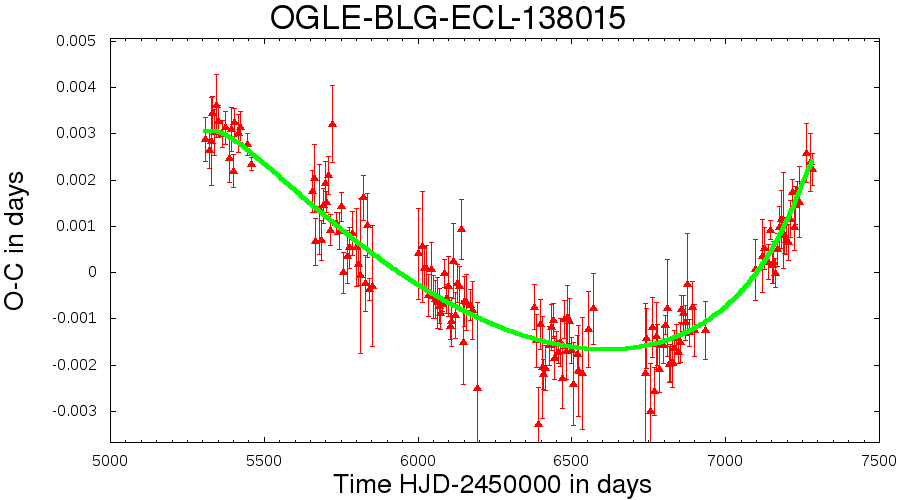}

\includegraphics[width=0.64\columnwidth]{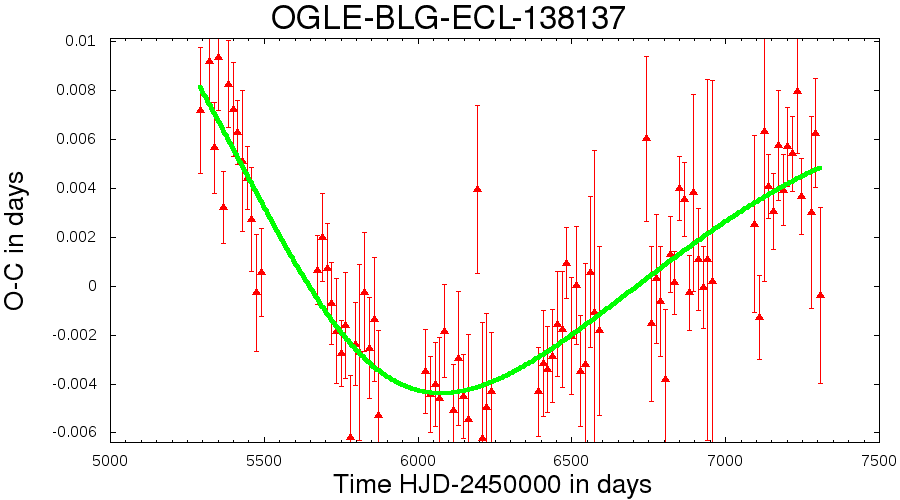}
\includegraphics[width=0.64\columnwidth]{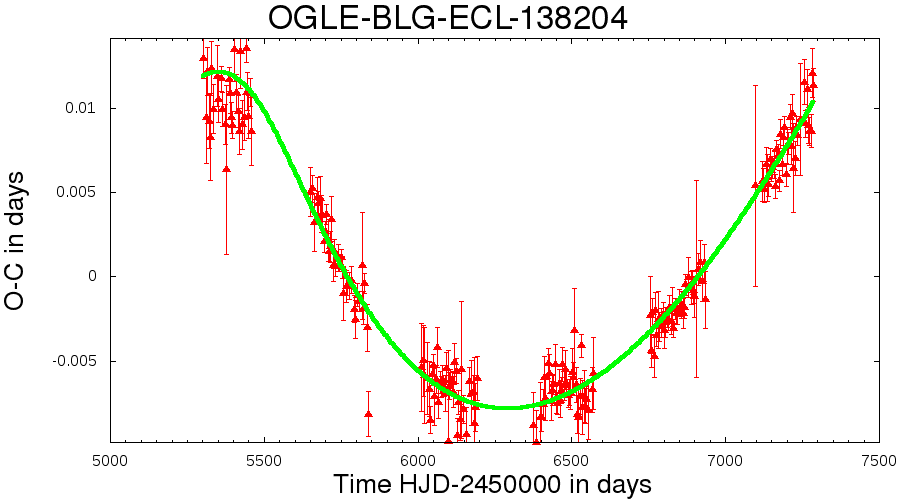}
\includegraphics[width=0.64\columnwidth]{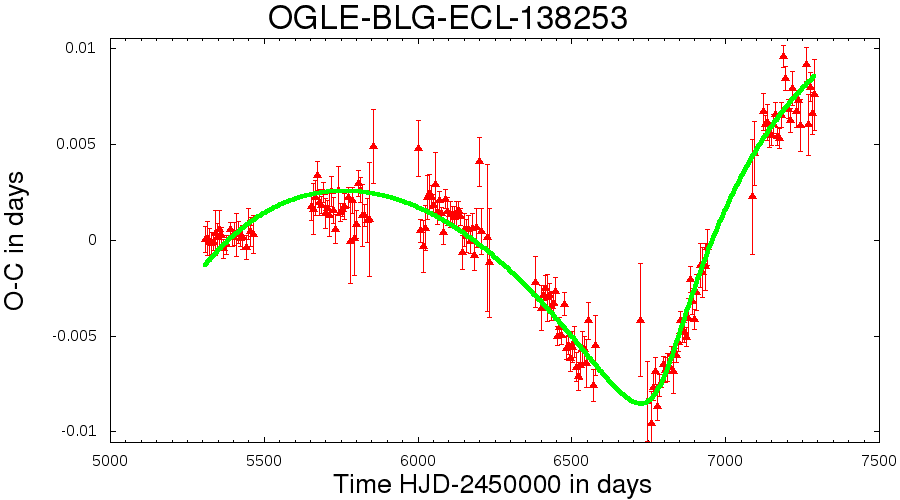}

\includegraphics[width=0.64\columnwidth]{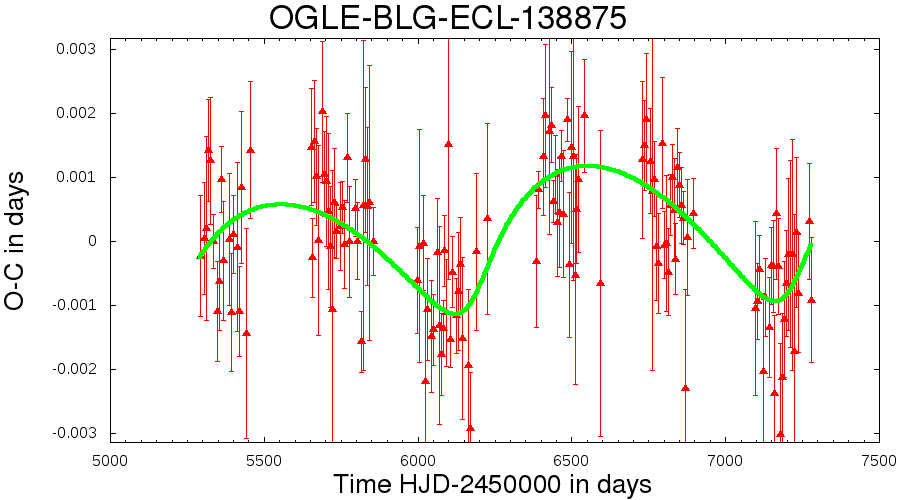}
\includegraphics[width=0.64\columnwidth]{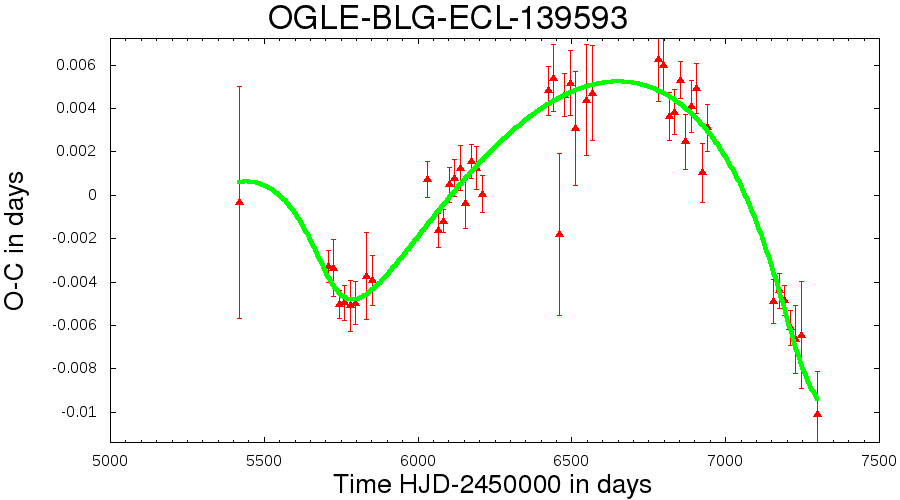}
\includegraphics[width=0.64\columnwidth]{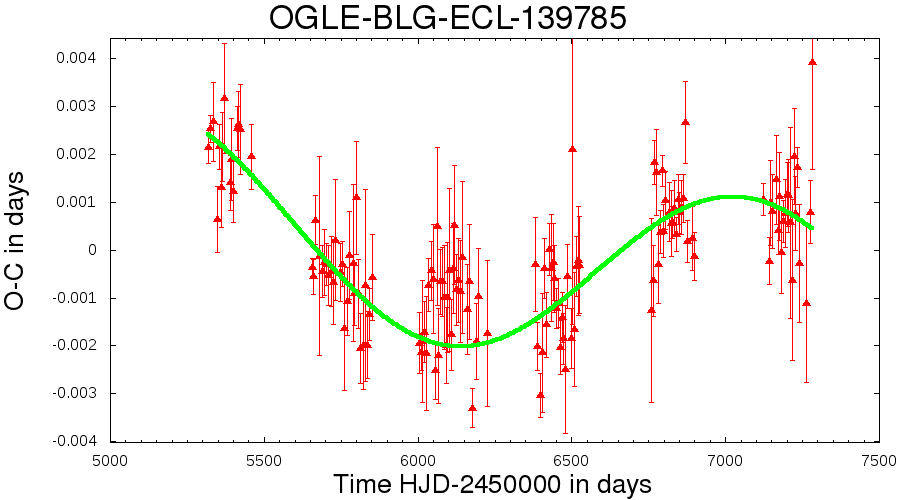}

\includegraphics[width=0.64\columnwidth]{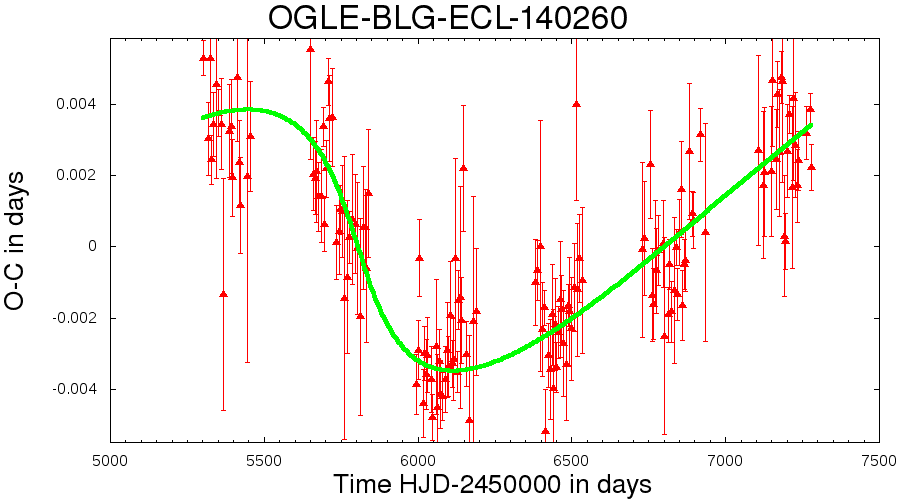}
\includegraphics[width=0.64\columnwidth]{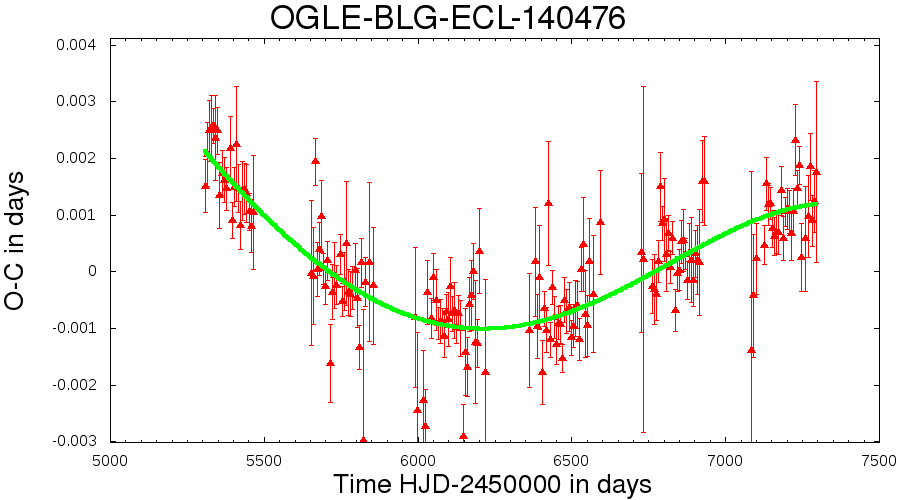}
\includegraphics[width=0.64\columnwidth]{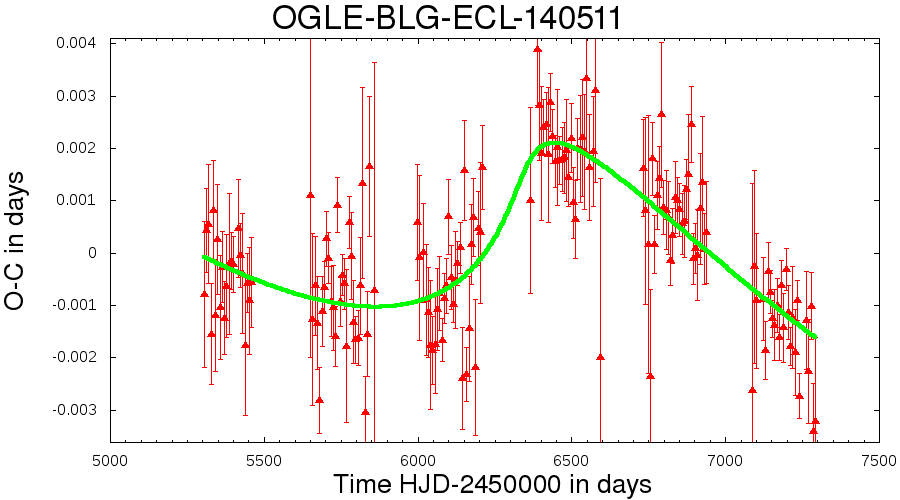}

\end{figure*}
\clearpage

\begin{figure*}
\includegraphics[width=0.64\columnwidth]{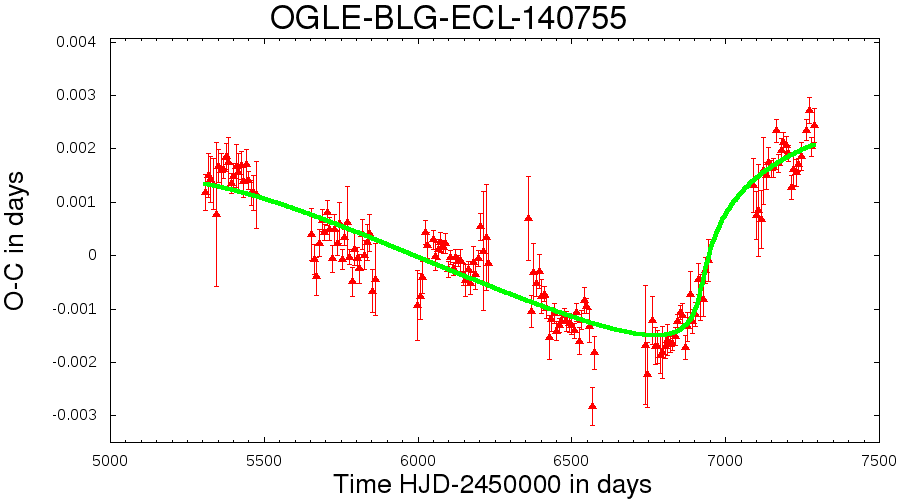}
\includegraphics[width=0.64\columnwidth]{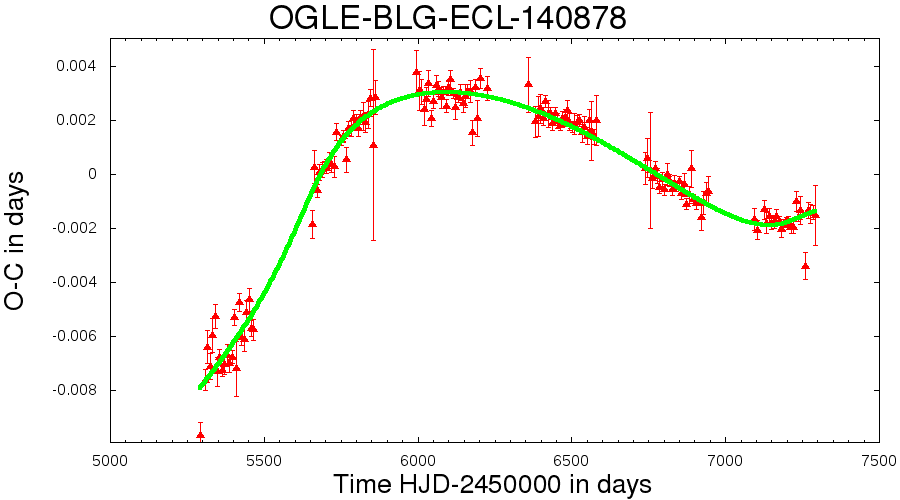}
\includegraphics[width=0.64\columnwidth]{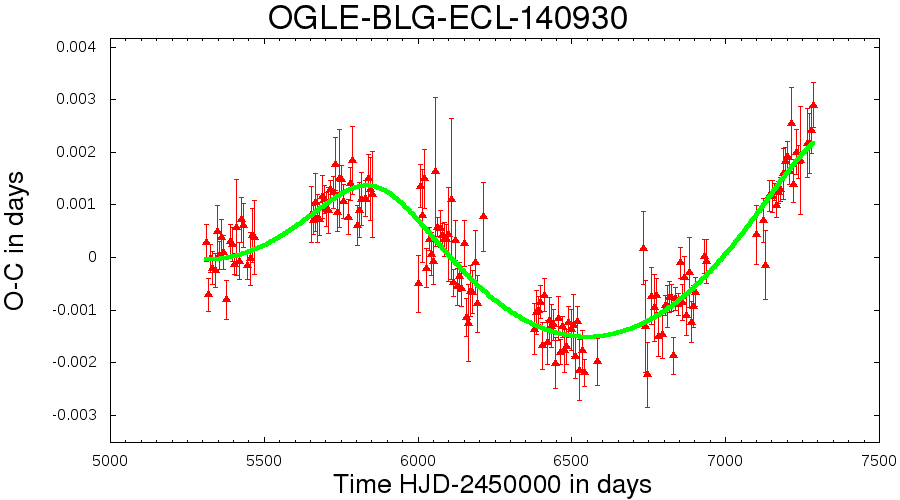}

\includegraphics[width=0.64\columnwidth]{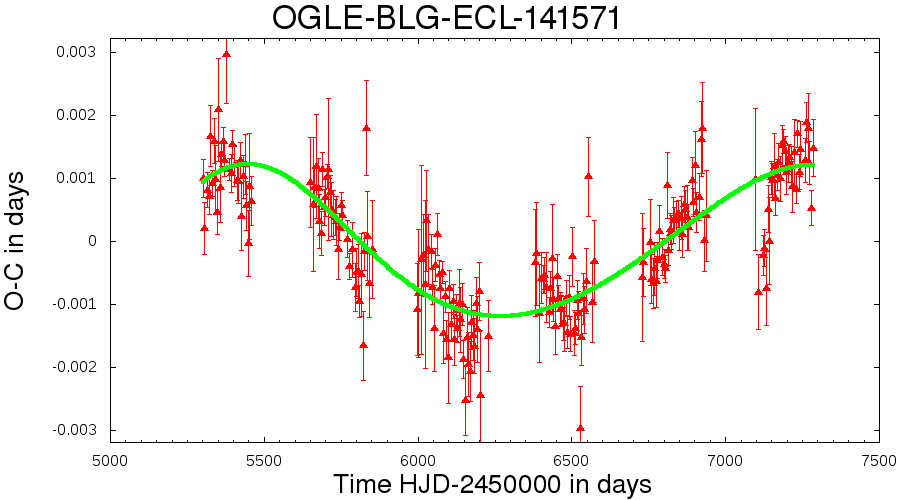}
\includegraphics[width=0.64\columnwidth]{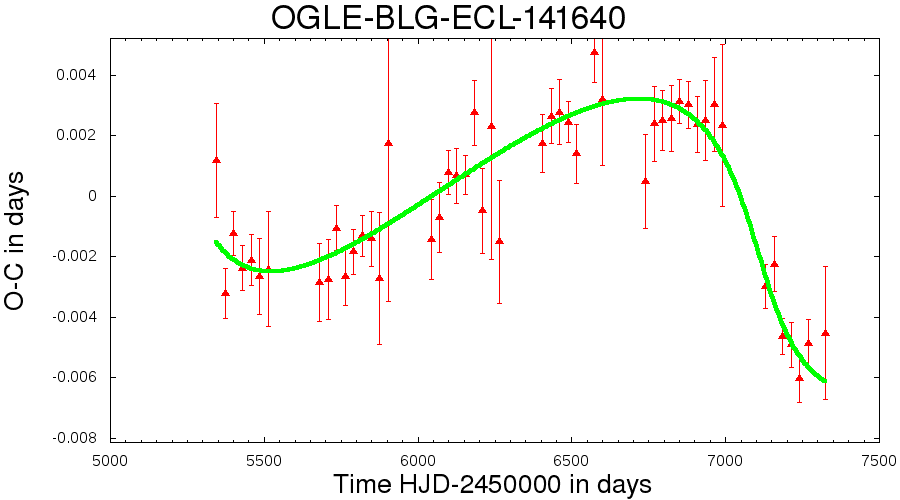}
\includegraphics[width=0.64\columnwidth]{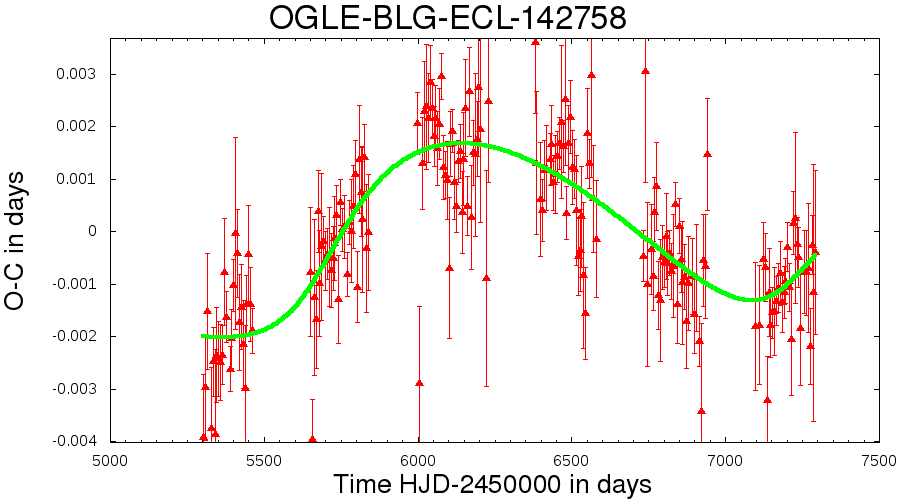}

\includegraphics[width=0.64\columnwidth]{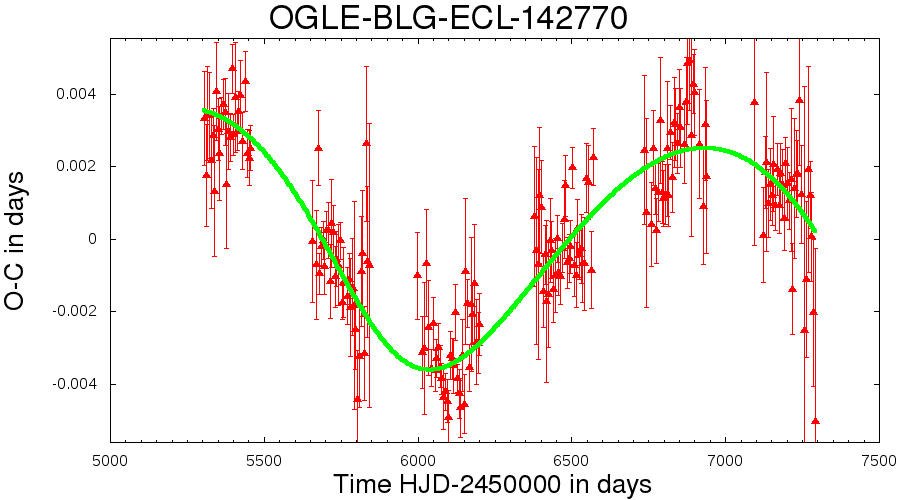}
\includegraphics[width=0.64\columnwidth]{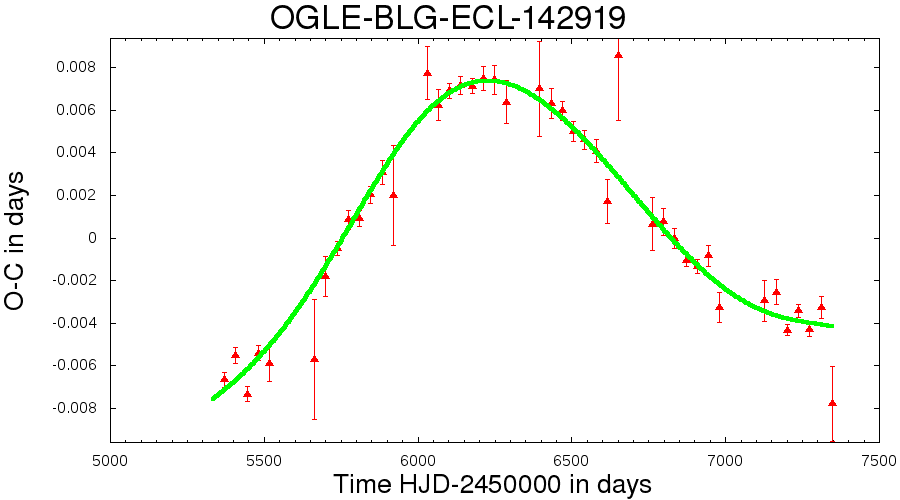}
\includegraphics[width=0.64\columnwidth]{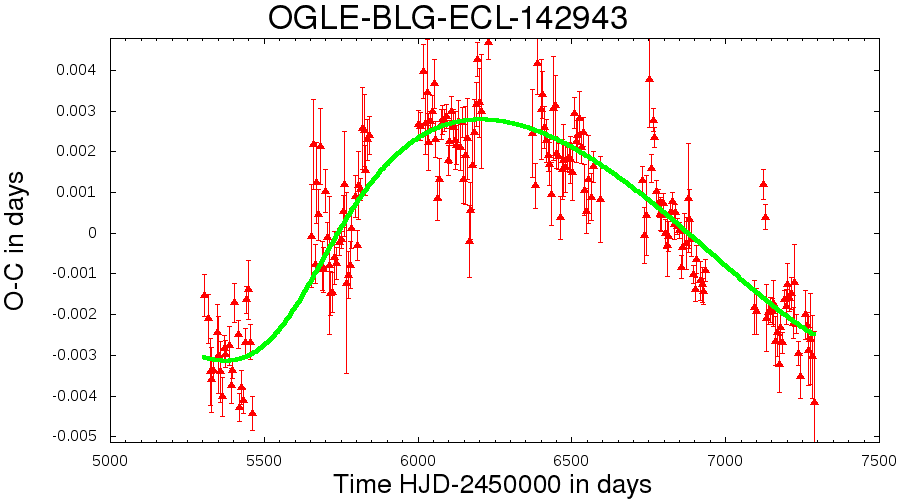}

\includegraphics[width=0.64\columnwidth]{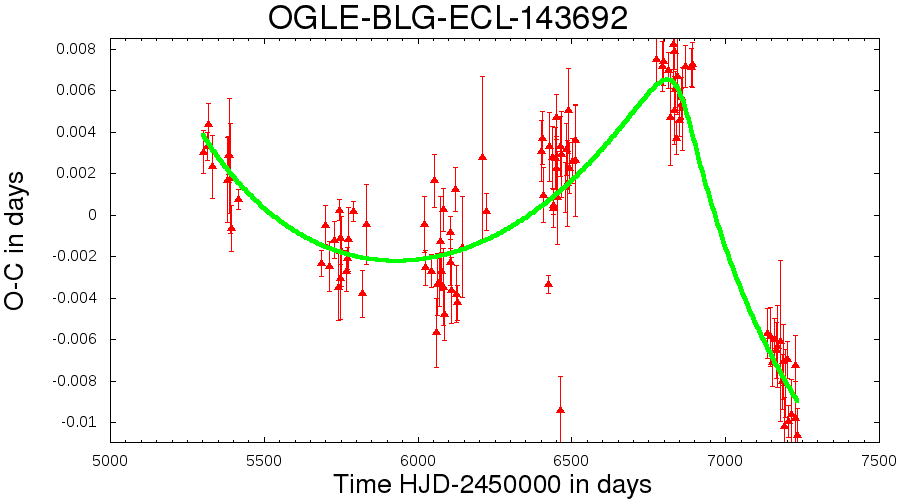}
\includegraphics[width=0.64\columnwidth]{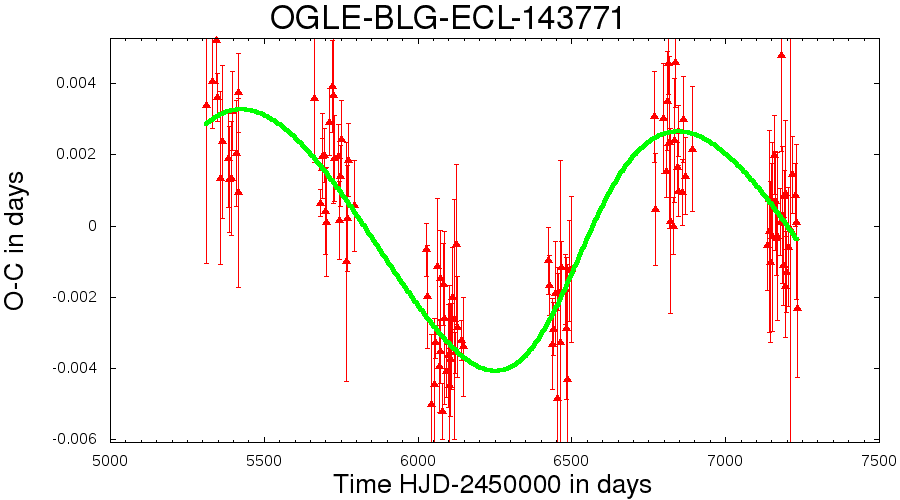}
\includegraphics[width=0.64\columnwidth]{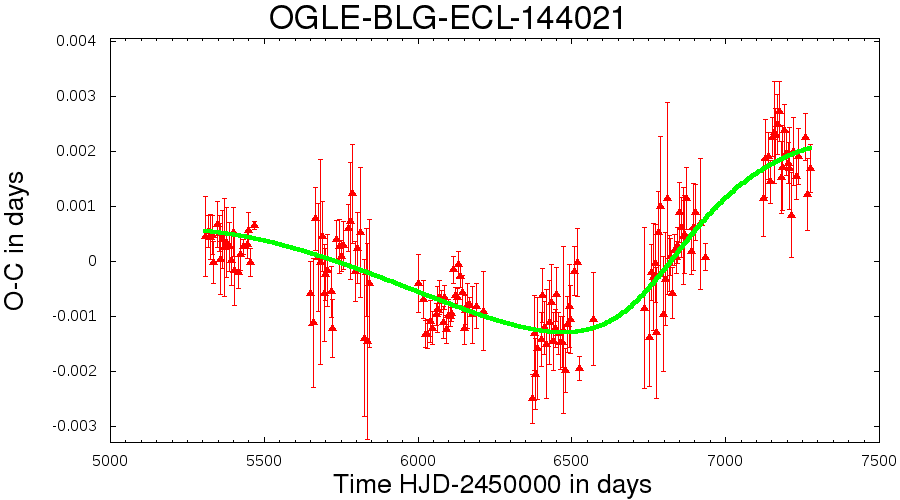}

\includegraphics[width=0.64\columnwidth]{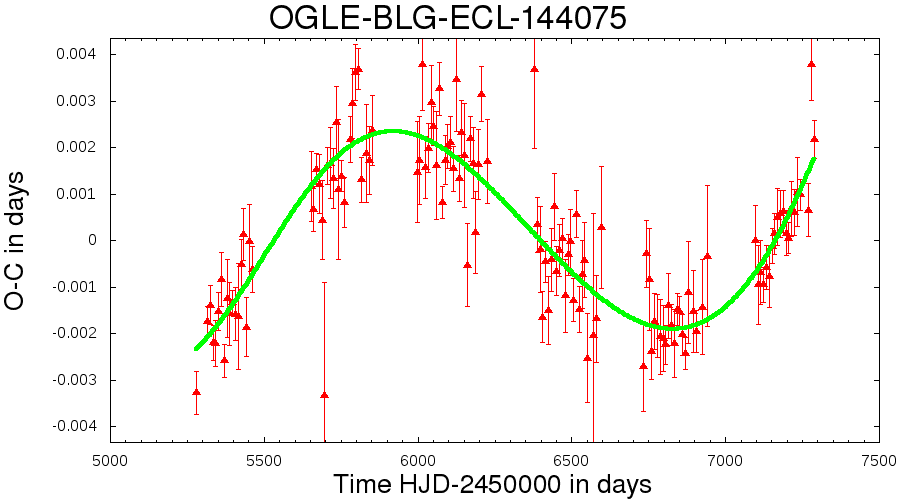}
\includegraphics[width=0.64\columnwidth]{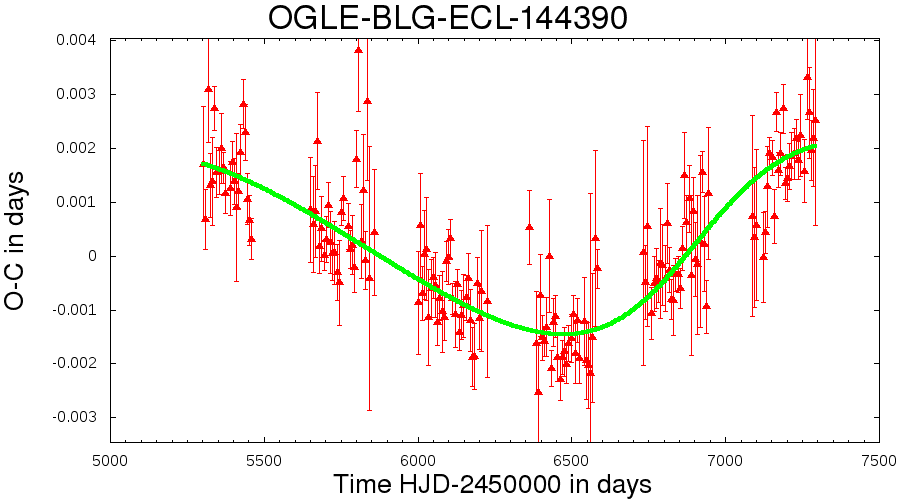}
\includegraphics[width=0.64\columnwidth]{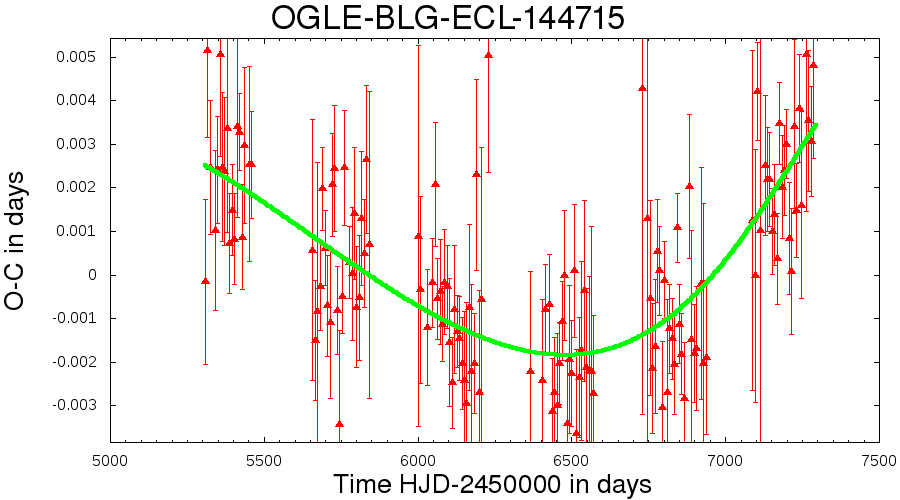}

\includegraphics[width=0.64\columnwidth]{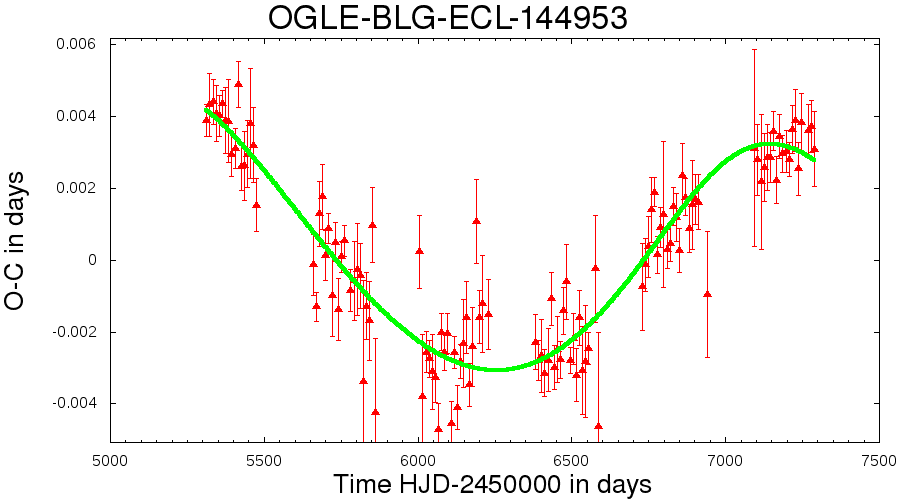}
\includegraphics[width=0.64\columnwidth]{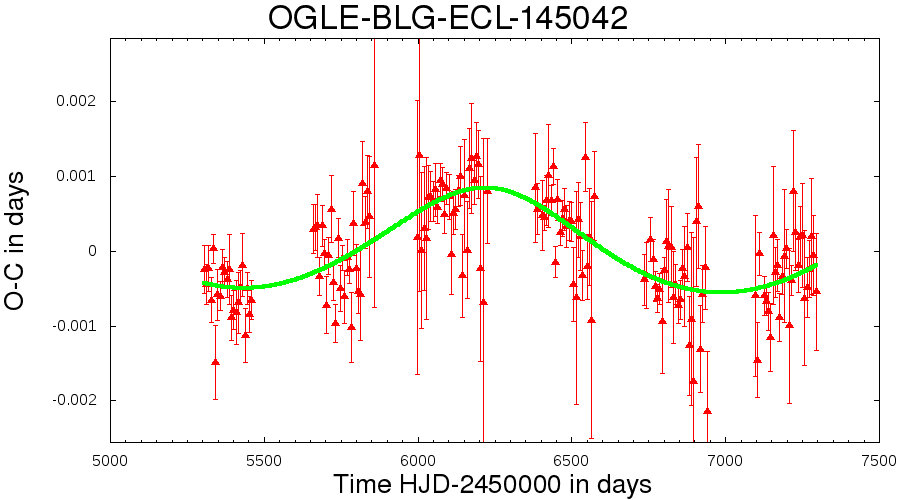}
\includegraphics[width=0.64\columnwidth]{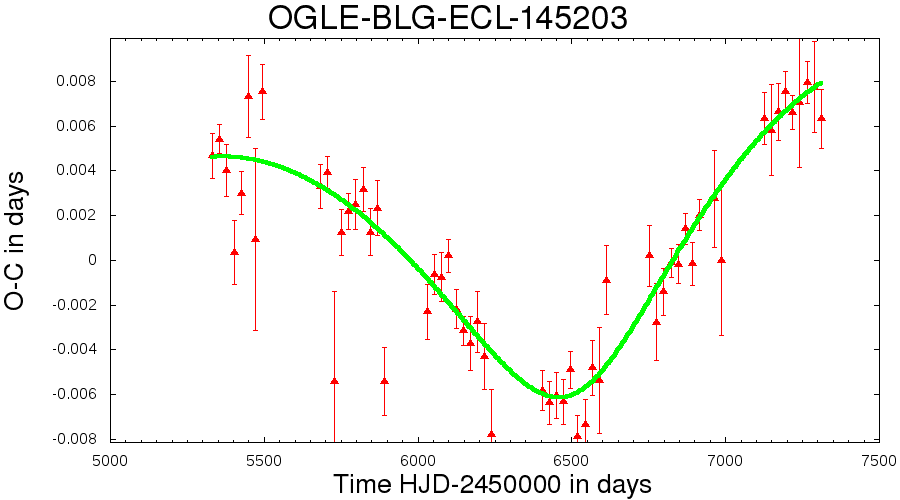}

\includegraphics[width=0.64\columnwidth]{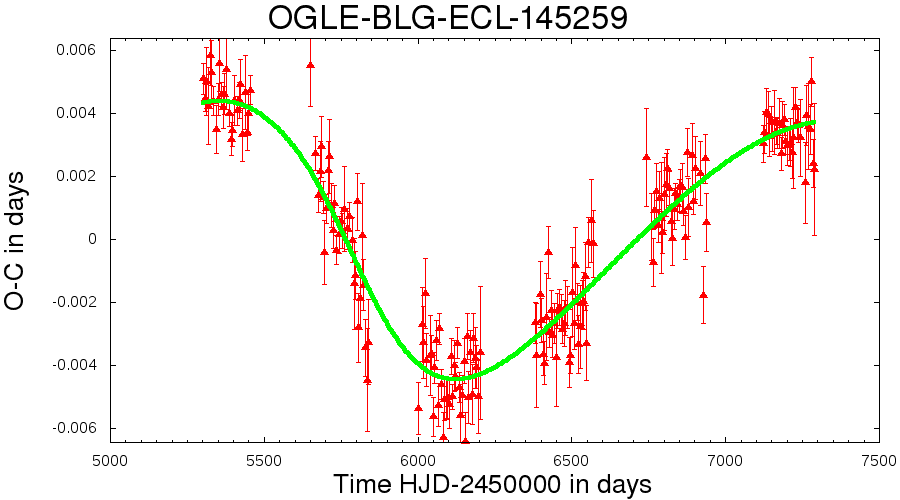}
\includegraphics[width=0.64\columnwidth]{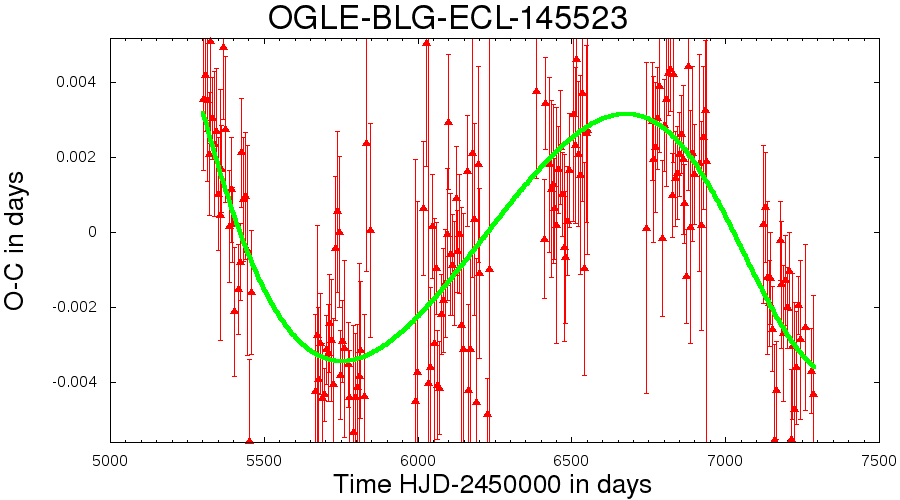}
\includegraphics[width=0.64\columnwidth]{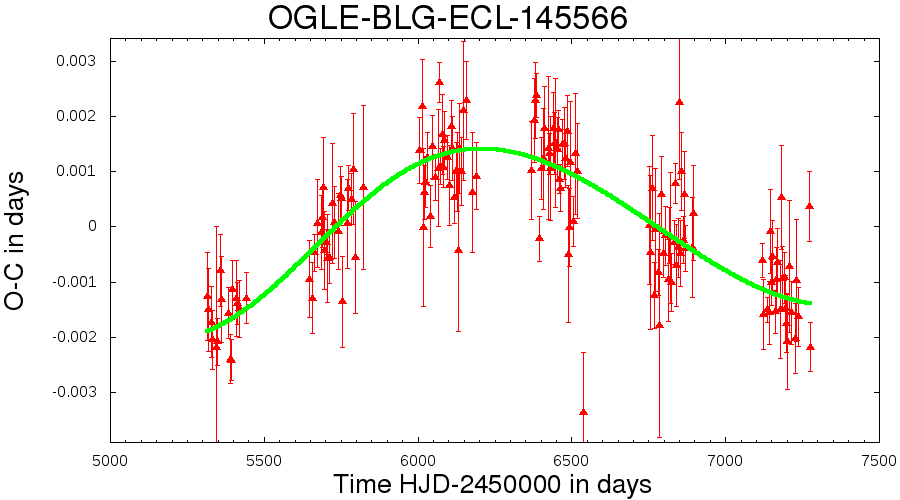}

\includegraphics[width=0.64\columnwidth]{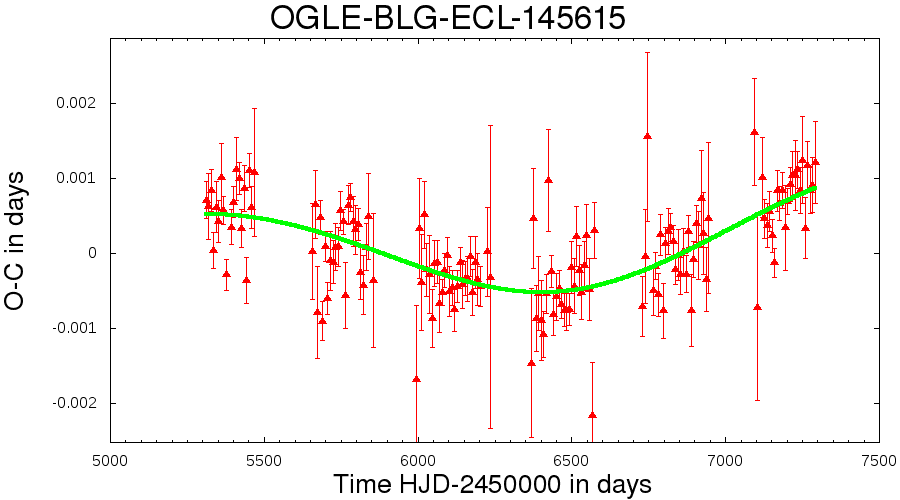}
\includegraphics[width=0.64\columnwidth]{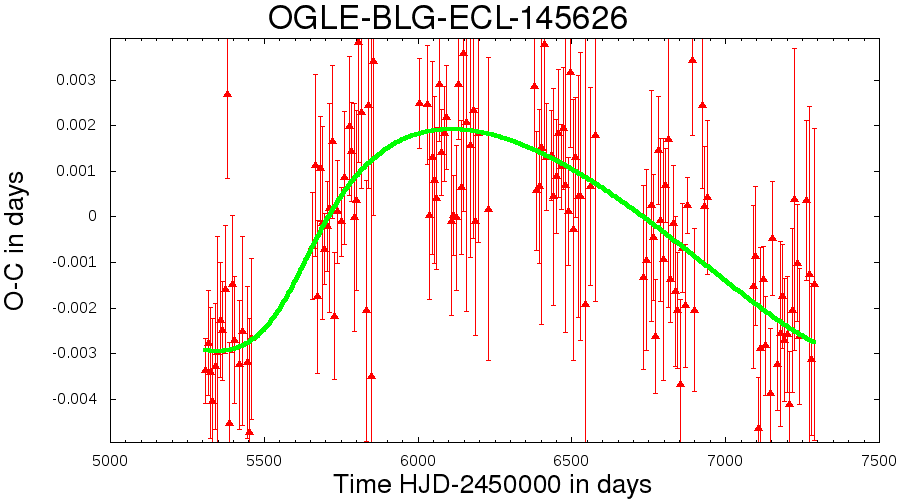}
\includegraphics[width=0.64\columnwidth]{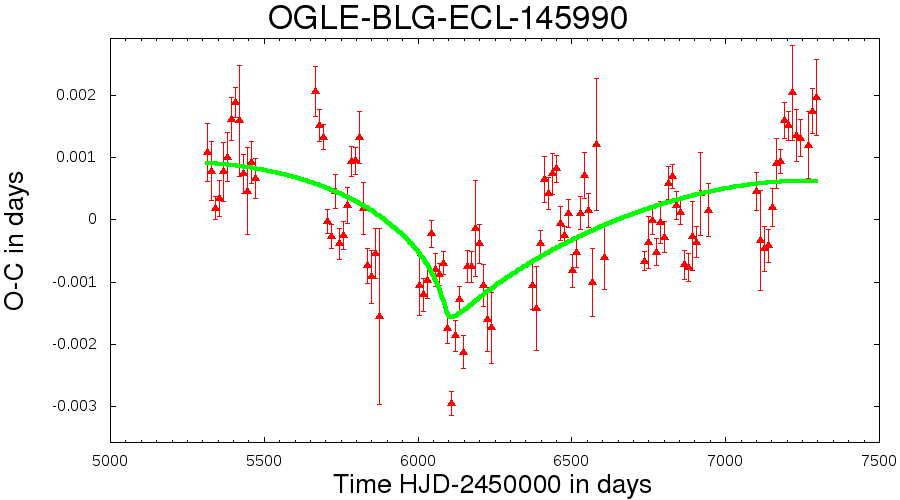}

\end{figure*}
\clearpage

\begin{figure*}
\includegraphics[width=0.64\columnwidth]{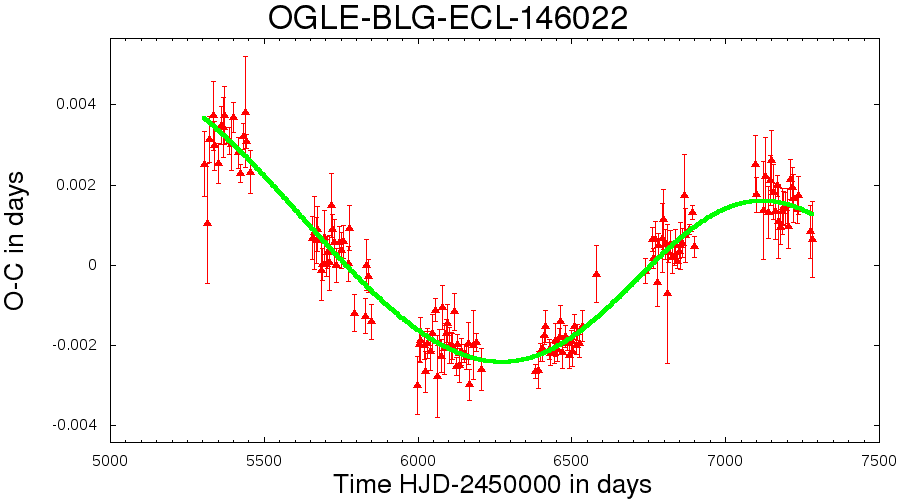}
\includegraphics[width=0.64\columnwidth]{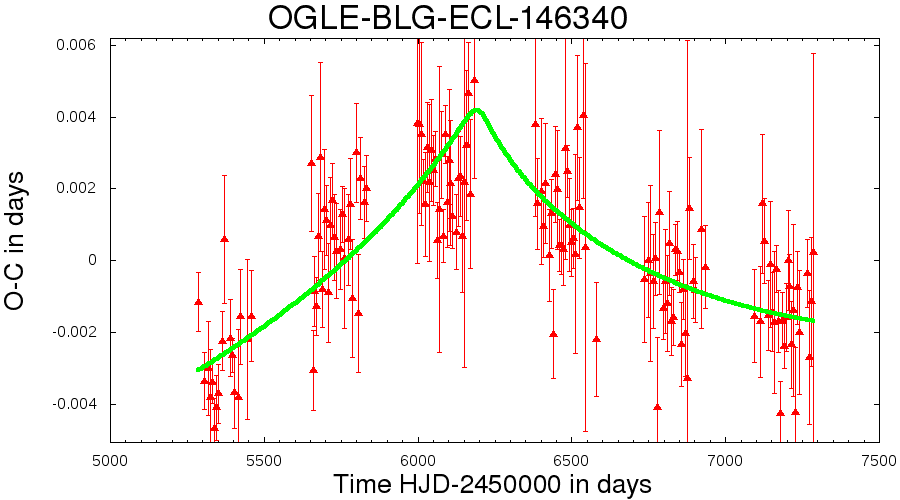}
\includegraphics[width=0.64\columnwidth]{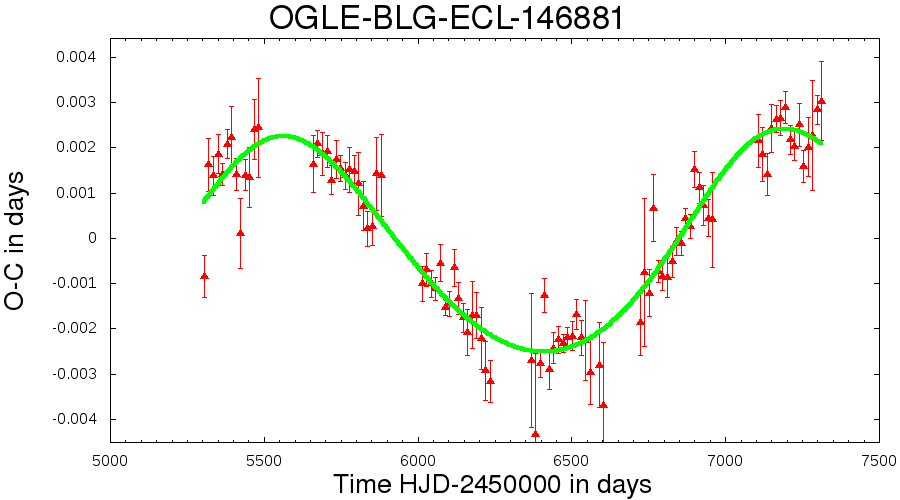}

\includegraphics[width=0.64\columnwidth]{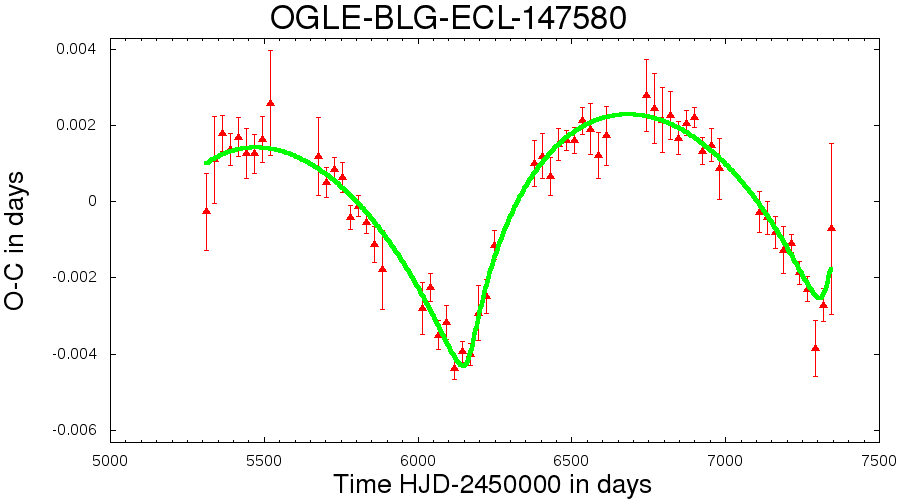}
\includegraphics[width=0.64\columnwidth]{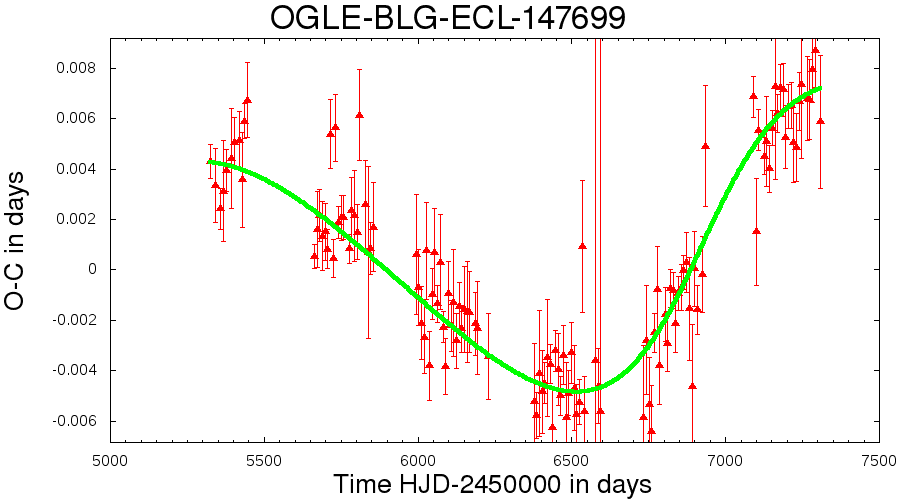}
\includegraphics[width=0.64\columnwidth]{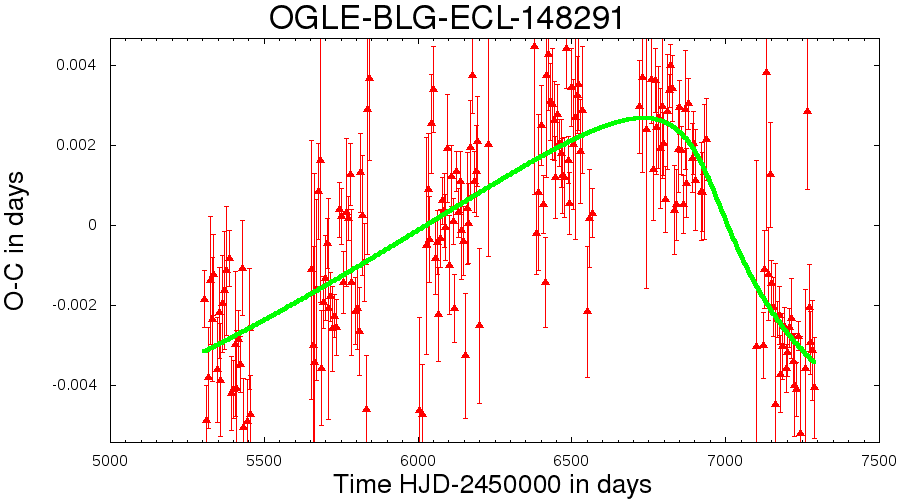}

\includegraphics[width=0.64\columnwidth]{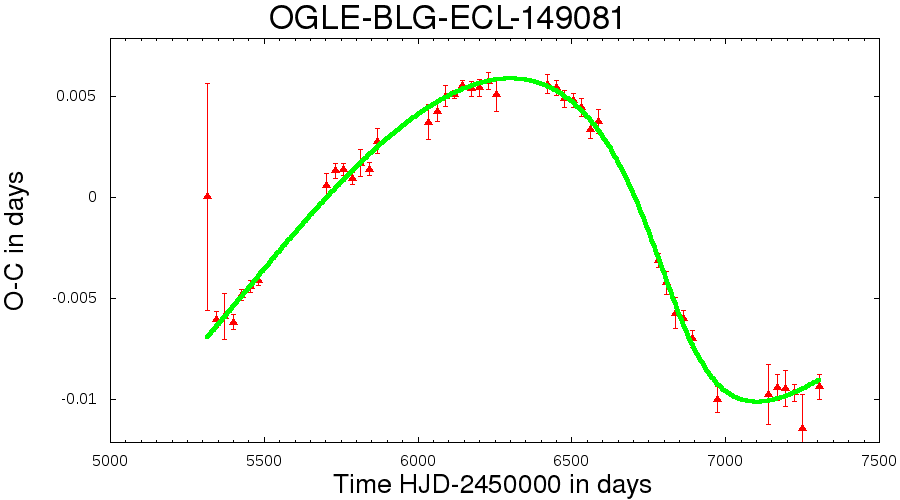}
\includegraphics[width=0.64\columnwidth]{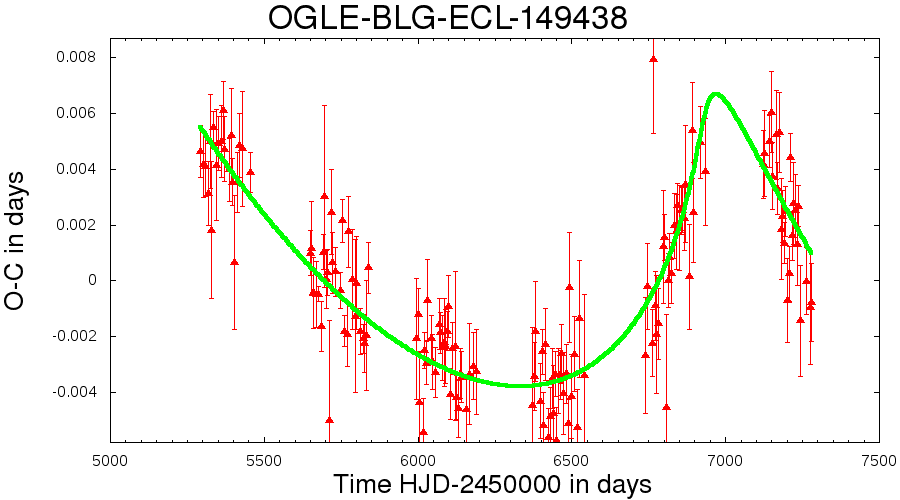}
\includegraphics[width=0.64\columnwidth]{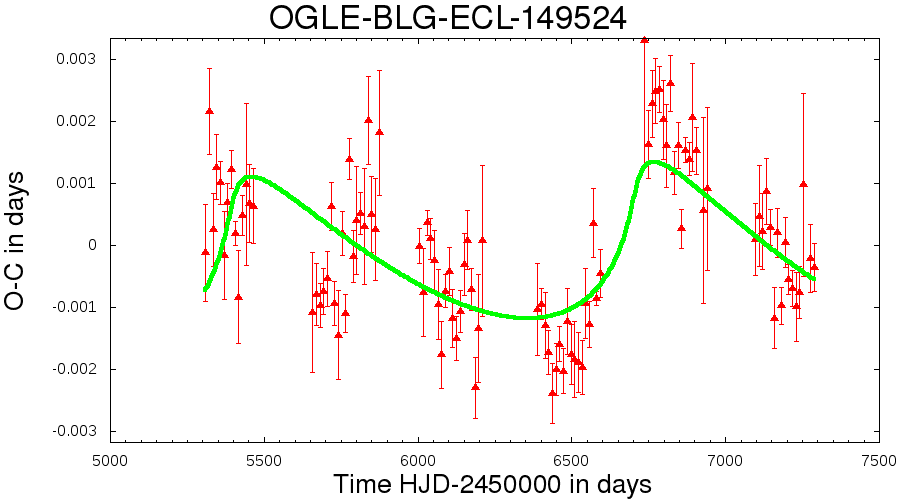}

\includegraphics[width=0.64\columnwidth]{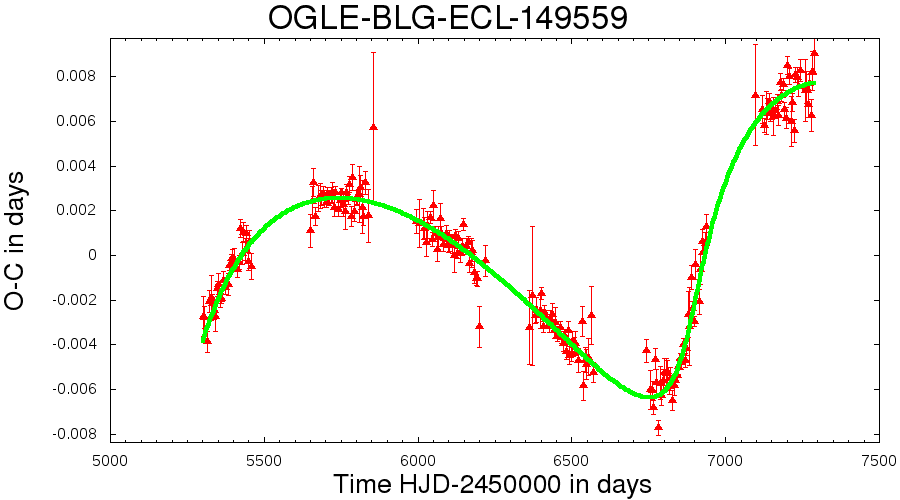}
\includegraphics[width=0.64\columnwidth]{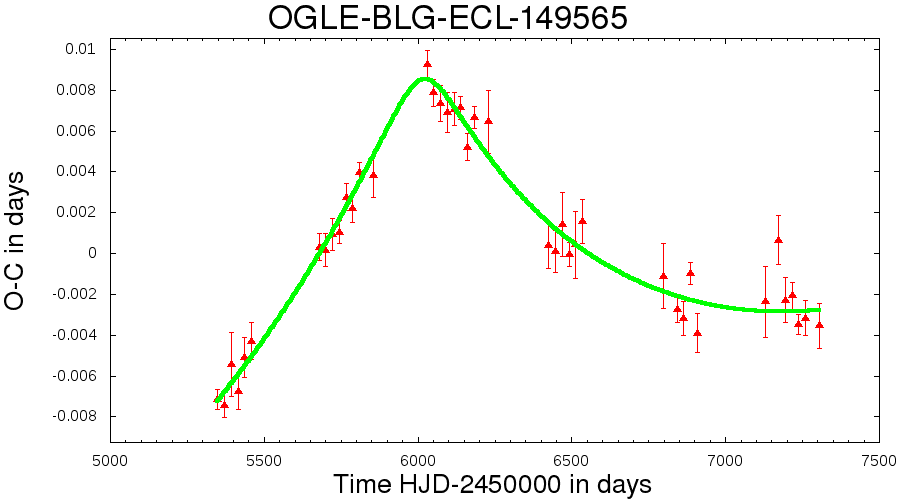}
\includegraphics[width=0.64\columnwidth]{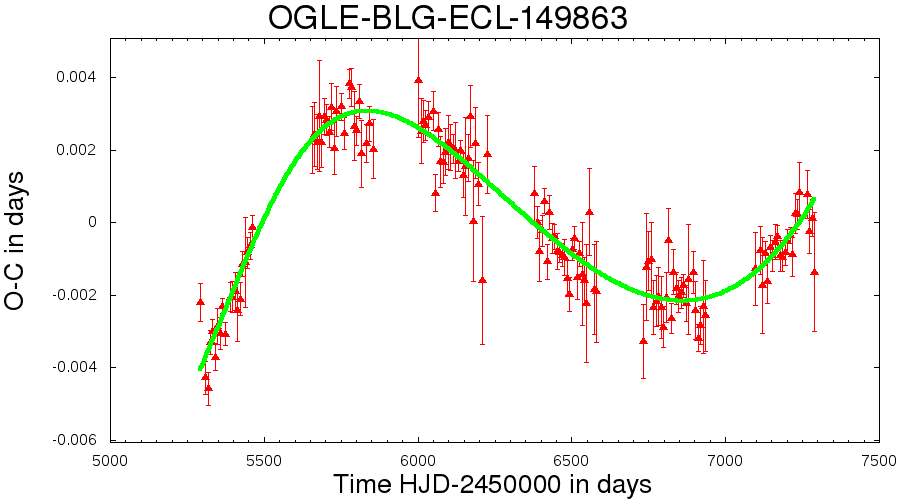}

\includegraphics[width=0.64\columnwidth]{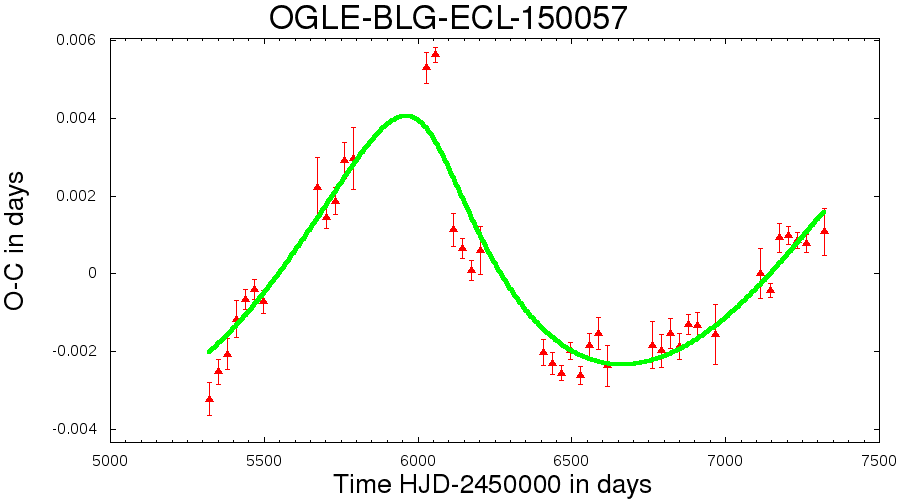}
\includegraphics[width=0.64\columnwidth]{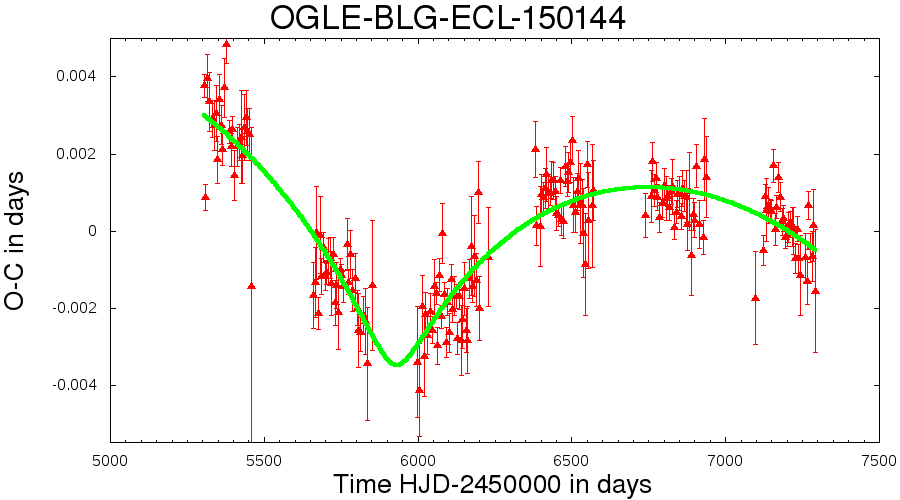}
\includegraphics[width=0.64\columnwidth]{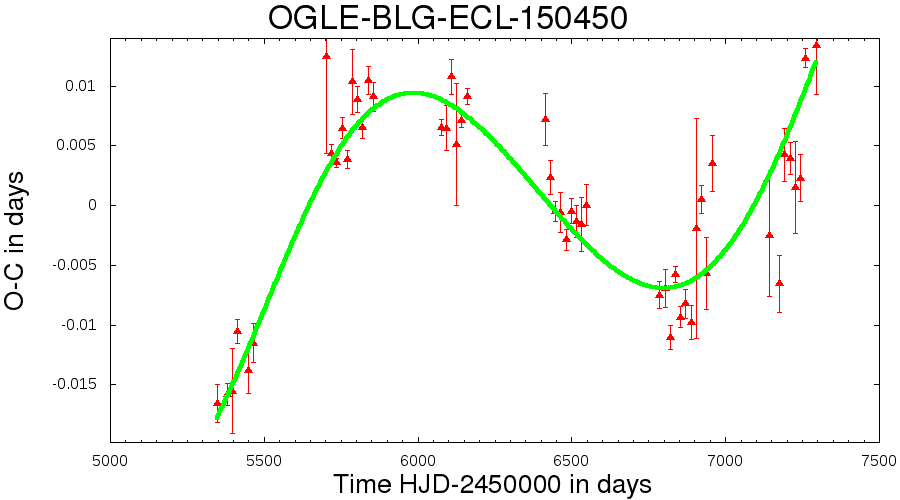}

\includegraphics[width=0.64\columnwidth]{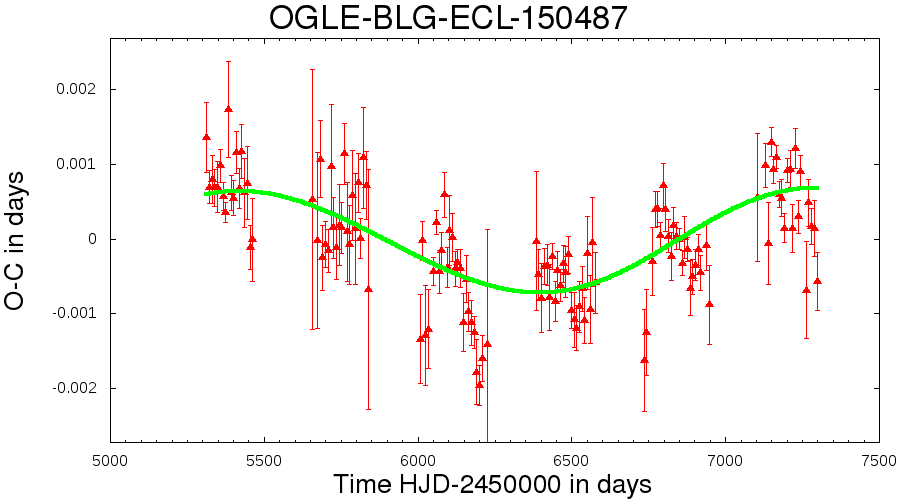}
\includegraphics[width=0.64\columnwidth]{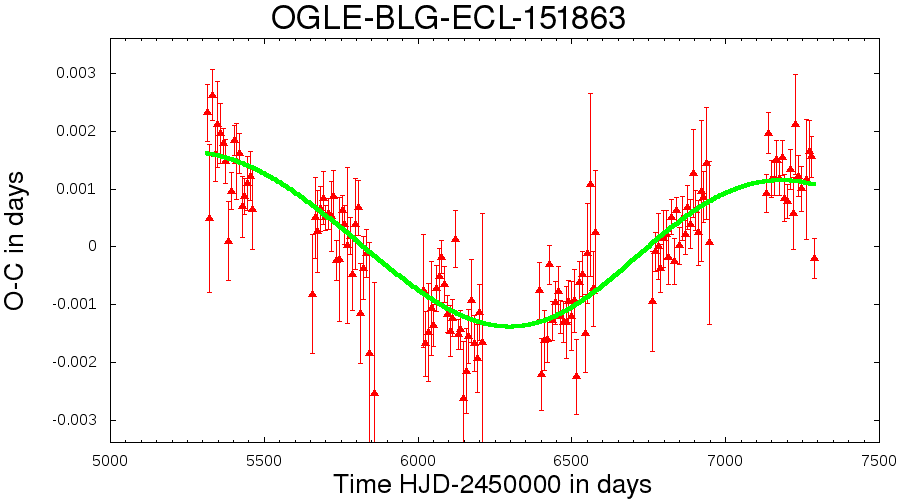}
\includegraphics[width=0.64\columnwidth]{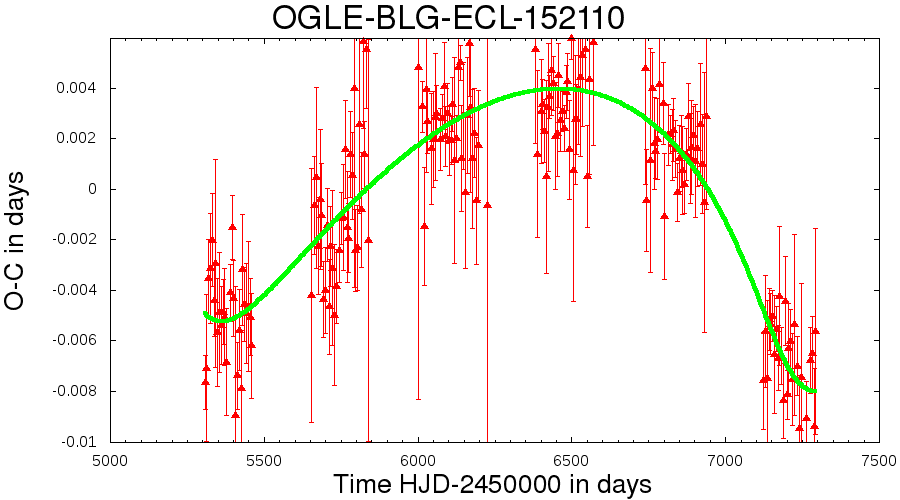}

\includegraphics[width=0.64\columnwidth]{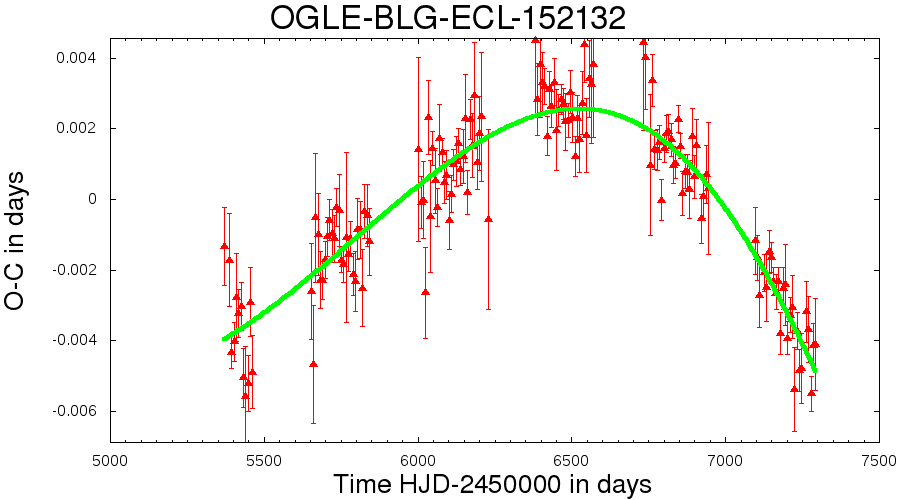}
\includegraphics[width=0.64\columnwidth]{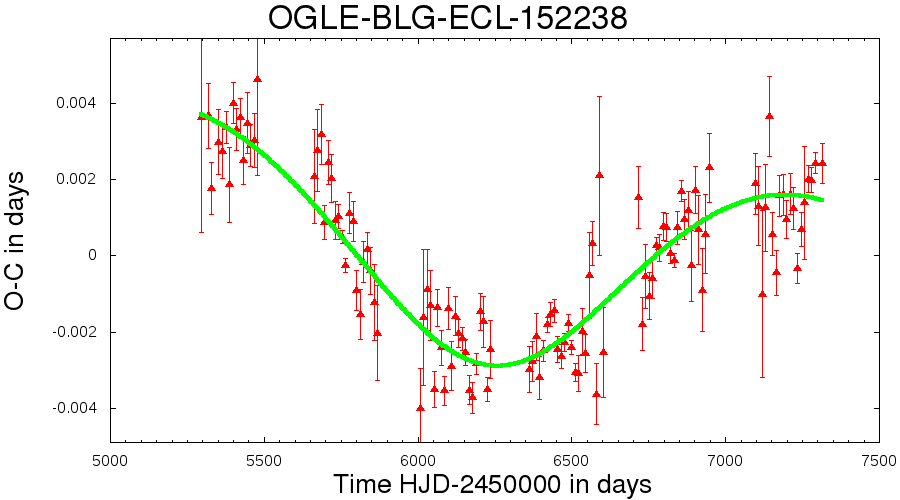}
\includegraphics[width=0.64\columnwidth]{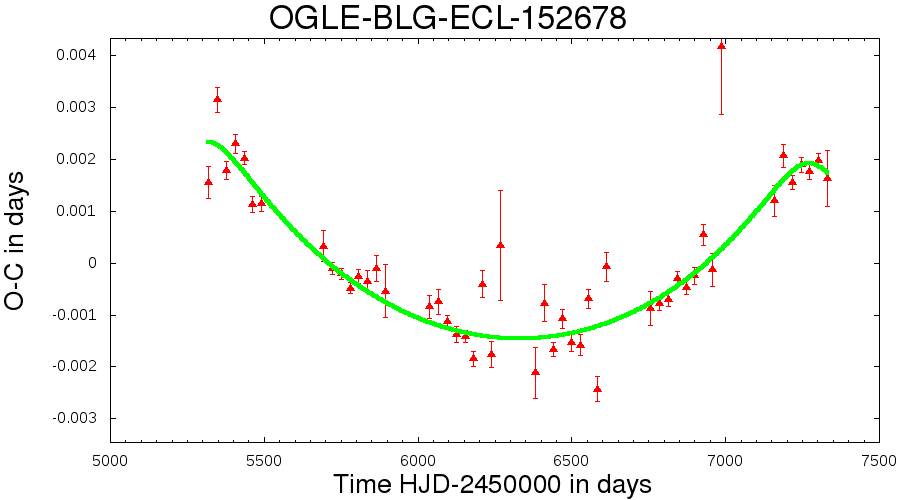}

\includegraphics[width=0.64\columnwidth]{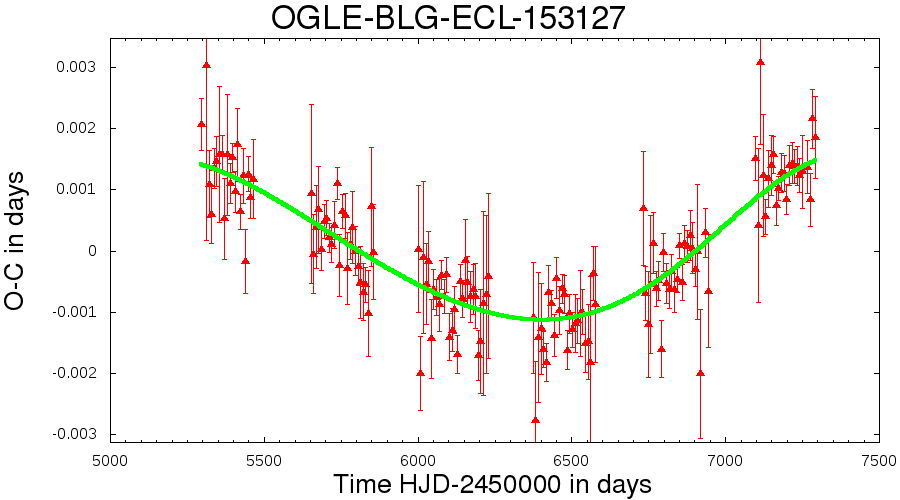}
\includegraphics[width=0.64\columnwidth]{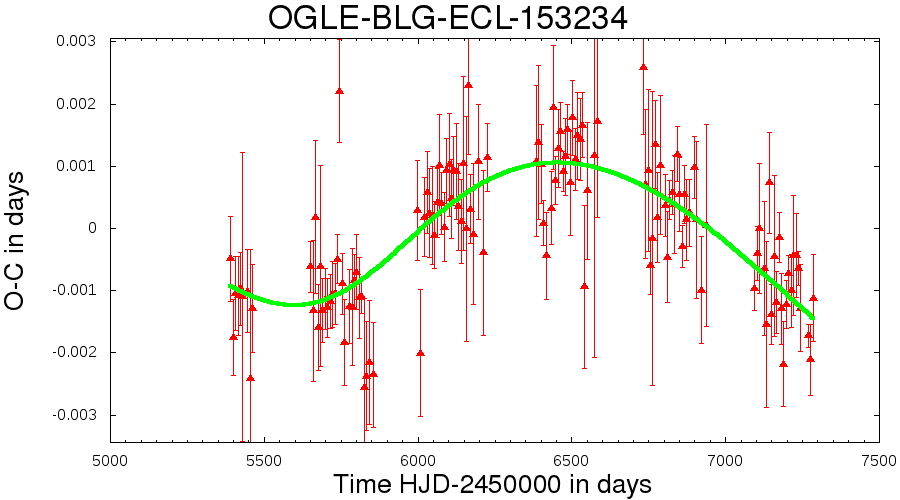}
\includegraphics[width=0.64\columnwidth]{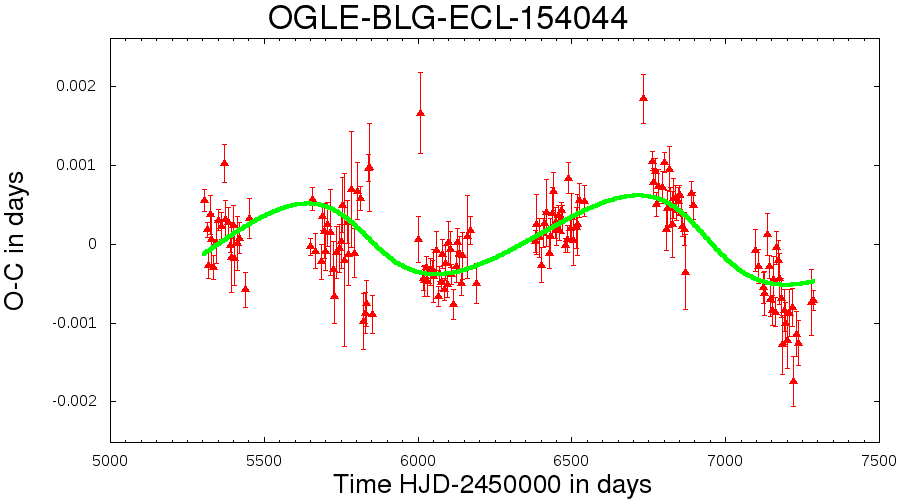}

\end{figure*}
\clearpage

\begin{figure*}
\includegraphics[width=0.64\columnwidth]{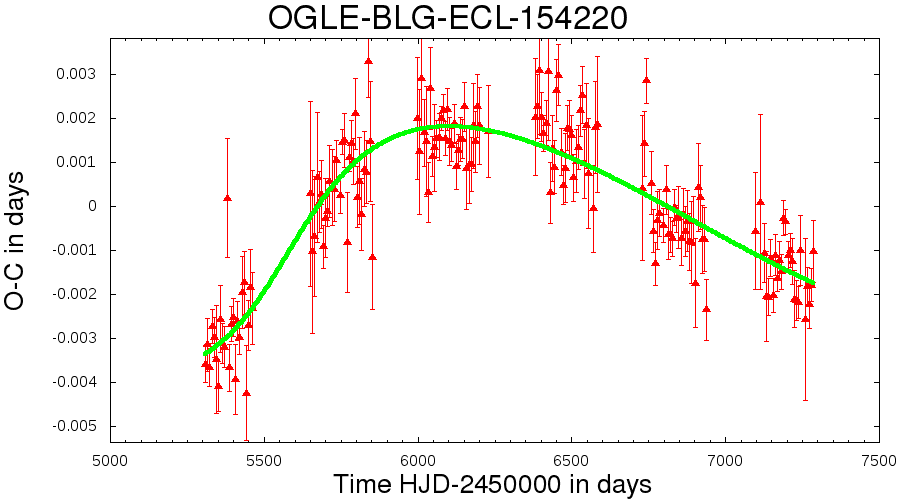}
\includegraphics[width=0.64\columnwidth]{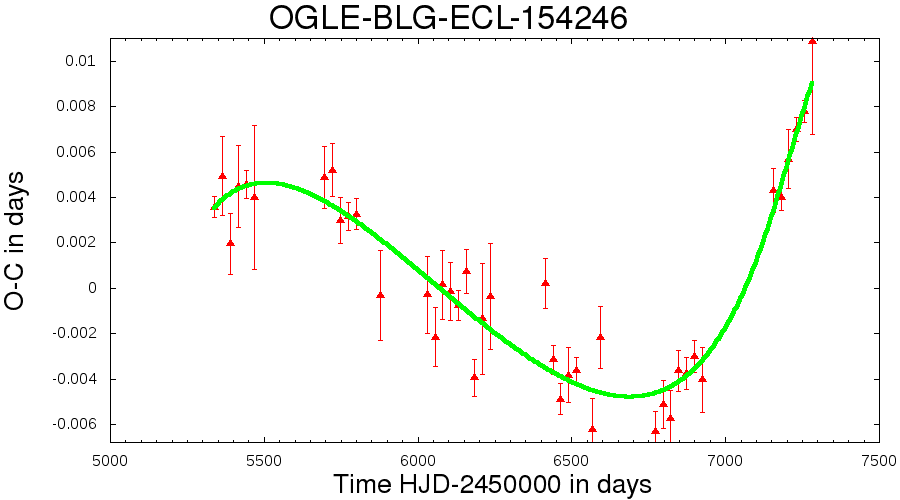}
\includegraphics[width=0.64\columnwidth]{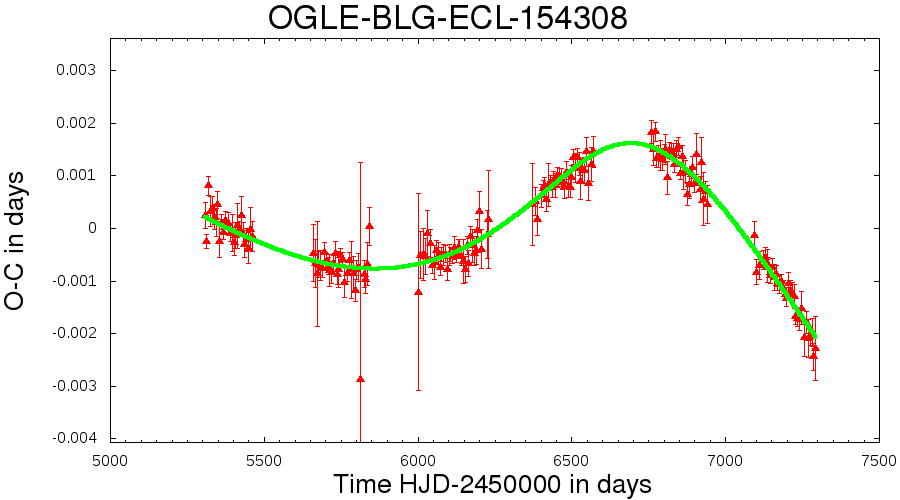}

\includegraphics[width=0.64\columnwidth]{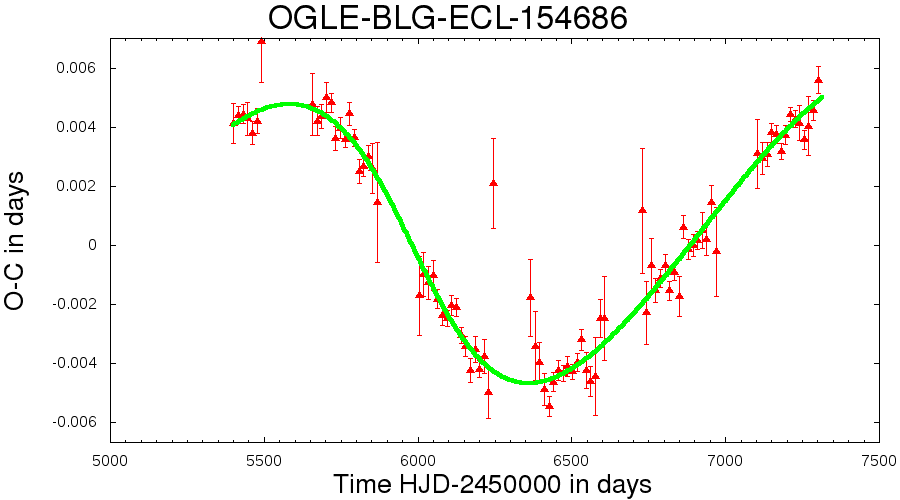}
\includegraphics[width=0.64\columnwidth]{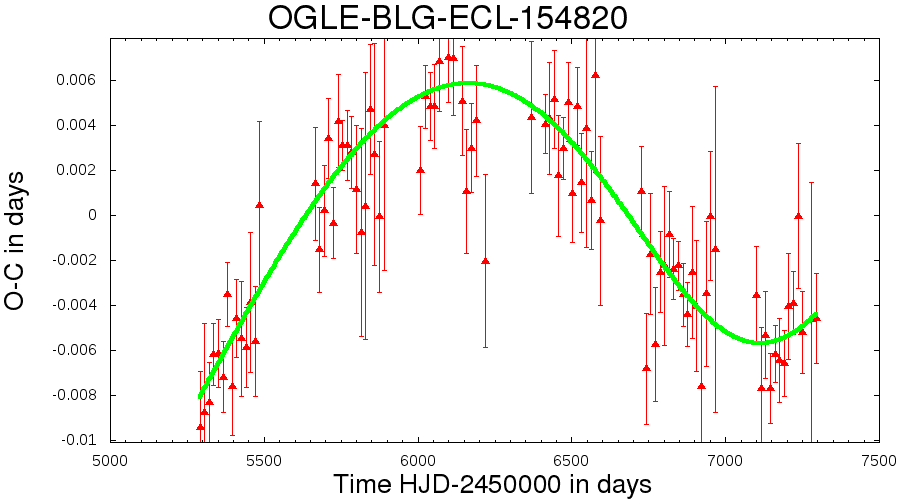}
\includegraphics[width=0.64\columnwidth]{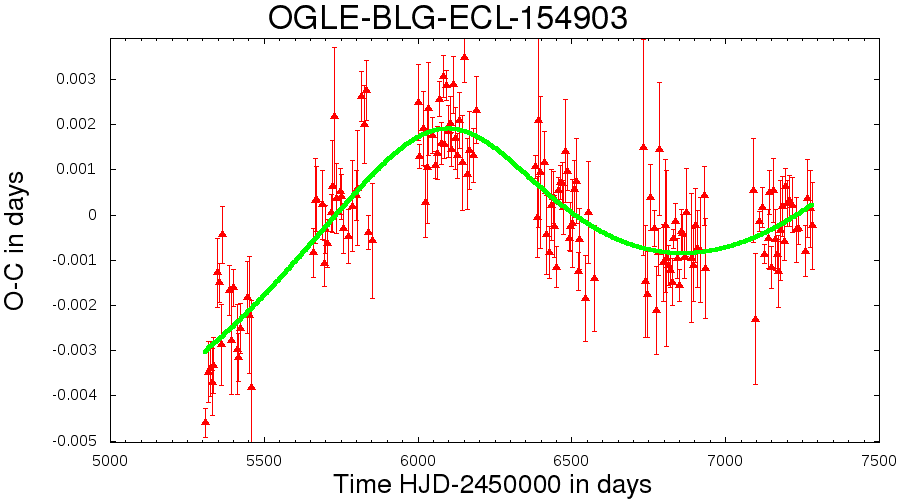}

\includegraphics[width=0.64\columnwidth]{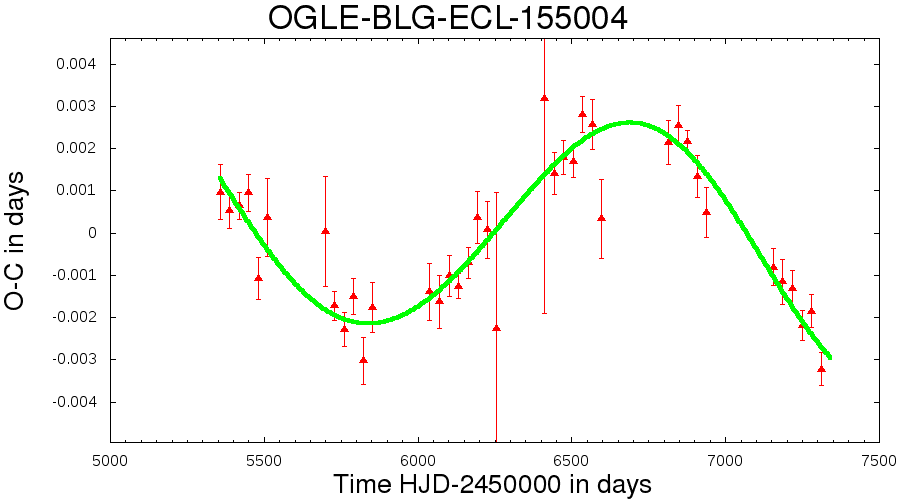}
\includegraphics[width=0.64\columnwidth]{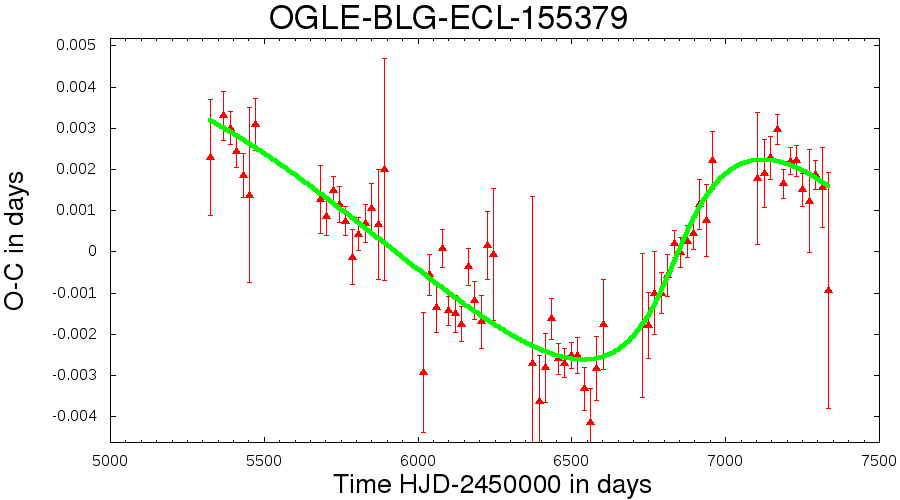}
\includegraphics[width=0.64\columnwidth]{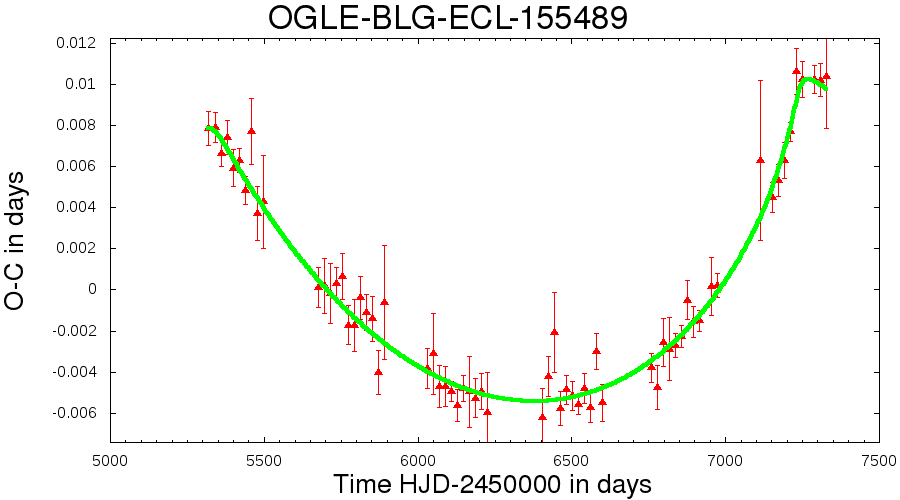}

\includegraphics[width=0.64\columnwidth]{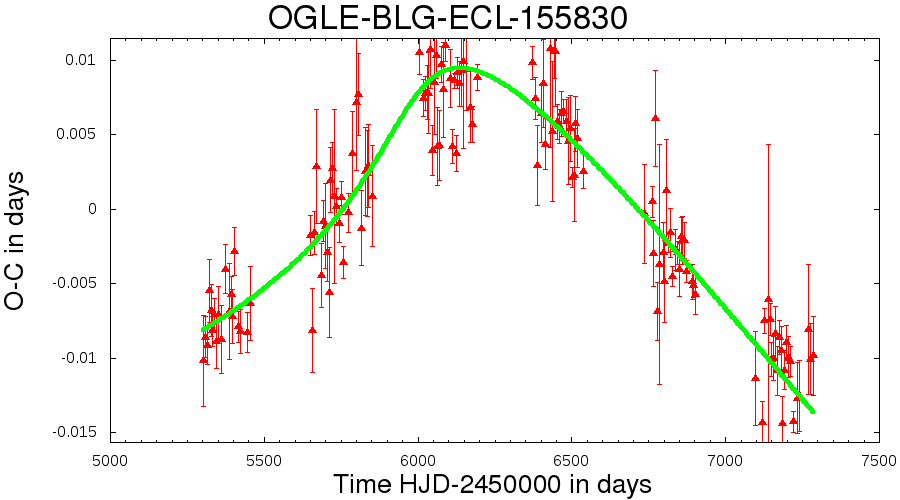}
\includegraphics[width=0.64\columnwidth]{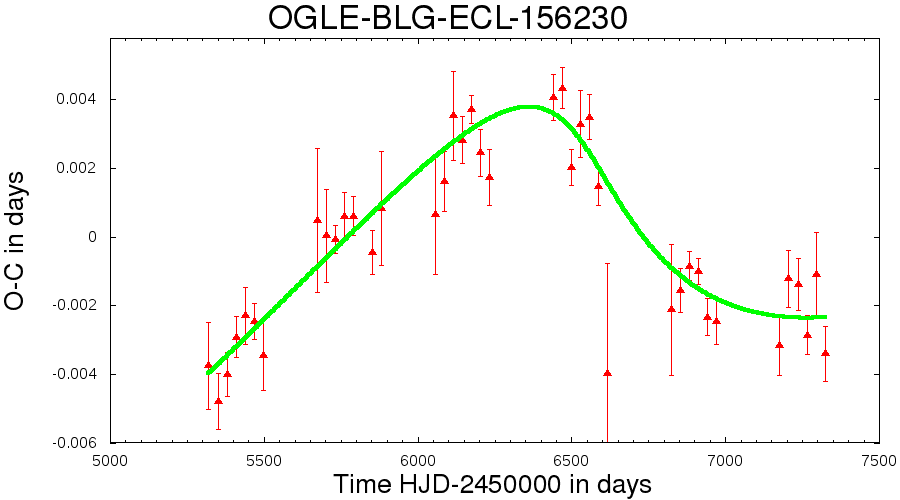}
\includegraphics[width=0.64\columnwidth]{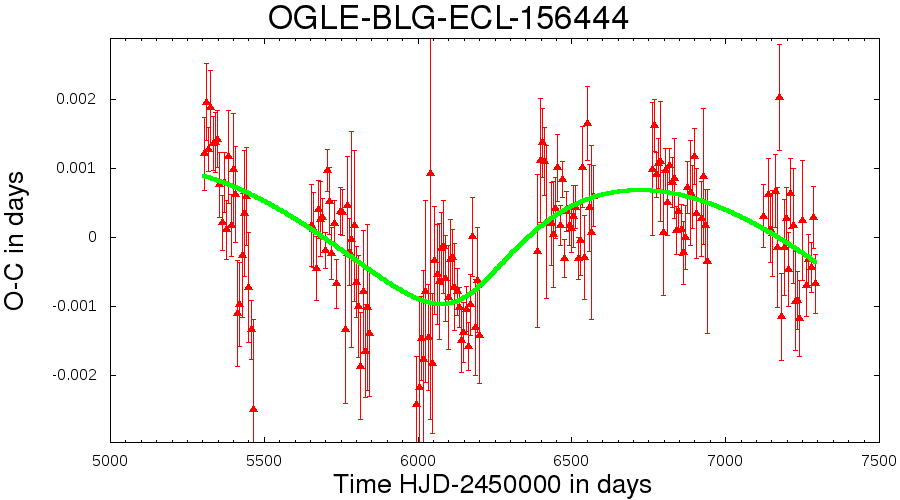}

\includegraphics[width=0.64\columnwidth]{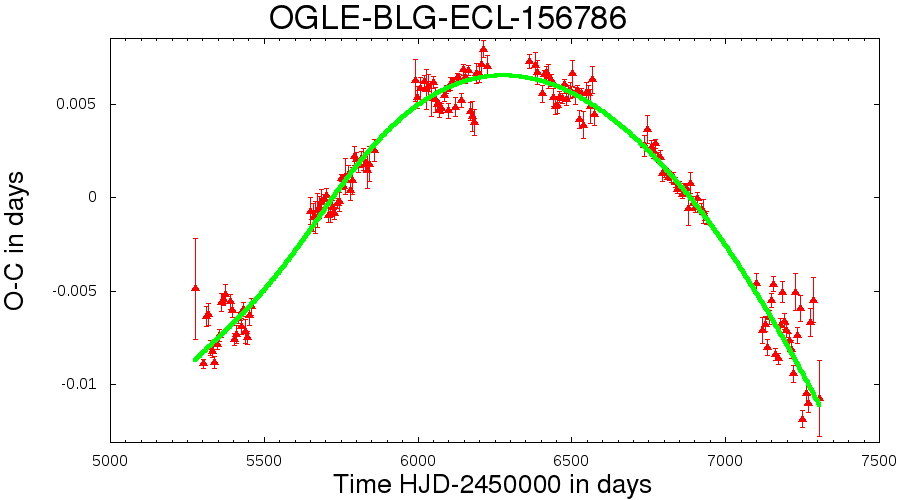}
\includegraphics[width=0.64\columnwidth]{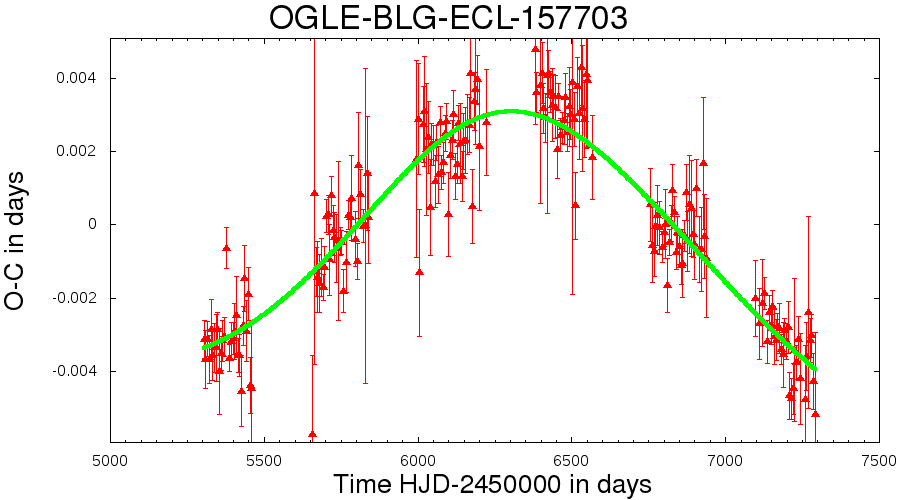}
\includegraphics[width=0.64\columnwidth]{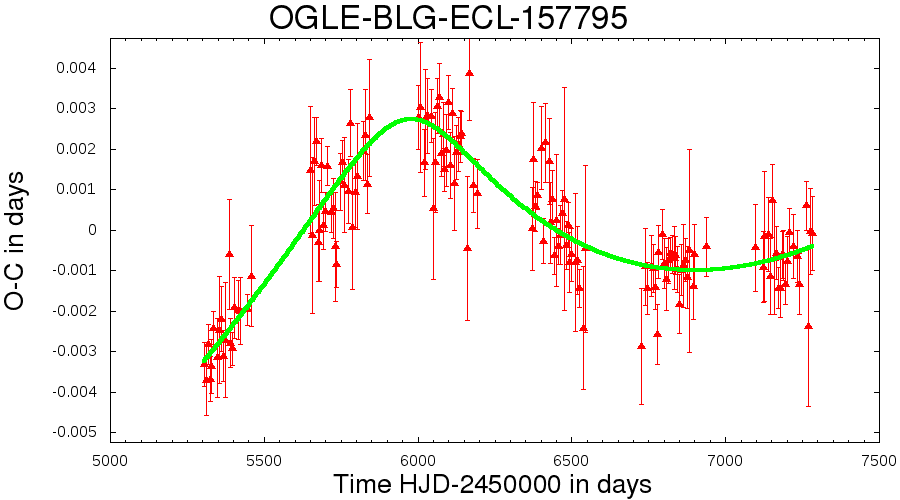}

\includegraphics[width=0.64\columnwidth]{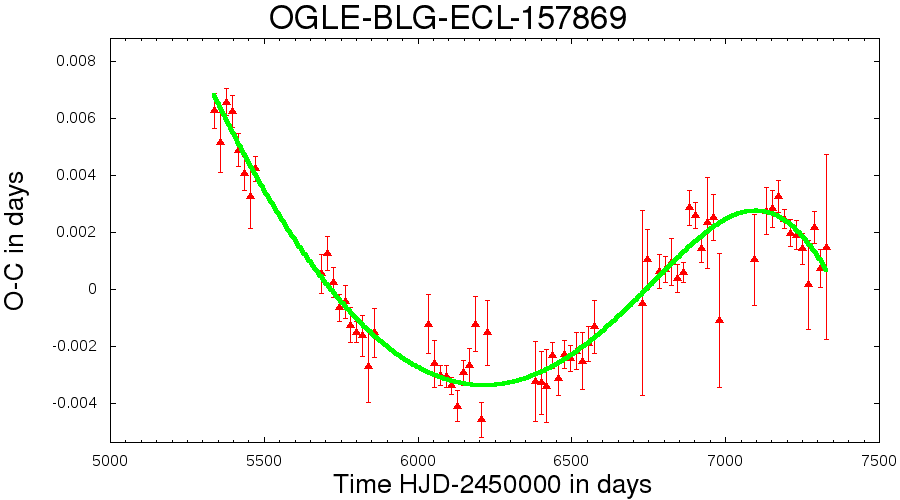}
\includegraphics[width=0.64\columnwidth]{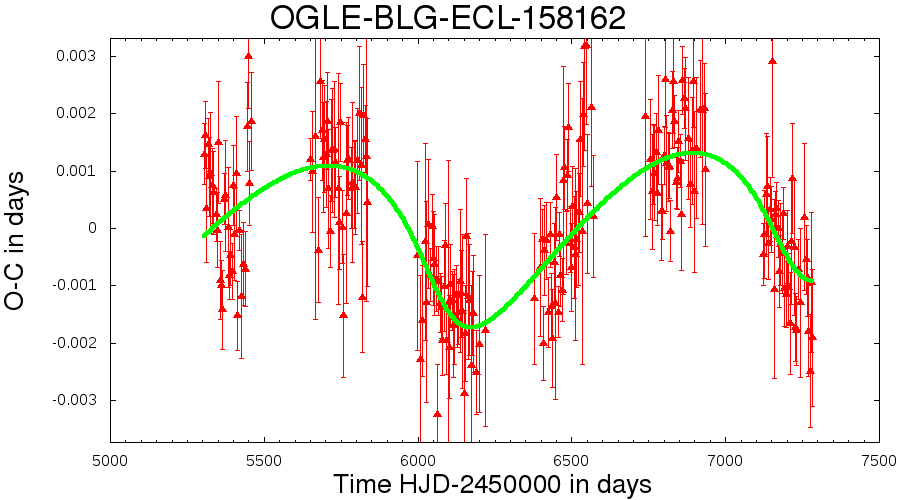}
\includegraphics[width=0.64\columnwidth]{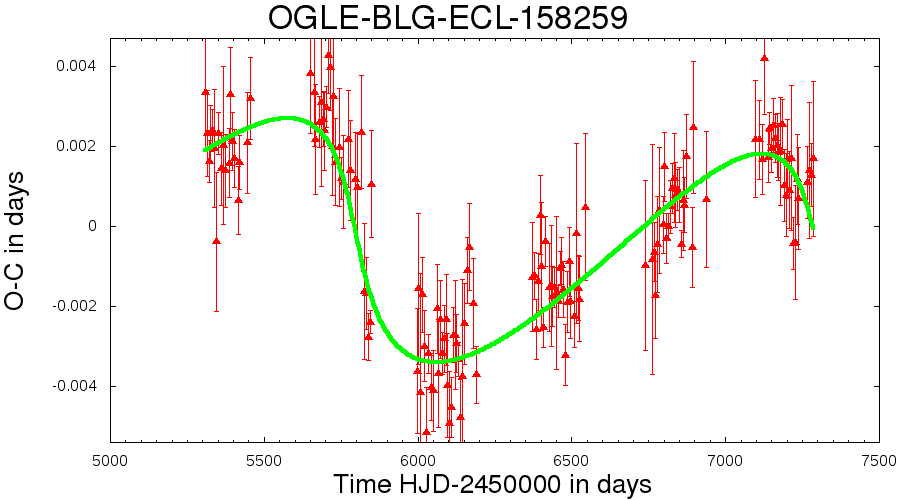}

\includegraphics[width=0.64\columnwidth]{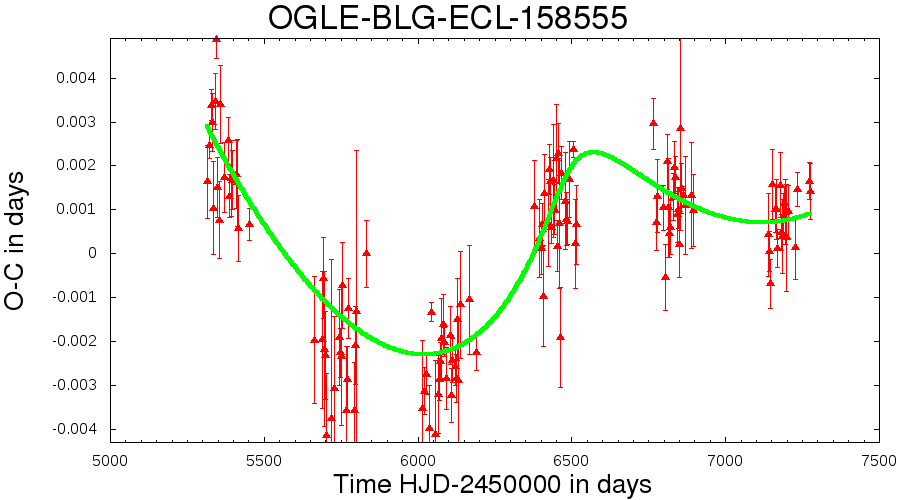}
\includegraphics[width=0.64\columnwidth]{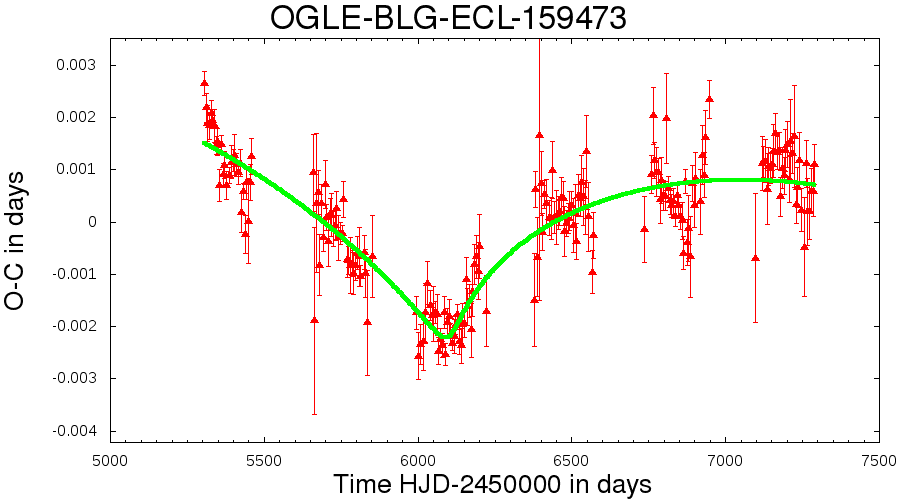}
\includegraphics[width=0.64\columnwidth]{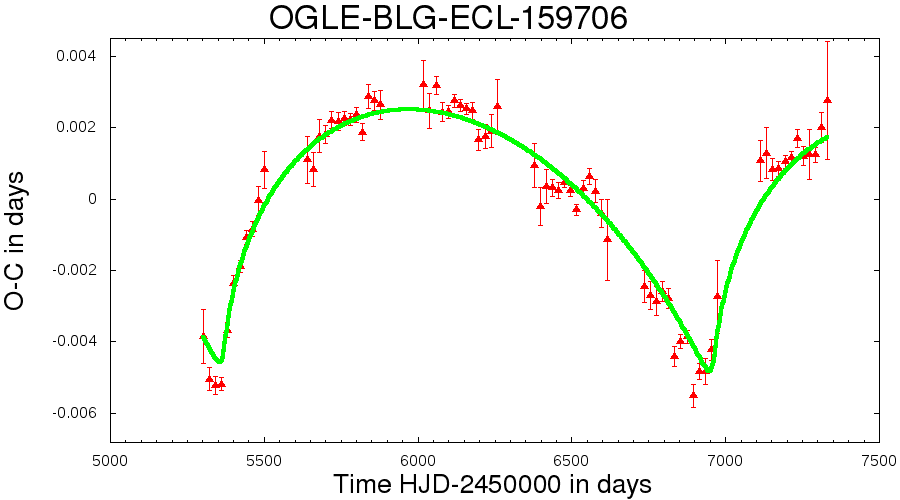}

\includegraphics[width=0.64\columnwidth]{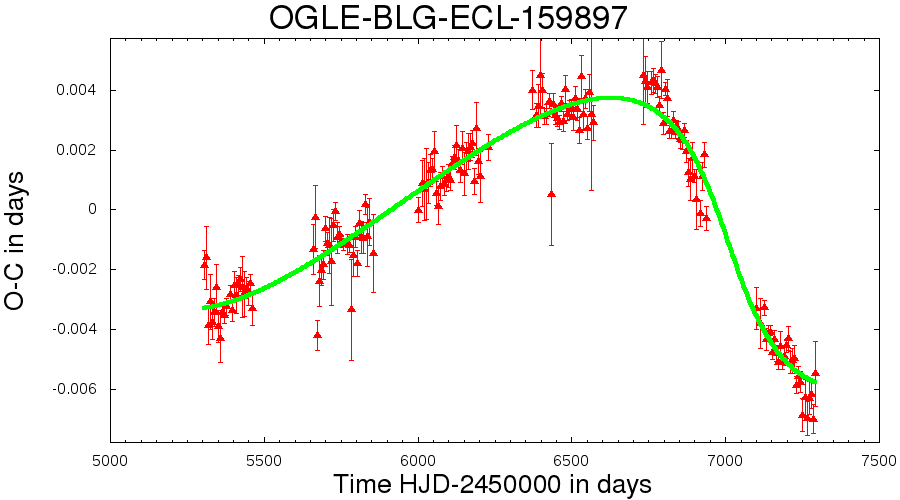}
\includegraphics[width=0.64\columnwidth]{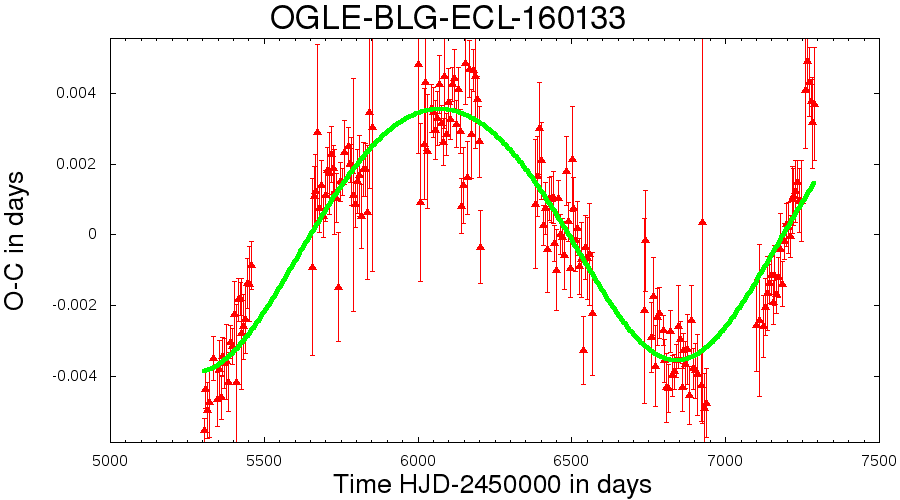}
\includegraphics[width=0.64\columnwidth]{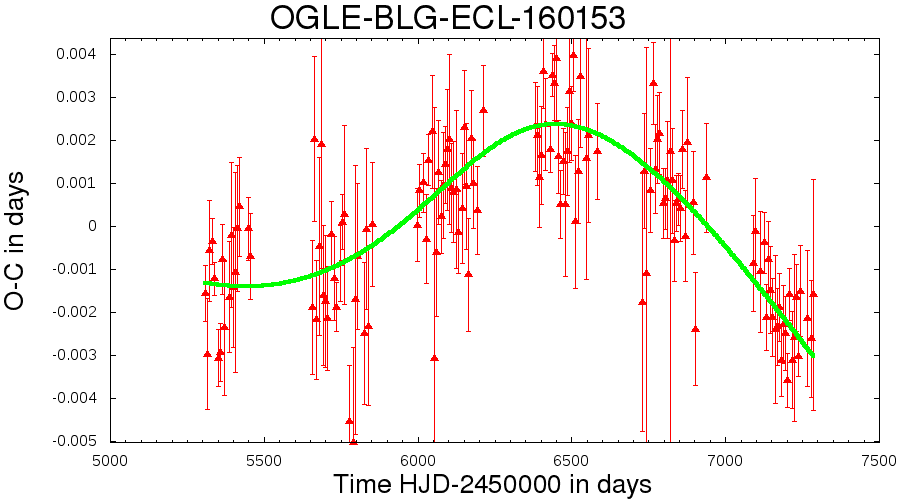}

\end{figure*}
\clearpage

\begin{figure*}
\includegraphics[width=0.64\columnwidth]{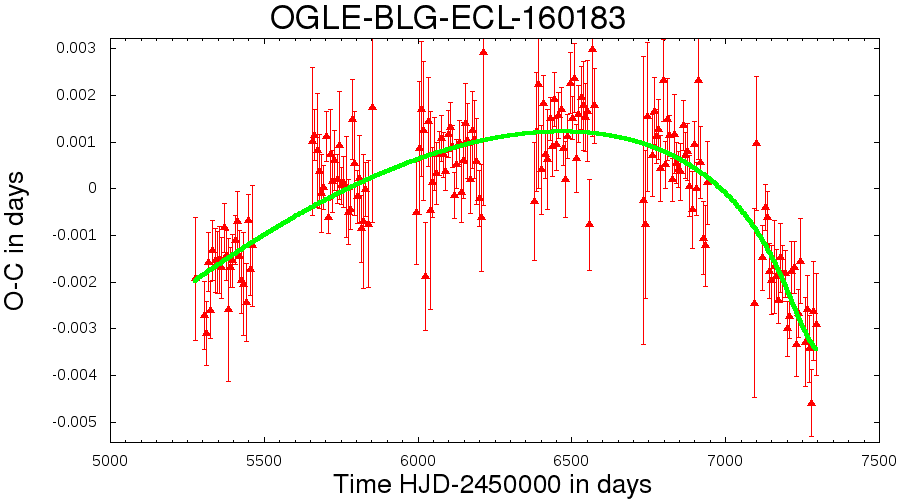}
\includegraphics[width=0.64\columnwidth]{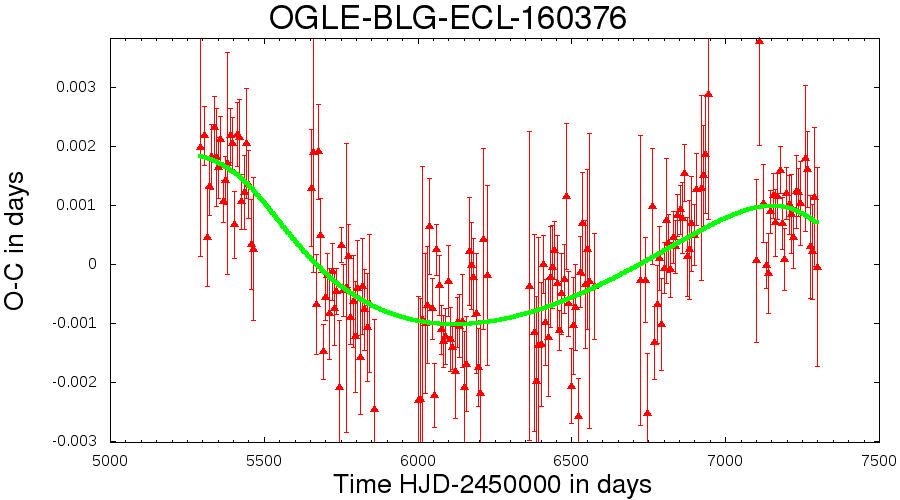}
\includegraphics[width=0.64\columnwidth]{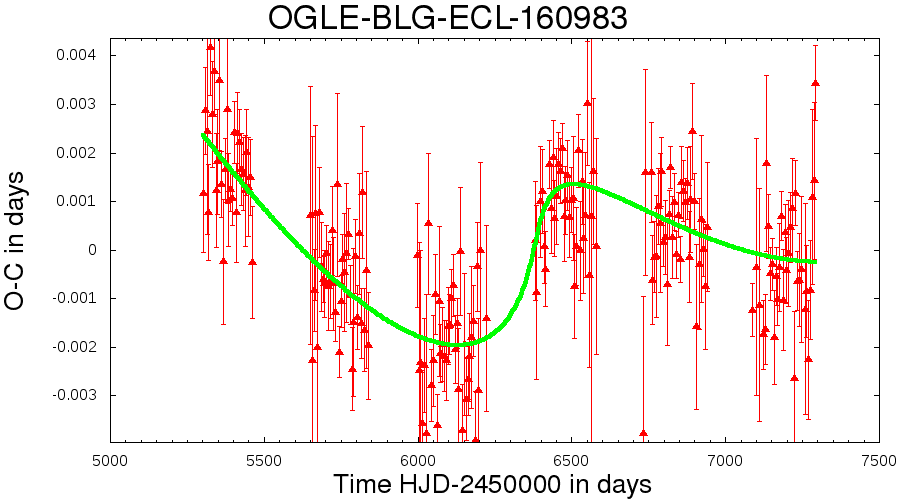}

\includegraphics[width=0.64\columnwidth]{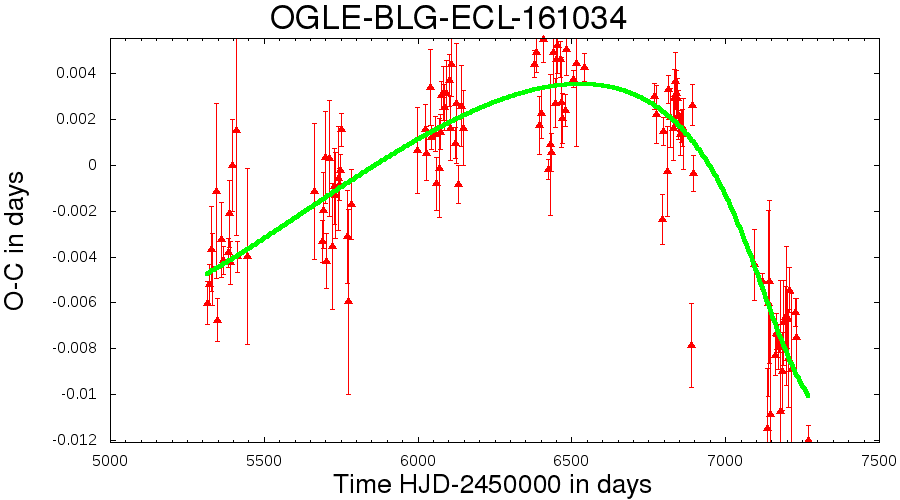}
\includegraphics[width=0.64\columnwidth]{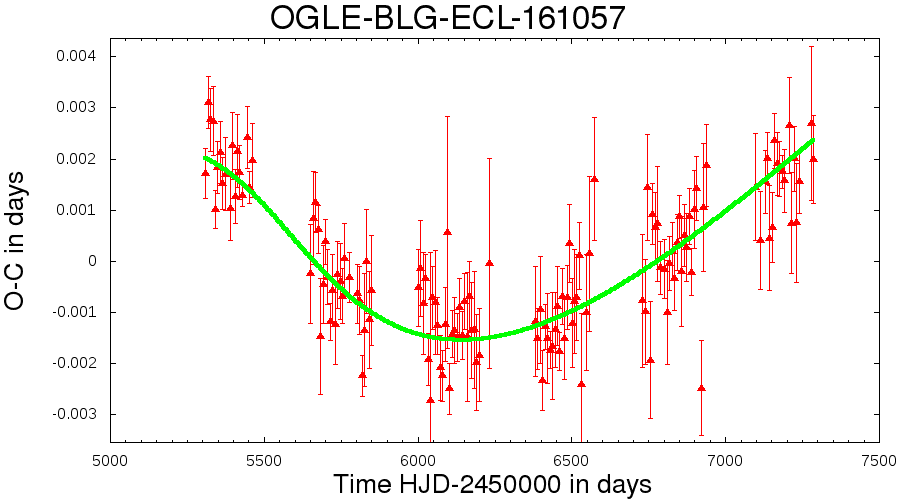}
\includegraphics[width=0.64\columnwidth]{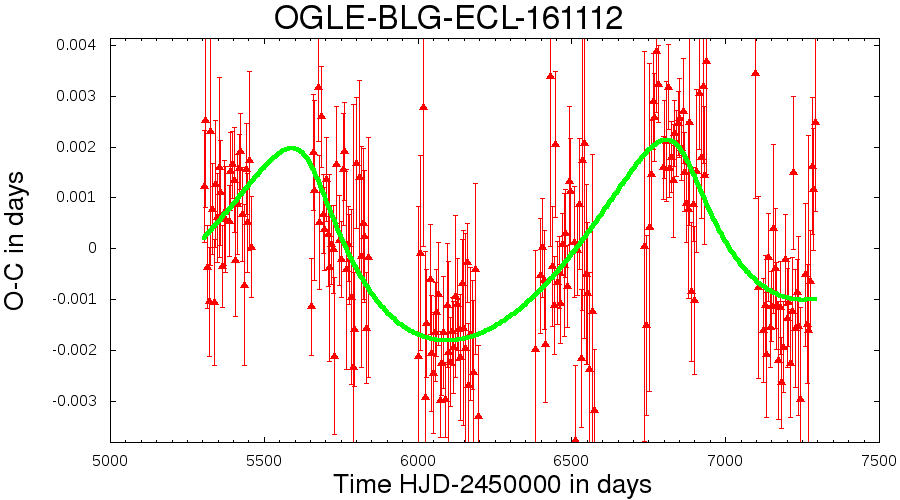}

\includegraphics[width=0.64\columnwidth]{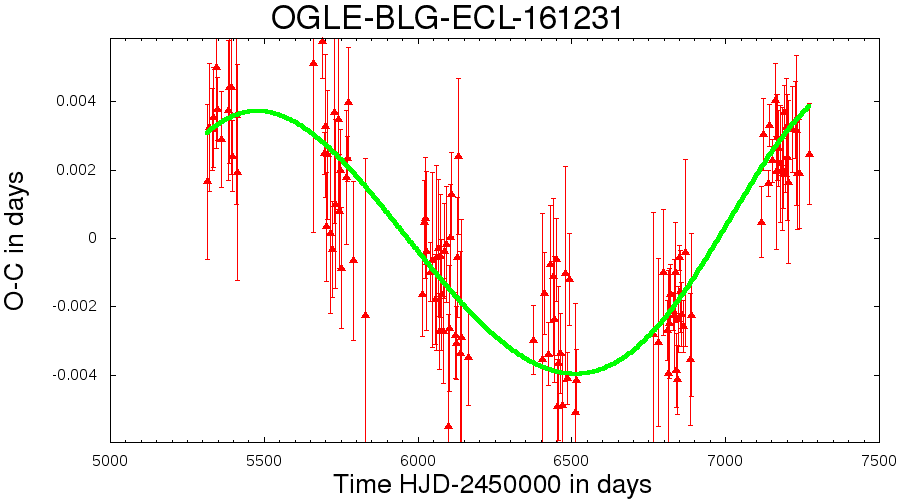}
\includegraphics[width=0.64\columnwidth]{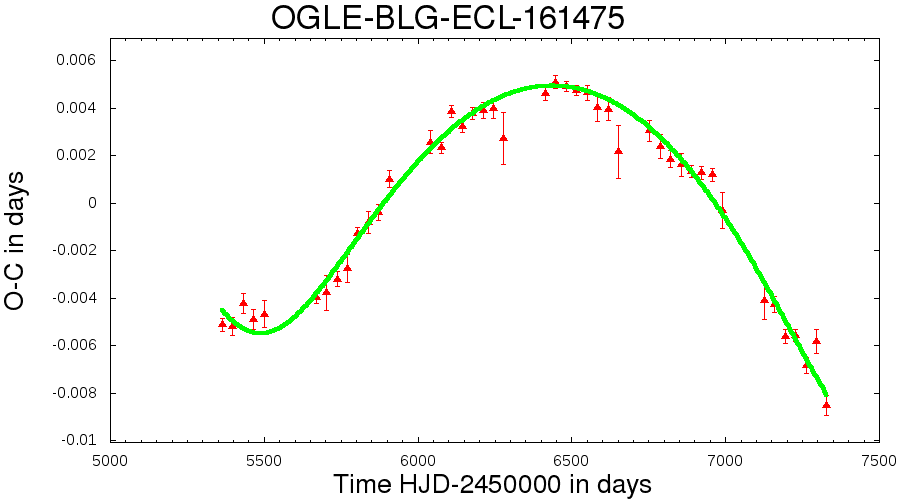}
\includegraphics[width=0.64\columnwidth]{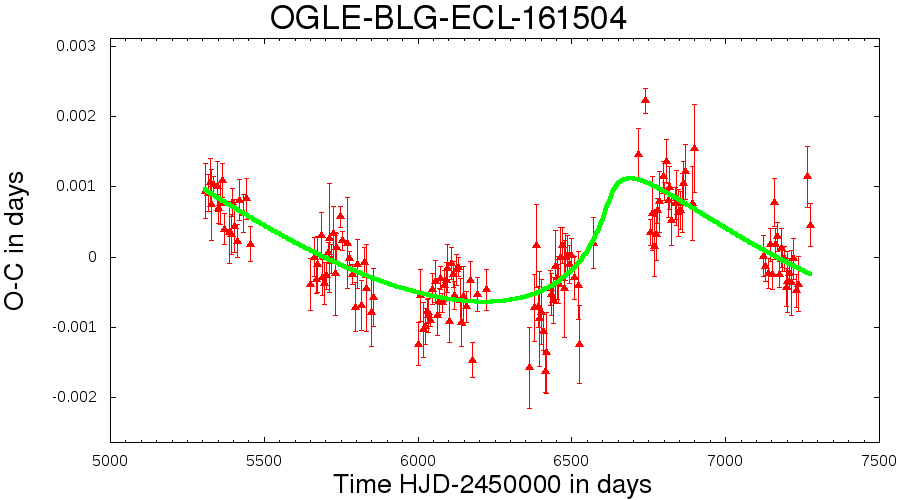}

\includegraphics[width=0.64\columnwidth]{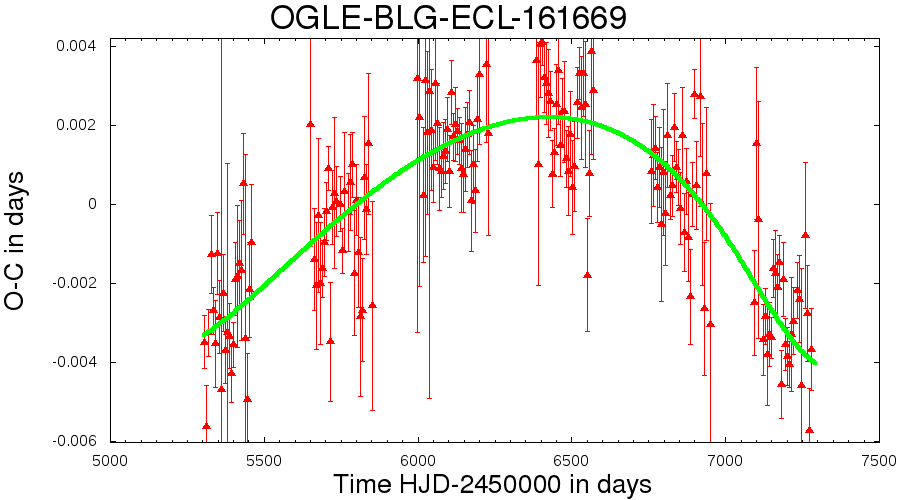}
\includegraphics[width=0.64\columnwidth]{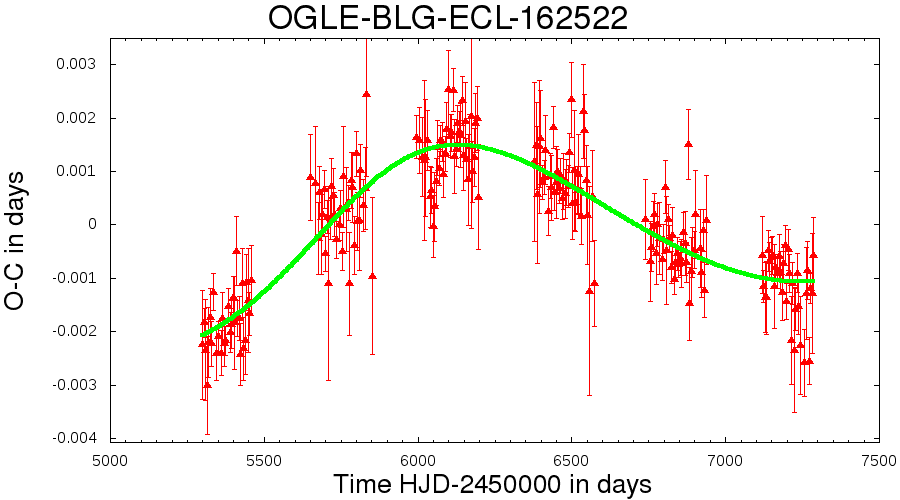}
\includegraphics[width=0.64\columnwidth]{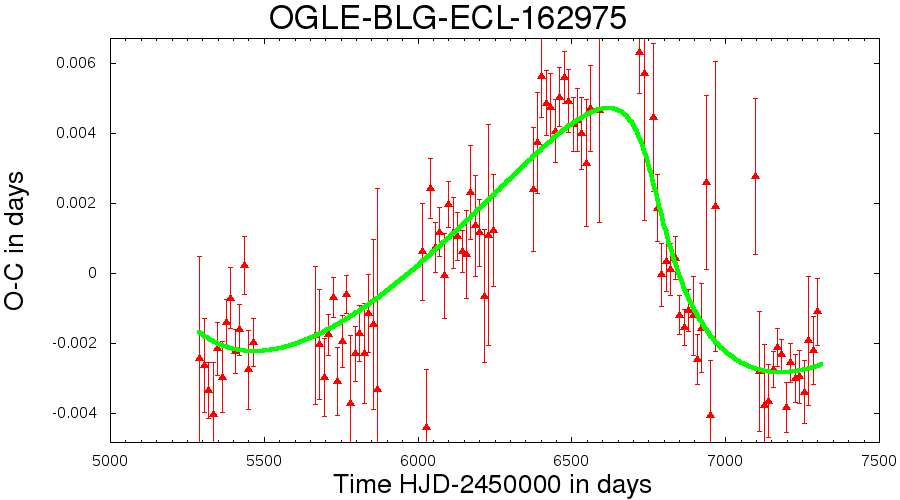}

\includegraphics[width=0.64\columnwidth]{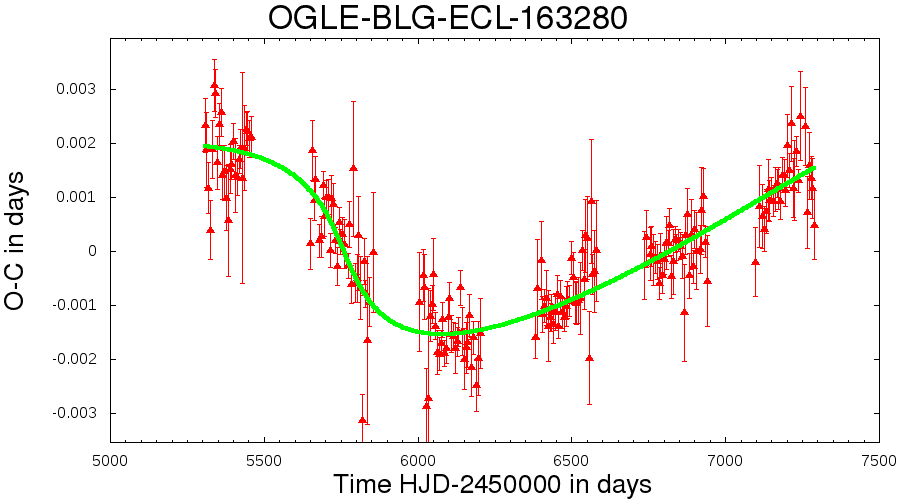}
\includegraphics[width=0.64\columnwidth]{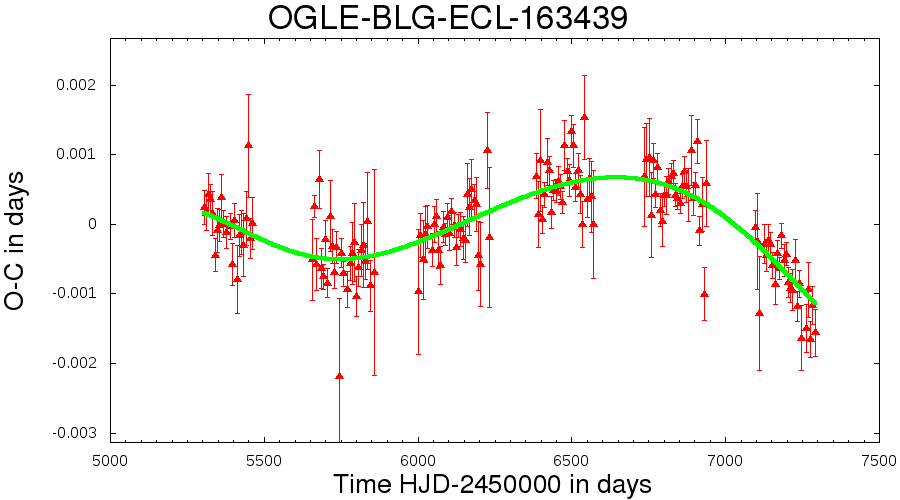}
\includegraphics[width=0.64\columnwidth]{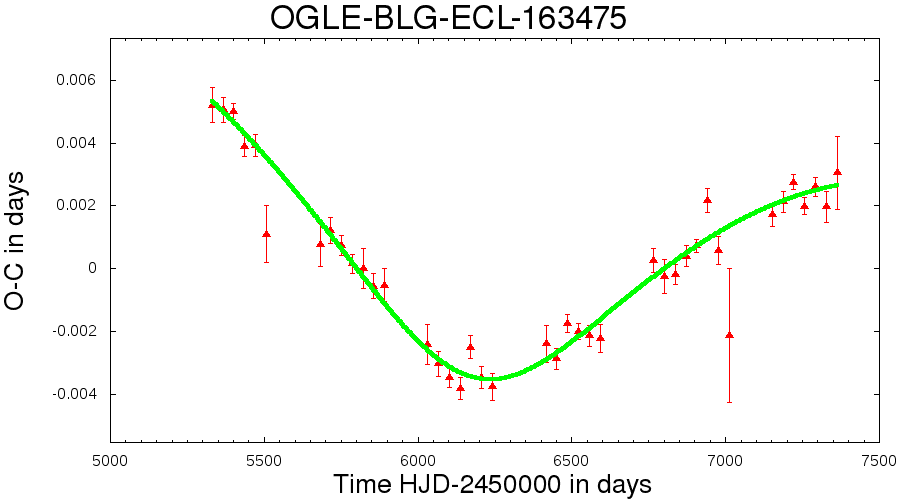}

\includegraphics[width=0.64\columnwidth]{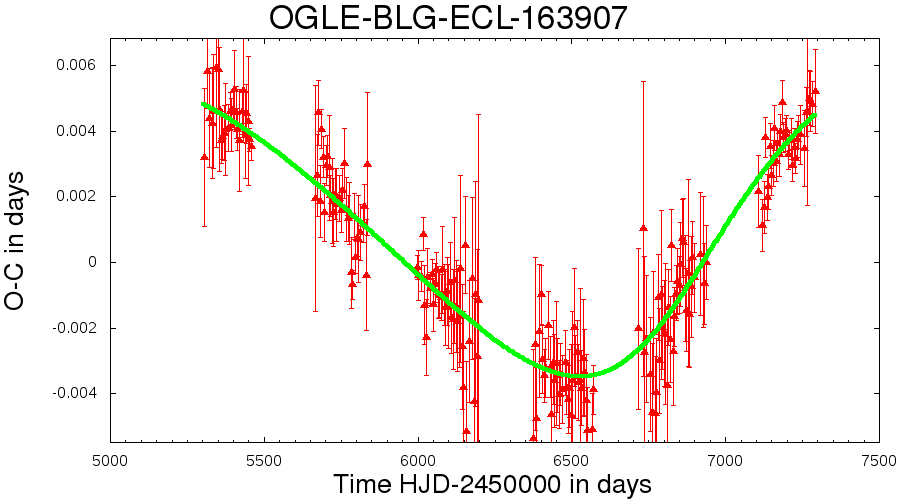}
\includegraphics[width=0.64\columnwidth]{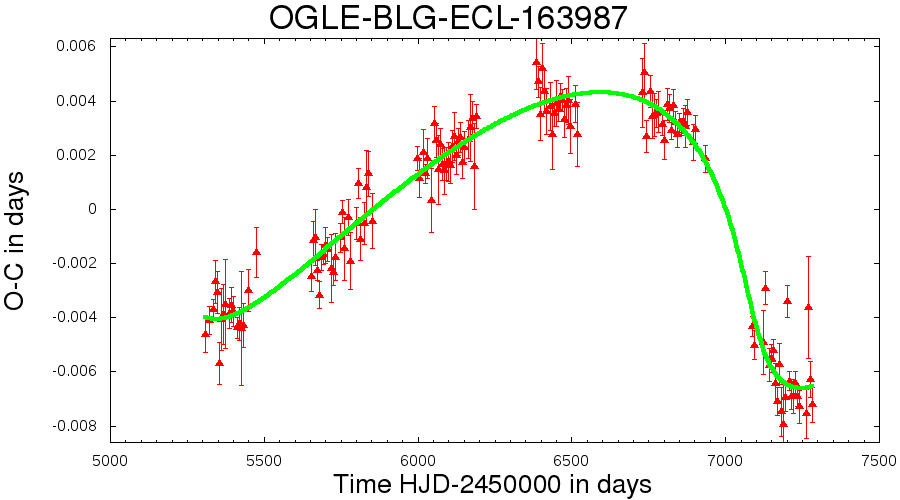}
\includegraphics[width=0.64\columnwidth]{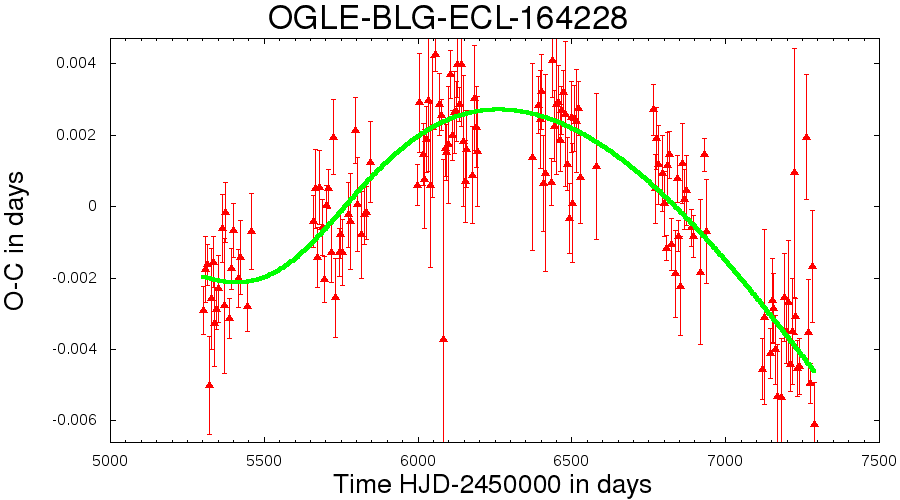}

\includegraphics[width=0.64\columnwidth]{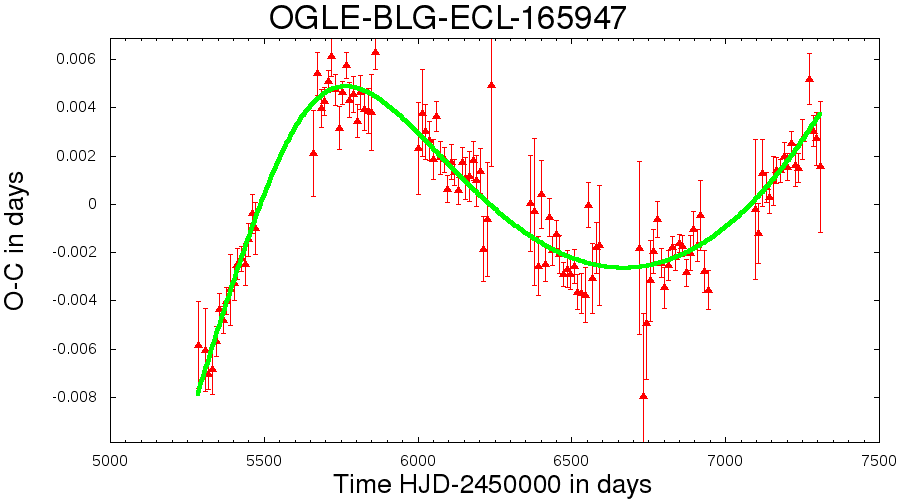}
\includegraphics[width=0.64\columnwidth]{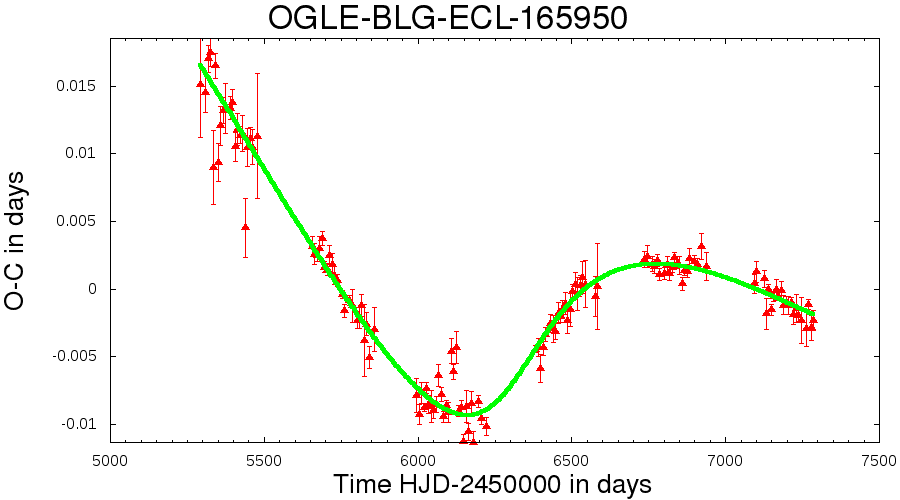}
\includegraphics[width=0.64\columnwidth]{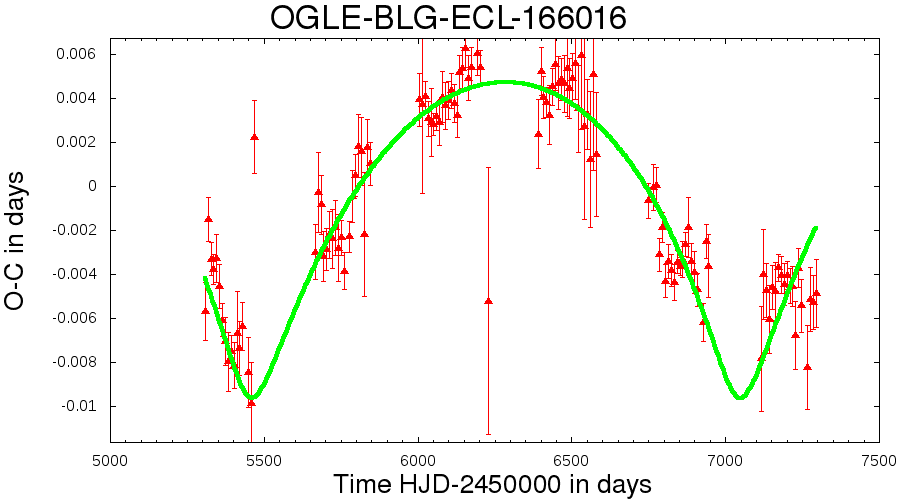}

\includegraphics[width=0.64\columnwidth]{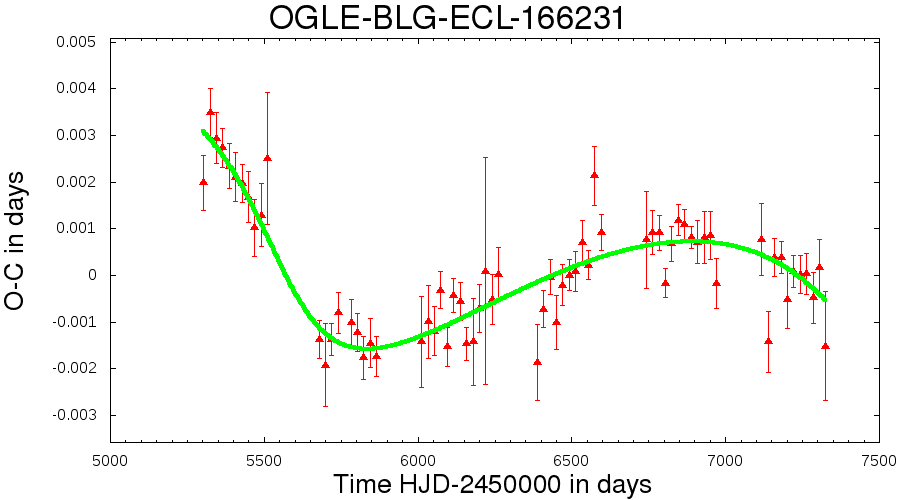}
\includegraphics[width=0.64\columnwidth]{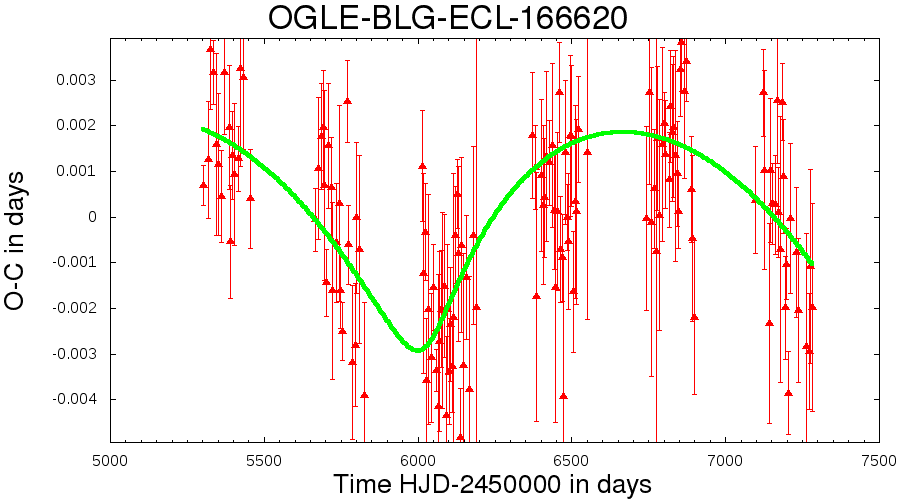}
\includegraphics[width=0.64\columnwidth]{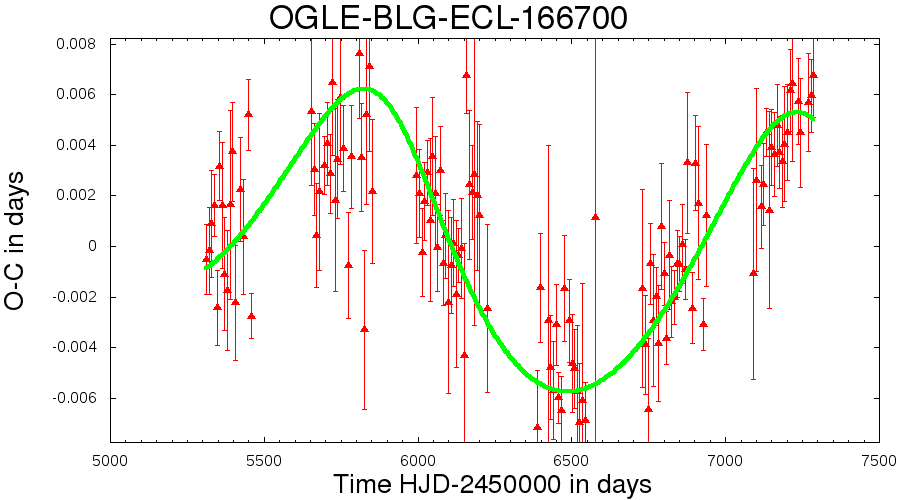}

\end{figure*}
\clearpage

\begin{figure*}
\includegraphics[width=0.64\columnwidth]{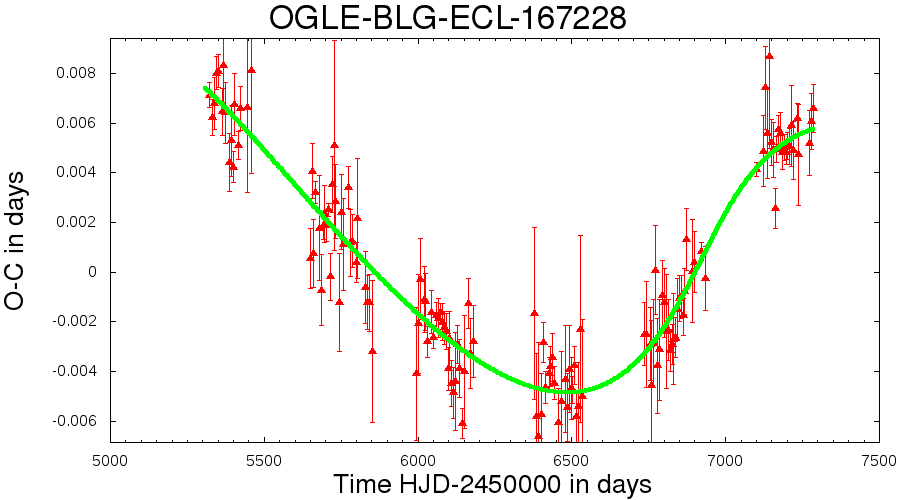}
\includegraphics[width=0.64\columnwidth]{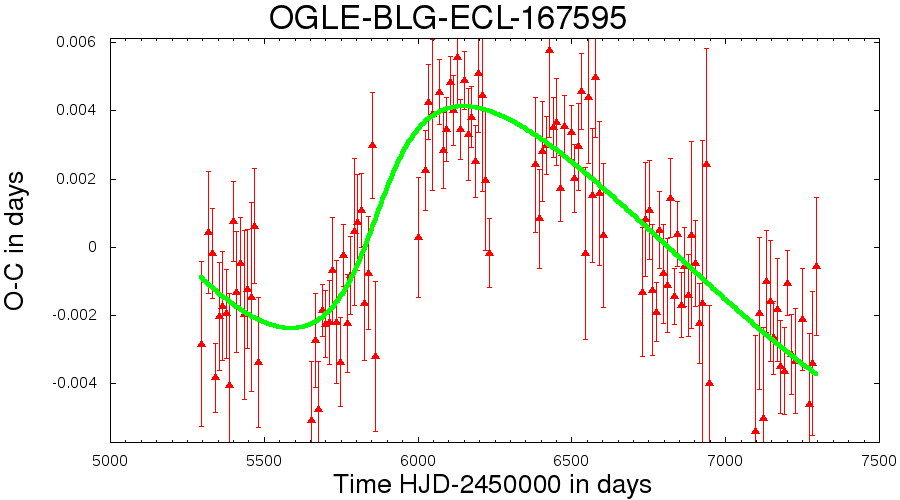}
\includegraphics[width=0.64\columnwidth]{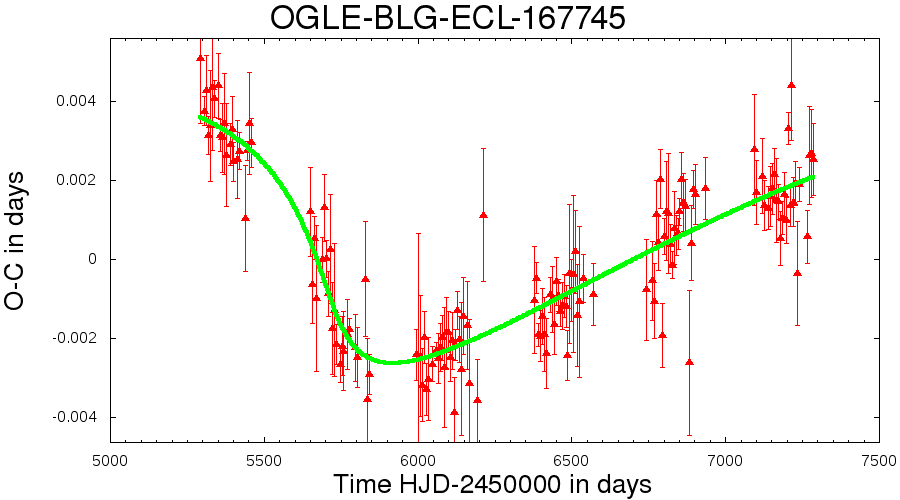}

\includegraphics[width=0.64\columnwidth]{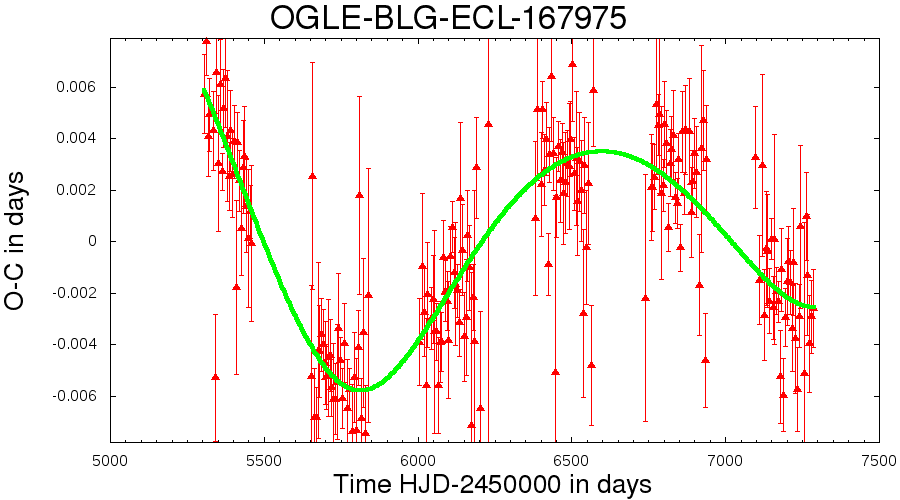}
\includegraphics[width=0.64\columnwidth]{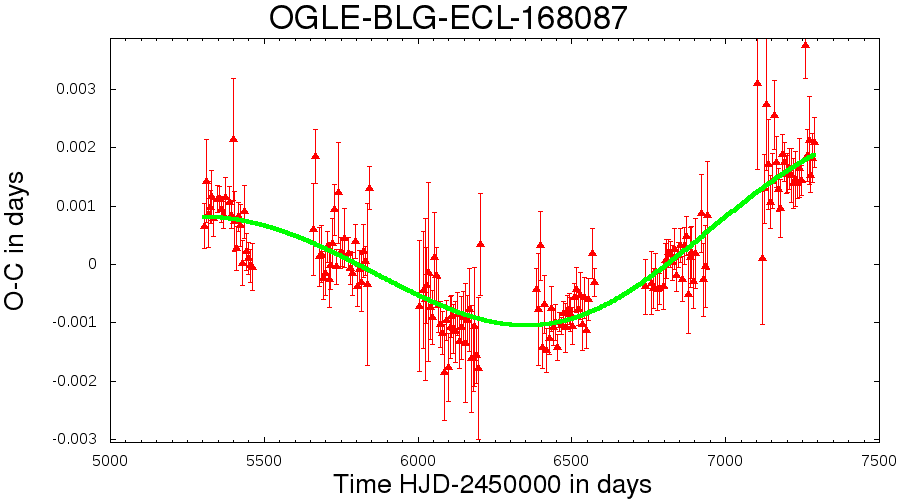}
\includegraphics[width=0.64\columnwidth]{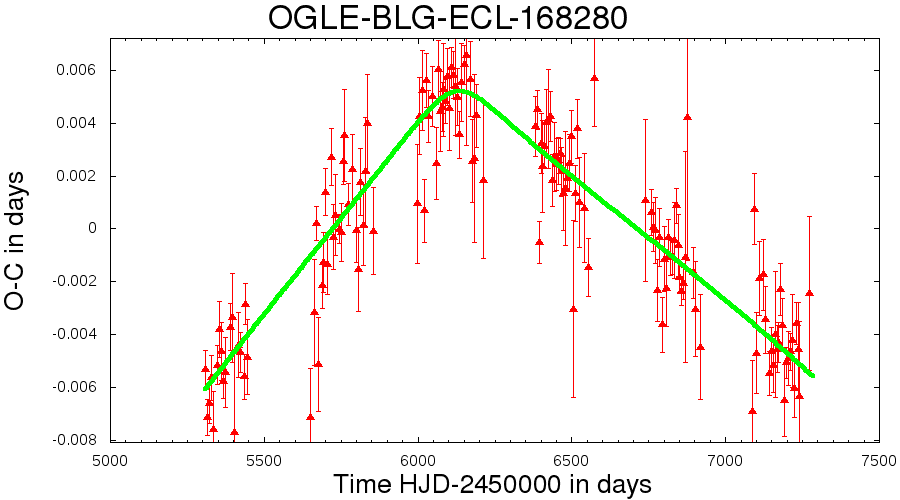}

\includegraphics[width=0.64\columnwidth]{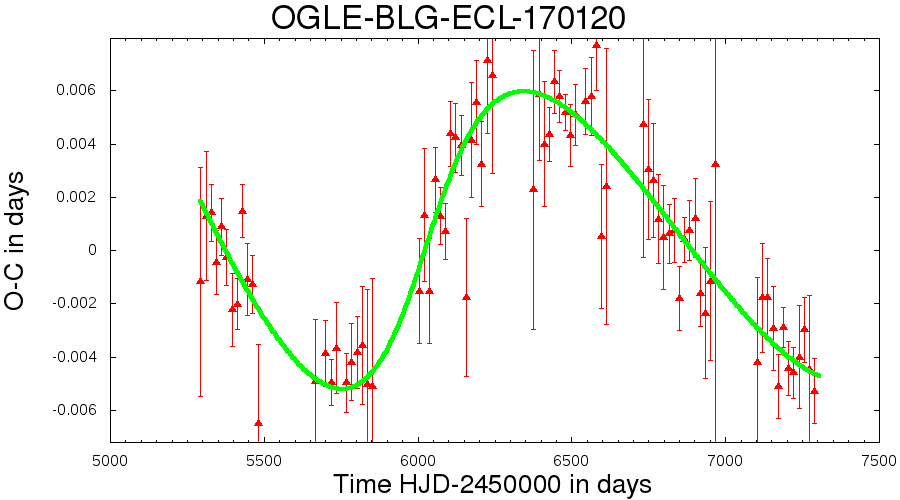}
\includegraphics[width=0.64\columnwidth]{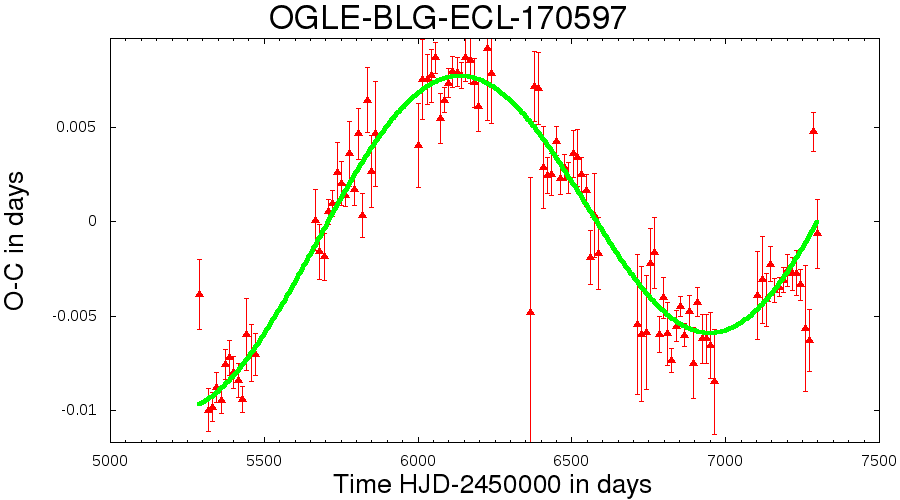}
\includegraphics[width=0.64\columnwidth]{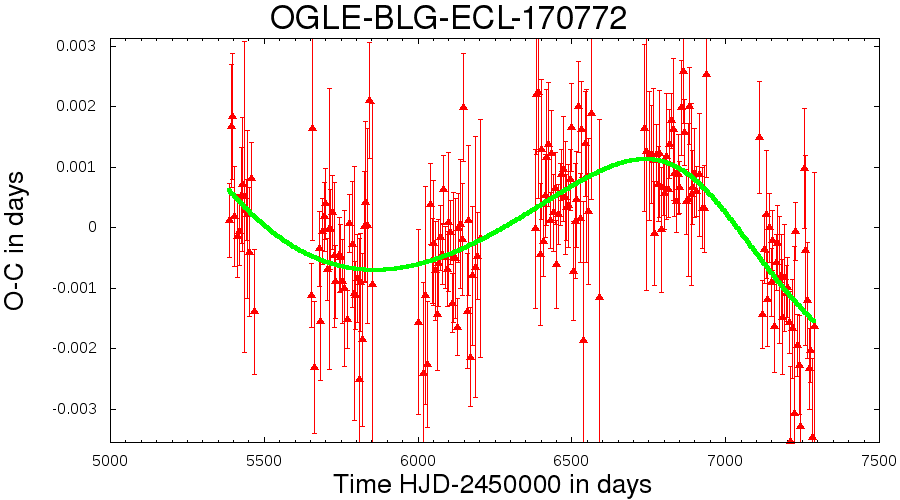}

\includegraphics[width=0.64\columnwidth]{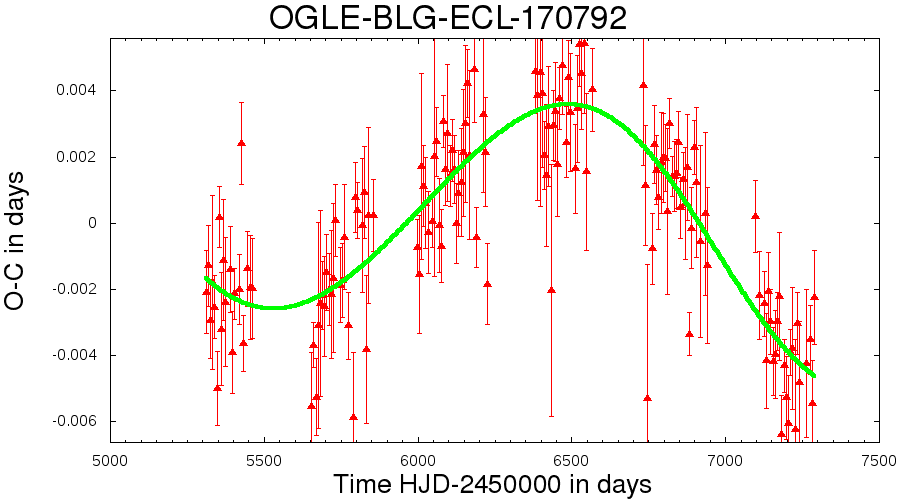}
\includegraphics[width=0.64\columnwidth]{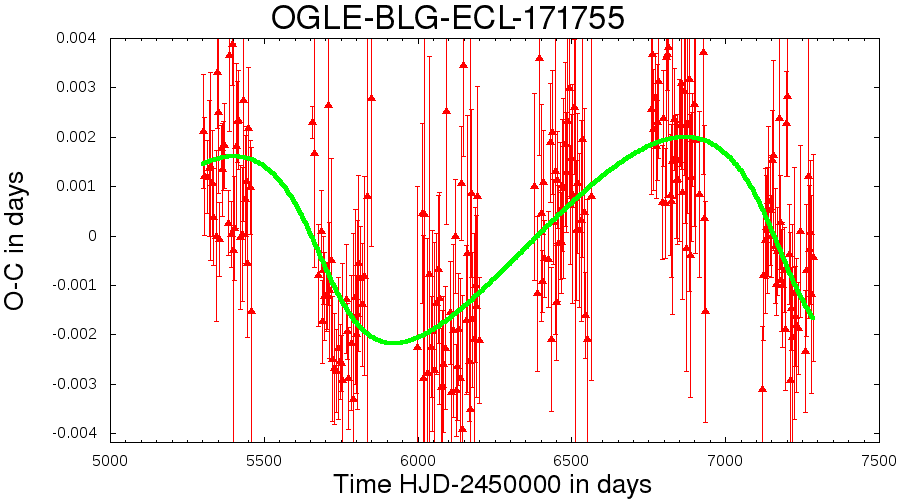}
\includegraphics[width=0.64\columnwidth]{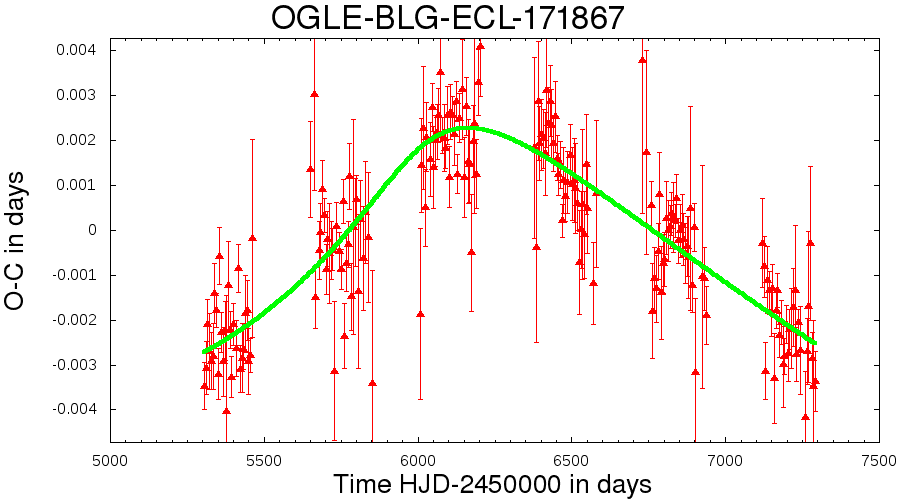}

\includegraphics[width=0.64\columnwidth]{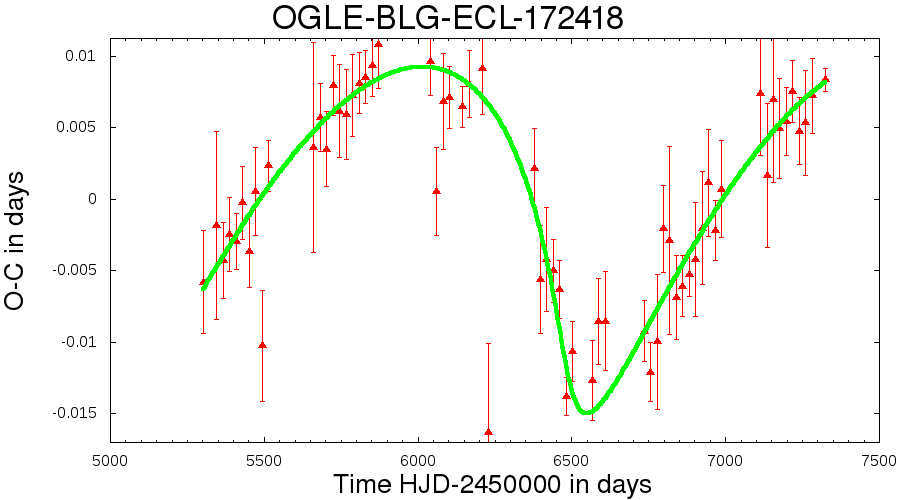}
\includegraphics[width=0.64\columnwidth]{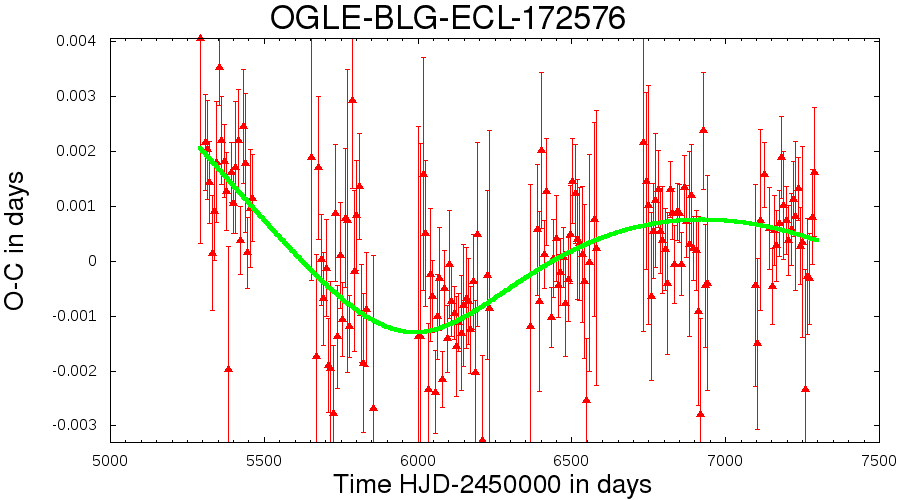}
\includegraphics[width=0.64\columnwidth]{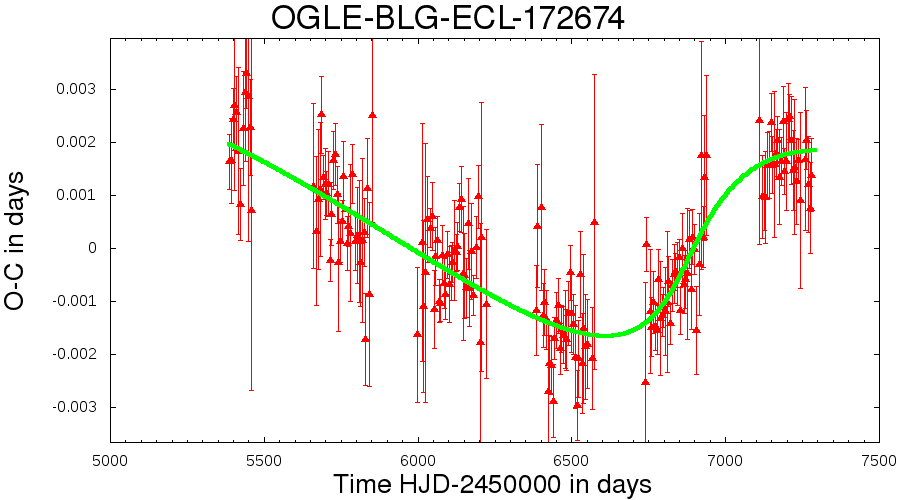}

\includegraphics[width=0.64\columnwidth]{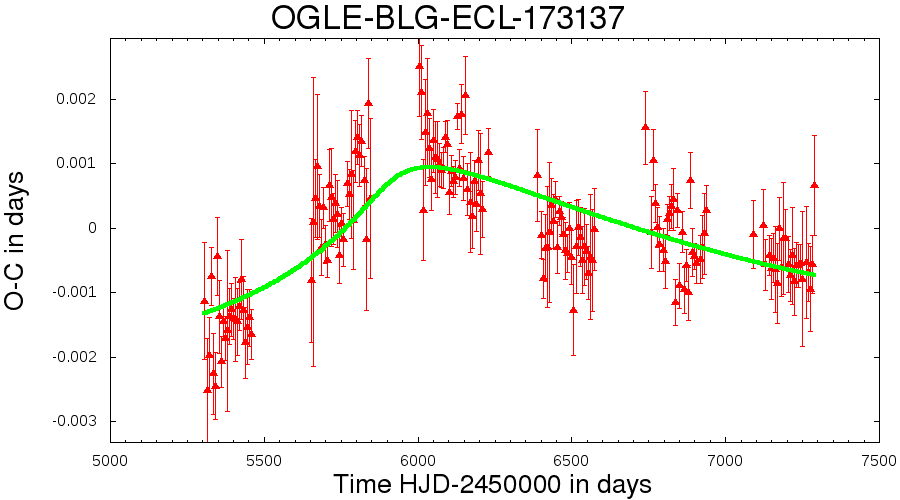}
\includegraphics[width=0.64\columnwidth]{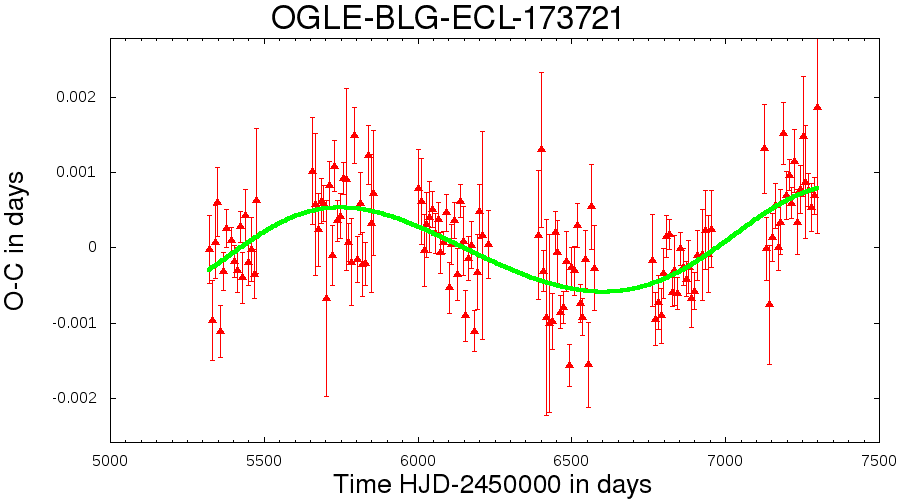}
\includegraphics[width=0.64\columnwidth]{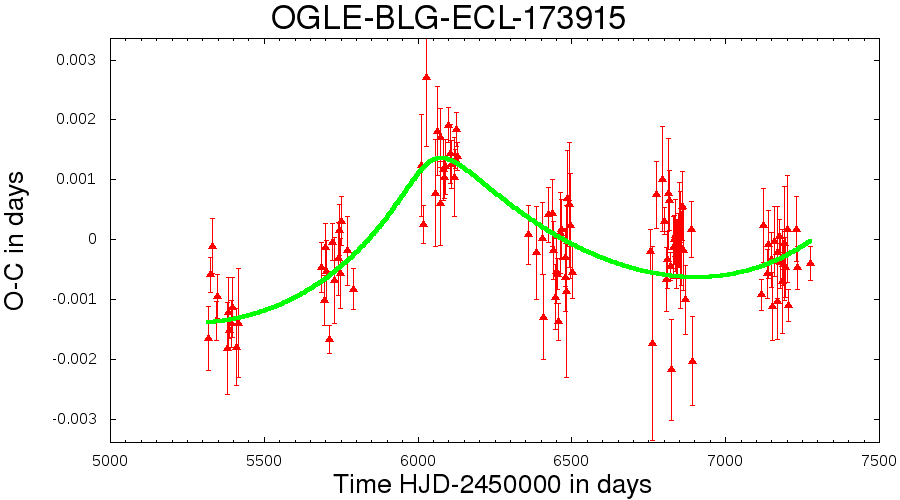}

\includegraphics[width=0.64\columnwidth]{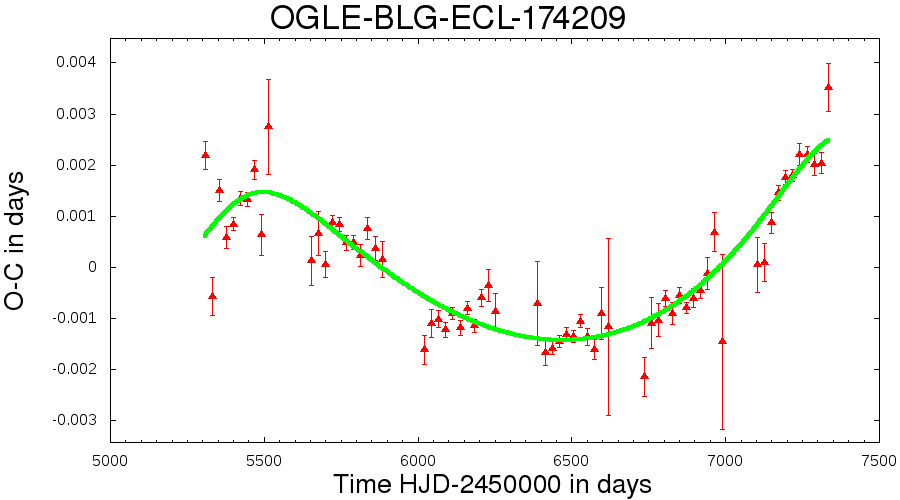}
\includegraphics[width=0.64\columnwidth]{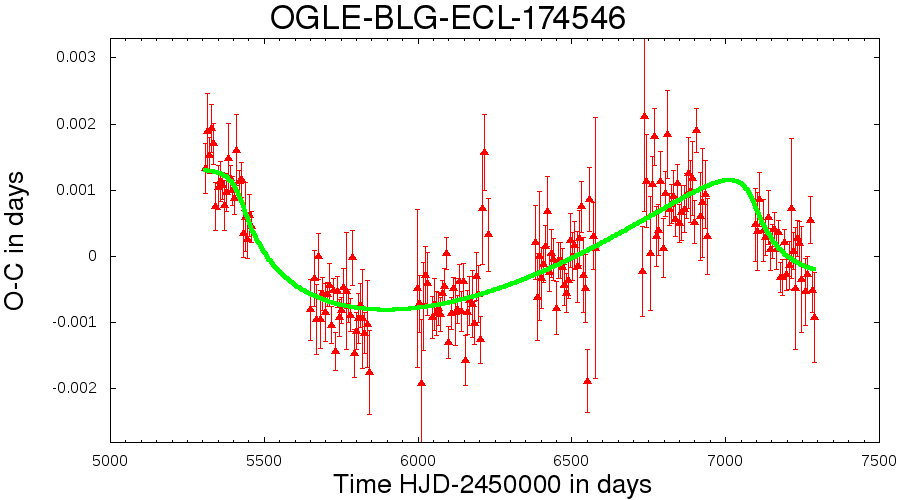}
\includegraphics[width=0.64\columnwidth]{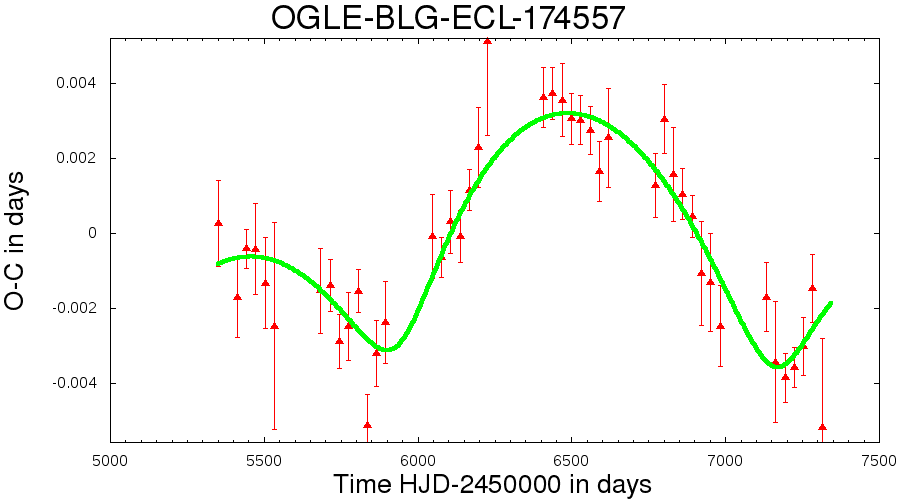}

\includegraphics[width=0.64\columnwidth]{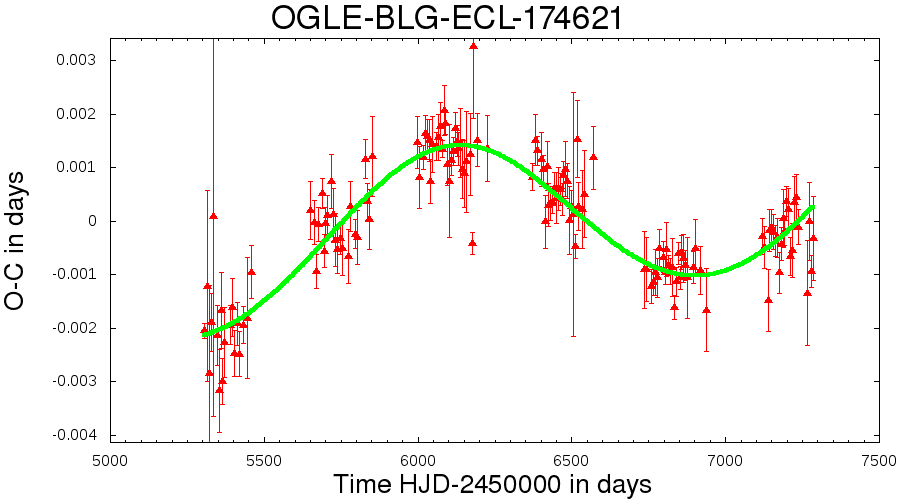}
\includegraphics[width=0.64\columnwidth]{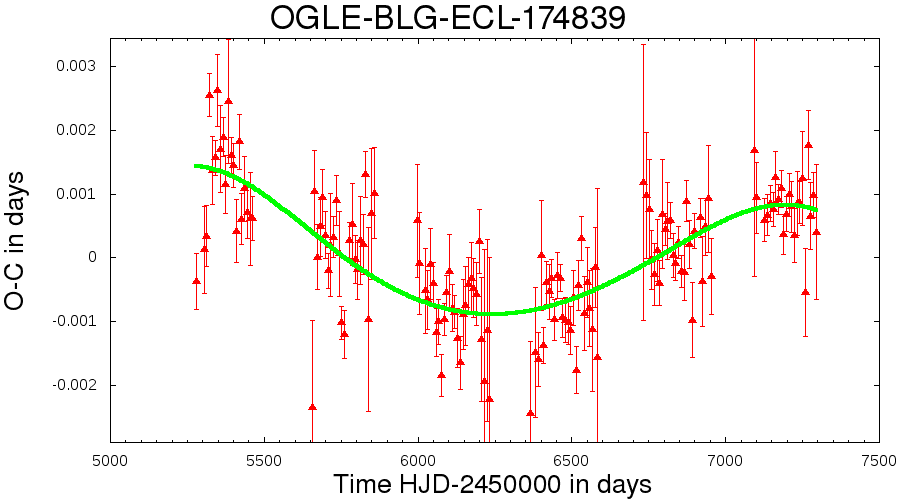}
\includegraphics[width=0.64\columnwidth]{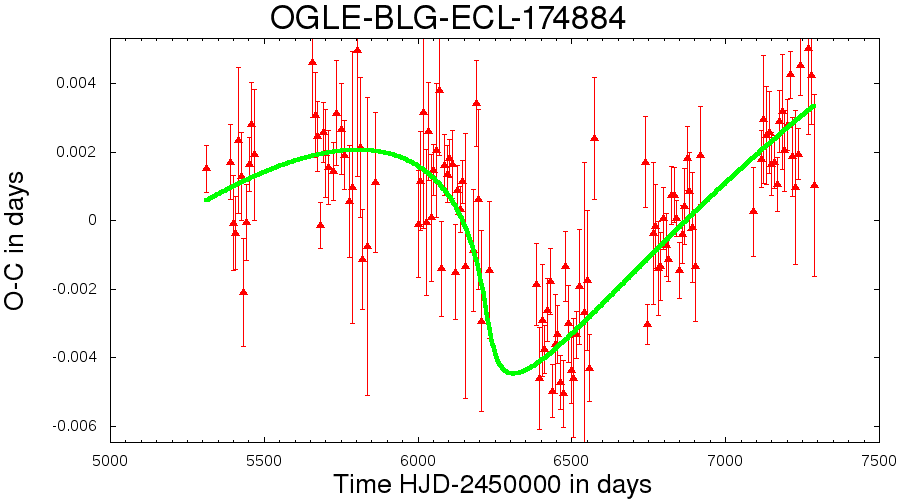}

\end{figure*}
\clearpage

\begin{figure*}
\includegraphics[width=0.64\columnwidth]{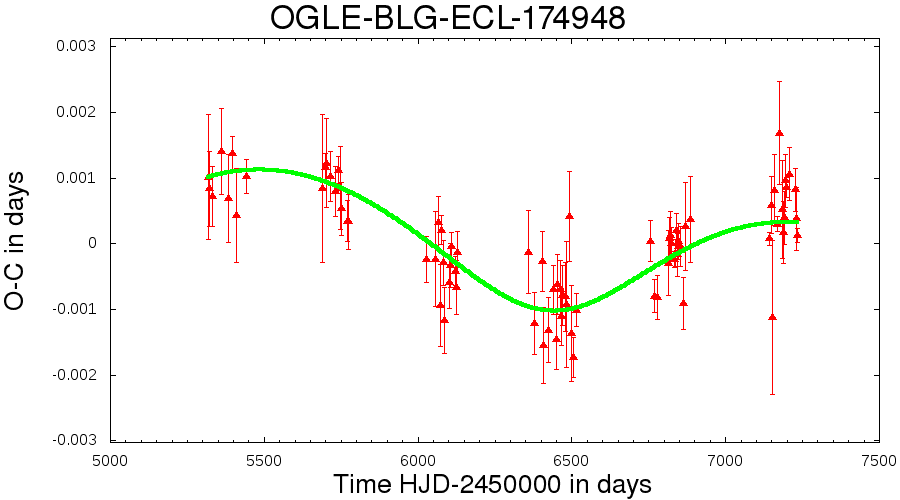}
\includegraphics[width=0.64\columnwidth]{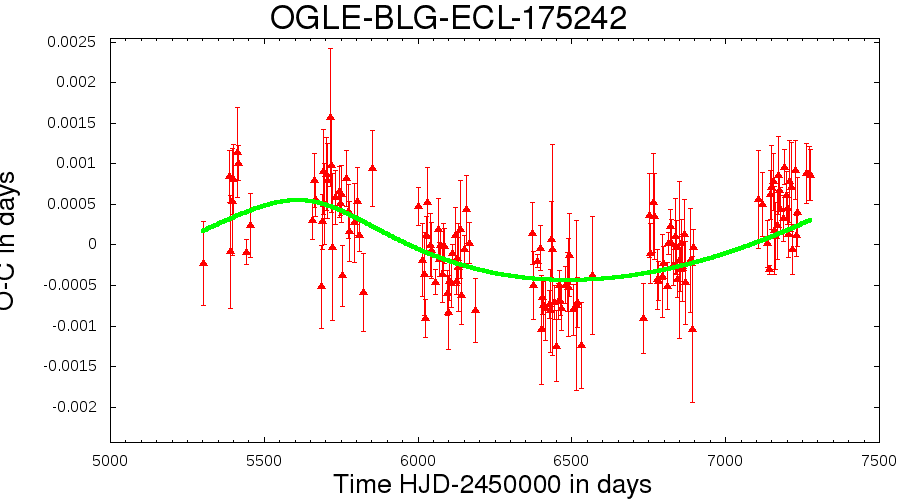}
\includegraphics[width=0.64\columnwidth]{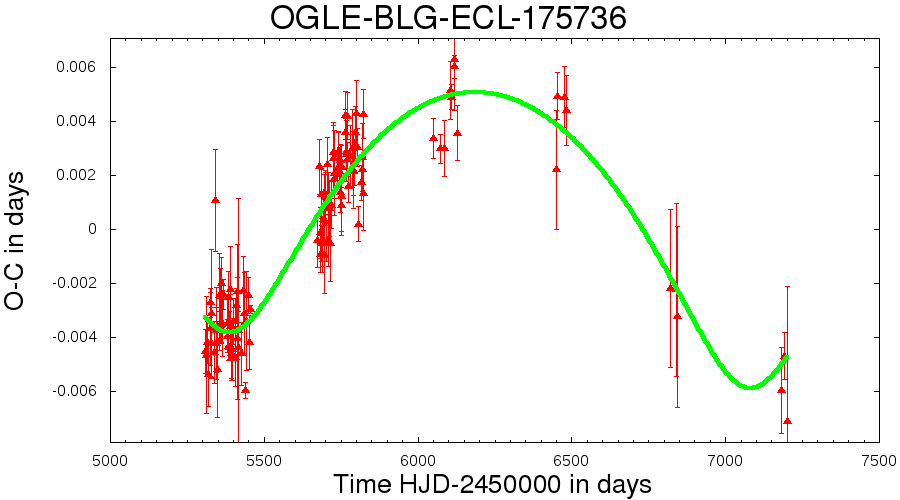}

\includegraphics[width=0.64\columnwidth]{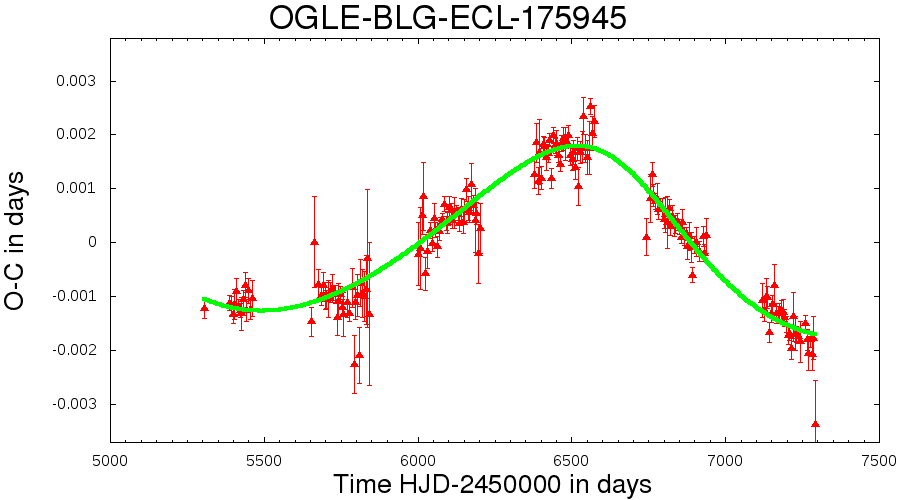}
\includegraphics[width=0.64\columnwidth]{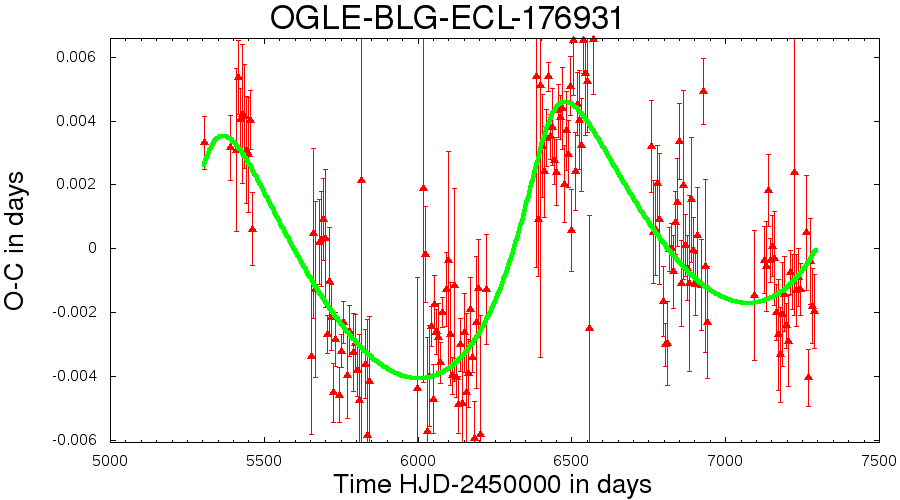}
\includegraphics[width=0.64\columnwidth]{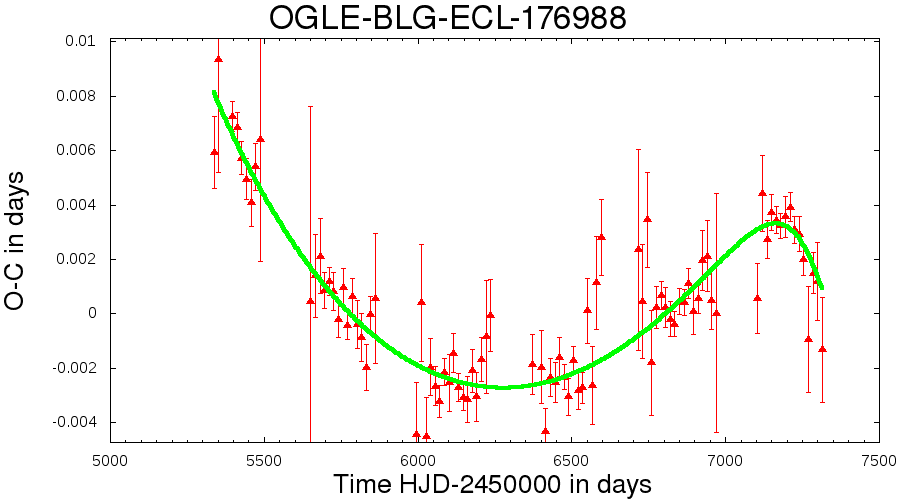}

\includegraphics[width=0.64\columnwidth]{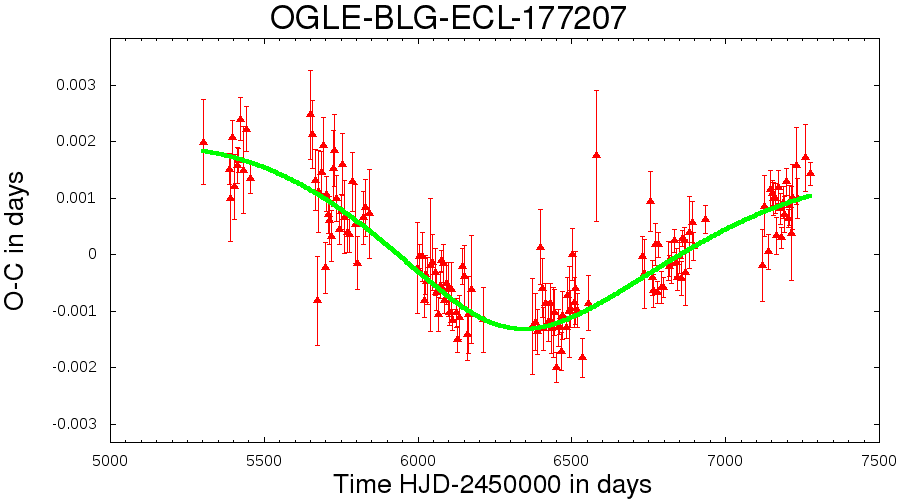}
\includegraphics[width=0.64\columnwidth]{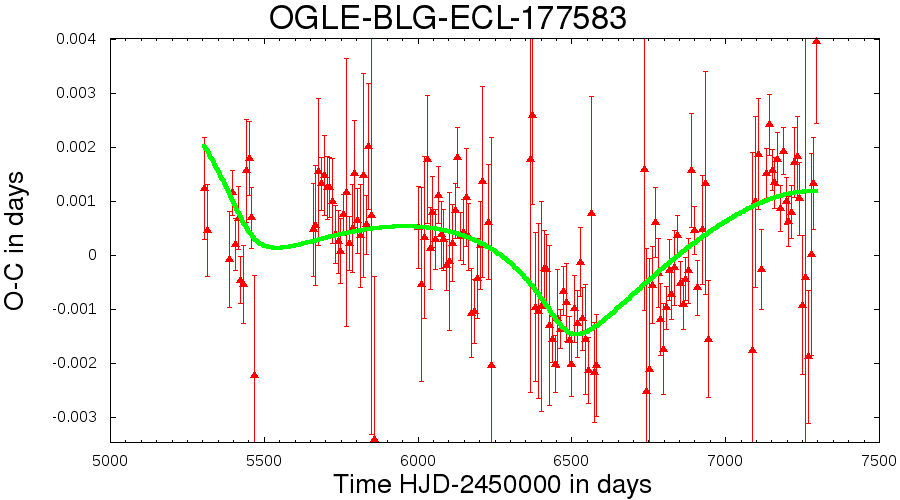}
\includegraphics[width=0.64\columnwidth]{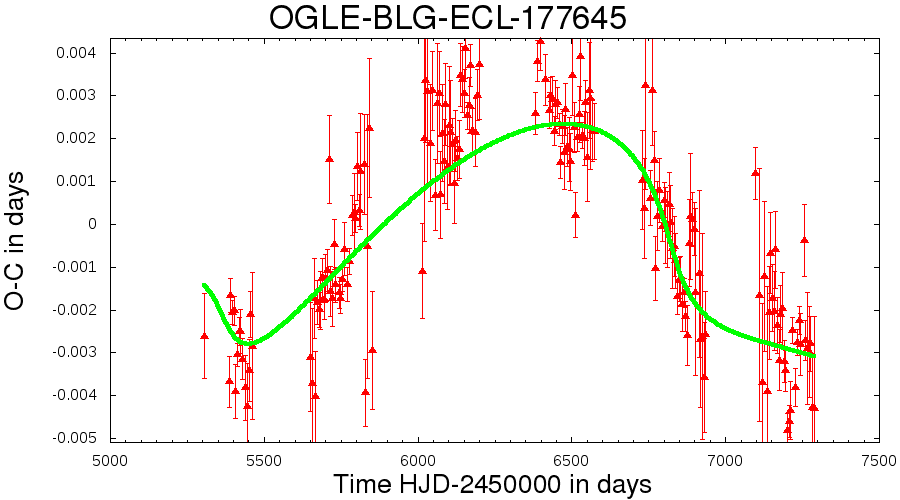}

\includegraphics[width=0.64\columnwidth]{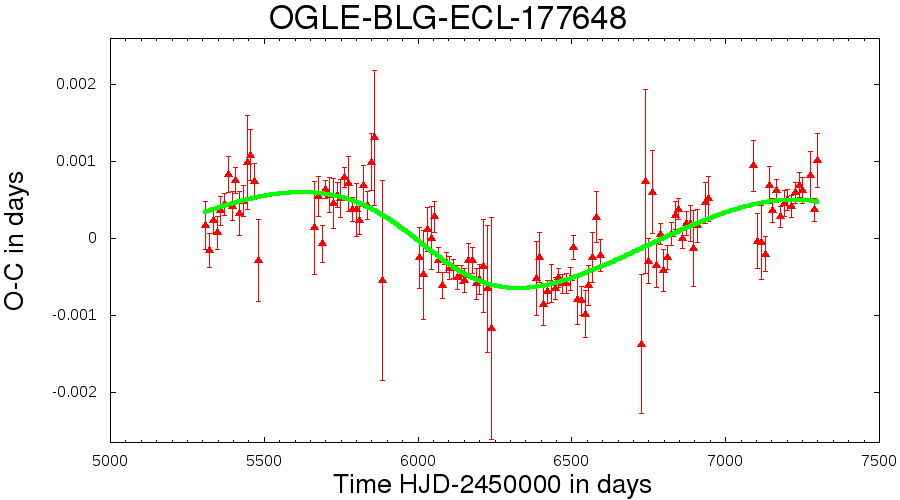}
\includegraphics[width=0.64\columnwidth]{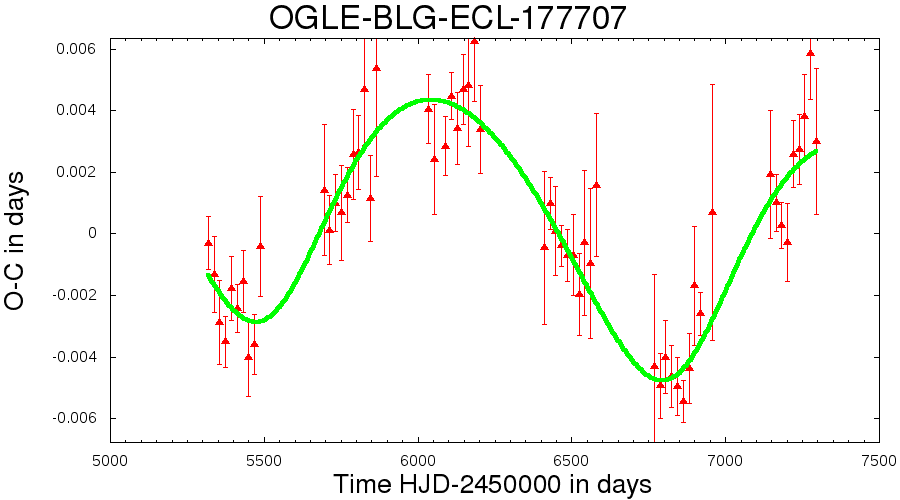}
\includegraphics[width=0.64\columnwidth]{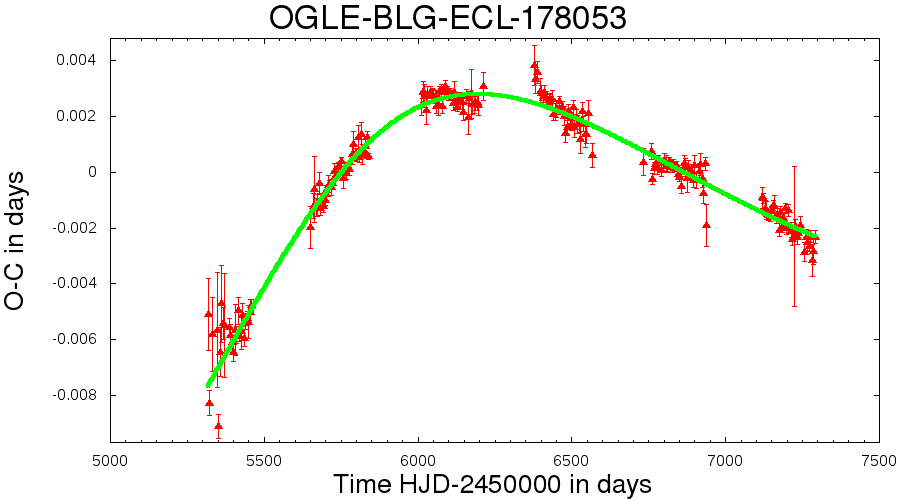}

\includegraphics[width=0.64\columnwidth]{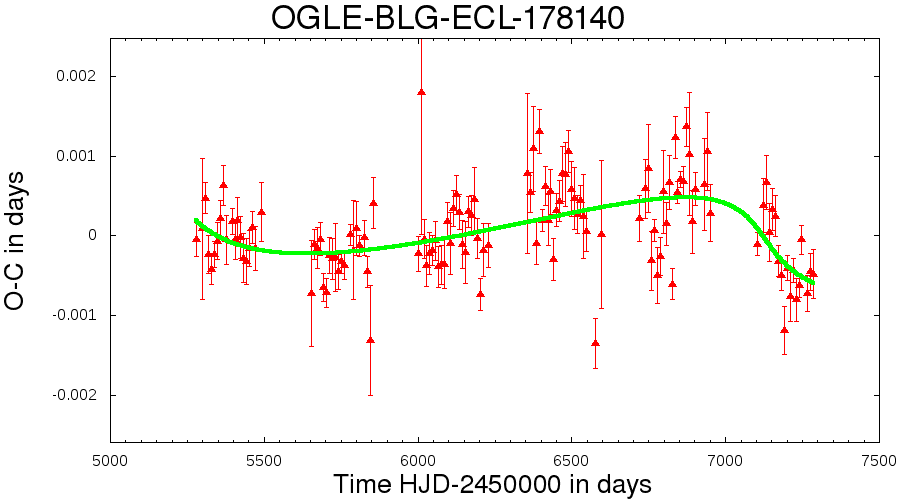}
\includegraphics[width=0.64\columnwidth]{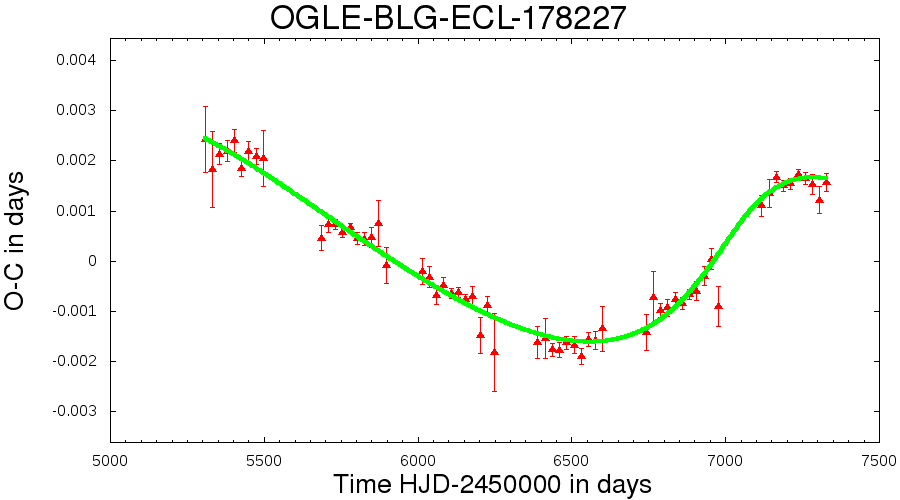}
\includegraphics[width=0.64\columnwidth]{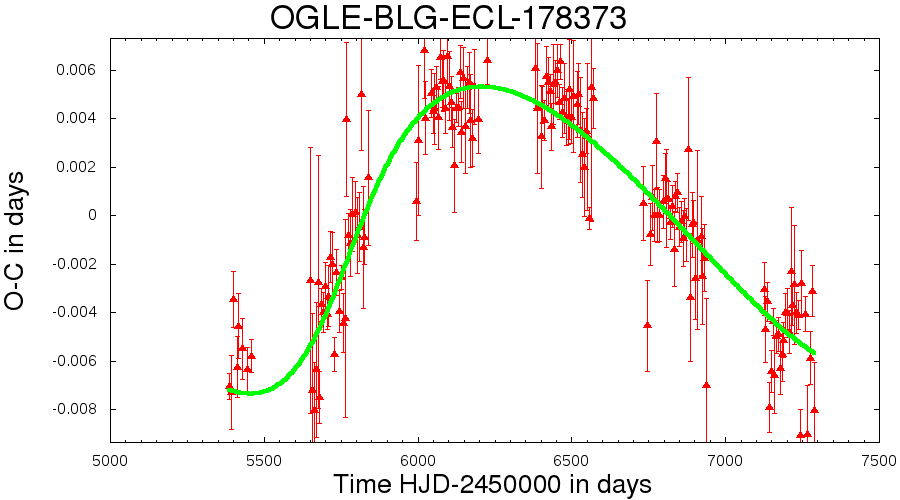}

\includegraphics[width=0.64\columnwidth]{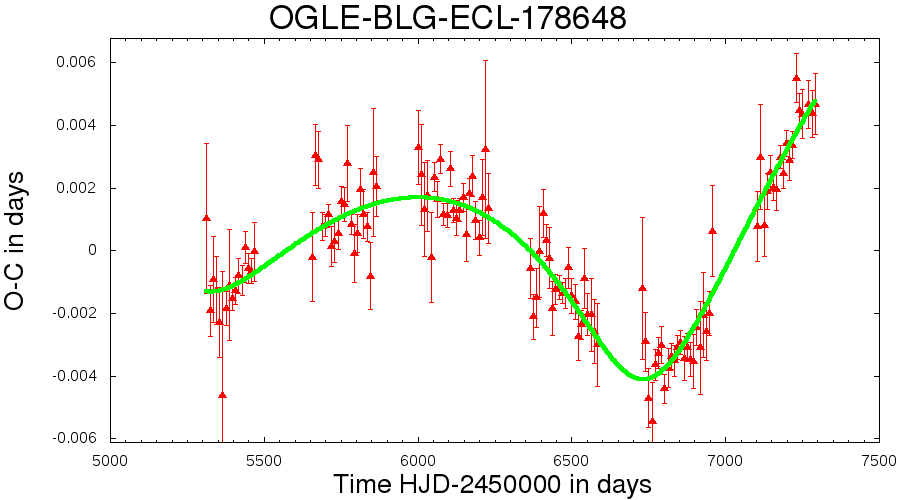}
\includegraphics[width=0.64\columnwidth]{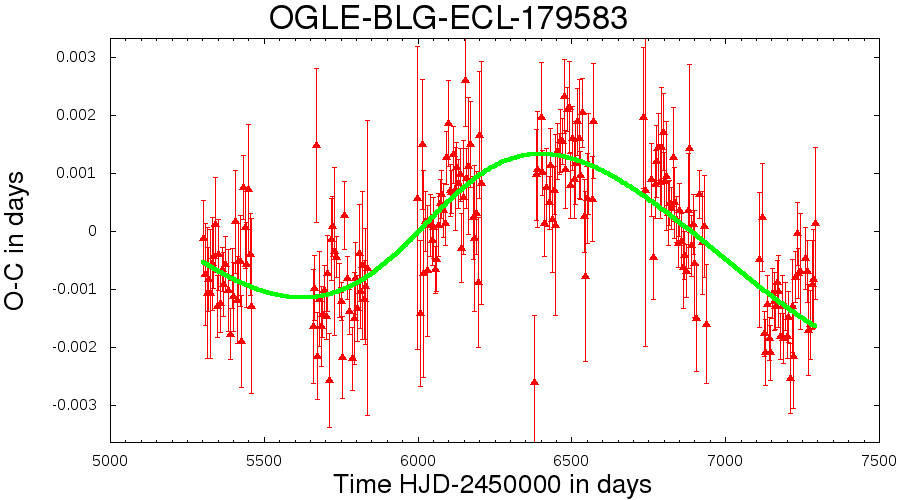}
\includegraphics[width=0.64\columnwidth]{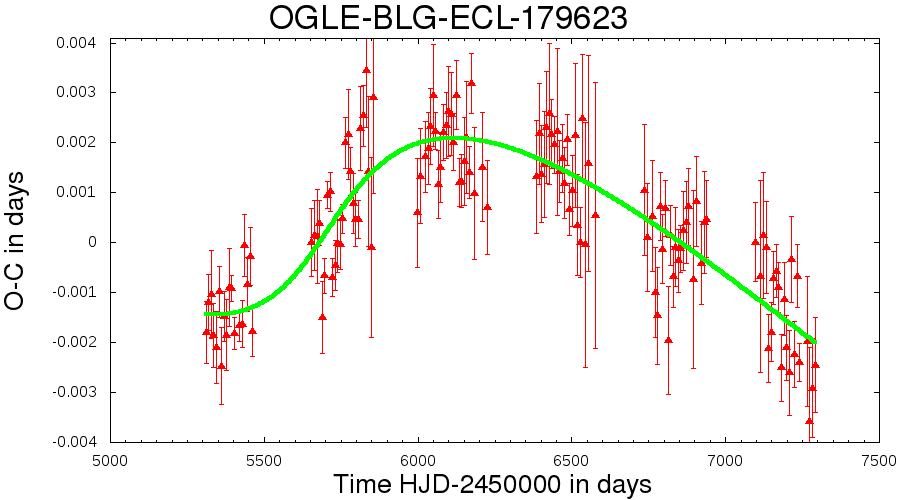}

\includegraphics[width=0.64\columnwidth]{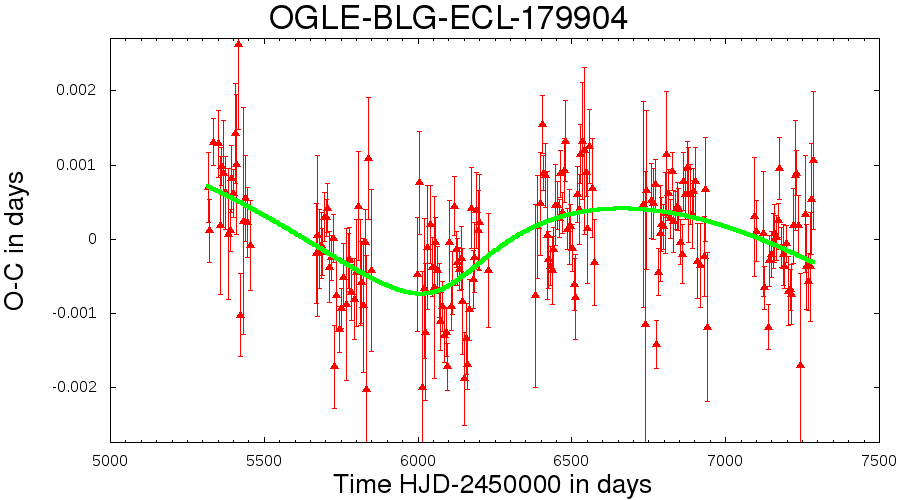}
\includegraphics[width=0.64\columnwidth]{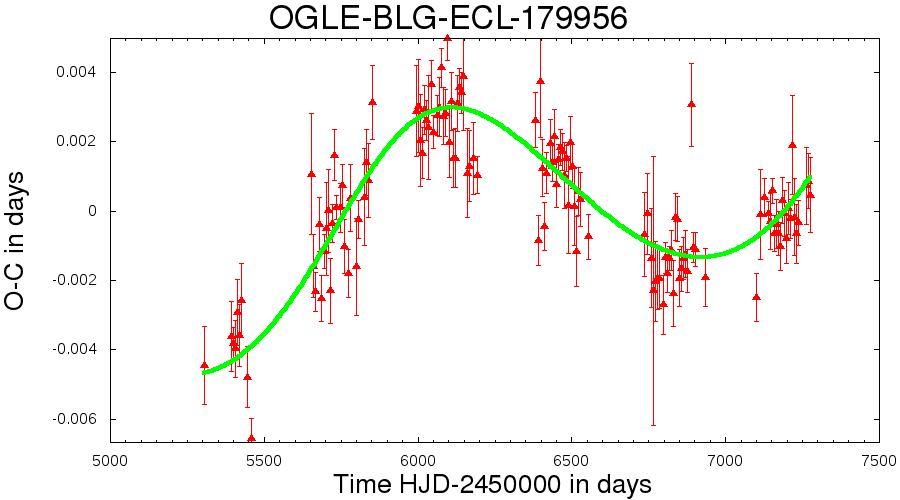}
\includegraphics[width=0.64\columnwidth]{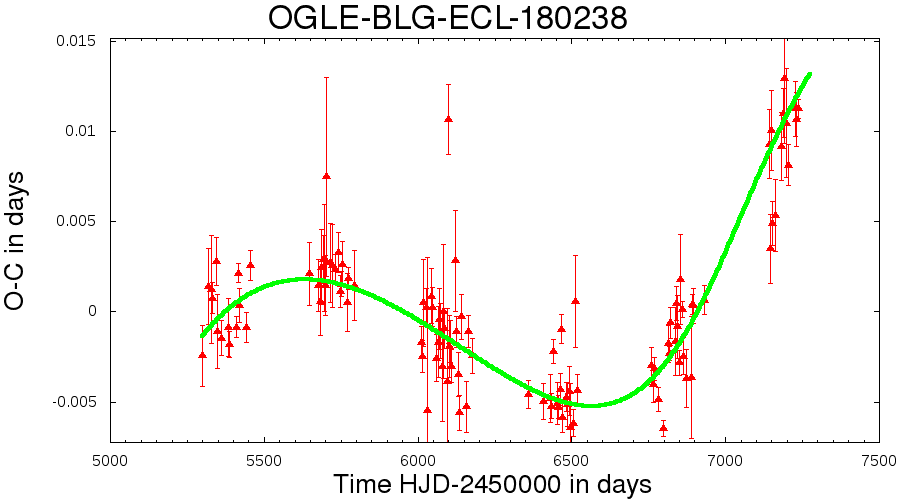}

\includegraphics[width=0.64\columnwidth]{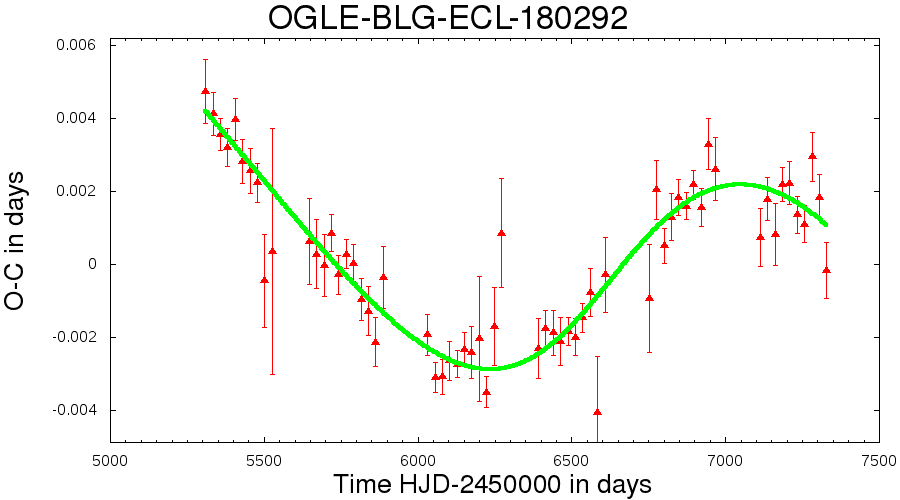}
\includegraphics[width=0.64\columnwidth]{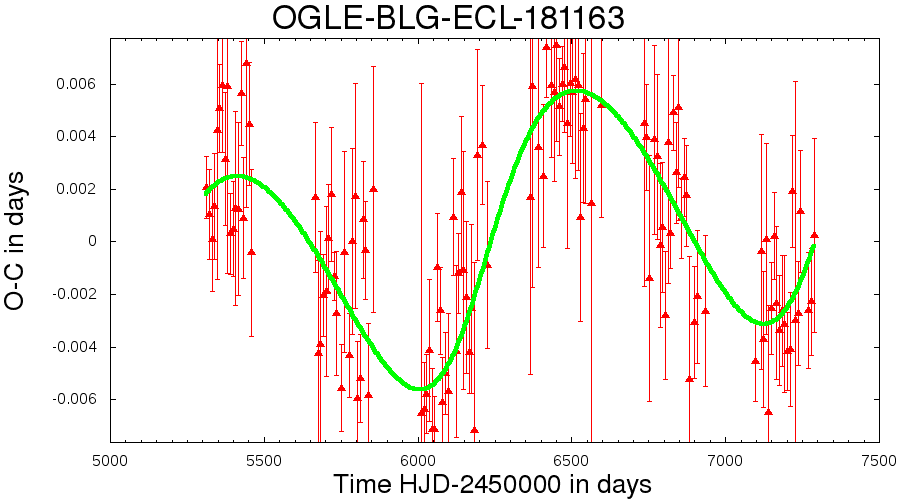}
\includegraphics[width=0.64\columnwidth]{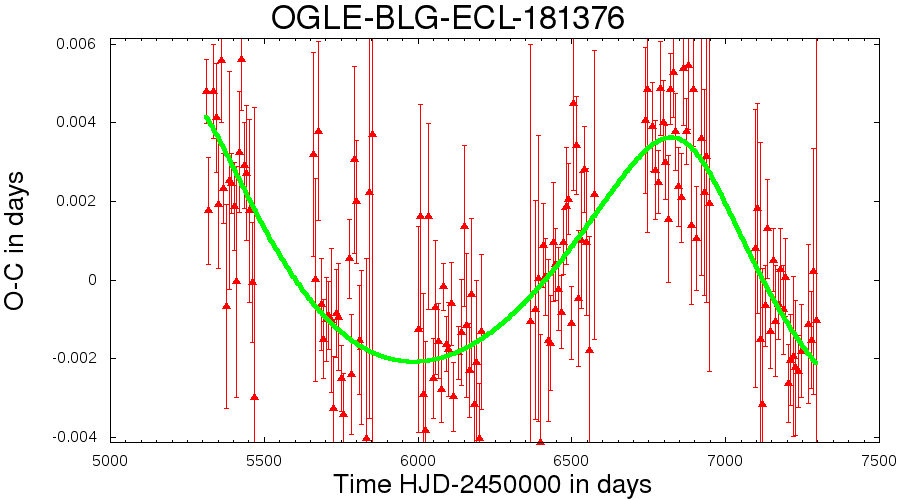}

\end{figure*}
\clearpage

\begin{figure*}
\includegraphics[width=0.64\columnwidth]{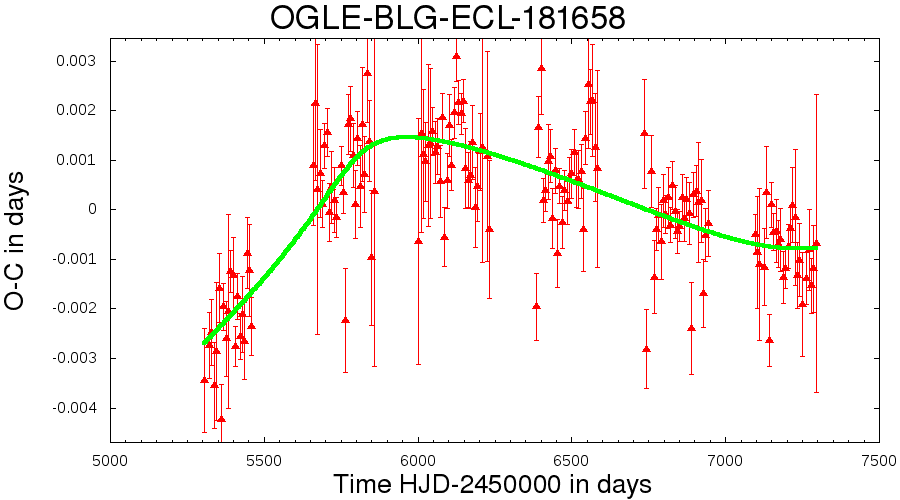}
\includegraphics[width=0.64\columnwidth]{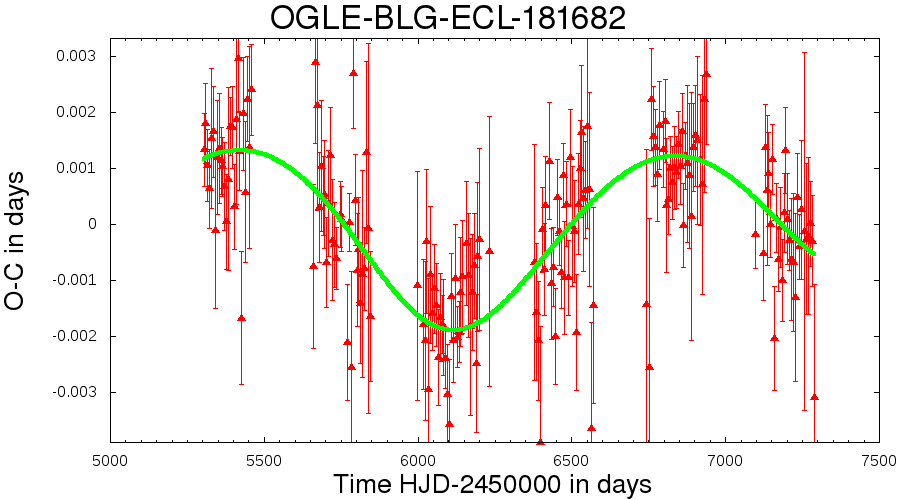}
\includegraphics[width=0.64\columnwidth]{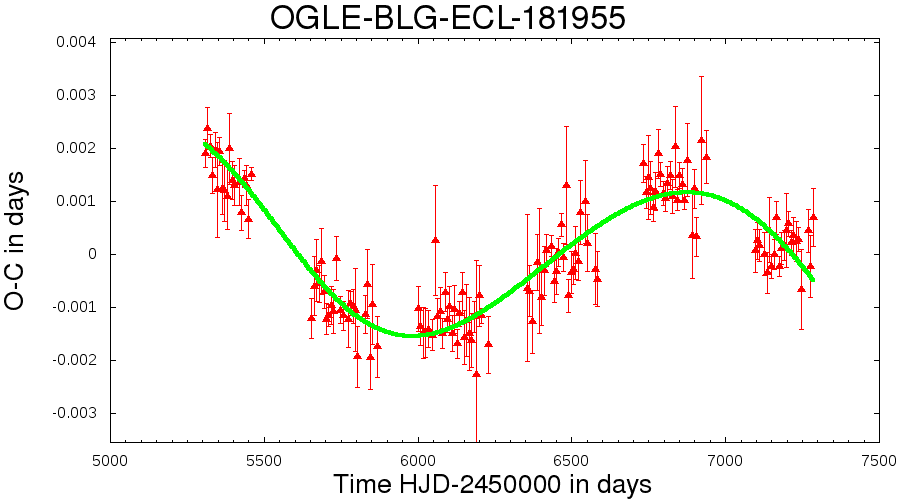}

\includegraphics[width=0.64\columnwidth]{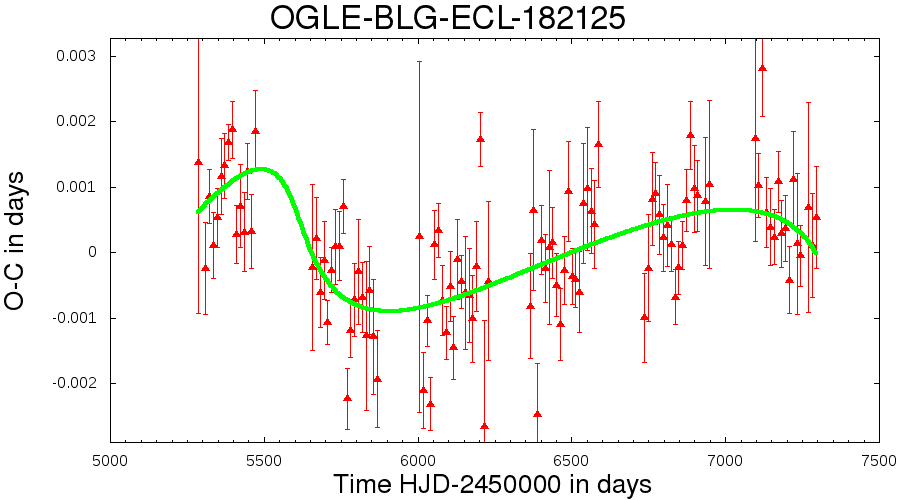}
\includegraphics[width=0.64\columnwidth]{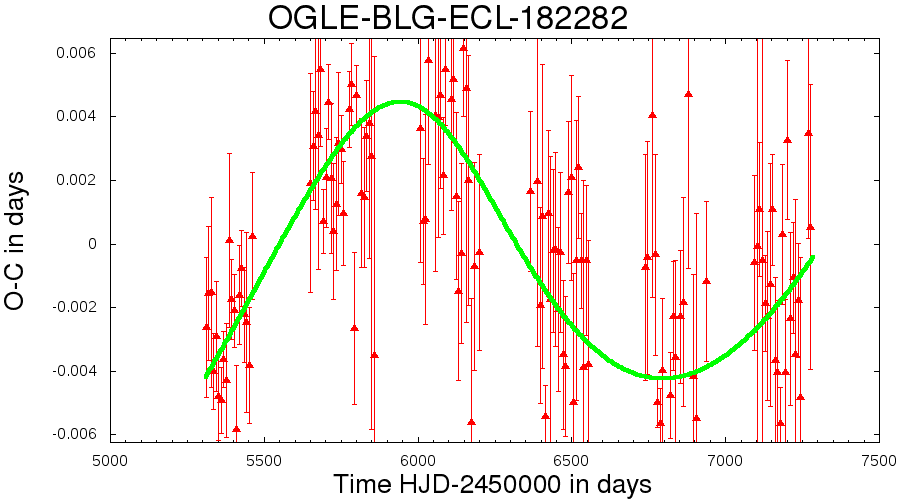}
\includegraphics[width=0.64\columnwidth]{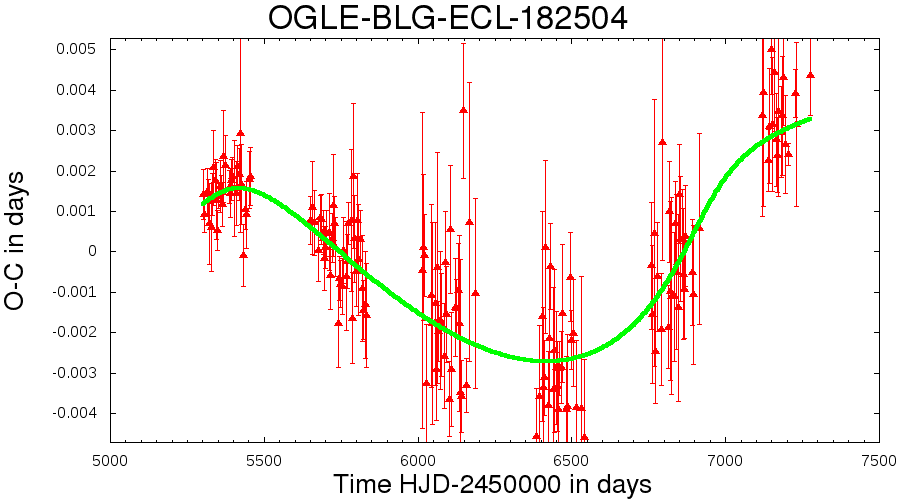}

\includegraphics[width=0.64\columnwidth]{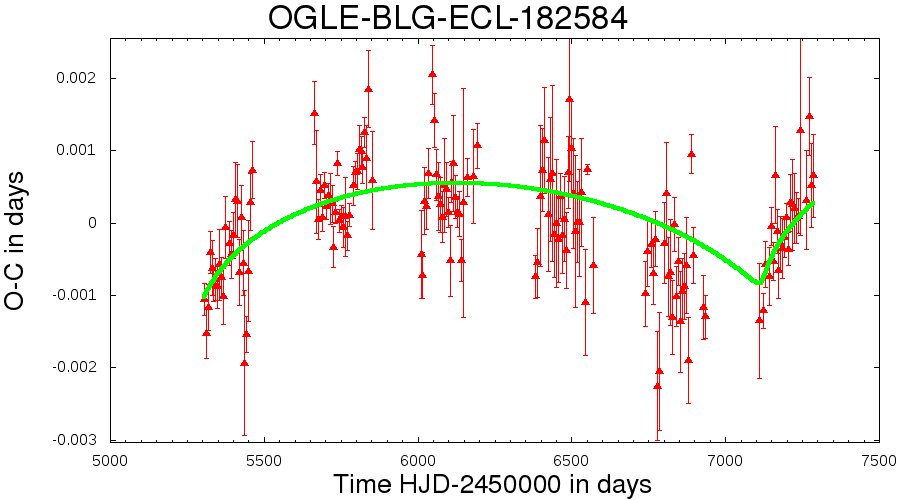}
\includegraphics[width=0.64\columnwidth]{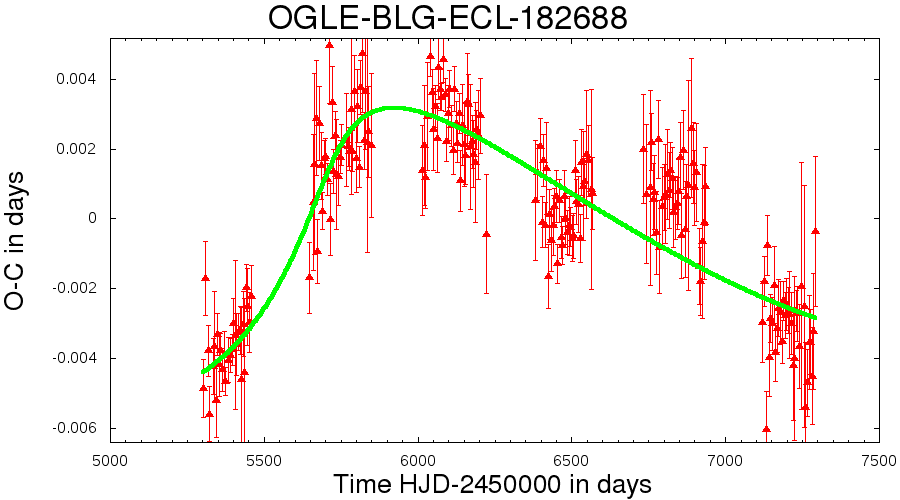}
\includegraphics[width=0.64\columnwidth]{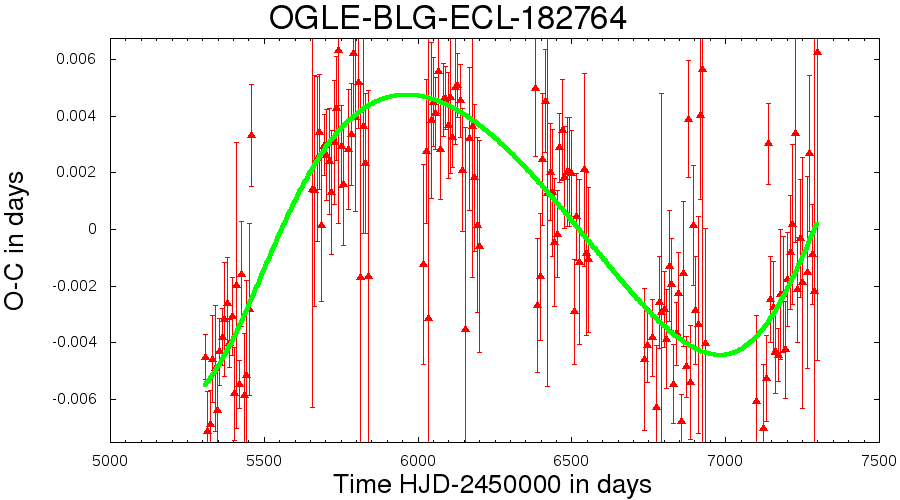}

\includegraphics[width=0.64\columnwidth]{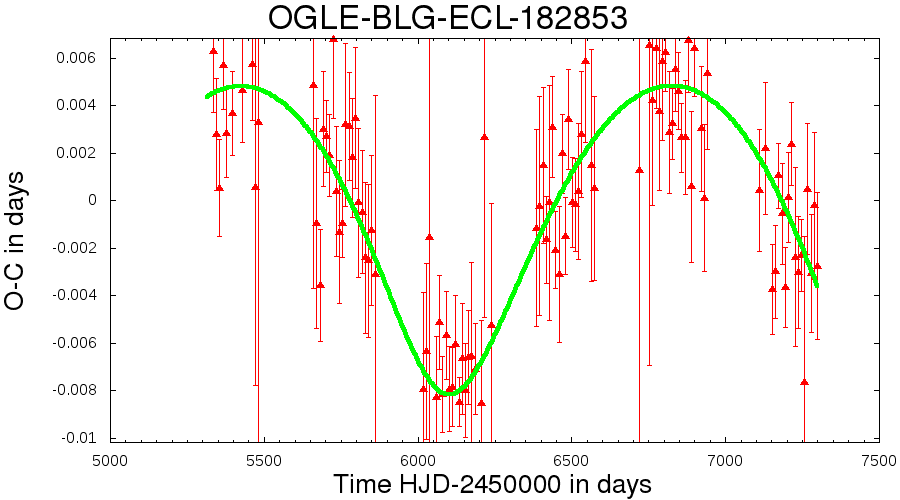}
\includegraphics[width=0.64\columnwidth]{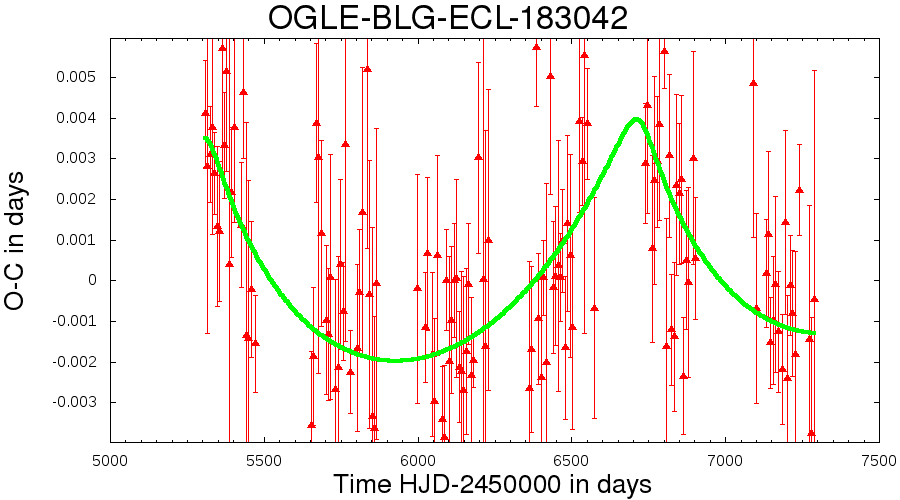}
\includegraphics[width=0.64\columnwidth]{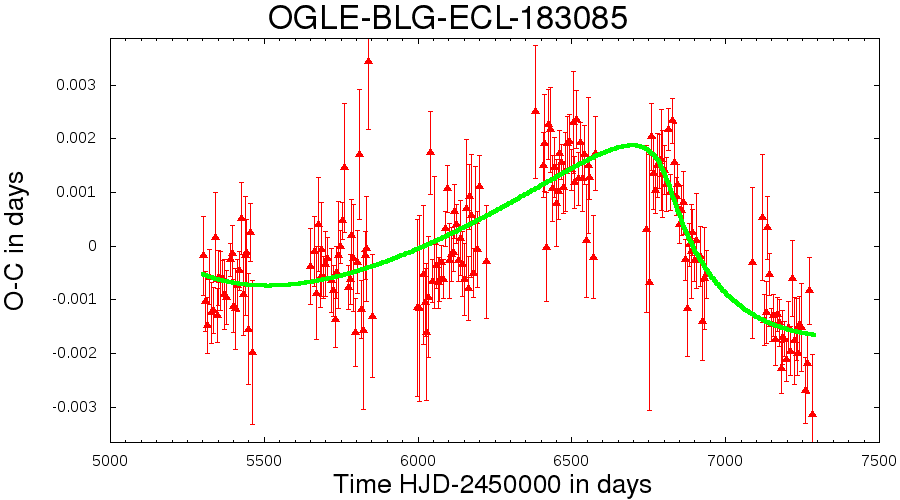}

\includegraphics[width=0.64\columnwidth]{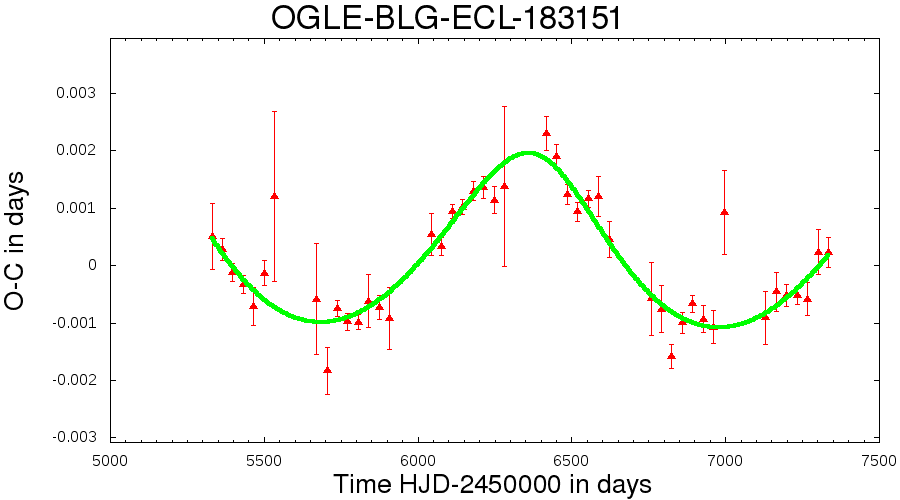}
\includegraphics[width=0.64\columnwidth]{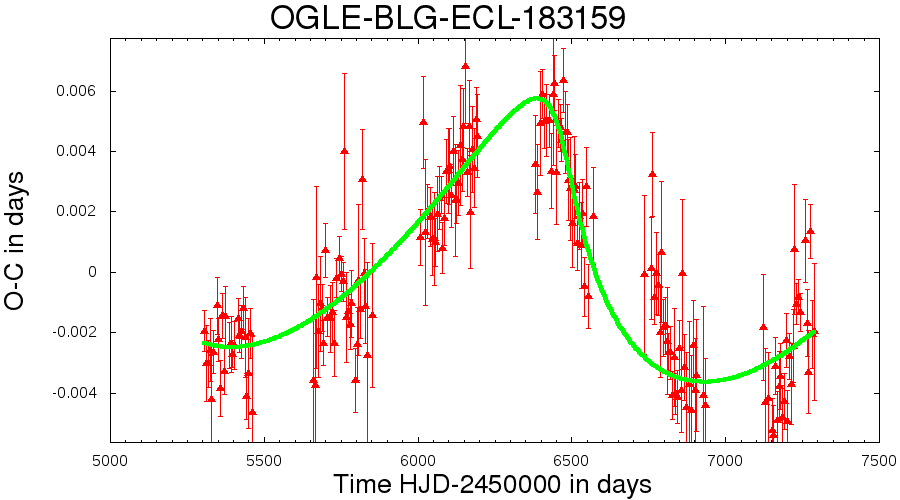}
\includegraphics[width=0.64\columnwidth]{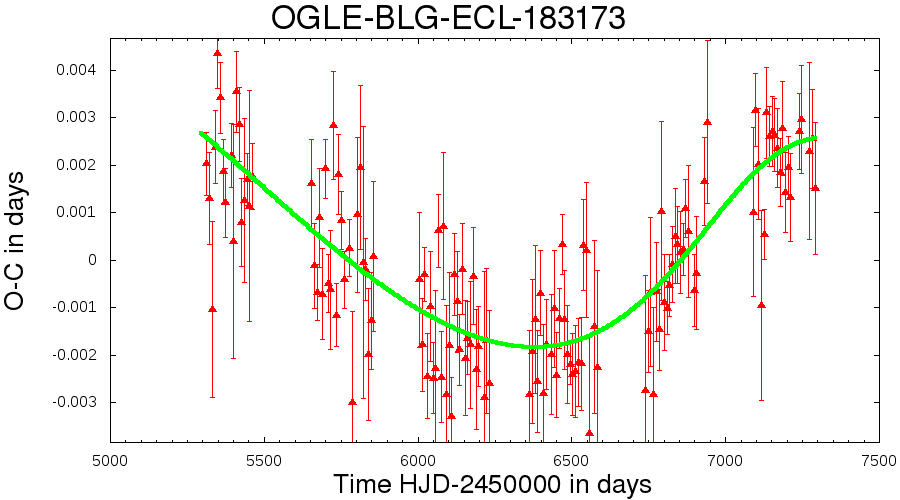}

\includegraphics[width=0.64\columnwidth]{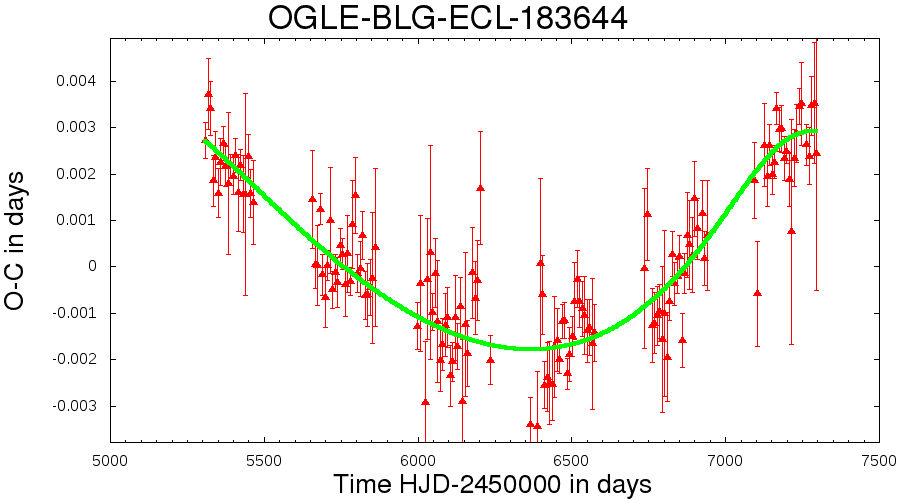}
\includegraphics[width=0.64\columnwidth]{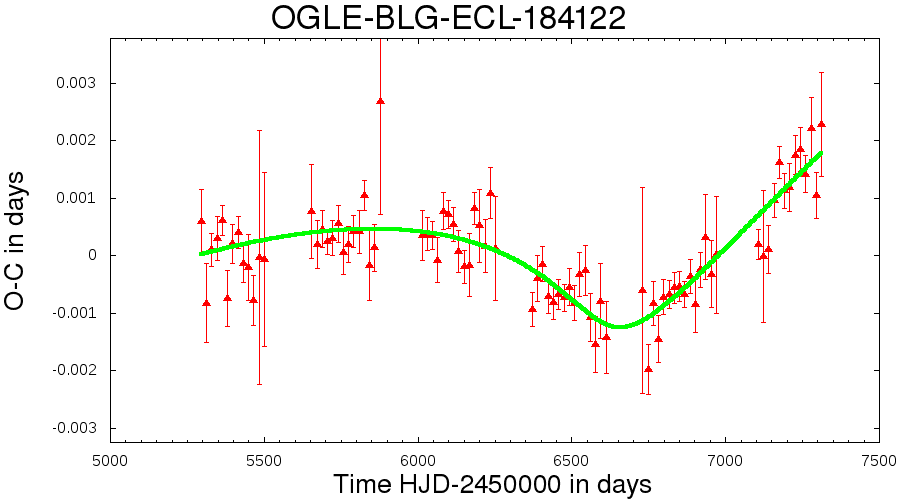}
\includegraphics[width=0.64\columnwidth]{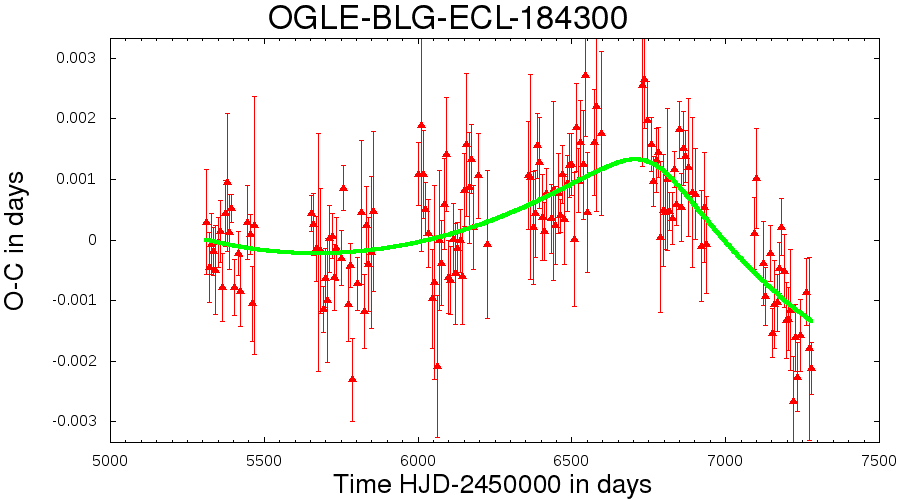}

\includegraphics[width=0.64\columnwidth]{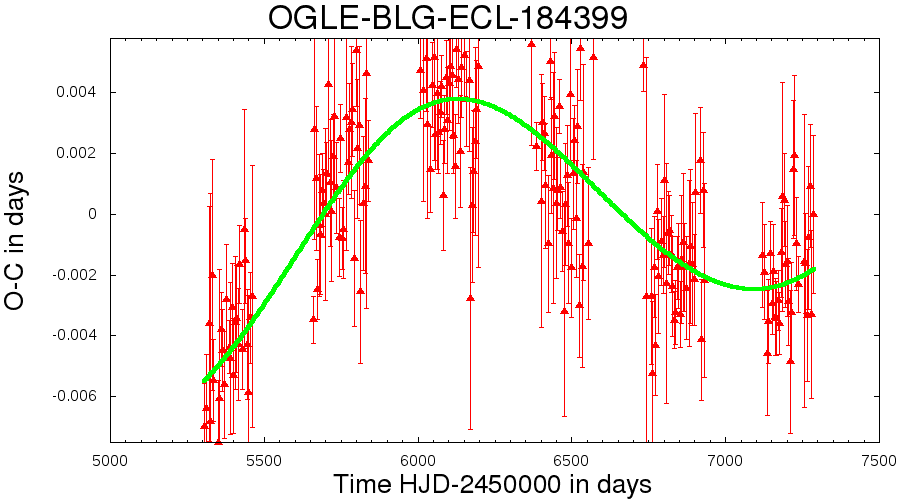}
\includegraphics[width=0.64\columnwidth]{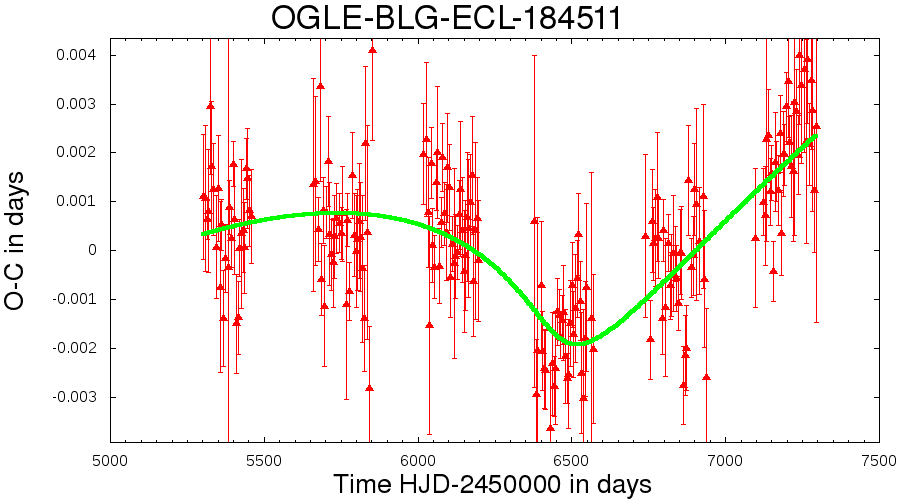}
\includegraphics[width=0.64\columnwidth]{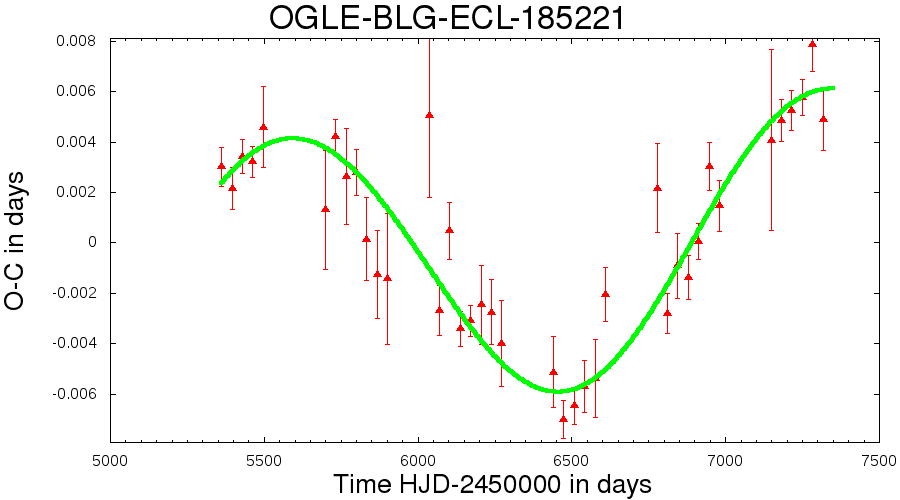}

\includegraphics[width=0.64\columnwidth]{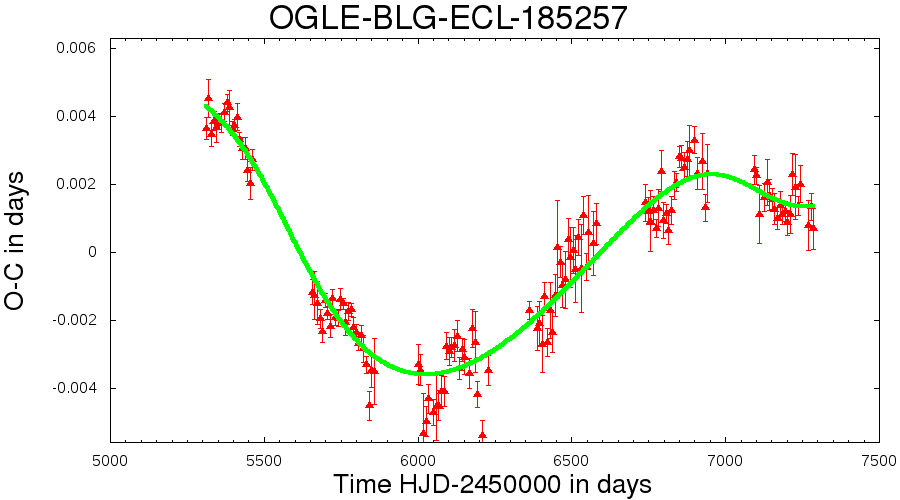}
\includegraphics[width=0.64\columnwidth]{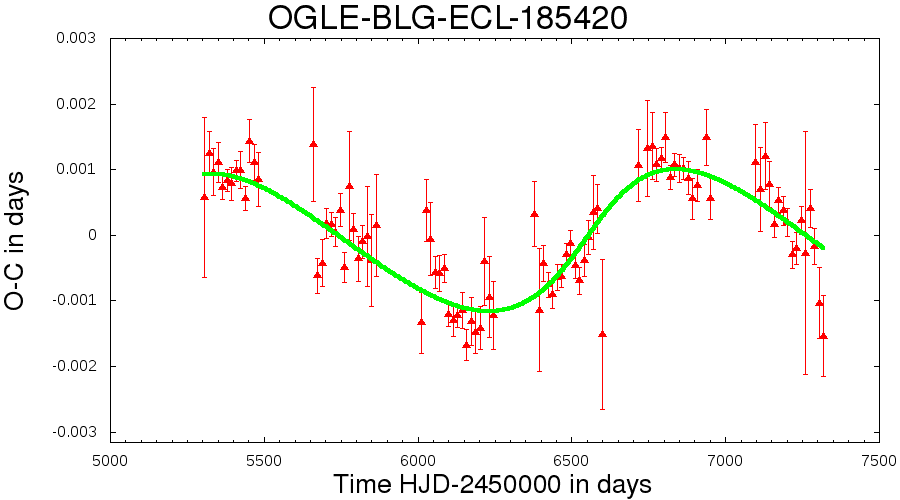}
\includegraphics[width=0.64\columnwidth]{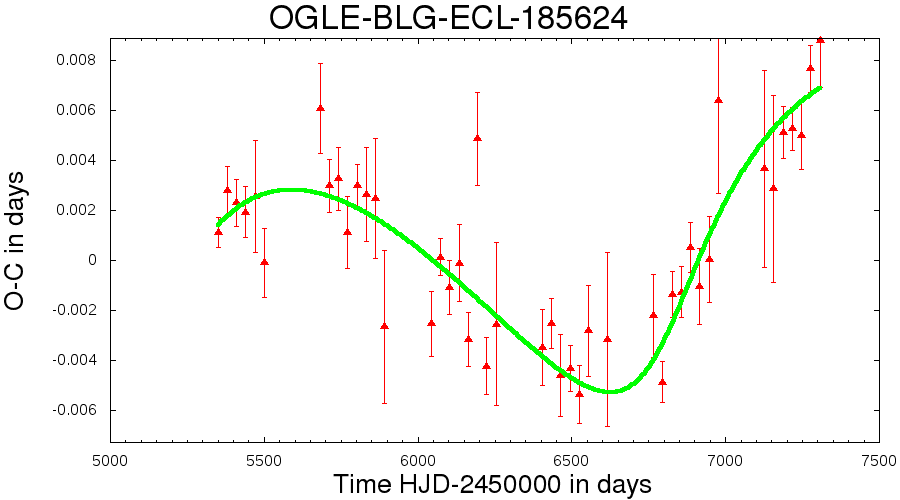}

\end{figure*}
\clearpage

\begin{figure*}
\includegraphics[width=0.64\columnwidth]{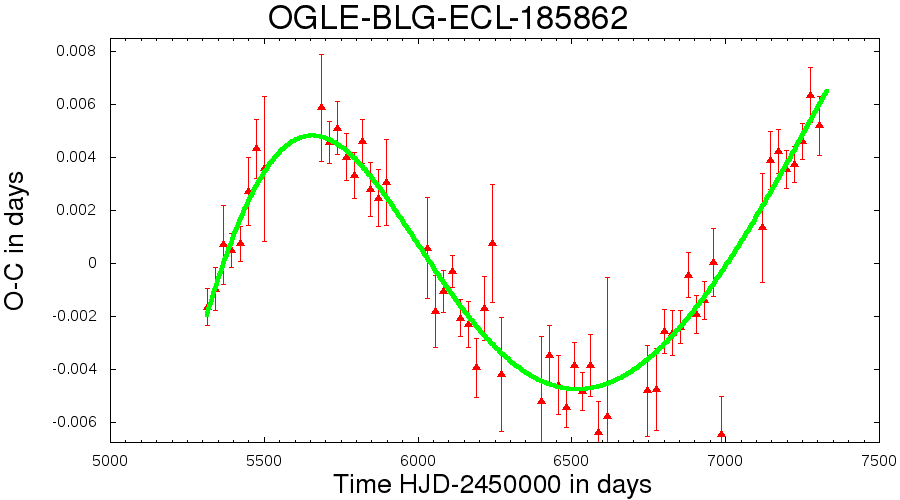}
\includegraphics[width=0.64\columnwidth]{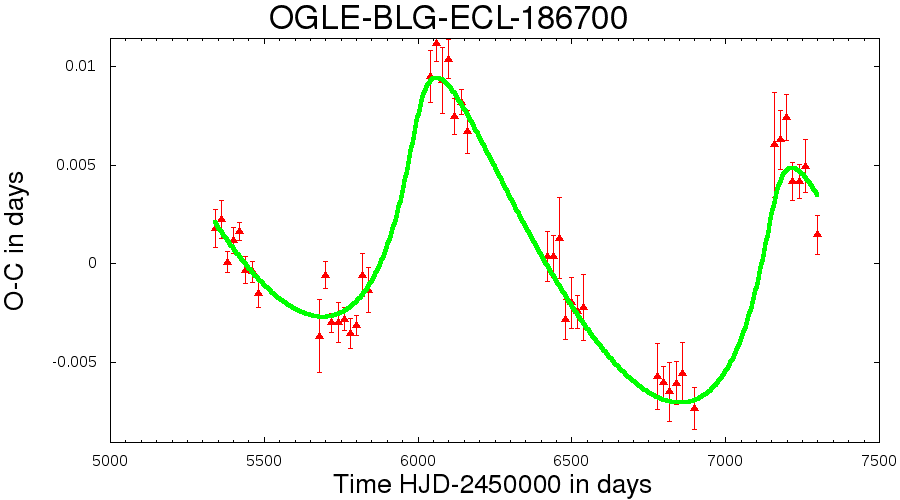}
\includegraphics[width=0.64\columnwidth]{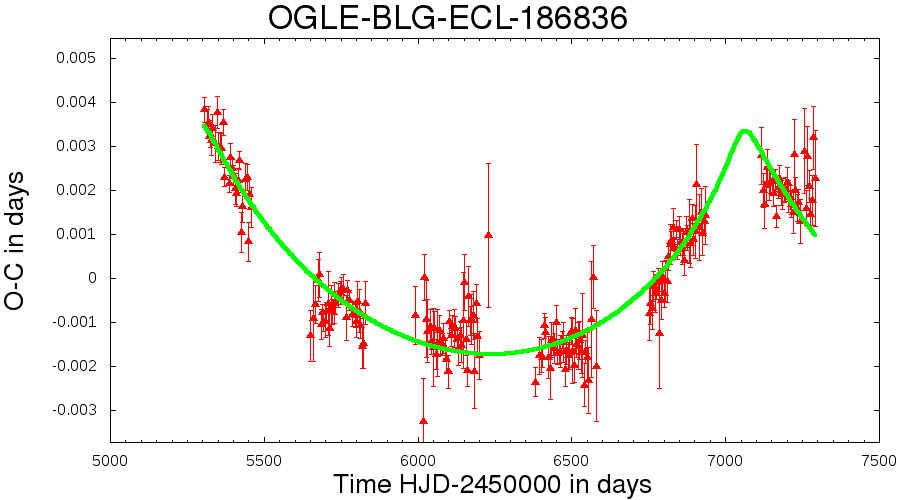}

\includegraphics[width=0.64\columnwidth]{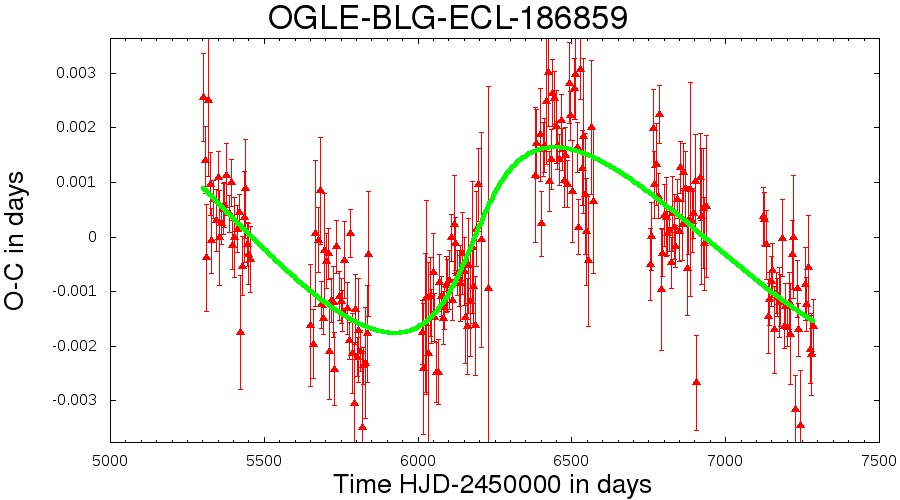}
\includegraphics[width=0.64\columnwidth]{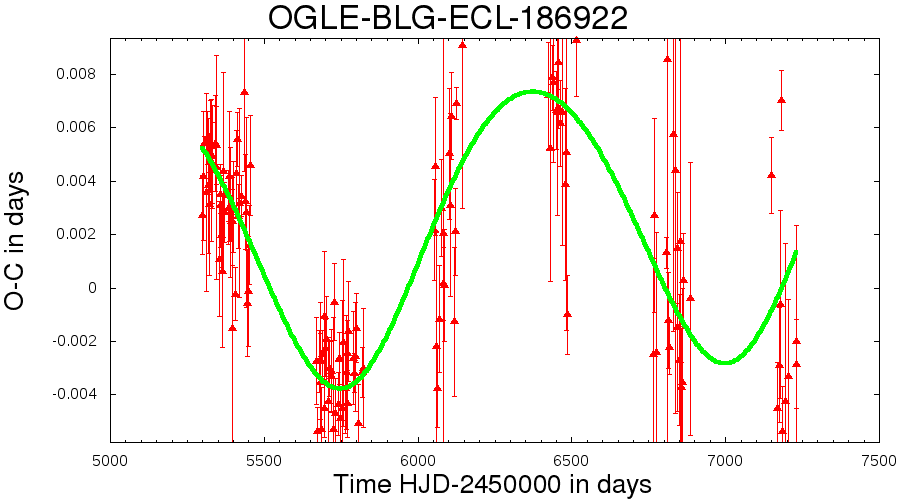}
\includegraphics[width=0.64\columnwidth]{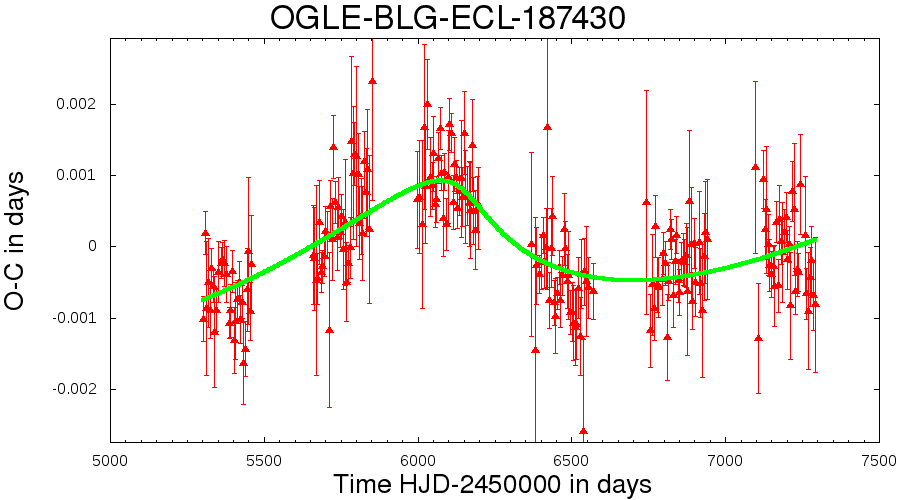}

\includegraphics[width=0.64\columnwidth]{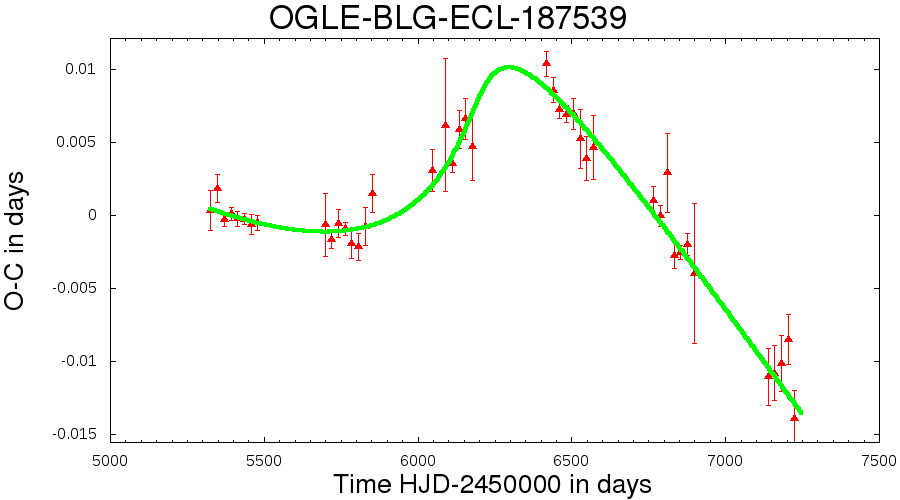}
\includegraphics[width=0.64\columnwidth]{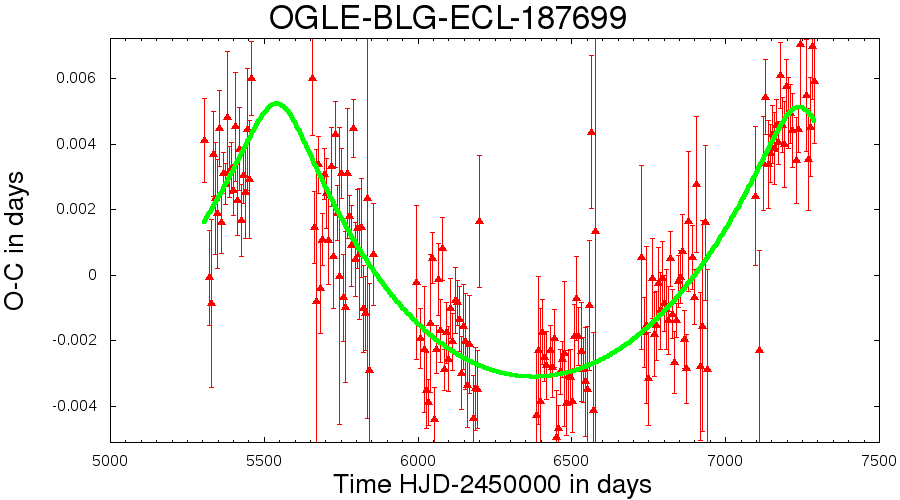}
\includegraphics[width=0.64\columnwidth]{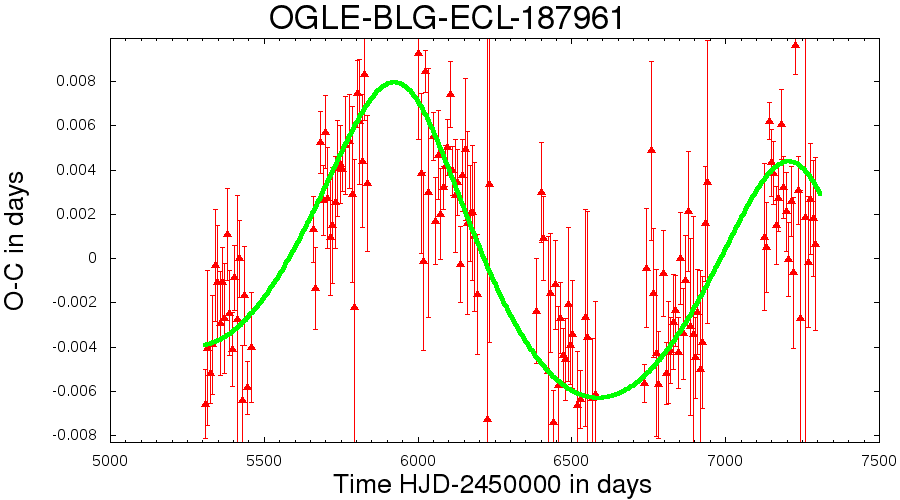}

\includegraphics[width=0.64\columnwidth]{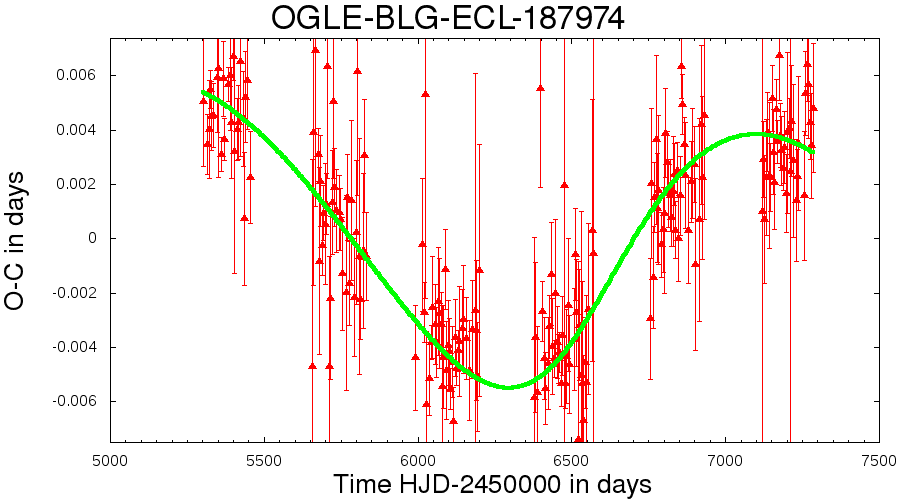}
\includegraphics[width=0.64\columnwidth]{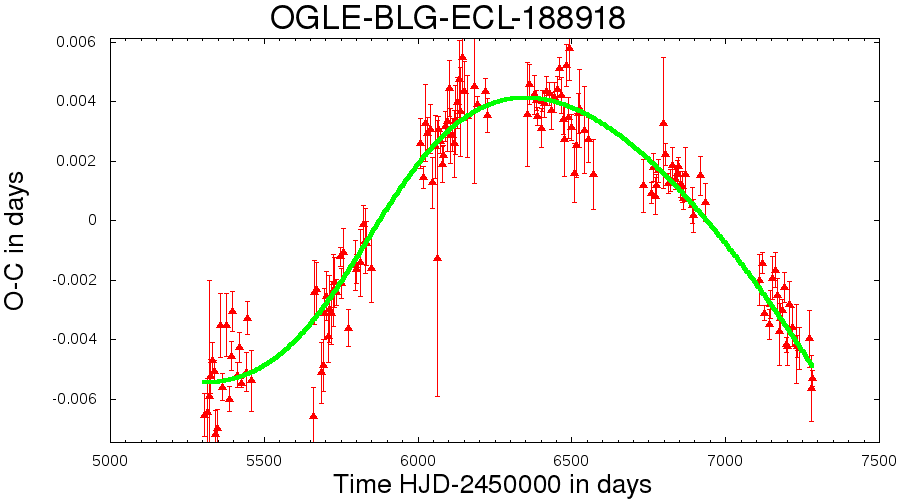}
\includegraphics[width=0.64\columnwidth]{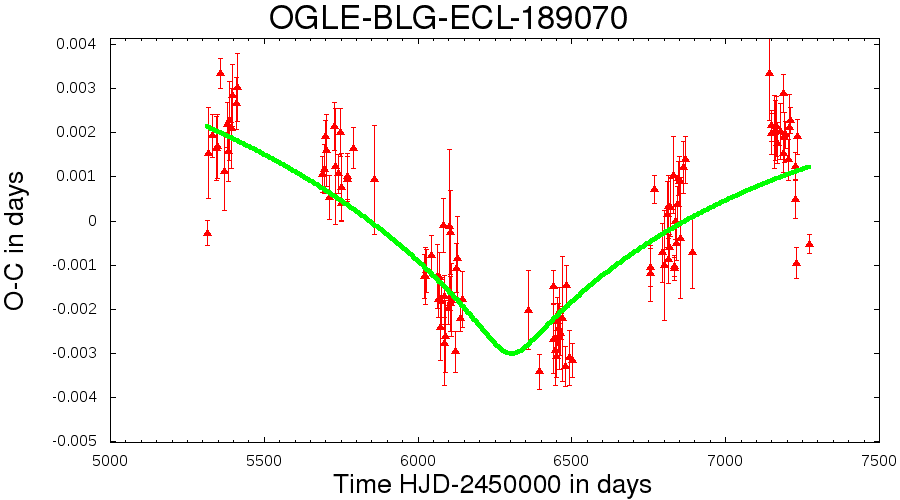}

\includegraphics[width=0.64\columnwidth]{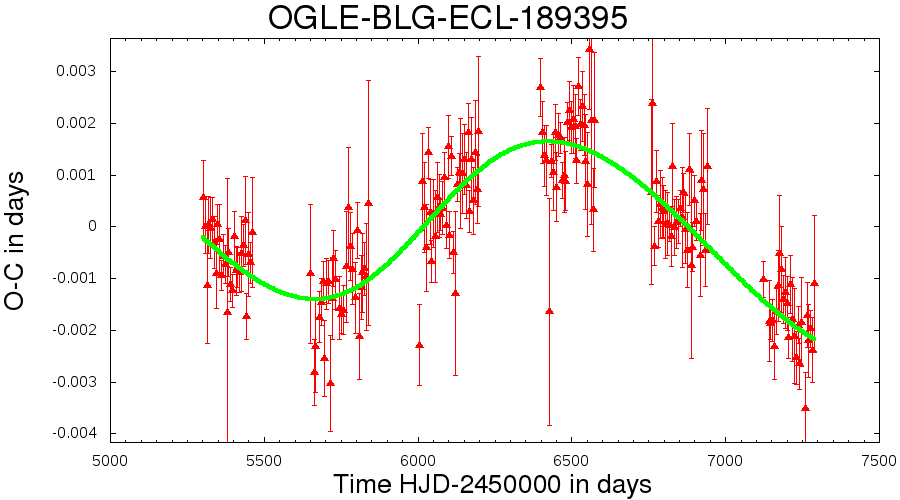}
\includegraphics[width=0.64\columnwidth]{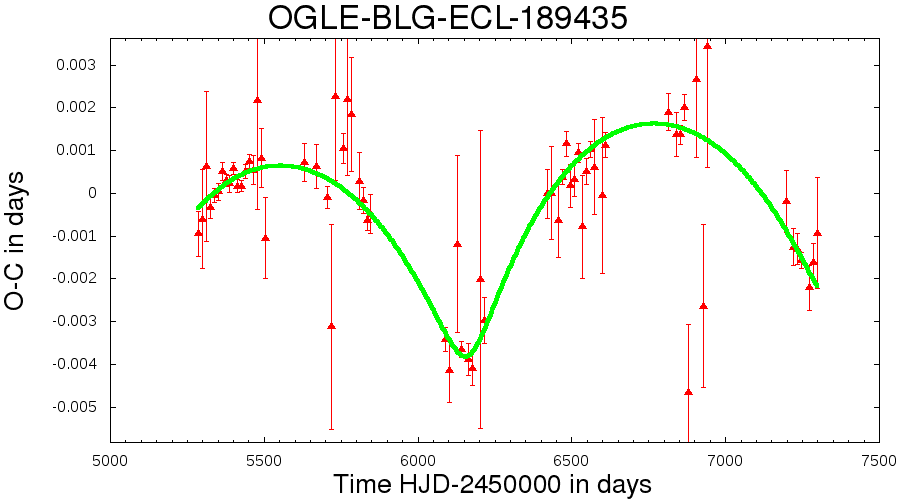}
\includegraphics[width=0.64\columnwidth]{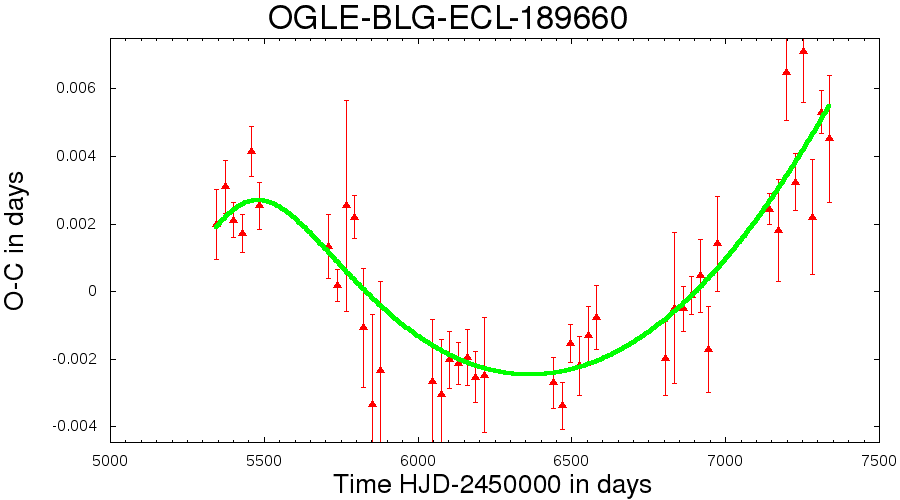}

\includegraphics[width=0.64\columnwidth]{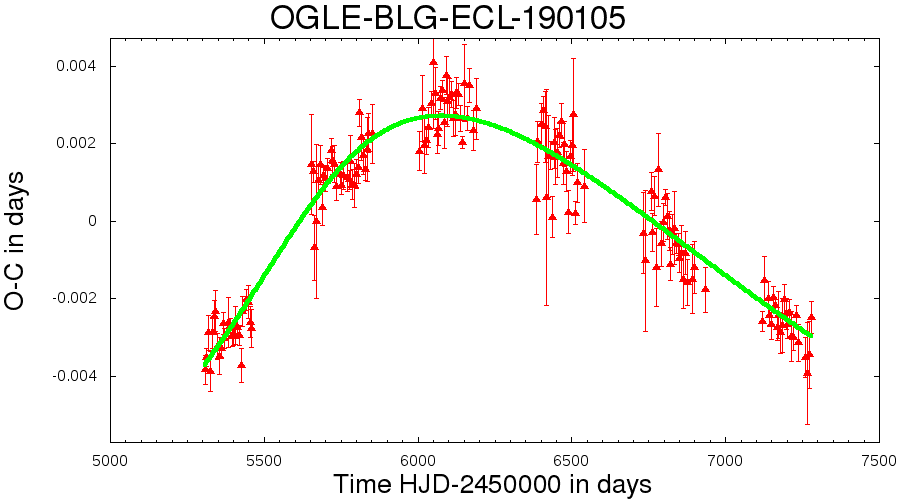}
\includegraphics[width=0.64\columnwidth]{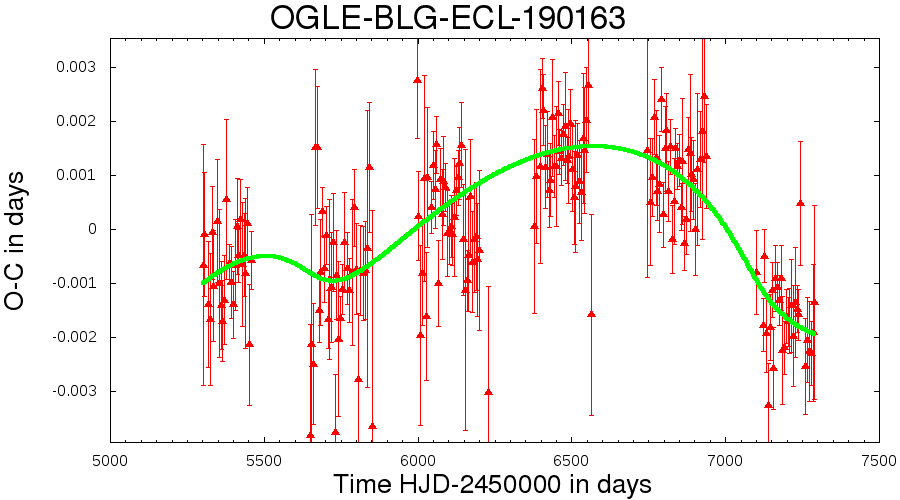}
\includegraphics[width=0.64\columnwidth]{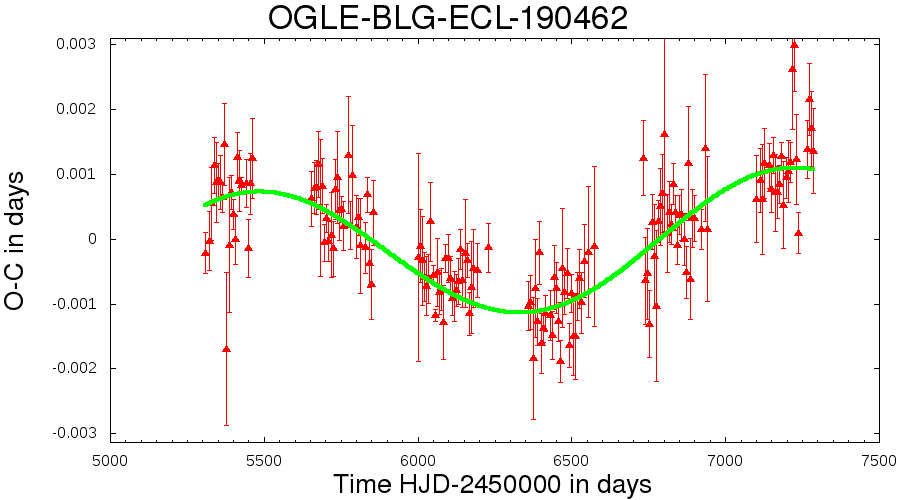}

\includegraphics[width=0.64\columnwidth]{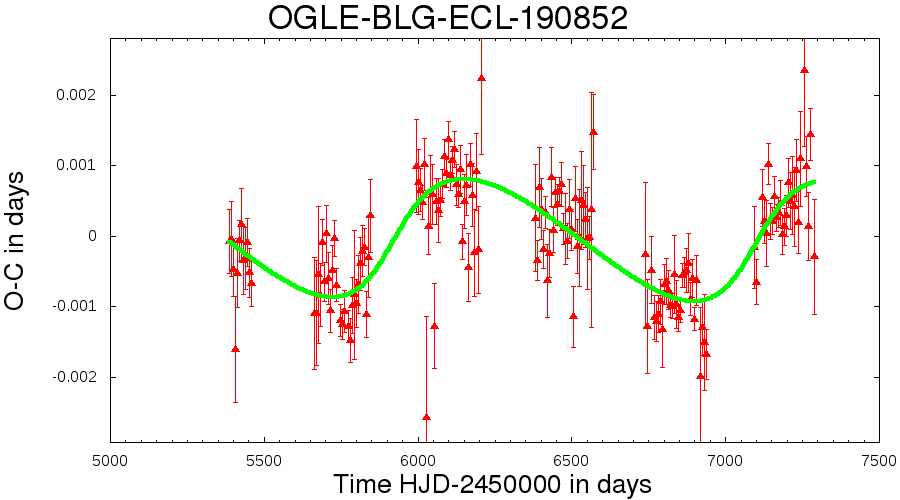}
\includegraphics[width=0.64\columnwidth]{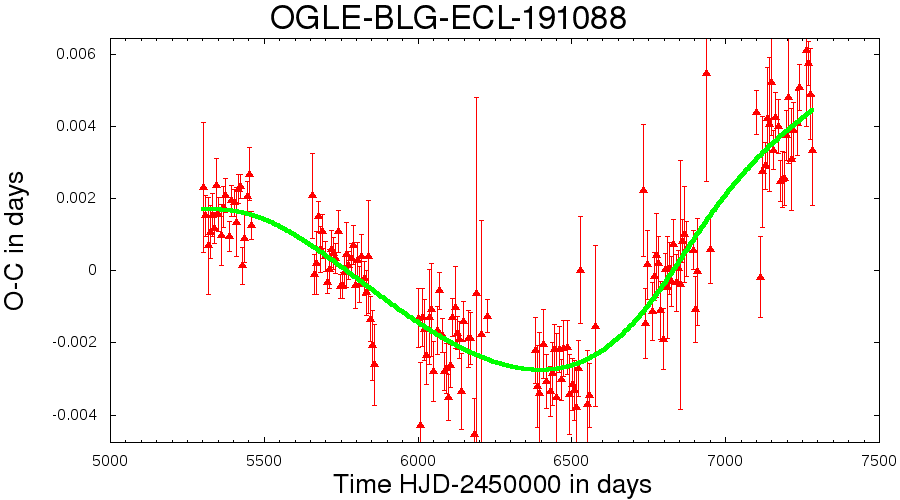}
\includegraphics[width=0.64\columnwidth]{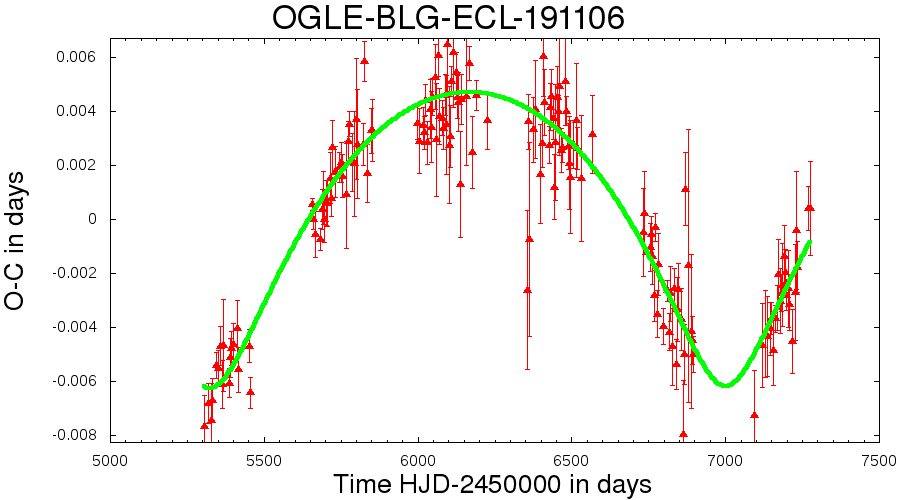}

\includegraphics[width=0.64\columnwidth]{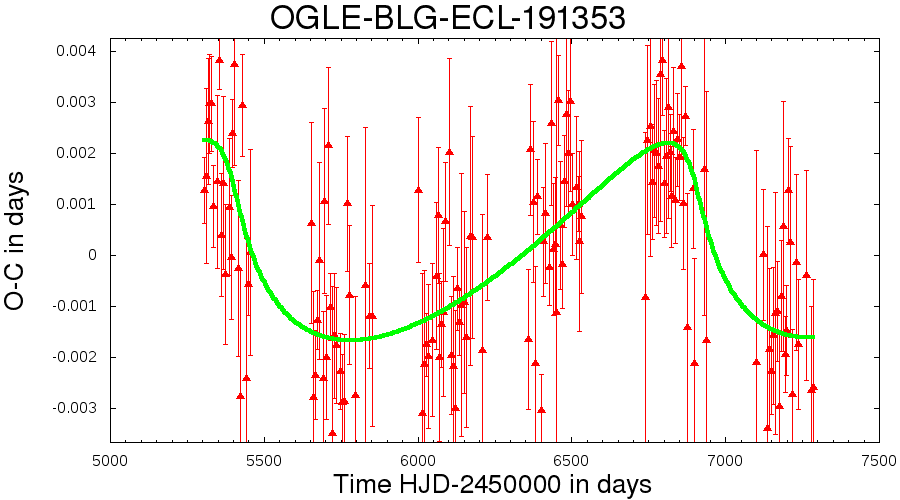}
\includegraphics[width=0.64\columnwidth]{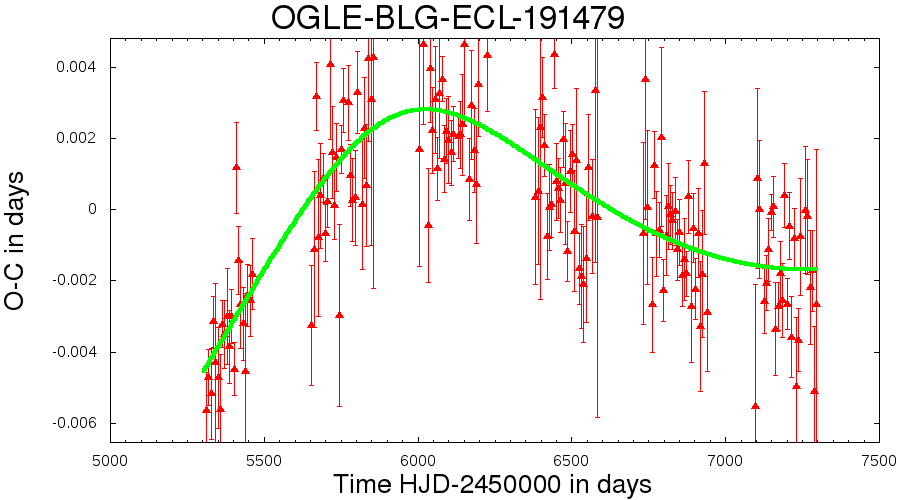}
\includegraphics[width=0.64\columnwidth]{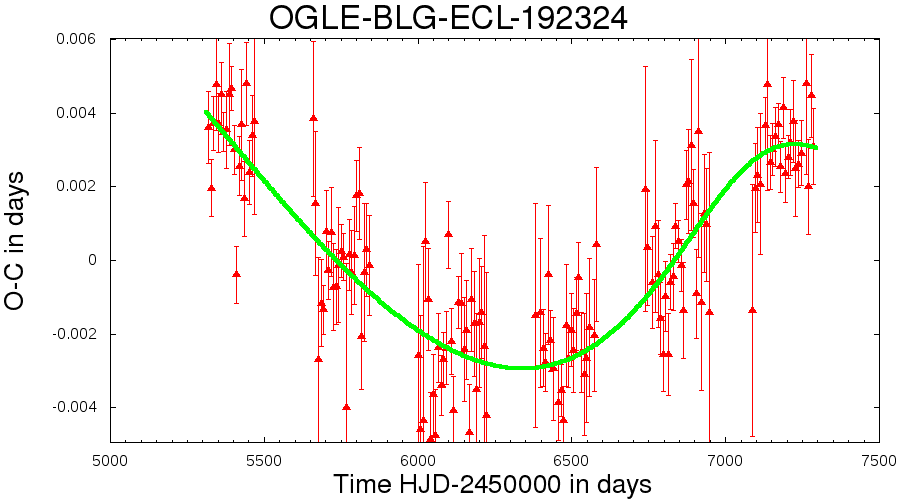}

\end{figure*}
\clearpage

\begin{figure*}
\includegraphics[width=0.64\columnwidth]{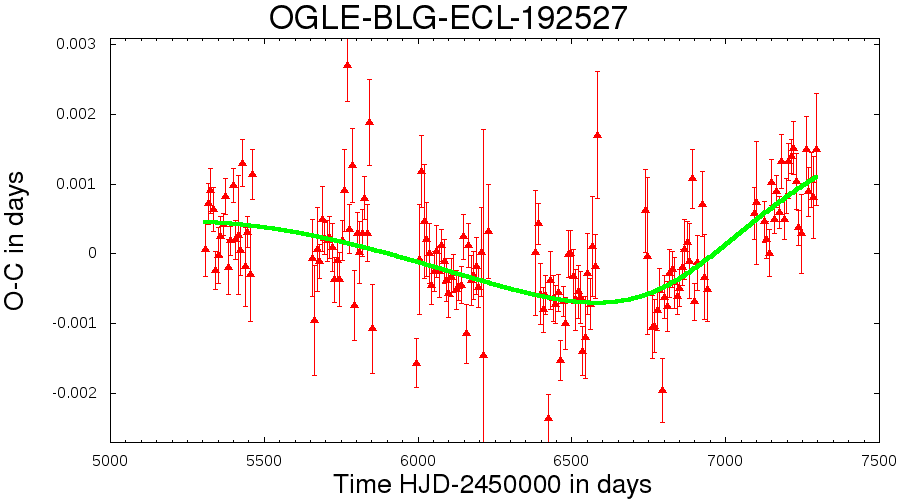}
\includegraphics[width=0.64\columnwidth]{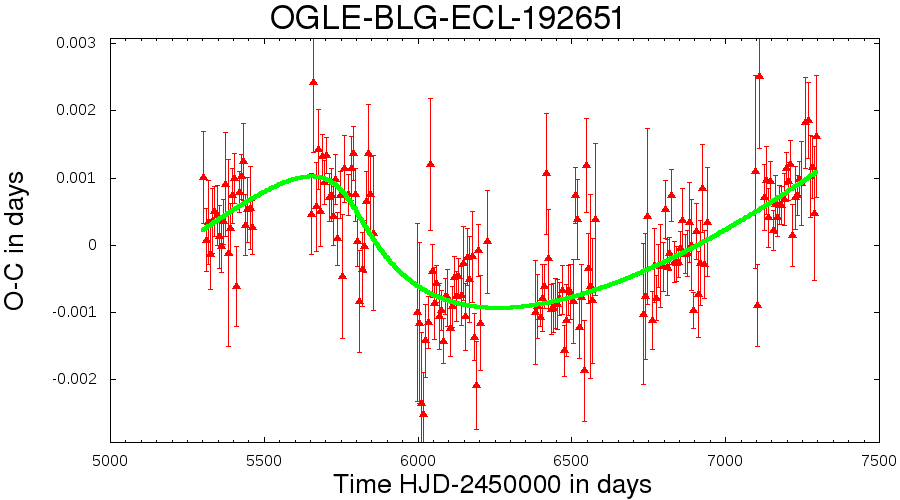}
\includegraphics[width=0.64\columnwidth]{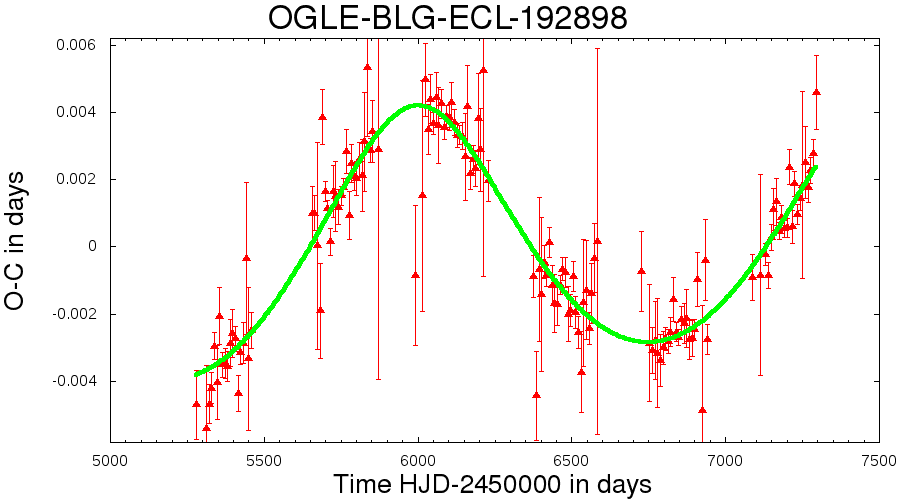}

\includegraphics[width=0.64\columnwidth]{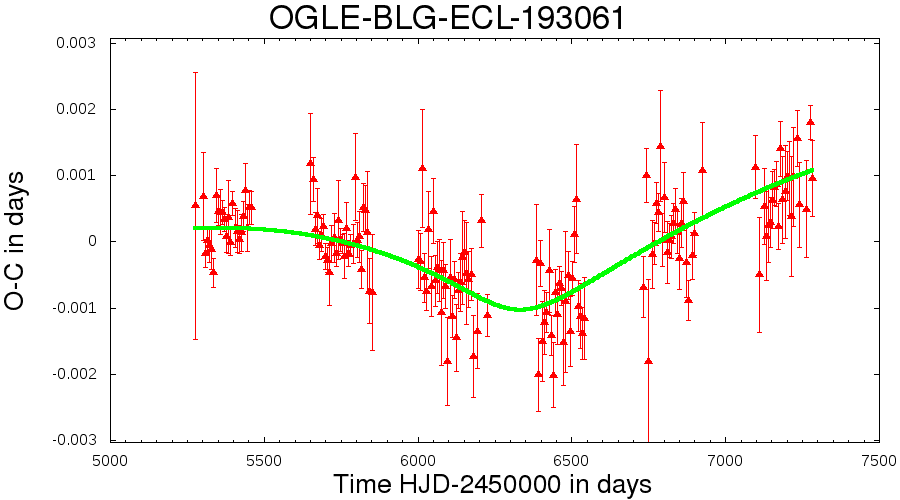}
\includegraphics[width=0.64\columnwidth]{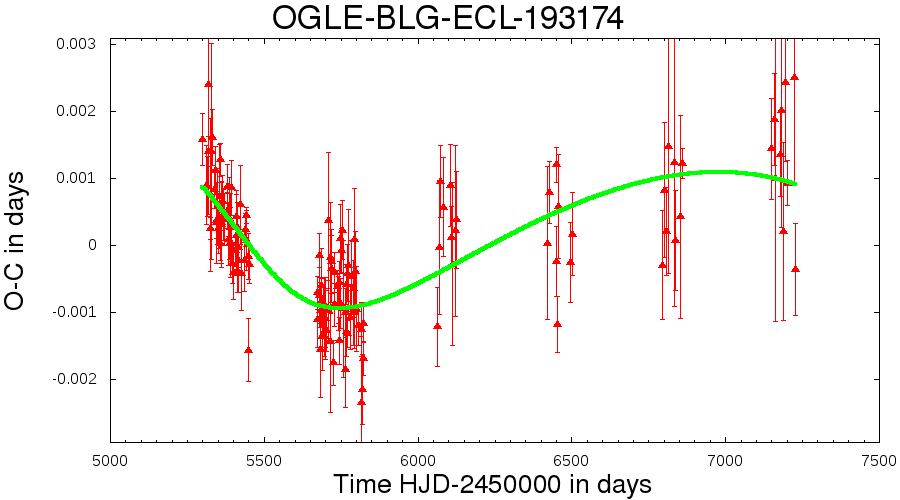}
\includegraphics[width=0.64\columnwidth]{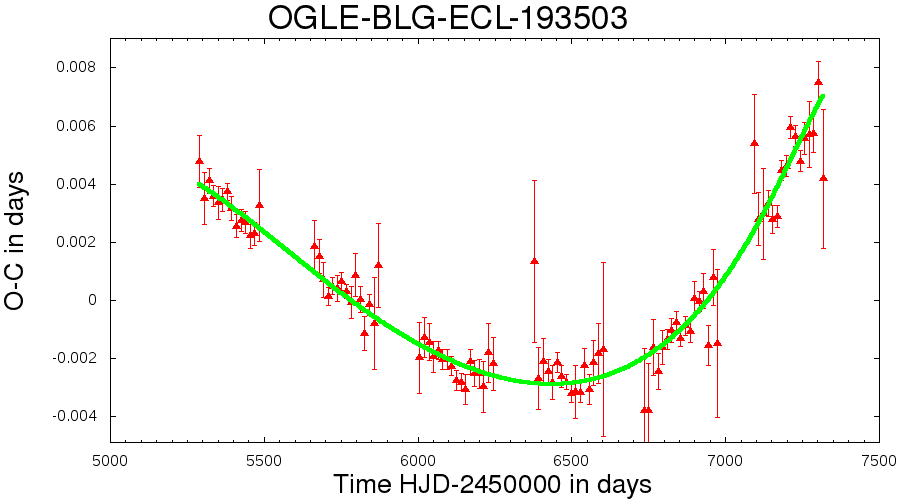}

\includegraphics[width=0.64\columnwidth]{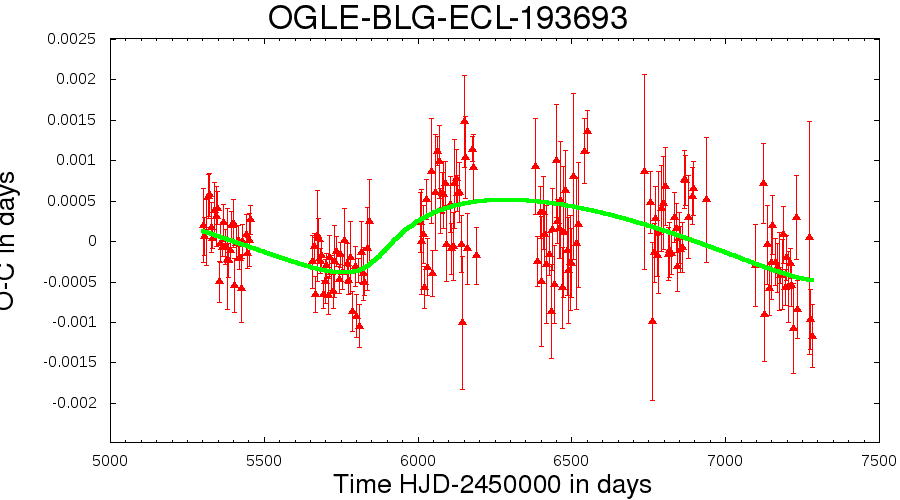}
\includegraphics[width=0.64\columnwidth]{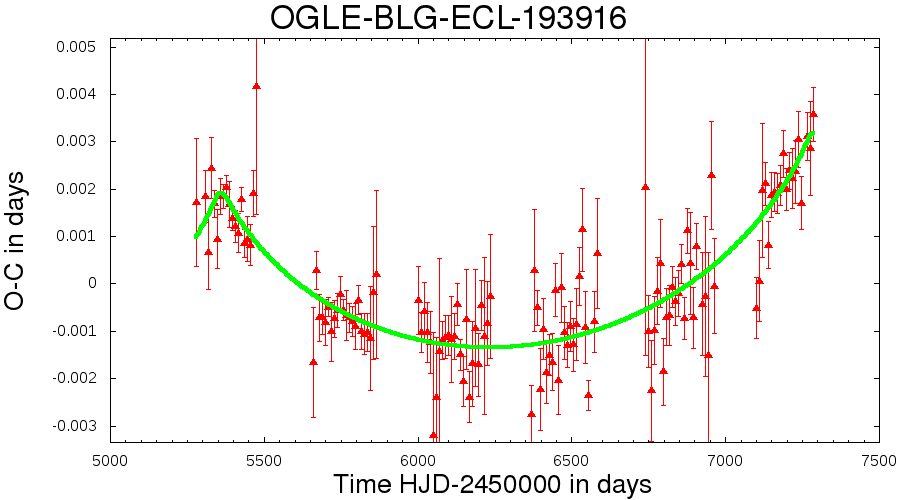}
\includegraphics[width=0.64\columnwidth]{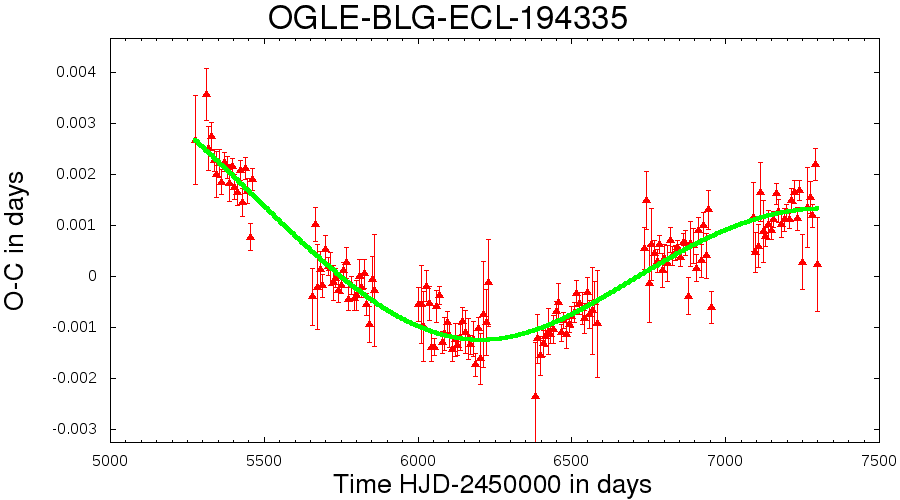}

\includegraphics[width=0.64\columnwidth]{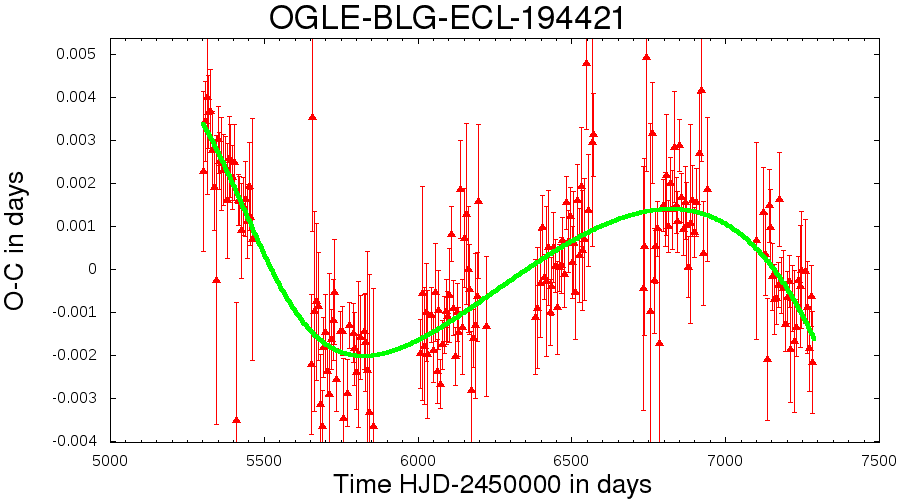}
\includegraphics[width=0.64\columnwidth]{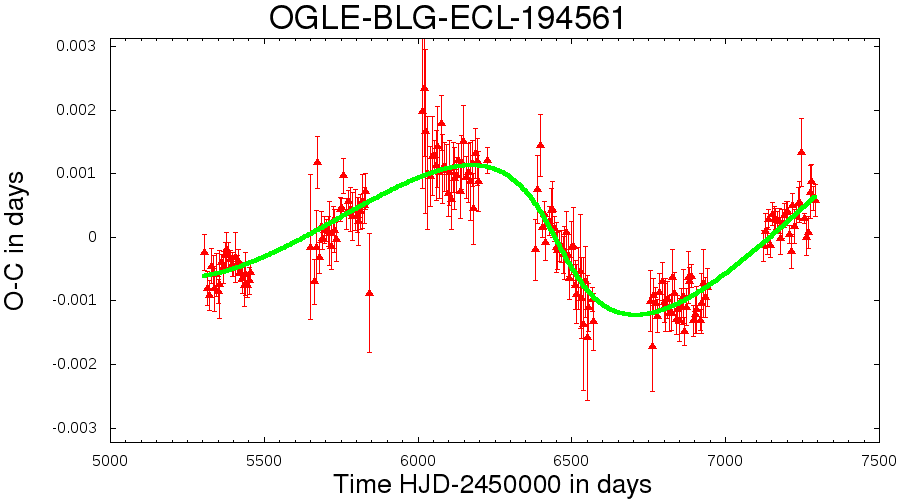}
\includegraphics[width=0.64\columnwidth]{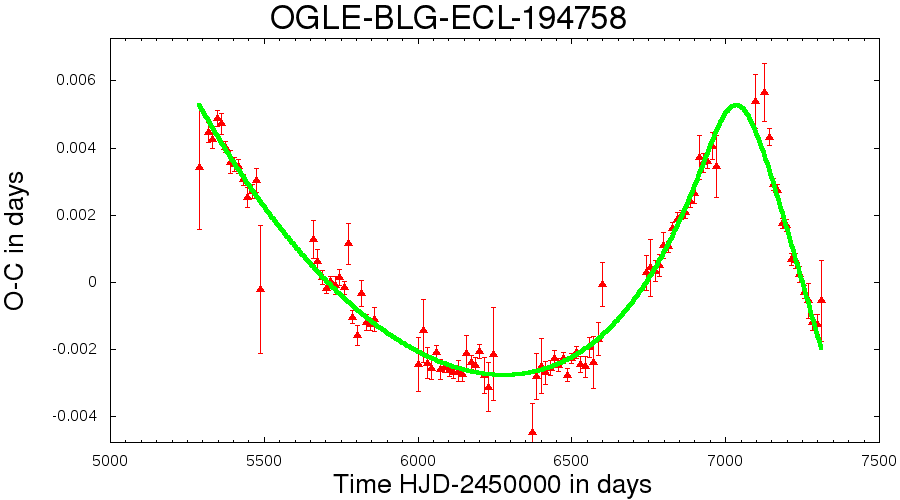}

\includegraphics[width=0.64\columnwidth]{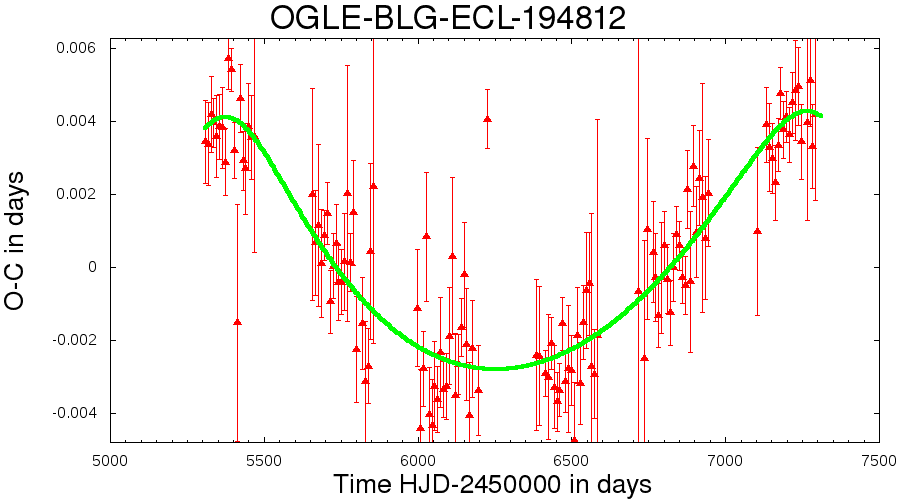}
\includegraphics[width=0.64\columnwidth]{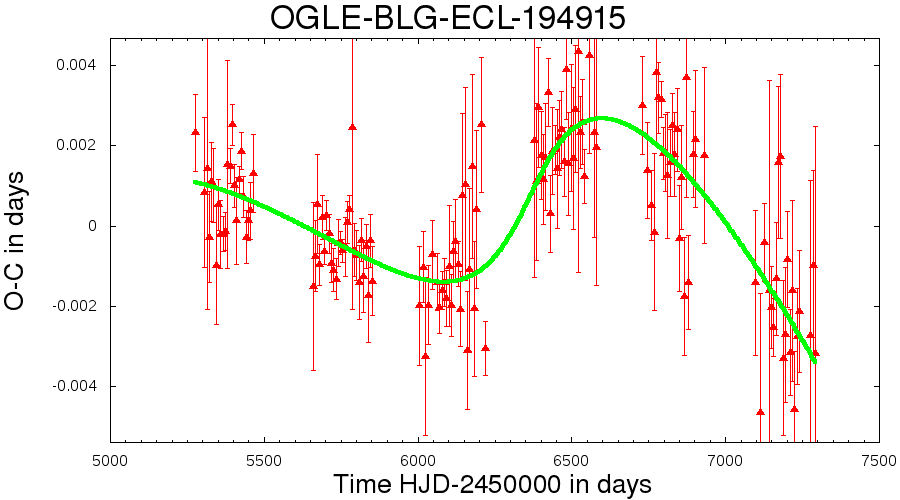}
\includegraphics[width=0.64\columnwidth]{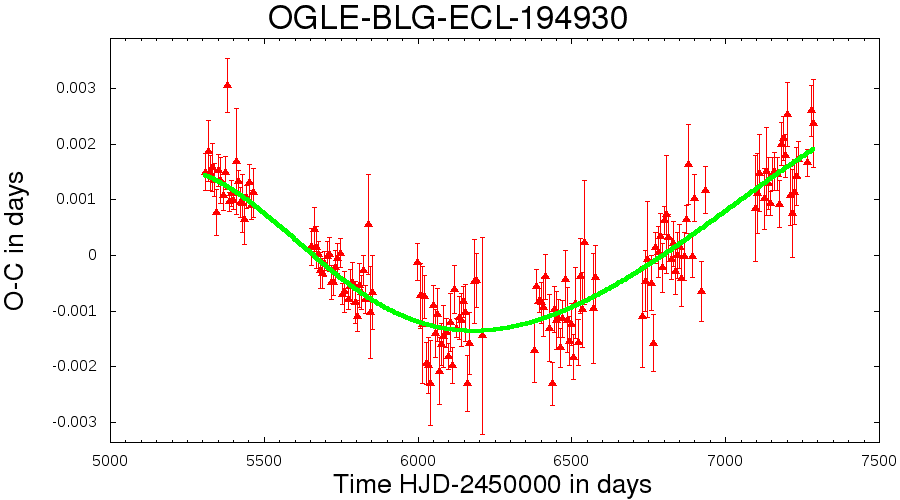}

\includegraphics[width=0.64\columnwidth]{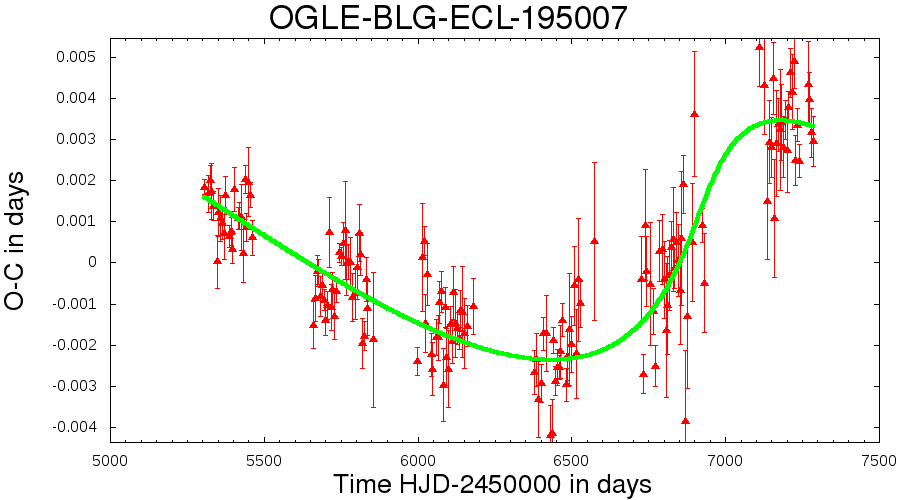}
\includegraphics[width=0.64\columnwidth]{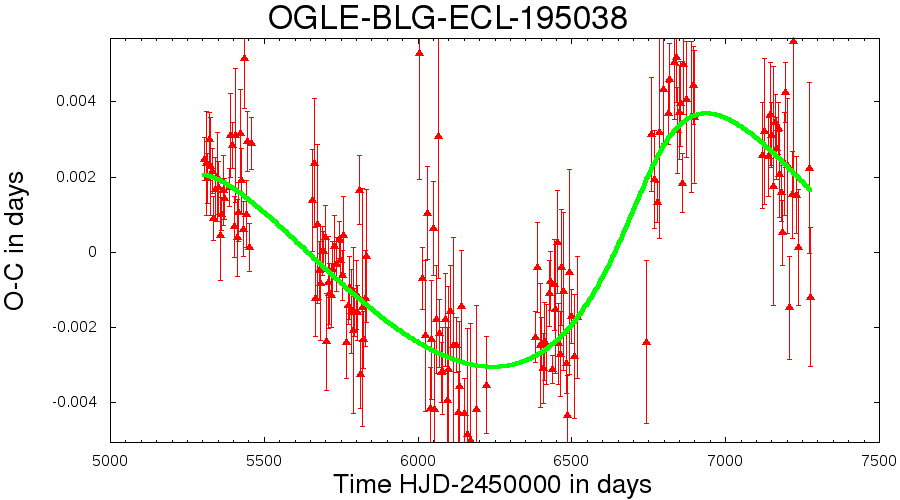}
\includegraphics[width=0.64\columnwidth]{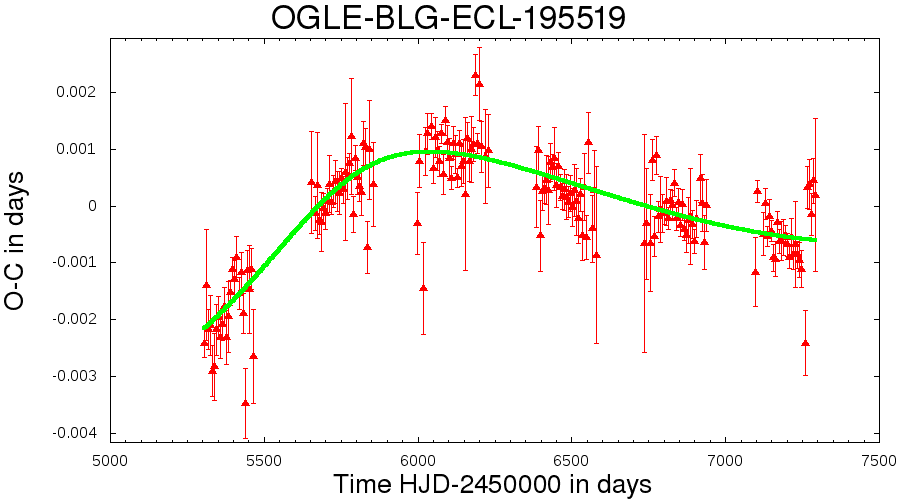}

\includegraphics[width=0.64\columnwidth]{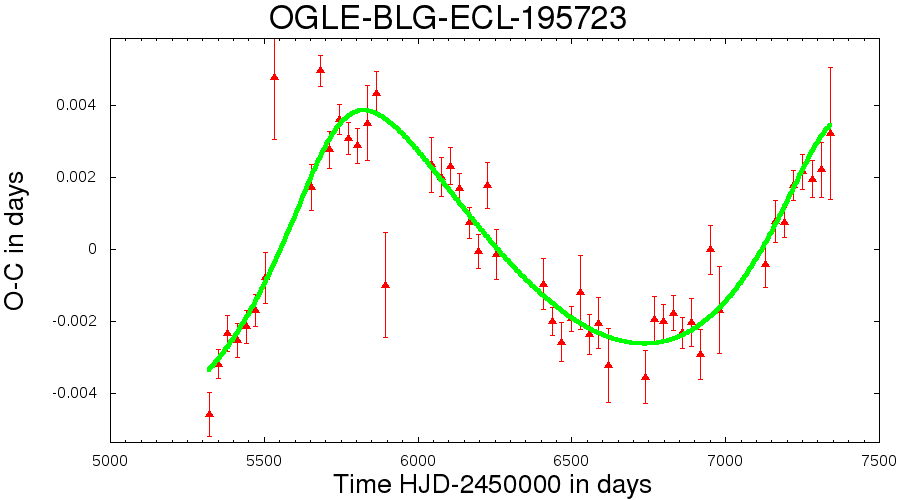}
\includegraphics[width=0.64\columnwidth]{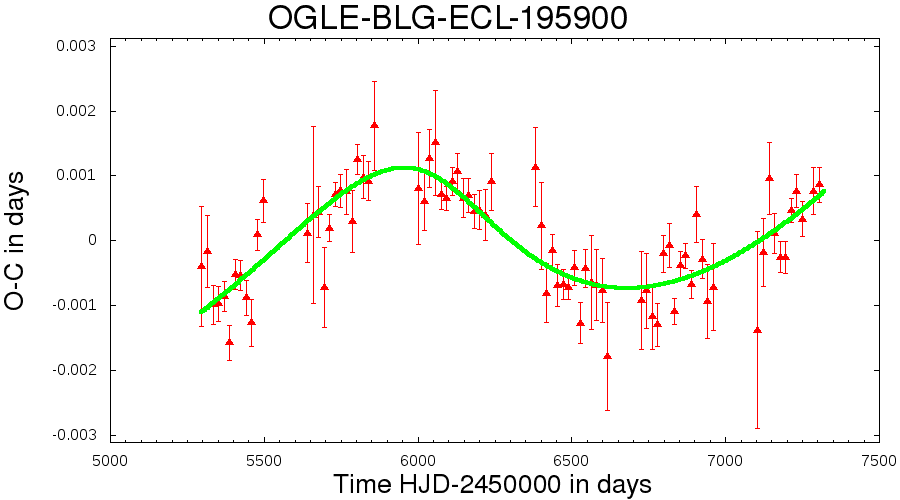}
\includegraphics[width=0.64\columnwidth]{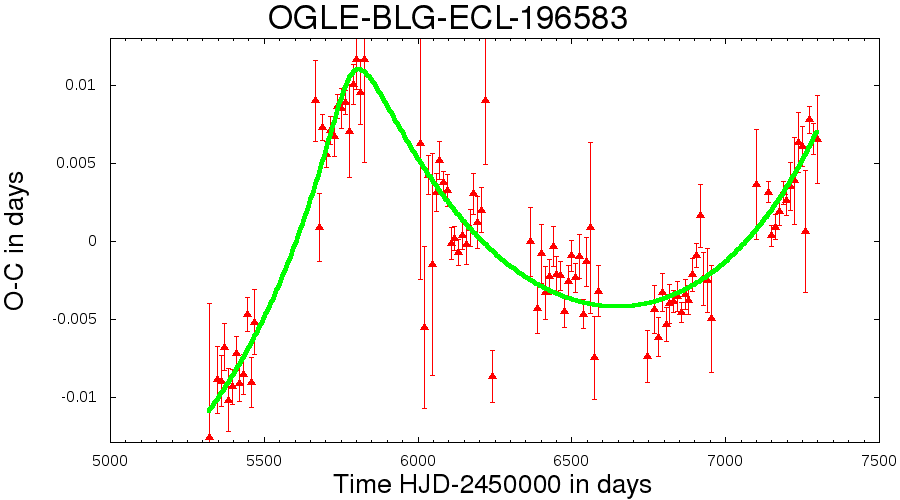}

\includegraphics[width=0.64\columnwidth]{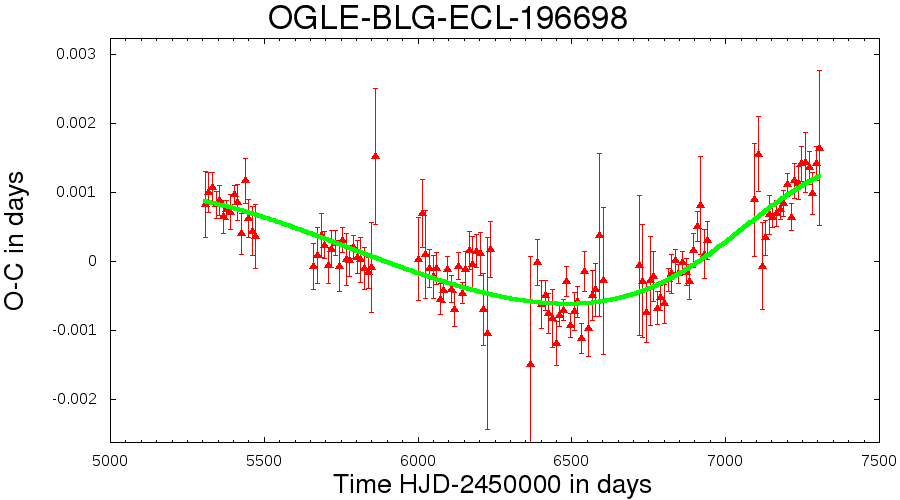}
\includegraphics[width=0.64\columnwidth]{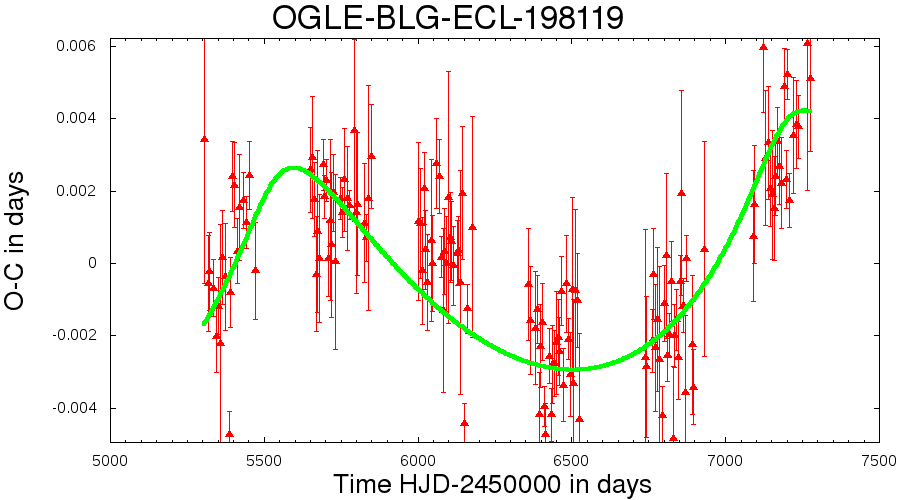}
\includegraphics[width=0.64\columnwidth]{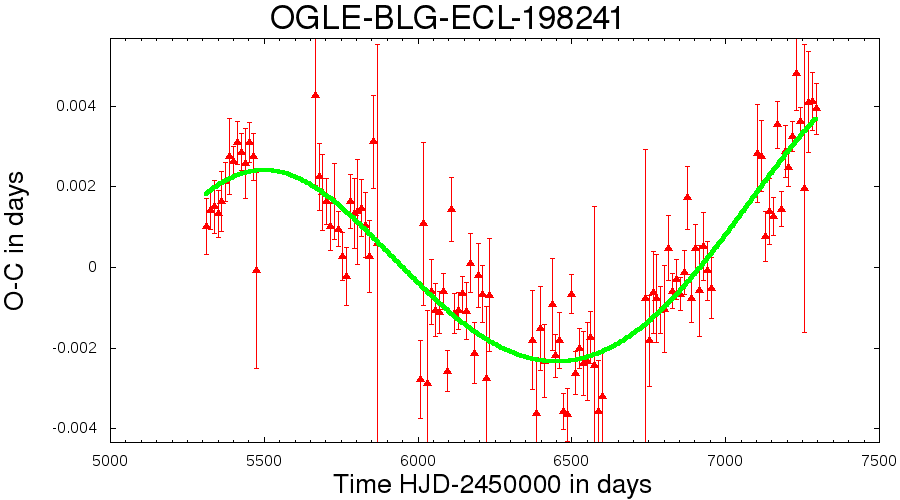}

\end{figure*}
\clearpage

\begin{figure*}
\includegraphics[width=0.64\columnwidth]{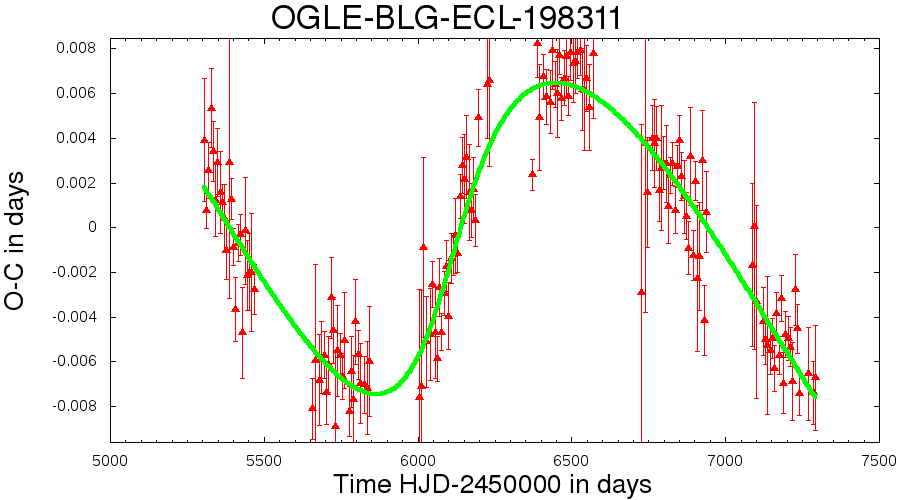}
\includegraphics[width=0.64\columnwidth]{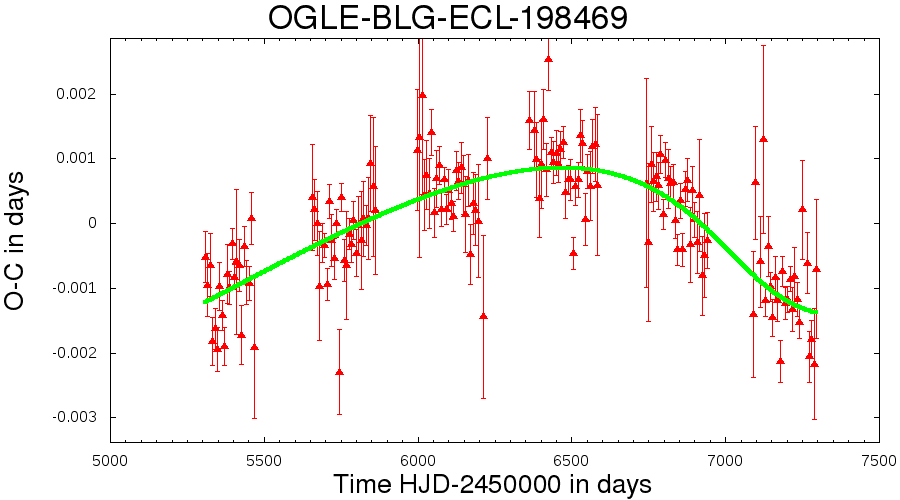}
\includegraphics[width=0.64\columnwidth]{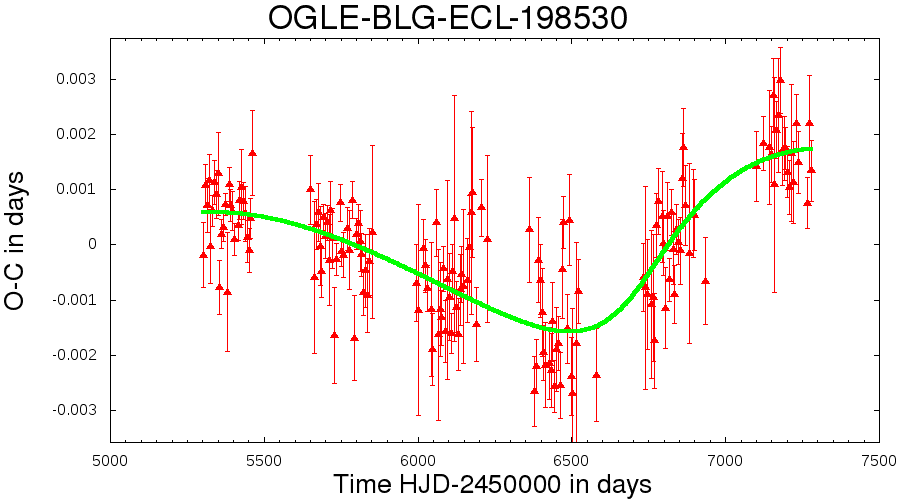}

\includegraphics[width=0.64\columnwidth]{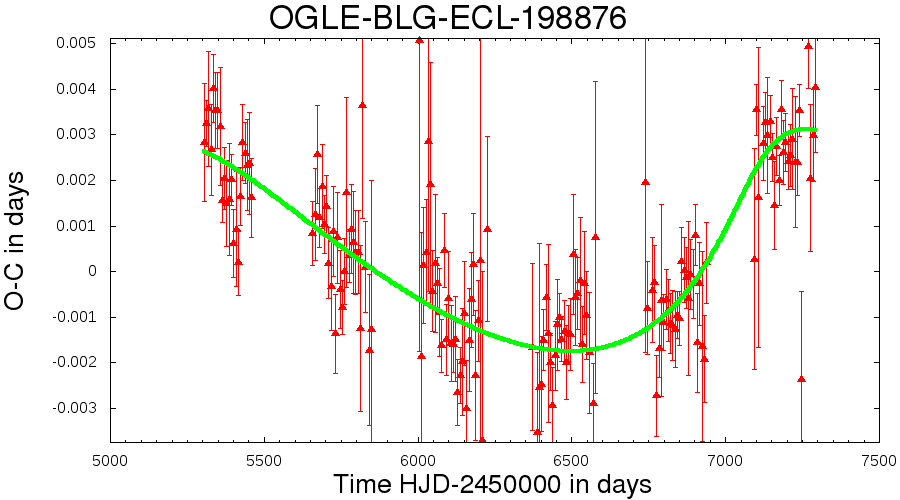}
\includegraphics[width=0.64\columnwidth]{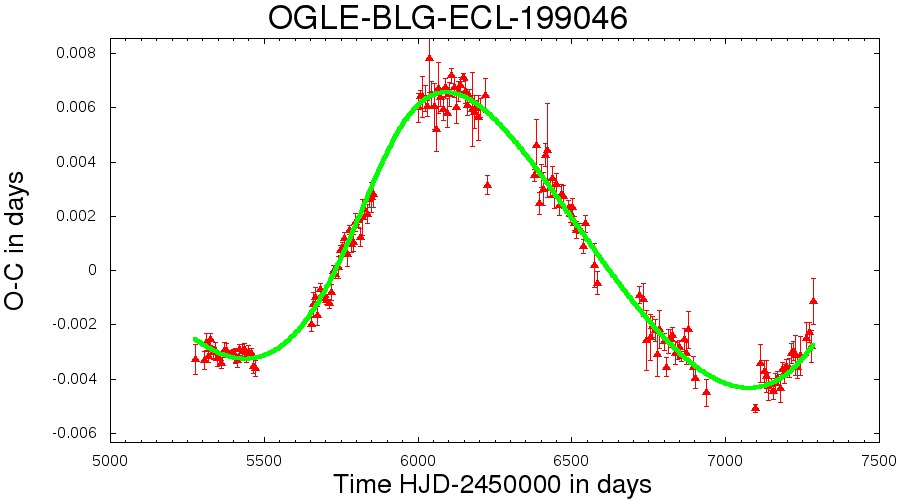}
\includegraphics[width=0.64\columnwidth]{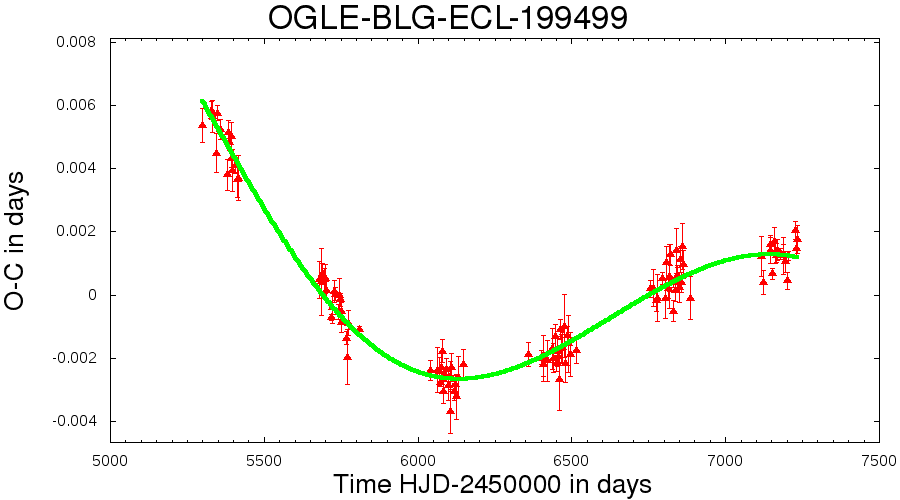}

\includegraphics[width=0.64\columnwidth]{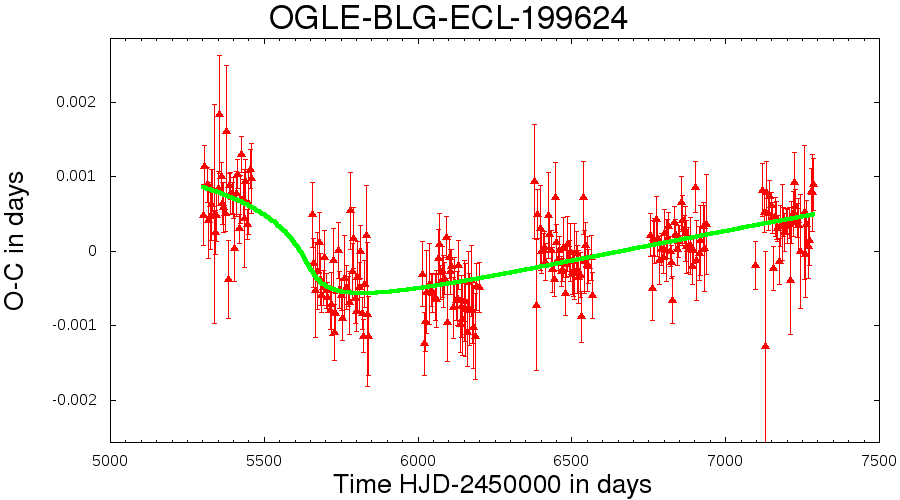}
\includegraphics[width=0.64\columnwidth]{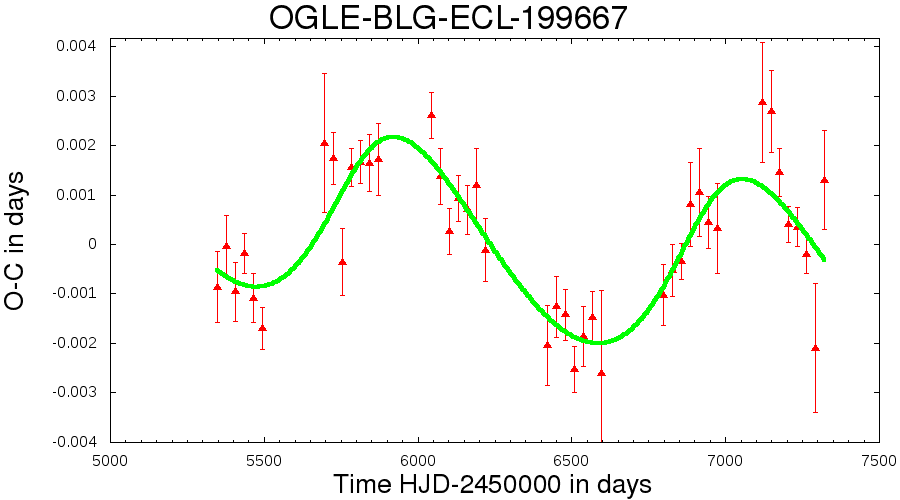}
\includegraphics[width=0.64\columnwidth]{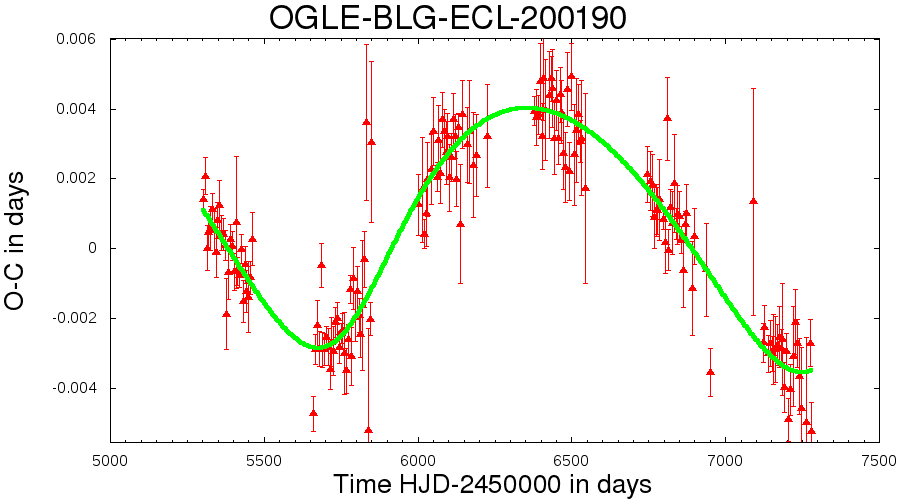}

\includegraphics[width=0.64\columnwidth]{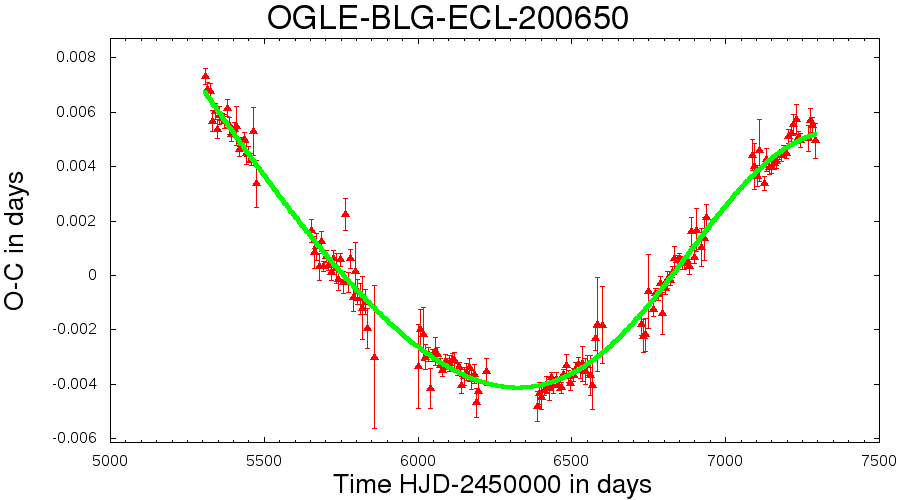}
\includegraphics[width=0.64\columnwidth]{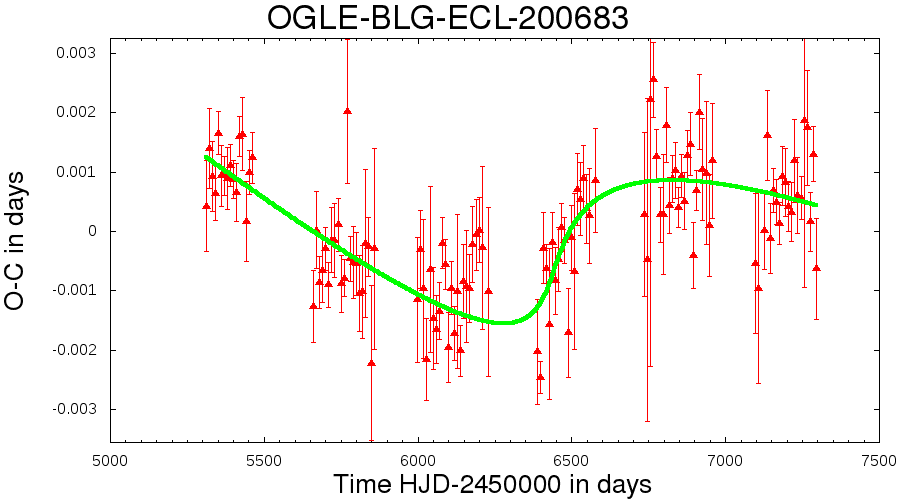}
\includegraphics[width=0.64\columnwidth]{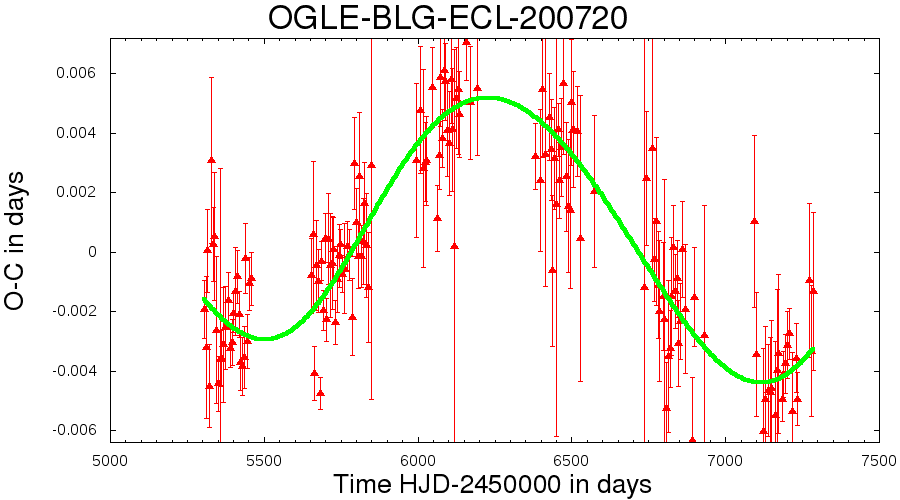}

\includegraphics[width=0.64\columnwidth]{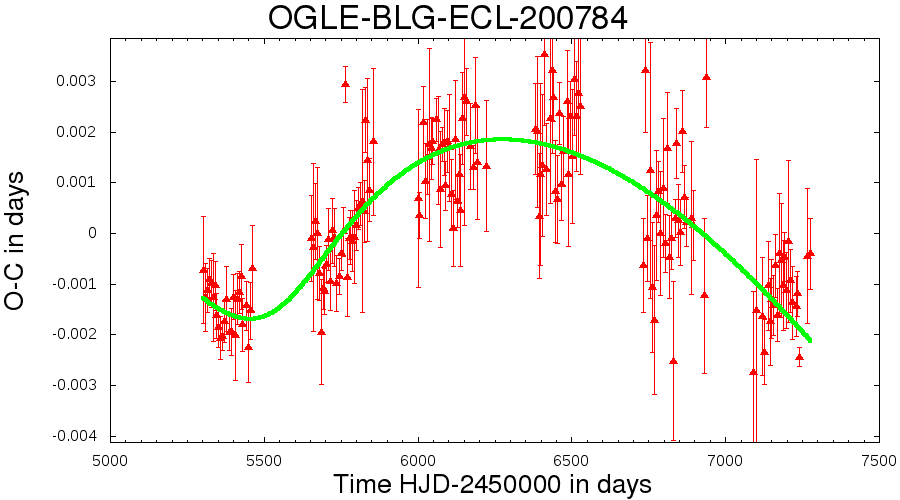}
\includegraphics[width=0.64\columnwidth]{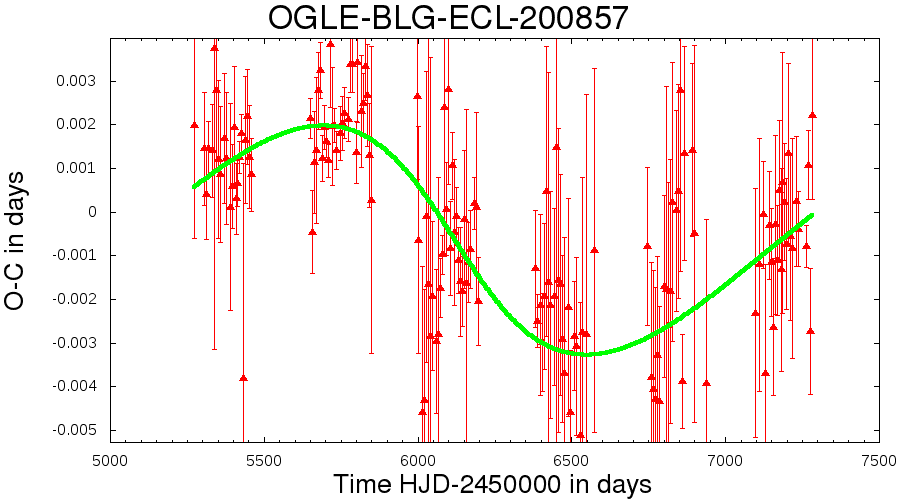}
\includegraphics[width=0.64\columnwidth]{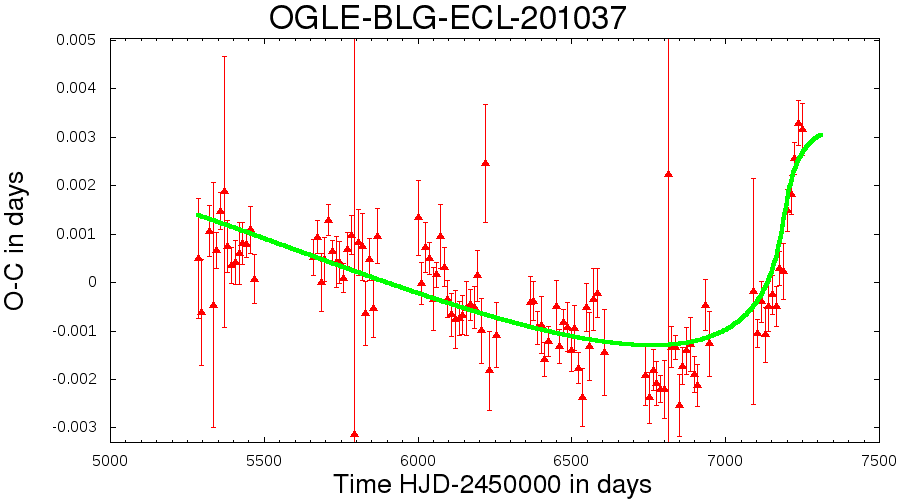}

\includegraphics[width=0.64\columnwidth]{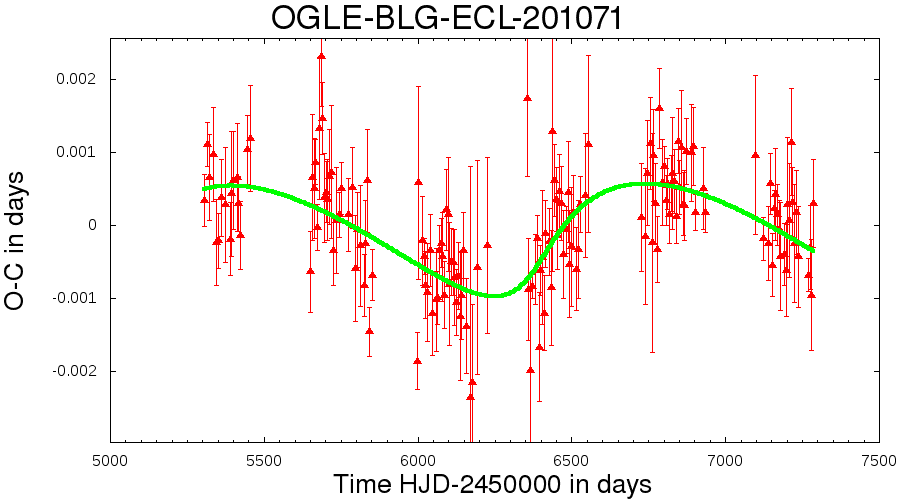}
\includegraphics[width=0.64\columnwidth]{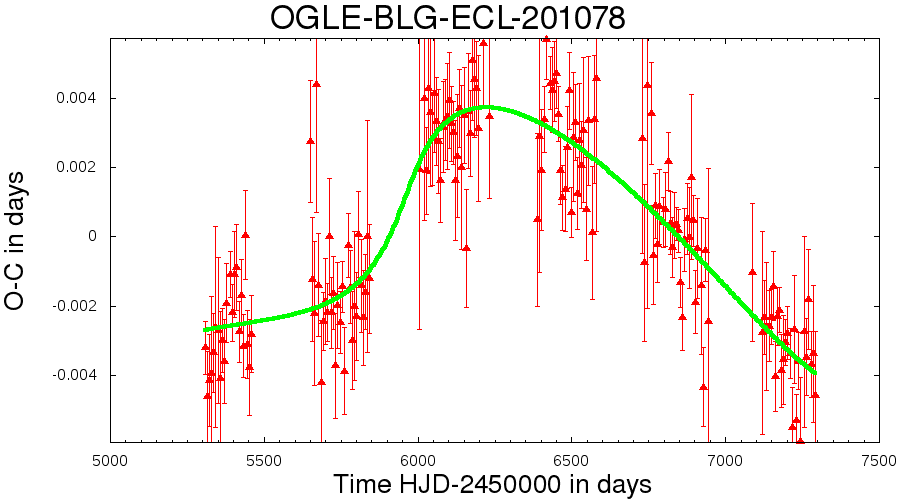}
\includegraphics[width=0.64\columnwidth]{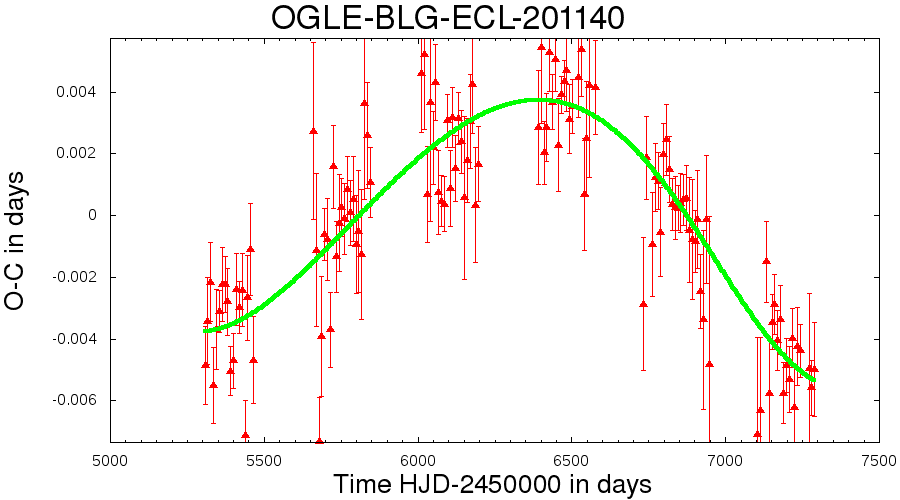}

\includegraphics[width=0.64\columnwidth]{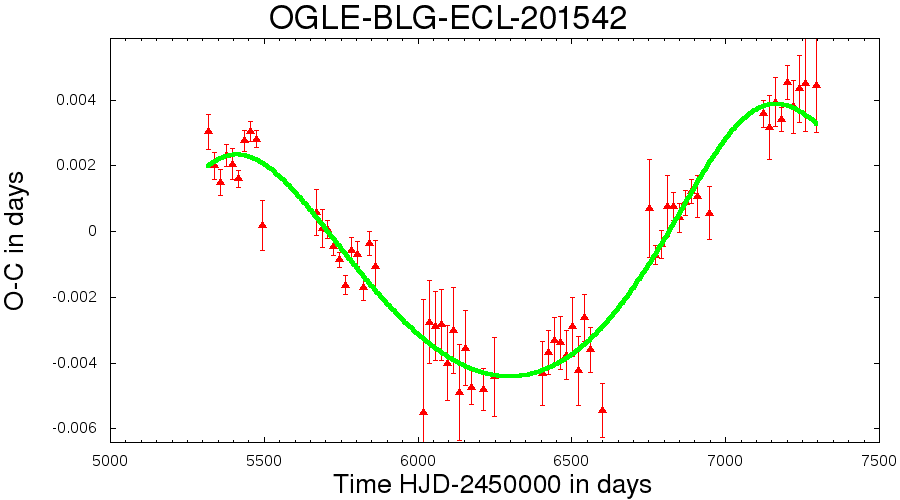}
\includegraphics[width=0.64\columnwidth]{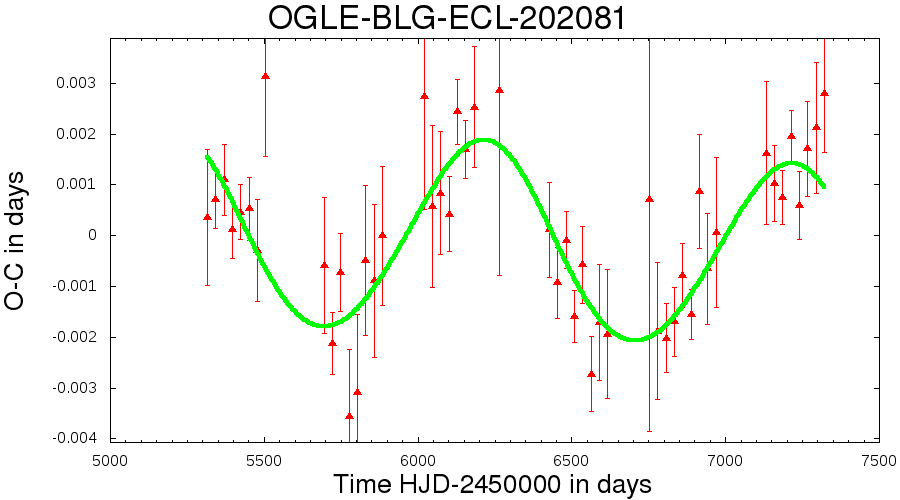}
\includegraphics[width=0.64\columnwidth]{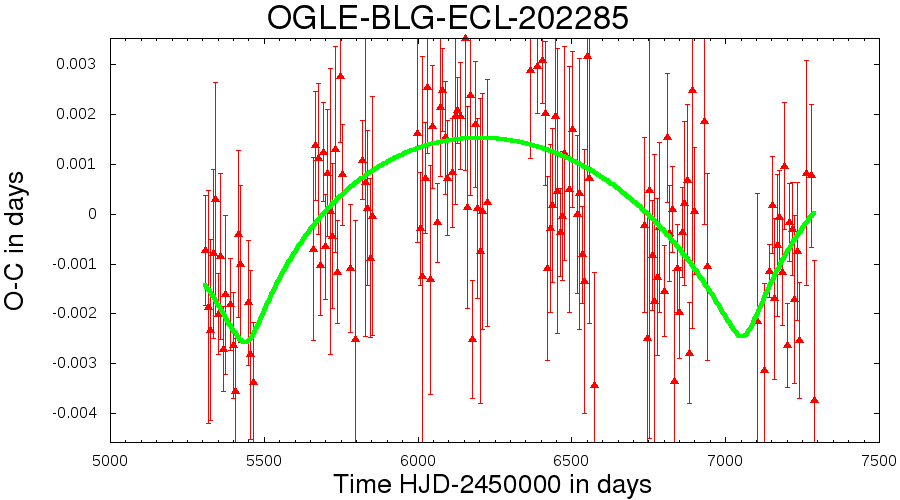}

\includegraphics[width=0.64\columnwidth]{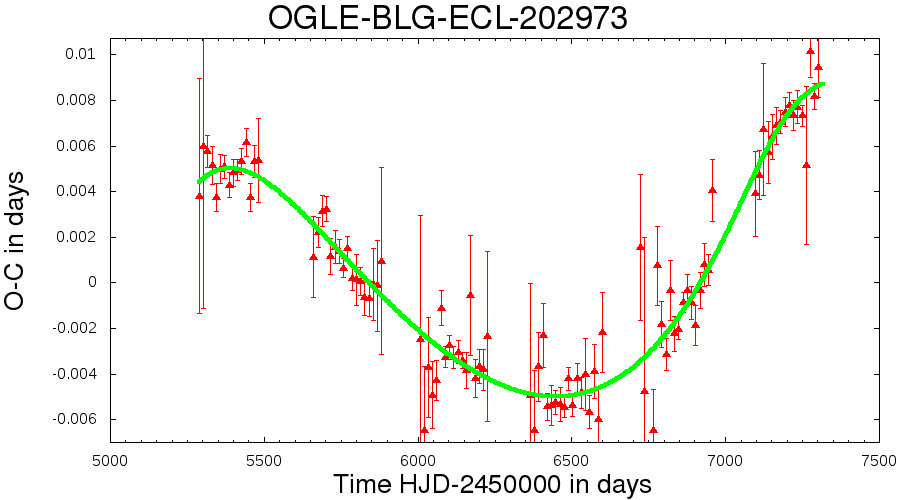}
\includegraphics[width=0.64\columnwidth]{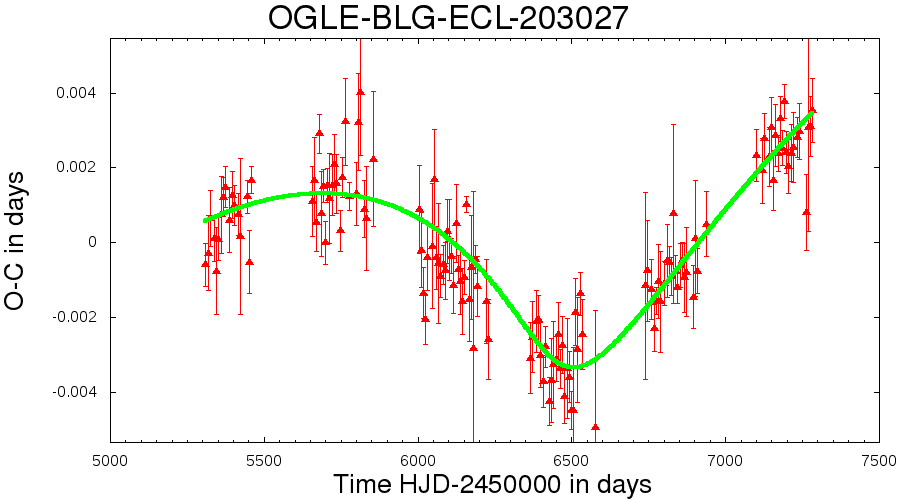}
\includegraphics[width=0.64\columnwidth]{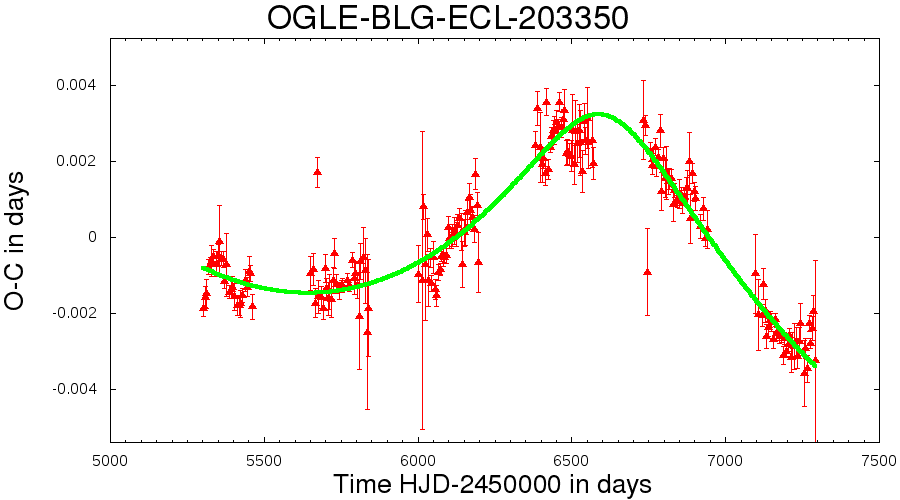}

\end{figure*}
\clearpage

\begin{figure*}
\includegraphics[width=0.64\columnwidth]{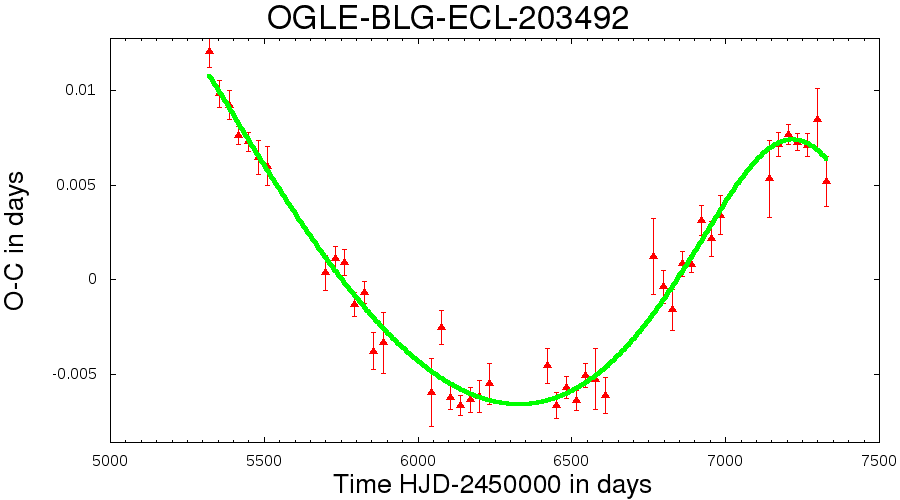}
\includegraphics[width=0.64\columnwidth]{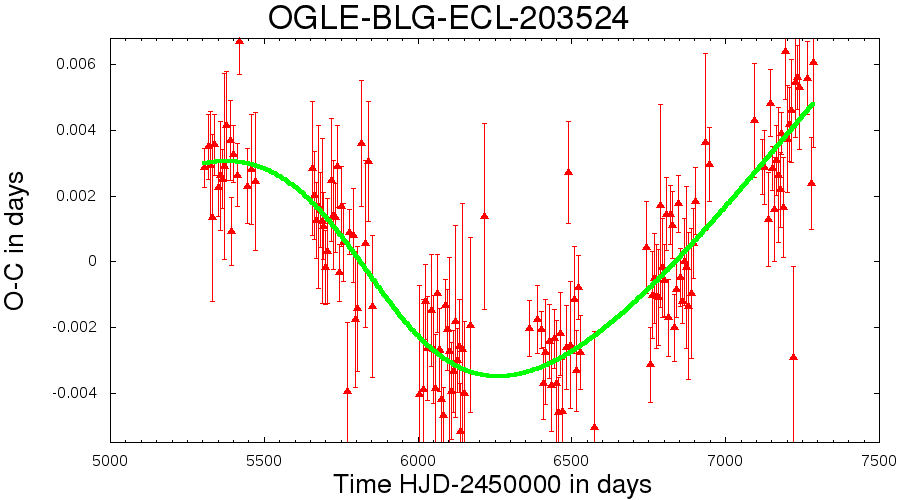}
\includegraphics[width=0.64\columnwidth]{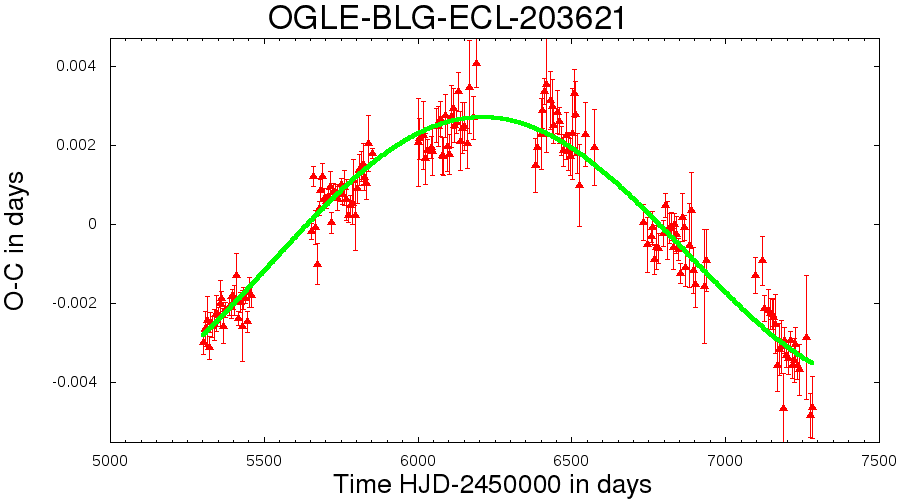}

\includegraphics[width=0.64\columnwidth]{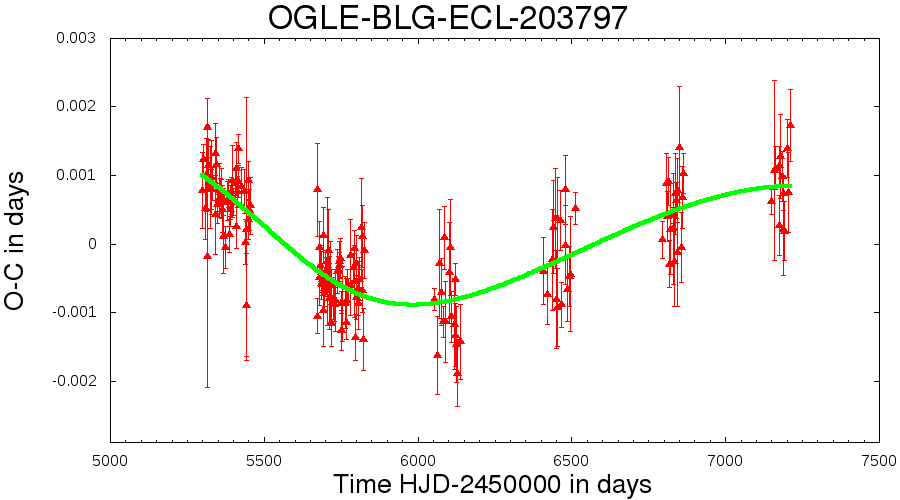}
\includegraphics[width=0.64\columnwidth]{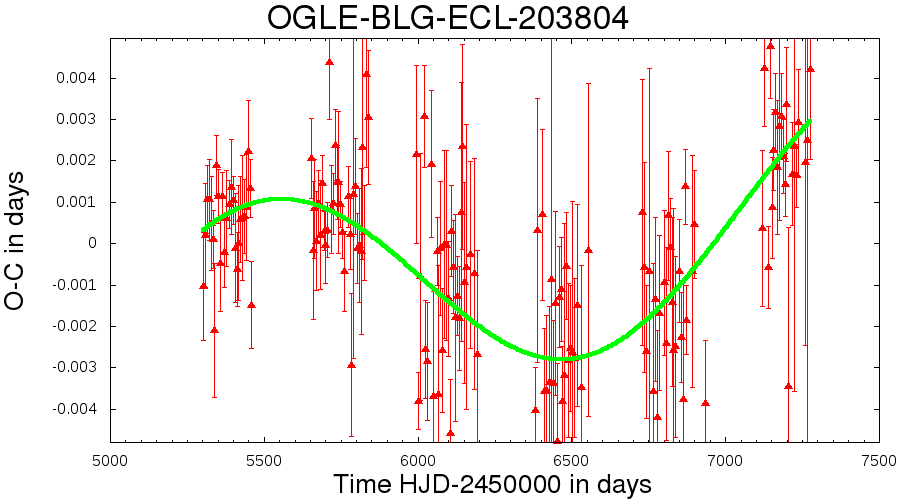}
\includegraphics[width=0.64\columnwidth]{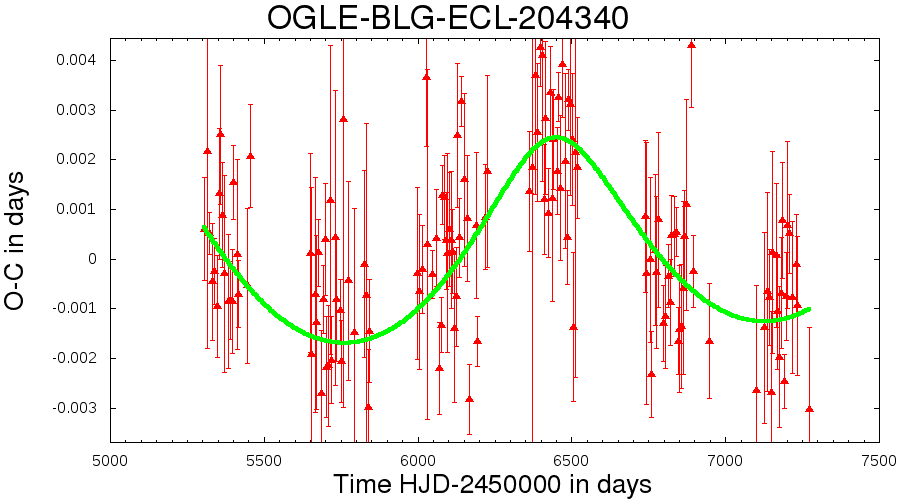}

\includegraphics[width=0.64\columnwidth]{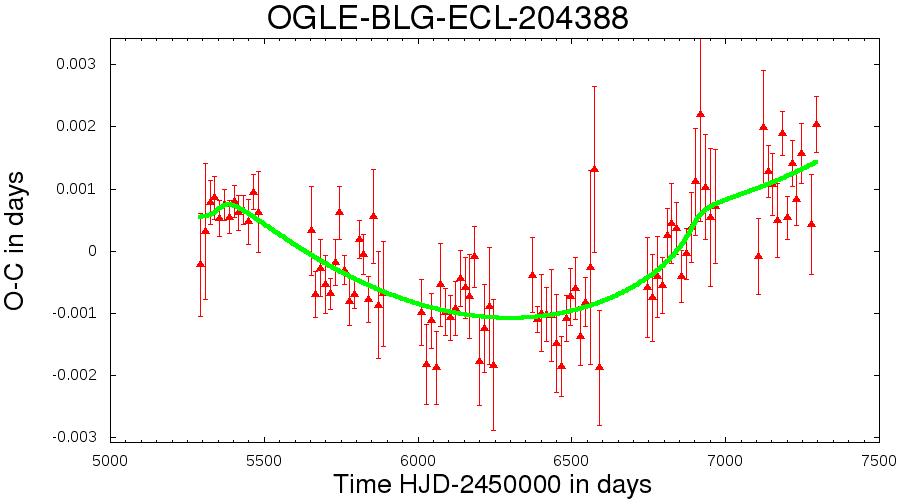}
\includegraphics[width=0.64\columnwidth]{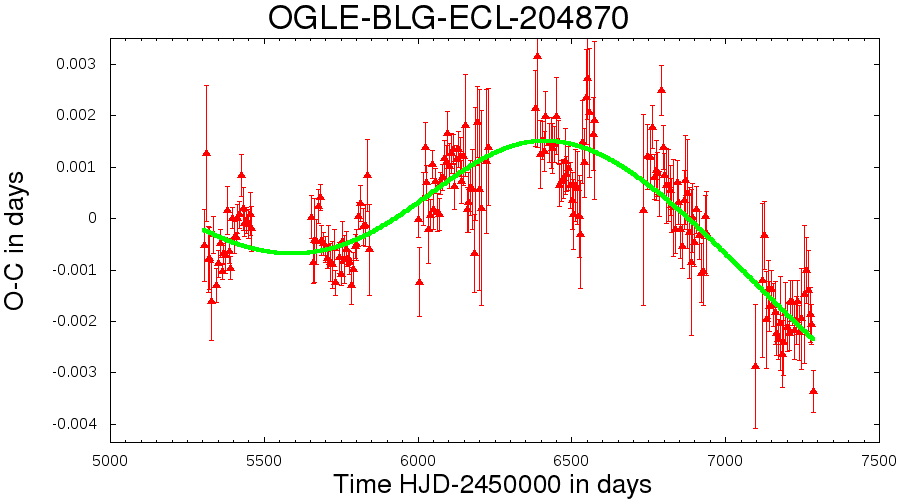}
\includegraphics[width=0.64\columnwidth]{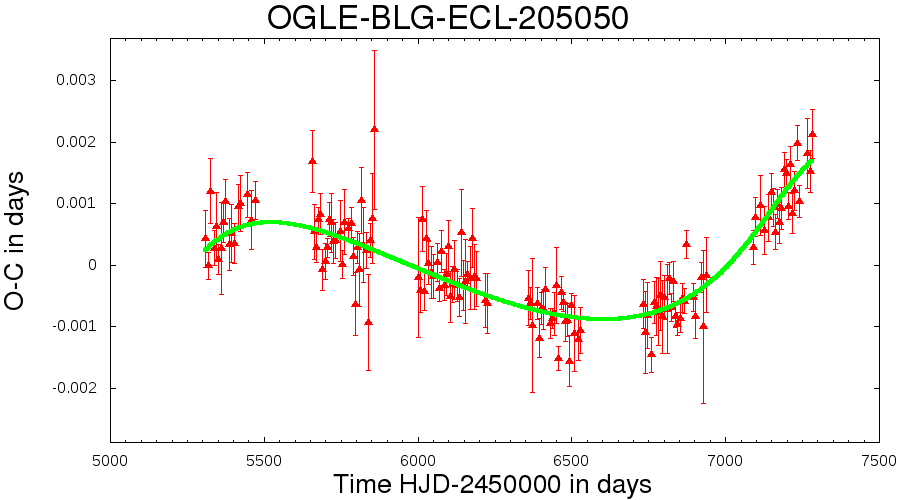}

\includegraphics[width=0.64\columnwidth]{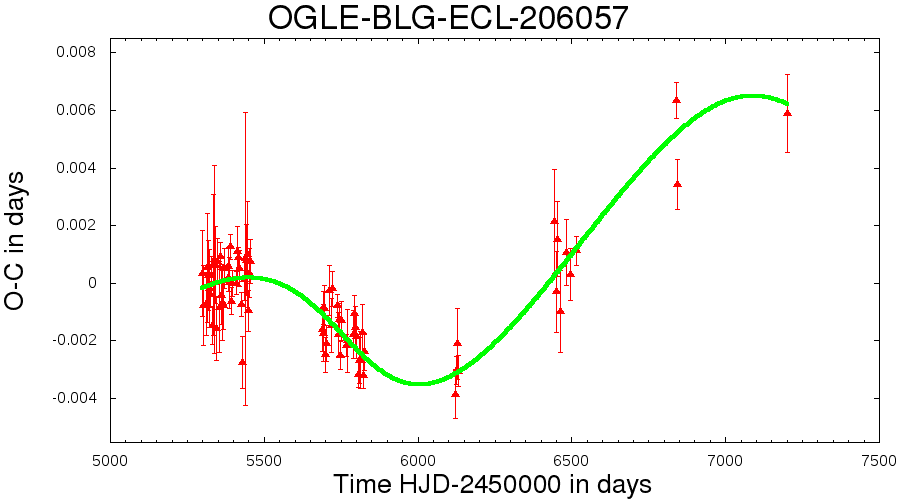}
\includegraphics[width=0.64\columnwidth]{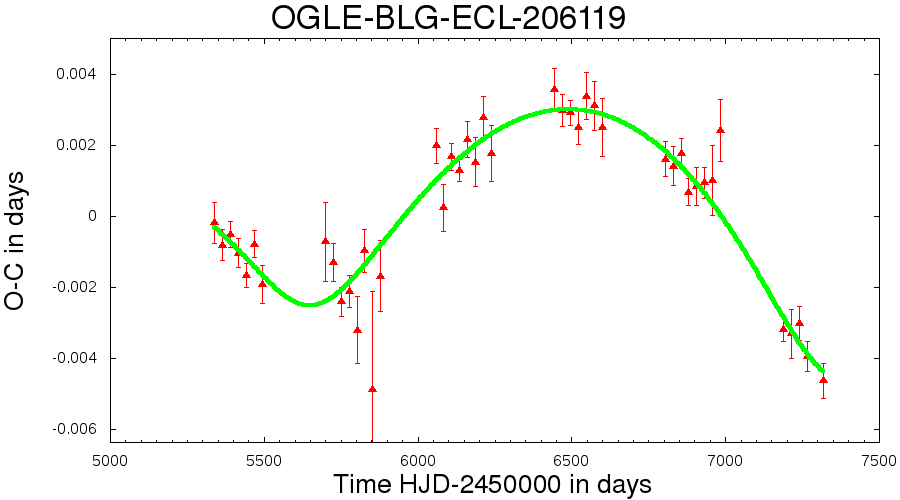}
\includegraphics[width=0.64\columnwidth]{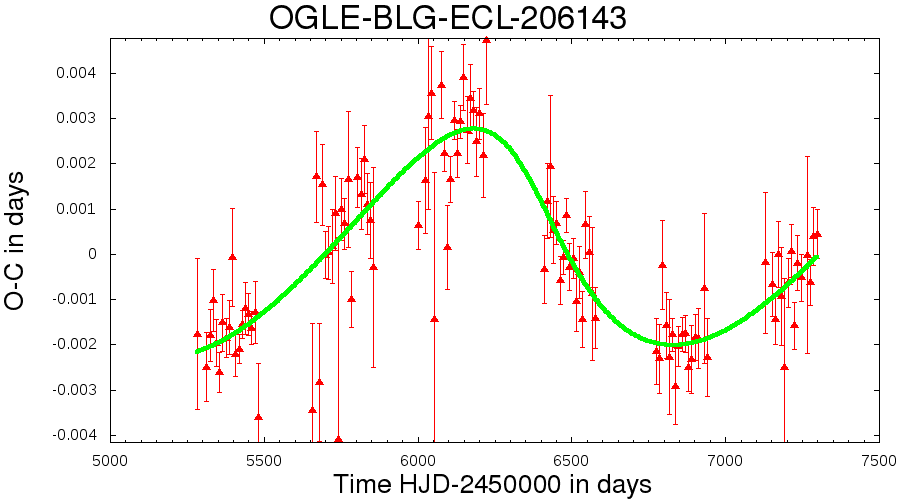}

\includegraphics[width=0.64\columnwidth]{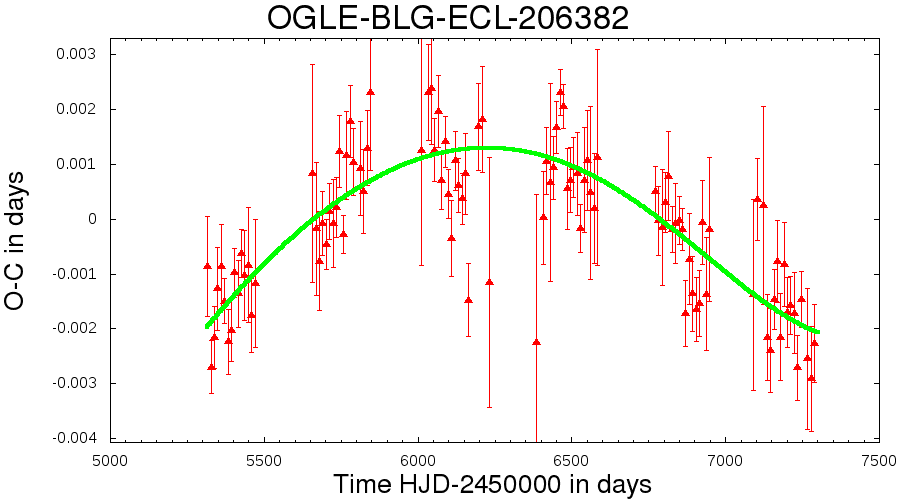}
\includegraphics[width=0.64\columnwidth]{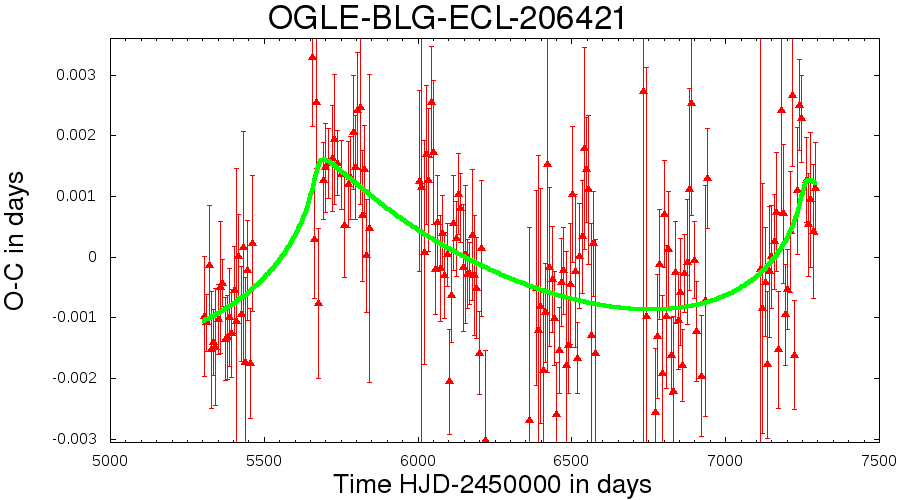}
\includegraphics[width=0.64\columnwidth]{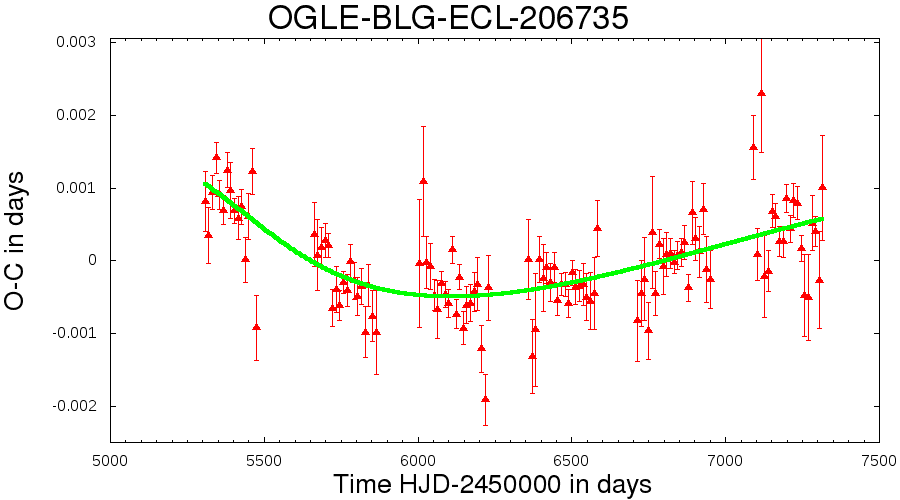}

\includegraphics[width=0.64\columnwidth]{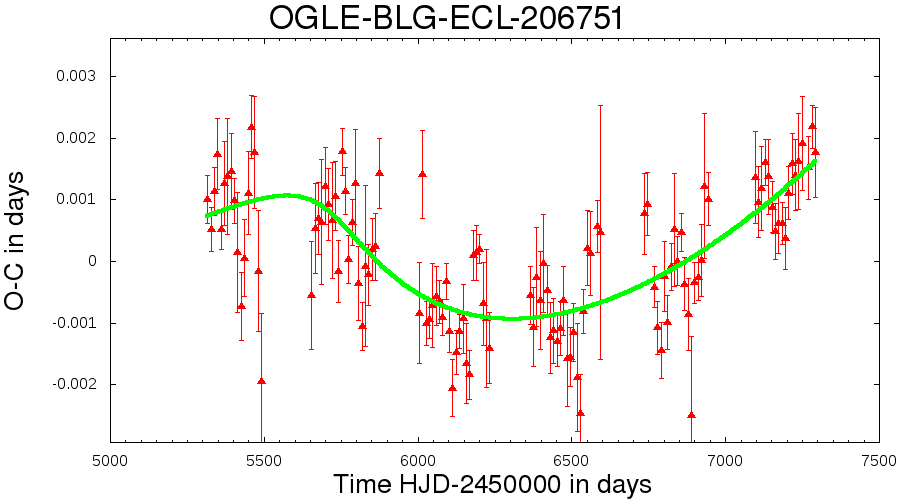}
\includegraphics[width=0.64\columnwidth]{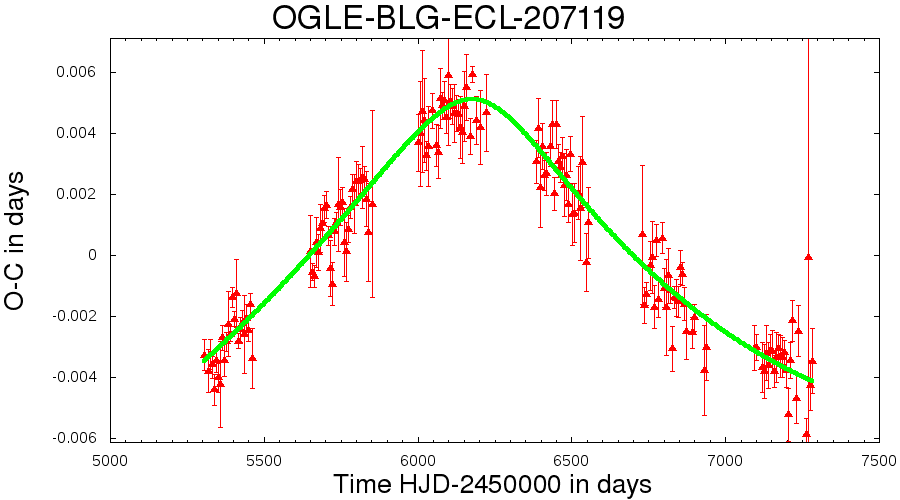}
\includegraphics[width=0.64\columnwidth]{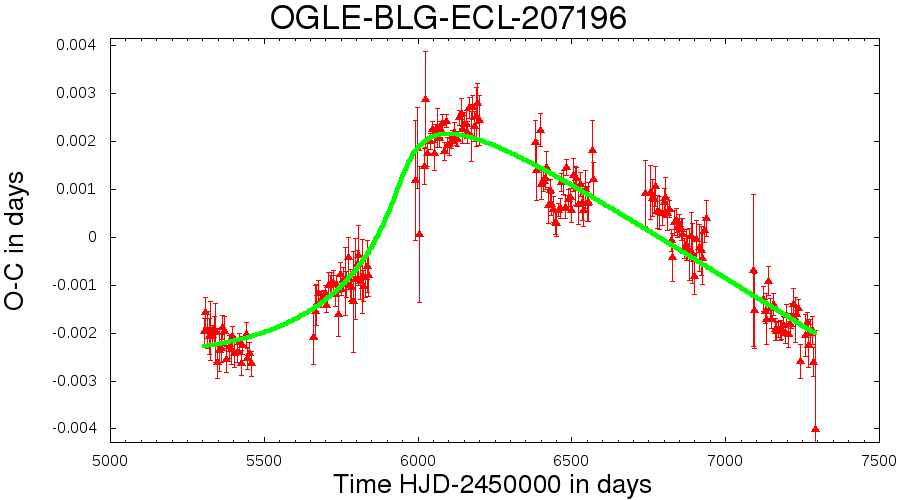}

\includegraphics[width=0.64\columnwidth]{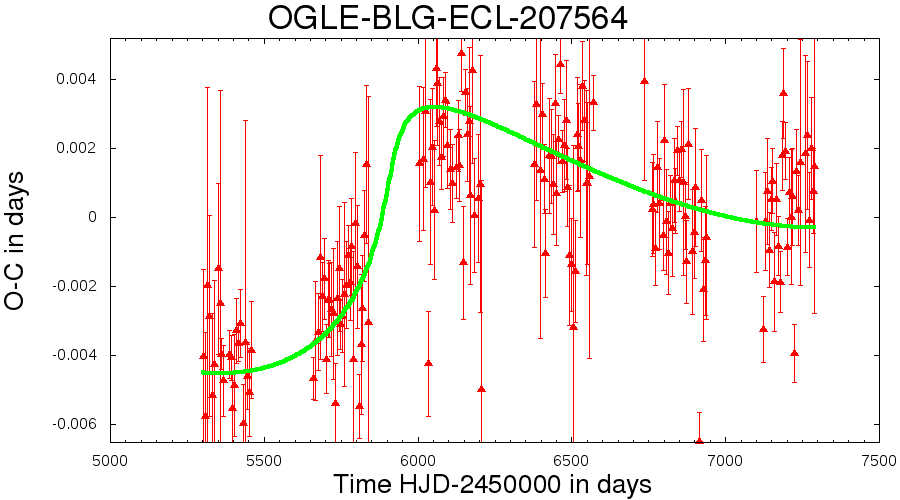}
\includegraphics[width=0.64\columnwidth]{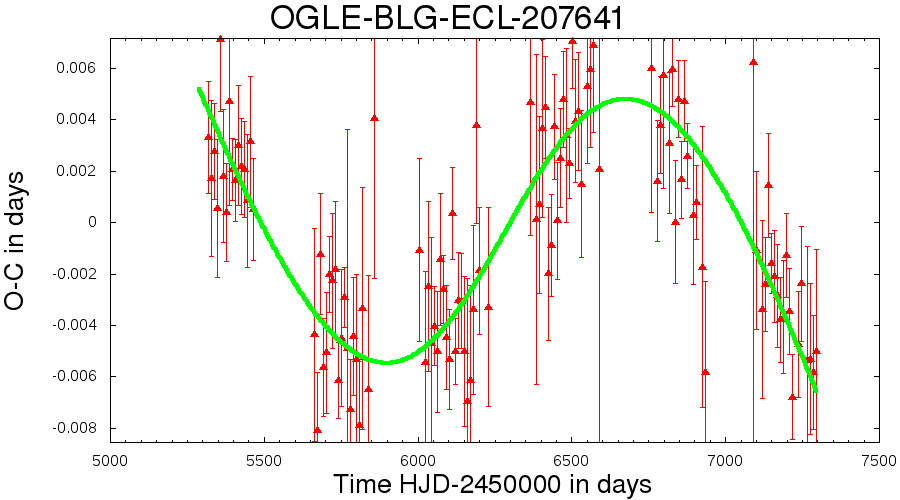}
\includegraphics[width=0.64\columnwidth]{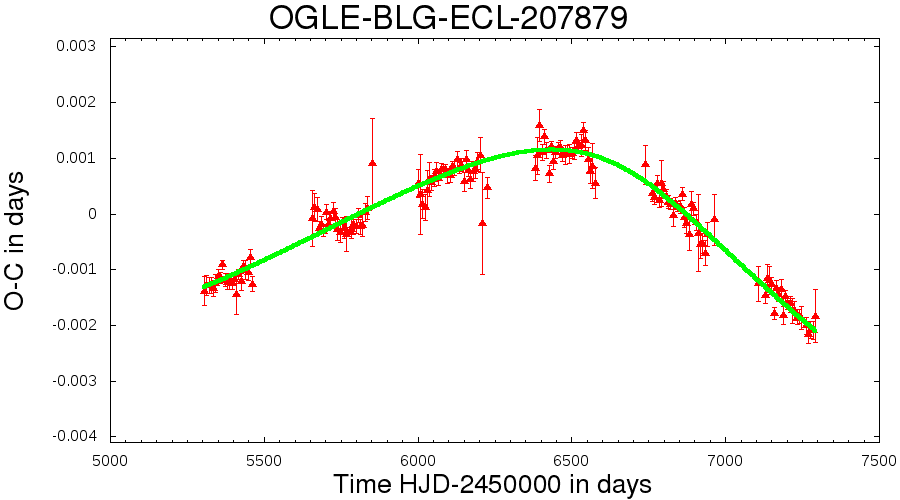}

\includegraphics[width=0.64\columnwidth]{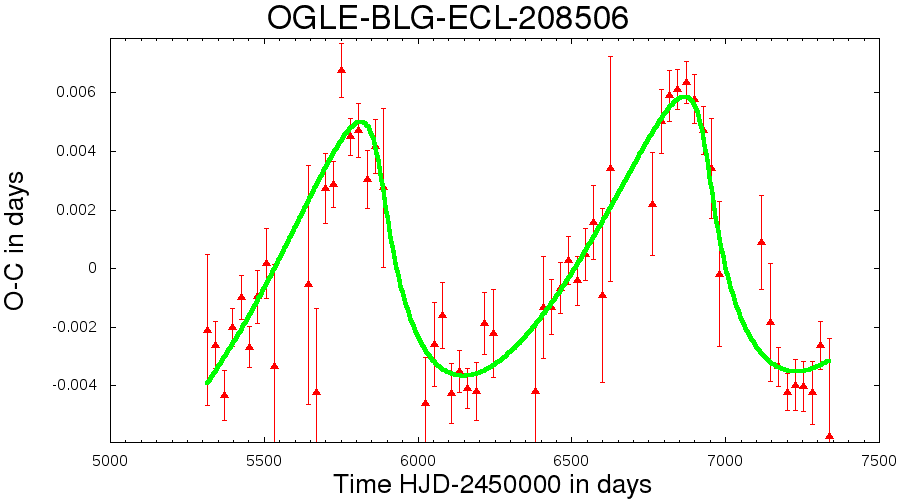}
\includegraphics[width=0.64\columnwidth]{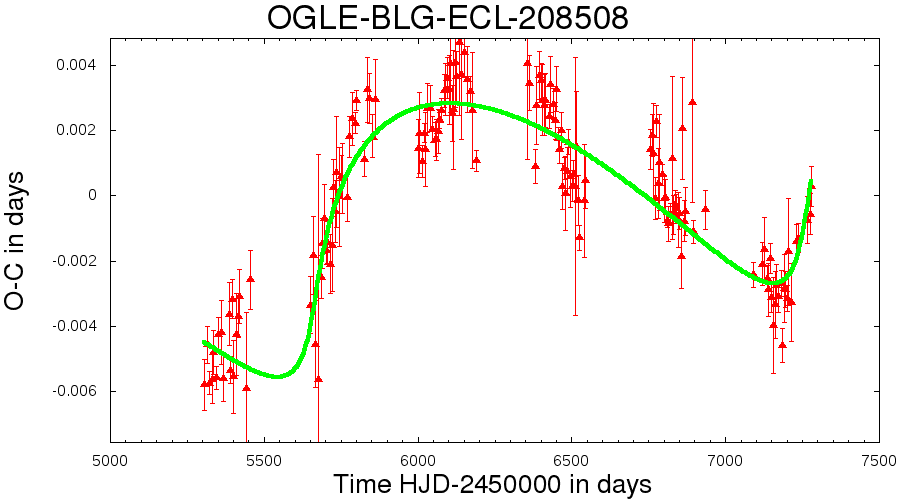}
\includegraphics[width=0.64\columnwidth]{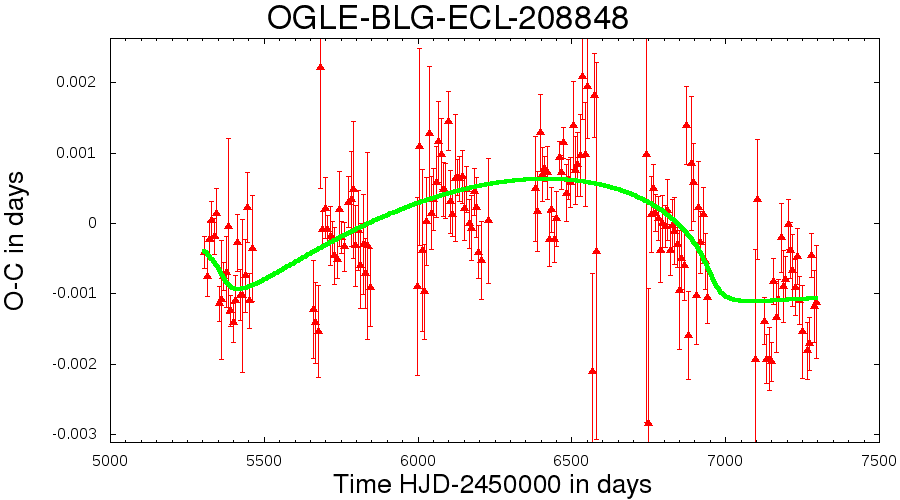}

\end{figure*}
\clearpage

\begin{figure*}
\includegraphics[width=0.64\columnwidth]{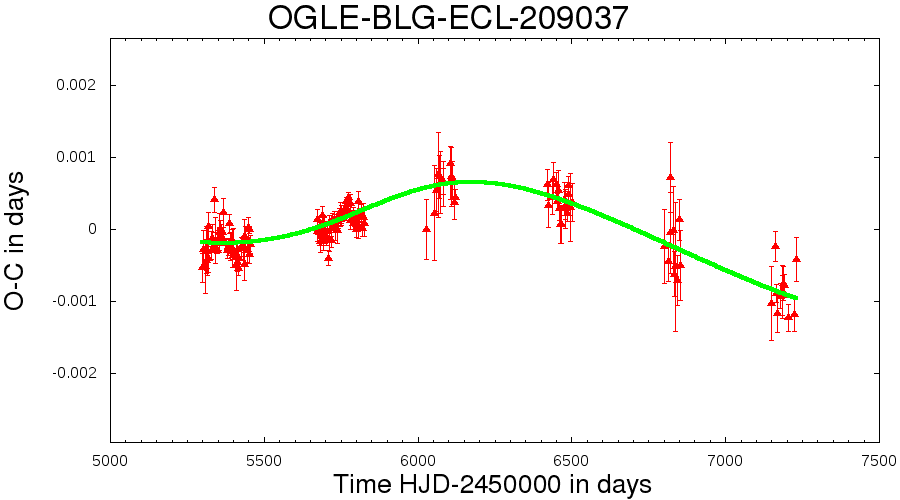}
\includegraphics[width=0.64\columnwidth]{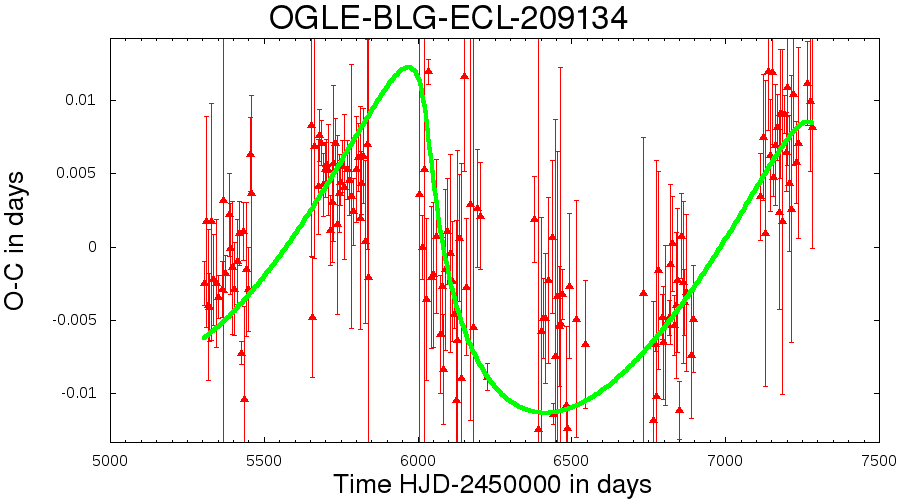}
\includegraphics[width=0.64\columnwidth]{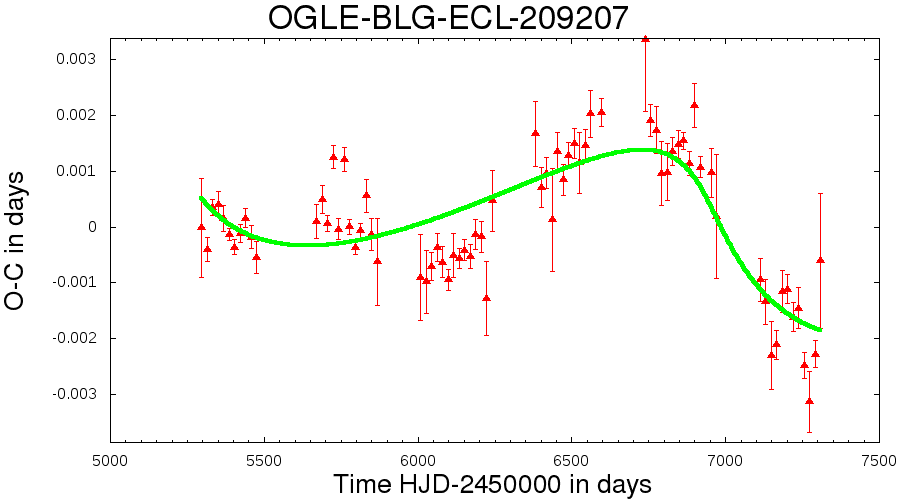}

\includegraphics[width=0.64\columnwidth]{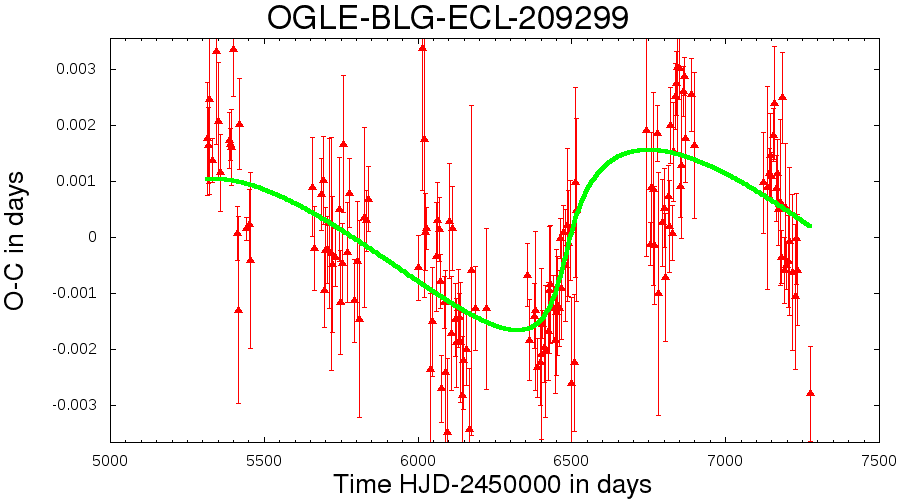}
\includegraphics[width=0.64\columnwidth]{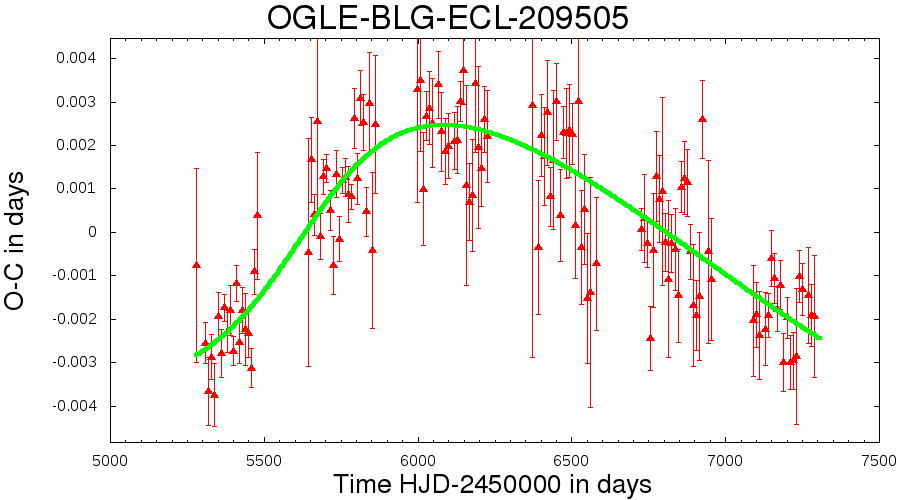}
\includegraphics[width=0.64\columnwidth]{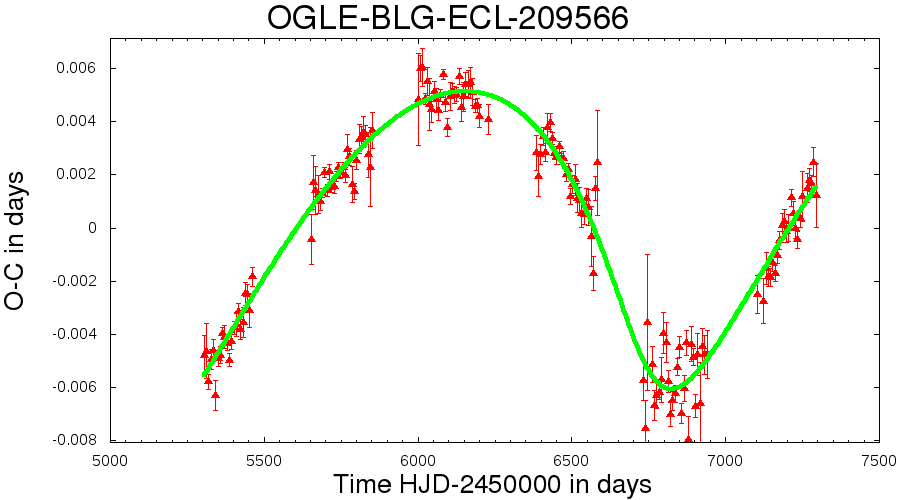}

\includegraphics[width=0.64\columnwidth]{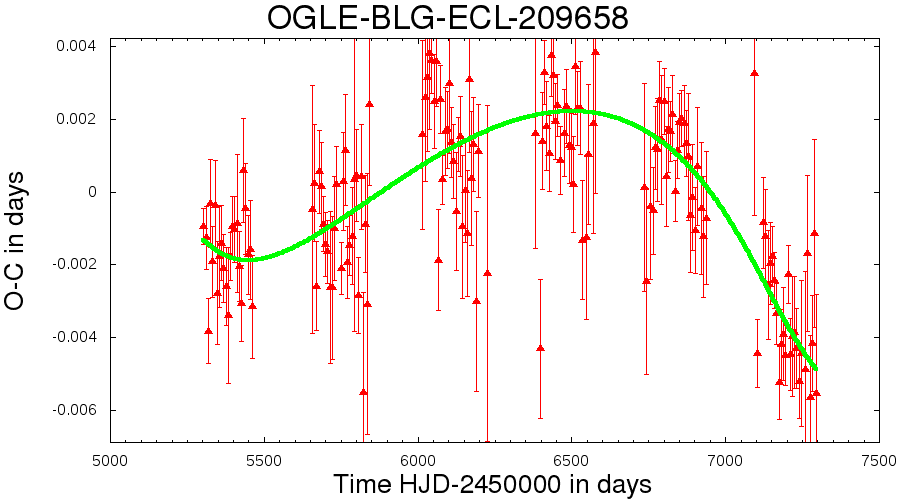}
\includegraphics[width=0.64\columnwidth]{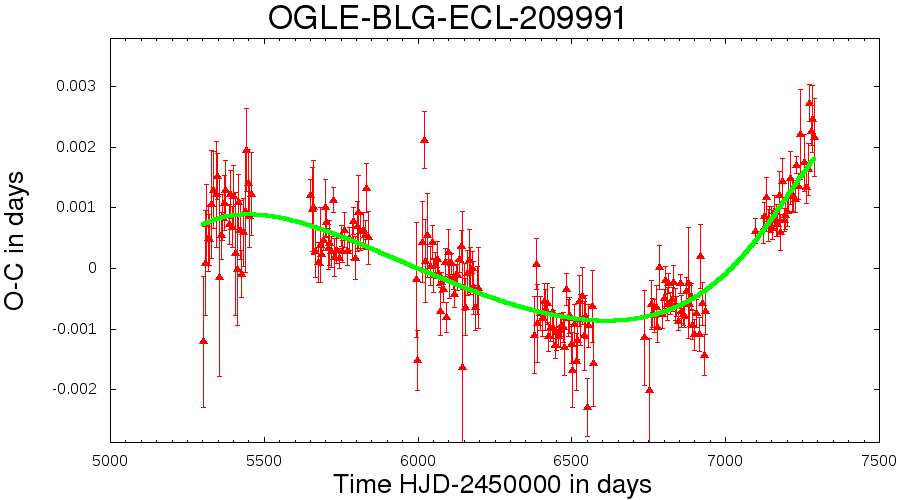}
\includegraphics[width=0.64\columnwidth]{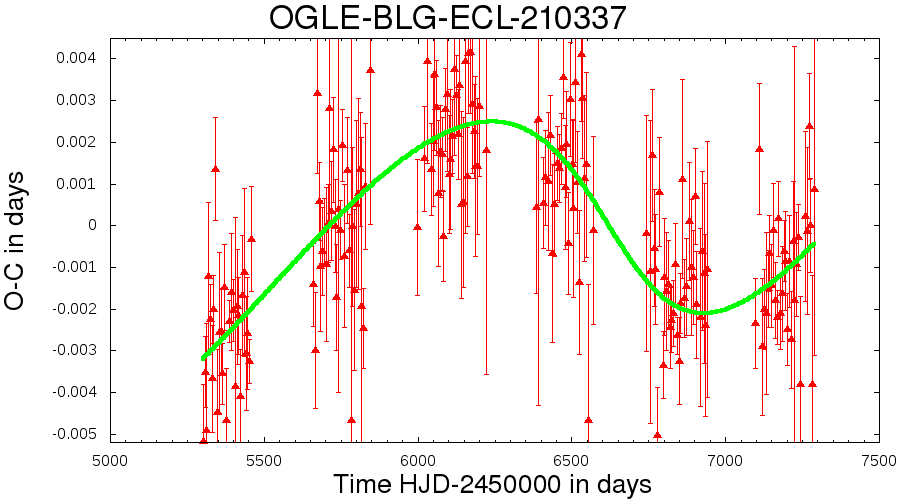}

\includegraphics[width=0.64\columnwidth]{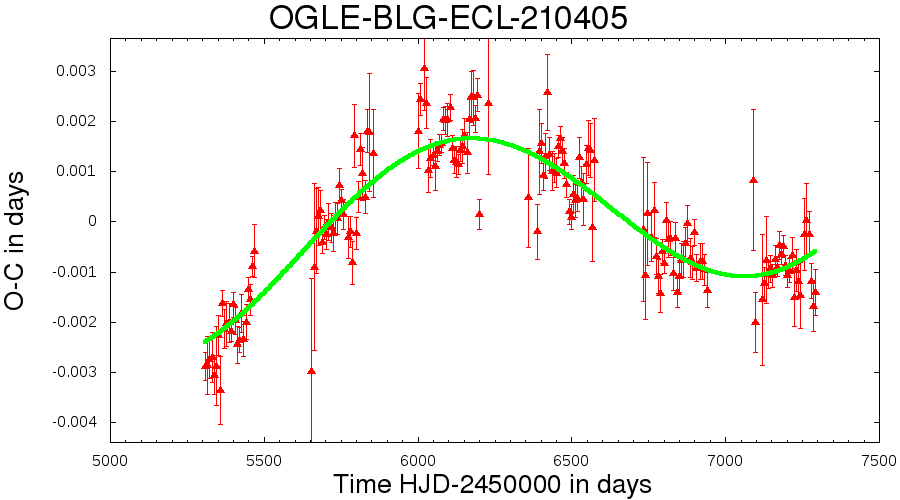}
\includegraphics[width=0.64\columnwidth]{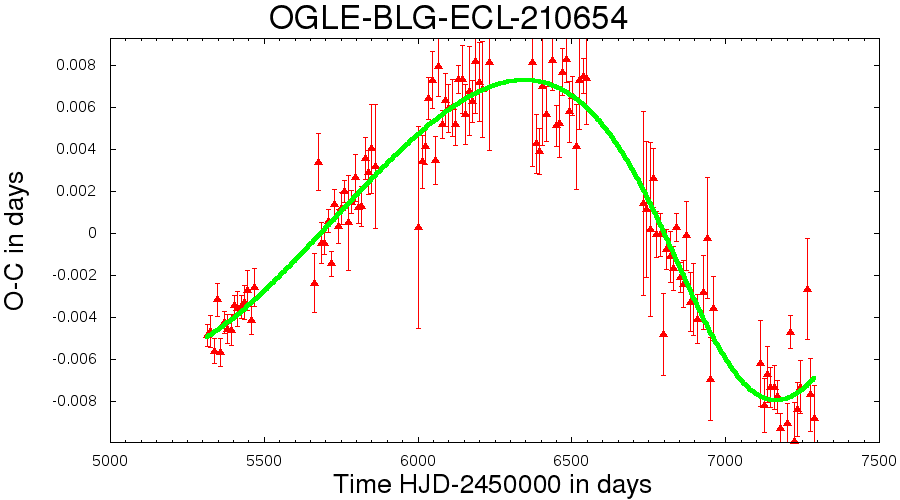}
\includegraphics[width=0.64\columnwidth]{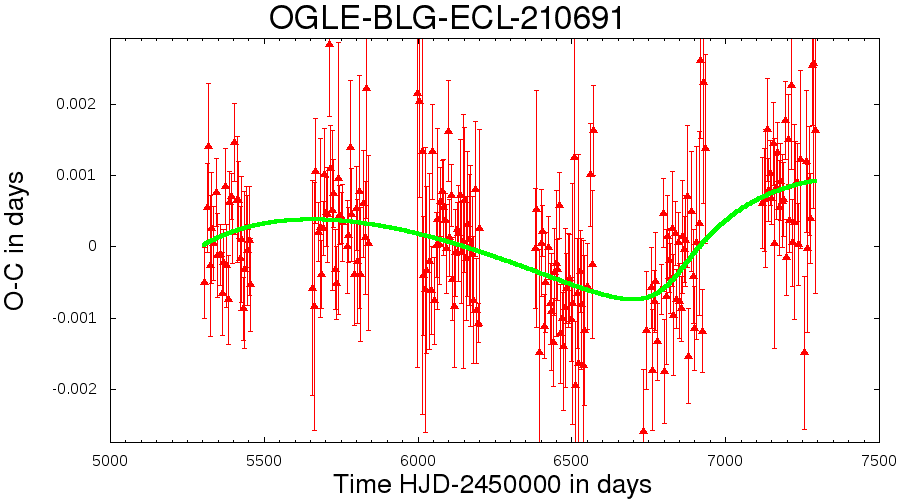}

\includegraphics[width=0.64\columnwidth]{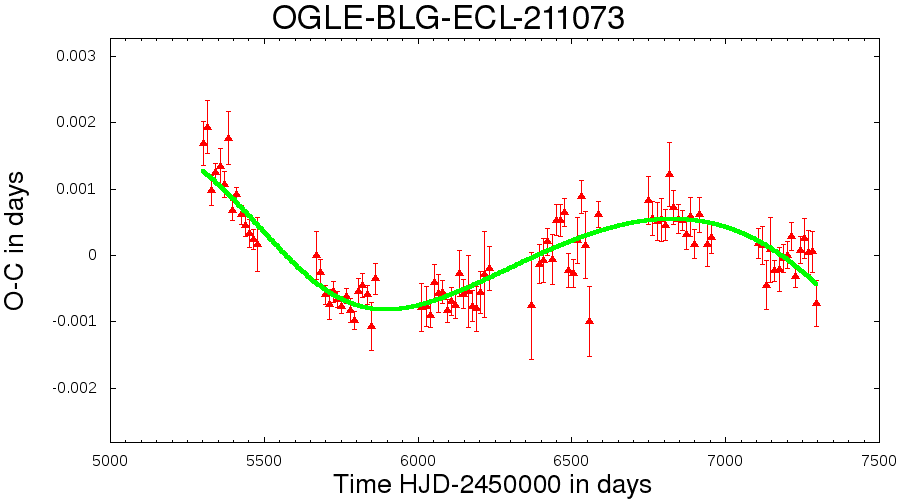}
\includegraphics[width=0.64\columnwidth]{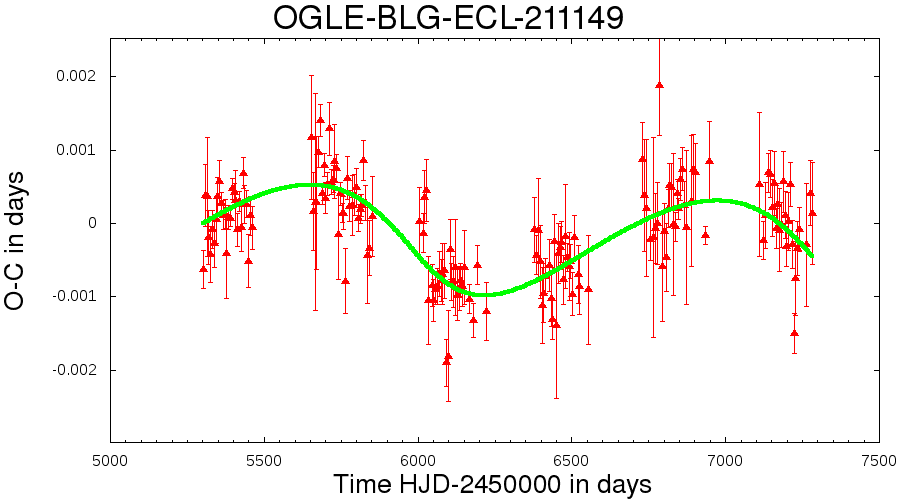}
\includegraphics[width=0.64\columnwidth]{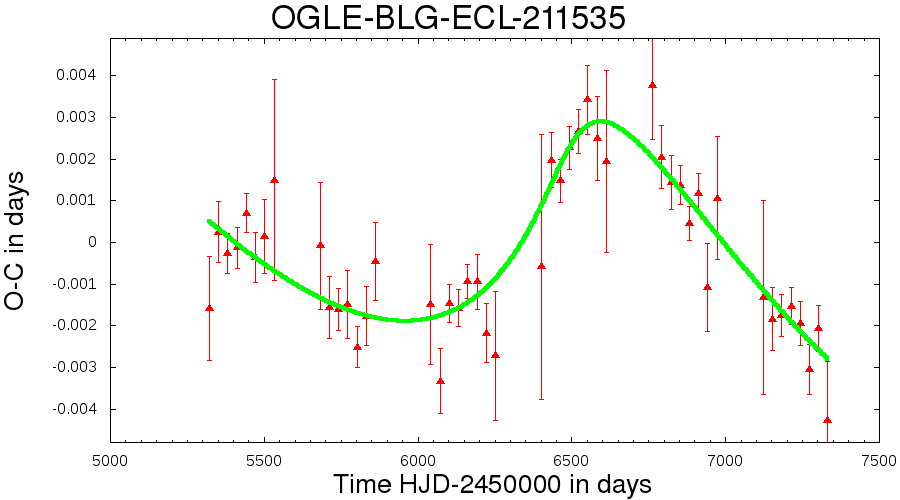}

\includegraphics[width=0.64\columnwidth]{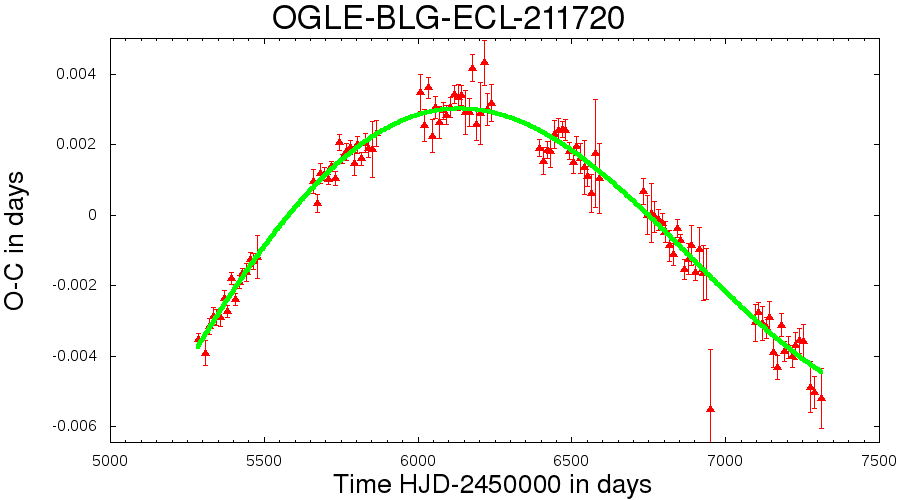}
\includegraphics[width=0.64\columnwidth]{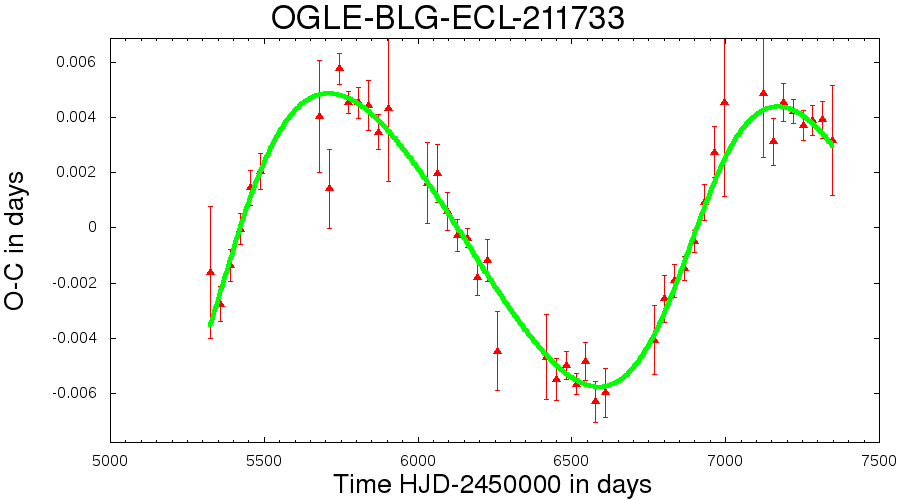}
\includegraphics[width=0.64\columnwidth]{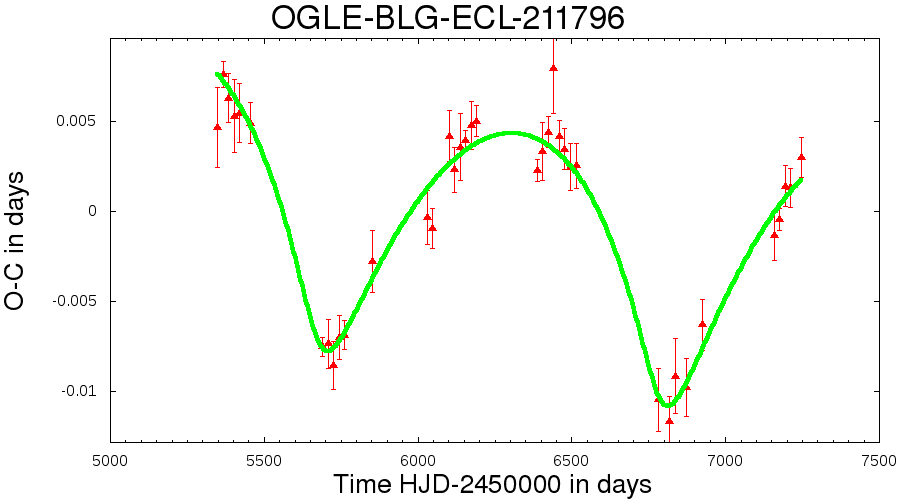}

\includegraphics[width=0.64\columnwidth]{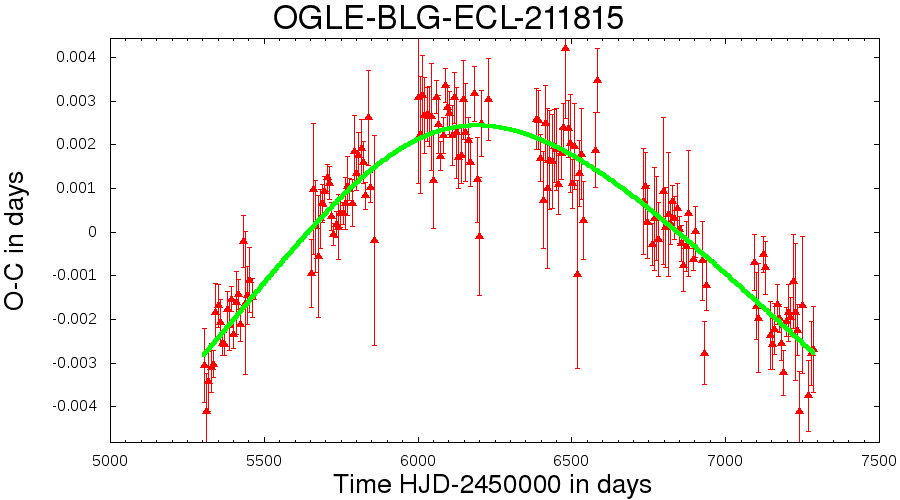}
\includegraphics[width=0.64\columnwidth]{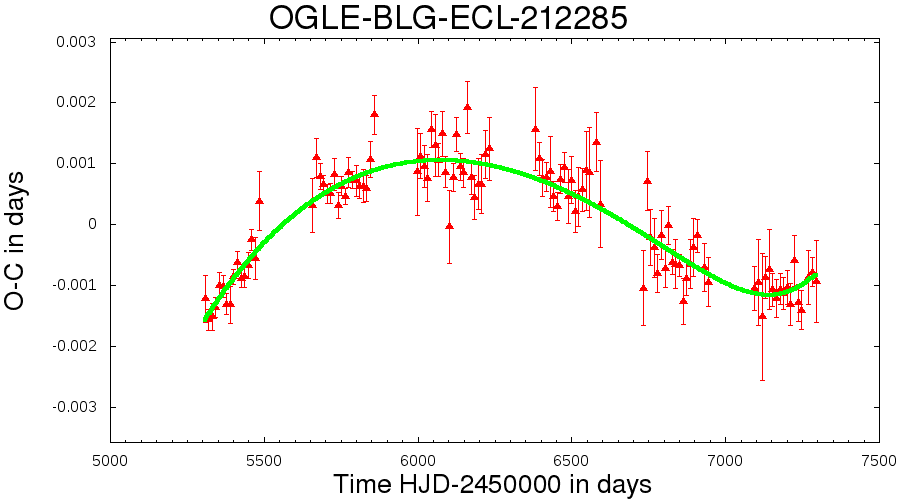}
\includegraphics[width=0.64\columnwidth]{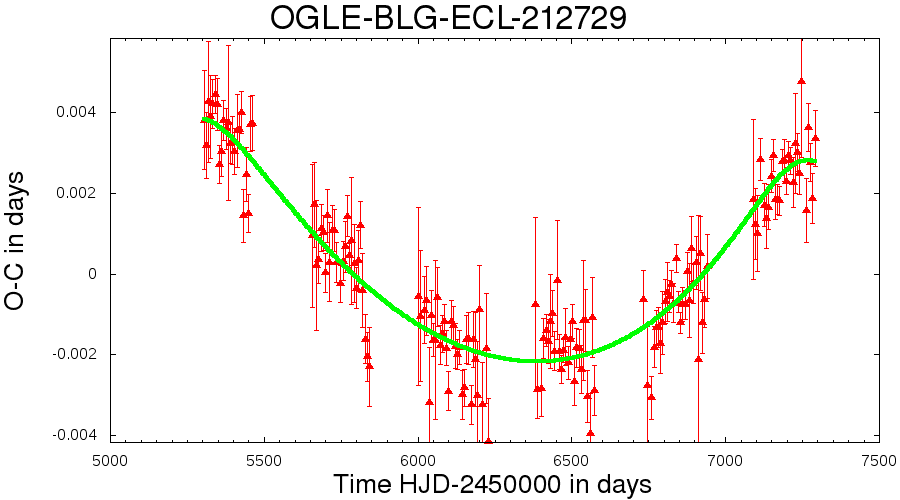}

\includegraphics[width=0.64\columnwidth]{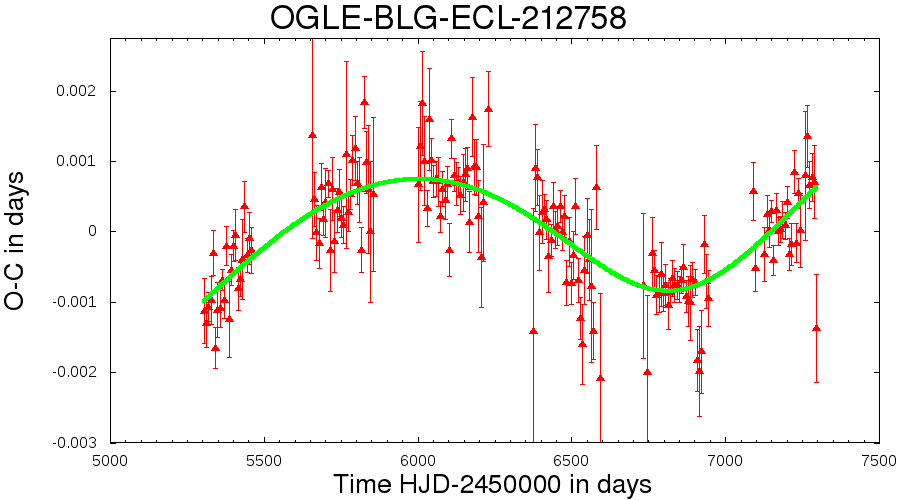}
\includegraphics[width=0.64\columnwidth]{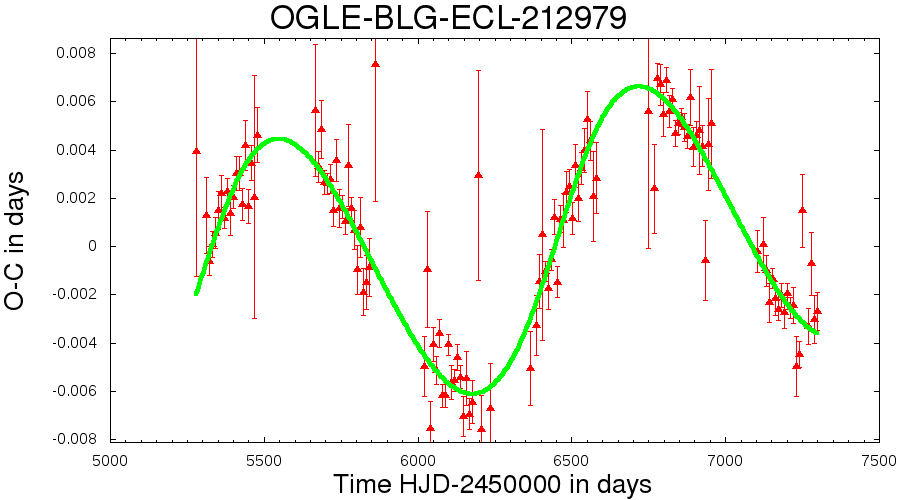}
\includegraphics[width=0.64\columnwidth]{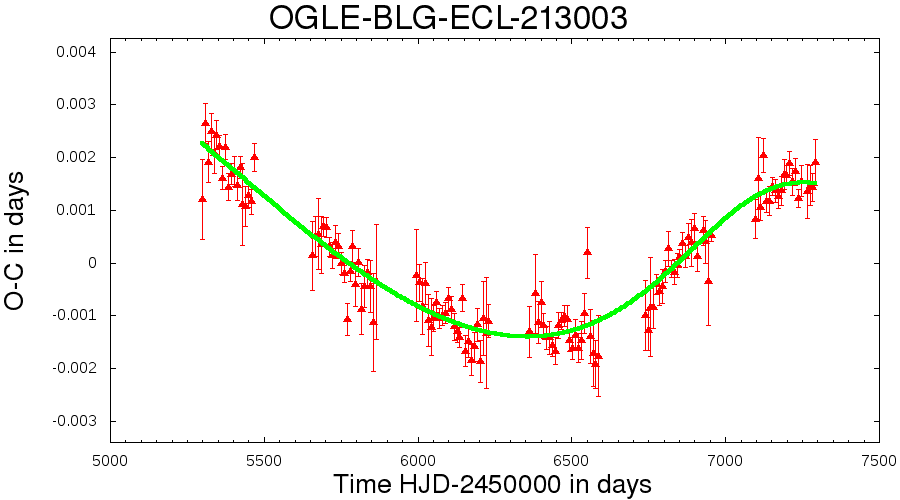}

\end{figure*}
\clearpage

\begin{figure*}
\includegraphics[width=0.64\columnwidth]{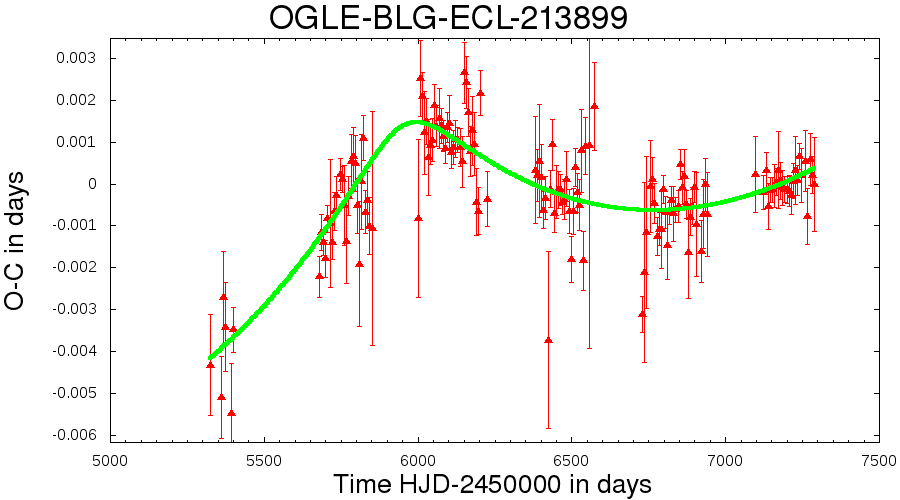}
\includegraphics[width=0.64\columnwidth]{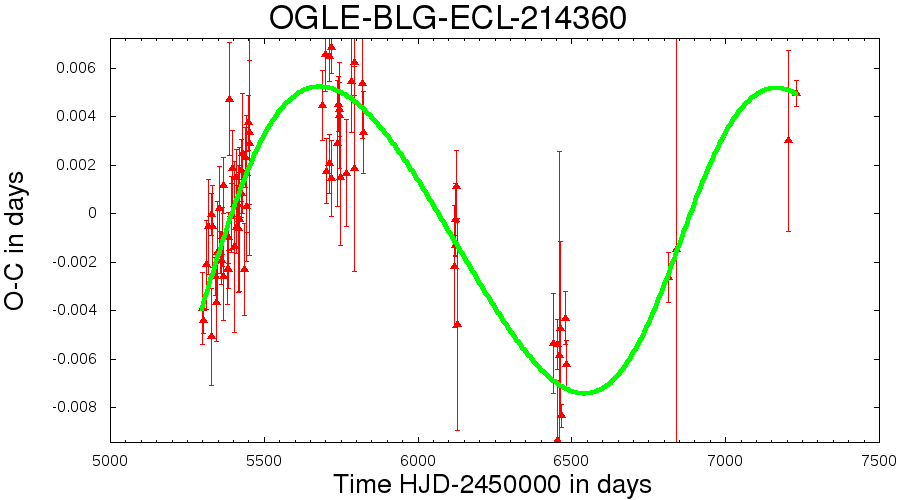}
\includegraphics[width=0.64\columnwidth]{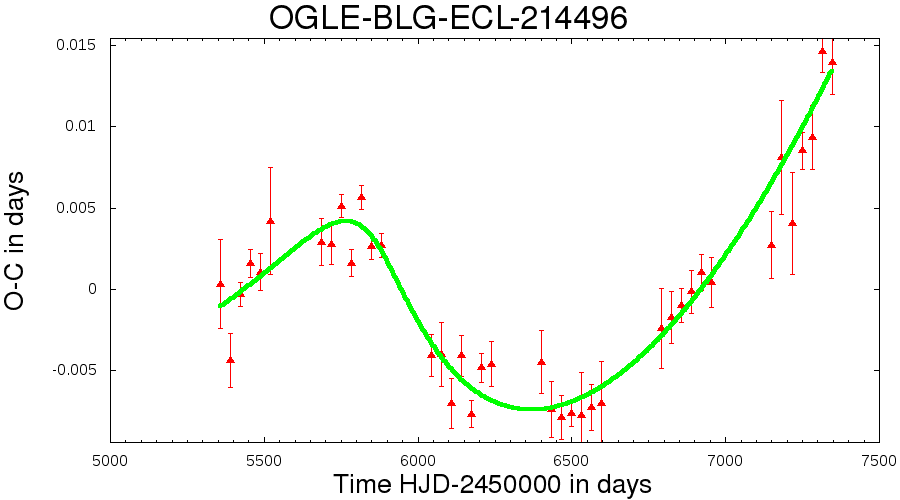}

\includegraphics[width=0.64\columnwidth]{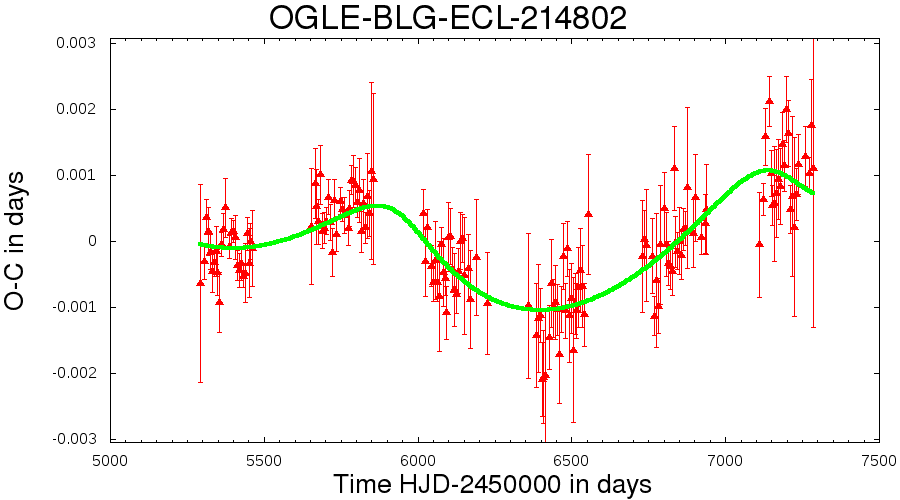}
\includegraphics[width=0.64\columnwidth]{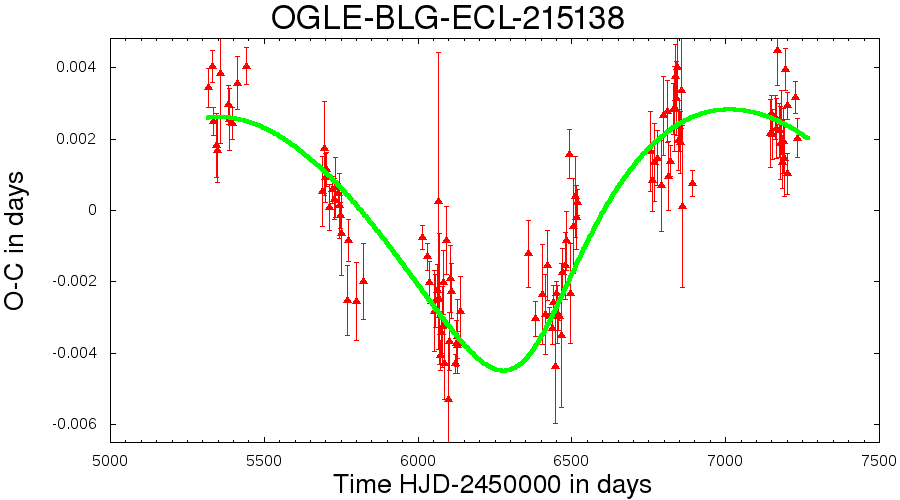}
\includegraphics[width=0.64\columnwidth]{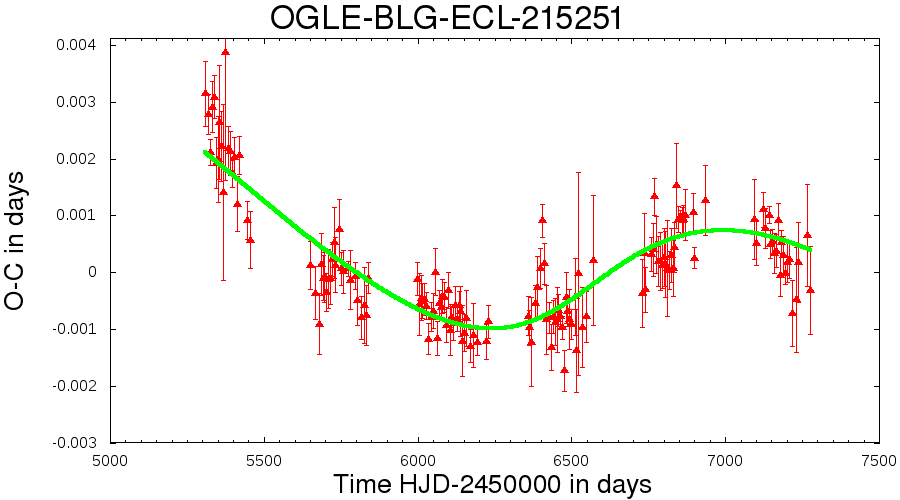}

\includegraphics[width=0.64\columnwidth]{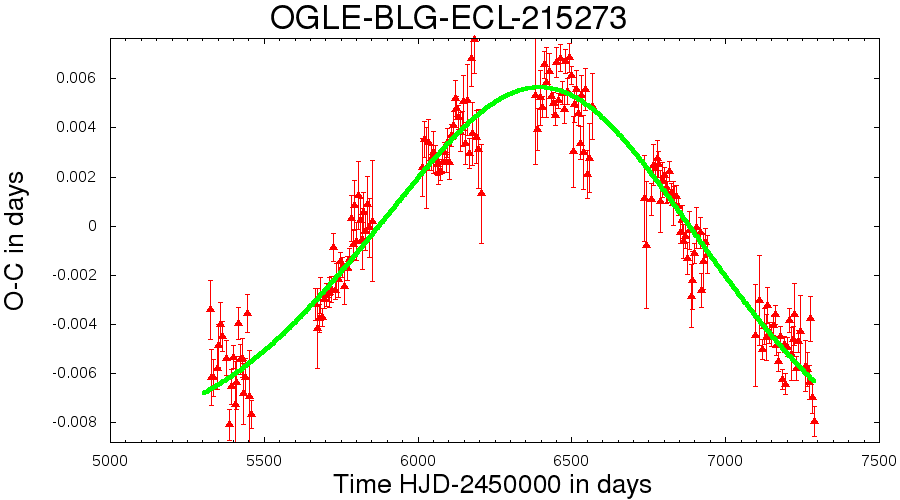}
\includegraphics[width=0.64\columnwidth]{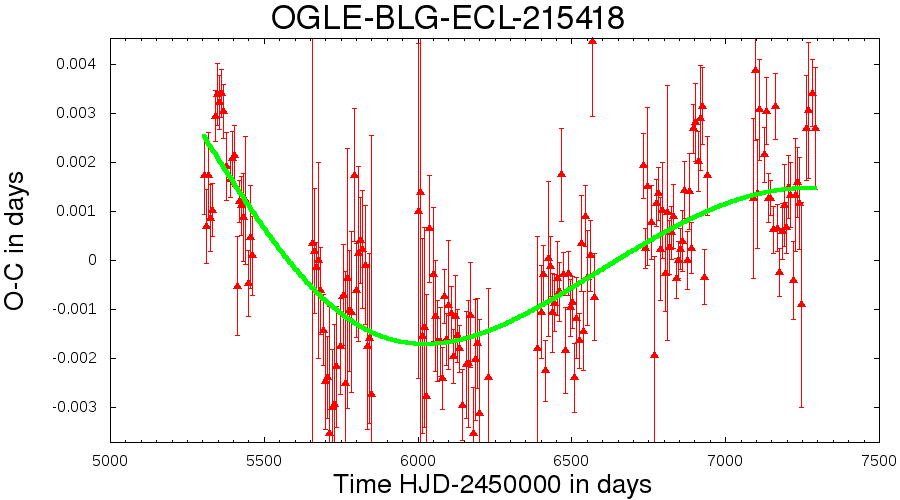}
\includegraphics[width=0.64\columnwidth]{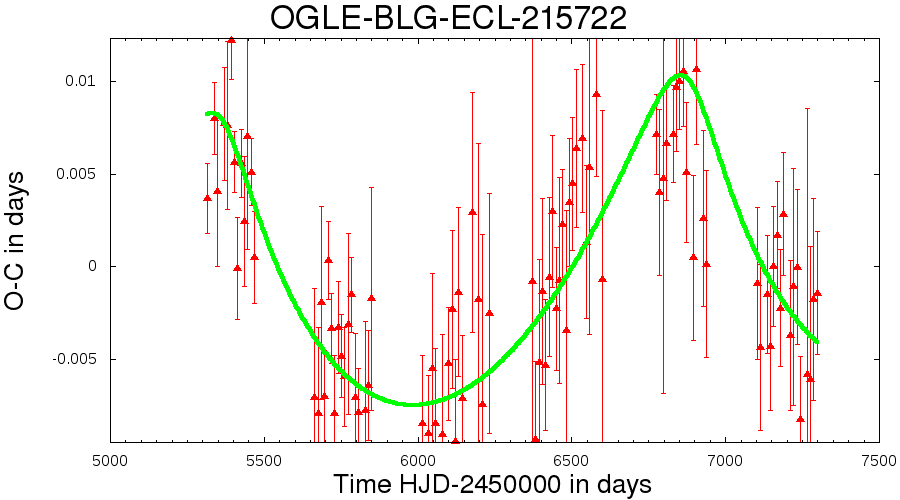}

\includegraphics[width=0.64\columnwidth]{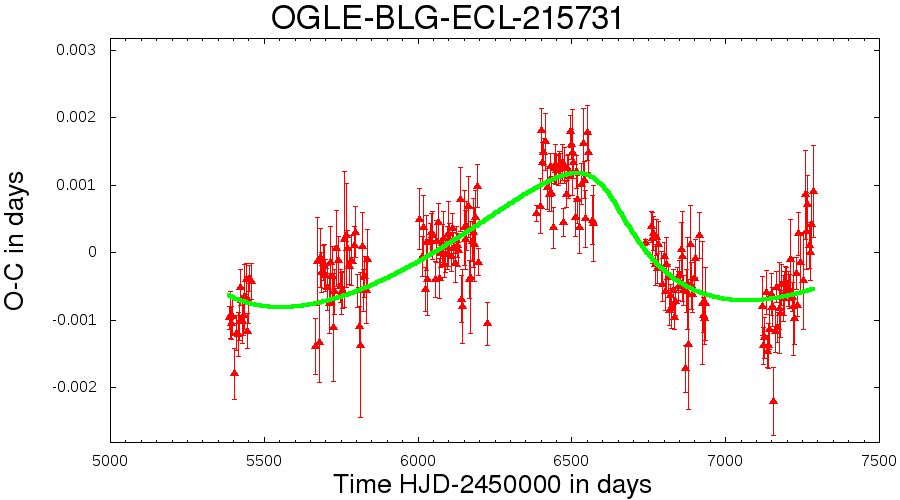}
\includegraphics[width=0.64\columnwidth]{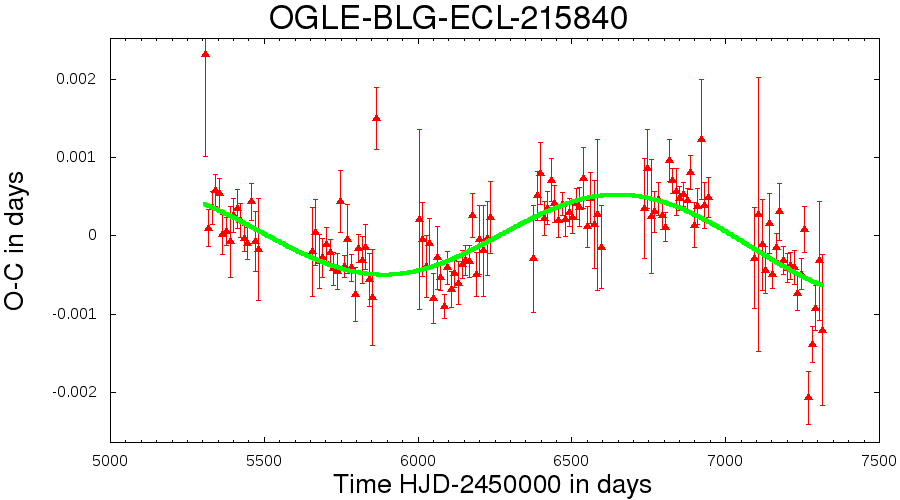}
\includegraphics[width=0.64\columnwidth]{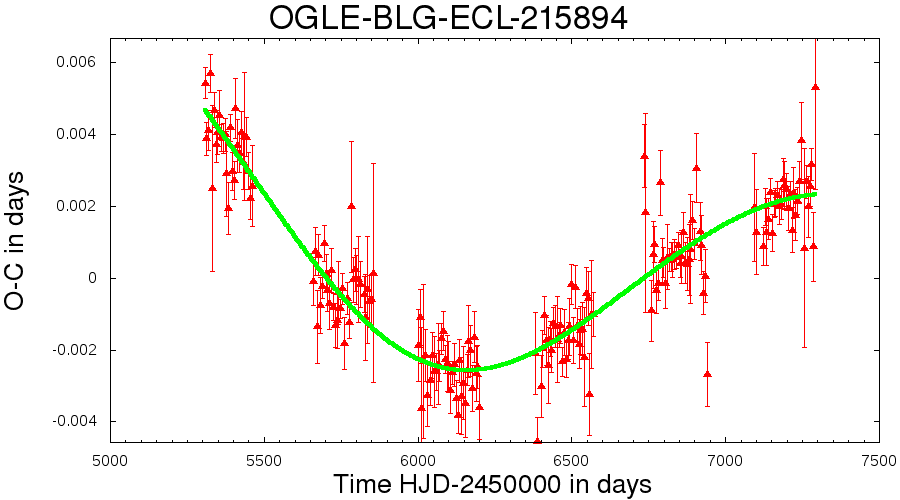}

\includegraphics[width=0.64\columnwidth]{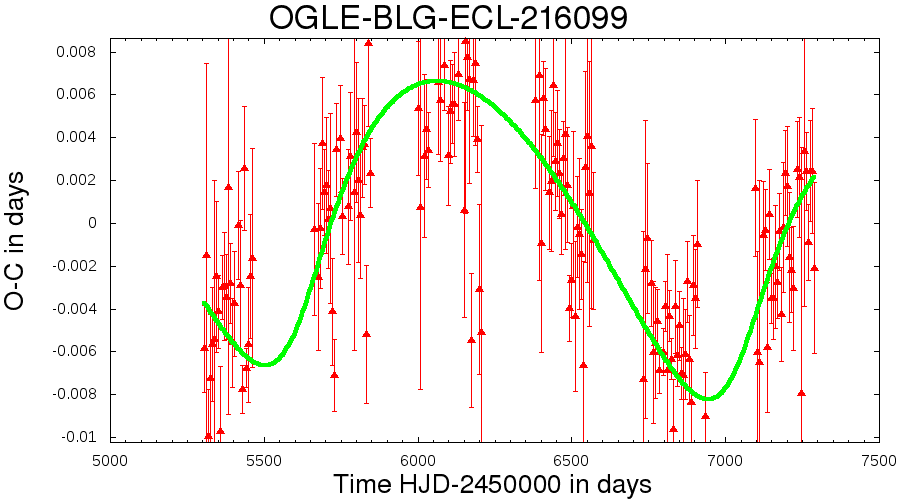}
\includegraphics[width=0.64\columnwidth]{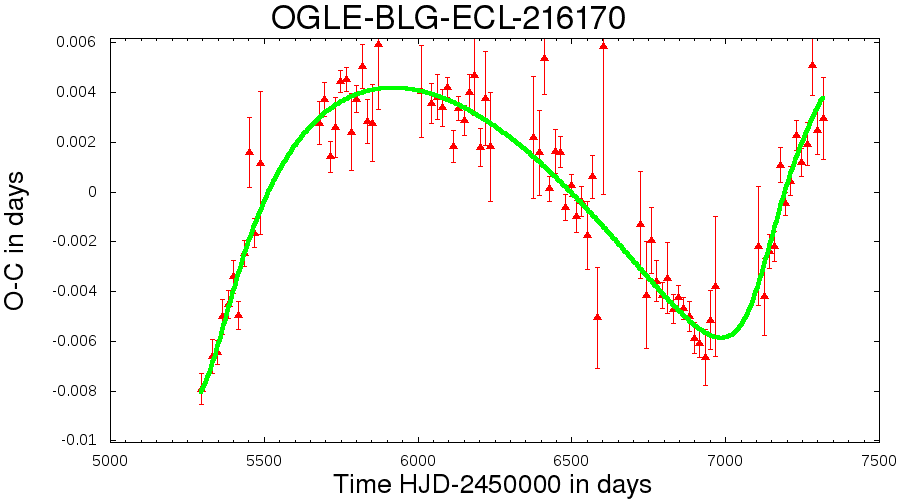}
\includegraphics[width=0.64\columnwidth]{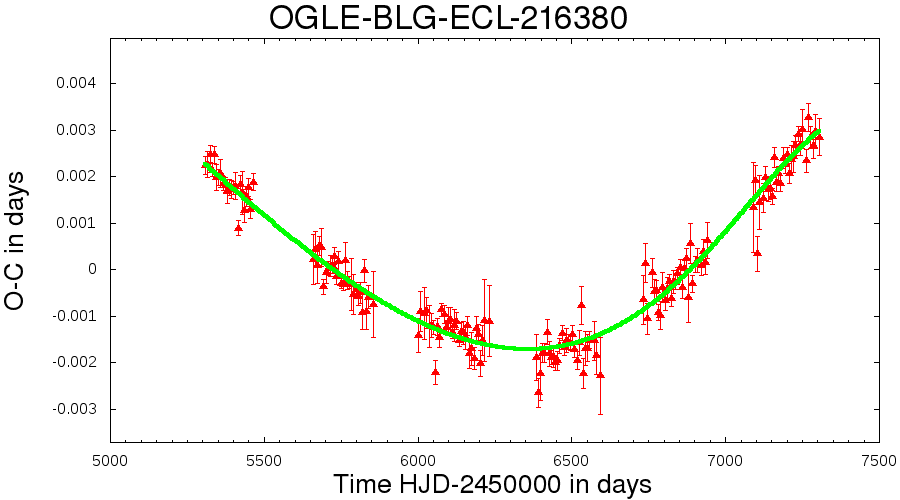}

\includegraphics[width=0.64\columnwidth]{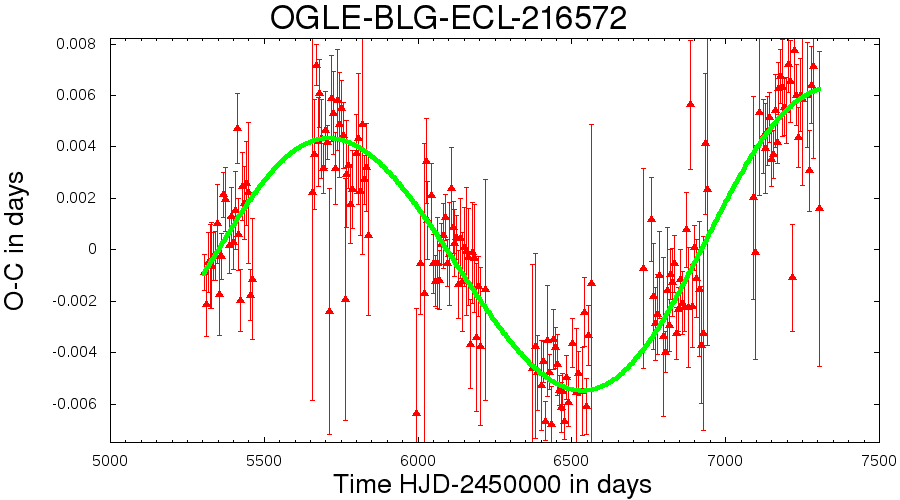}
\includegraphics[width=0.64\columnwidth]{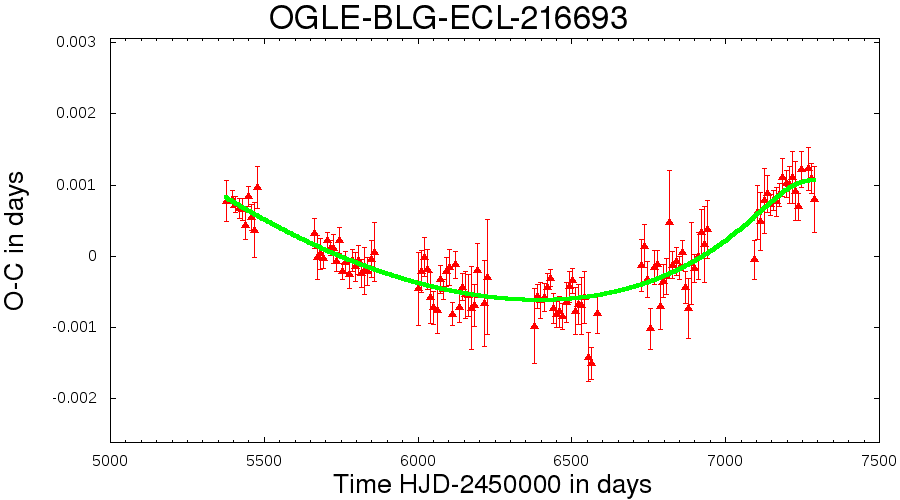}
\includegraphics[width=0.64\columnwidth]{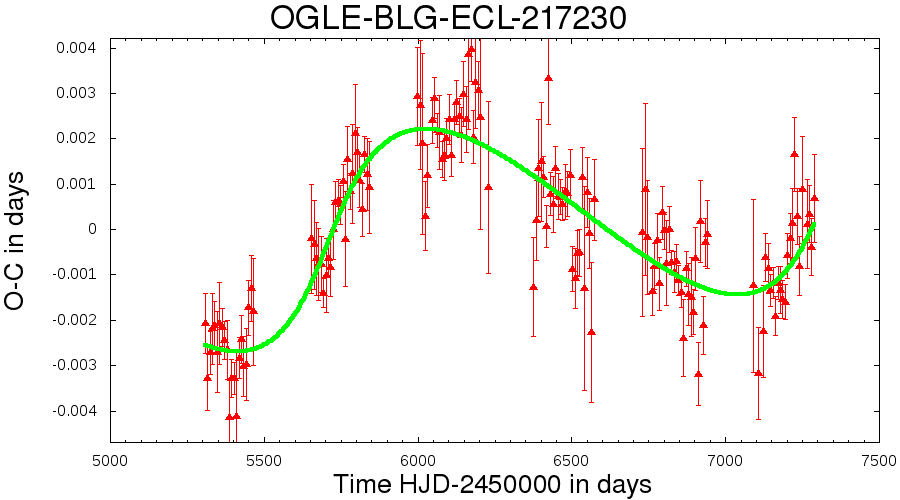}

\includegraphics[width=0.64\columnwidth]{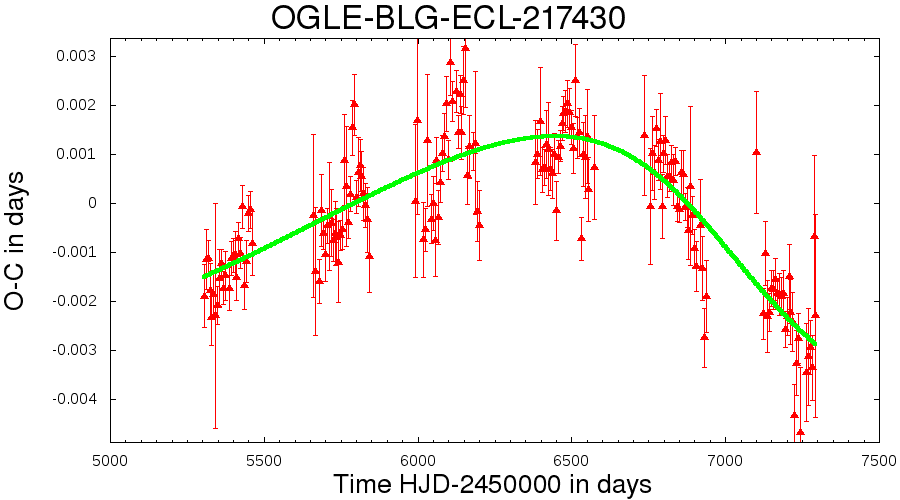}
\includegraphics[width=0.64\columnwidth]{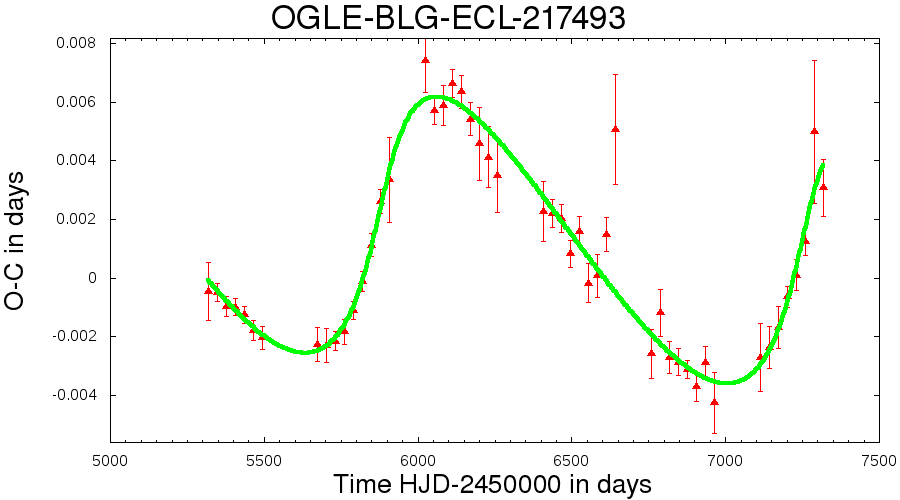}
\includegraphics[width=0.64\columnwidth]{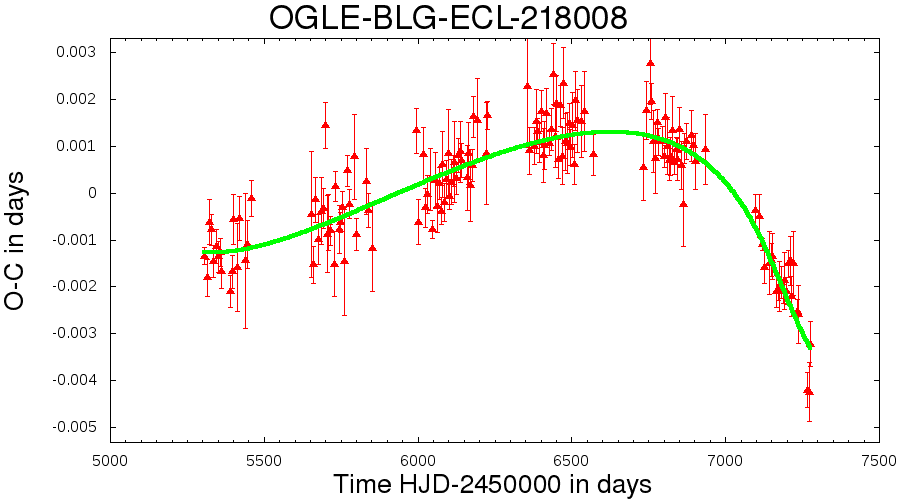}

\includegraphics[width=0.64\columnwidth]{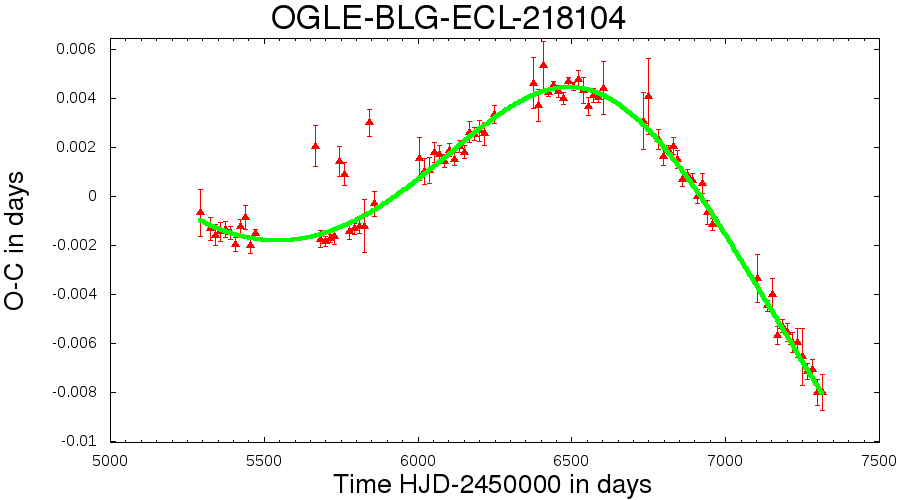}
\includegraphics[width=0.64\columnwidth]{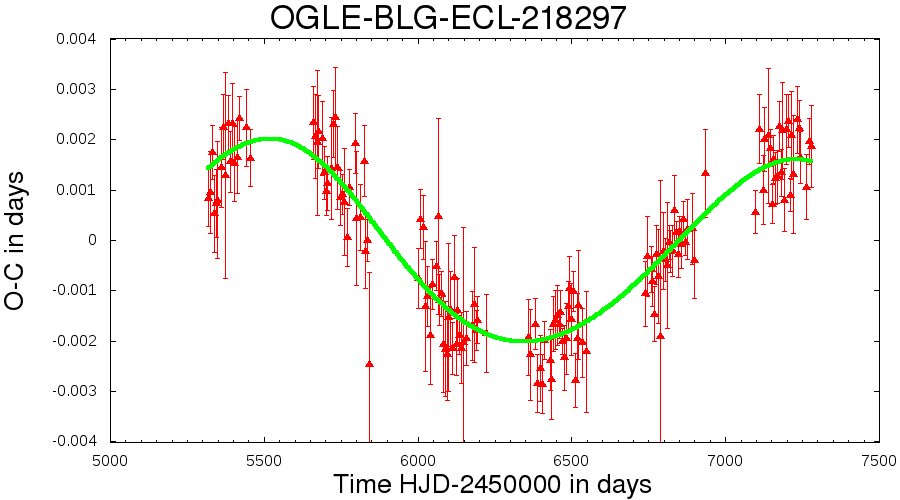}
\includegraphics[width=0.64\columnwidth]{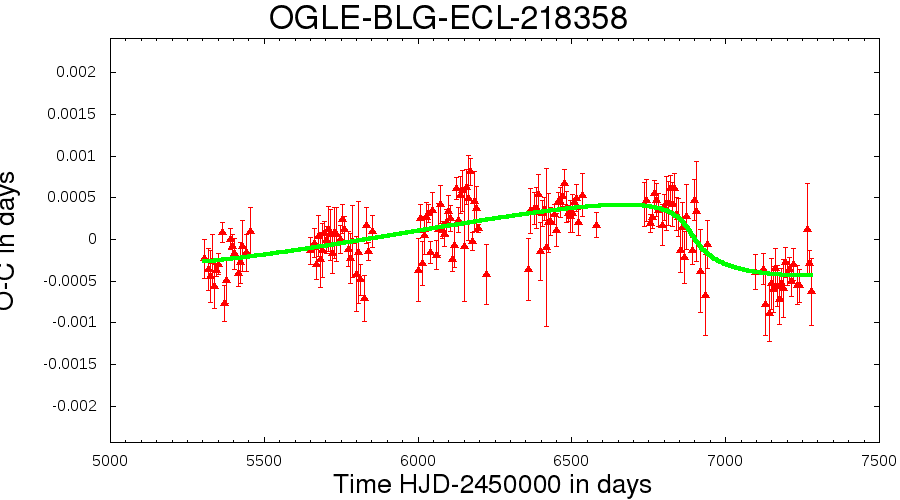}

\end{figure*}
\clearpage

\begin{figure*}
\includegraphics[width=0.64\columnwidth]{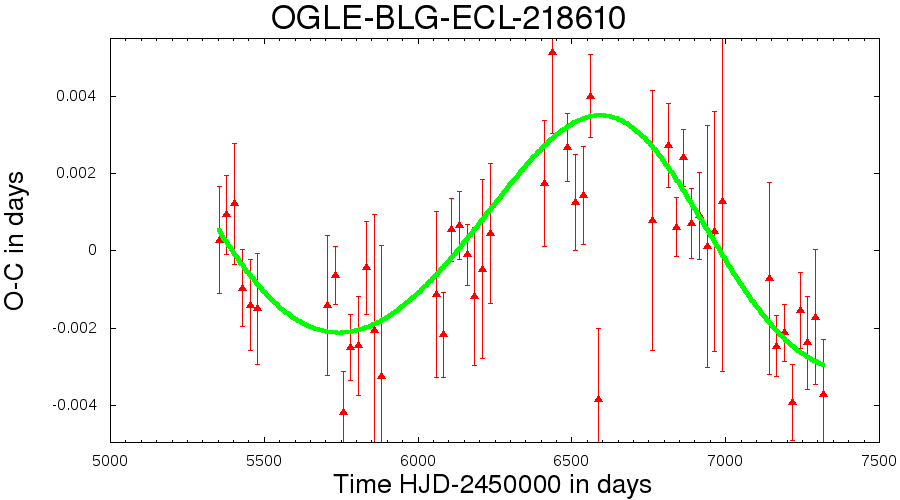}
\includegraphics[width=0.64\columnwidth]{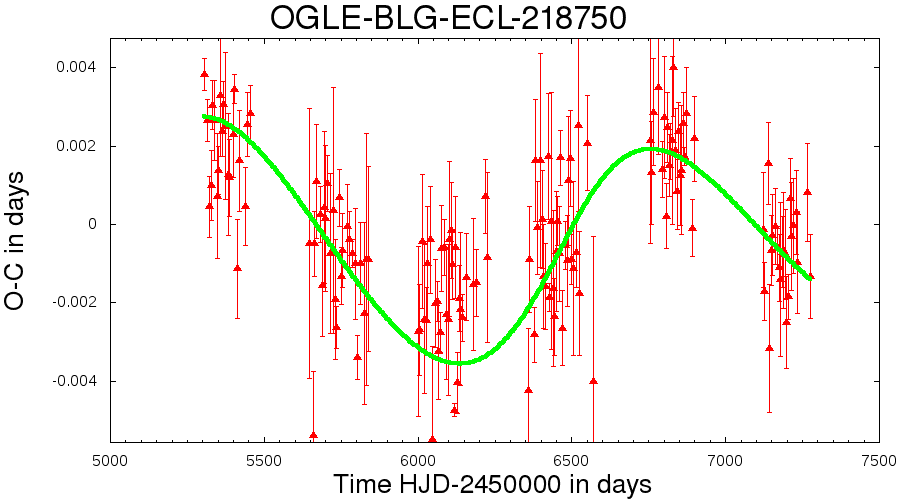}
\includegraphics[width=0.64\columnwidth]{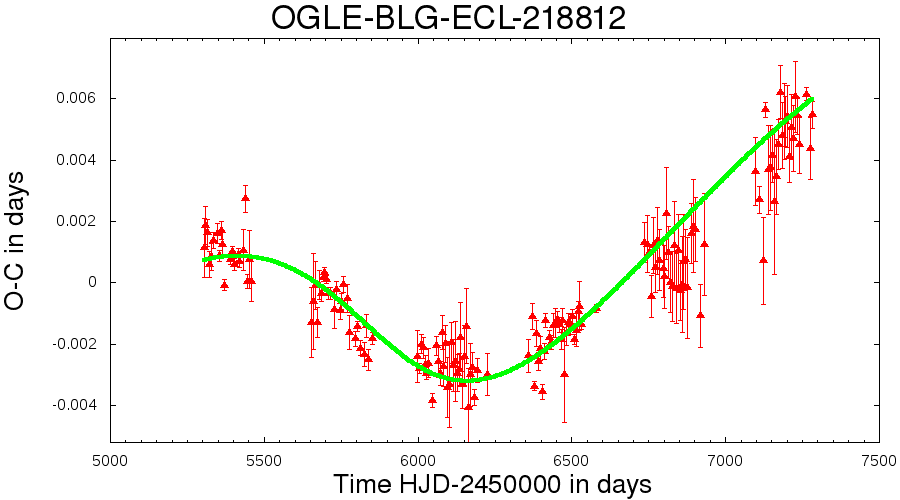}

\includegraphics[width=0.64\columnwidth]{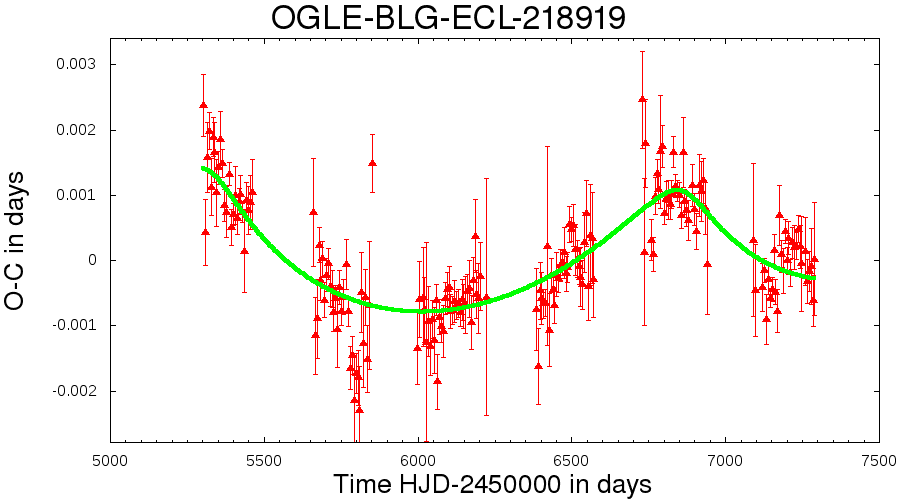}
\includegraphics[width=0.64\columnwidth]{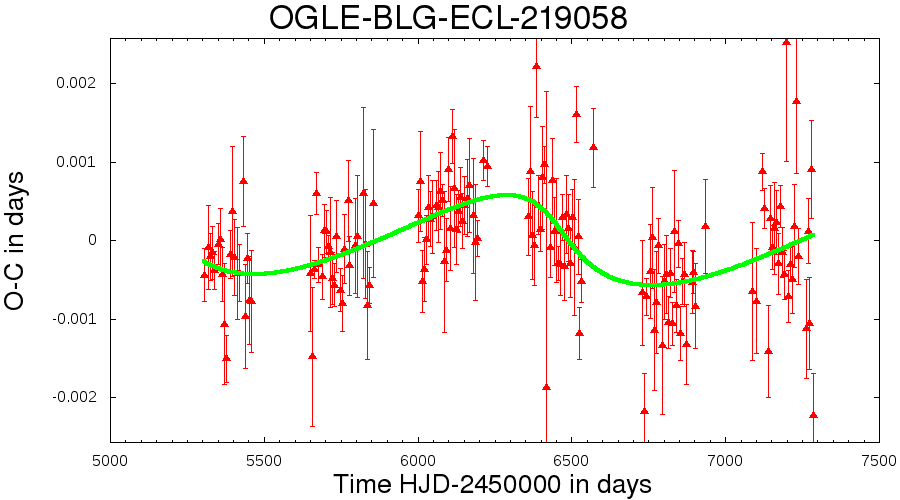}
\includegraphics[width=0.64\columnwidth]{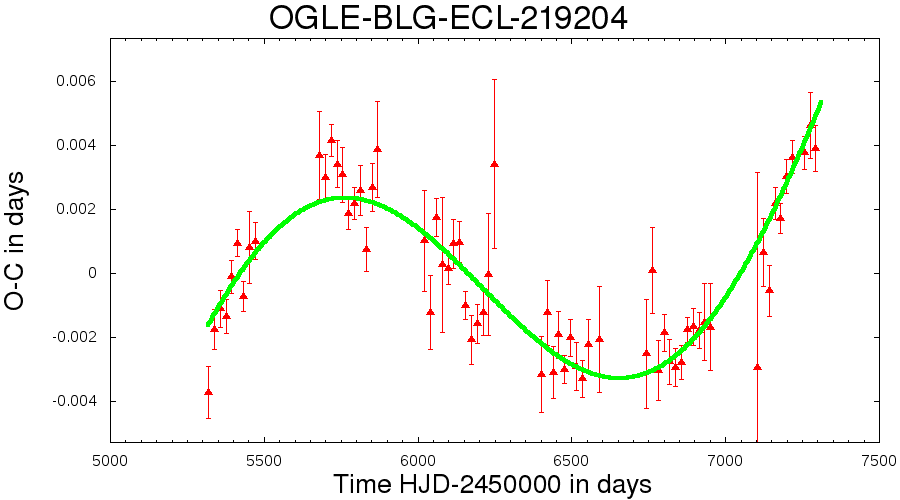}

\includegraphics[width=0.64\columnwidth]{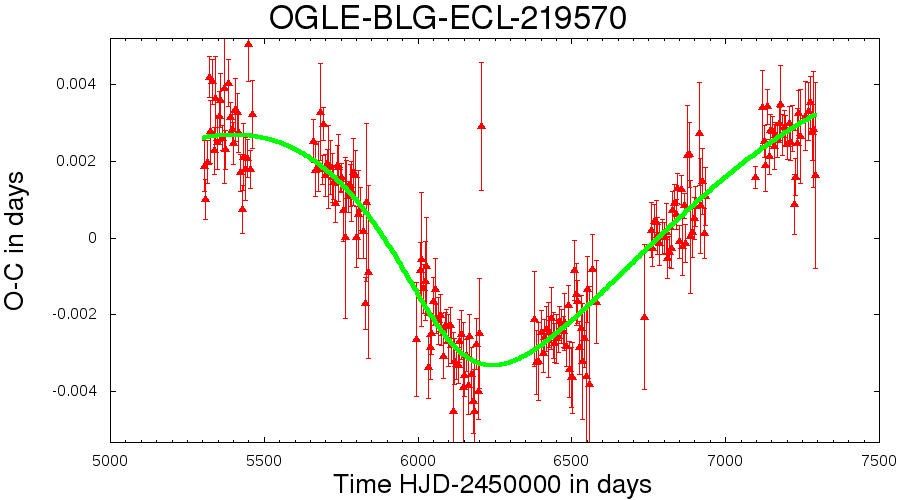}
\includegraphics[width=0.64\columnwidth]{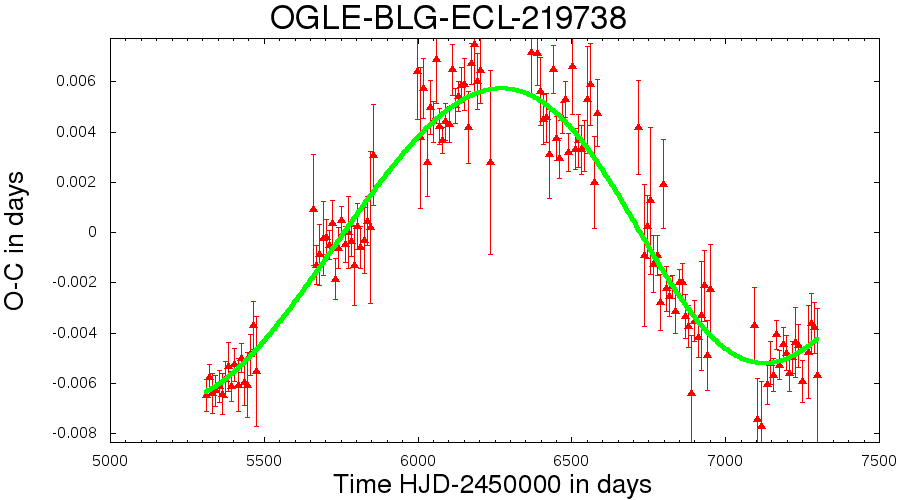}
\includegraphics[width=0.64\columnwidth]{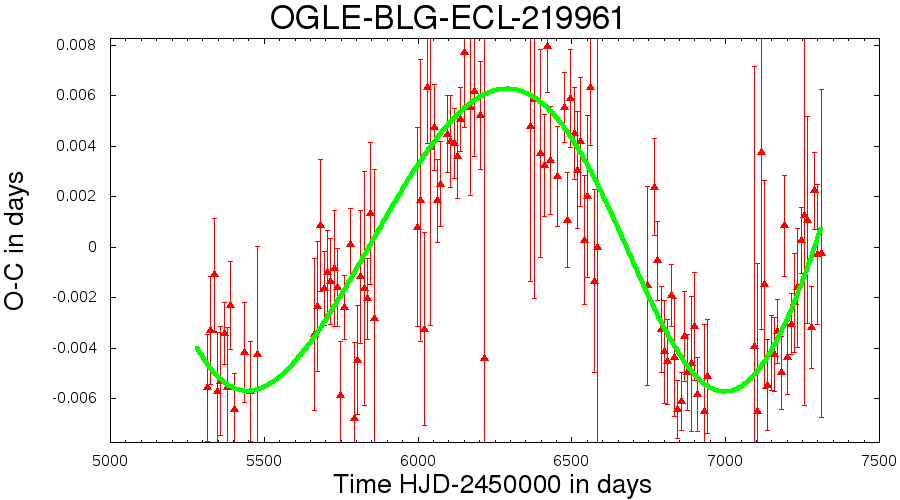}

\includegraphics[width=0.64\columnwidth]{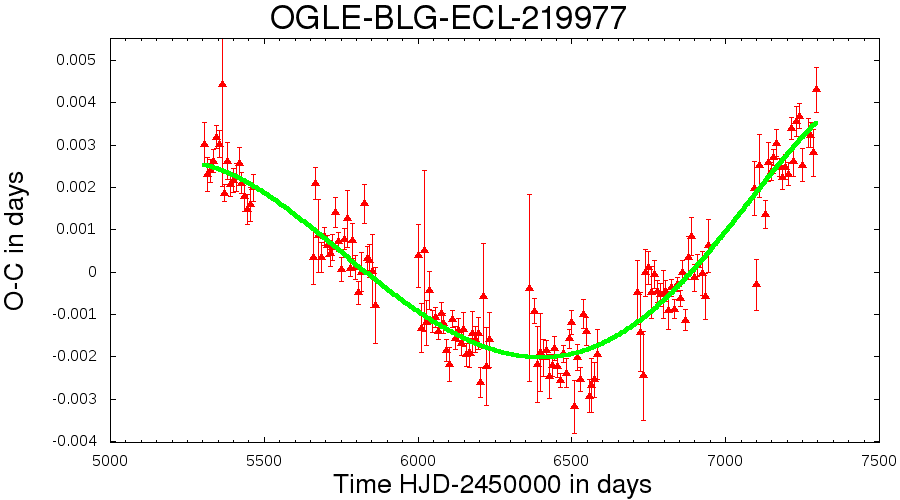}
\includegraphics[width=0.64\columnwidth]{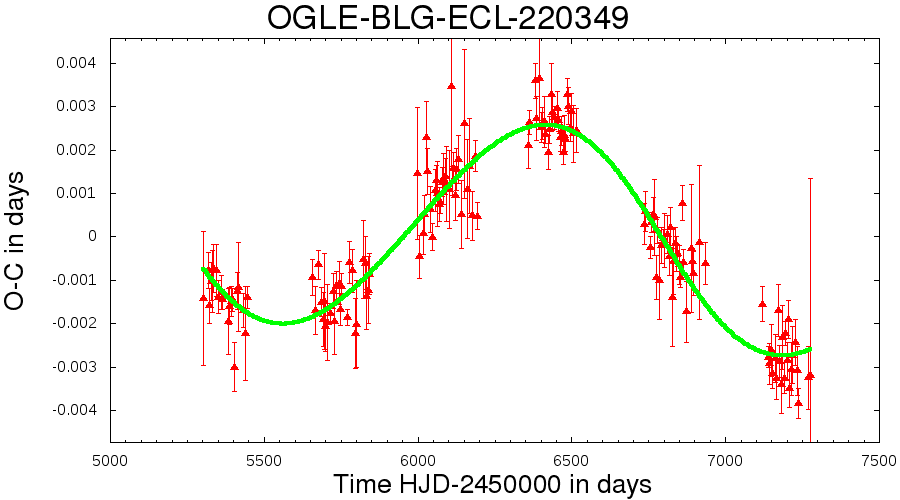}
\includegraphics[width=0.64\columnwidth]{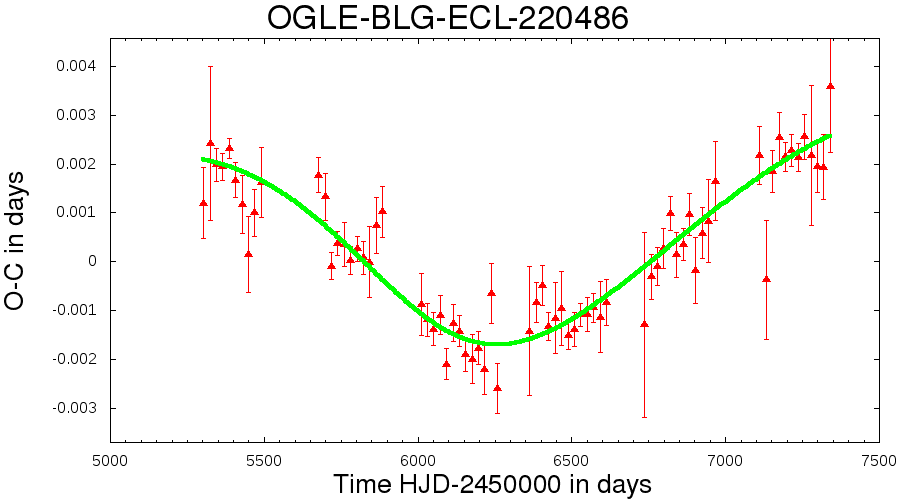}

\includegraphics[width=0.64\columnwidth]{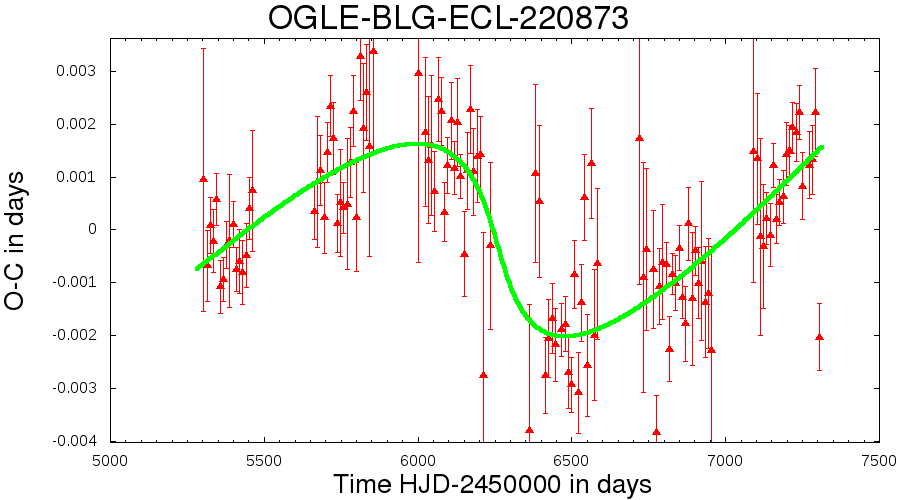}
\includegraphics[width=0.64\columnwidth]{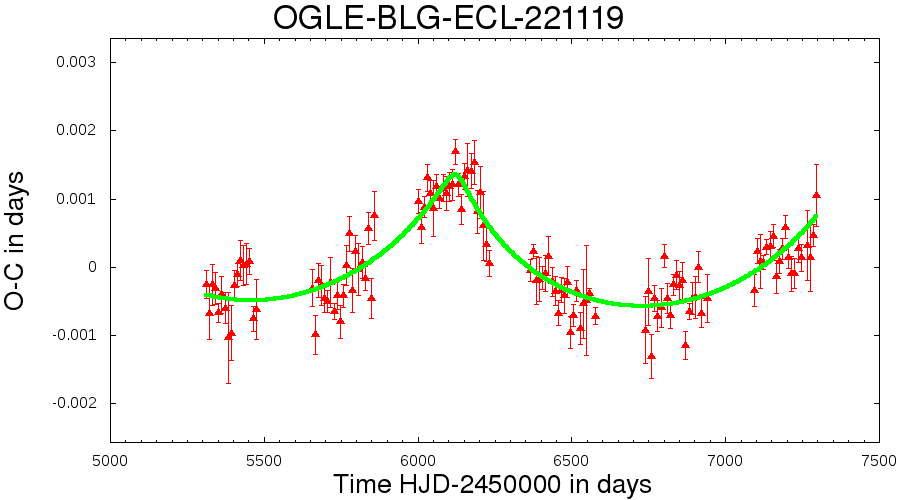}
\includegraphics[width=0.64\columnwidth]{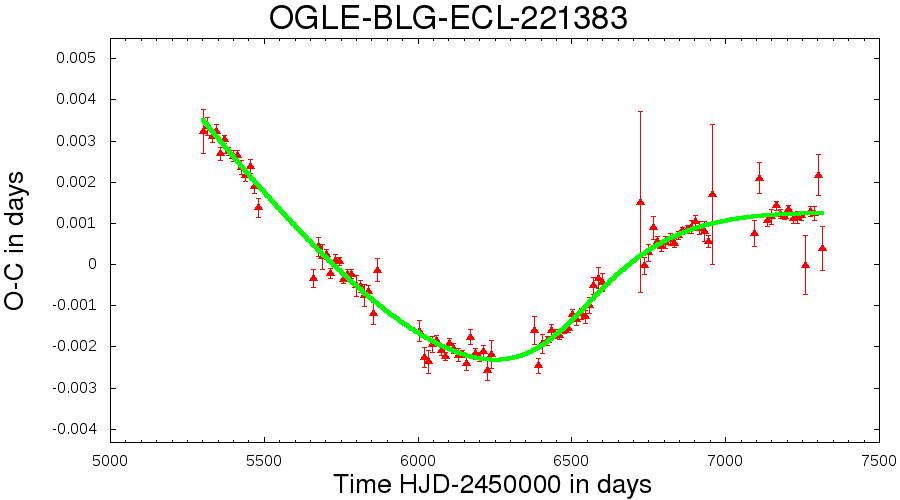}

\includegraphics[width=0.64\columnwidth]{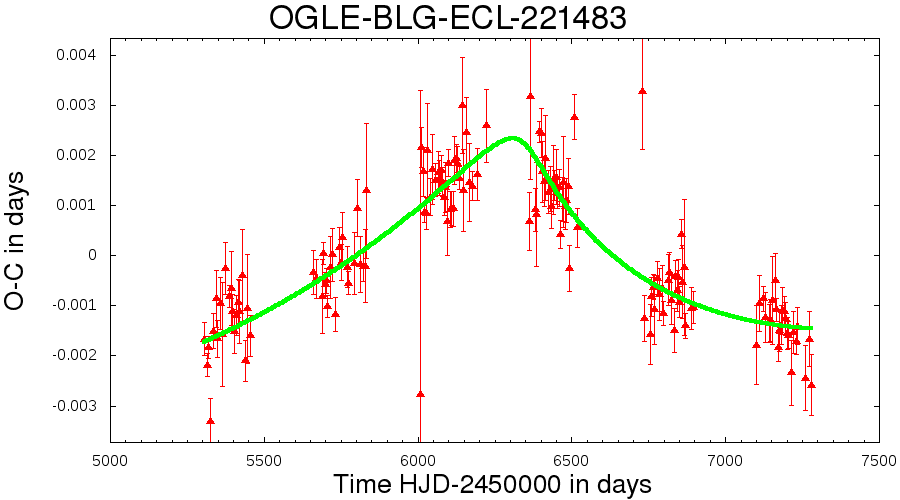}
\includegraphics[width=0.64\columnwidth]{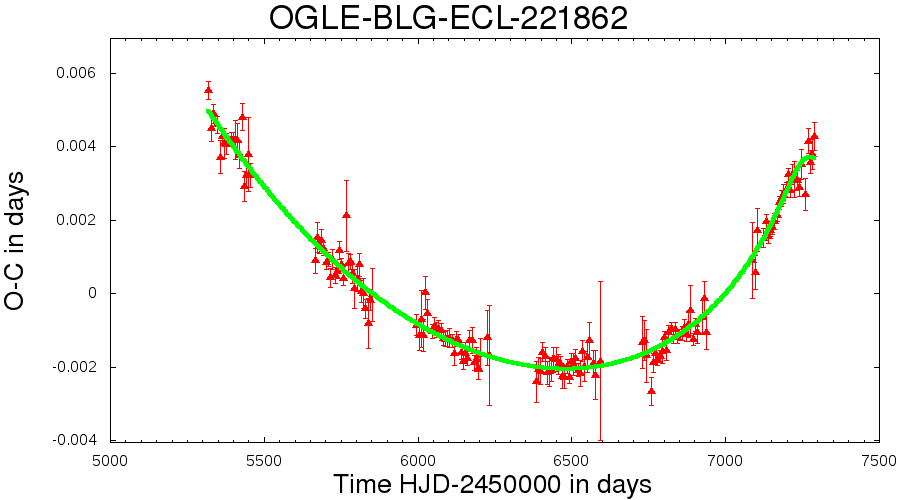}
\includegraphics[width=0.64\columnwidth]{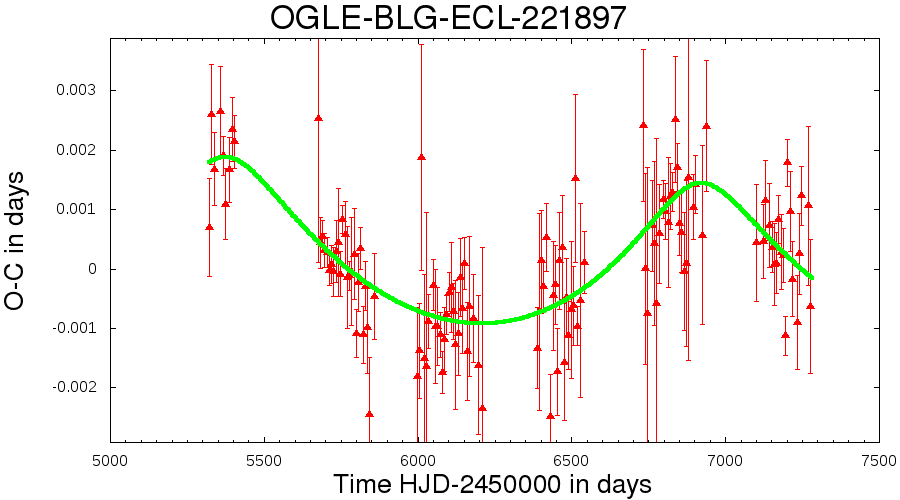}

\includegraphics[width=0.64\columnwidth]{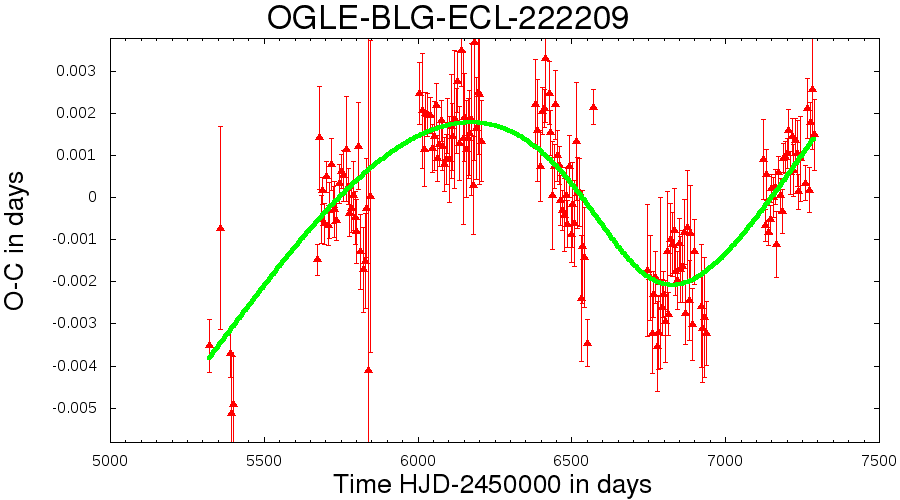}
\includegraphics[width=0.64\columnwidth]{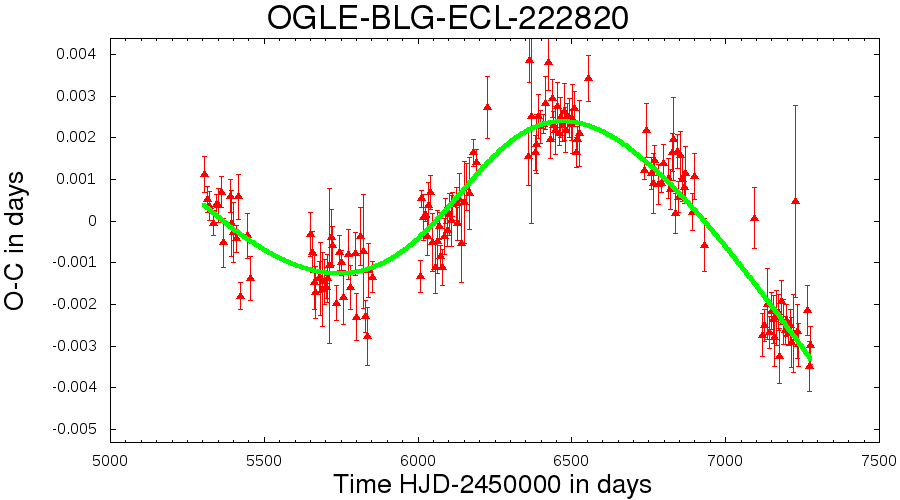}
\includegraphics[width=0.64\columnwidth]{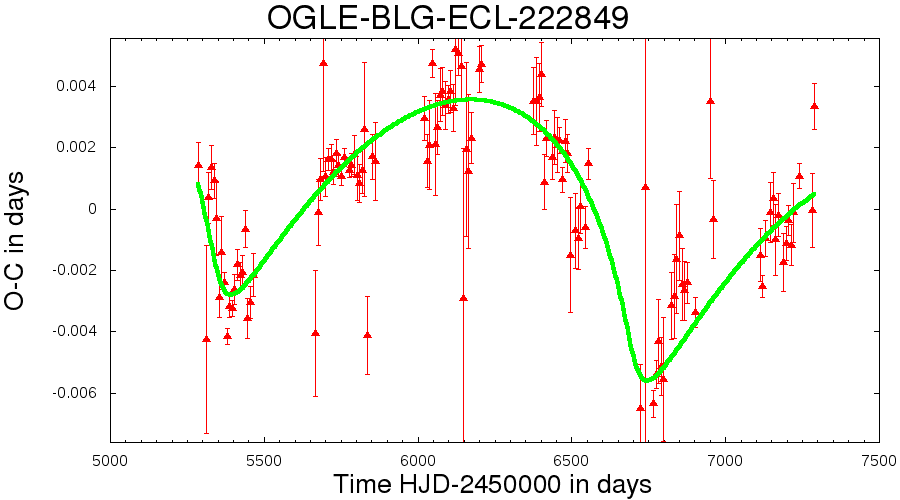}

\includegraphics[width=0.64\columnwidth]{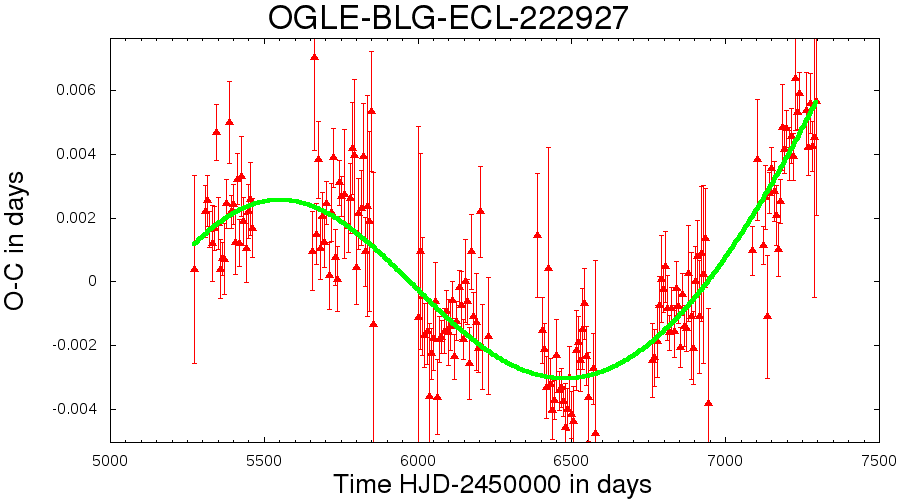}
\includegraphics[width=0.64\columnwidth]{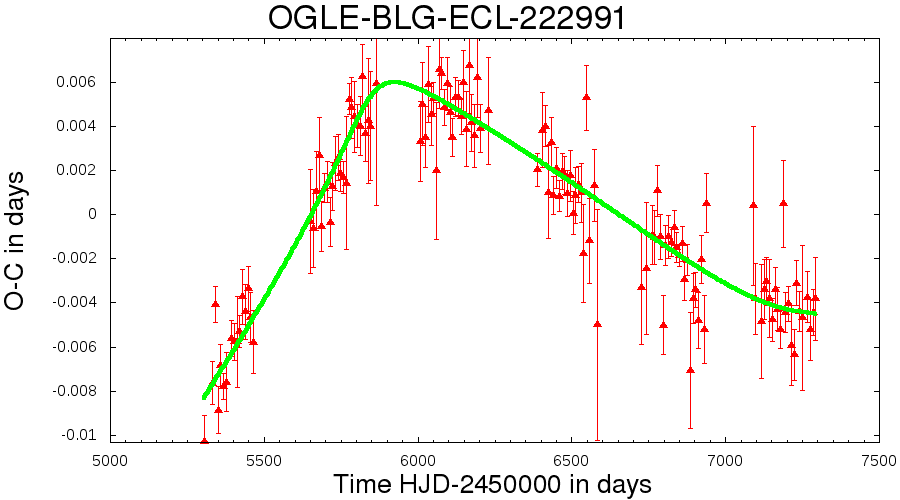}
\includegraphics[width=0.64\columnwidth]{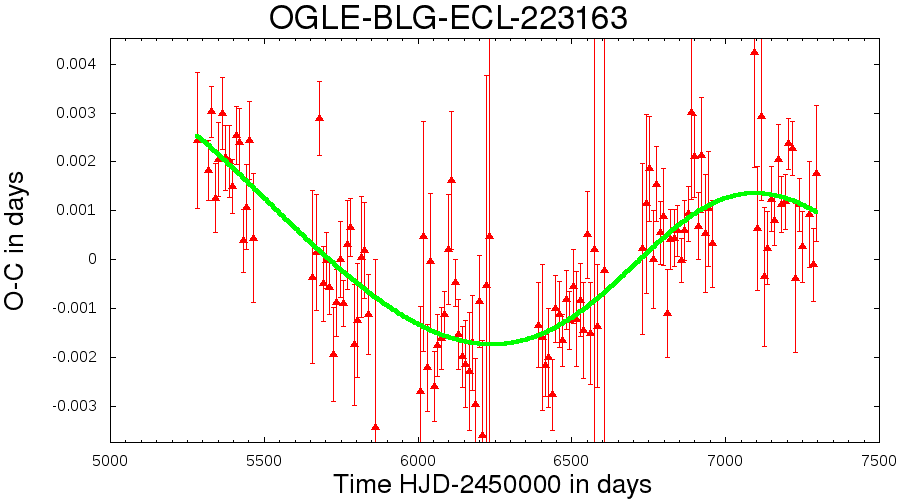}

\end{figure*}
\clearpage

\begin{figure*}
\includegraphics[width=0.64\columnwidth]{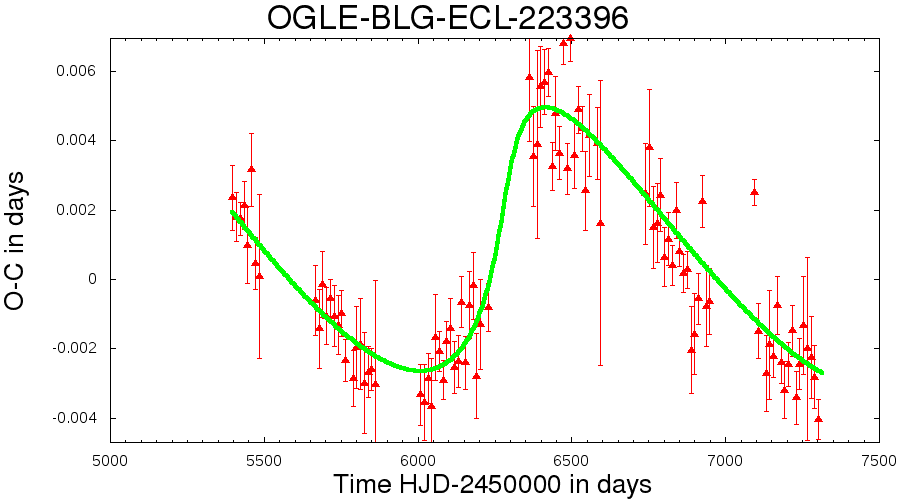}
\includegraphics[width=0.64\columnwidth]{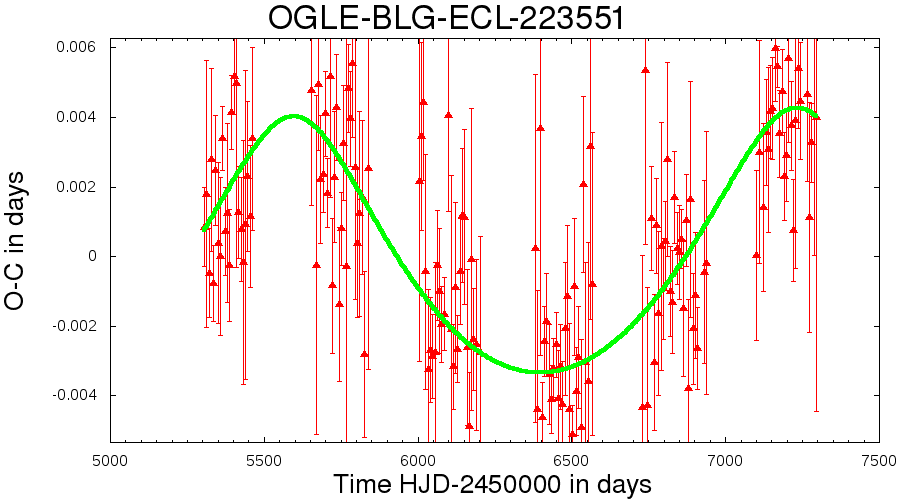}
\includegraphics[width=0.64\columnwidth]{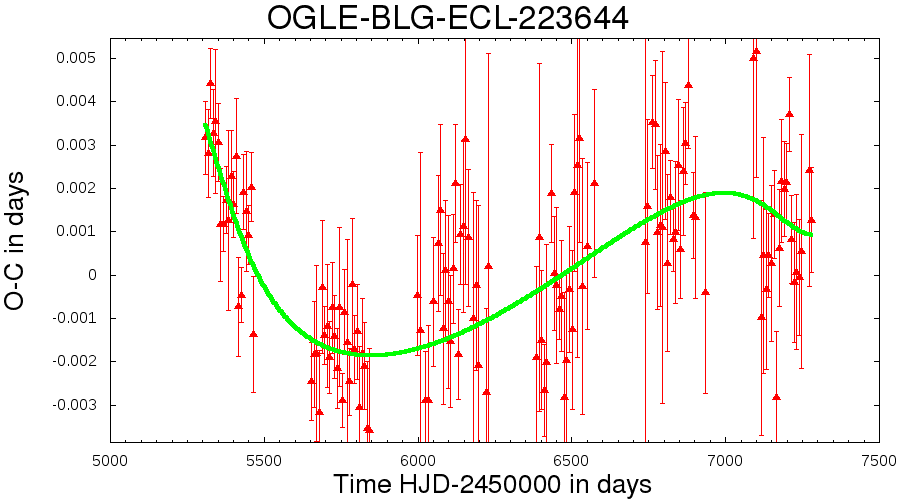}

\includegraphics[width=0.64\columnwidth]{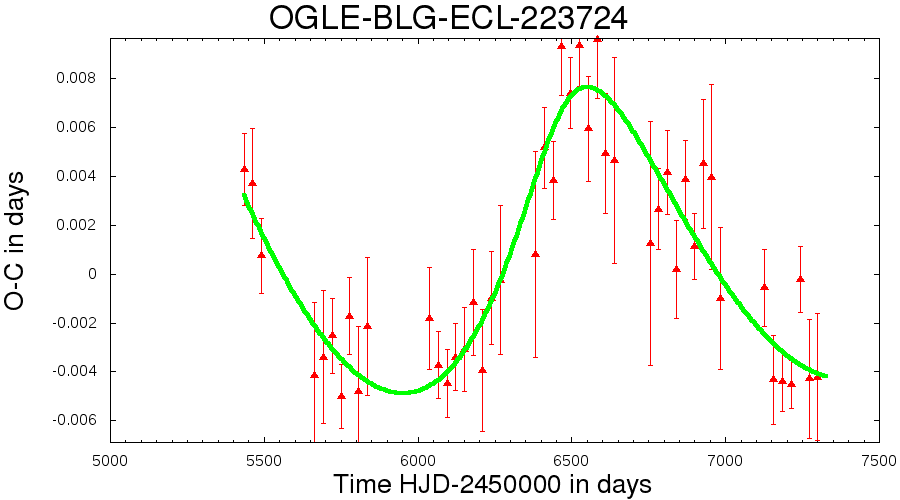}
\includegraphics[width=0.64\columnwidth]{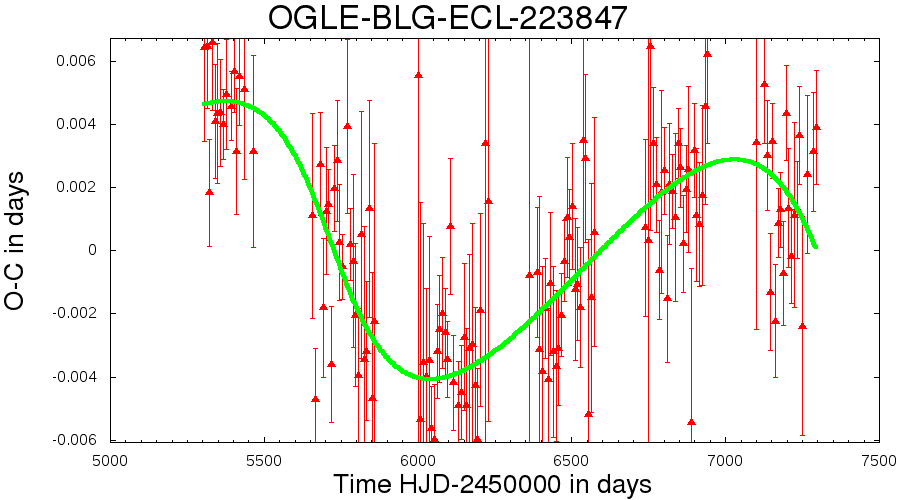}
\includegraphics[width=0.64\columnwidth]{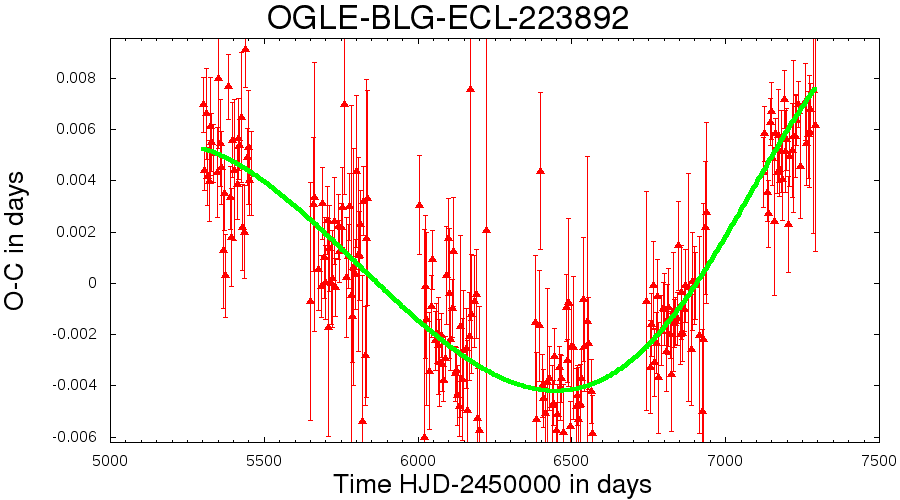}

\includegraphics[width=0.64\columnwidth]{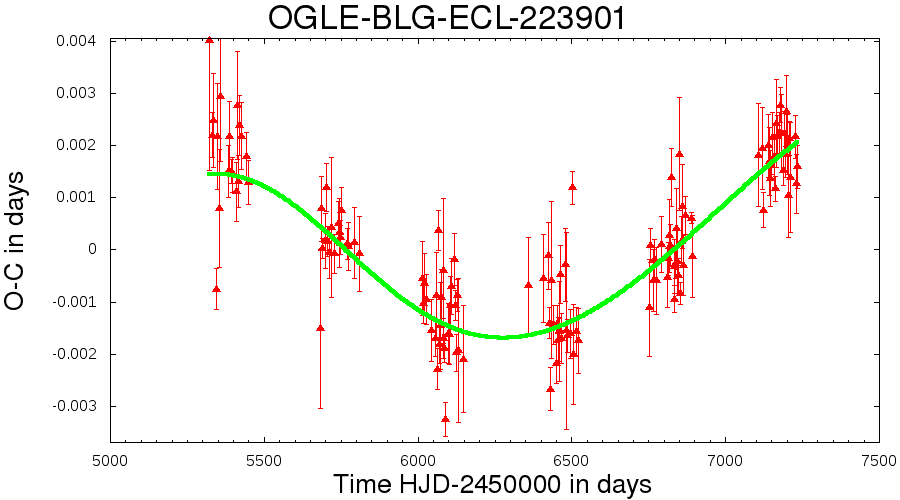}
\includegraphics[width=0.64\columnwidth]{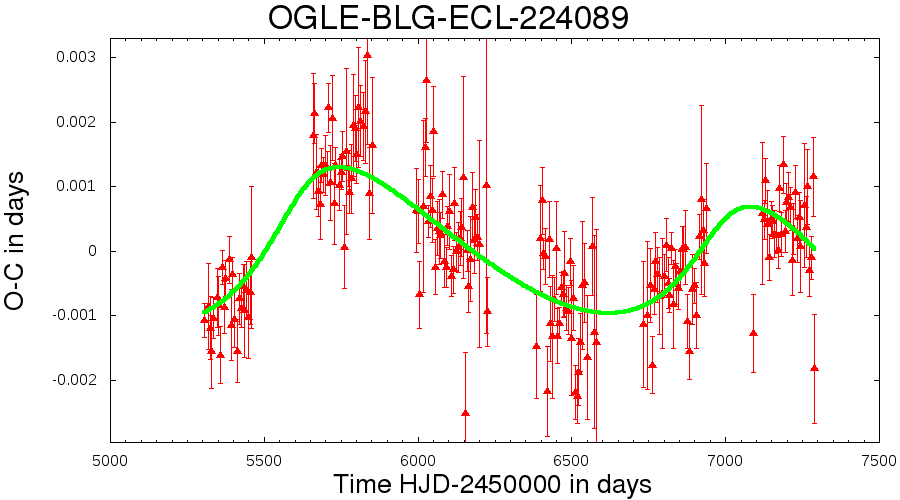}
\includegraphics[width=0.64\columnwidth]{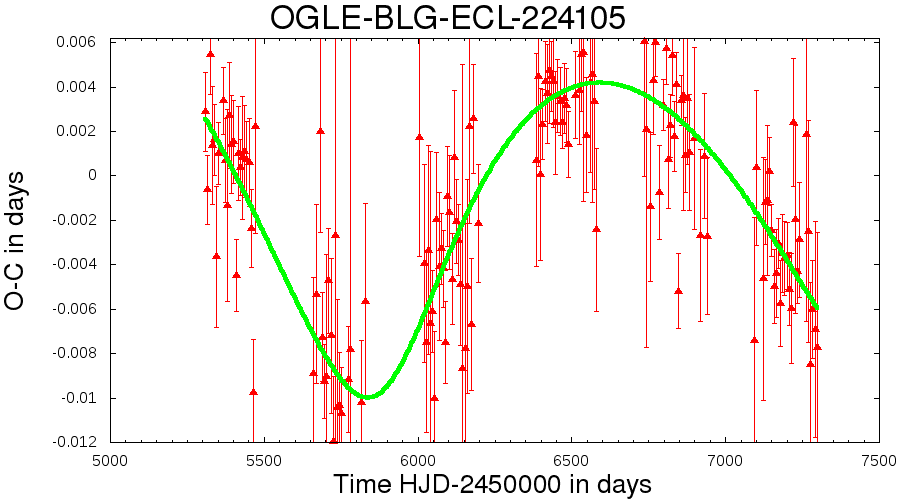}

\includegraphics[width=0.64\columnwidth]{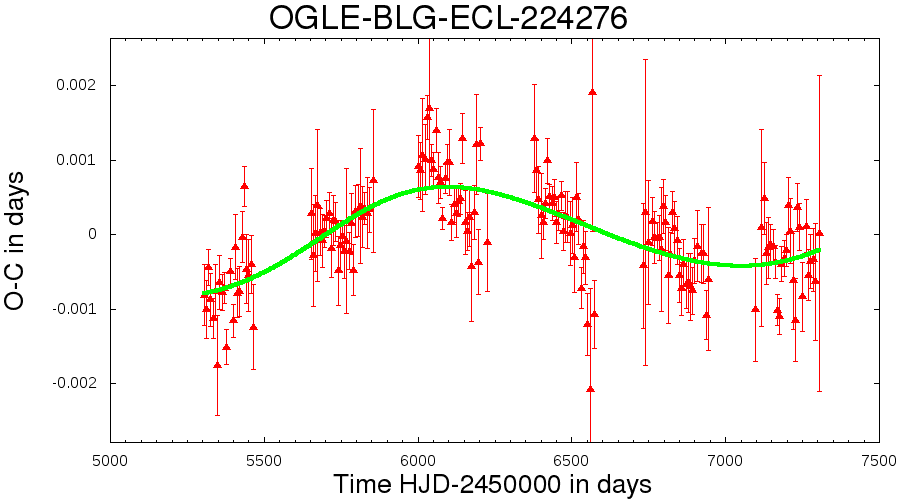}
\includegraphics[width=0.64\columnwidth]{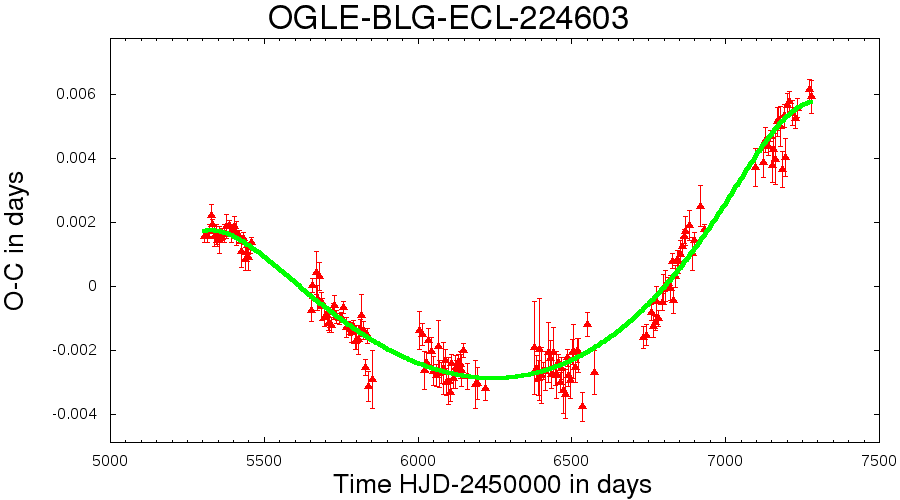}
\includegraphics[width=0.64\columnwidth]{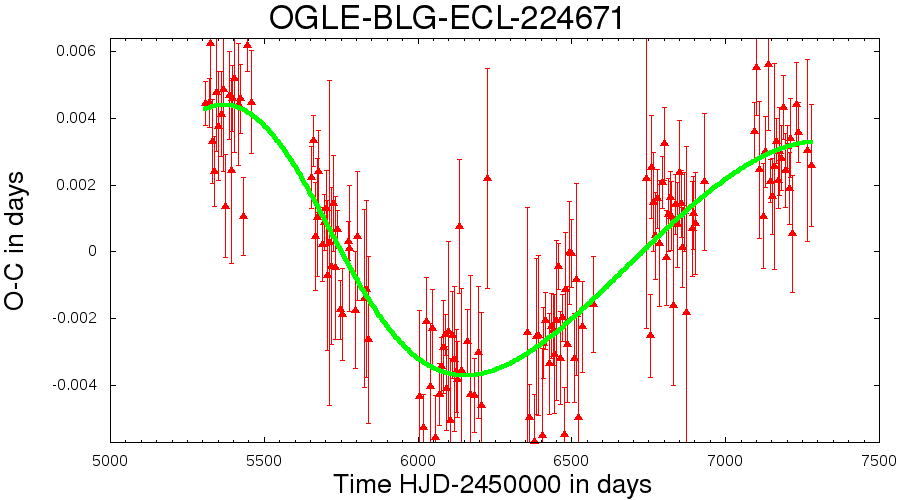}

\includegraphics[width=0.64\columnwidth]{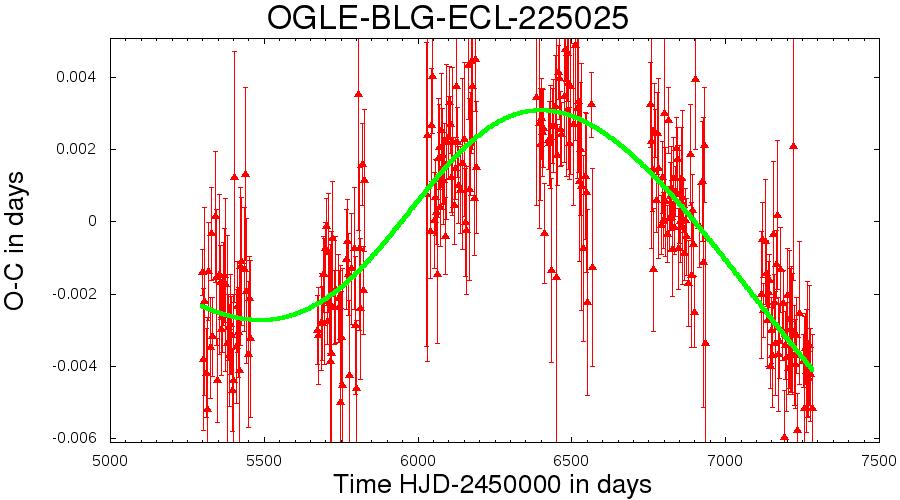}
\includegraphics[width=0.64\columnwidth]{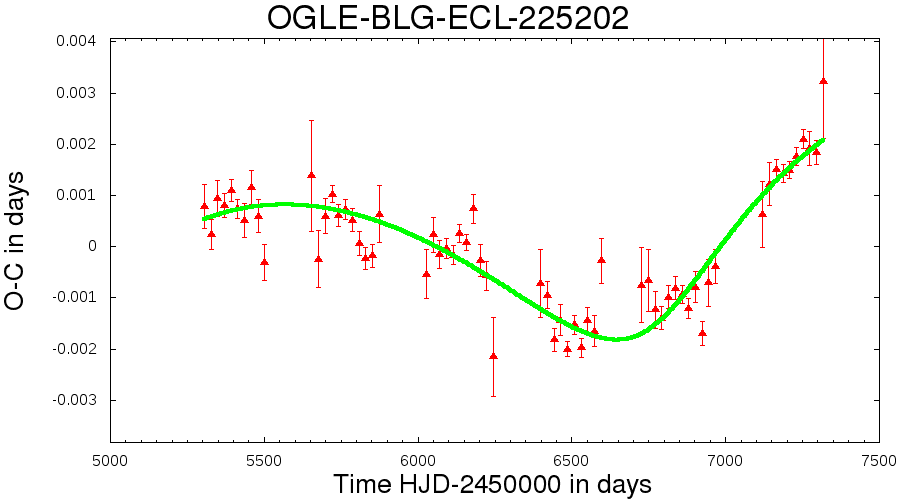}
\includegraphics[width=0.64\columnwidth]{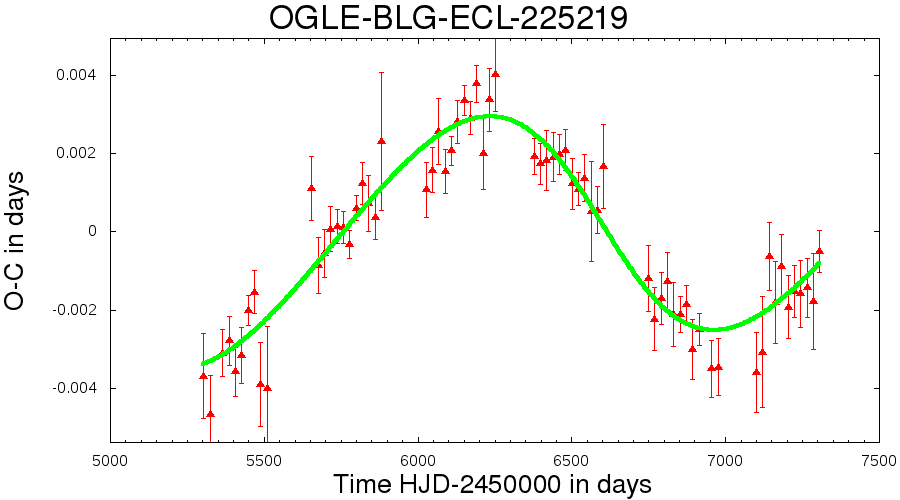}

\includegraphics[width=0.64\columnwidth]{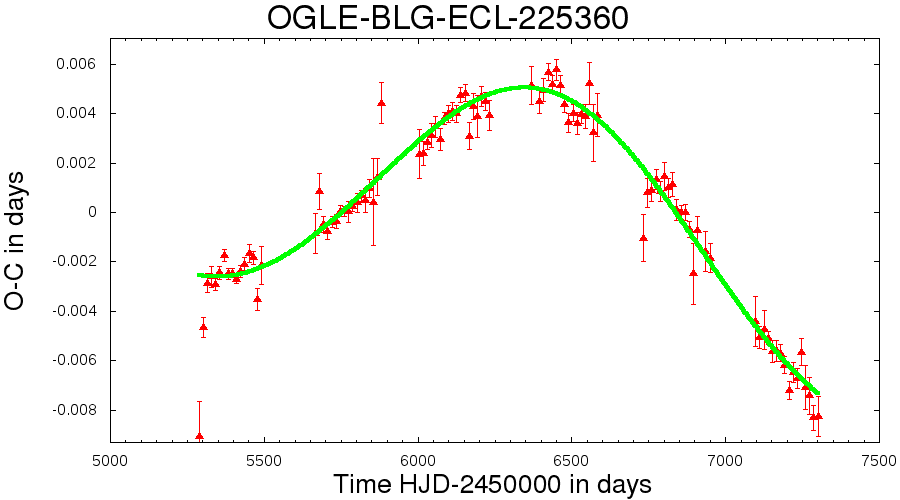}
\includegraphics[width=0.64\columnwidth]{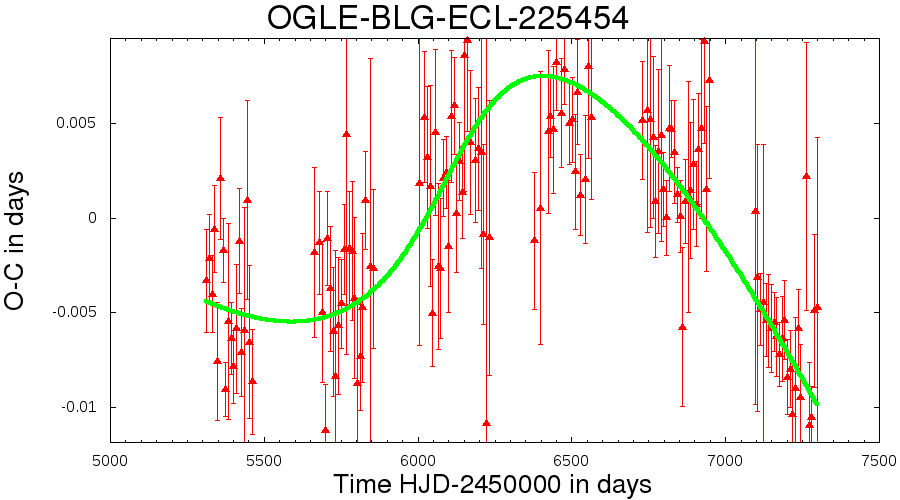}
\includegraphics[width=0.64\columnwidth]{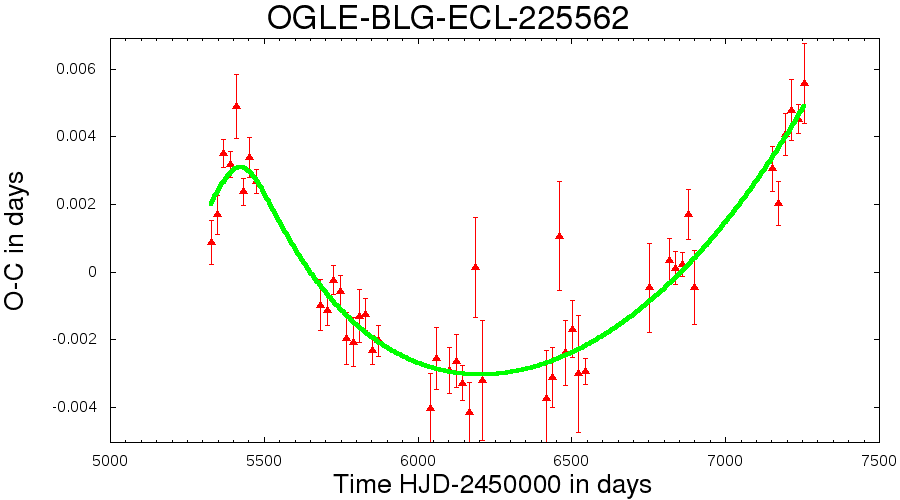}

\includegraphics[width=0.64\columnwidth]{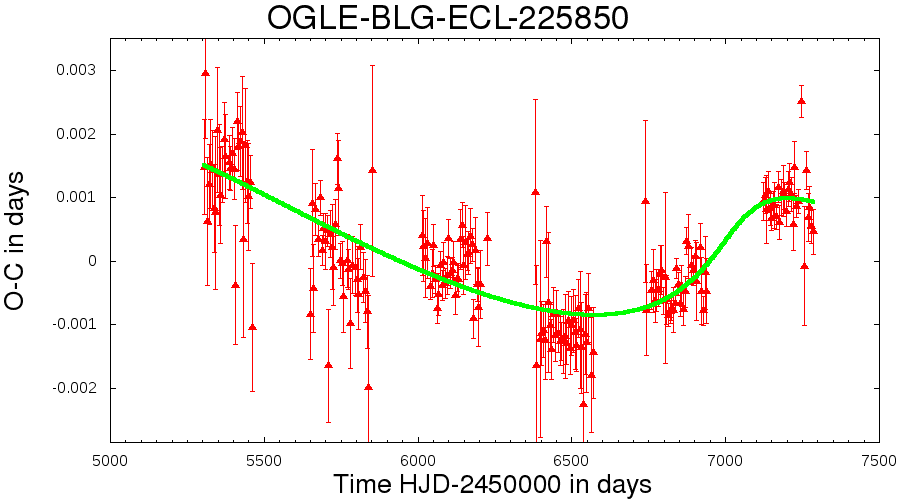}
\includegraphics[width=0.64\columnwidth]{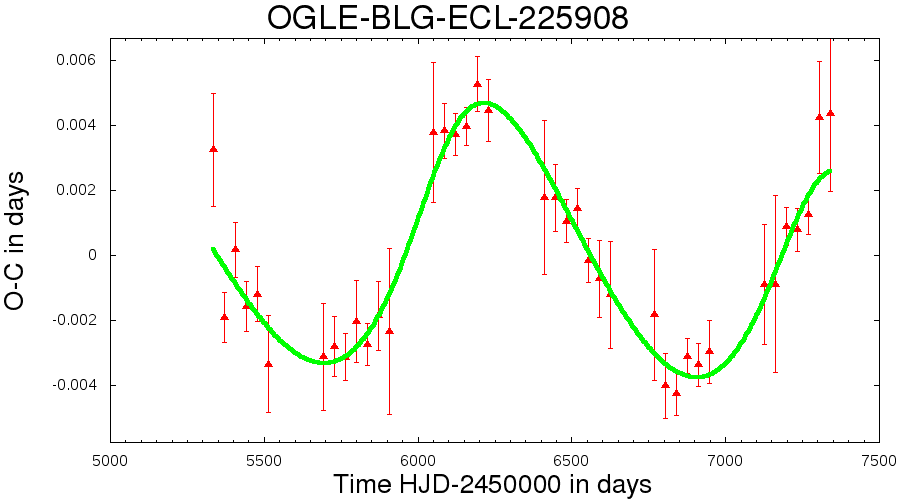}
\includegraphics[width=0.64\columnwidth]{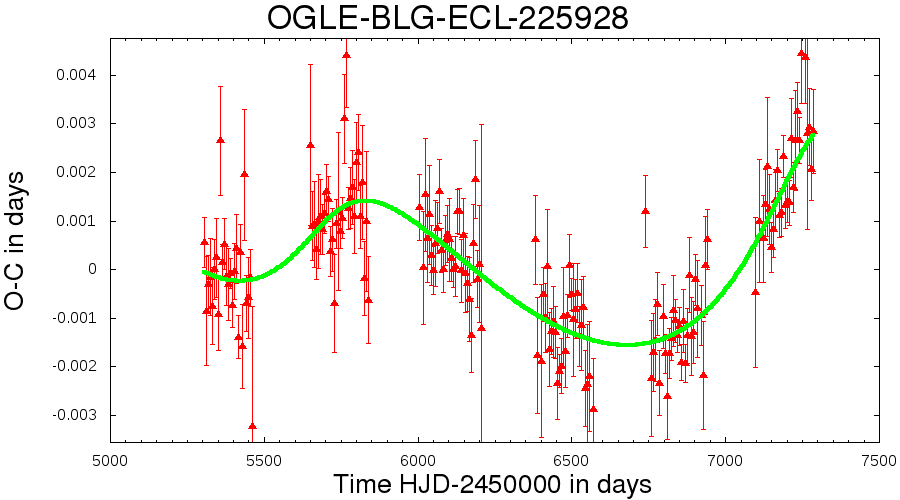}

\includegraphics[width=0.64\columnwidth]{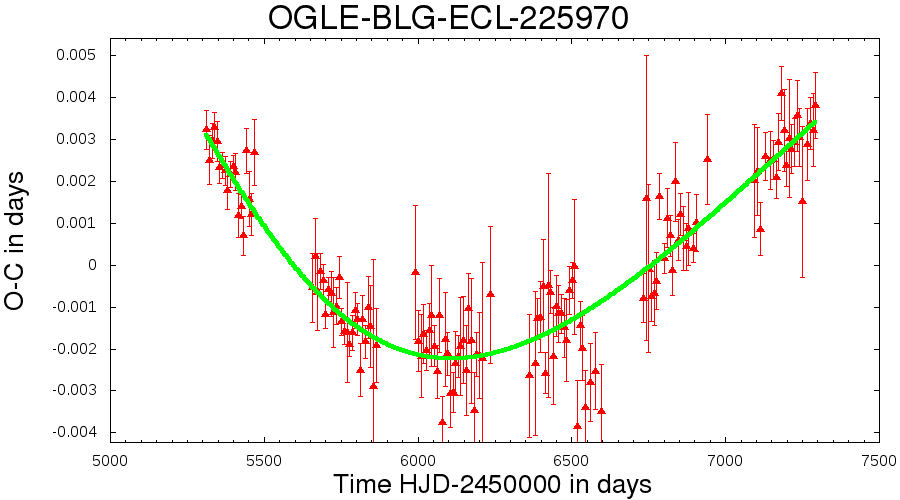}
\includegraphics[width=0.64\columnwidth]{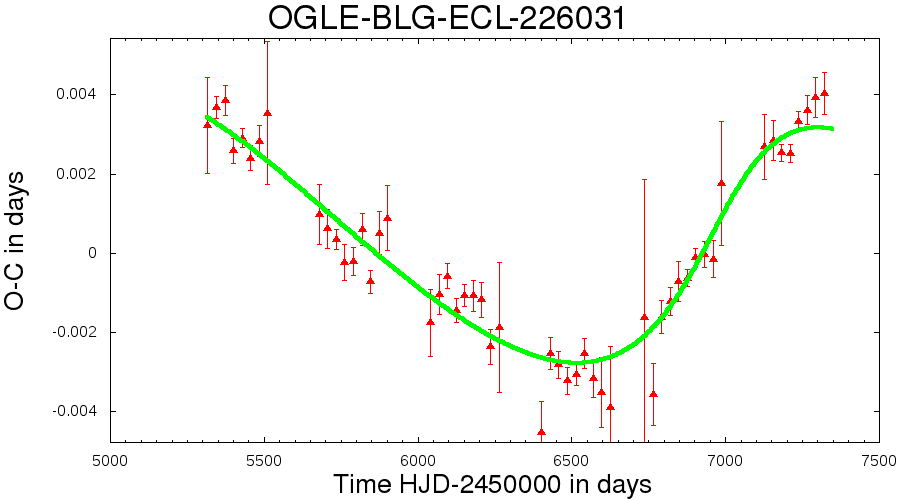}
\includegraphics[width=0.64\columnwidth]{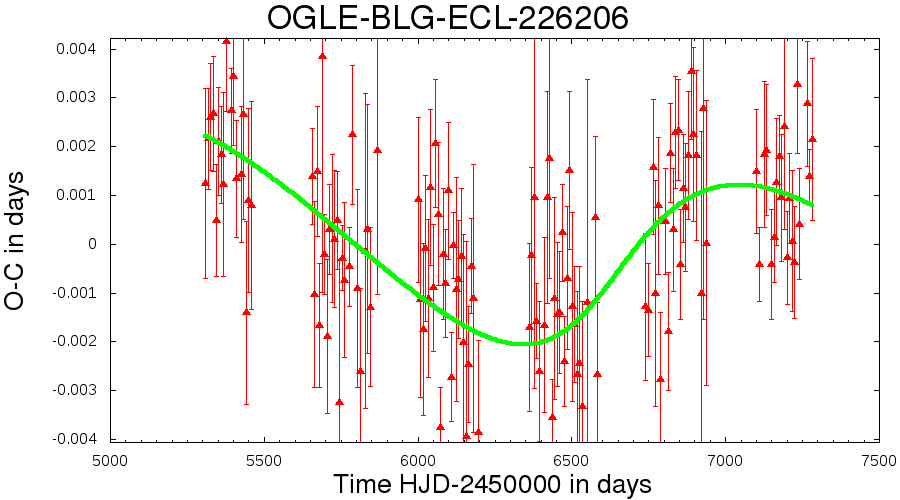}

\end{figure*}
\clearpage

\begin{figure*}
\includegraphics[width=0.64\columnwidth]{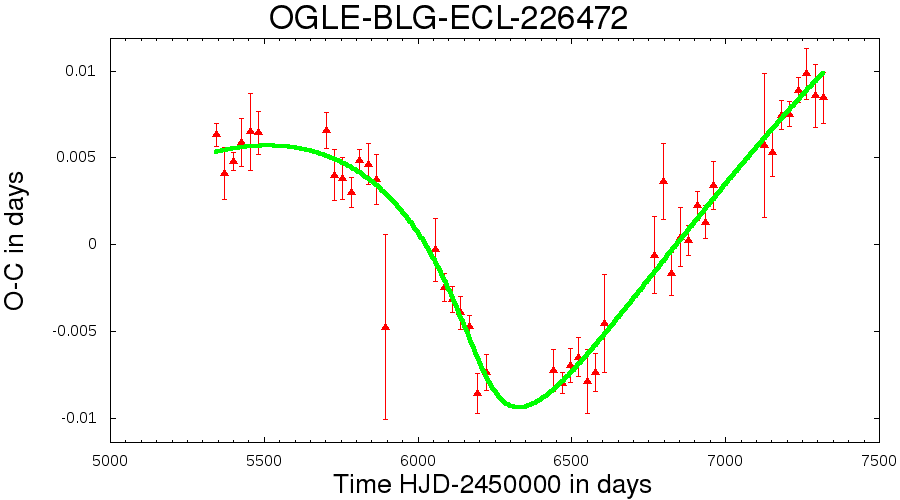}
\includegraphics[width=0.64\columnwidth]{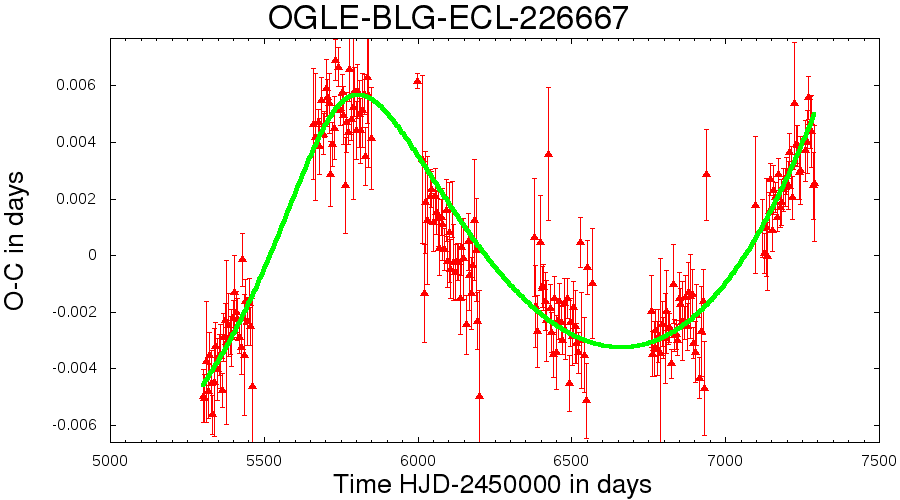}
\includegraphics[width=0.64\columnwidth]{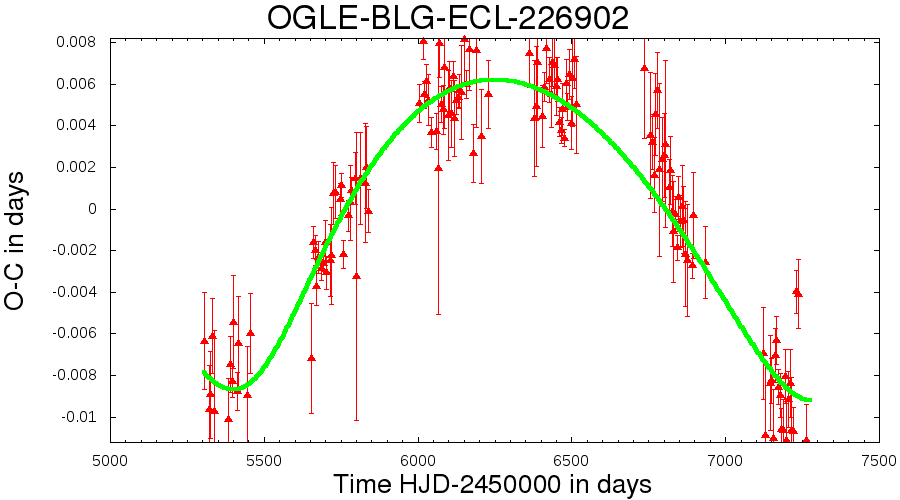}

\includegraphics[width=0.64\columnwidth]{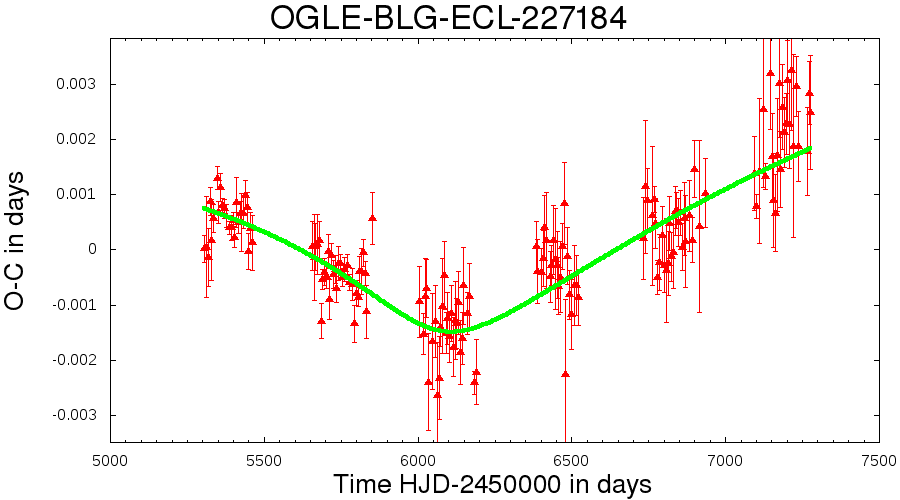}
\includegraphics[width=0.64\columnwidth]{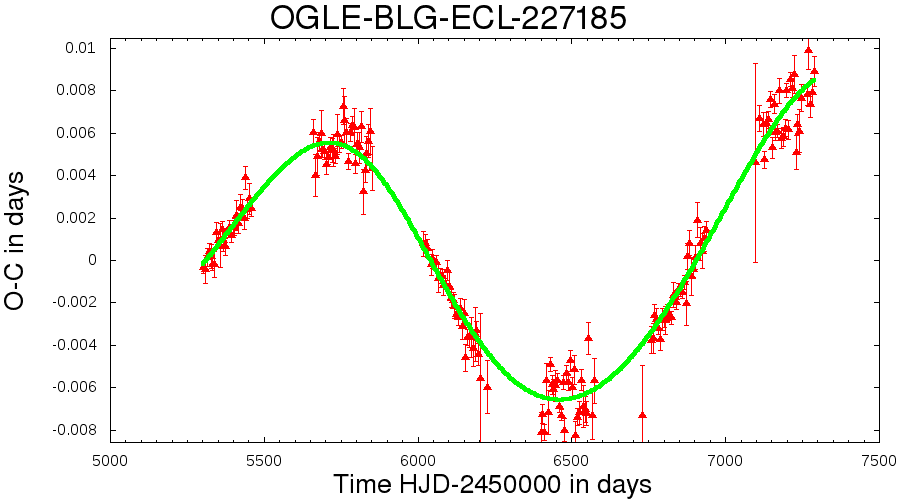}
\includegraphics[width=0.64\columnwidth]{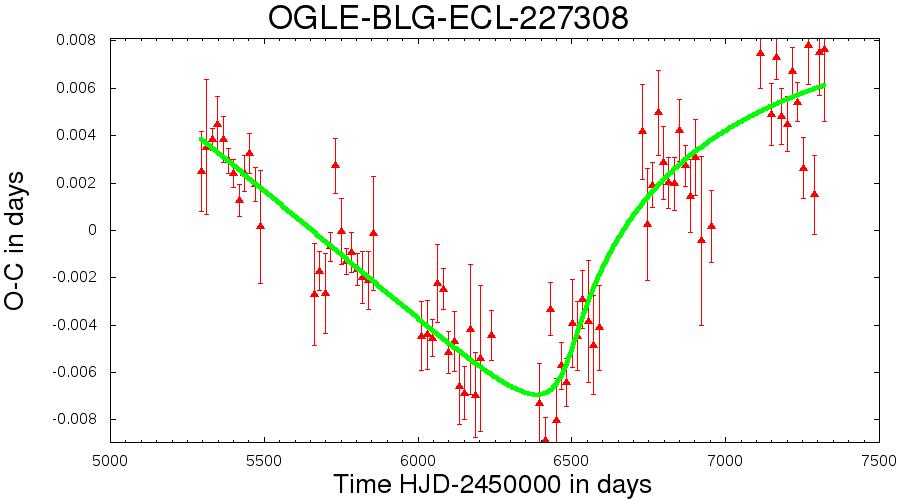}

\includegraphics[width=0.64\columnwidth]{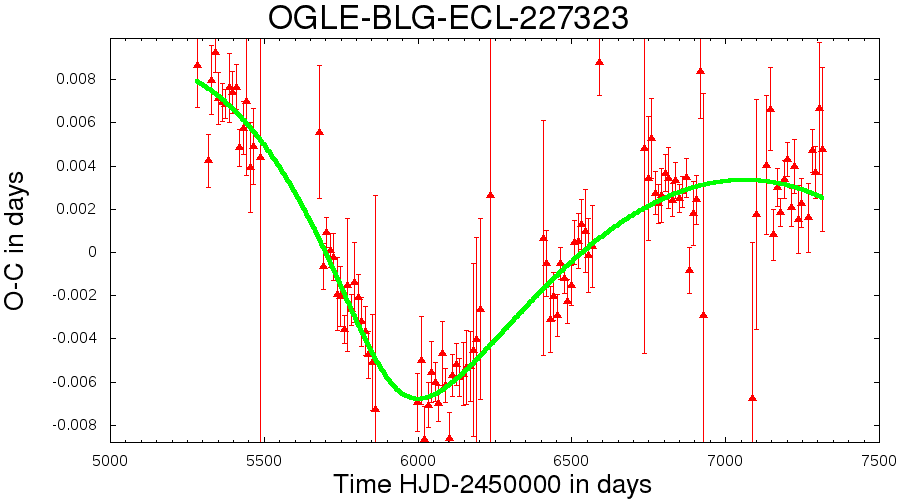}
\includegraphics[width=0.64\columnwidth]{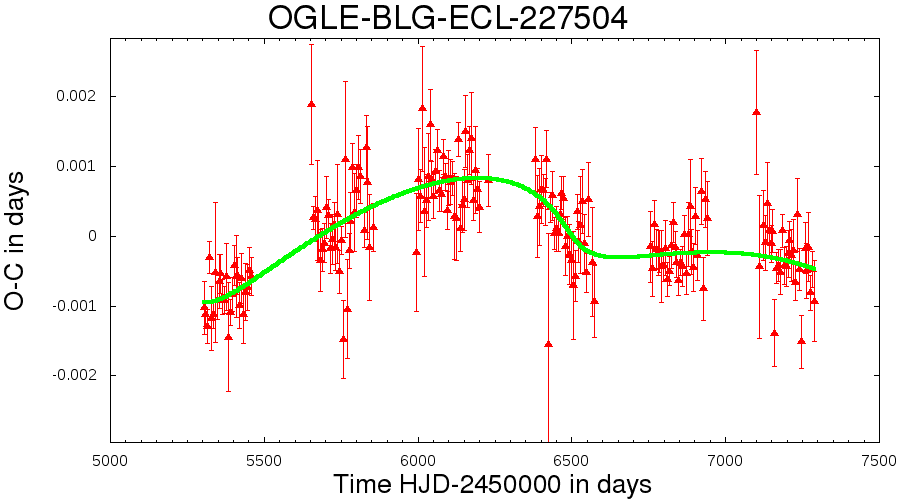}
\includegraphics[width=0.64\columnwidth]{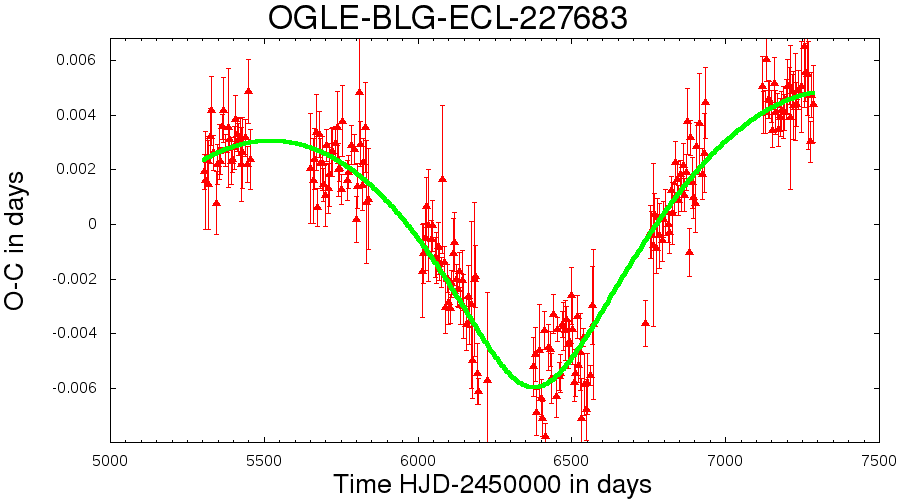}

\includegraphics[width=0.64\columnwidth]{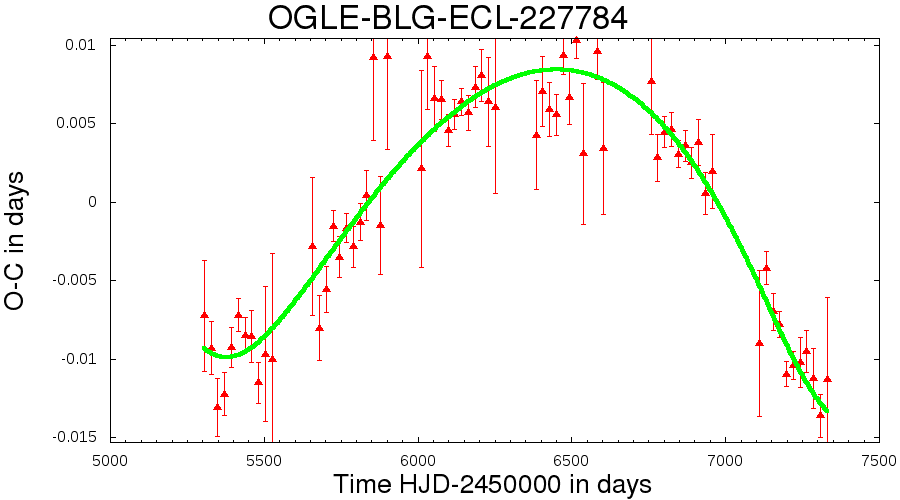}
\includegraphics[width=0.64\columnwidth]{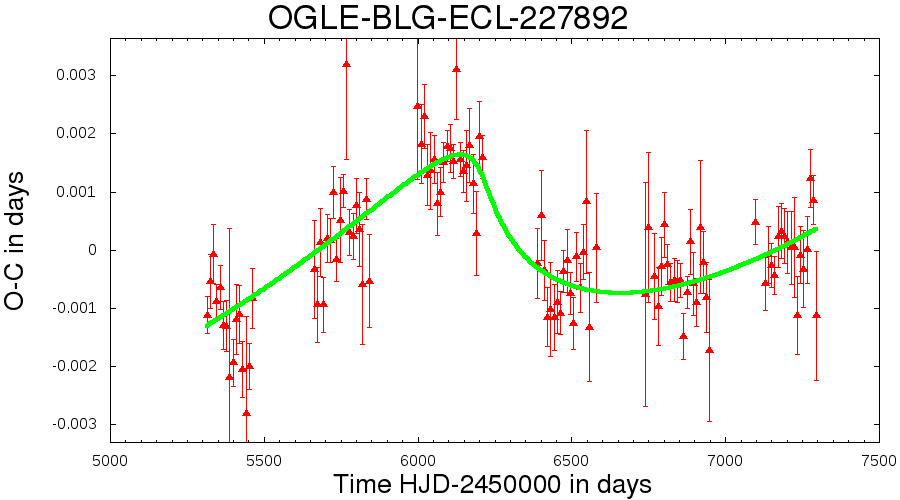}
\includegraphics[width=0.64\columnwidth]{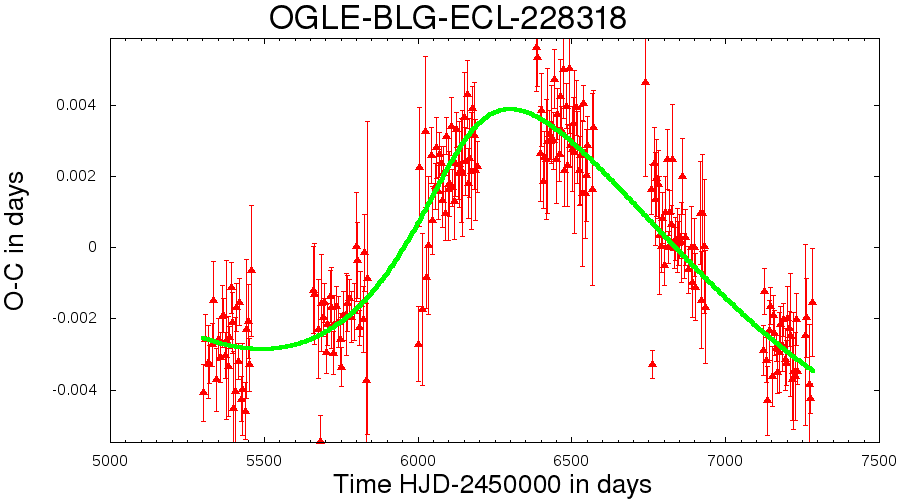}

\includegraphics[width=0.64\columnwidth]{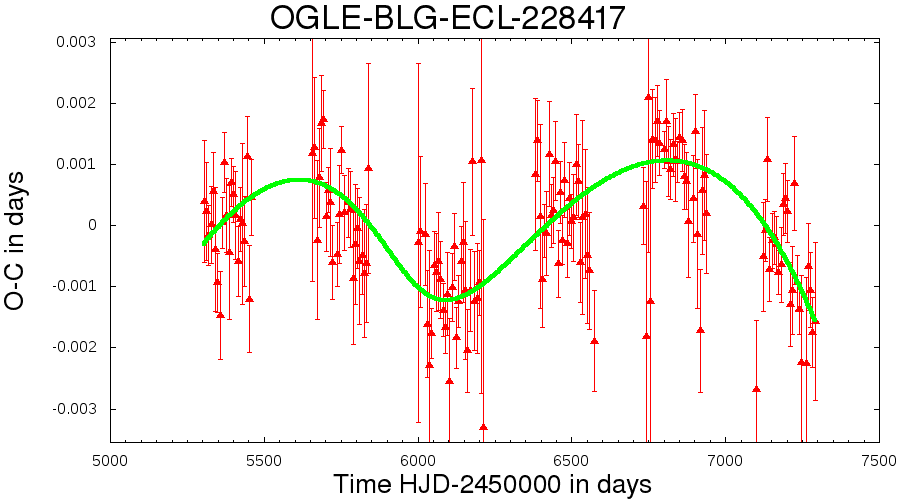}
\includegraphics[width=0.64\columnwidth]{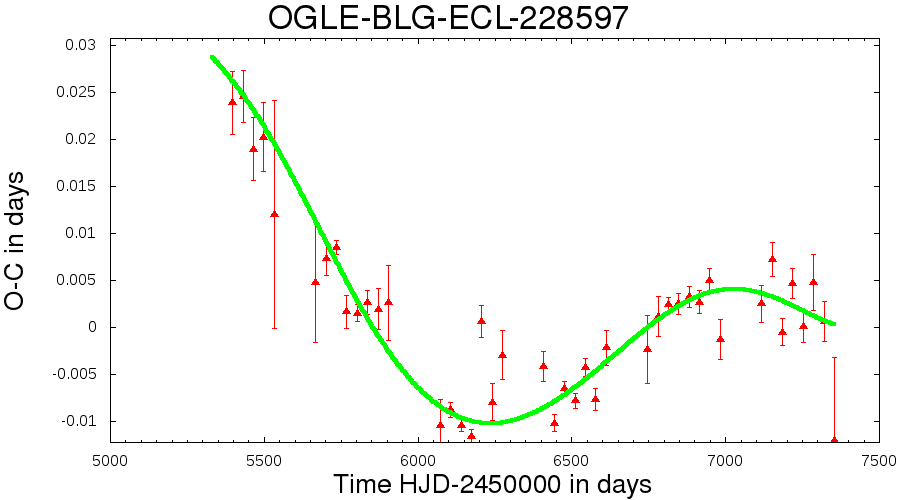}
\includegraphics[width=0.64\columnwidth]{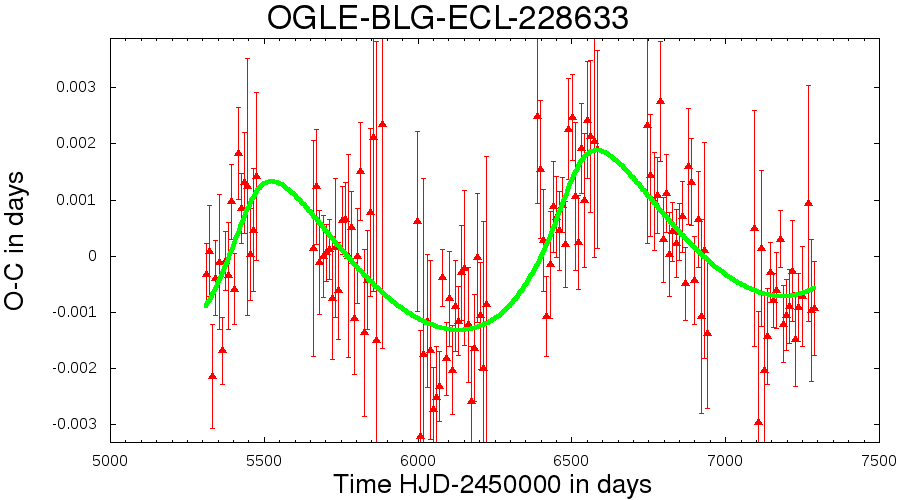}

\includegraphics[width=0.64\columnwidth]{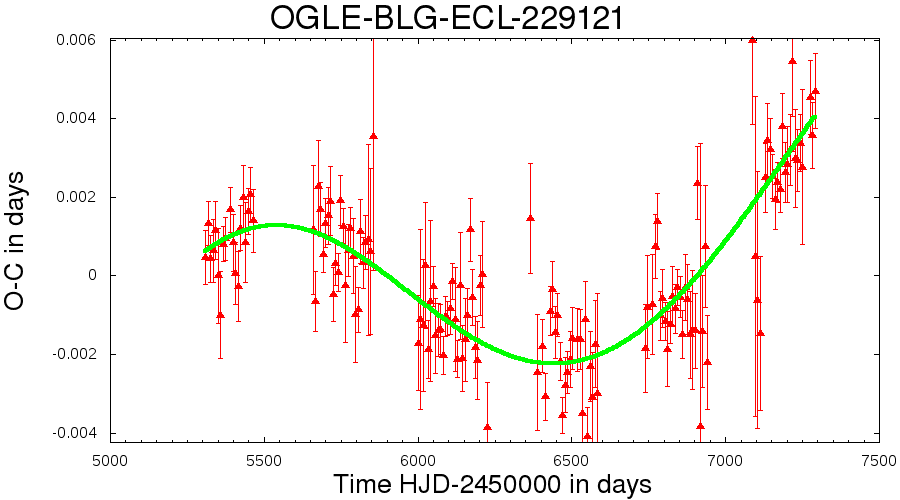}
\includegraphics[width=0.64\columnwidth]{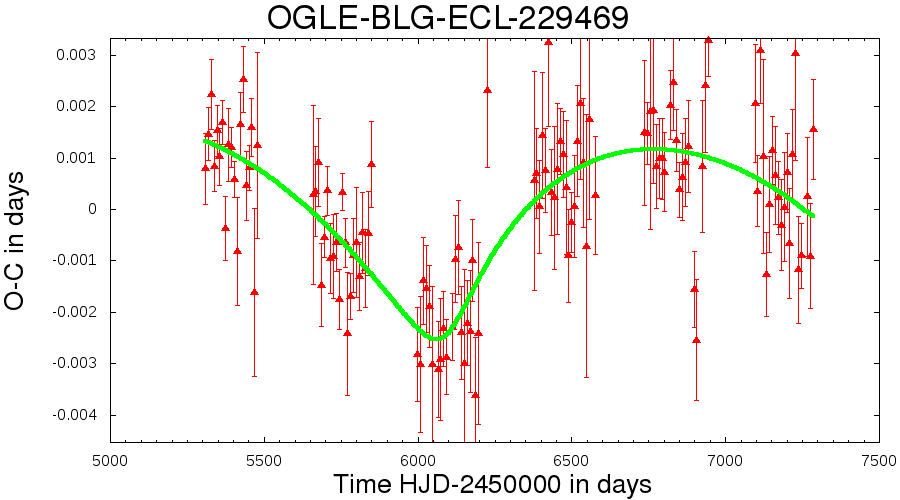}
\includegraphics[width=0.64\columnwidth]{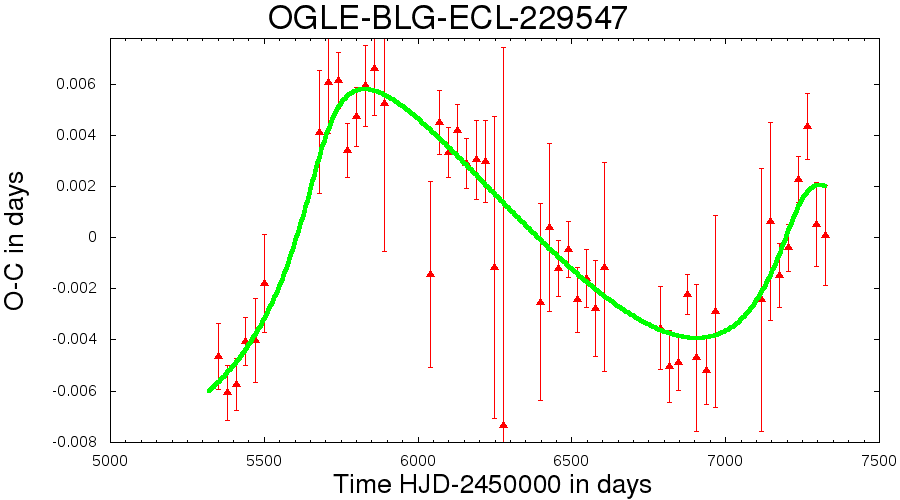}

\includegraphics[width=0.64\columnwidth]{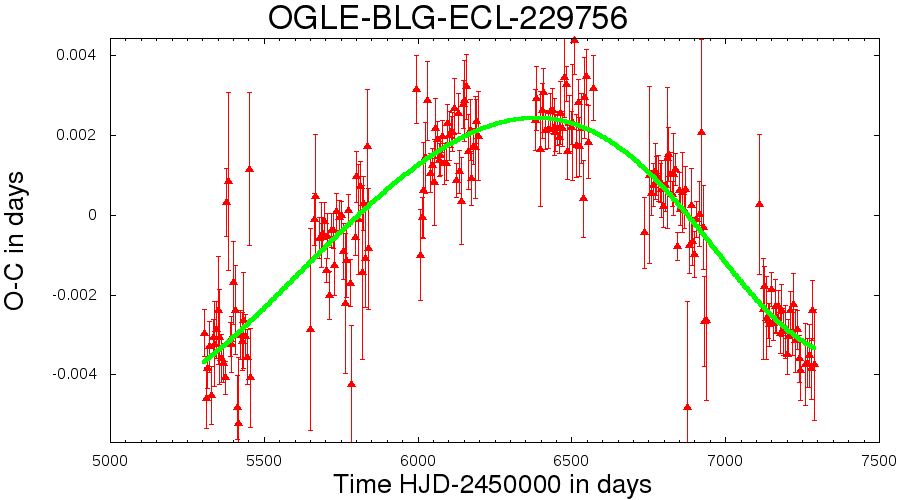}
\includegraphics[width=0.64\columnwidth]{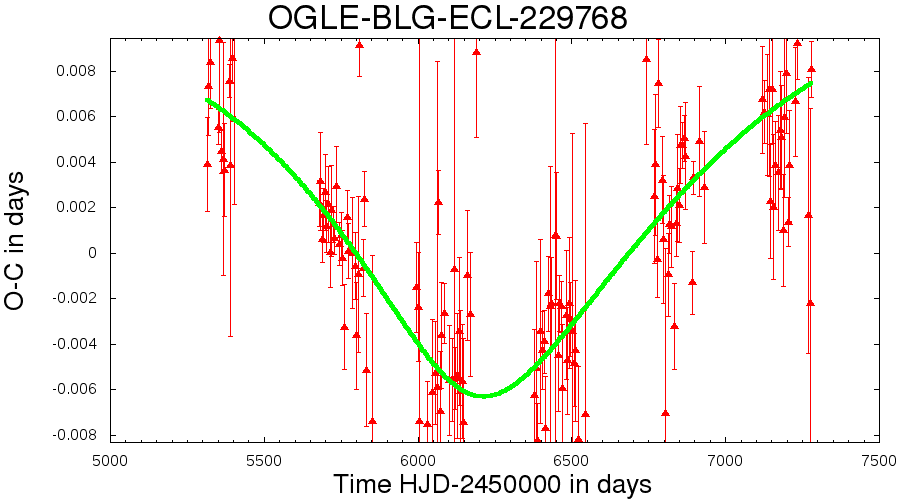}
\includegraphics[width=0.64\columnwidth]{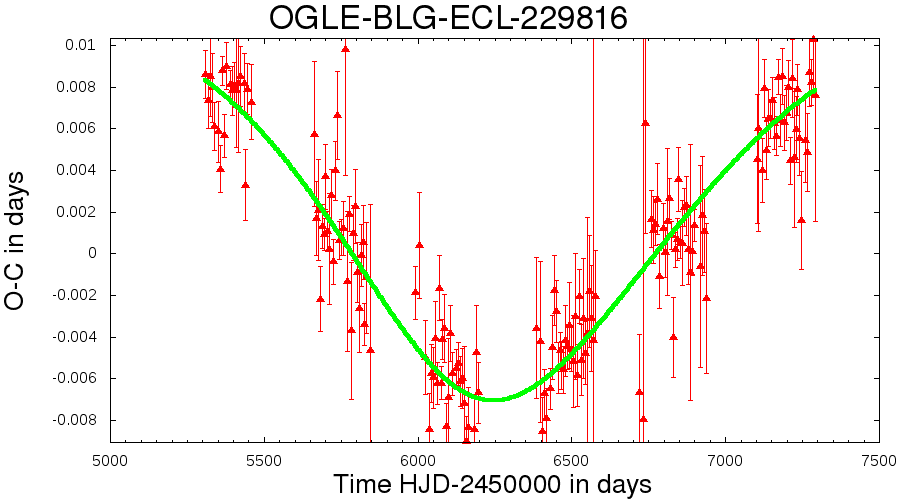}

\includegraphics[width=0.64\columnwidth]{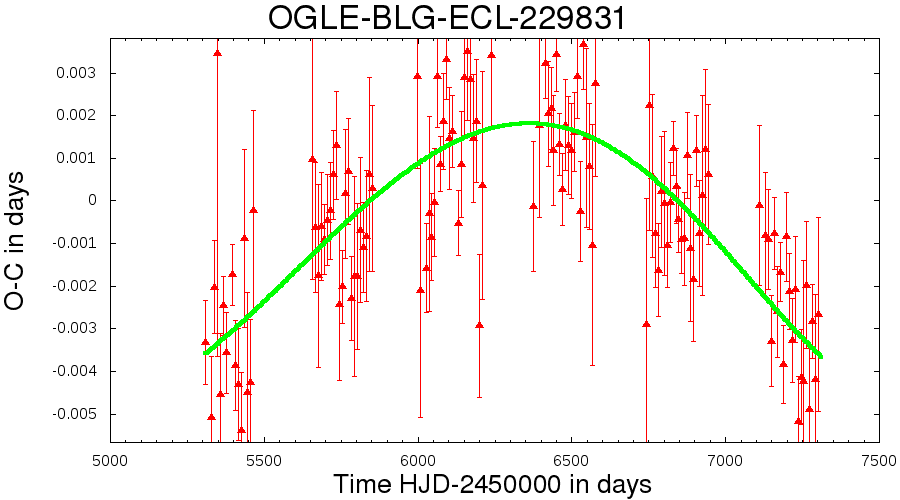}
\includegraphics[width=0.64\columnwidth]{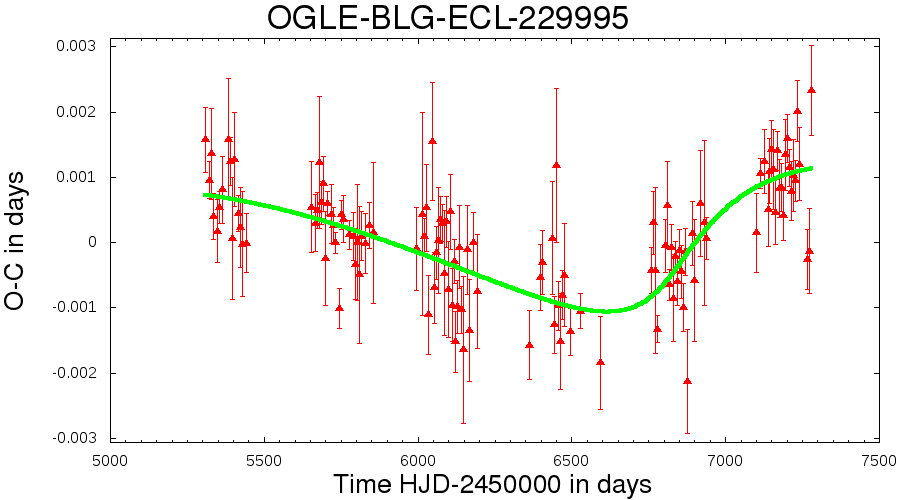}
\includegraphics[width=0.64\columnwidth]{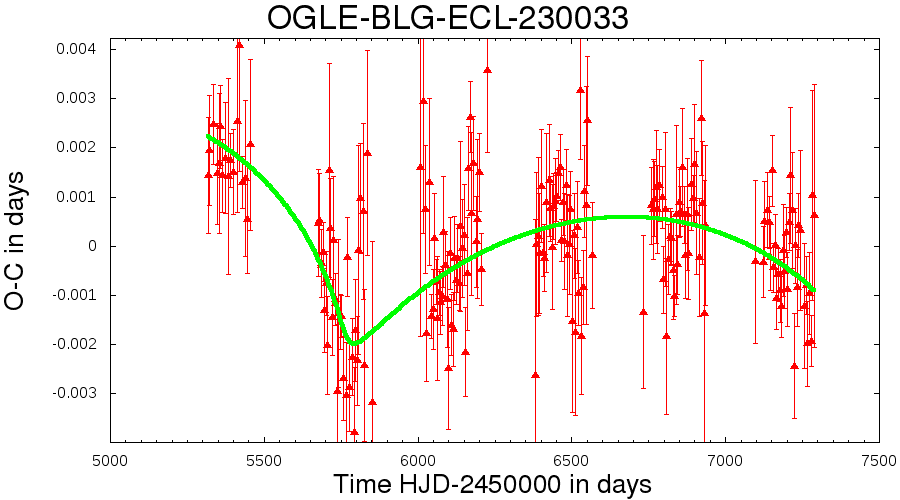}

\end{figure*}
\clearpage

\begin{figure*}
\includegraphics[width=0.64\columnwidth]{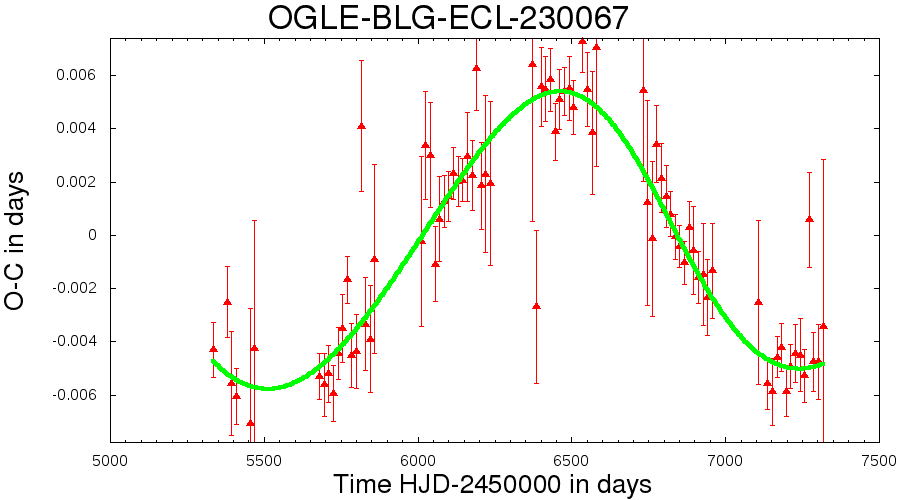}
\includegraphics[width=0.64\columnwidth]{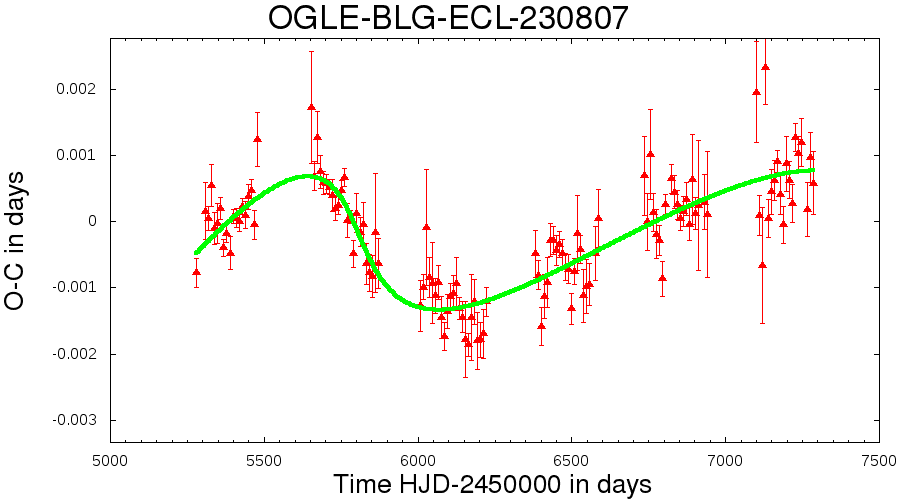}
\includegraphics[width=0.64\columnwidth]{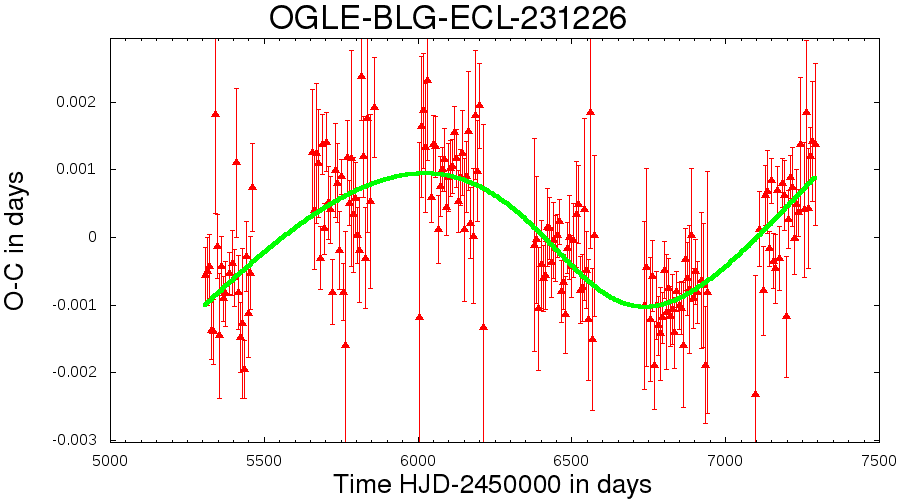}

\includegraphics[width=0.64\columnwidth]{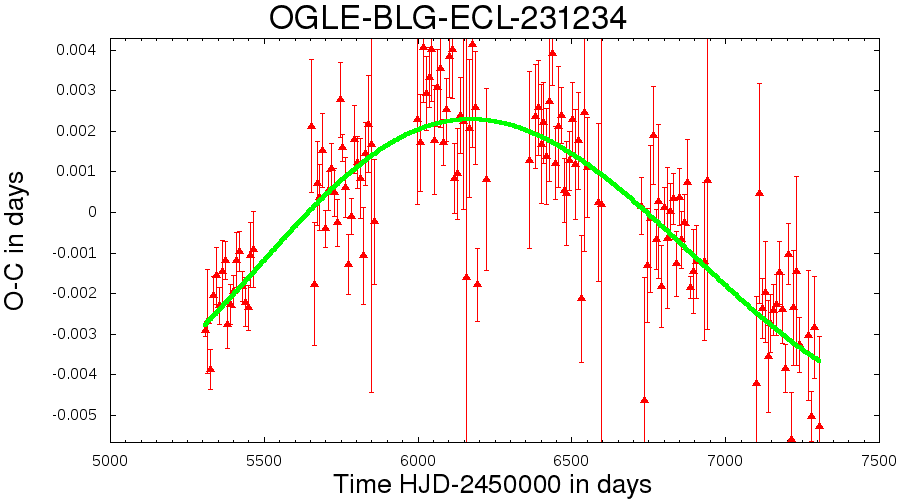}
\includegraphics[width=0.64\columnwidth]{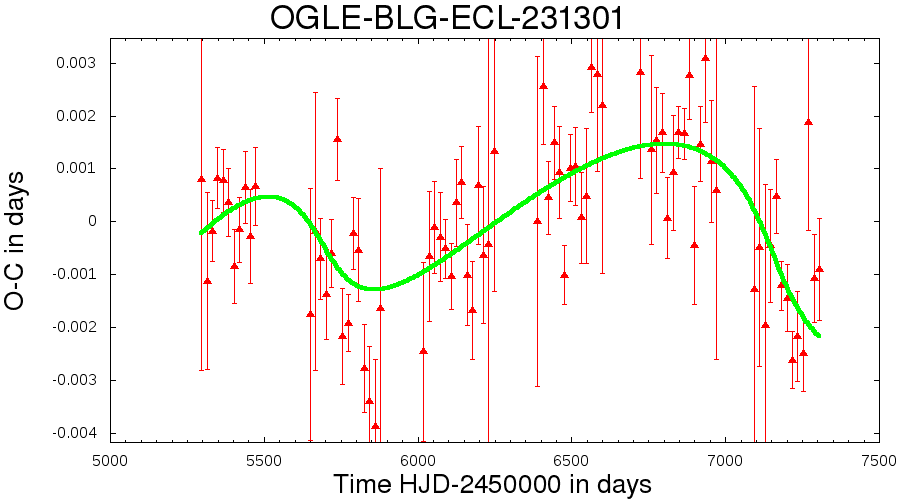}
\includegraphics[width=0.64\columnwidth]{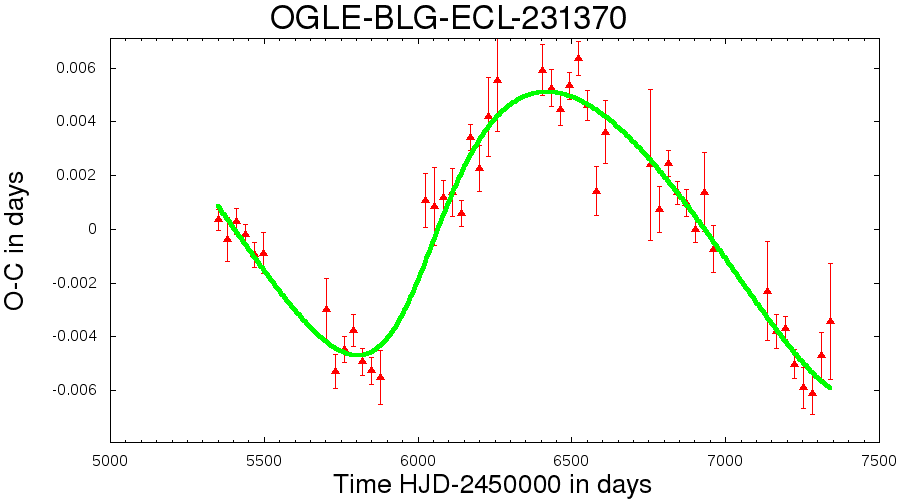}

\includegraphics[width=0.64\columnwidth]{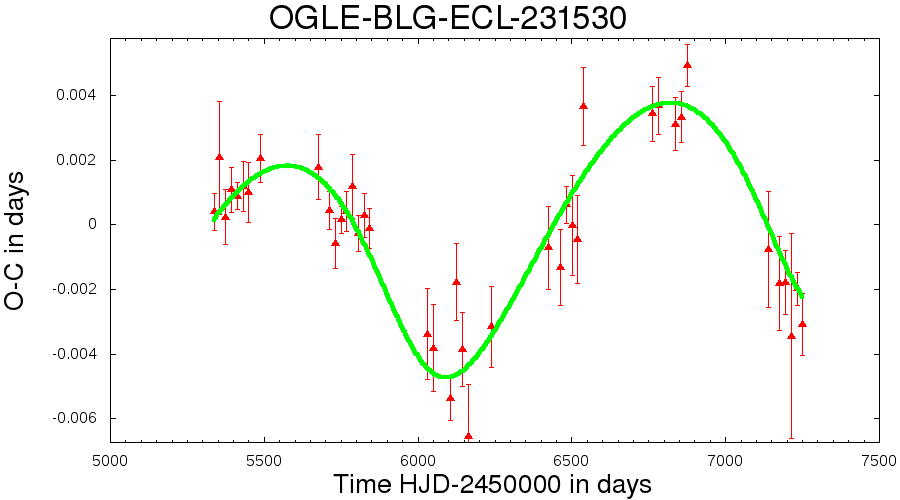}
\includegraphics[width=0.64\columnwidth]{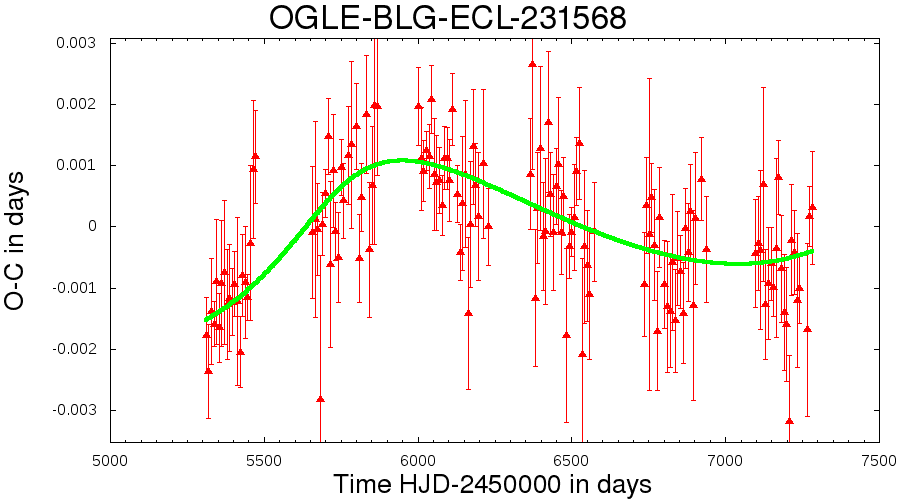}
\includegraphics[width=0.64\columnwidth]{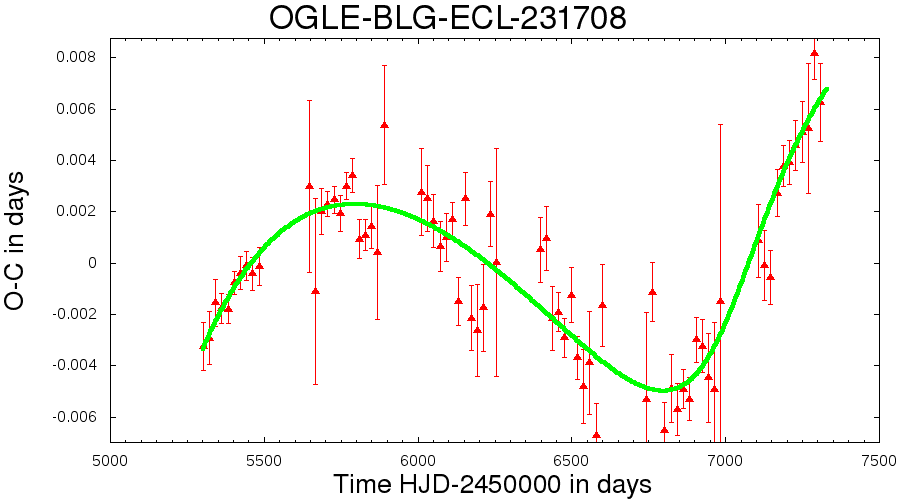}

\includegraphics[width=0.64\columnwidth]{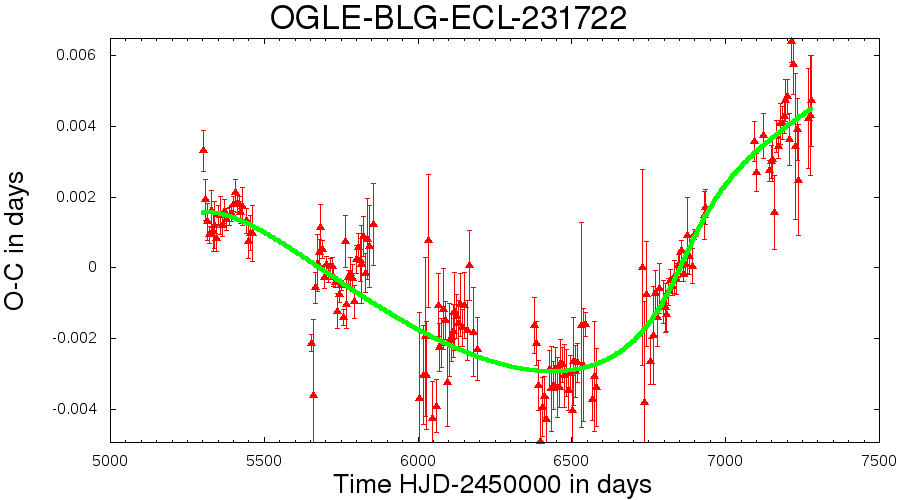}
\includegraphics[width=0.64\columnwidth]{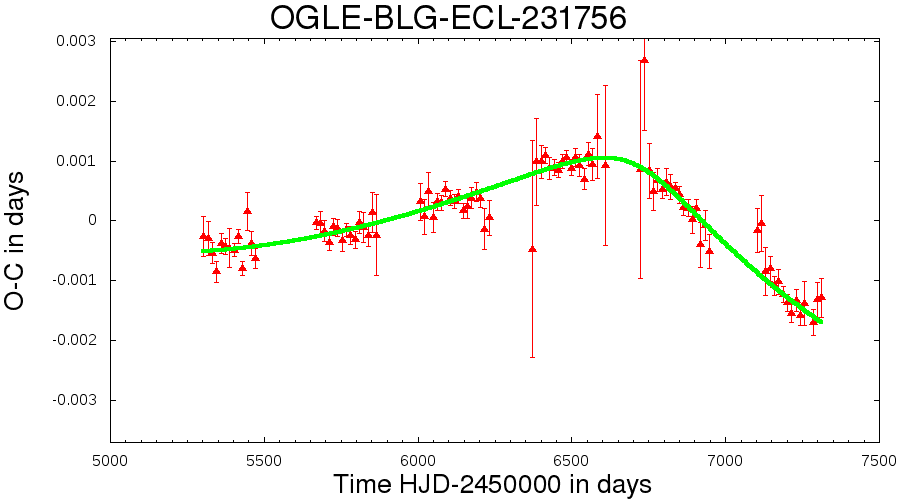}
\includegraphics[width=0.64\columnwidth]{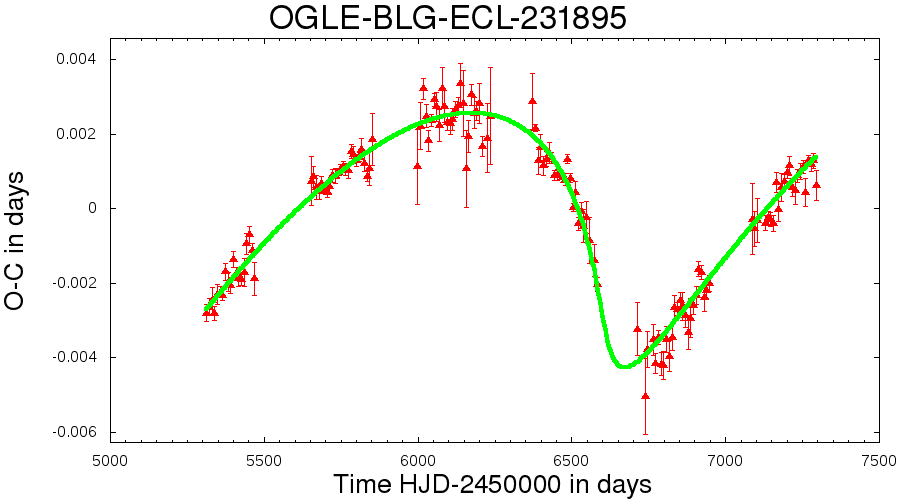}

\includegraphics[width=0.64\columnwidth]{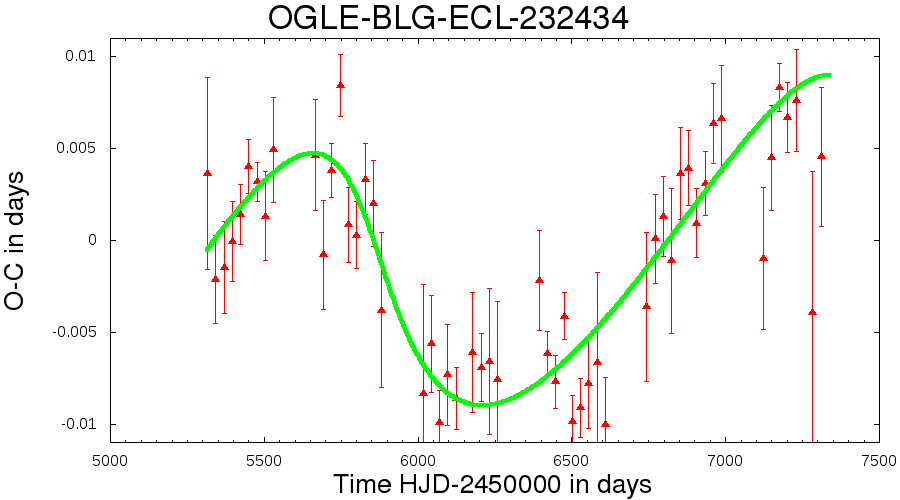}
\includegraphics[width=0.64\columnwidth]{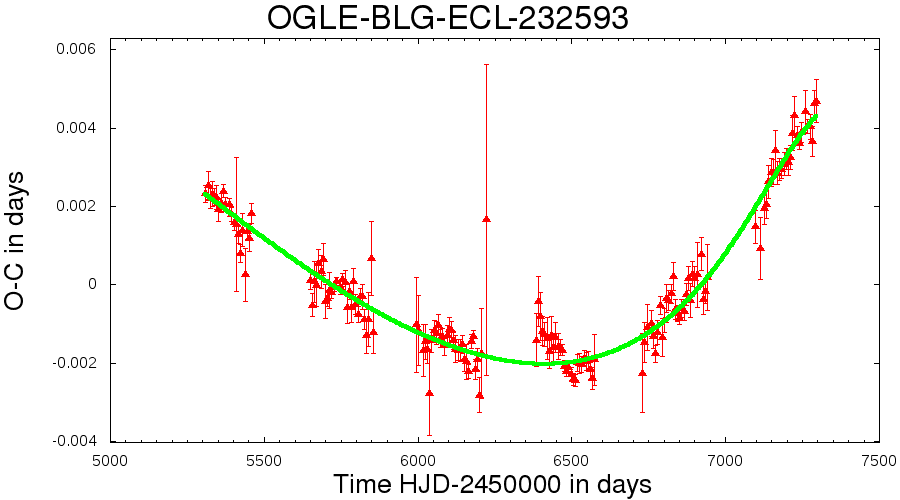}
\includegraphics[width=0.64\columnwidth]{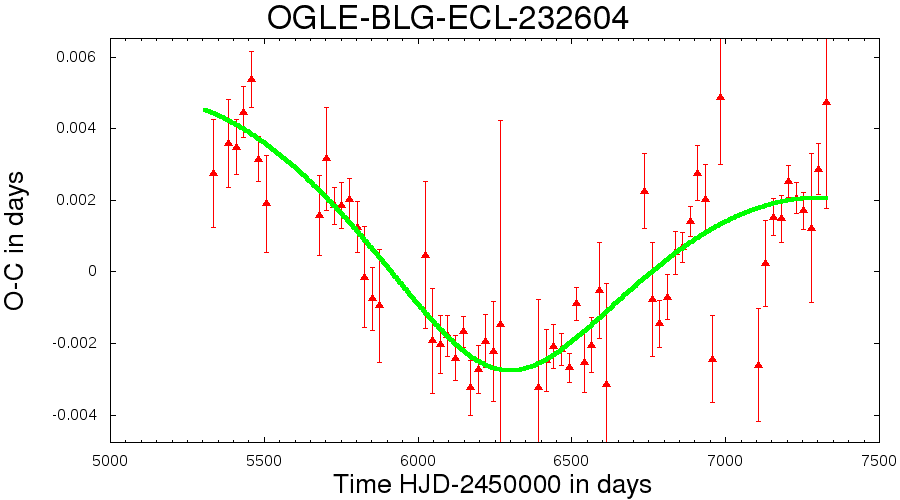}

\includegraphics[width=0.64\columnwidth]{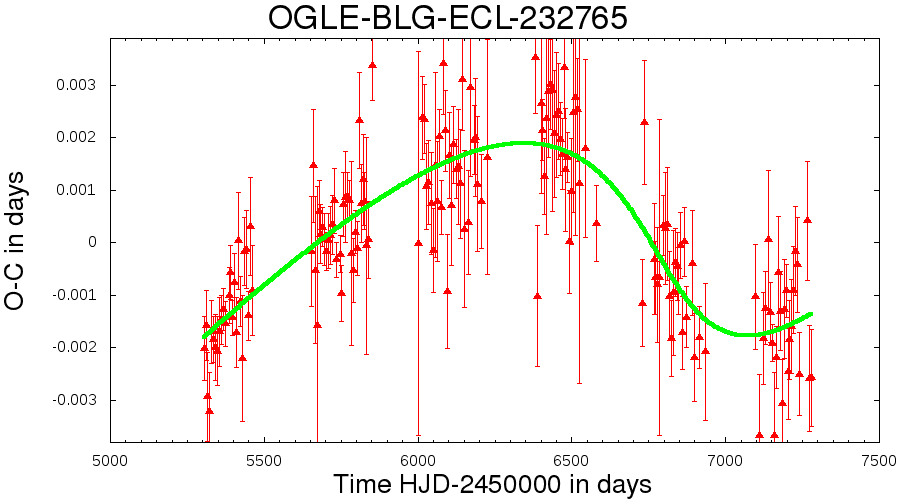}
\includegraphics[width=0.64\columnwidth]{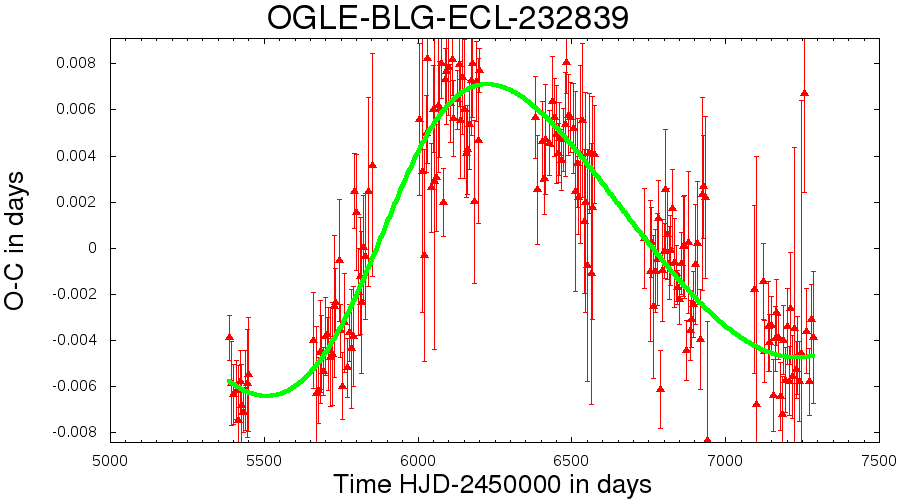}
\includegraphics[width=0.64\columnwidth]{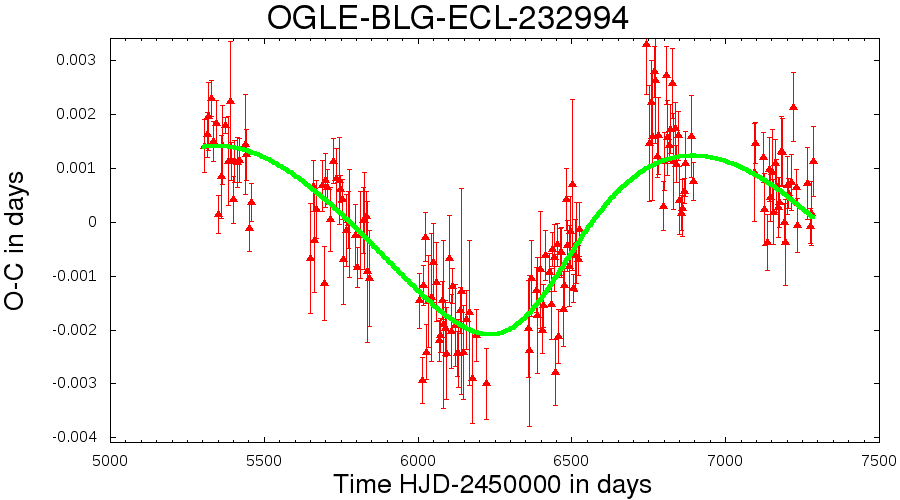}

\includegraphics[width=0.64\columnwidth]{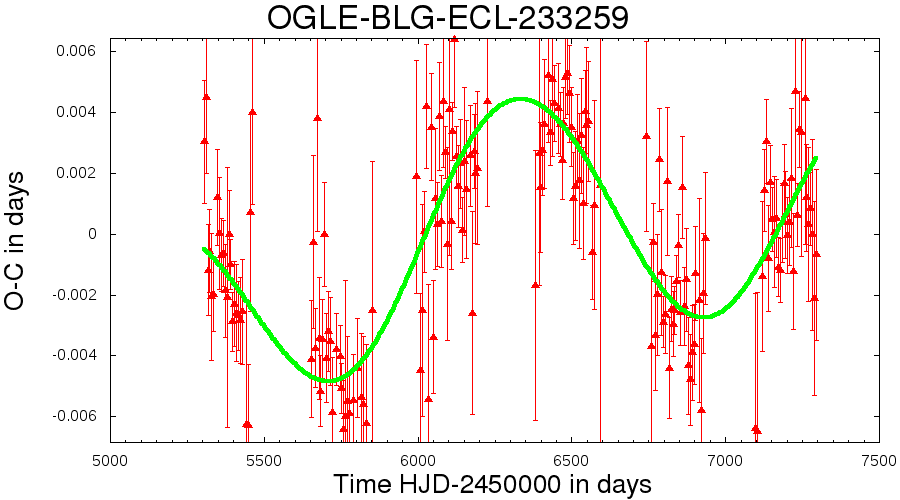}
\includegraphics[width=0.64\columnwidth]{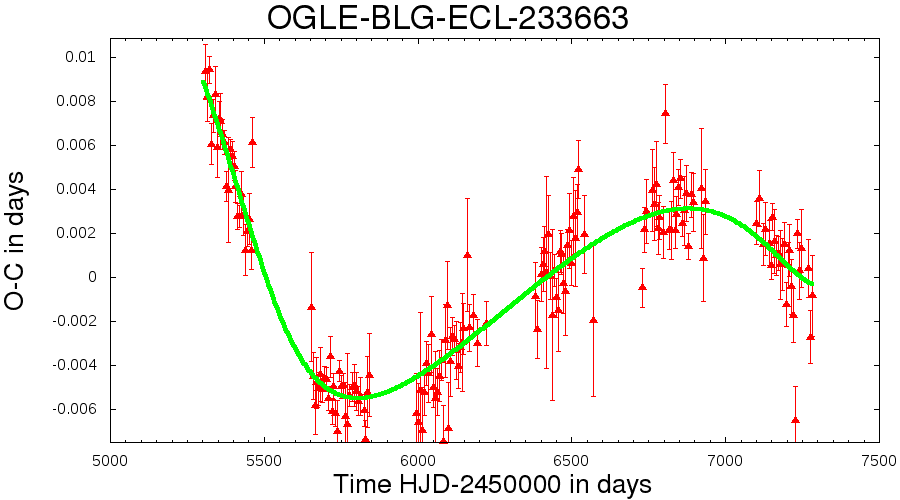}
\includegraphics[width=0.64\columnwidth]{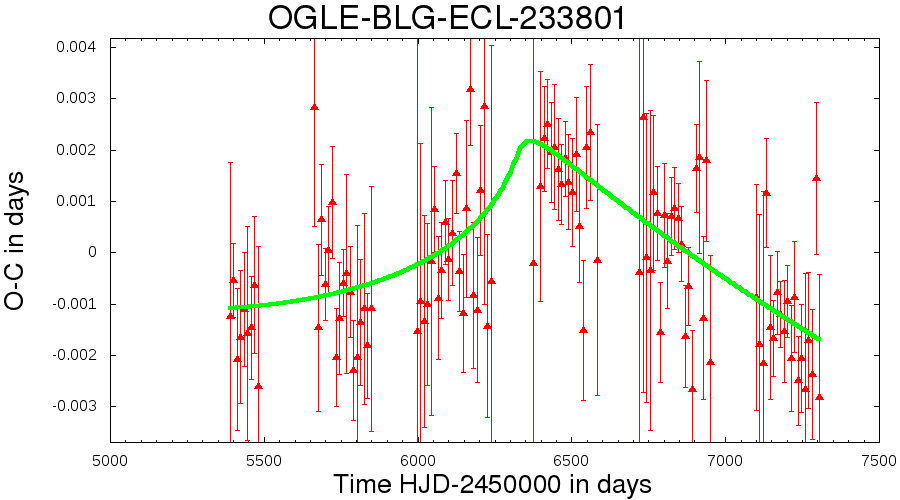}

\includegraphics[width=0.64\columnwidth]{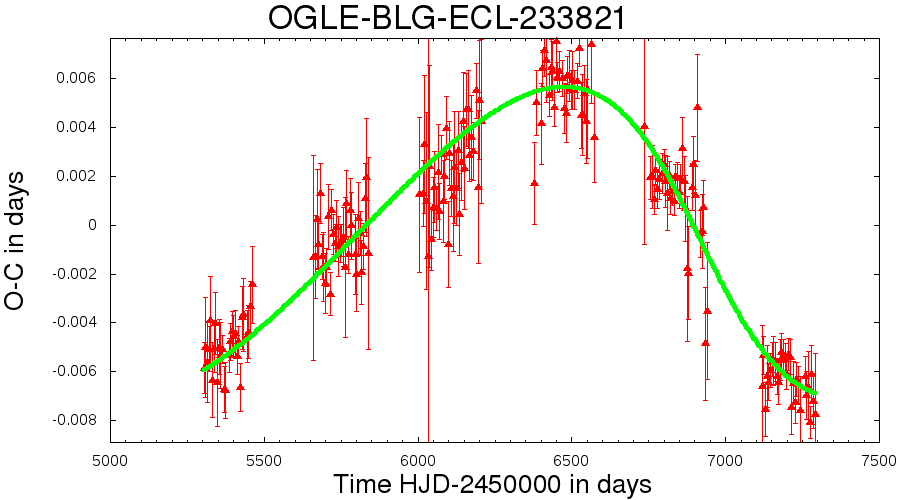}
\includegraphics[width=0.64\columnwidth]{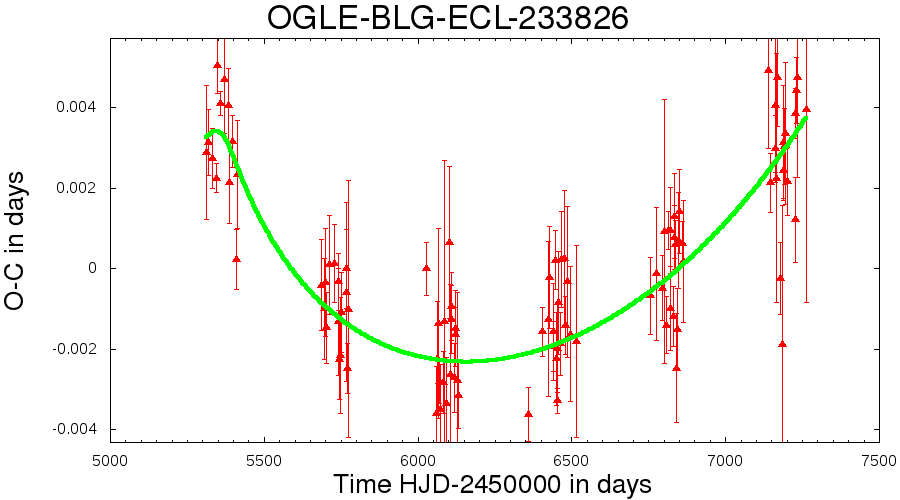}
\includegraphics[width=0.64\columnwidth]{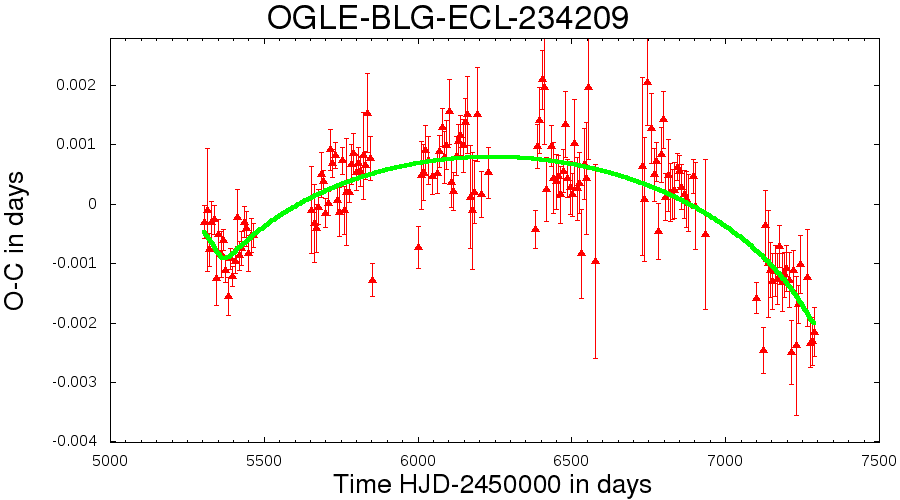}

\end{figure*}
\clearpage

\begin{figure*}
\includegraphics[width=0.64\columnwidth]{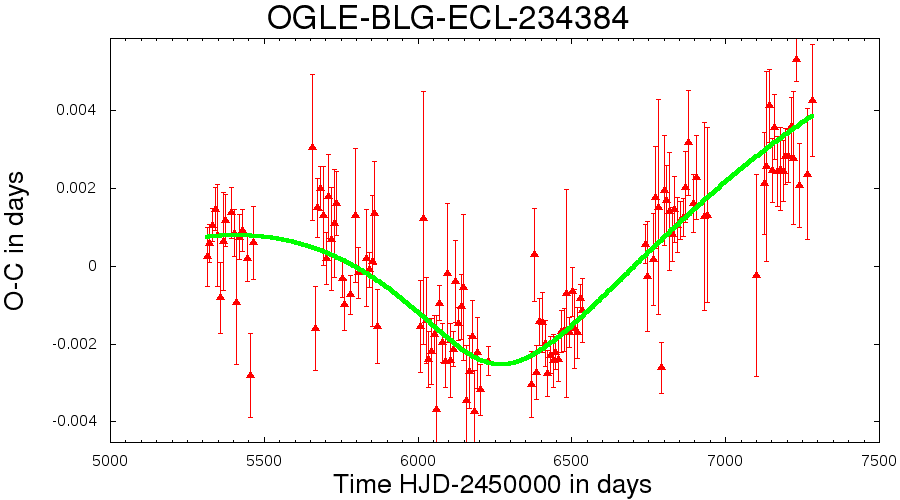}
\includegraphics[width=0.64\columnwidth]{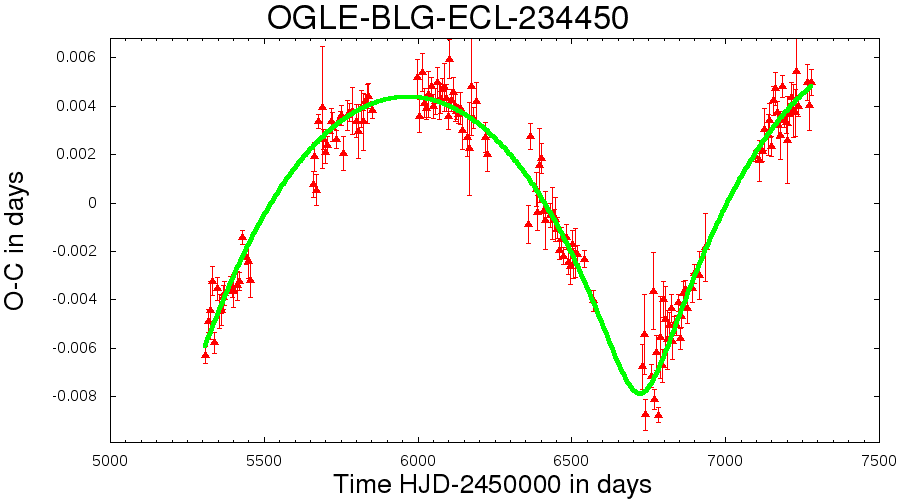}
\includegraphics[width=0.64\columnwidth]{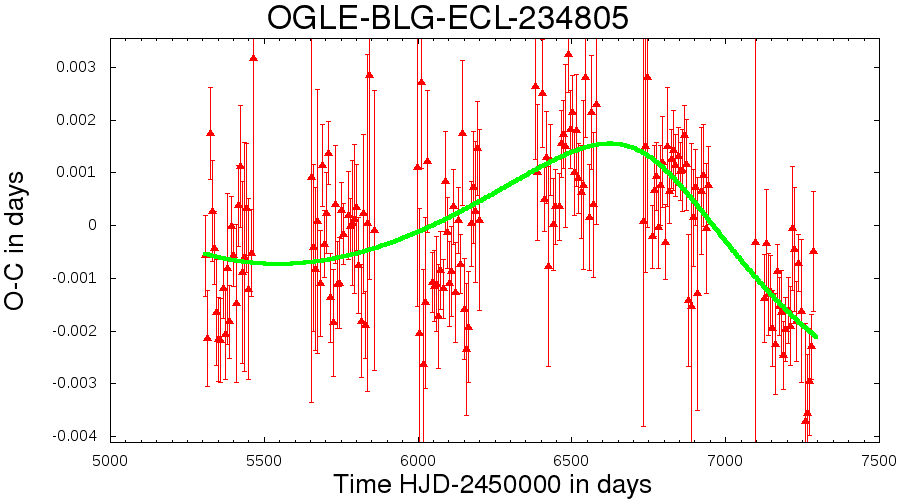}

\includegraphics[width=0.64\columnwidth]{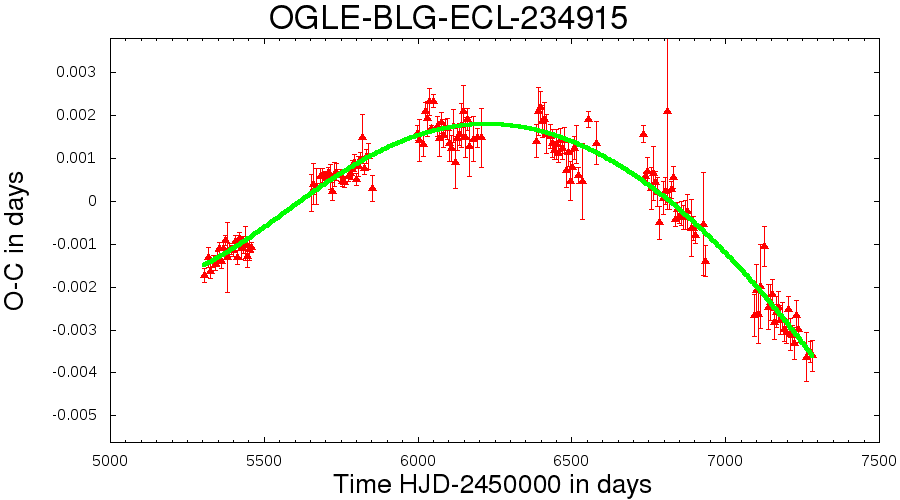}
\includegraphics[width=0.64\columnwidth]{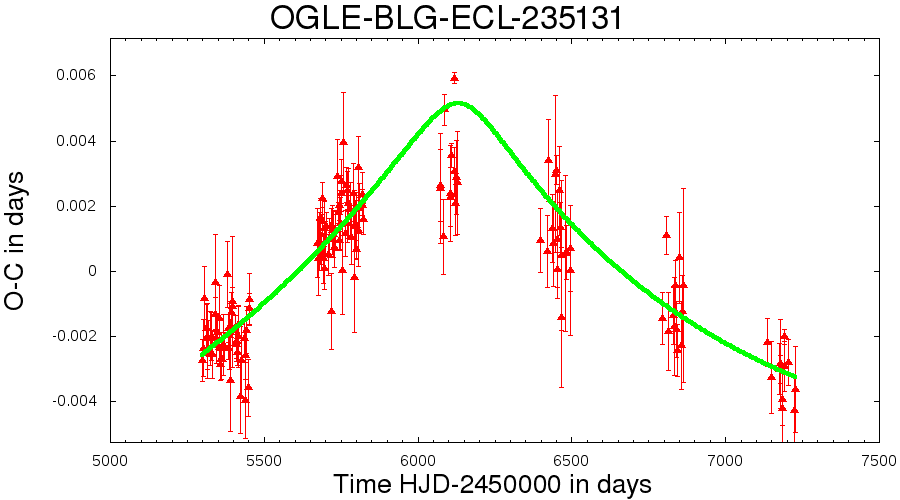}
\includegraphics[width=0.64\columnwidth]{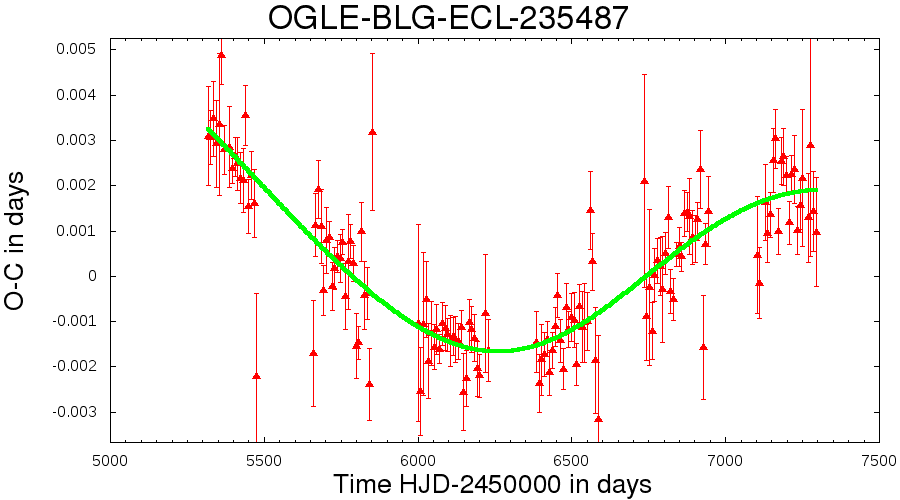}

\includegraphics[width=0.64\columnwidth]{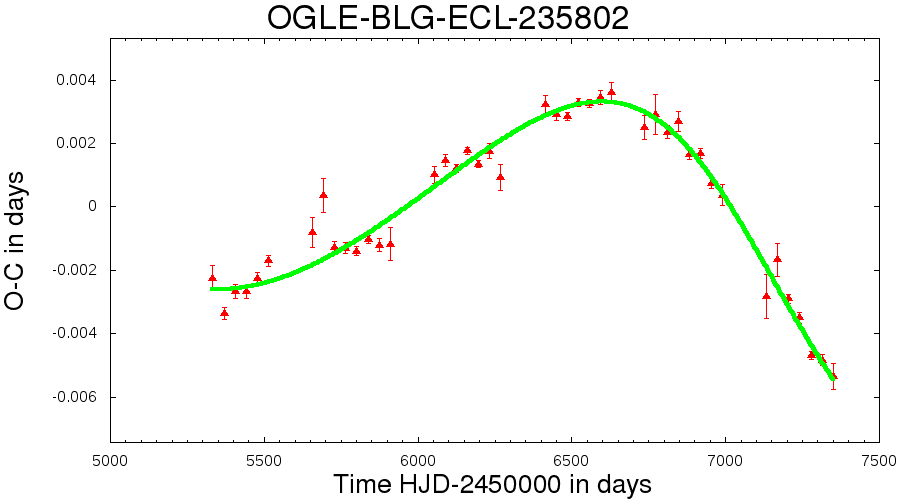}
\includegraphics[width=0.64\columnwidth]{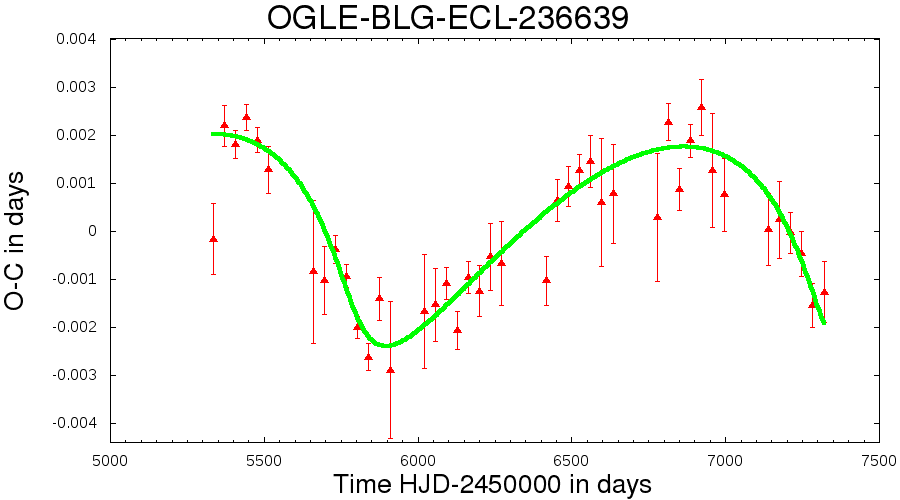}
\includegraphics[width=0.64\columnwidth]{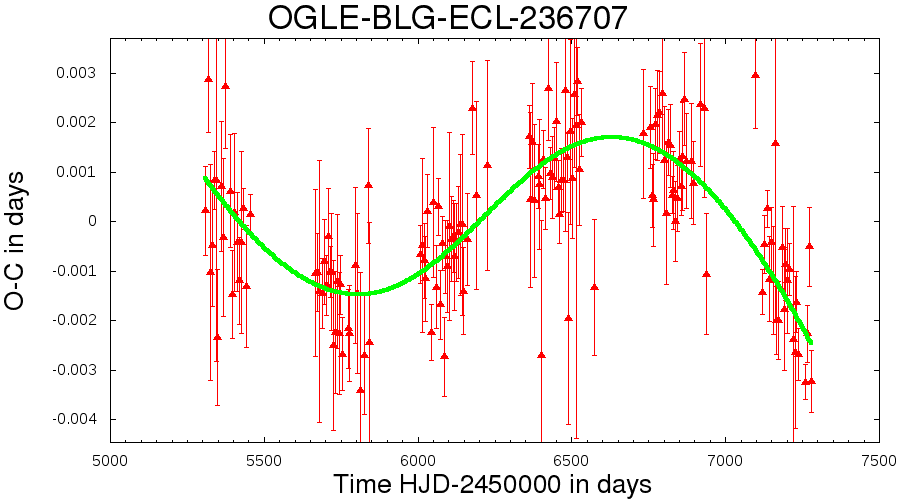}

\includegraphics[width=0.64\columnwidth]{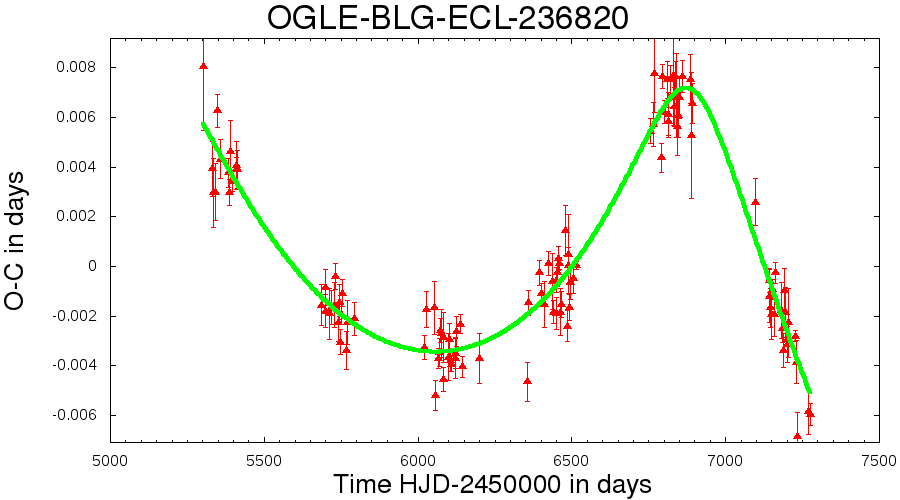}
\includegraphics[width=0.64\columnwidth]{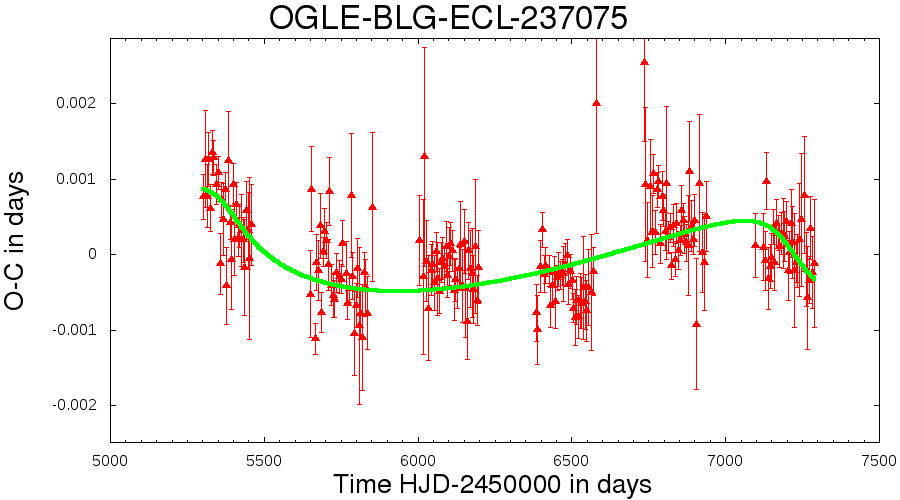}
\includegraphics[width=0.64\columnwidth]{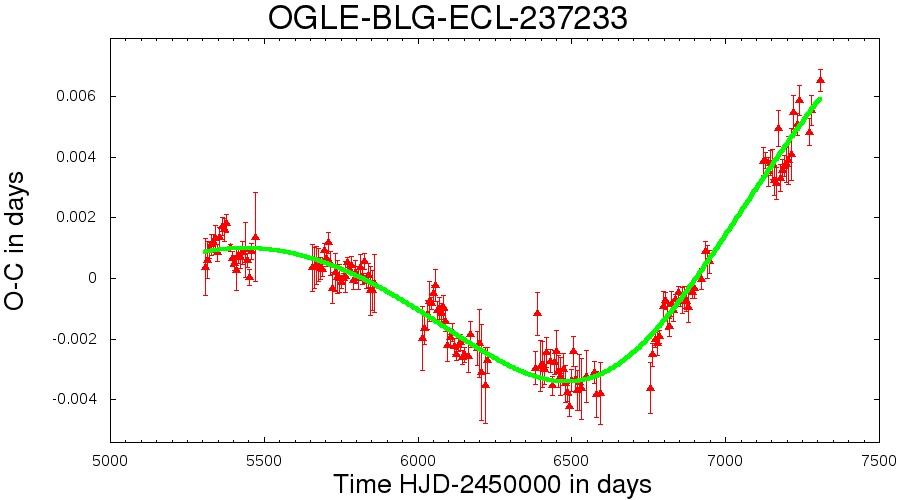}

\includegraphics[width=0.64\columnwidth]{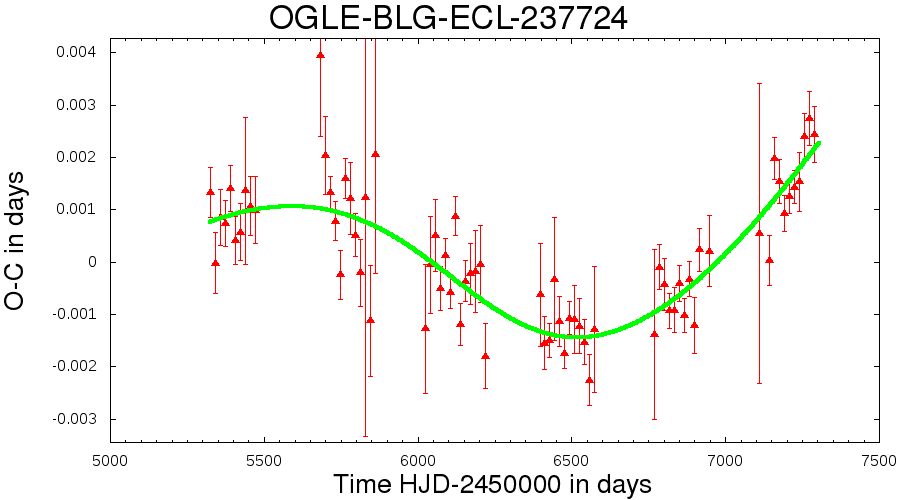}
\includegraphics[width=0.64\columnwidth]{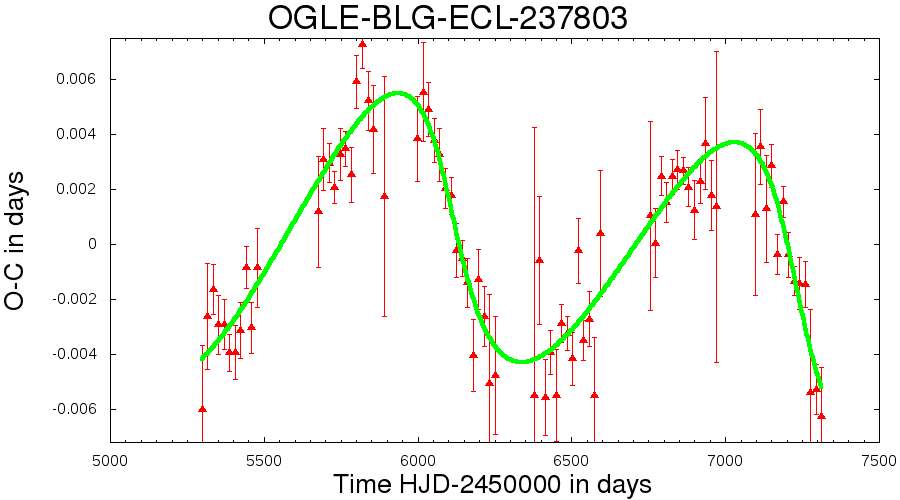}
\includegraphics[width=0.64\columnwidth]{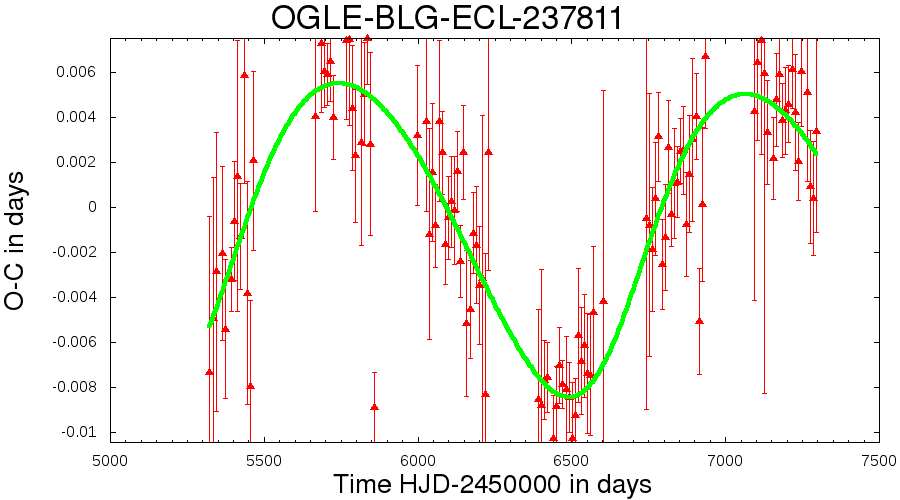}

\includegraphics[width=0.64\columnwidth]{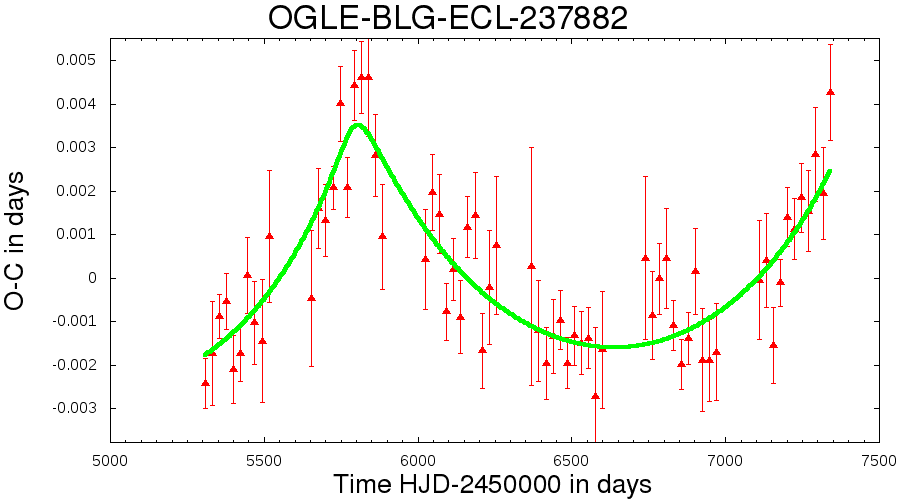}
\includegraphics[width=0.64\columnwidth]{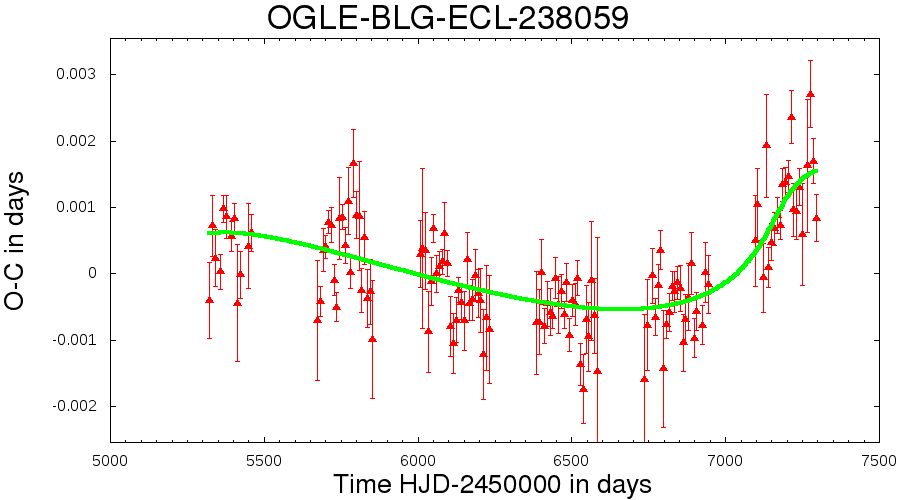}
\includegraphics[width=0.64\columnwidth]{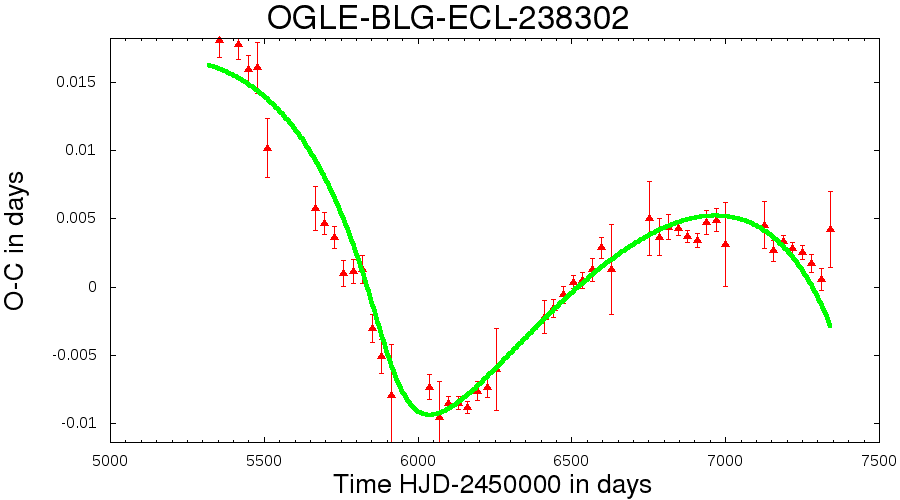}

\includegraphics[width=0.64\columnwidth]{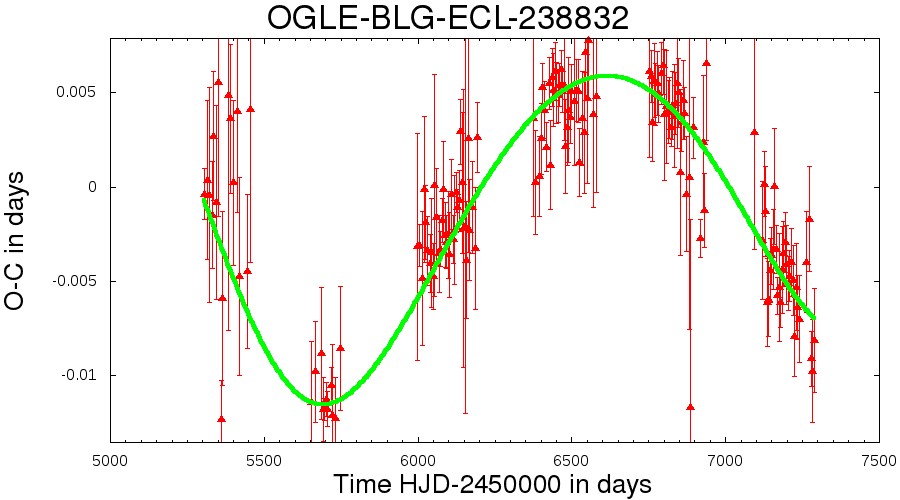}
\includegraphics[width=0.64\columnwidth]{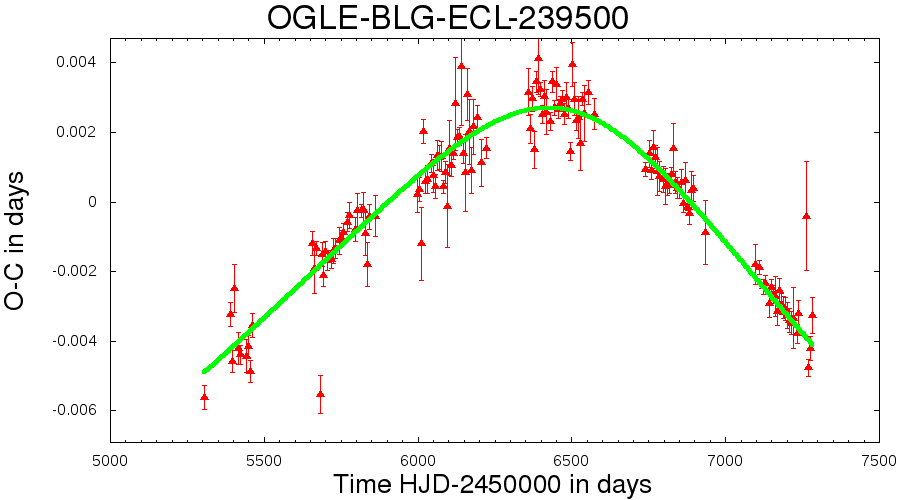}
\includegraphics[width=0.64\columnwidth]{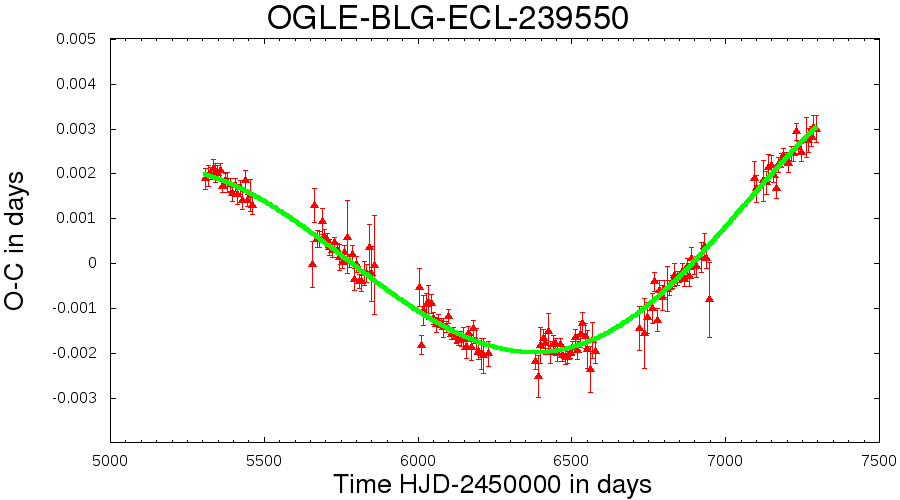}

\includegraphics[width=0.64\columnwidth]{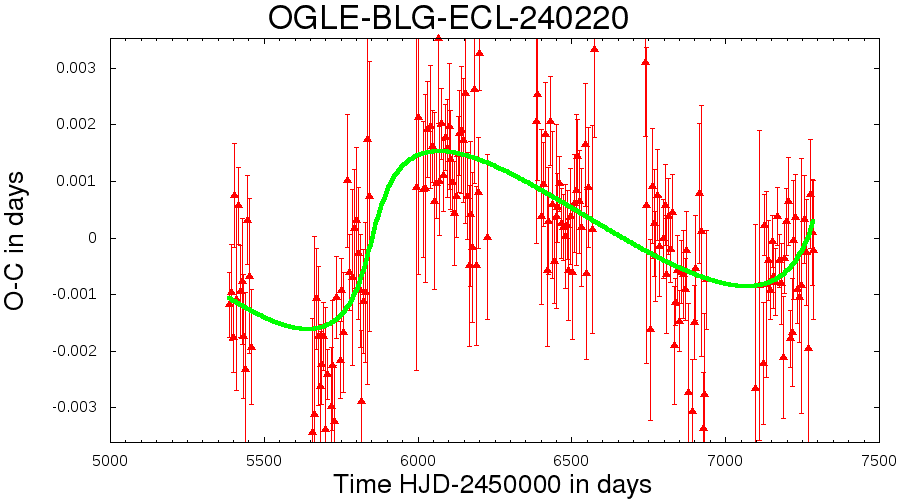}
\includegraphics[width=0.64\columnwidth]{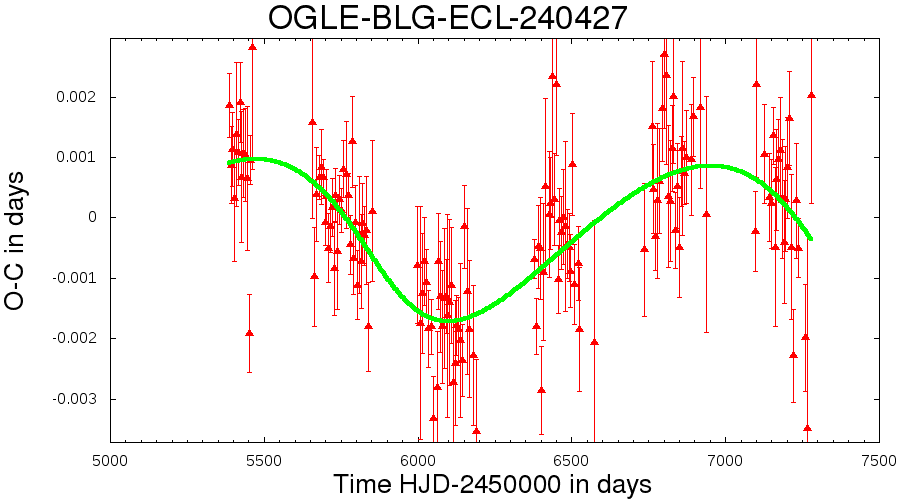}
\includegraphics[width=0.64\columnwidth]{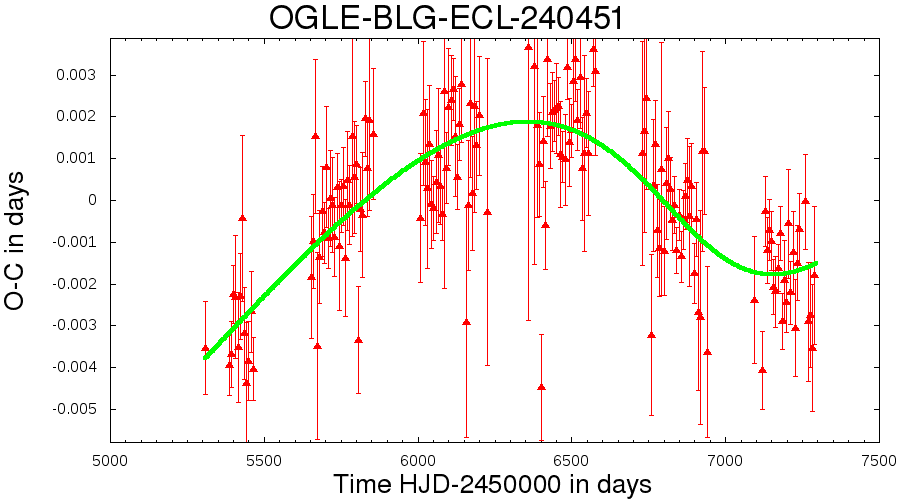}

\end{figure*}
\clearpage

\begin{figure*}
\includegraphics[width=0.64\columnwidth]{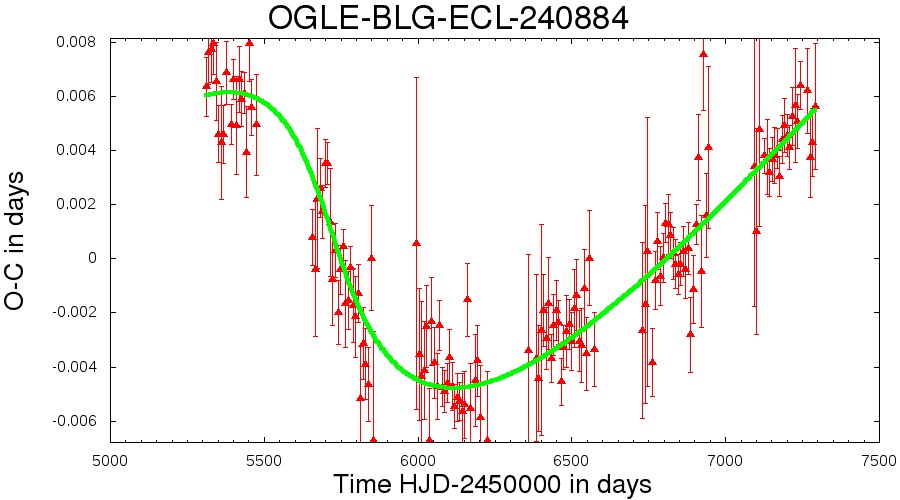}
\includegraphics[width=0.64\columnwidth]{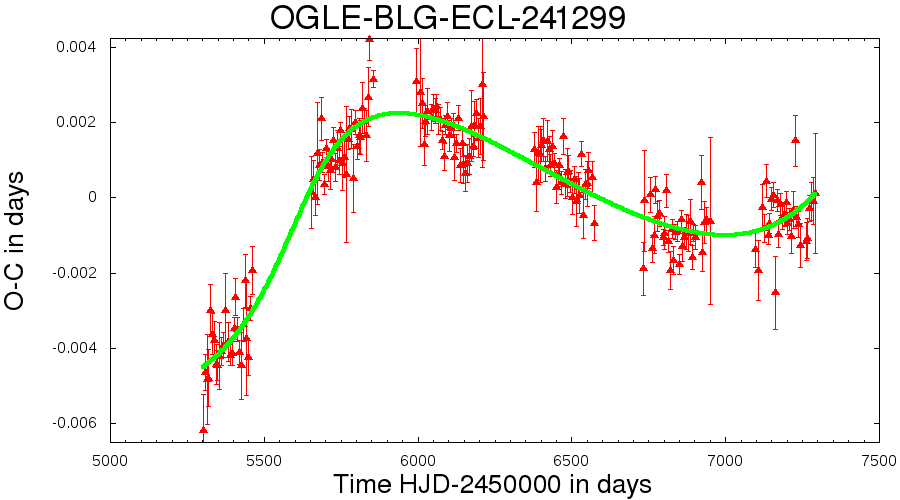}
\includegraphics[width=0.64\columnwidth]{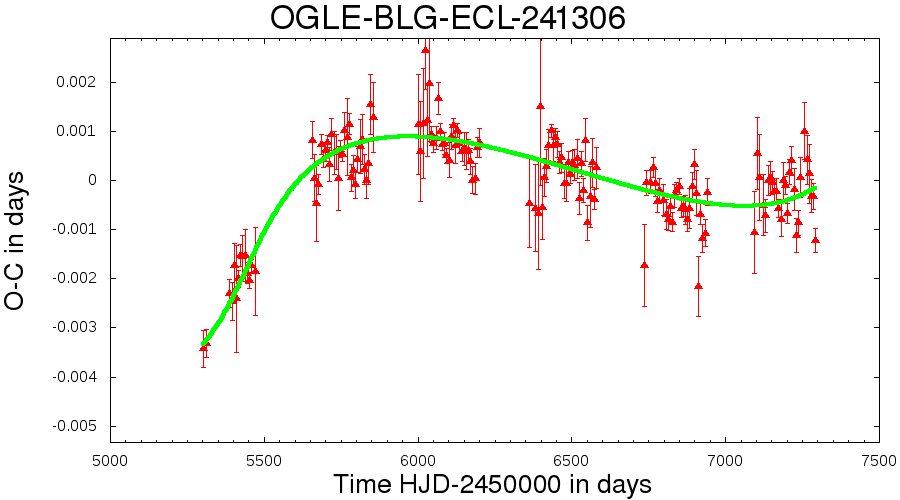}

\includegraphics[width=0.64\columnwidth]{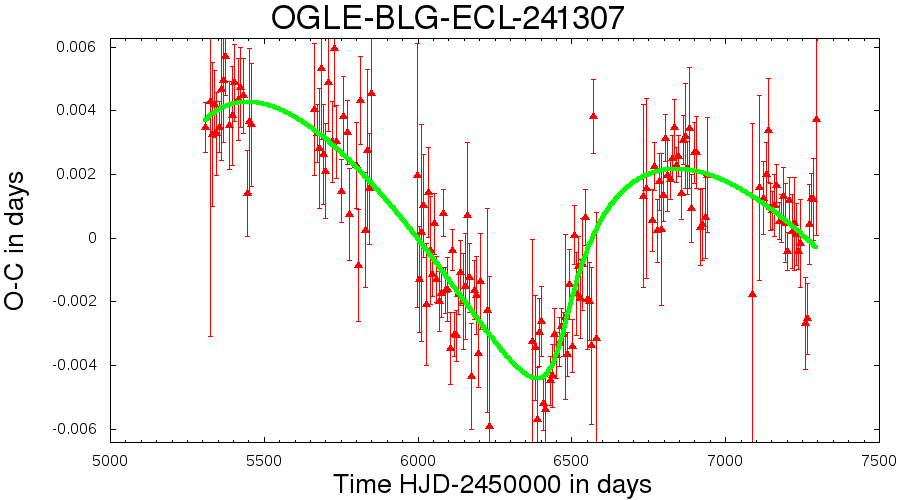}
\includegraphics[width=0.64\columnwidth]{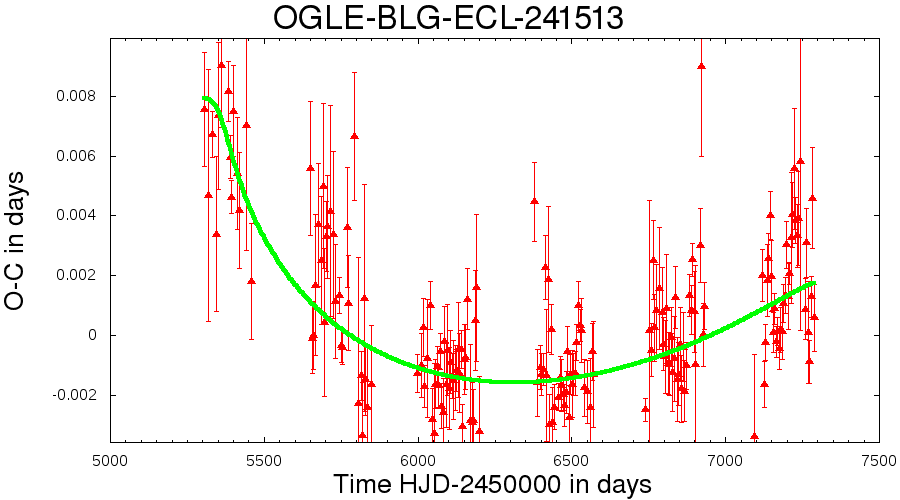}
\includegraphics[width=0.64\columnwidth]{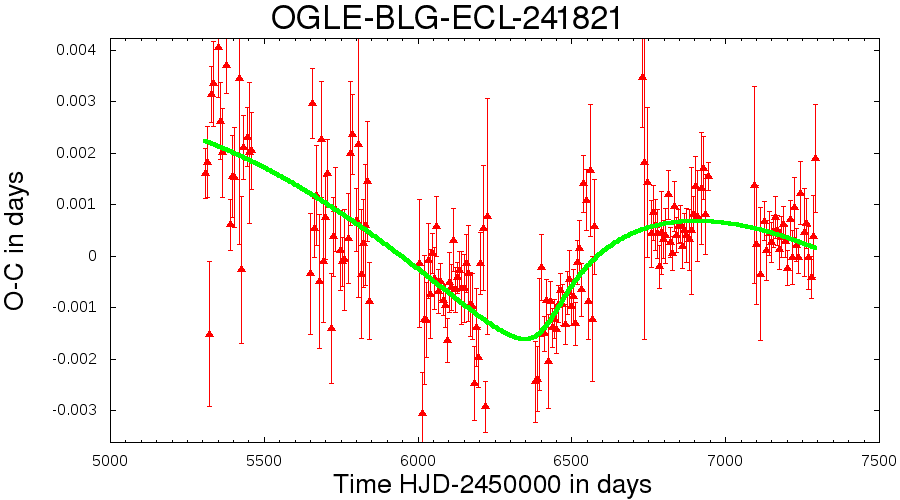}

\includegraphics[width=0.64\columnwidth]{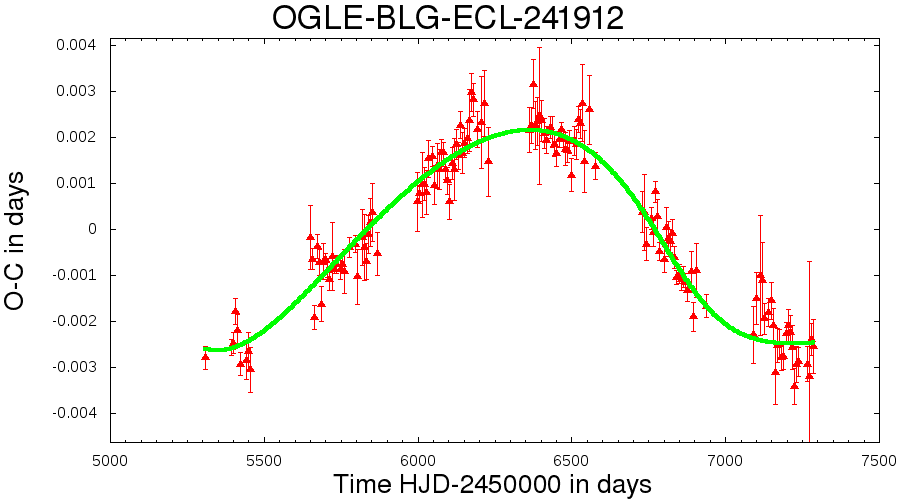}
\includegraphics[width=0.64\columnwidth]{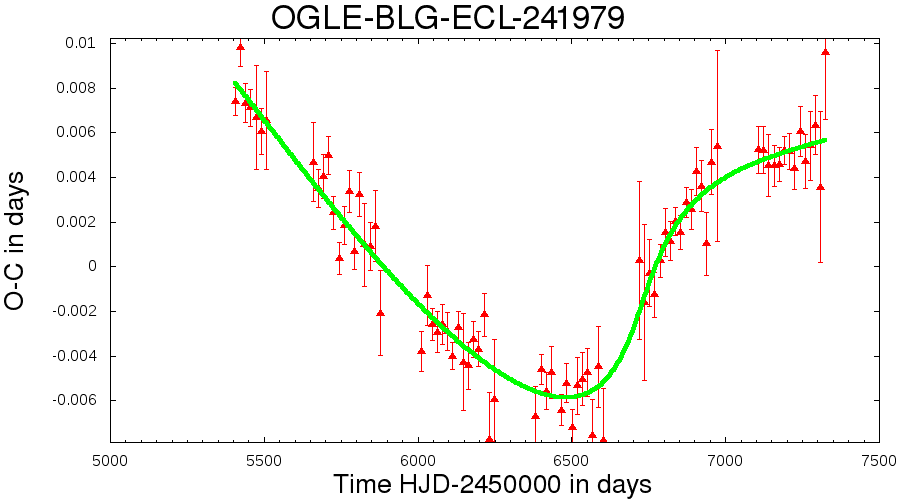}
\includegraphics[width=0.64\columnwidth]{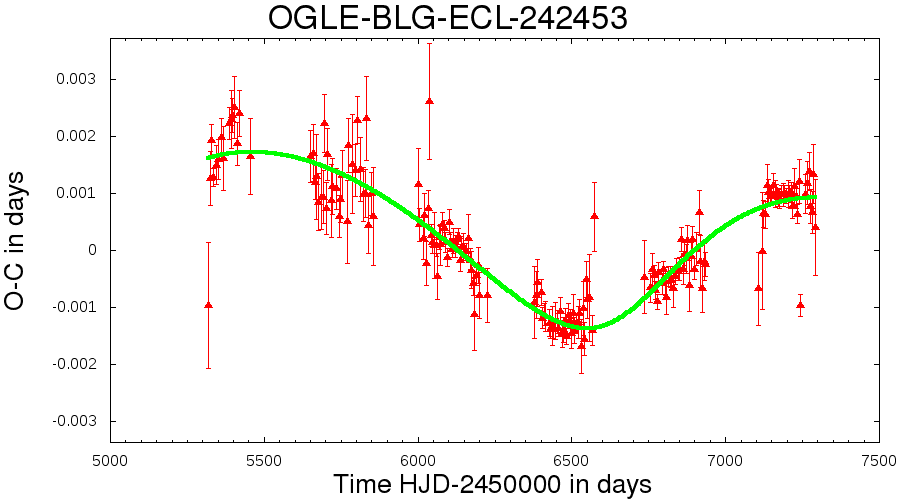}

\includegraphics[width=0.64\columnwidth]{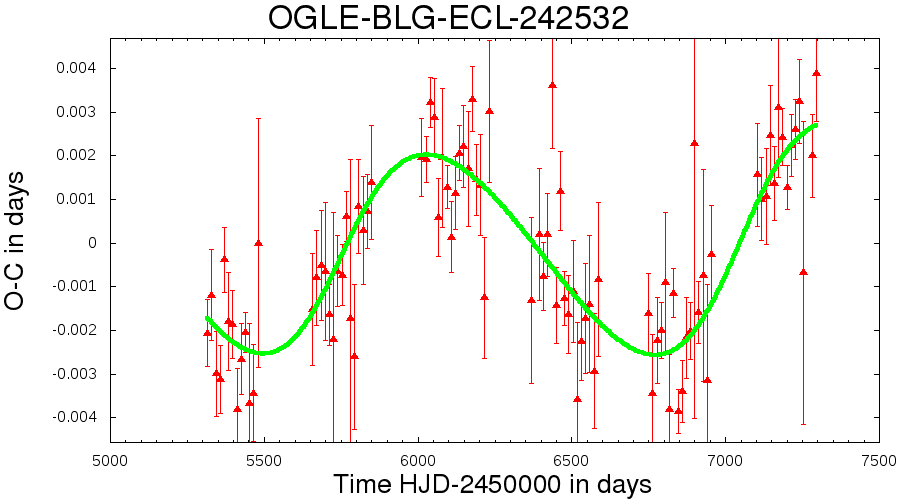}
\includegraphics[width=0.64\columnwidth]{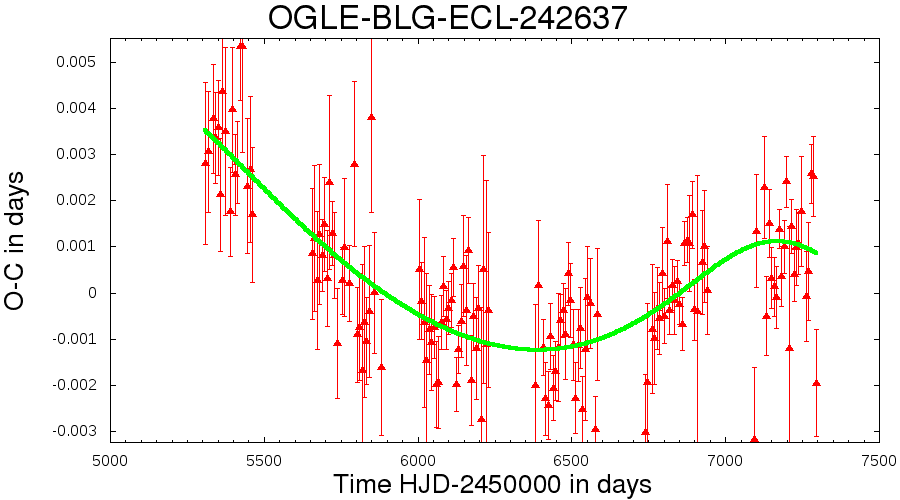}
\includegraphics[width=0.64\columnwidth]{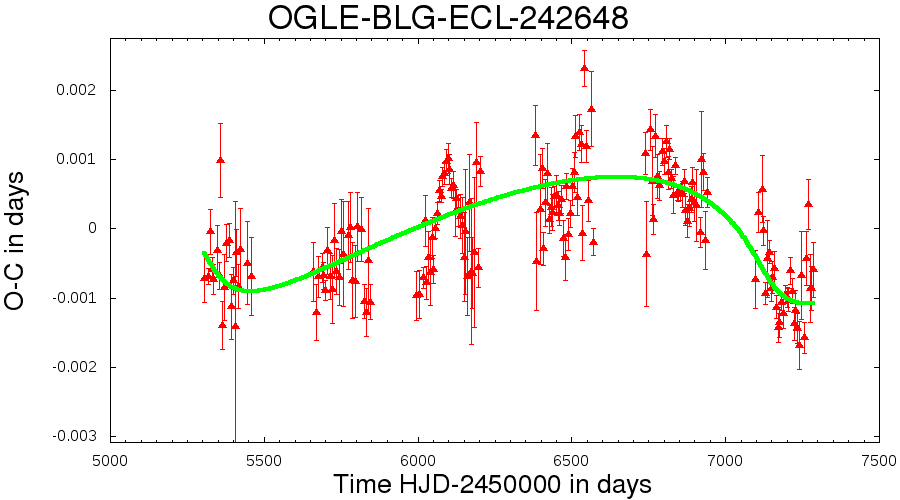}

\includegraphics[width=0.64\columnwidth]{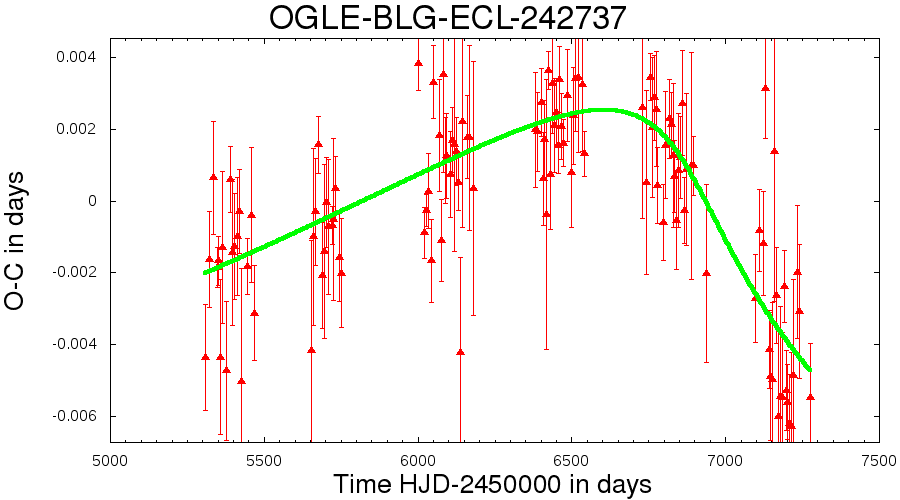}
\includegraphics[width=0.64\columnwidth]{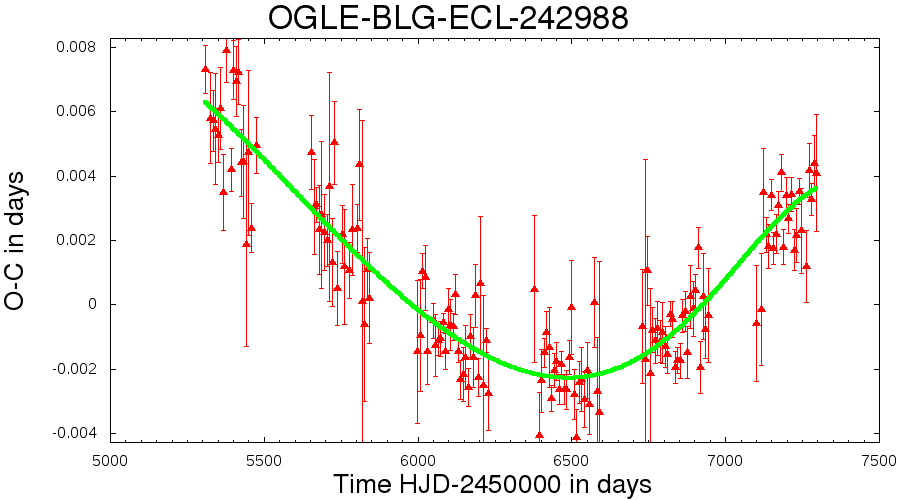}
\includegraphics[width=0.64\columnwidth]{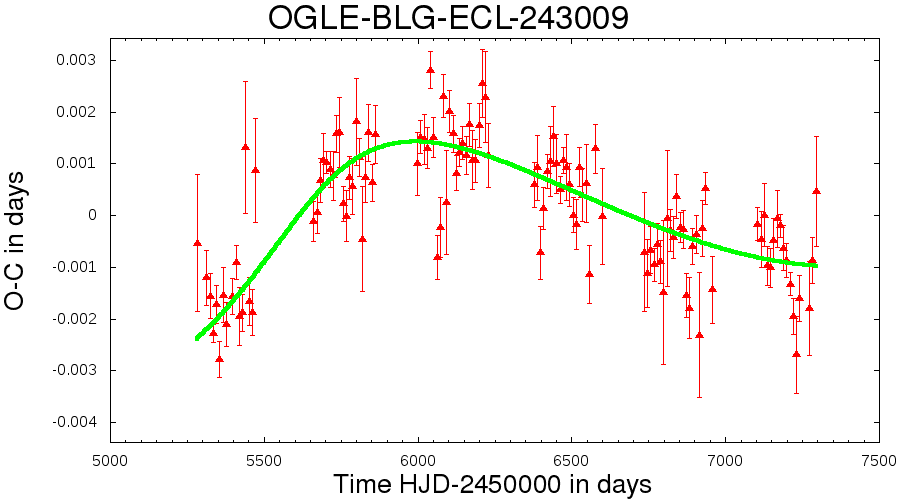}

\includegraphics[width=0.64\columnwidth]{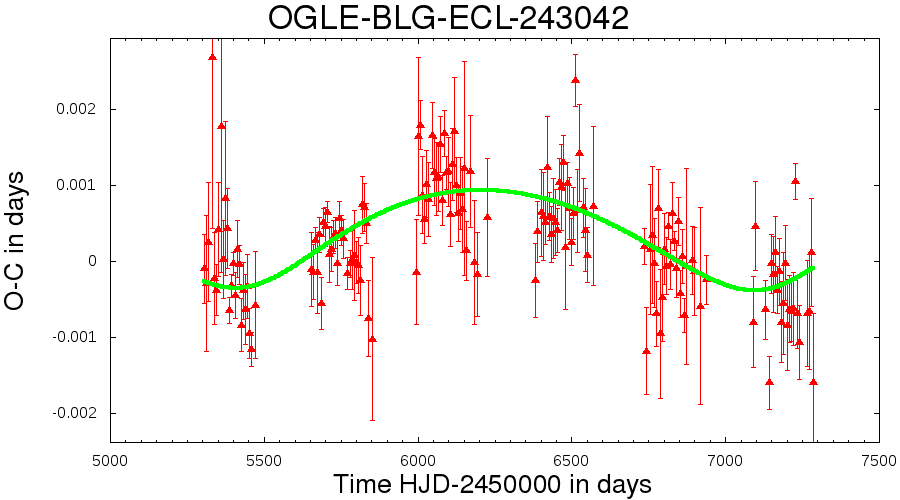}
\includegraphics[width=0.64\columnwidth]{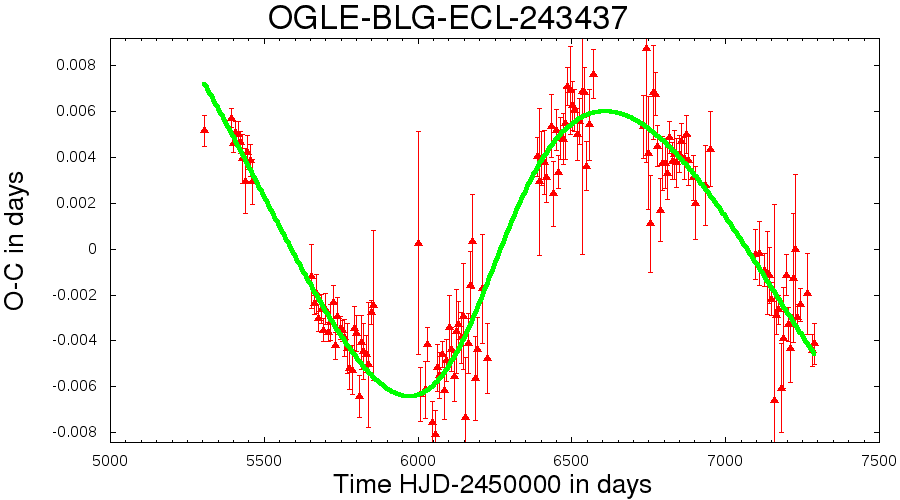}
\includegraphics[width=0.64\columnwidth]{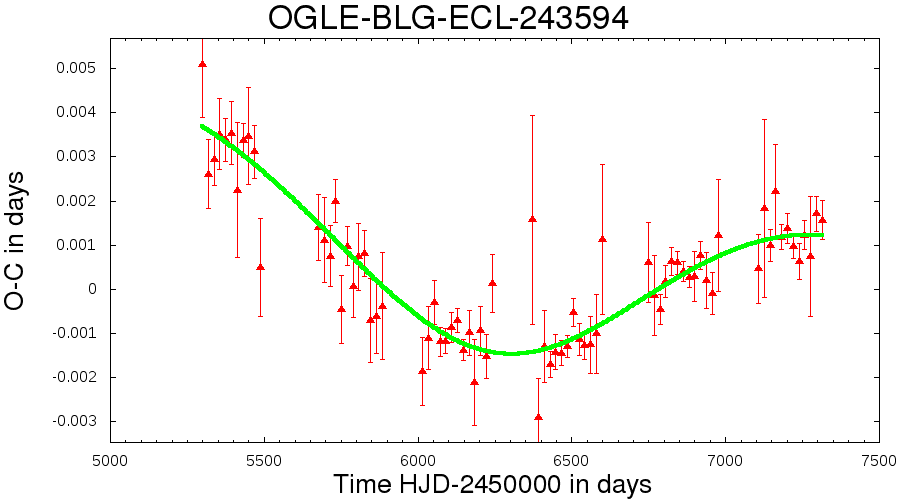}

\includegraphics[width=0.64\columnwidth]{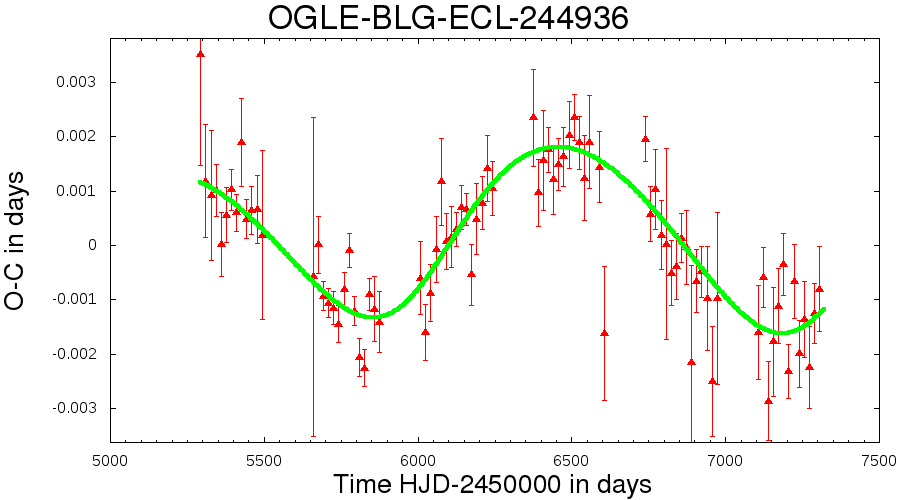}
\includegraphics[width=0.64\columnwidth]{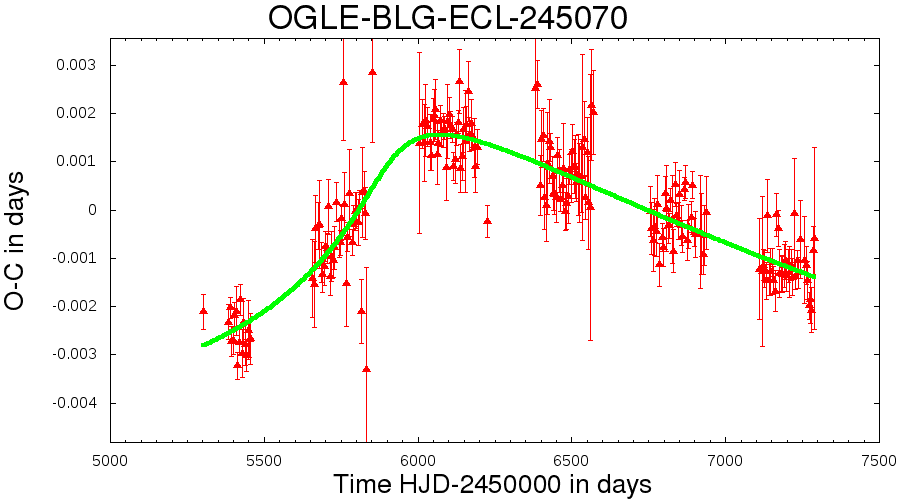}
\includegraphics[width=0.64\columnwidth]{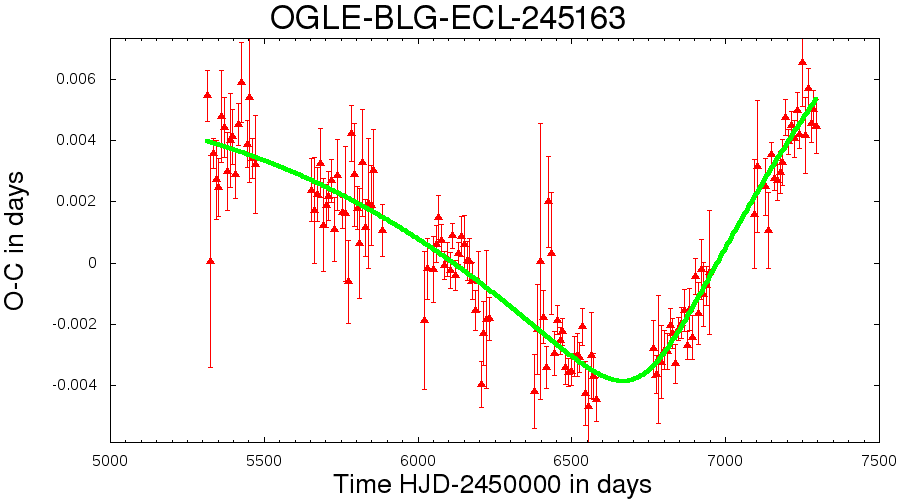}

\includegraphics[width=0.64\columnwidth]{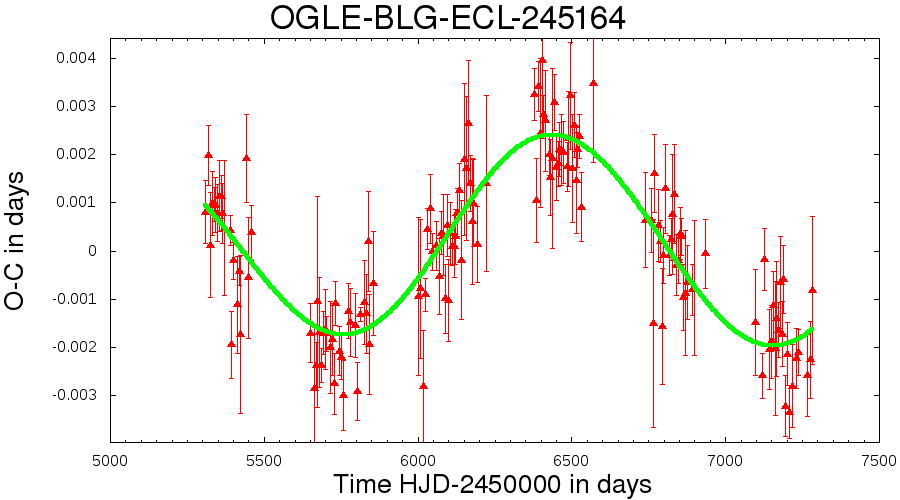}
\includegraphics[width=0.64\columnwidth]{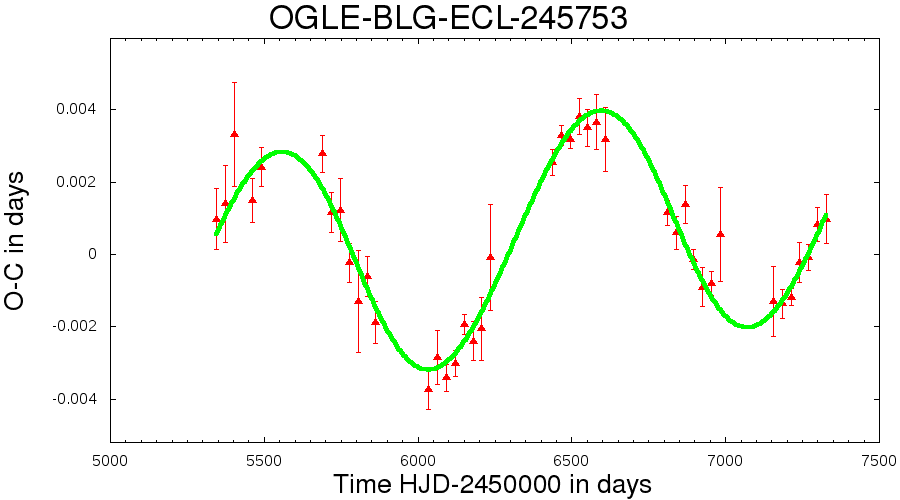}
\includegraphics[width=0.64\columnwidth]{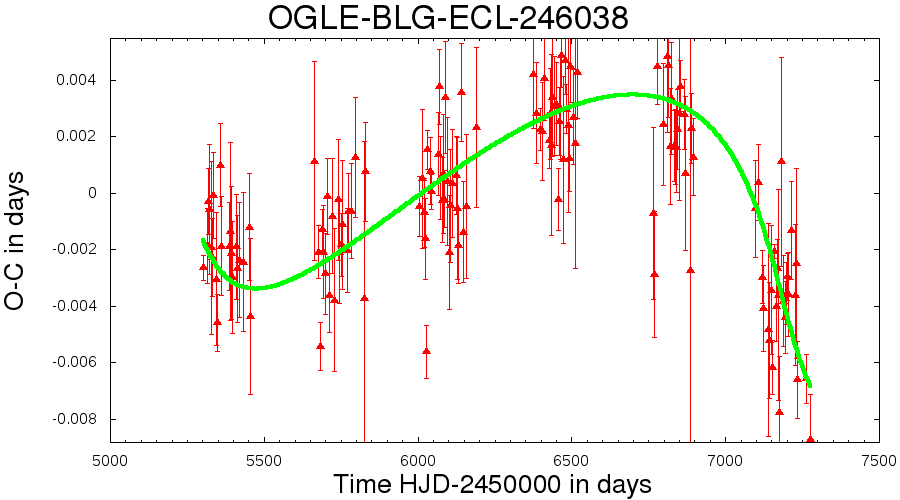}

\end{figure*}
\clearpage

\begin{figure*}
\includegraphics[width=0.64\columnwidth]{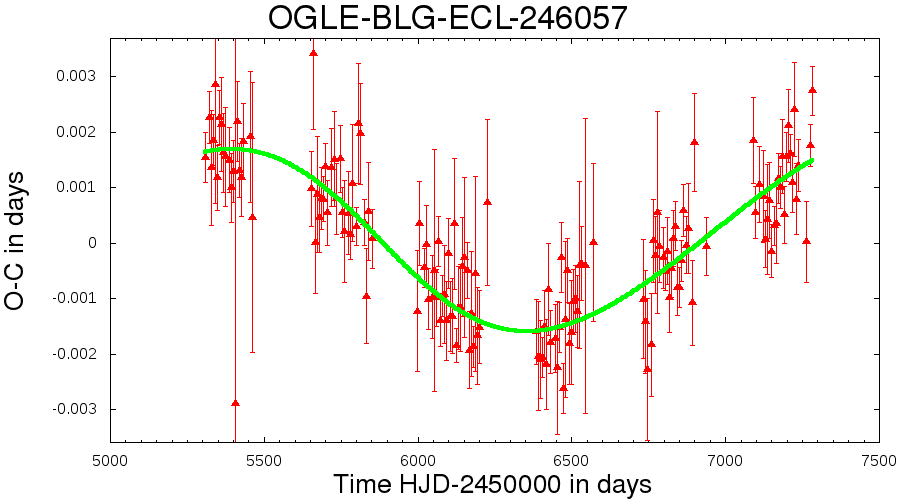}
\includegraphics[width=0.64\columnwidth]{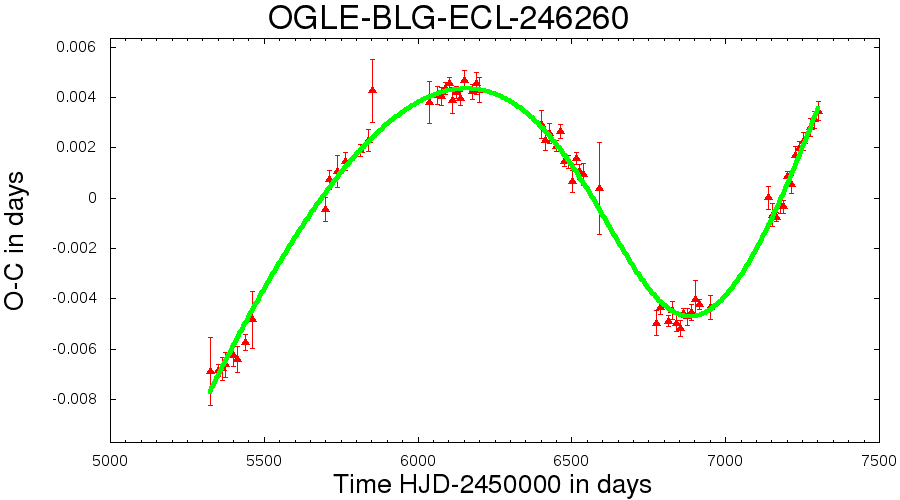}
\includegraphics[width=0.64\columnwidth]{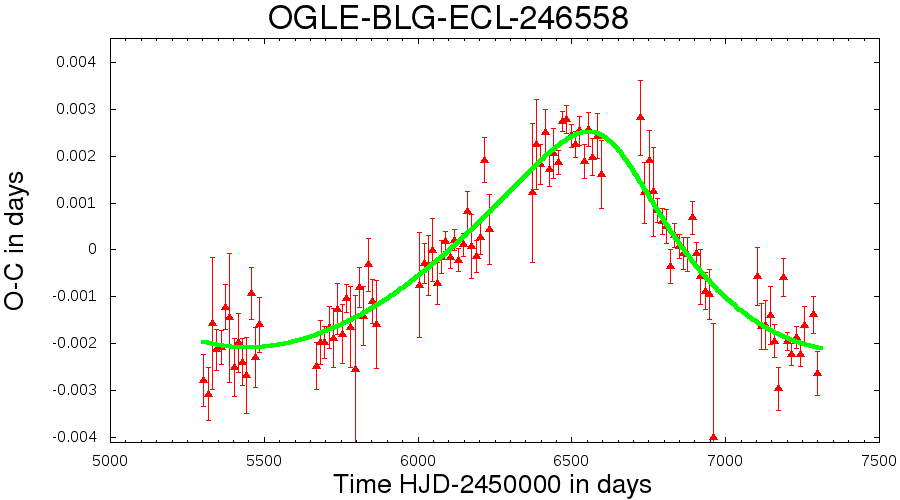}

\includegraphics[width=0.64\columnwidth]{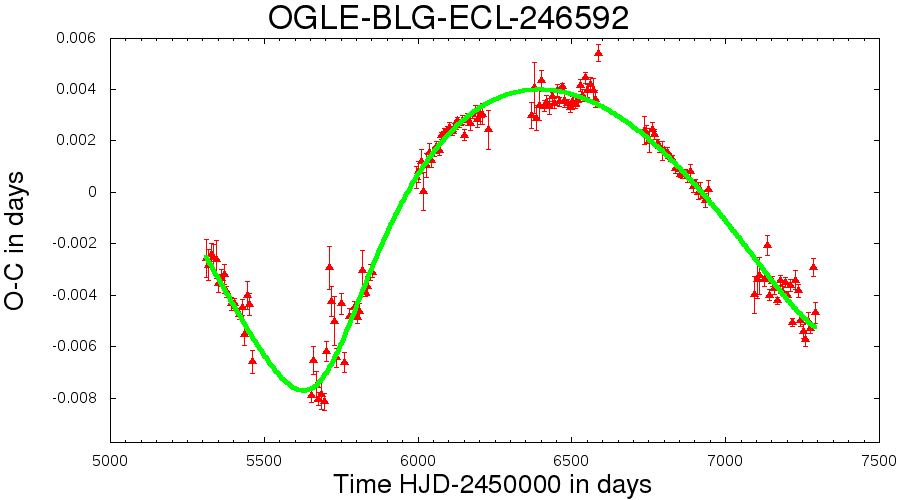}
\includegraphics[width=0.64\columnwidth]{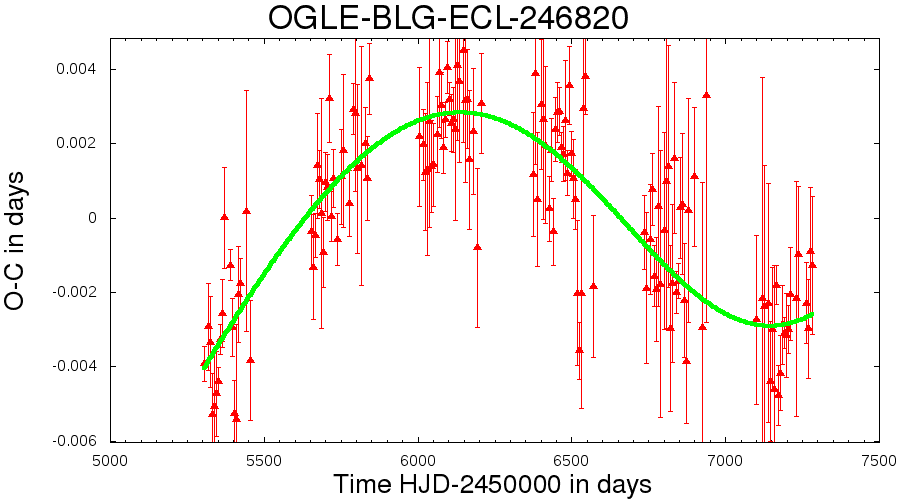}
\includegraphics[width=0.64\columnwidth]{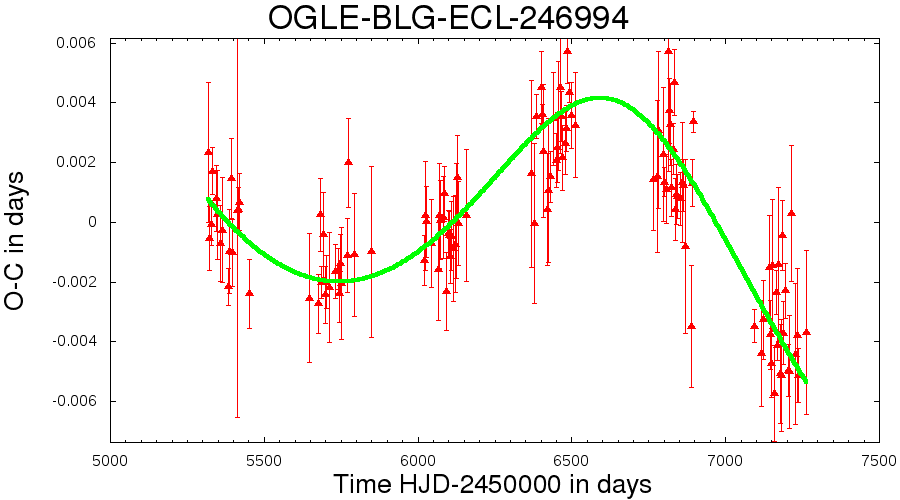}

\includegraphics[width=0.64\columnwidth]{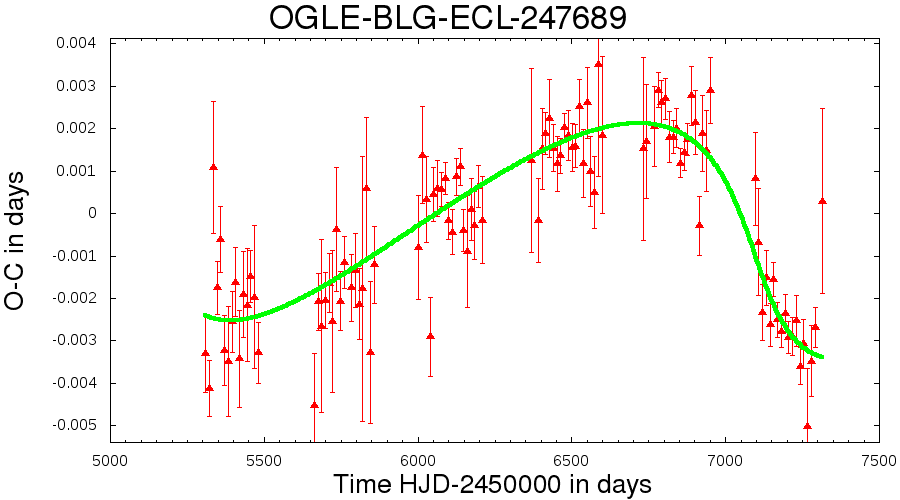}
\includegraphics[width=0.64\columnwidth]{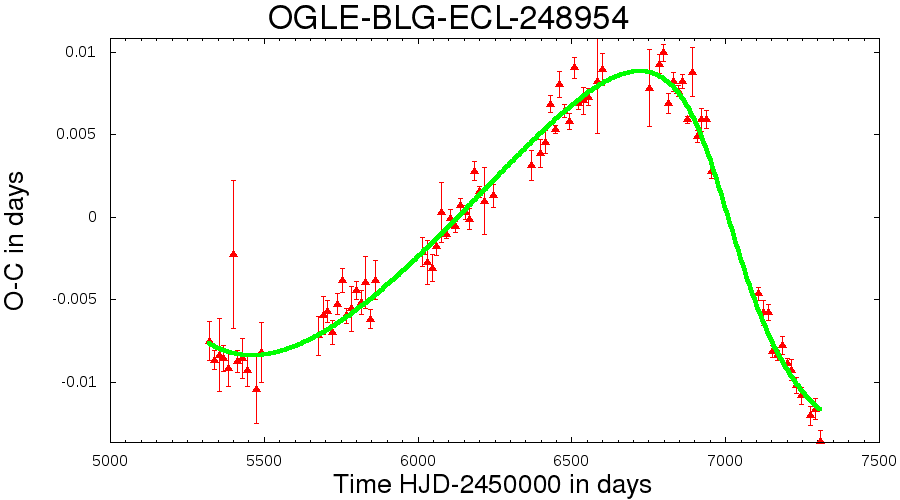}
\includegraphics[width=0.64\columnwidth]{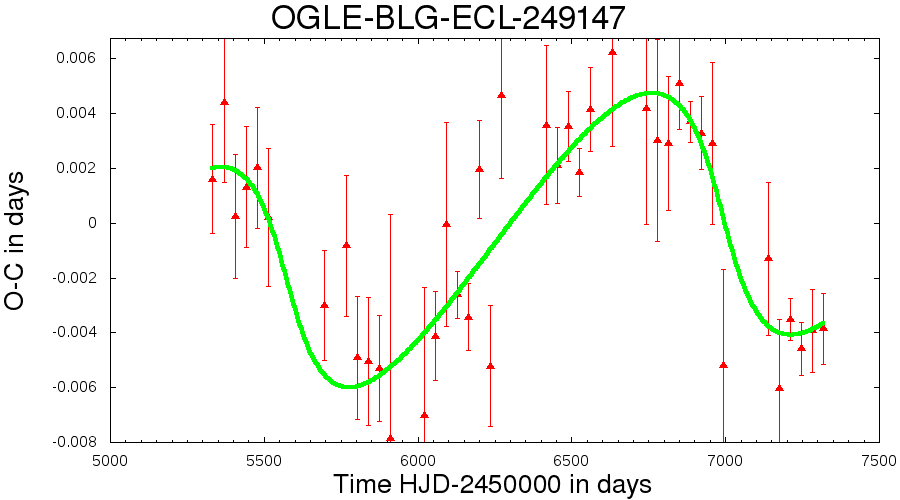}

\includegraphics[width=0.64\columnwidth]{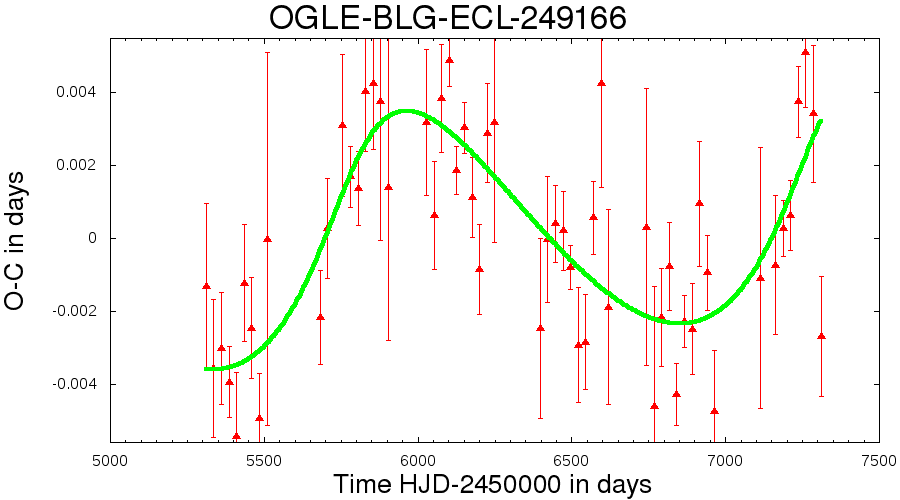}
\includegraphics[width=0.64\columnwidth]{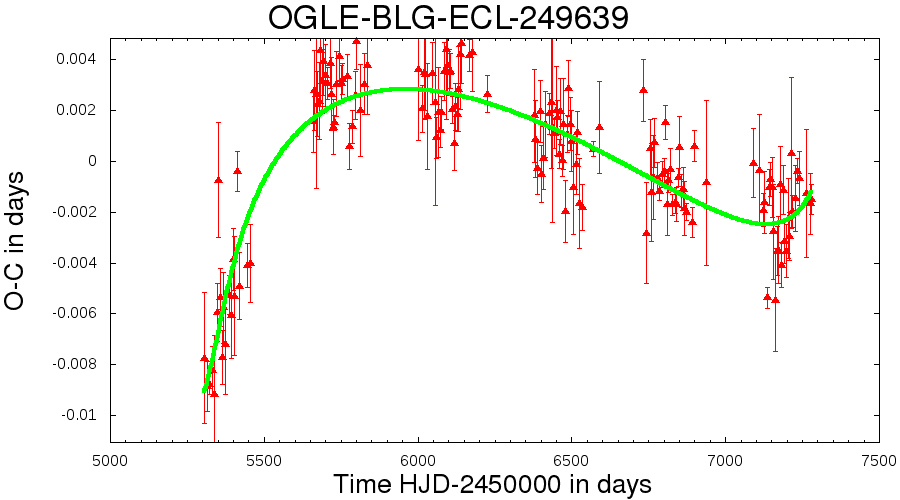}
\includegraphics[width=0.64\columnwidth]{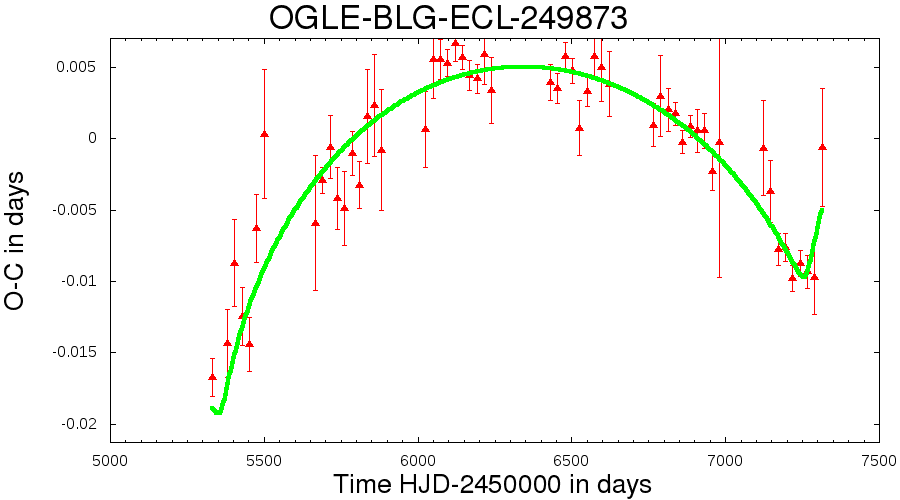}

\includegraphics[width=0.64\columnwidth]{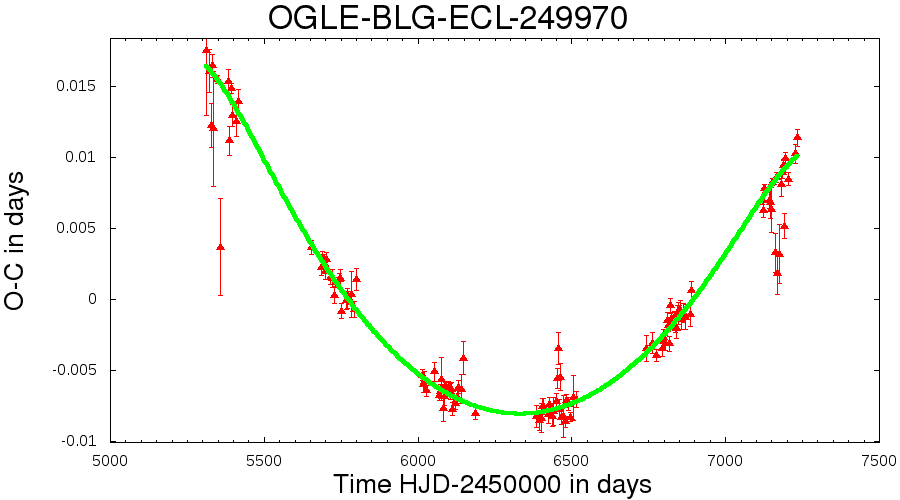}
\includegraphics[width=0.64\columnwidth]{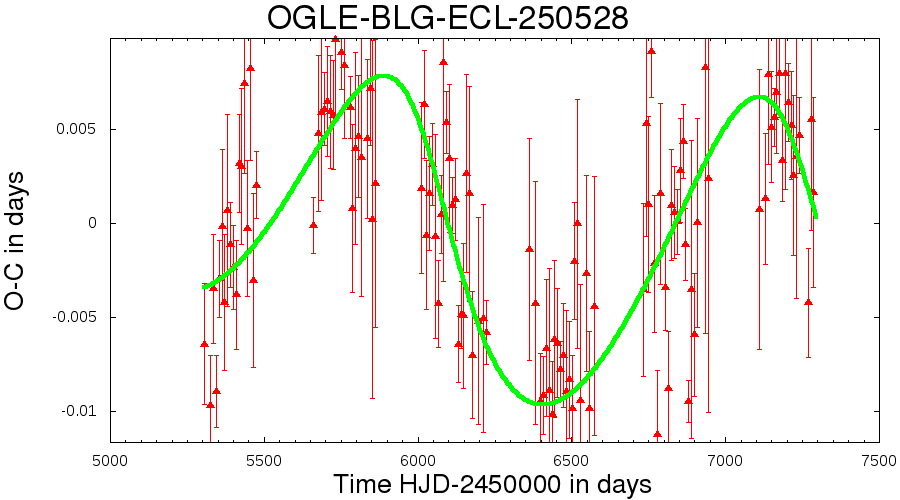}
\includegraphics[width=0.64\columnwidth]{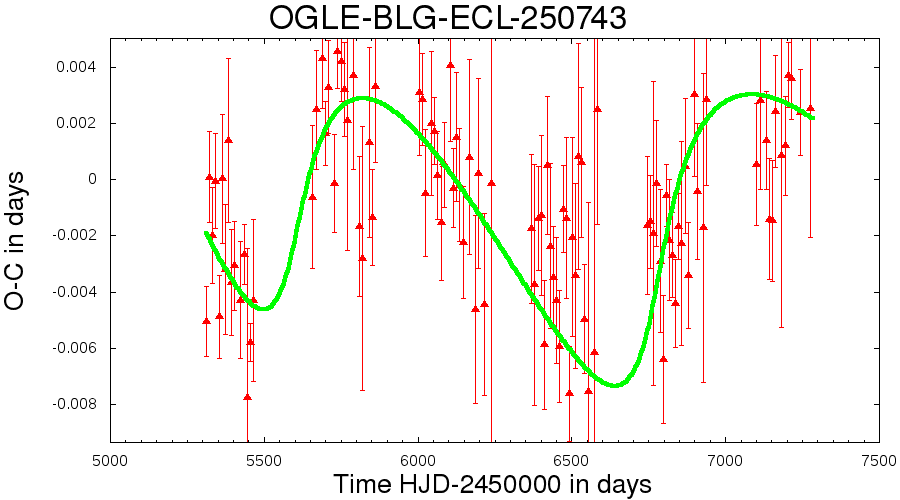}

\includegraphics[width=0.64\columnwidth]{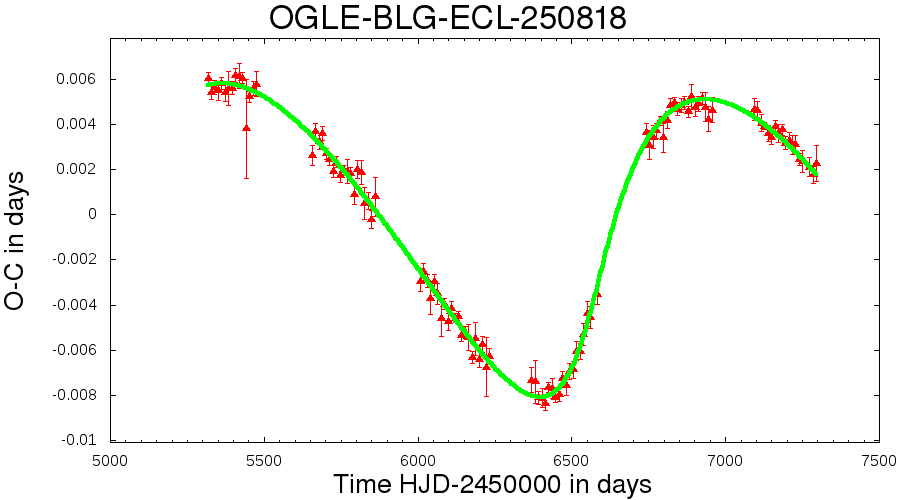}
\includegraphics[width=0.64\columnwidth]{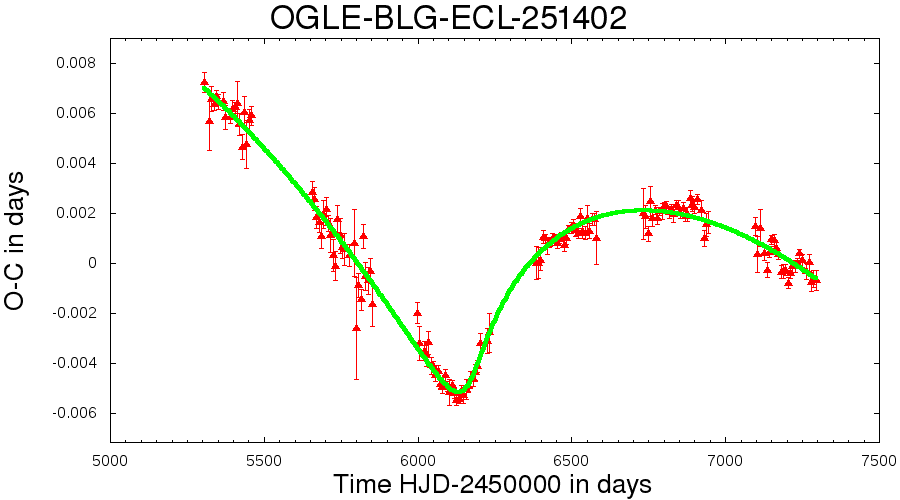}
\includegraphics[width=0.64\columnwidth]{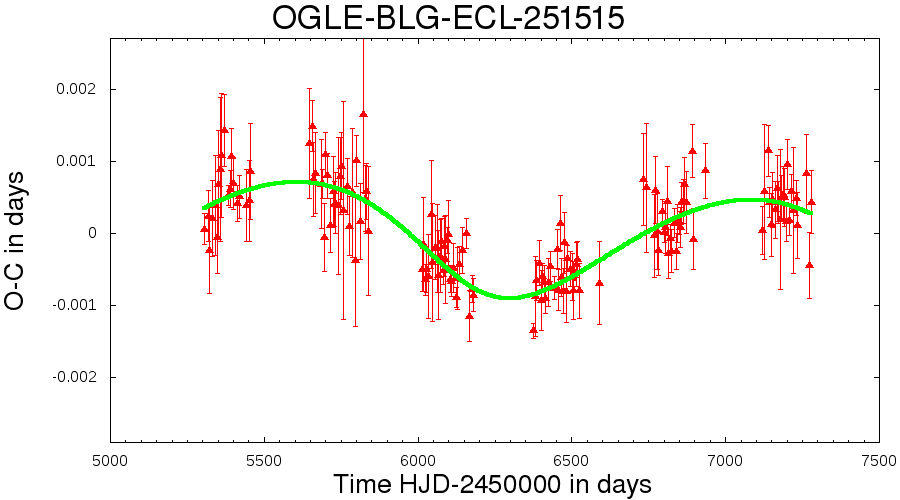}

\includegraphics[width=0.64\columnwidth]{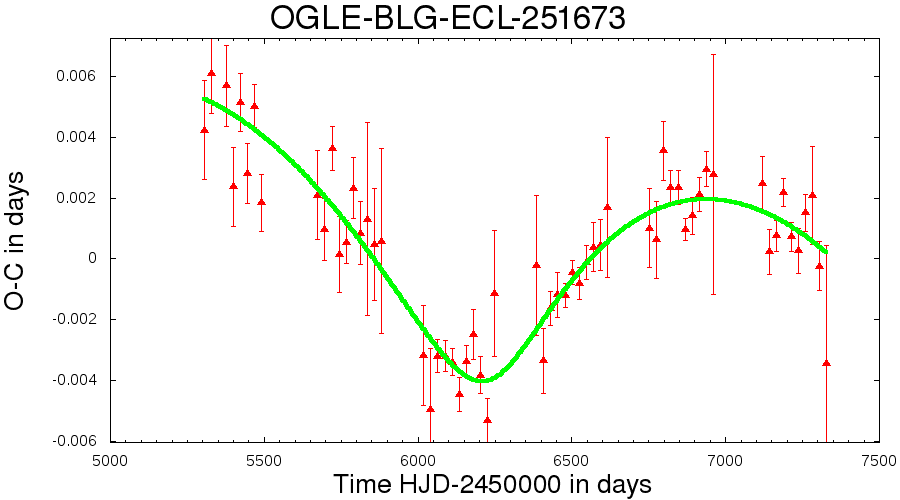}
\includegraphics[width=0.64\columnwidth]{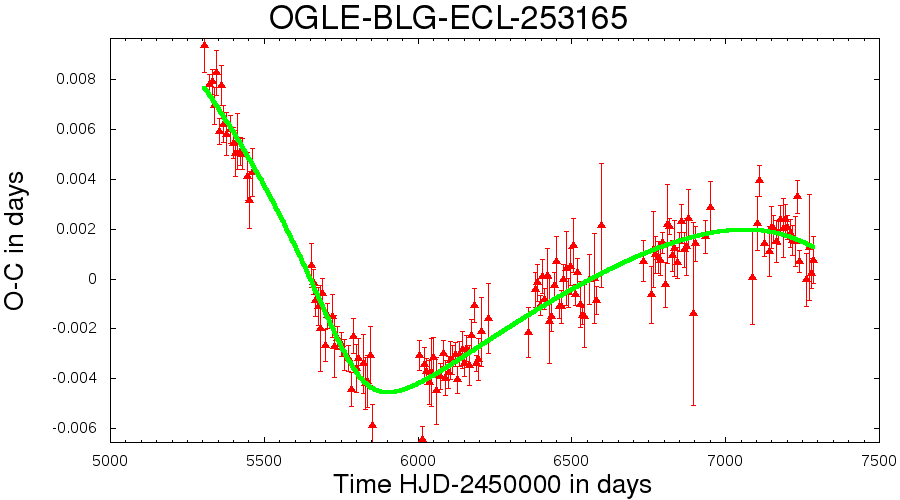}
\includegraphics[width=0.64\columnwidth]{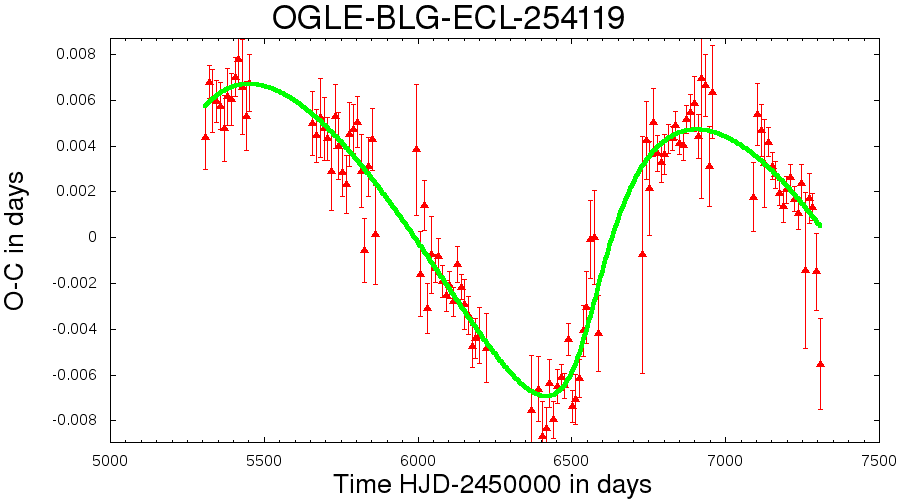}

\includegraphics[width=0.64\columnwidth]{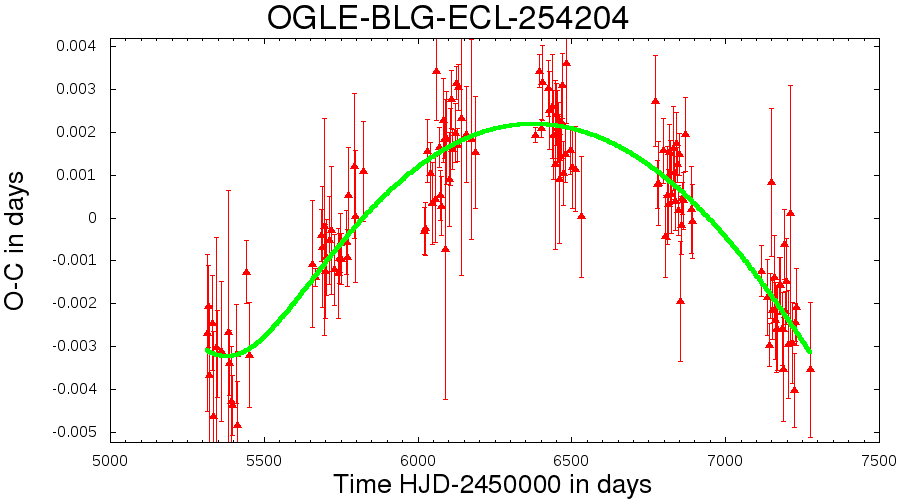}
\includegraphics[width=0.64\columnwidth]{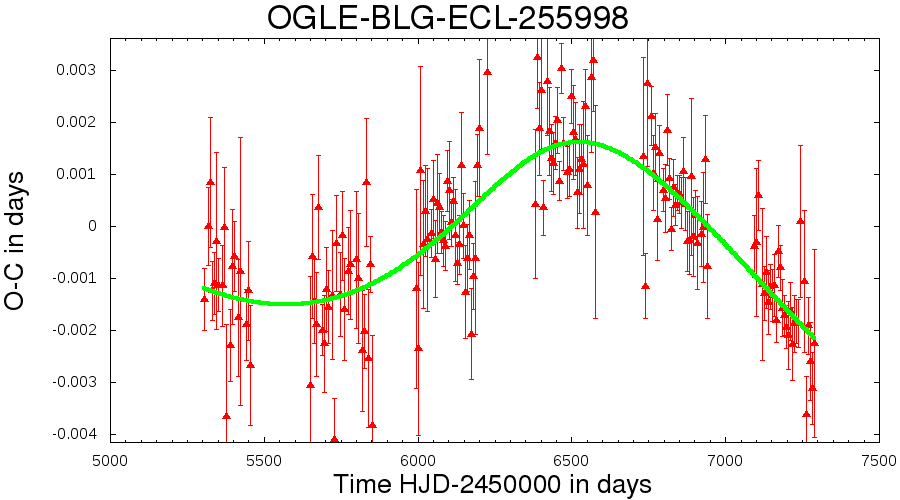}
\includegraphics[width=0.64\columnwidth]{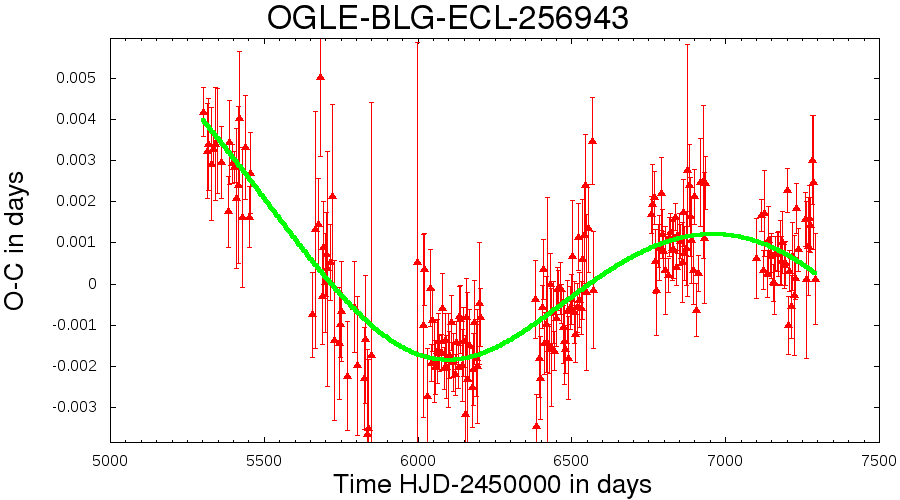}

\end{figure*}
\clearpage

\begin{figure*}
\includegraphics[width=0.64\columnwidth]{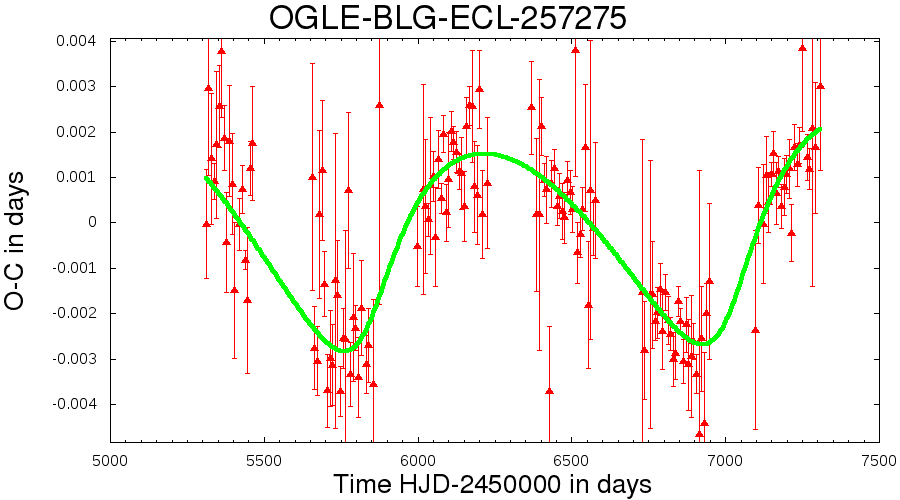}
\includegraphics[width=0.64\columnwidth]{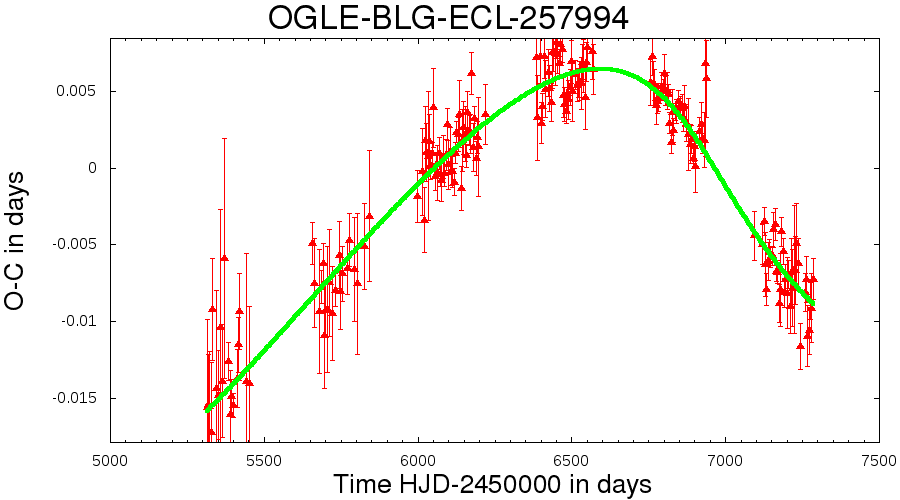}
\includegraphics[width=0.64\columnwidth]{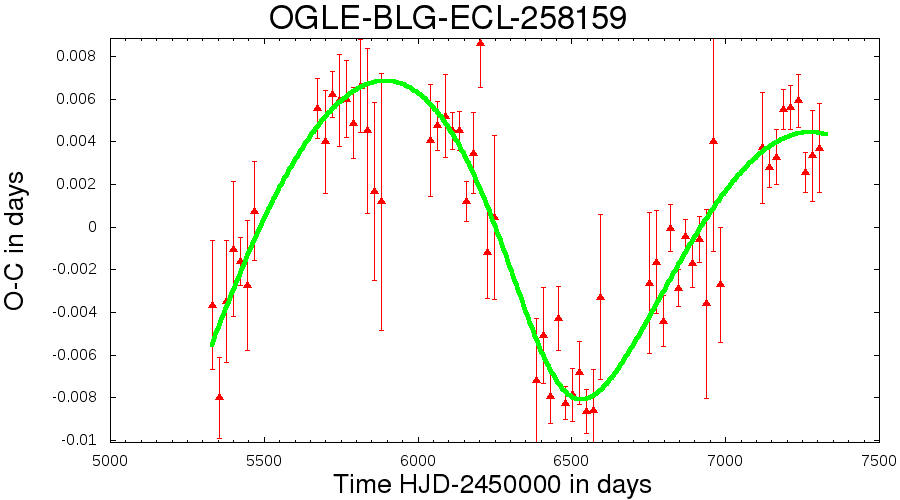}

\includegraphics[width=0.64\columnwidth]{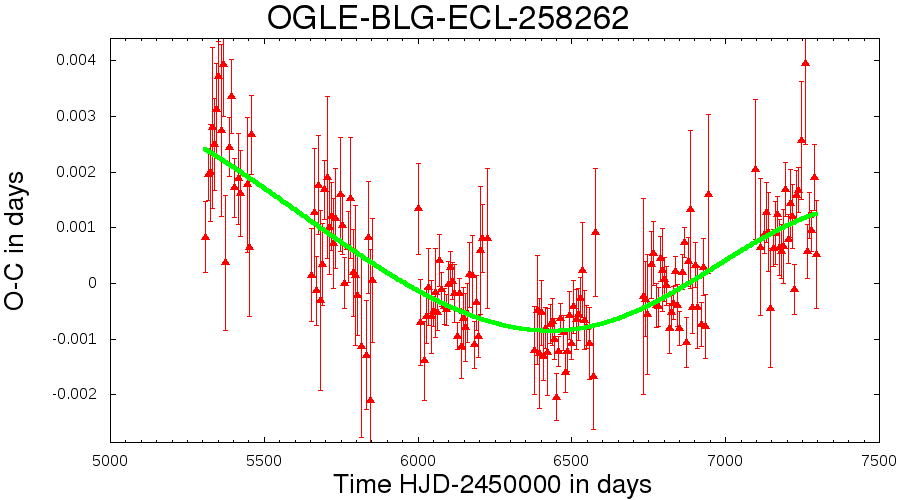}
\includegraphics[width=0.64\columnwidth]{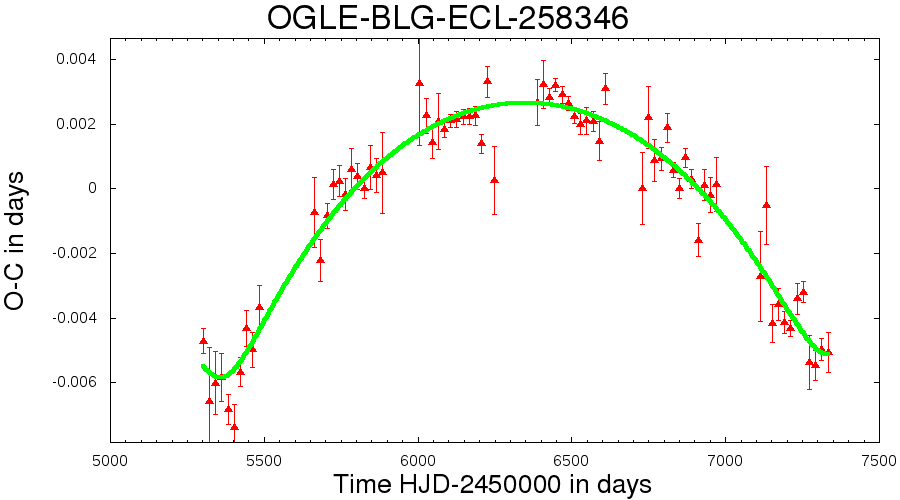}
\includegraphics[width=0.64\columnwidth]{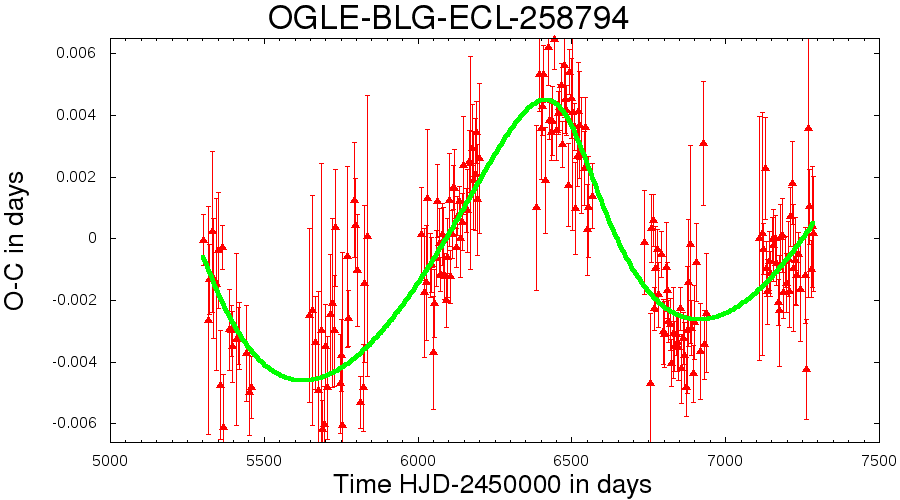}

\includegraphics[width=0.64\columnwidth]{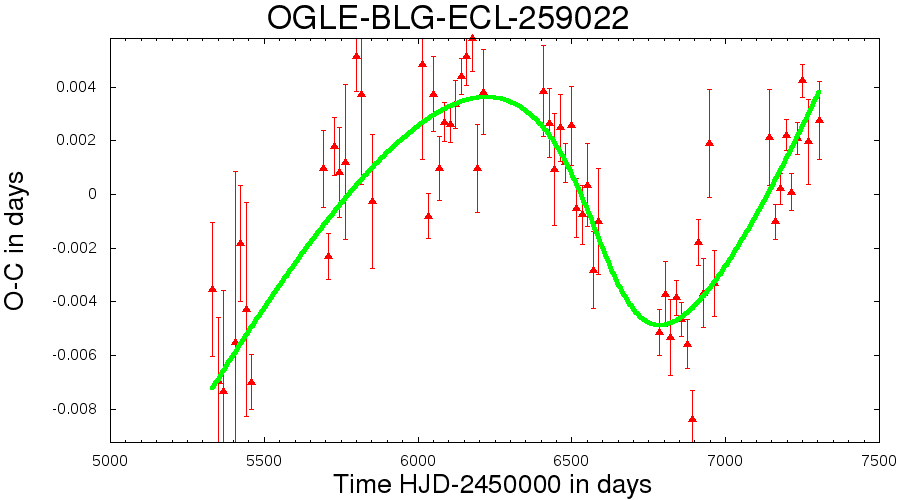}
\includegraphics[width=0.64\columnwidth]{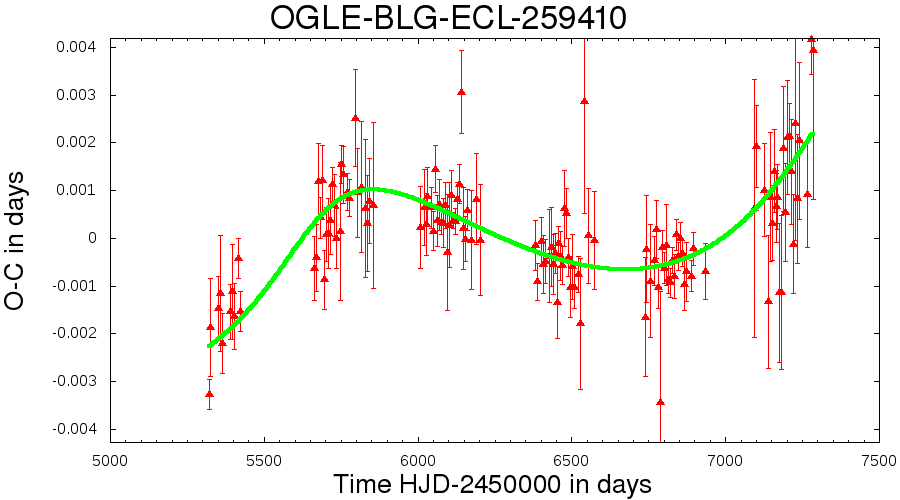}
\includegraphics[width=0.64\columnwidth]{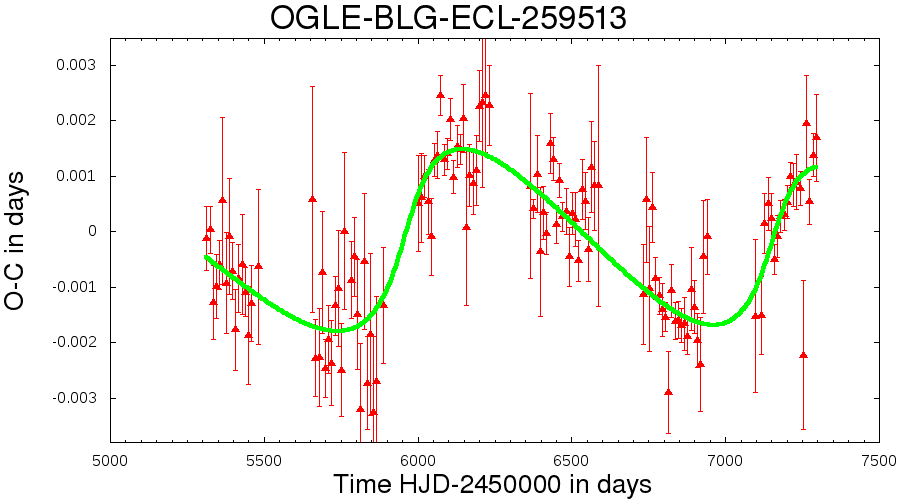}

\includegraphics[width=0.64\columnwidth]{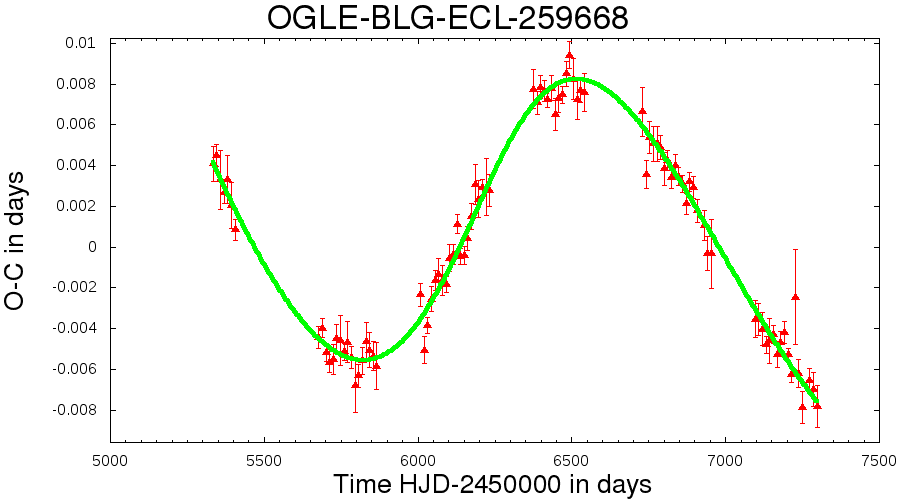}
\includegraphics[width=0.64\columnwidth]{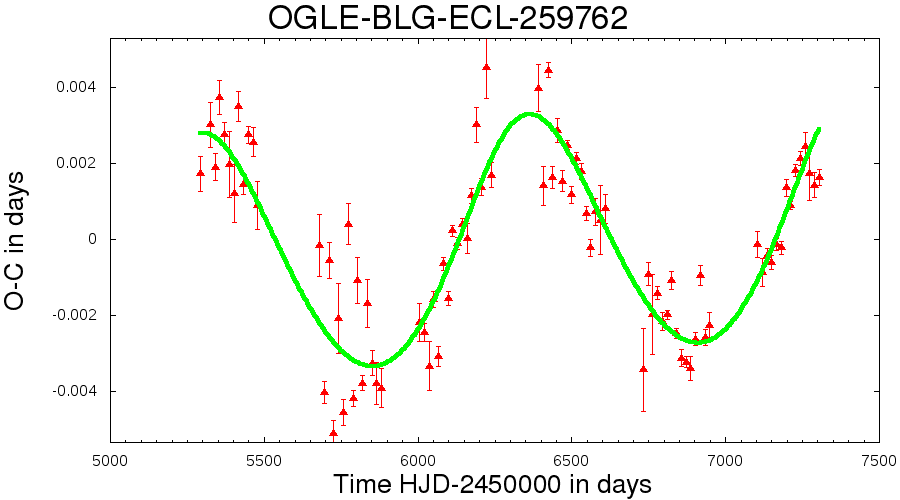}
\includegraphics[width=0.64\columnwidth]{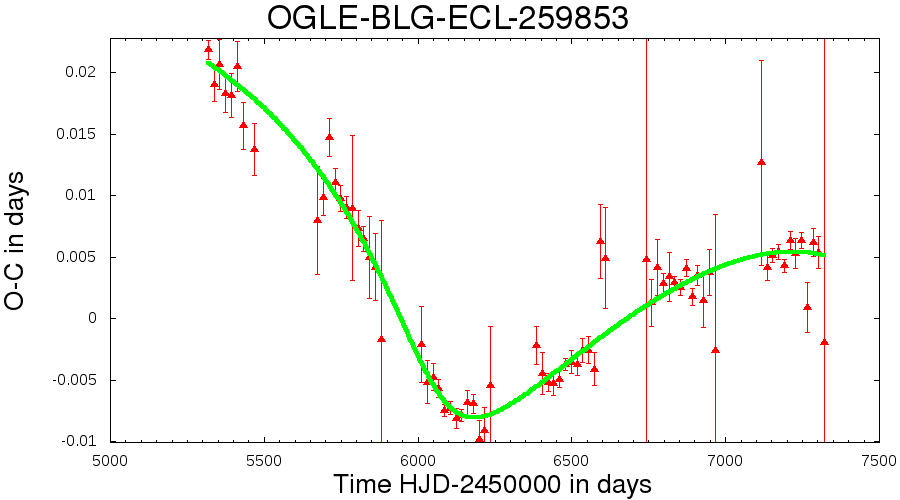}

\includegraphics[width=0.64\columnwidth]{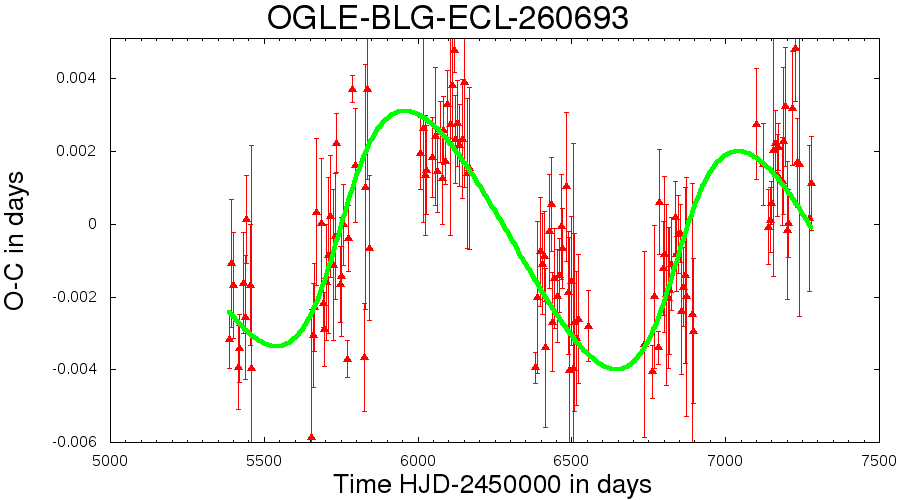}
\includegraphics[width=0.64\columnwidth]{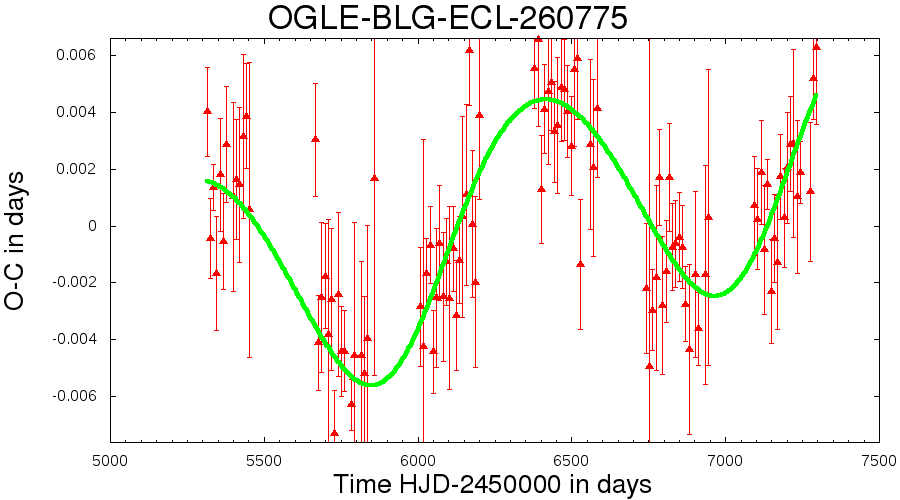}
\includegraphics[width=0.64\columnwidth]{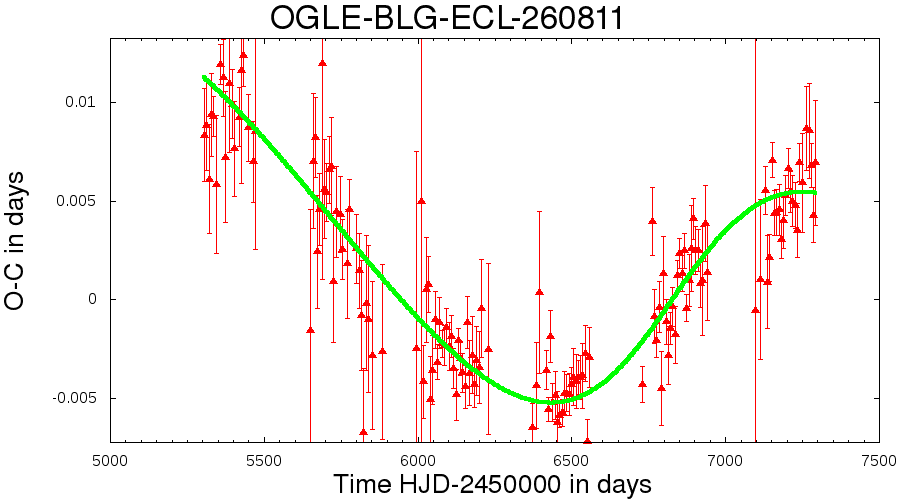}

\includegraphics[width=0.64\columnwidth]{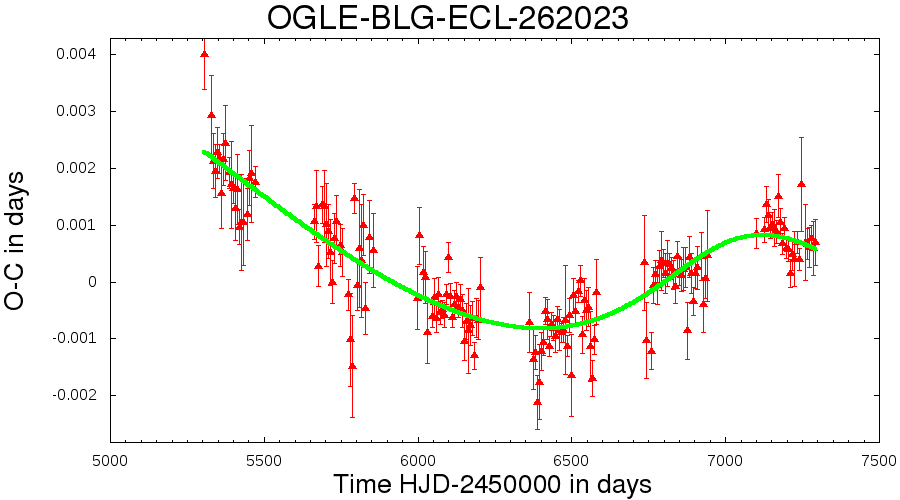}
\includegraphics[width=0.64\columnwidth]{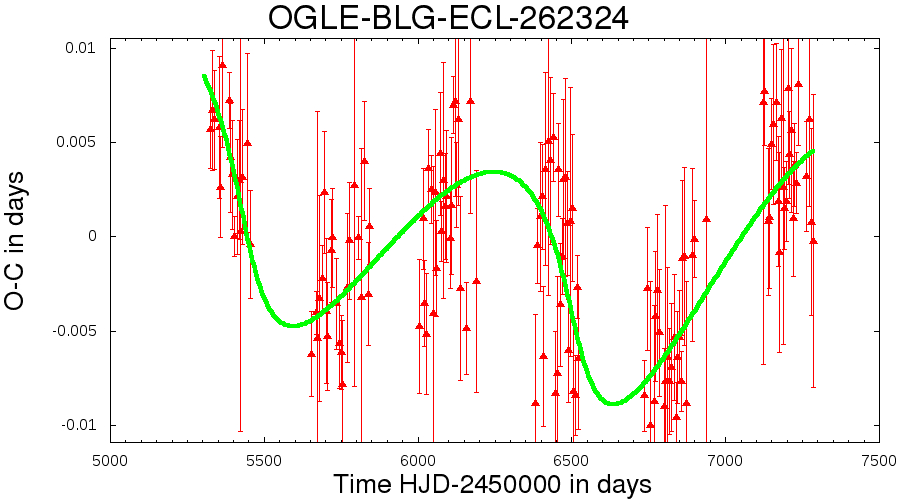}
\includegraphics[width=0.64\columnwidth]{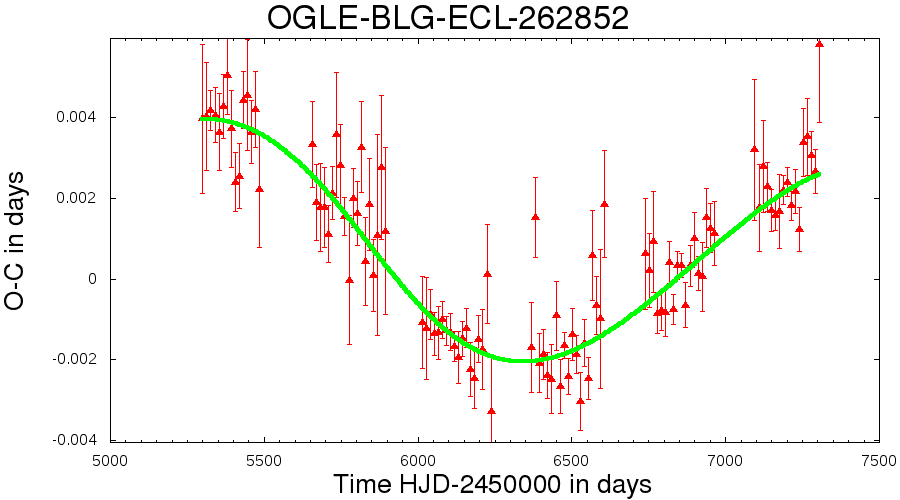}

\includegraphics[width=0.64\columnwidth]{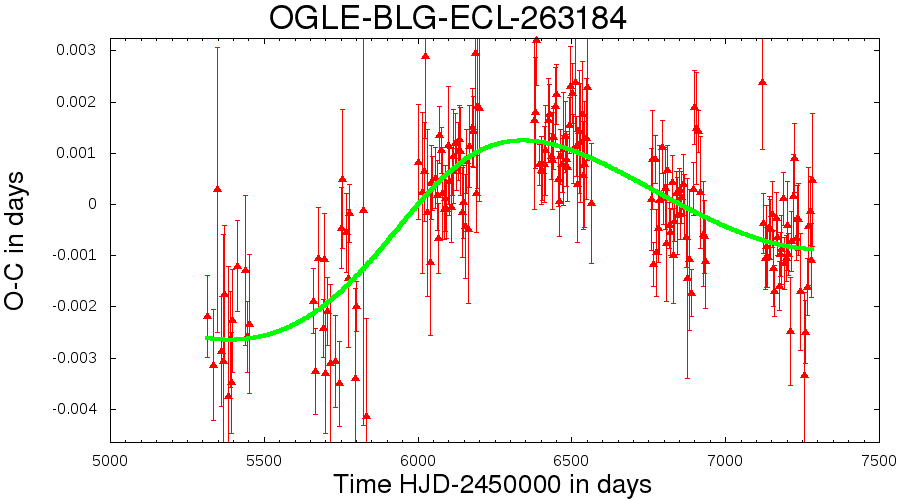}
\includegraphics[width=0.64\columnwidth]{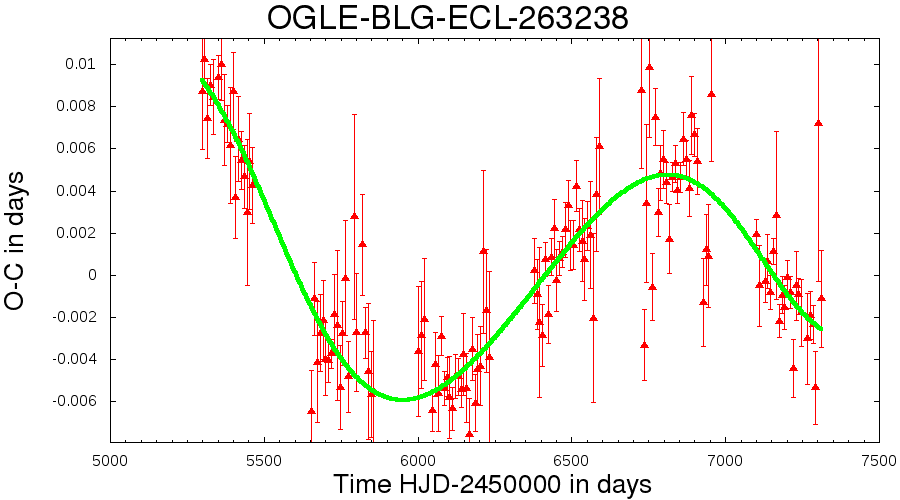}
\includegraphics[width=0.64\columnwidth]{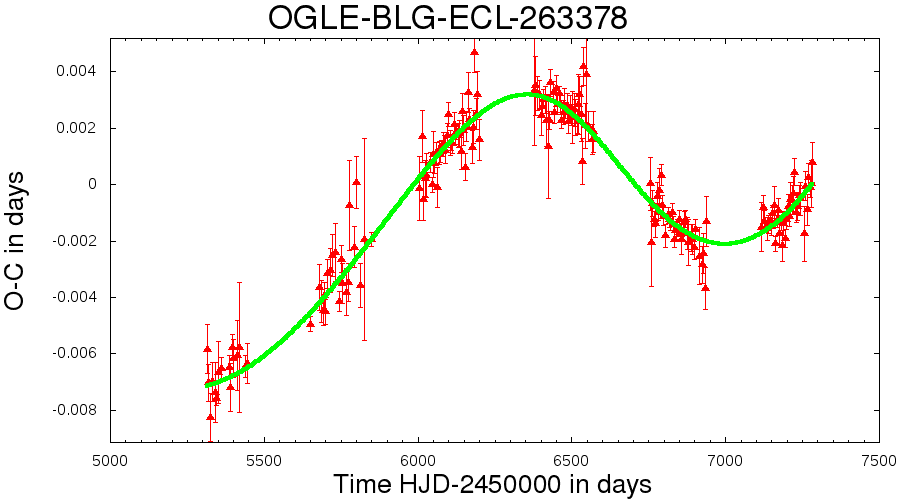}

\includegraphics[width=0.64\columnwidth]{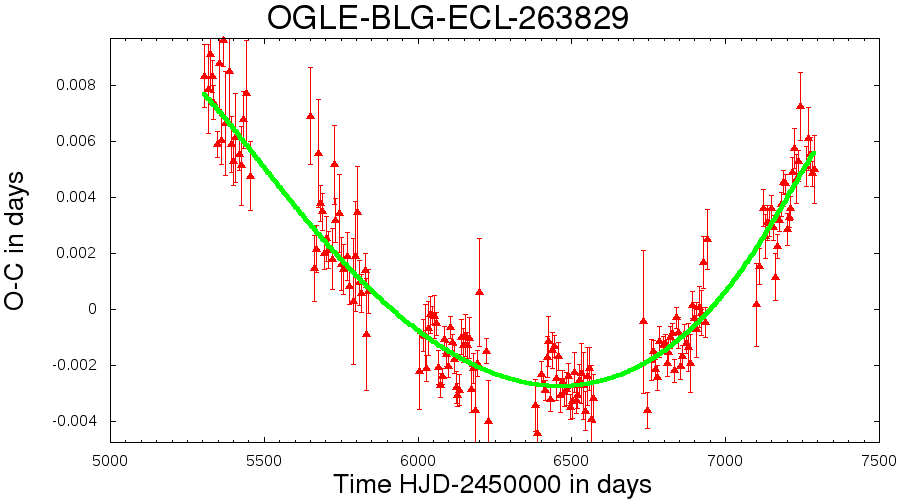}
\includegraphics[width=0.64\columnwidth]{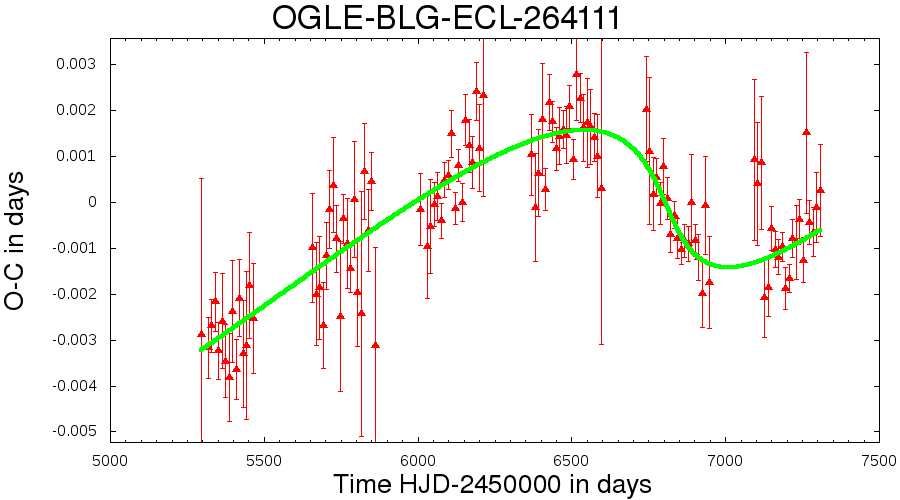}
\includegraphics[width=0.64\columnwidth]{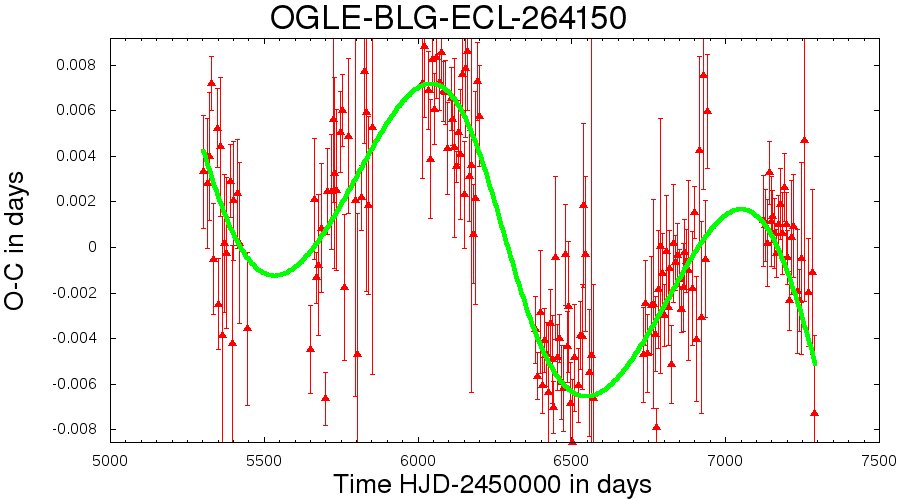}

\end{figure*}
\clearpage

\begin{figure*}
\includegraphics[width=0.64\columnwidth]{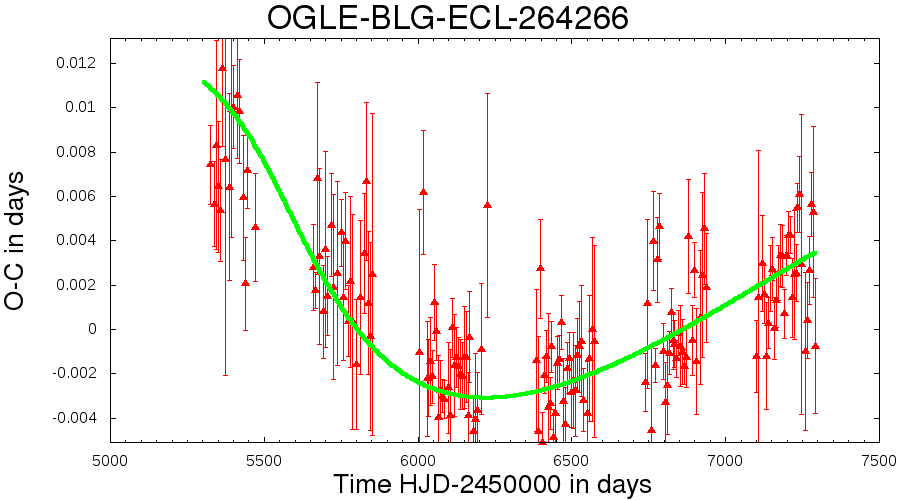}
\includegraphics[width=0.64\columnwidth]{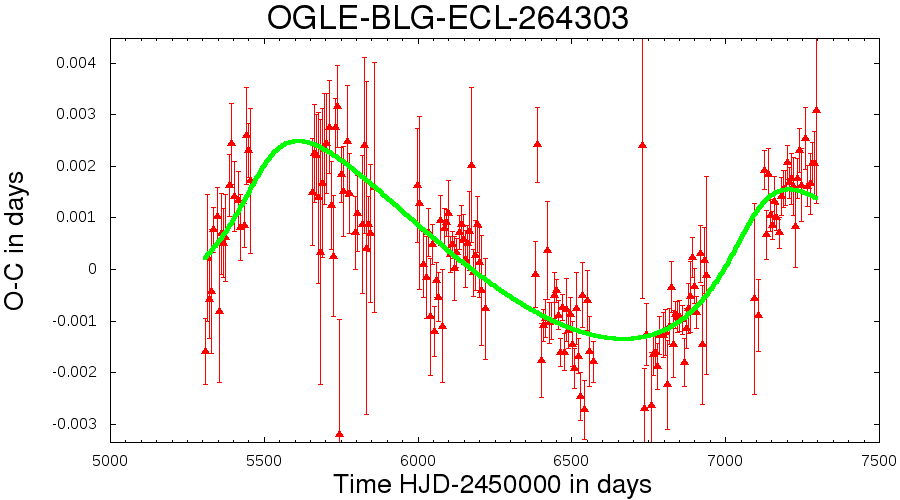}
\includegraphics[width=0.64\columnwidth]{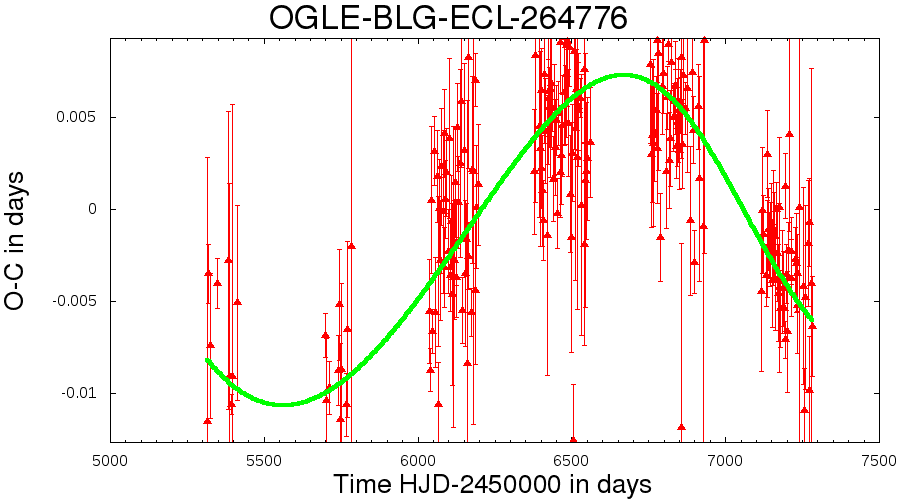}

\includegraphics[width=0.64\columnwidth]{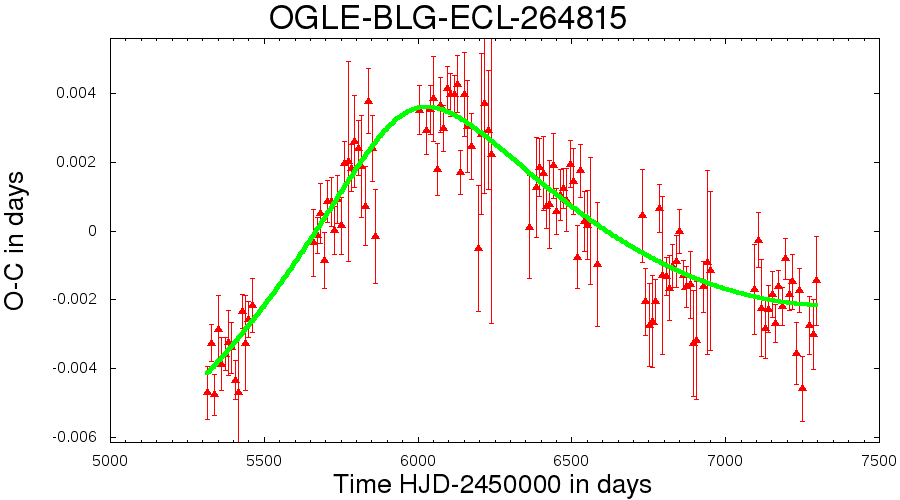}
\includegraphics[width=0.64\columnwidth]{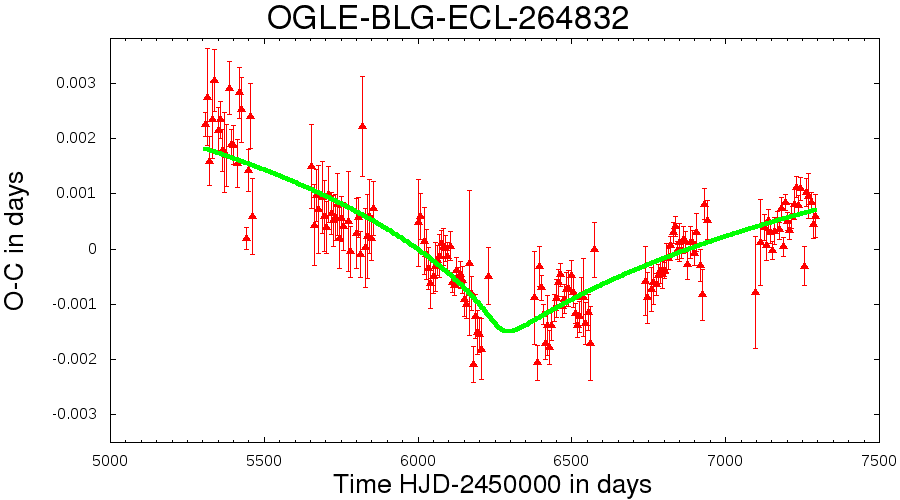}
\includegraphics[width=0.64\columnwidth]{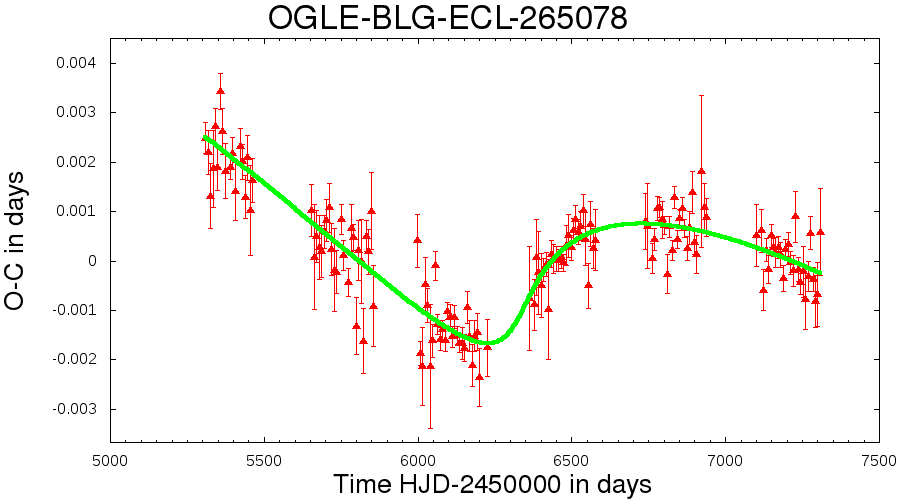}

\includegraphics[width=0.64\columnwidth]{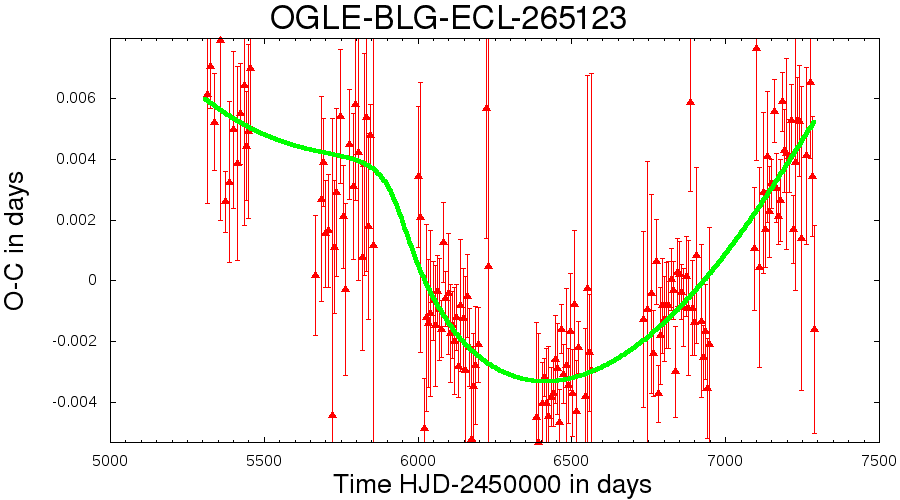}
\includegraphics[width=0.64\columnwidth]{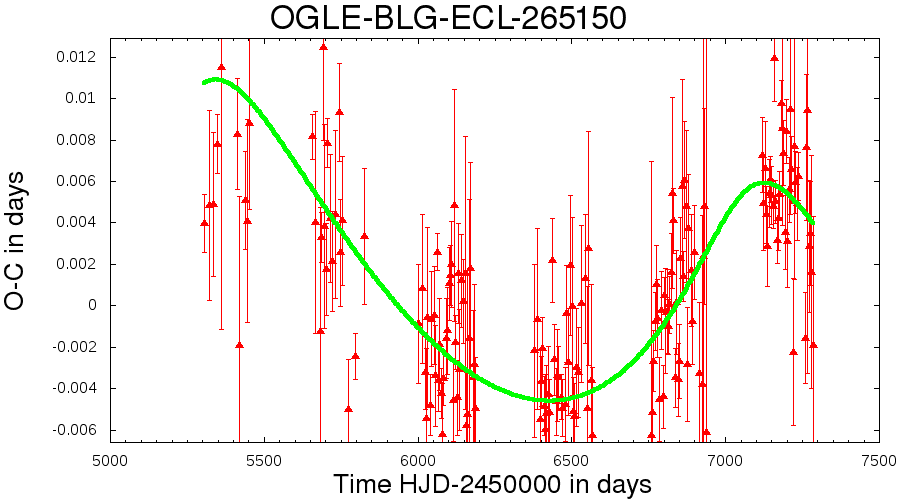}
\includegraphics[width=0.64\columnwidth]{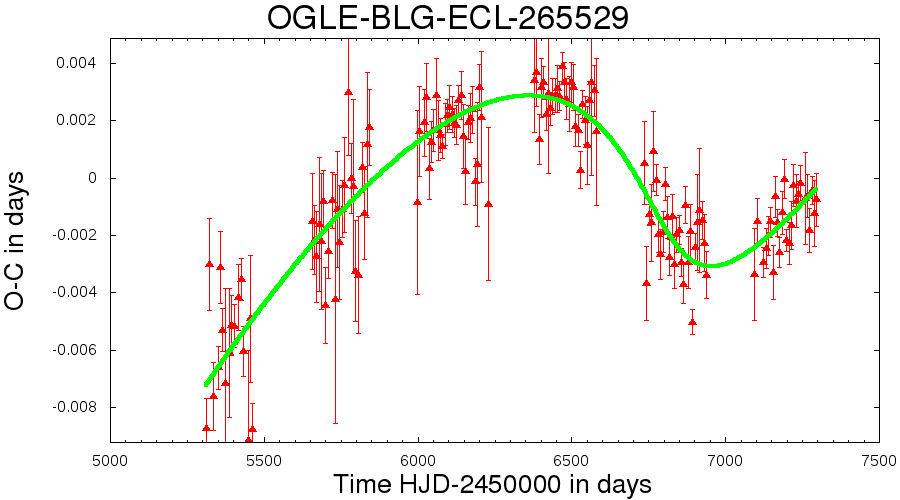}

\includegraphics[width=0.64\columnwidth]{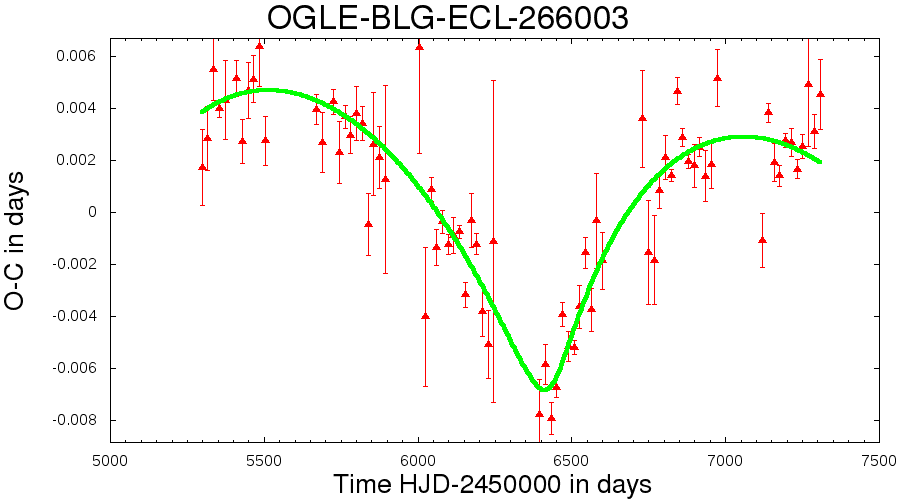}
\includegraphics[width=0.64\columnwidth]{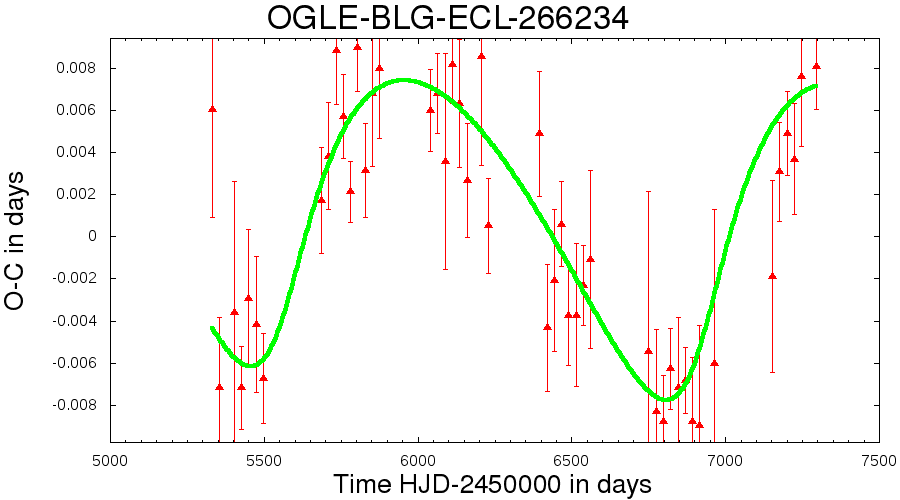}
\includegraphics[width=0.64\columnwidth]{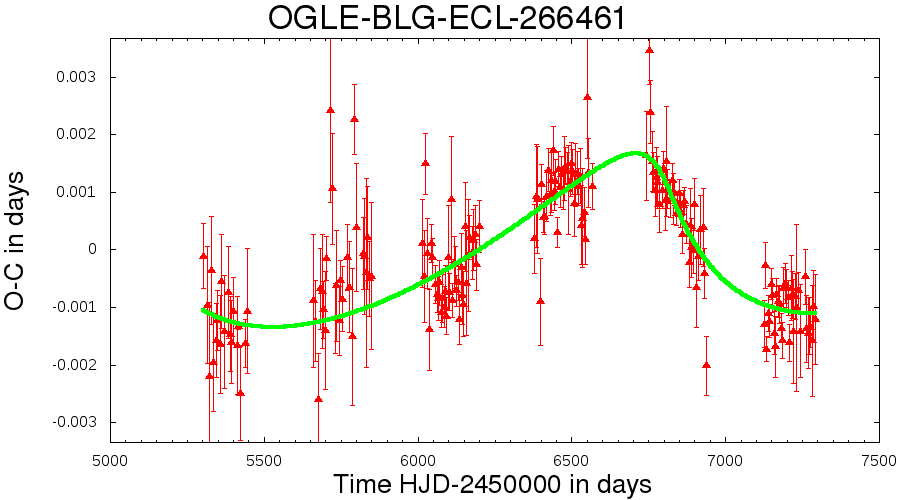}

\includegraphics[width=0.64\columnwidth]{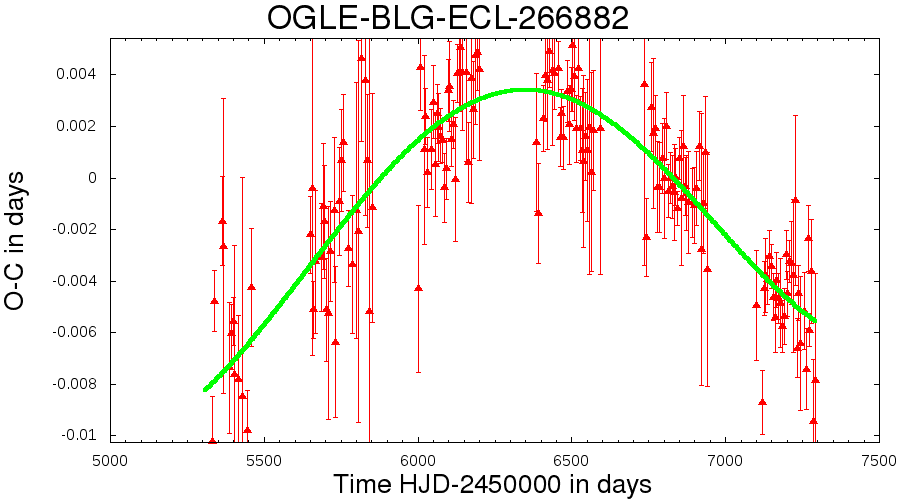}
\includegraphics[width=0.64\columnwidth]{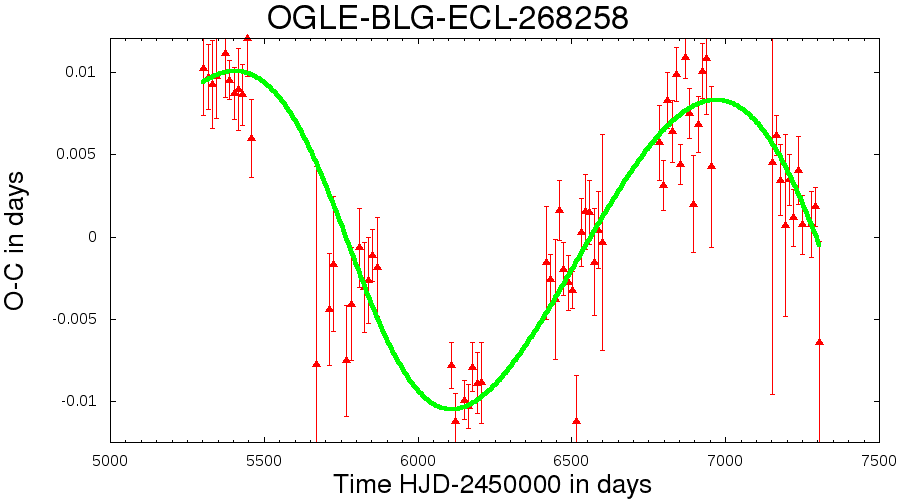}
\includegraphics[width=0.64\columnwidth]{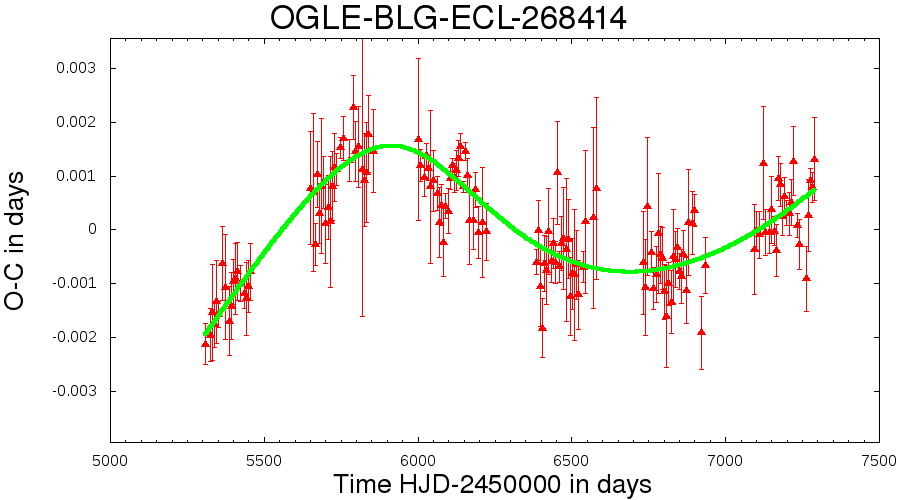}

\includegraphics[width=0.64\columnwidth]{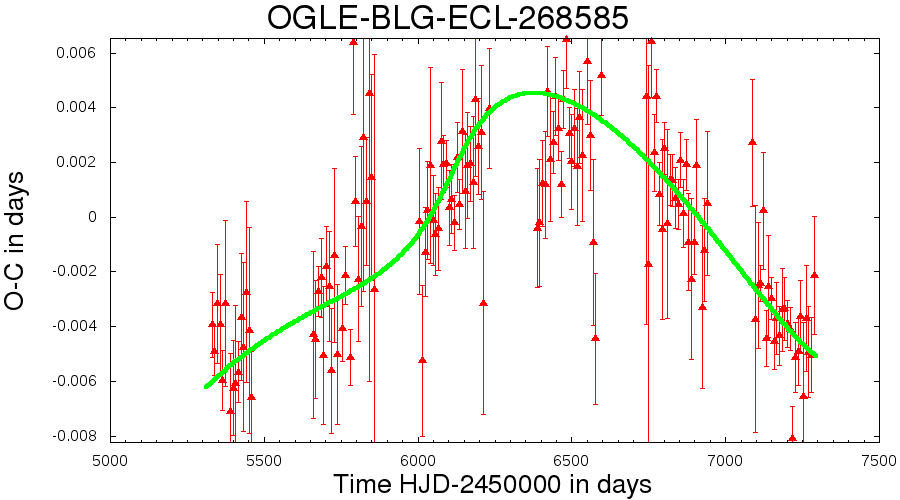}
\includegraphics[width=0.64\columnwidth]{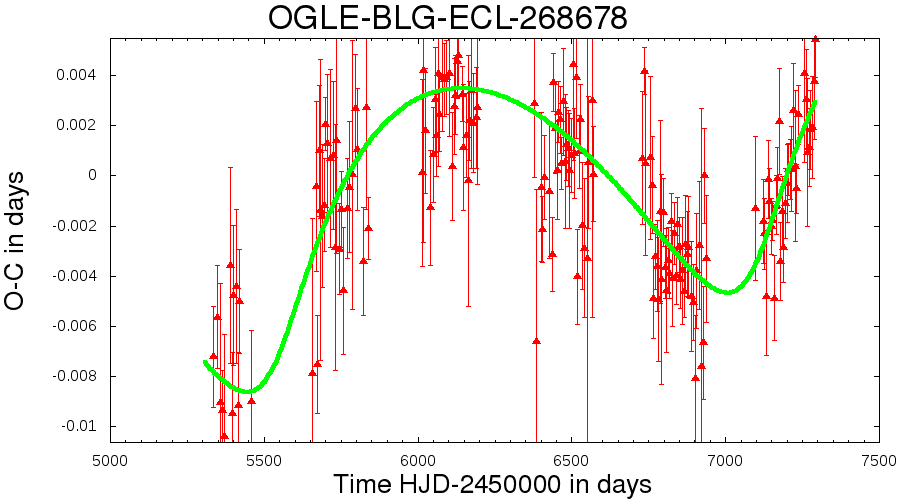}
\includegraphics[width=0.64\columnwidth]{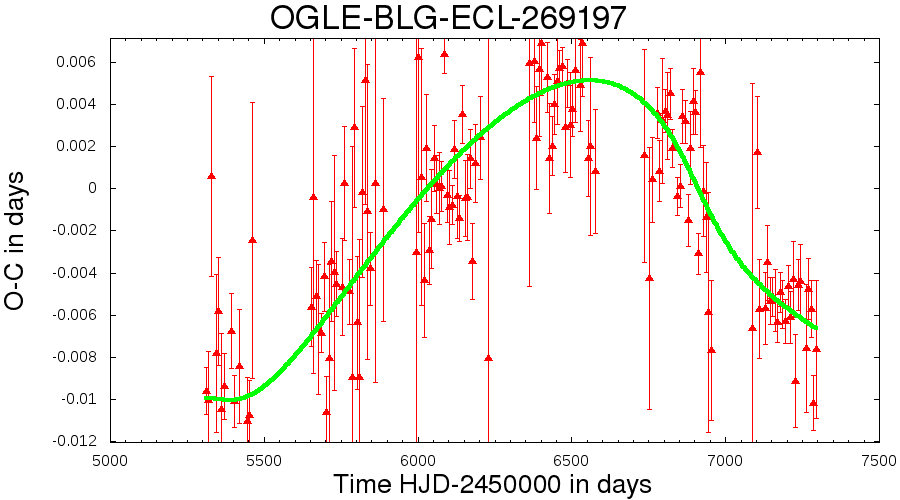}

\includegraphics[width=0.64\columnwidth]{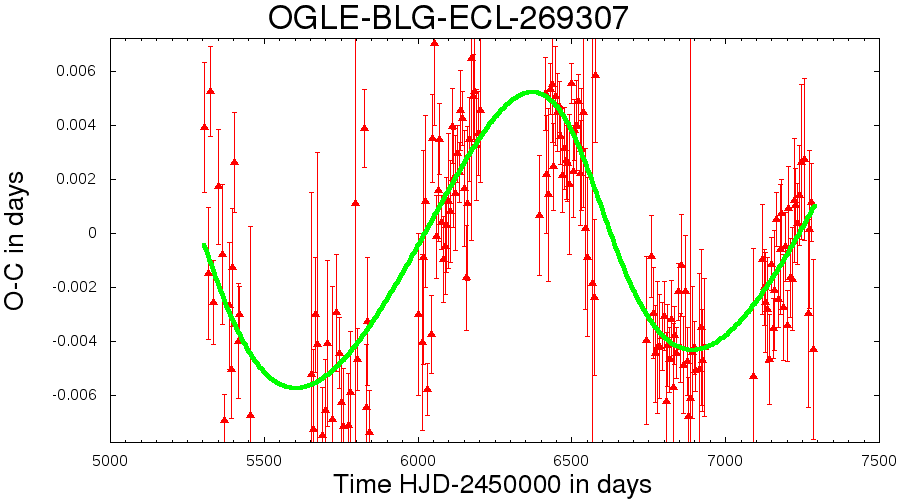}
\includegraphics[width=0.64\columnwidth]{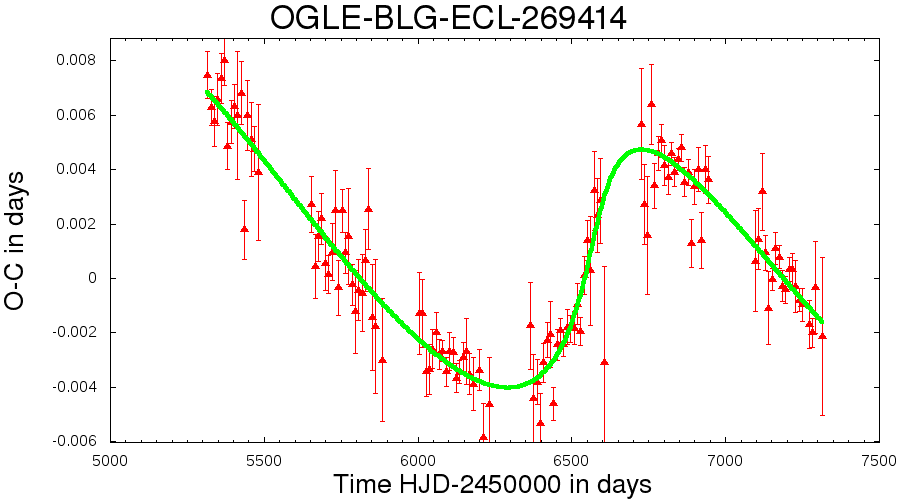}
\includegraphics[width=0.64\columnwidth]{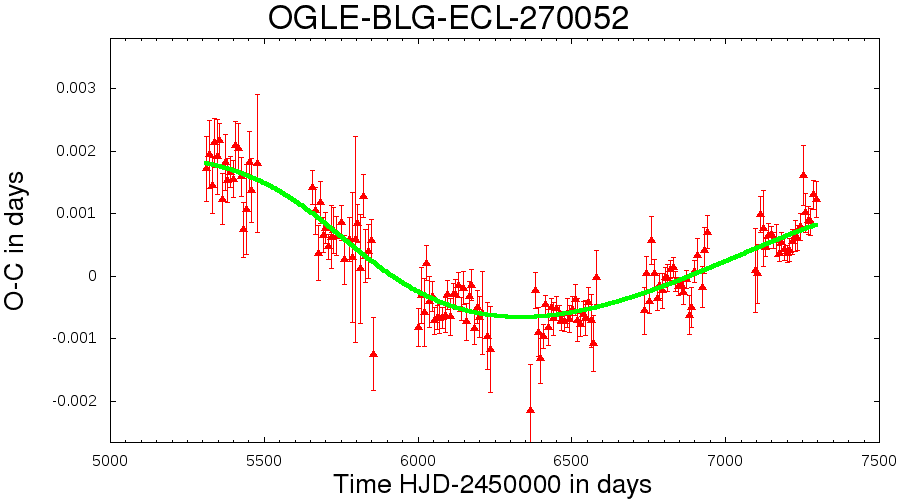}

\includegraphics[width=0.64\columnwidth]{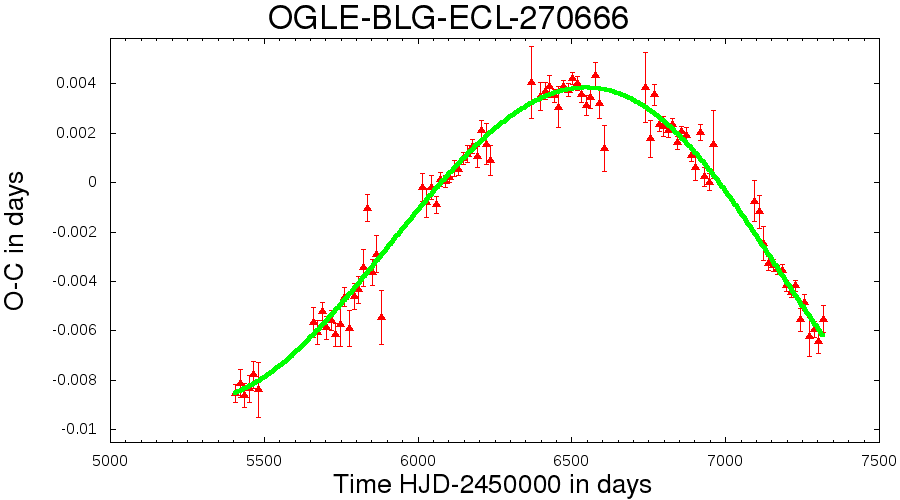}
\includegraphics[width=0.64\columnwidth]{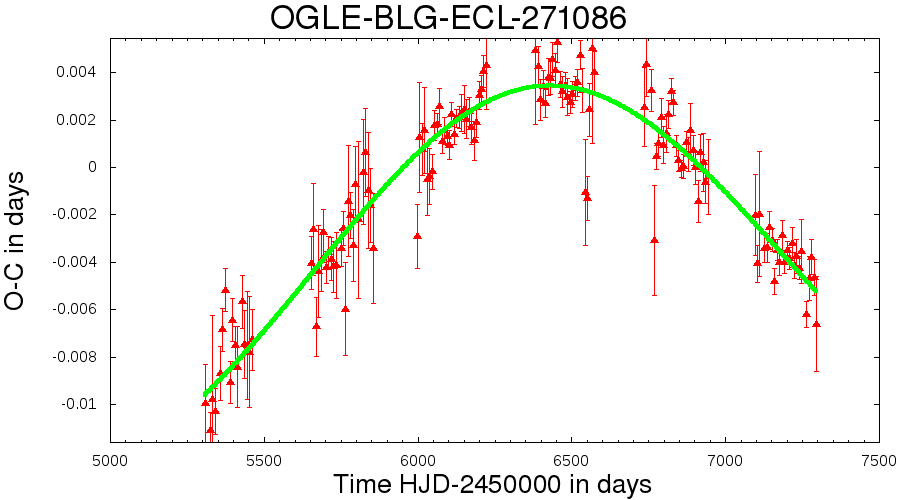}
\includegraphics[width=0.64\columnwidth]{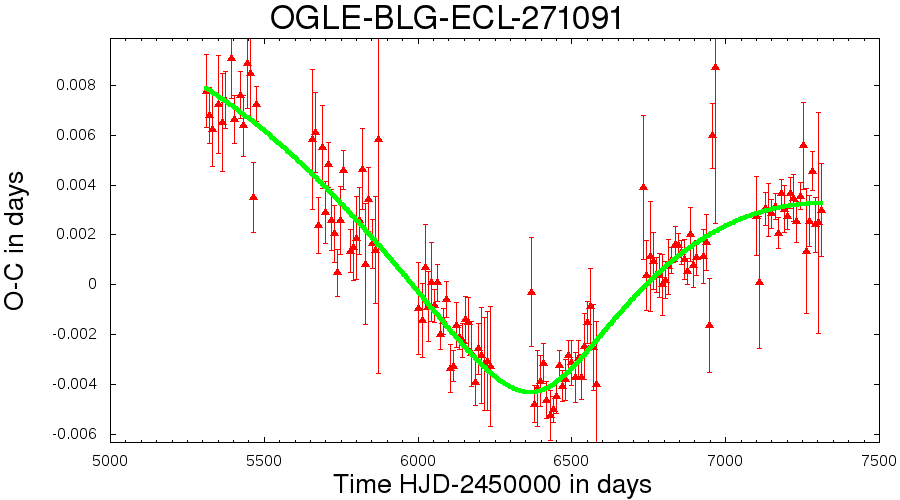}

\end{figure*}
\clearpage

\begin{figure*}
\includegraphics[width=0.64\columnwidth]{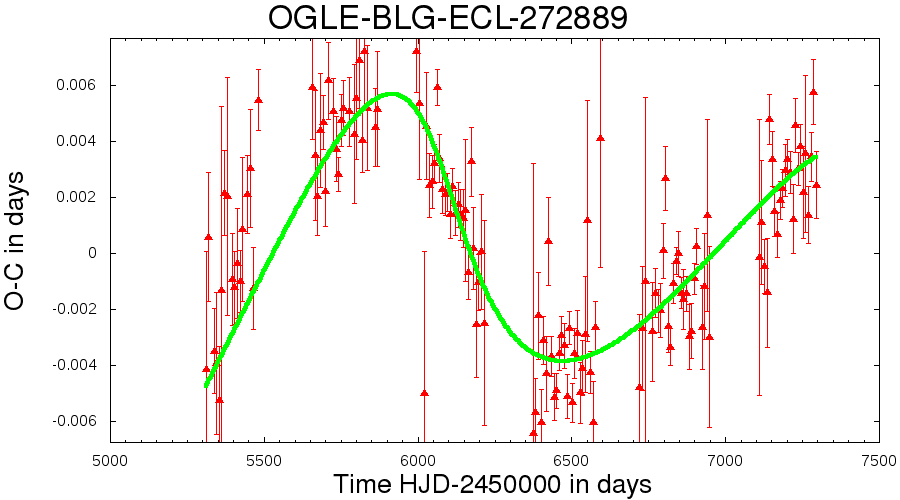}
\includegraphics[width=0.64\columnwidth]{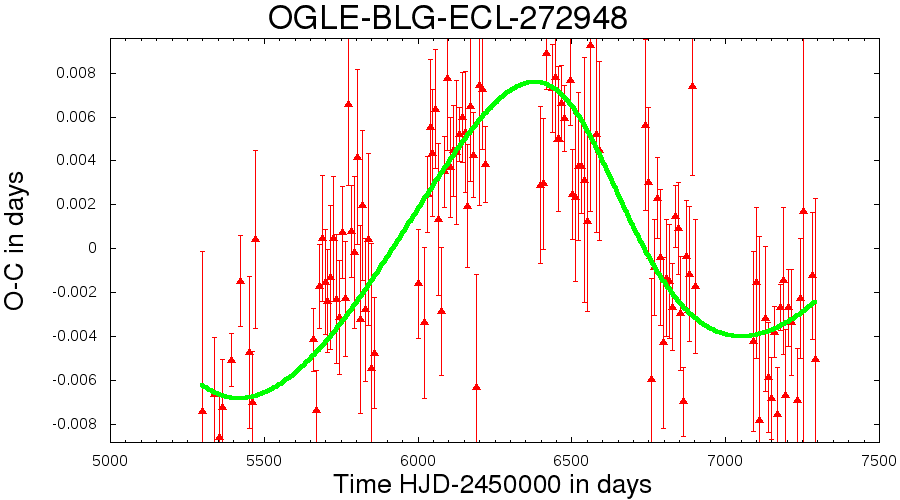}
\includegraphics[width=0.64\columnwidth]{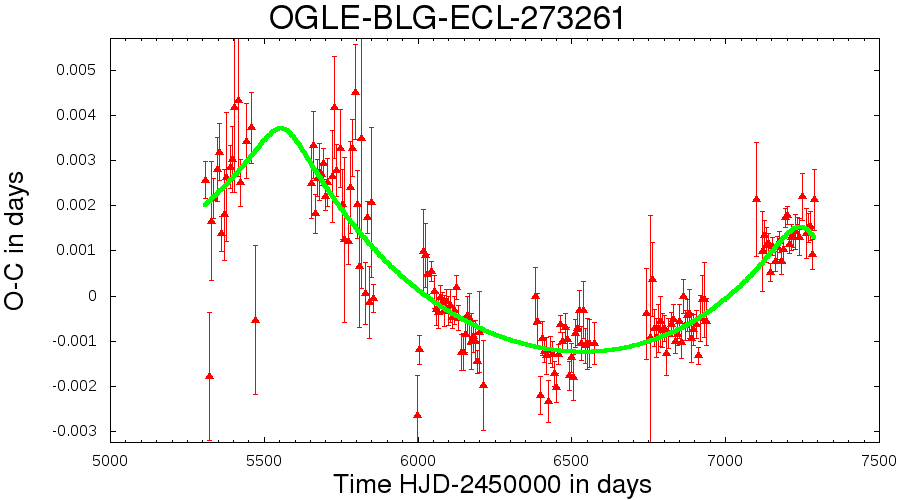}

\includegraphics[width=0.64\columnwidth]{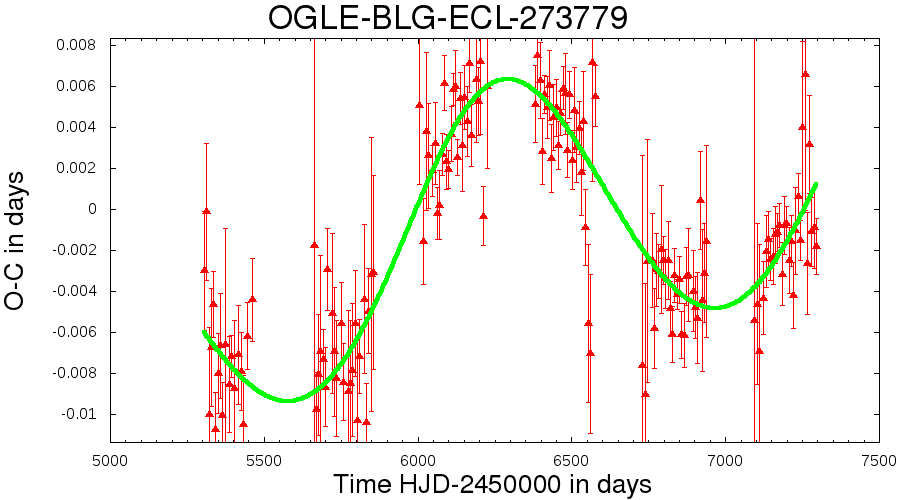}
\includegraphics[width=0.64\columnwidth]{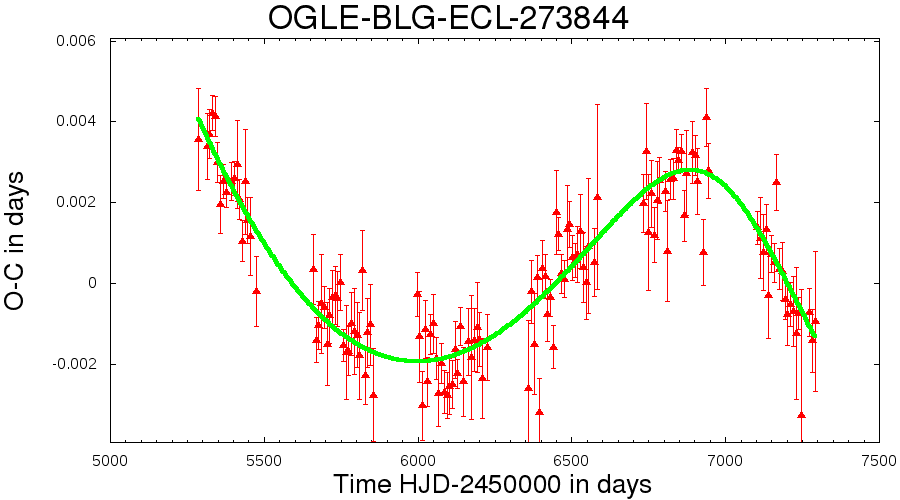}
\includegraphics[width=0.64\columnwidth]{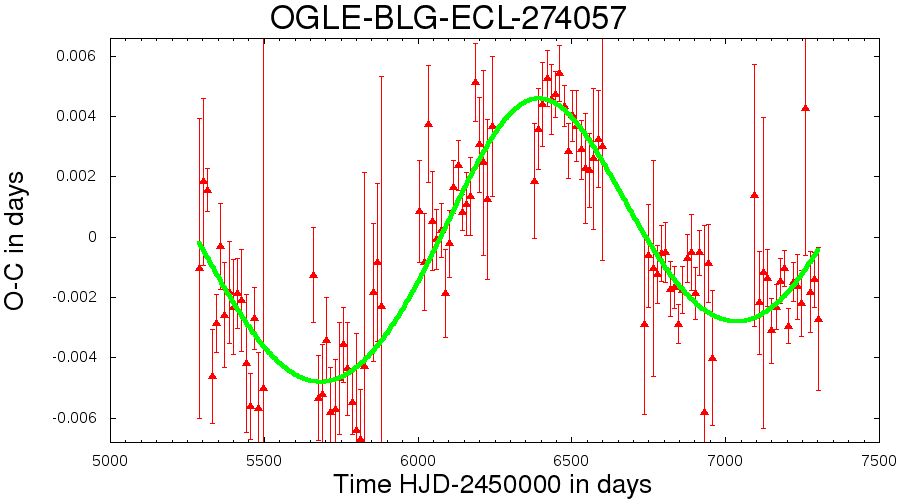}

\includegraphics[width=0.64\columnwidth]{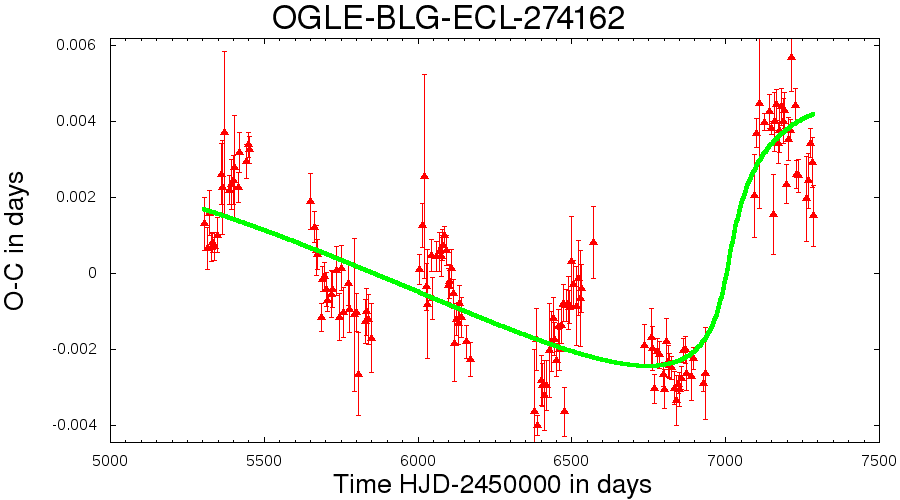}
\includegraphics[width=0.64\columnwidth]{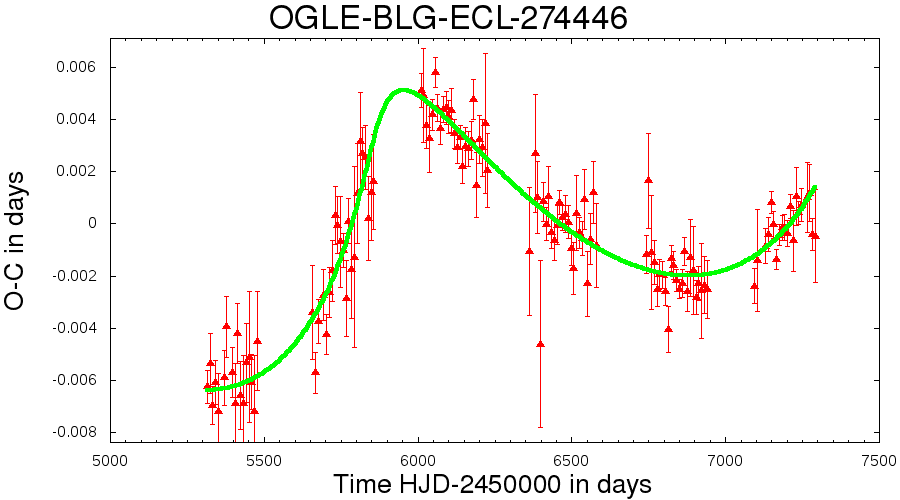}
\includegraphics[width=0.64\columnwidth]{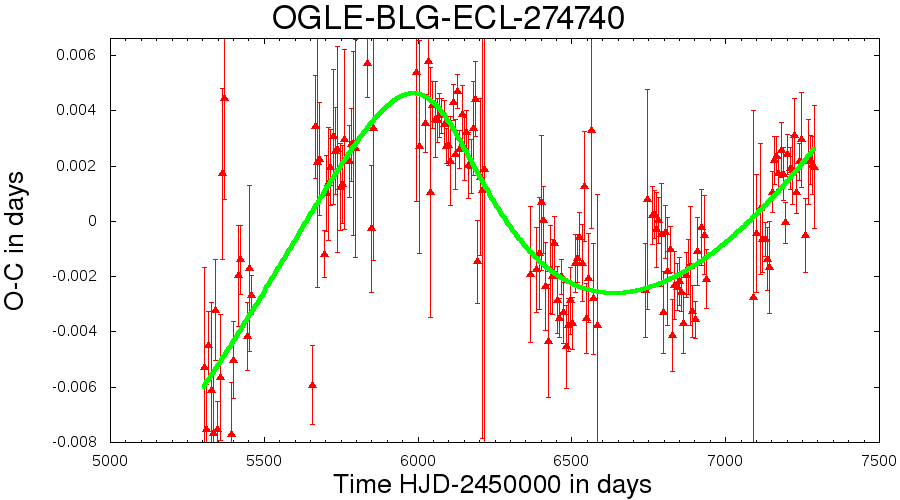}

\includegraphics[width=0.64\columnwidth]{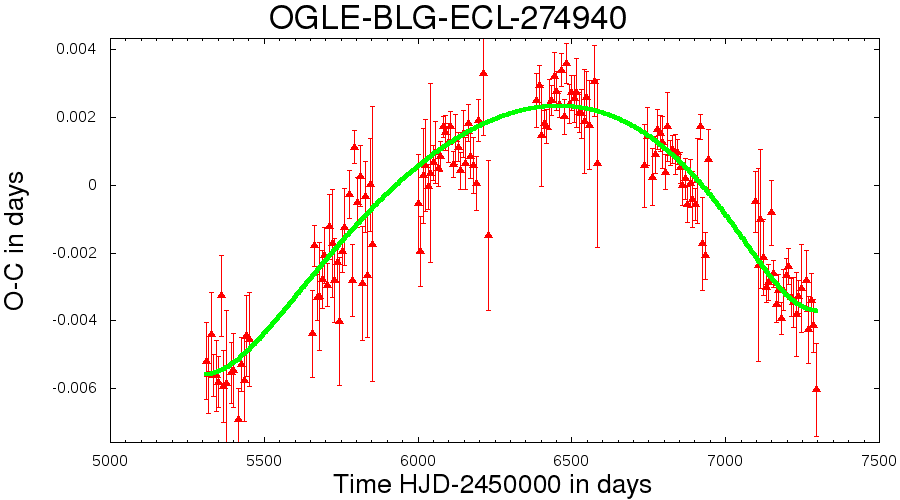}
\includegraphics[width=0.64\columnwidth]{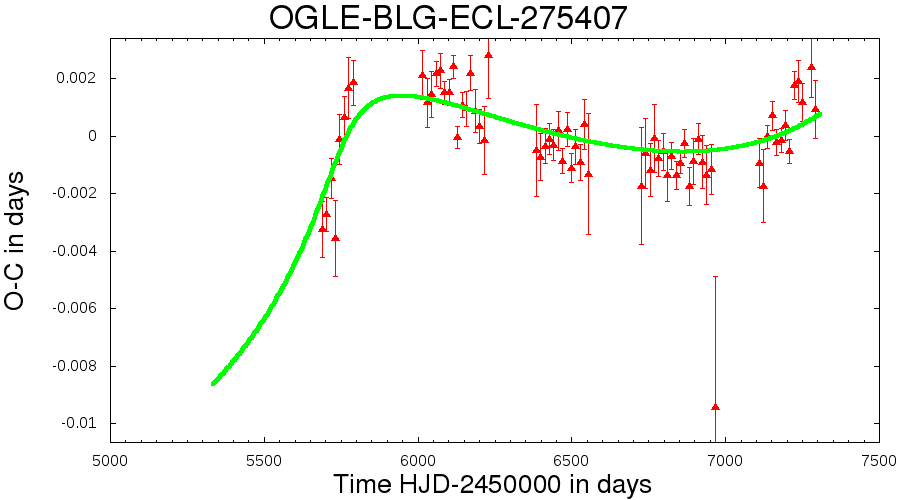}
\includegraphics[width=0.64\columnwidth]{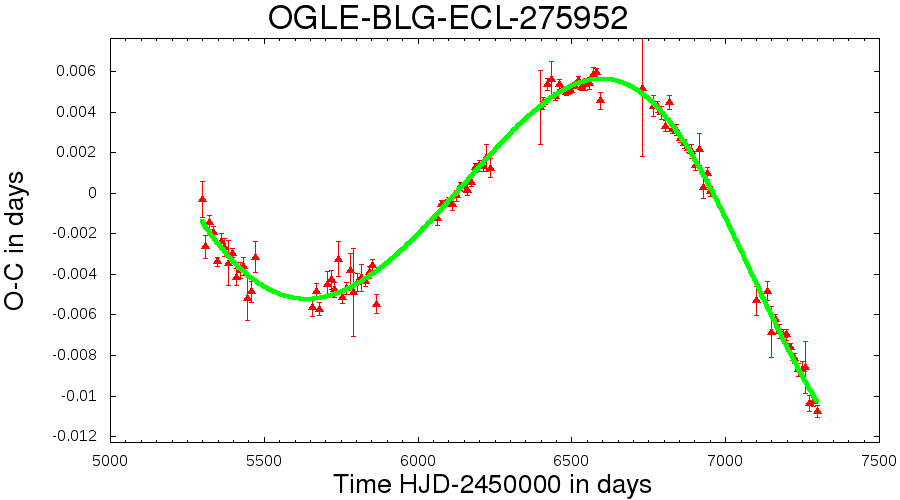}

\includegraphics[width=0.64\columnwidth]{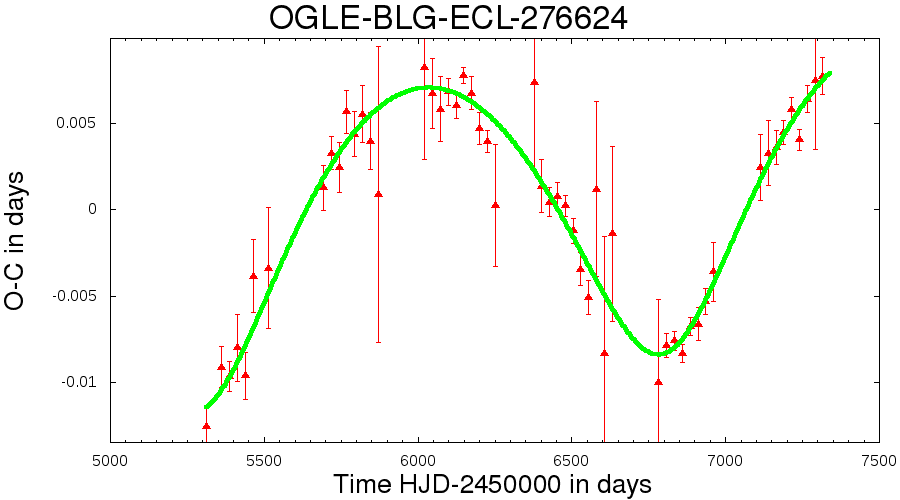}
\includegraphics[width=0.64\columnwidth]{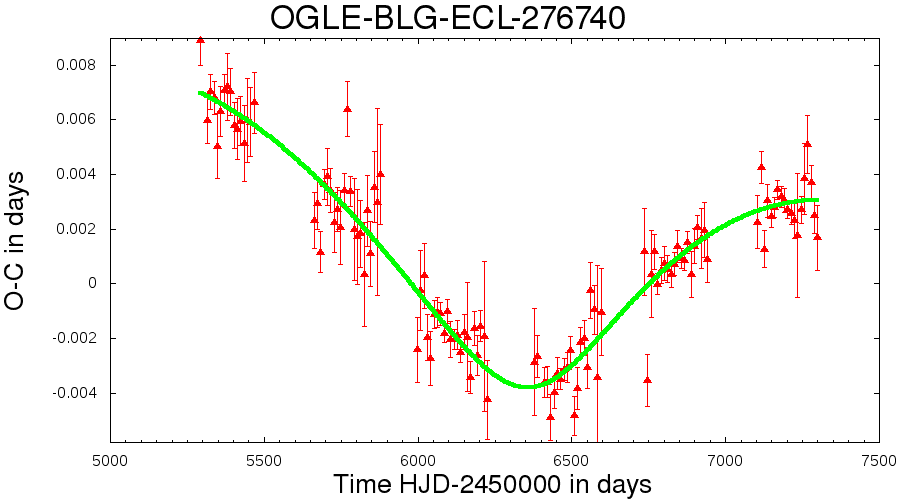}
\includegraphics[width=0.64\columnwidth]{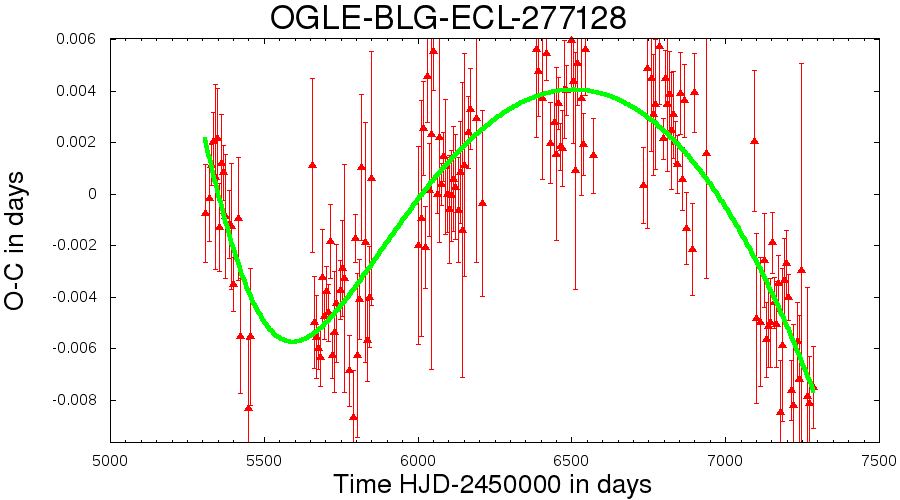}

\includegraphics[width=0.64\columnwidth]{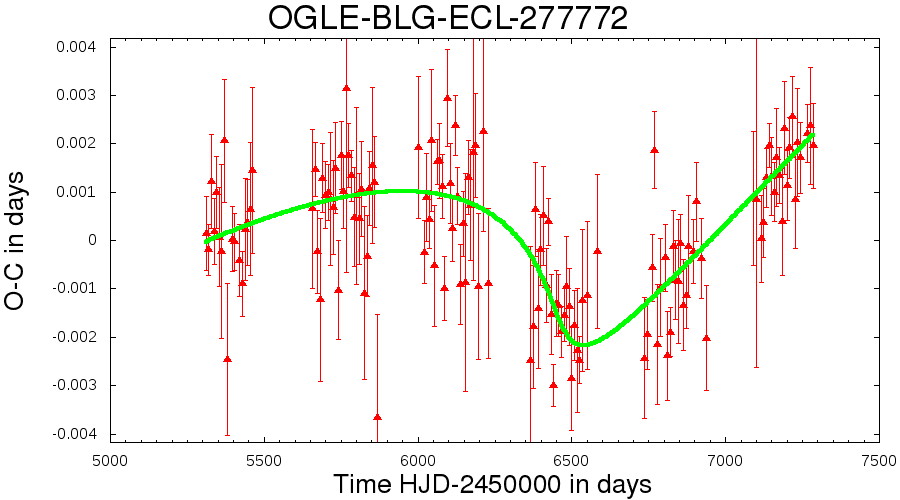}
\includegraphics[width=0.64\columnwidth]{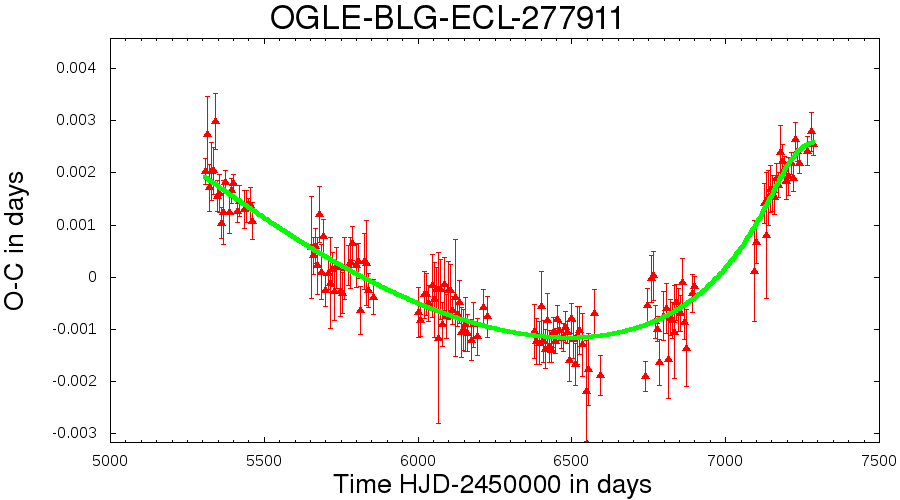}
\includegraphics[width=0.64\columnwidth]{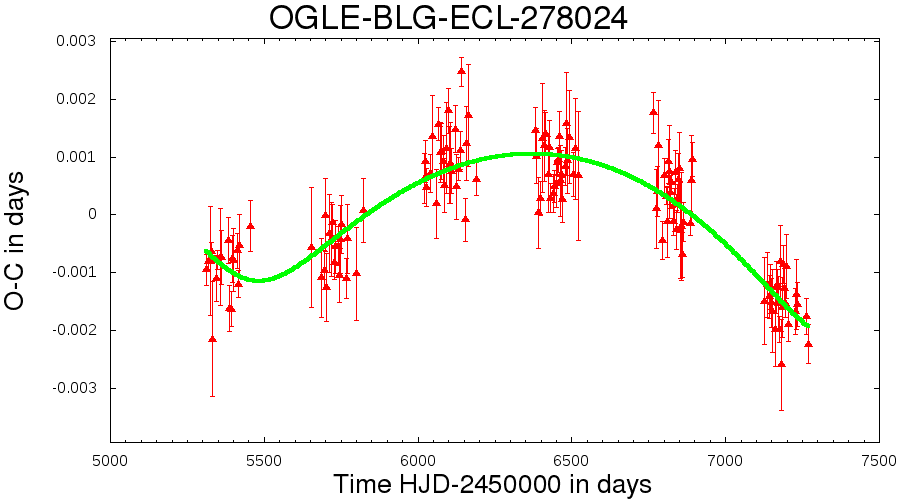}

\includegraphics[width=0.64\columnwidth]{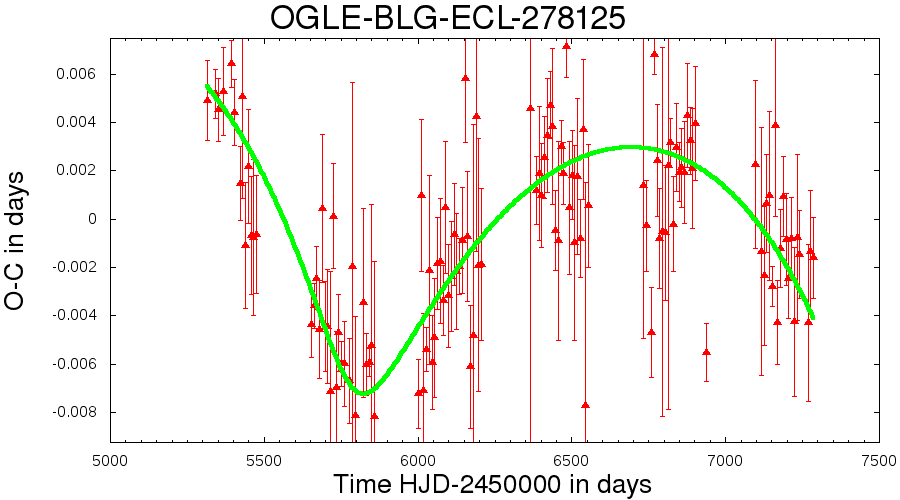}
\includegraphics[width=0.64\columnwidth]{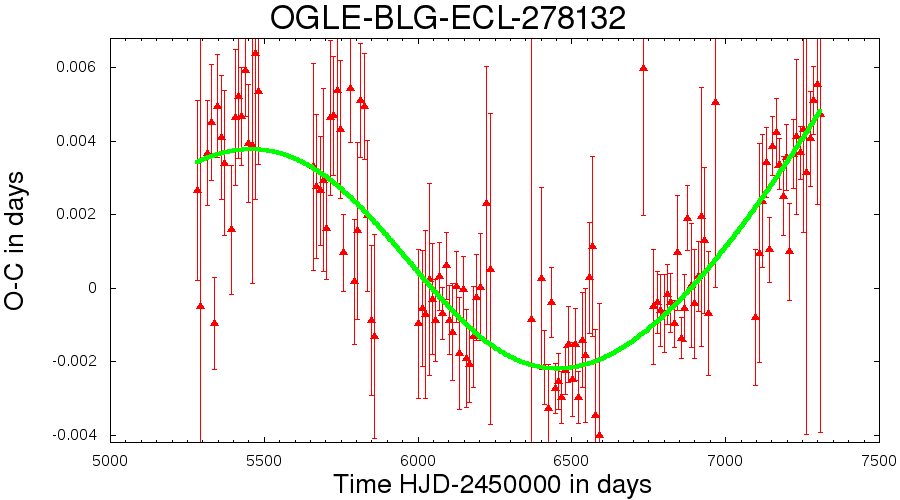}
\includegraphics[width=0.64\columnwidth]{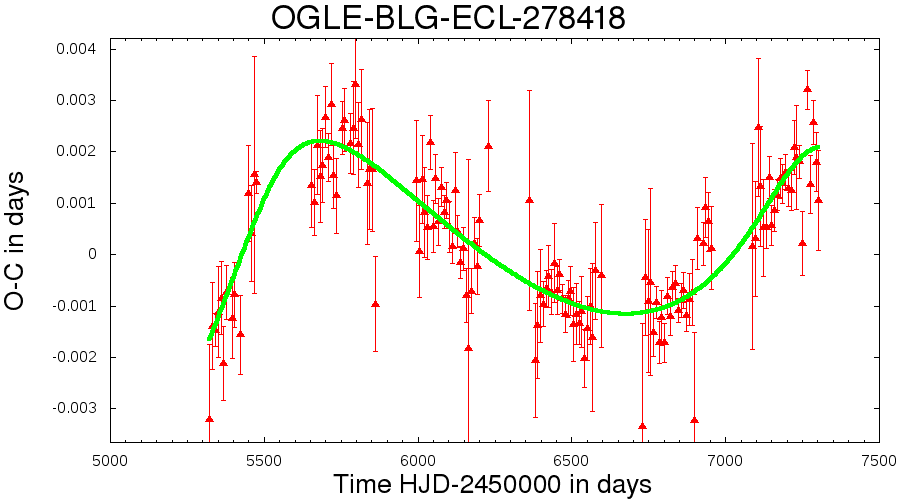}

\includegraphics[width=0.64\columnwidth]{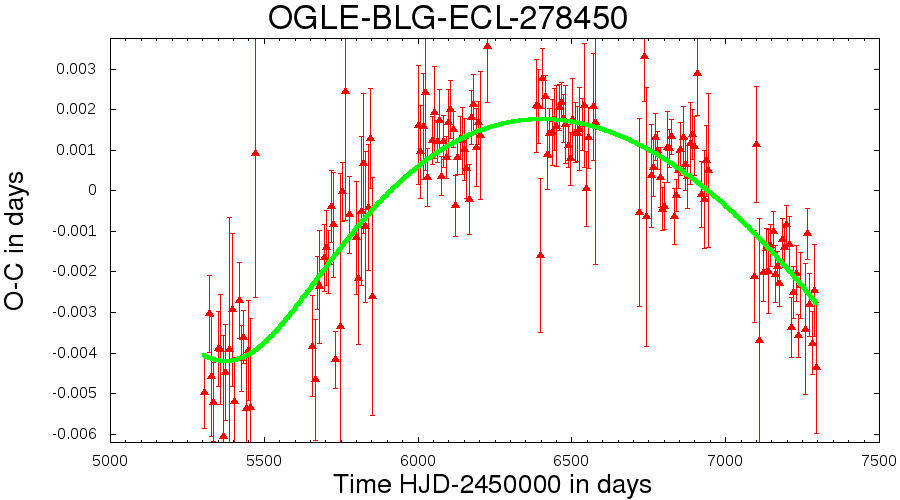}
\includegraphics[width=0.64\columnwidth]{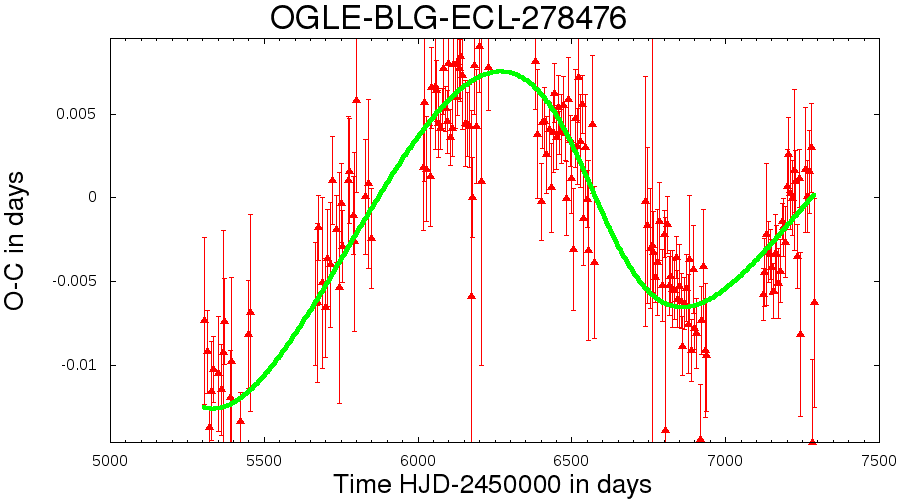}
\includegraphics[width=0.64\columnwidth]{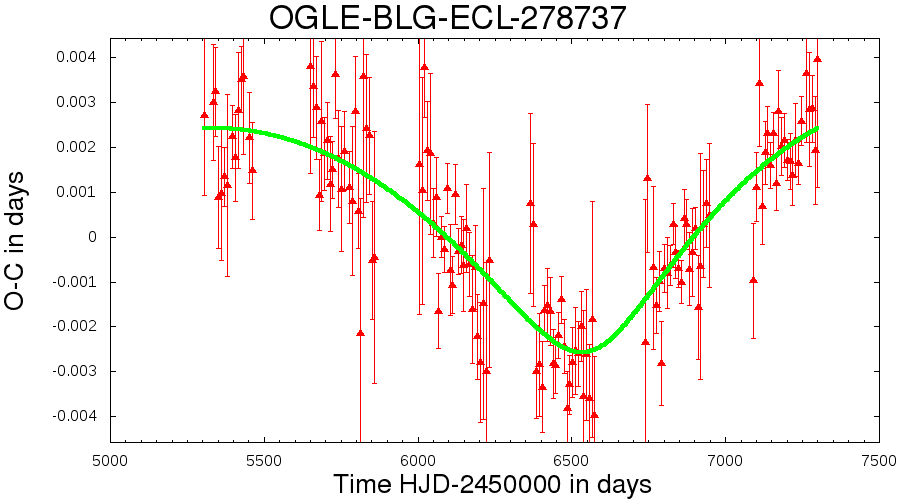}

\end{figure*}
\clearpage

\begin{figure*}
\includegraphics[width=0.64\columnwidth]{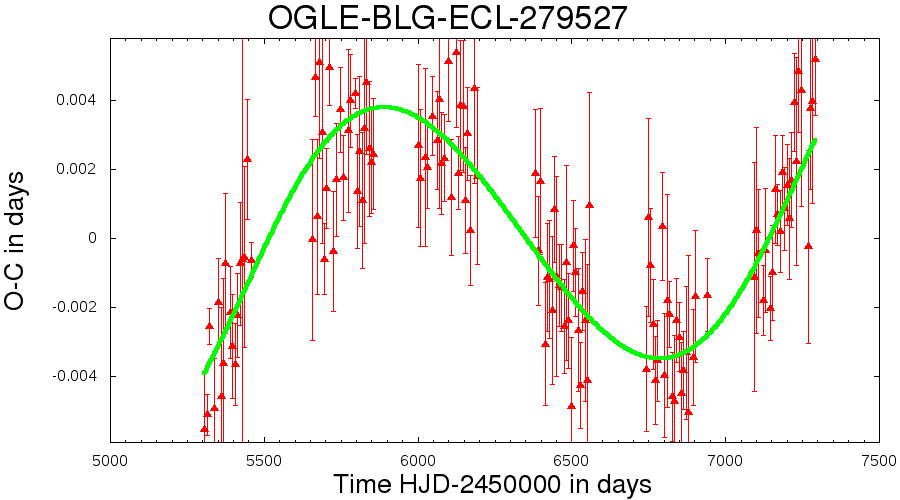}
\includegraphics[width=0.64\columnwidth]{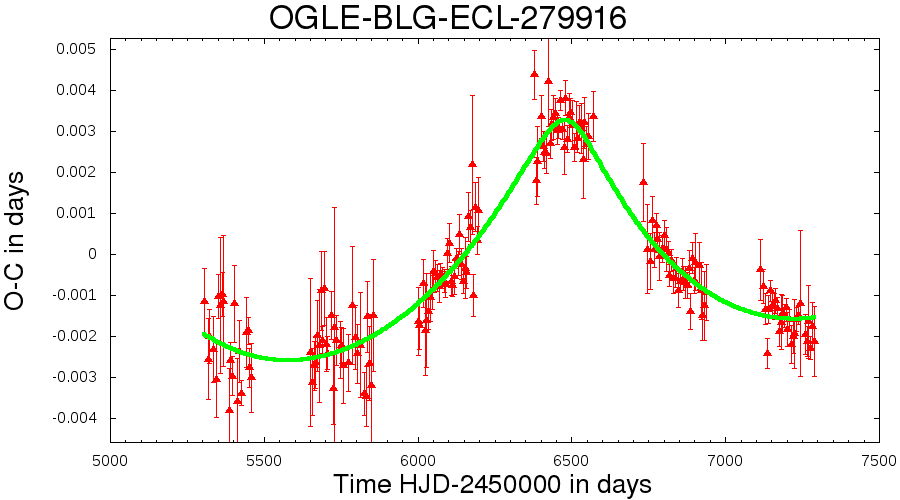}
\includegraphics[width=0.64\columnwidth]{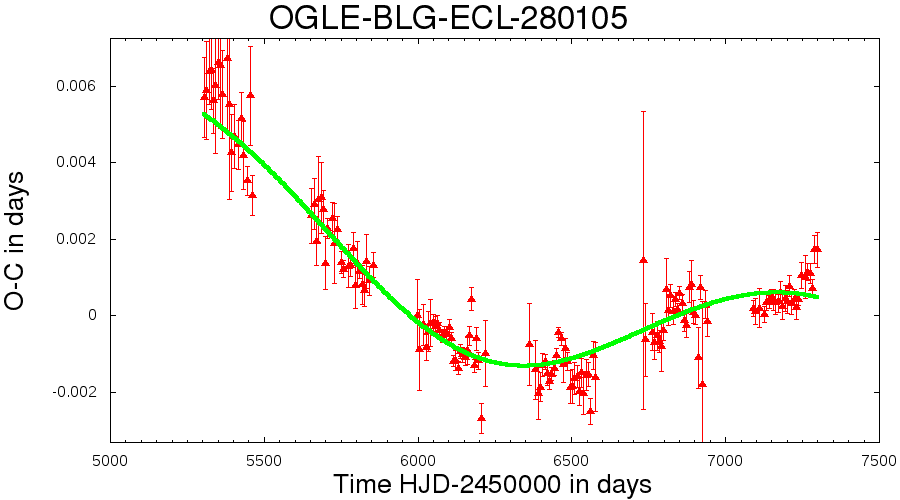}

\includegraphics[width=0.64\columnwidth]{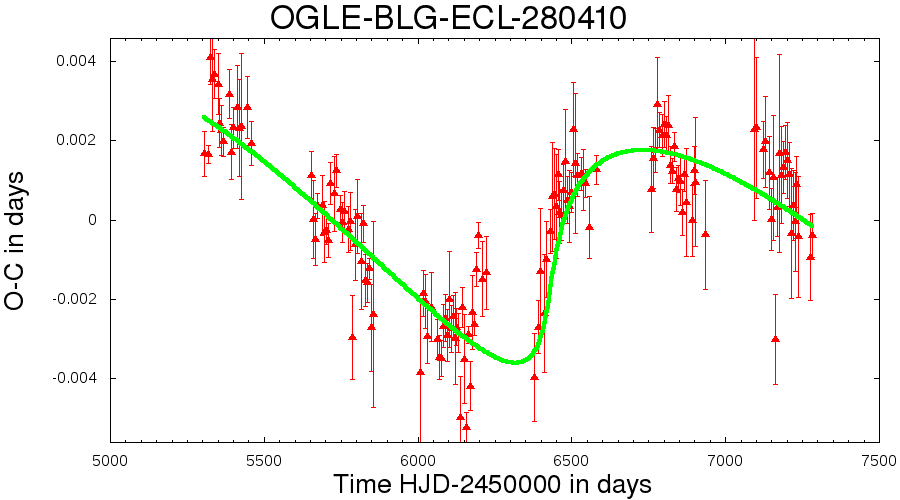}
\includegraphics[width=0.64\columnwidth]{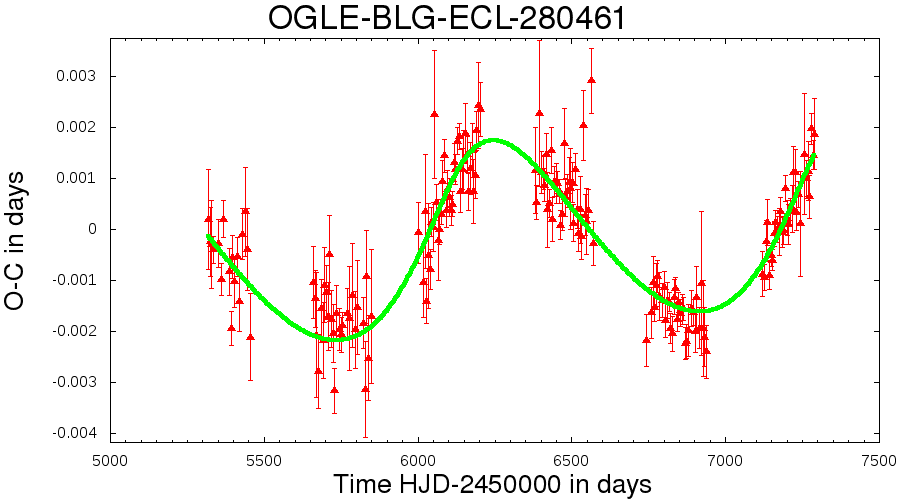}
\includegraphics[width=0.64\columnwidth]{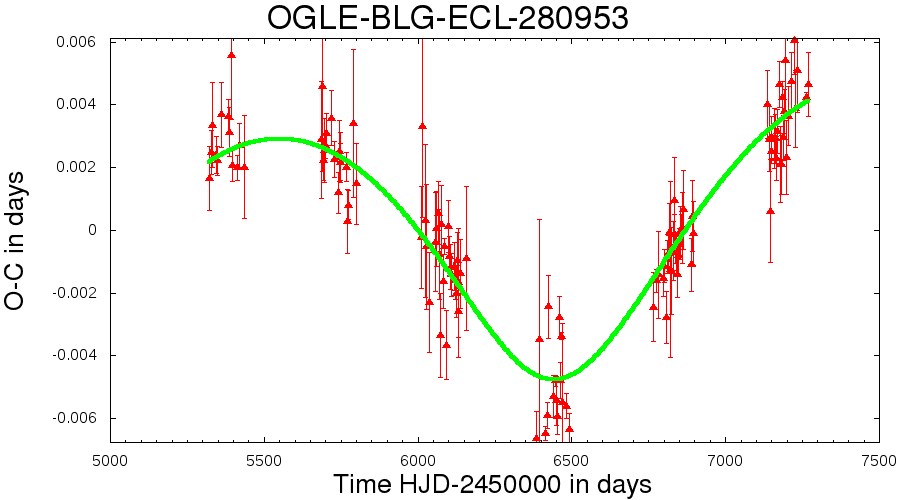}

\includegraphics[width=0.64\columnwidth]{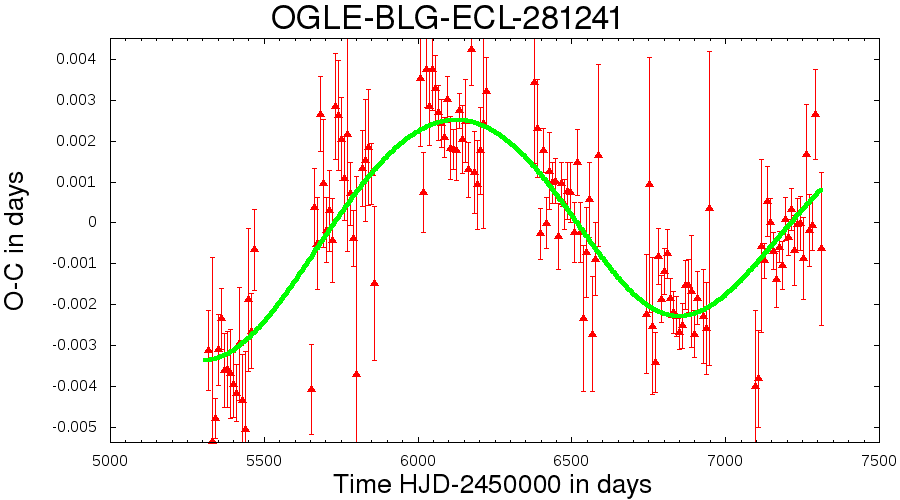}
\includegraphics[width=0.64\columnwidth]{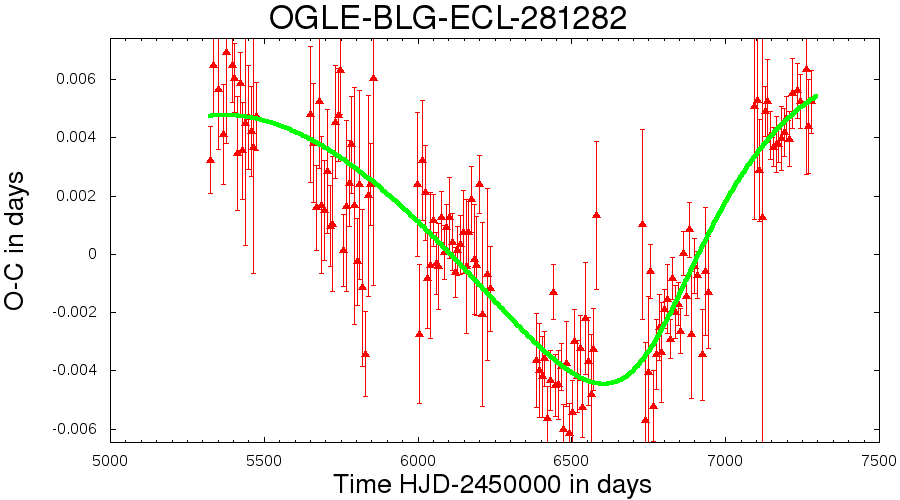}
\includegraphics[width=0.64\columnwidth]{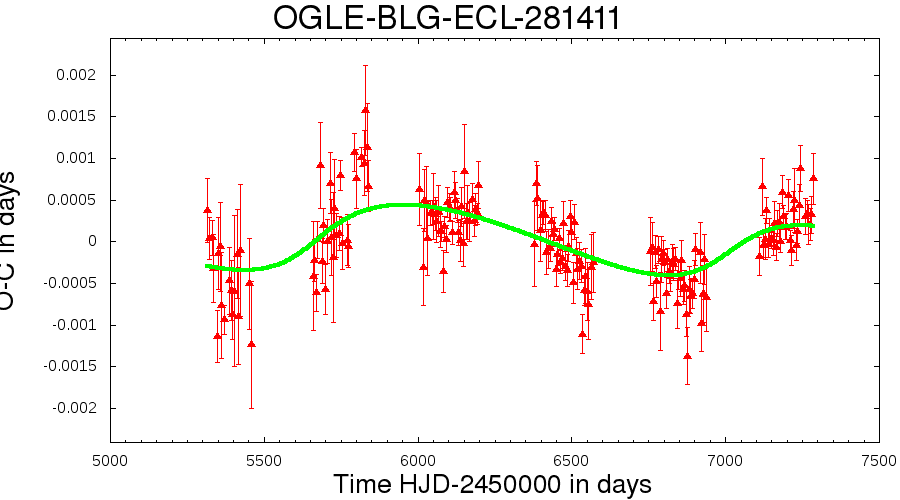}

\includegraphics[width=0.64\columnwidth]{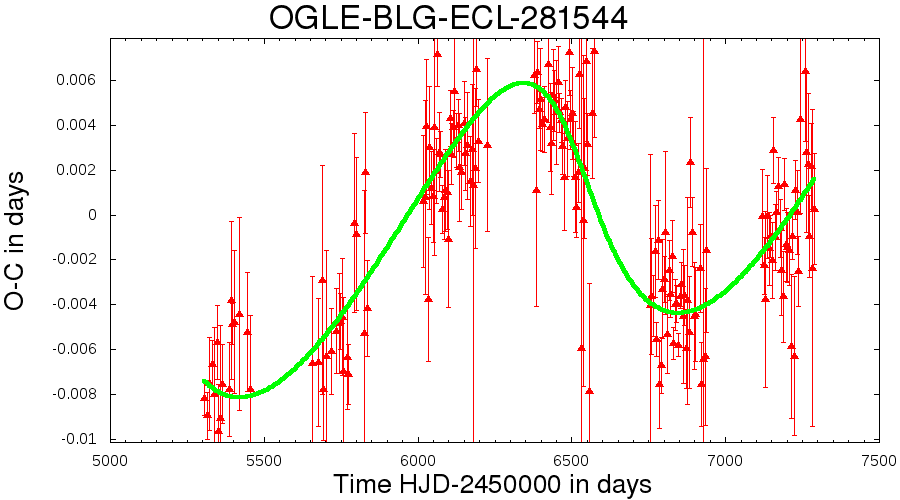}
\includegraphics[width=0.64\columnwidth]{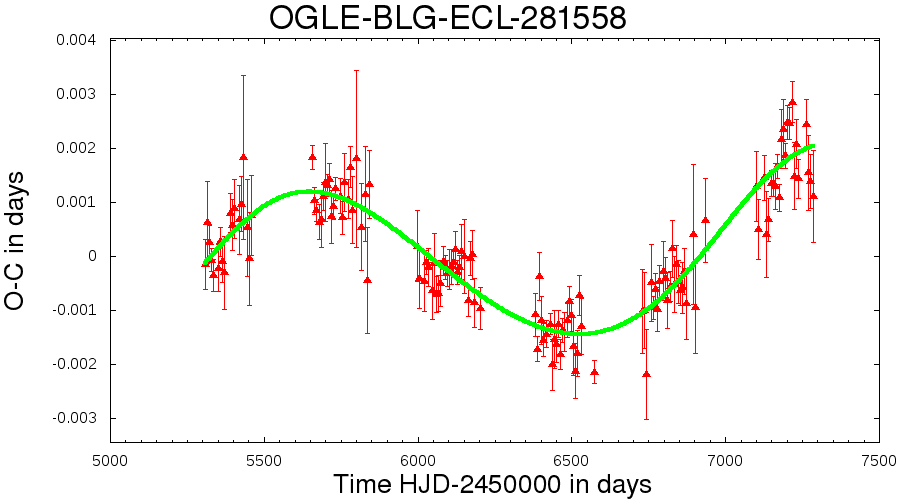}
\includegraphics[width=0.64\columnwidth]{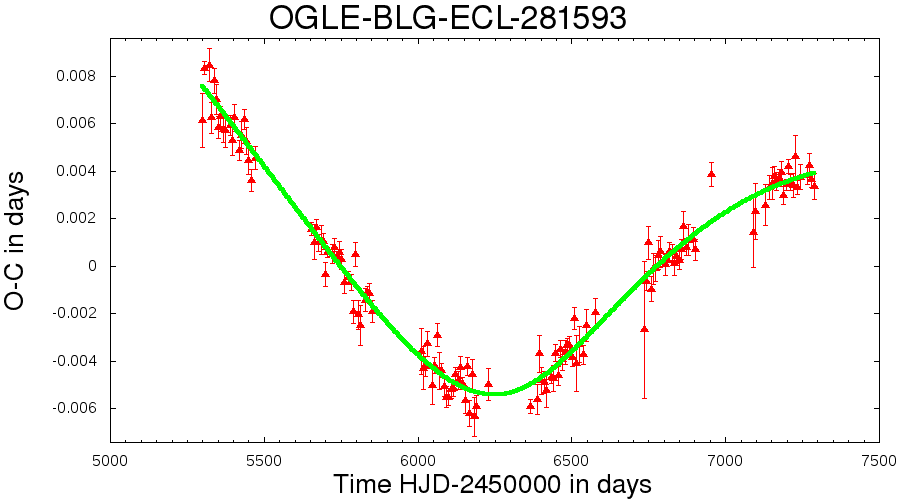}

\includegraphics[width=0.64\columnwidth]{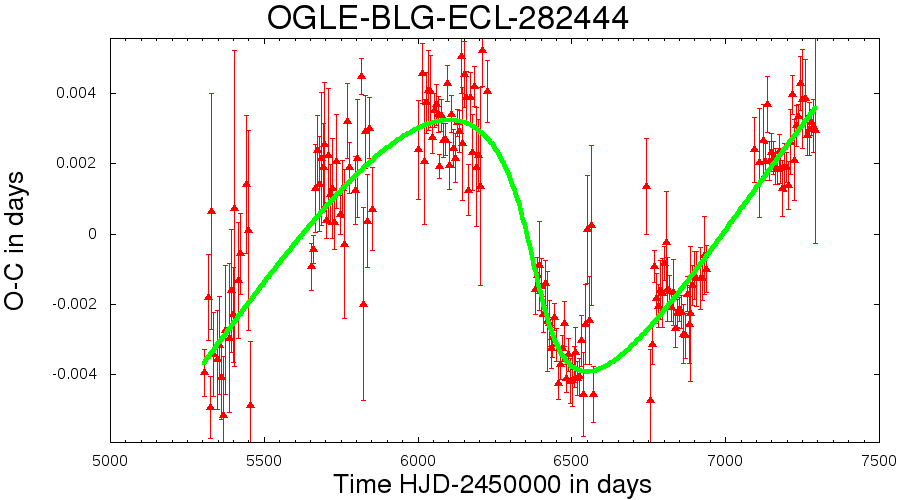}
\includegraphics[width=0.64\columnwidth]{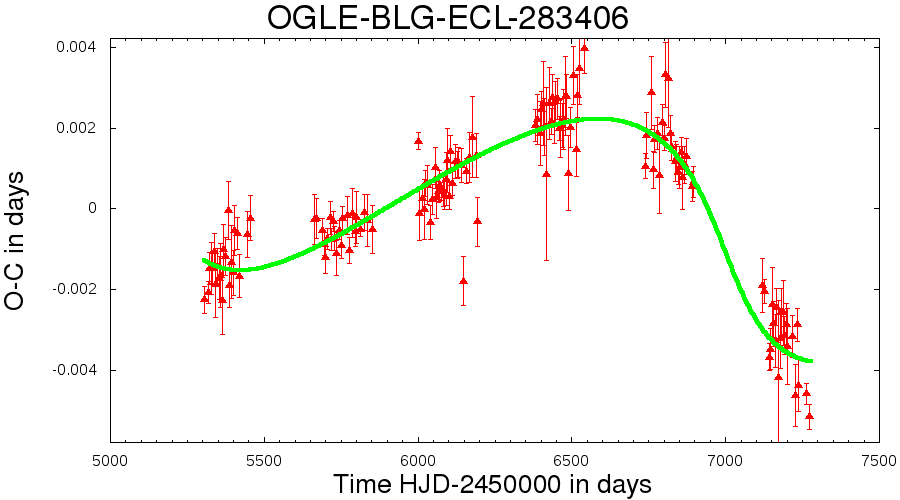}
\includegraphics[width=0.64\columnwidth]{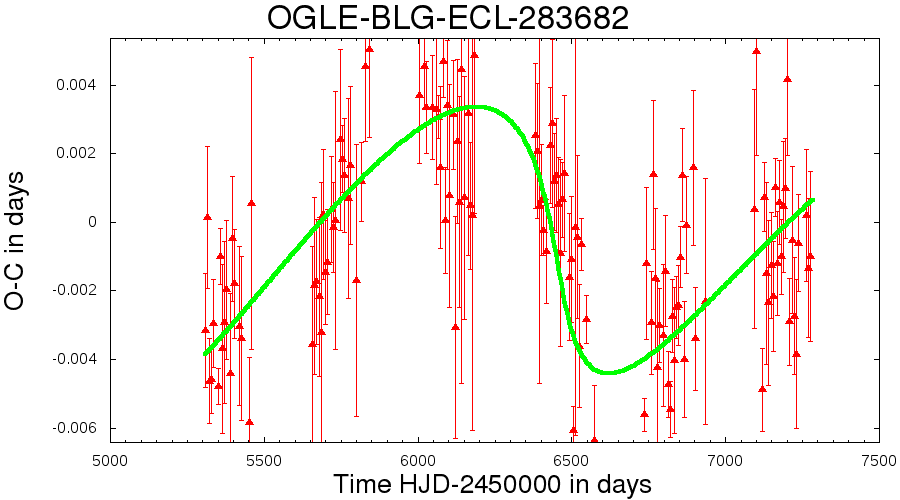}

\includegraphics[width=0.64\columnwidth]{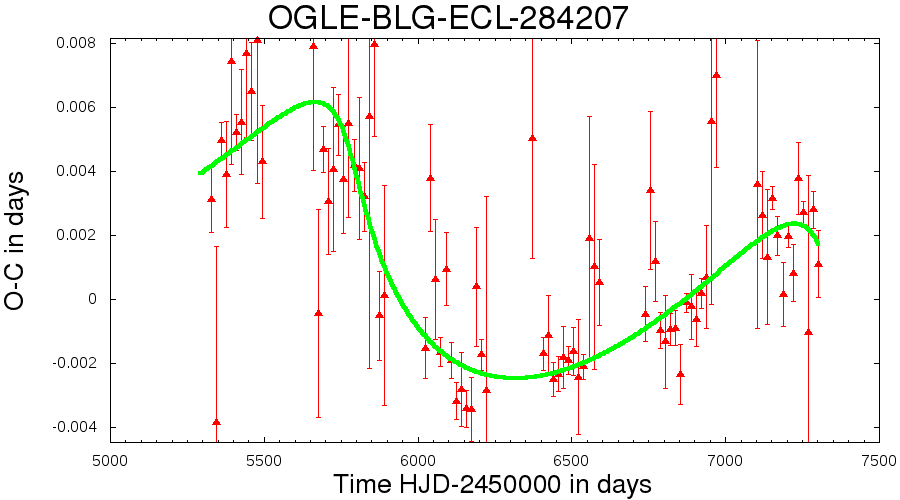}
\includegraphics[width=0.64\columnwidth]{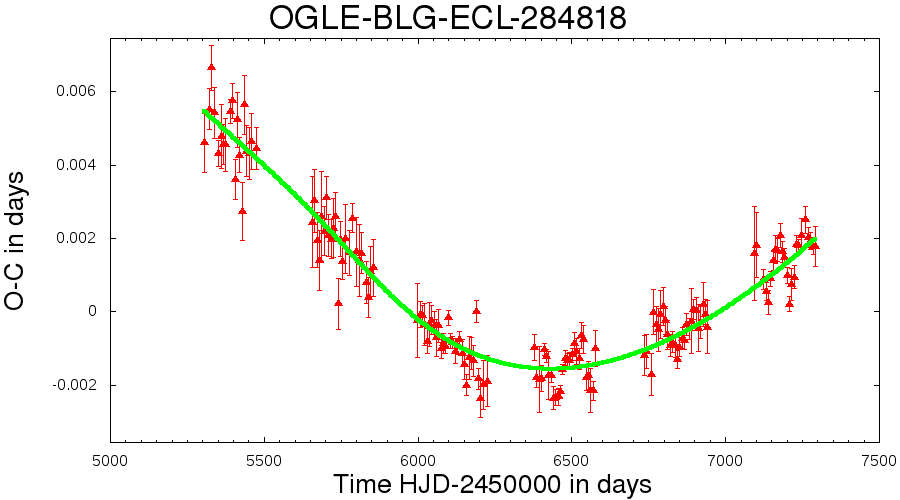}
\includegraphics[width=0.64\columnwidth]{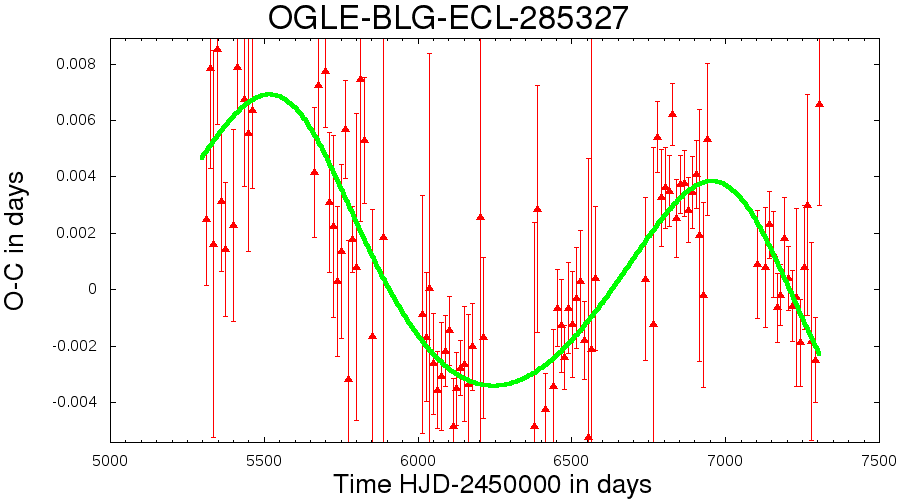}

\includegraphics[width=0.64\columnwidth]{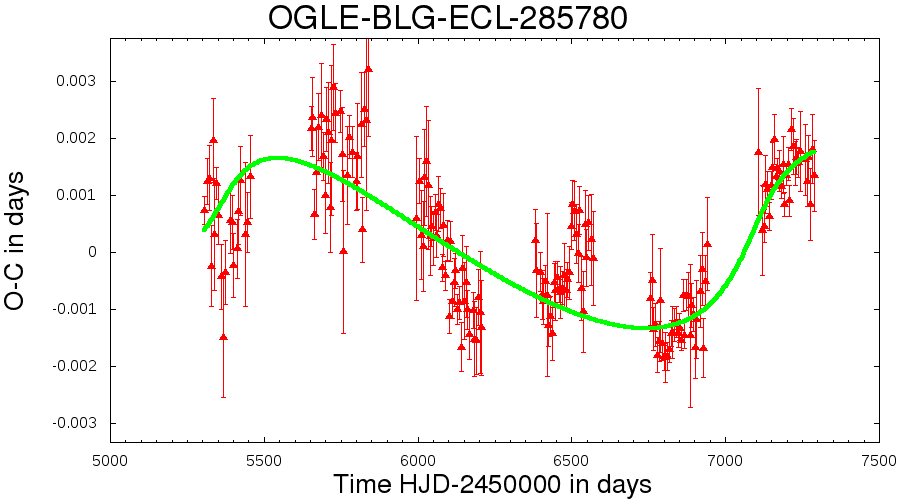}
\includegraphics[width=0.64\columnwidth]{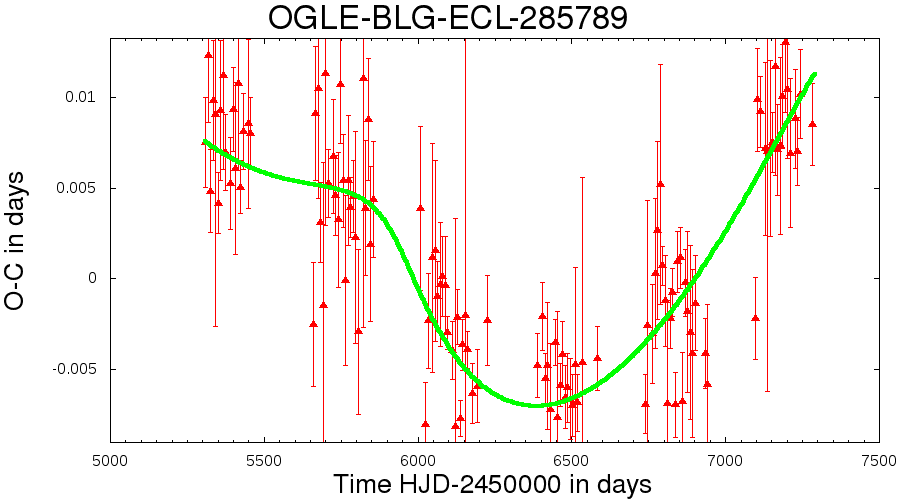}
\includegraphics[width=0.64\columnwidth]{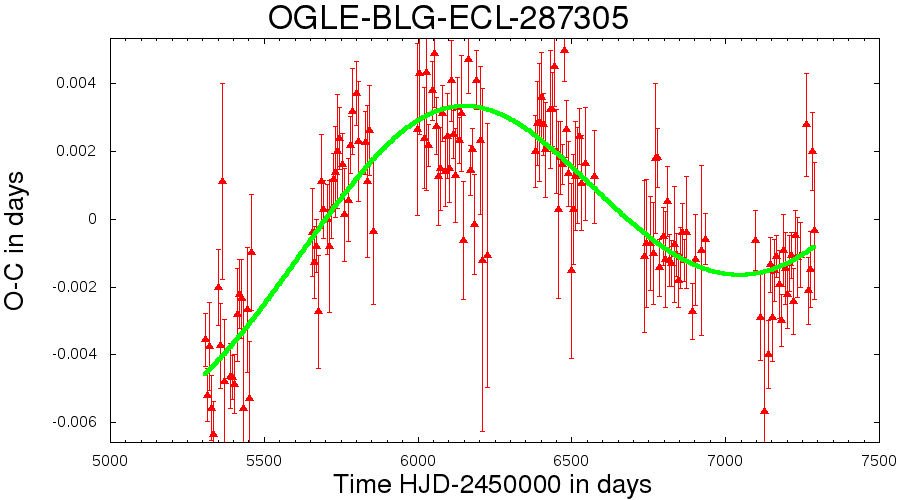}

\includegraphics[width=0.64\columnwidth]{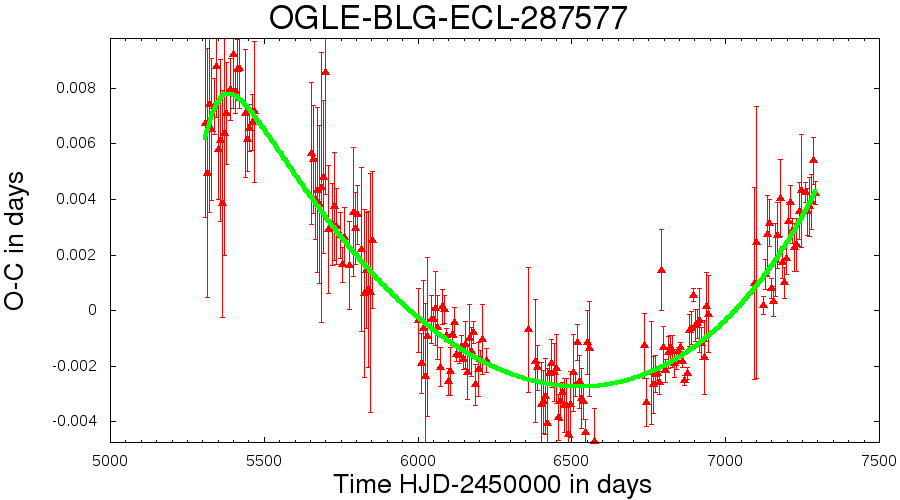}
\includegraphics[width=0.64\columnwidth]{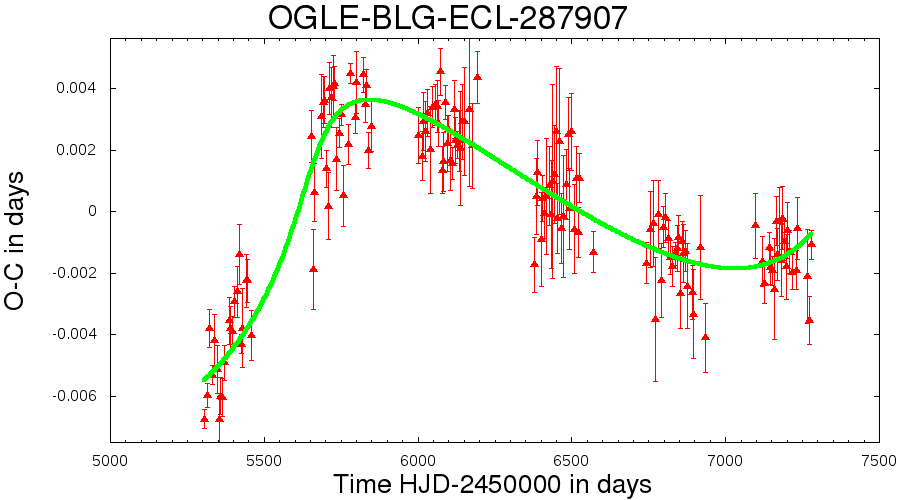}
\includegraphics[width=0.64\columnwidth]{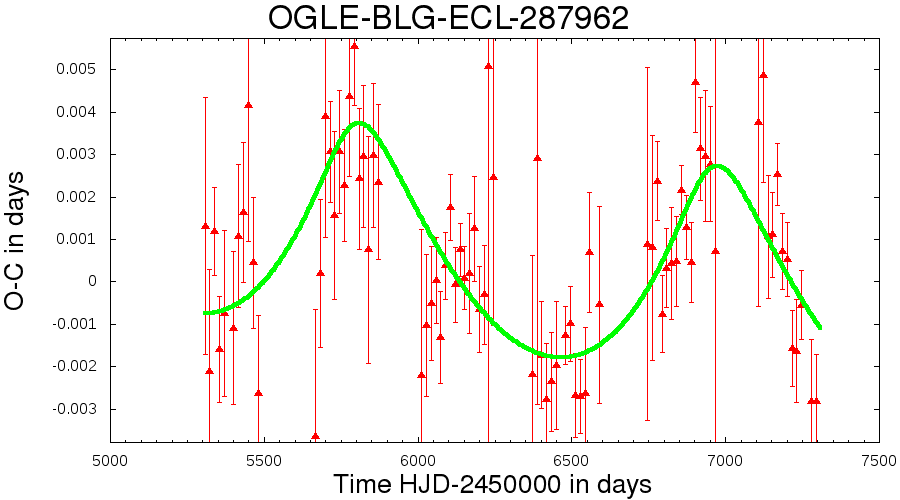}

\end{figure*}
\clearpage

\begin{figure*}
\includegraphics[width=0.64\columnwidth]{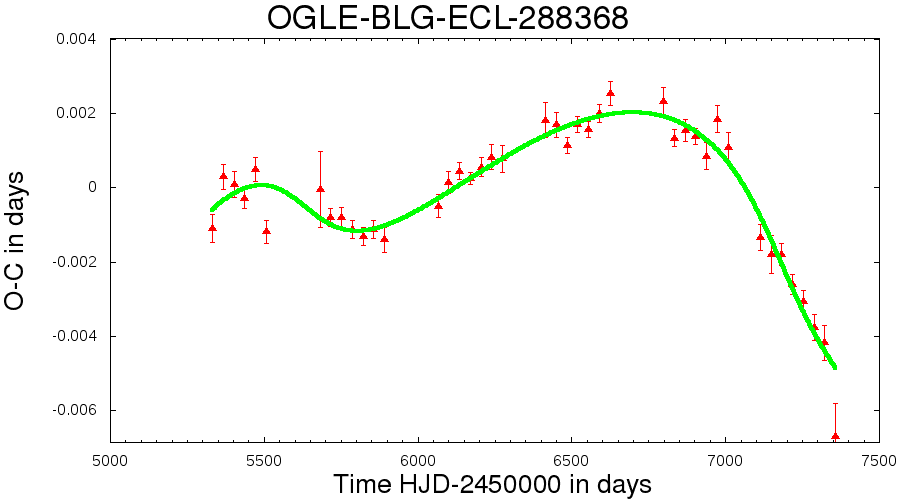}
\includegraphics[width=0.64\columnwidth]{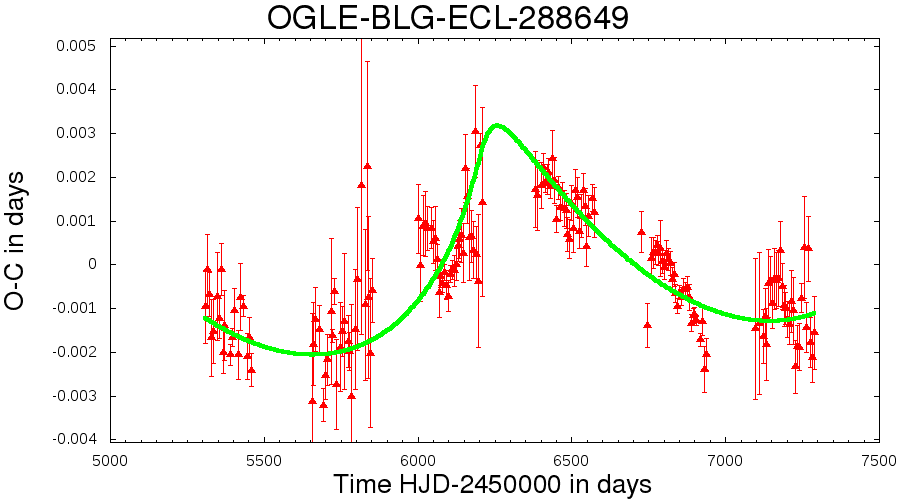}
\includegraphics[width=0.64\columnwidth]{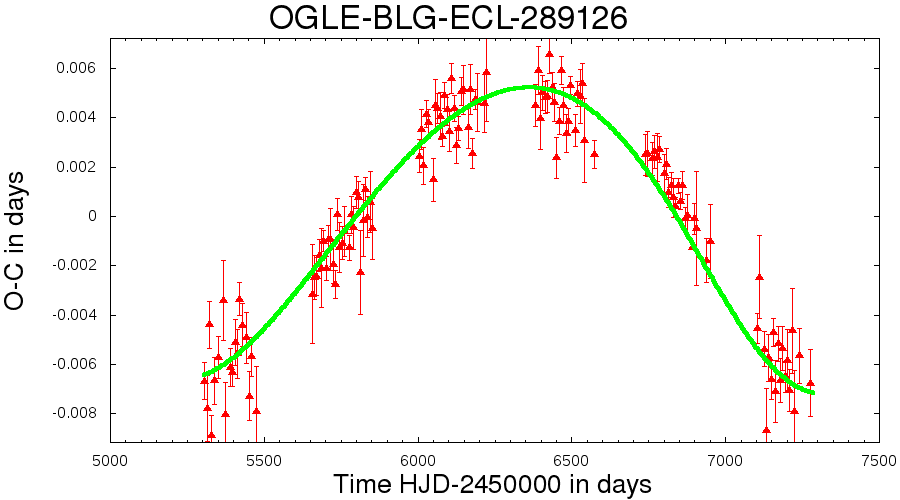}

\includegraphics[width=0.64\columnwidth]{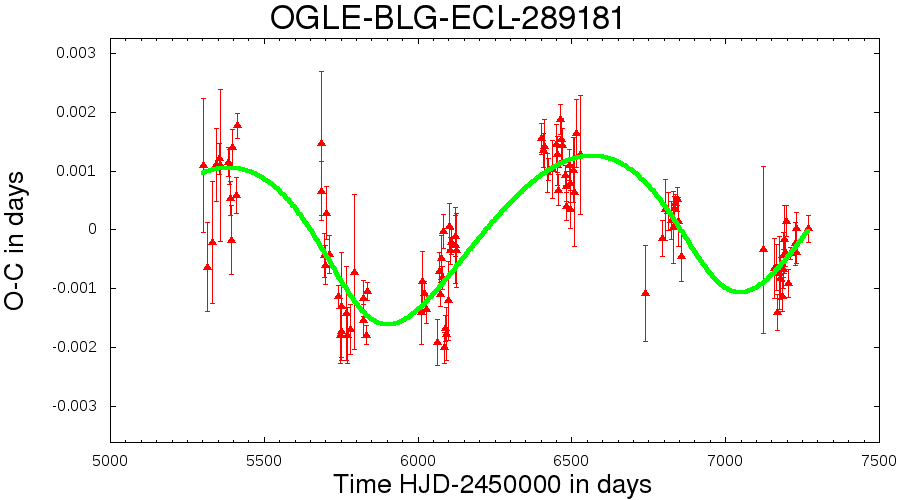}
\includegraphics[width=0.64\columnwidth]{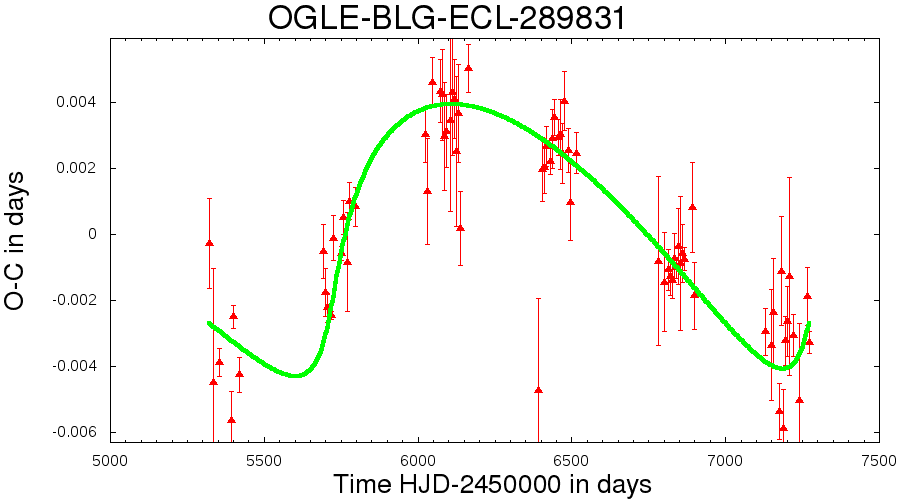}
\includegraphics[width=0.64\columnwidth]{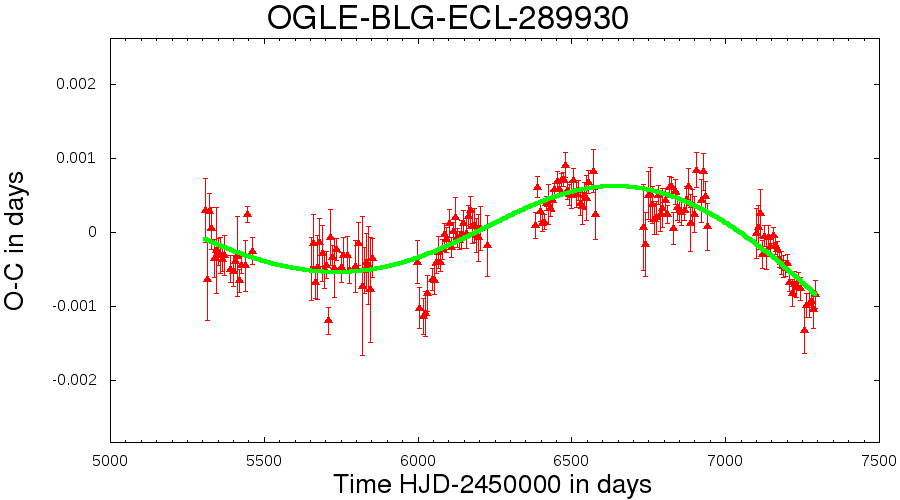}

\includegraphics[width=0.64\columnwidth]{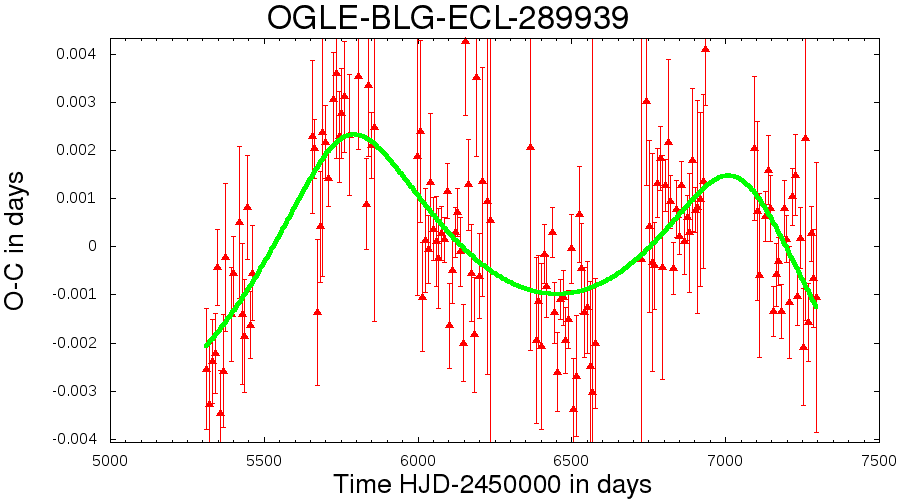}
\includegraphics[width=0.64\columnwidth]{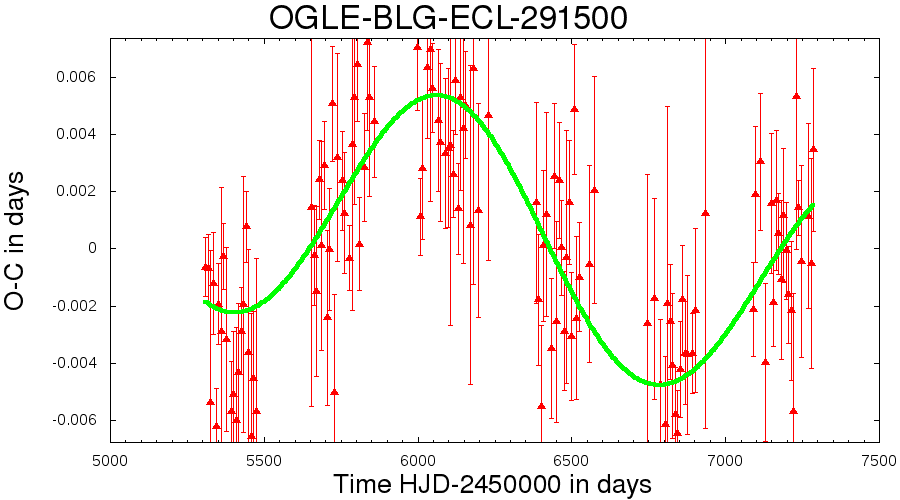}
\includegraphics[width=0.64\columnwidth]{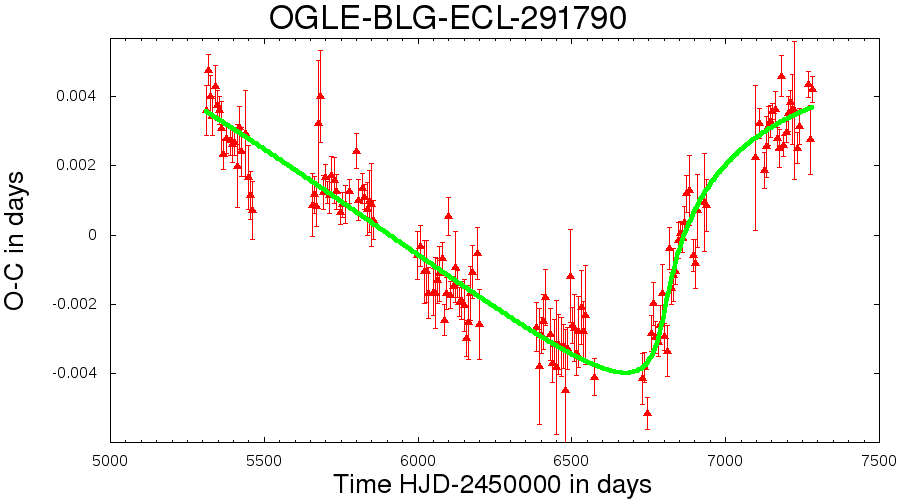}

\includegraphics[width=0.64\columnwidth]{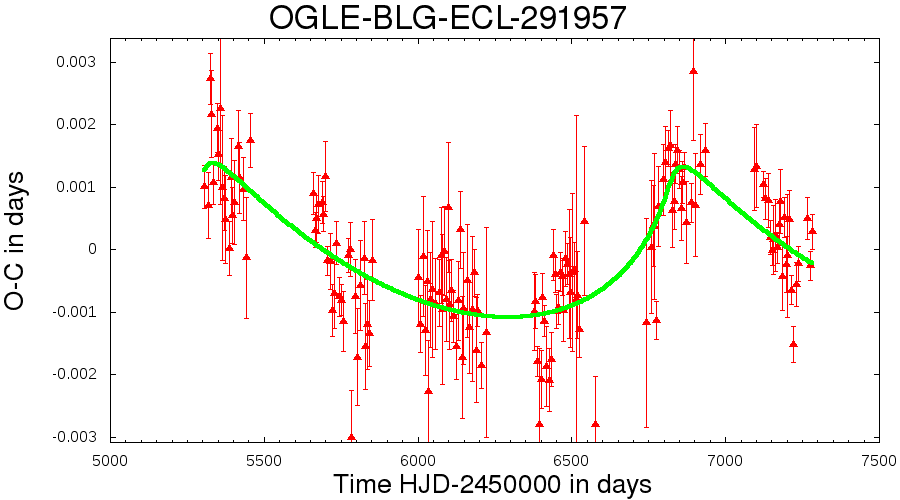}
\includegraphics[width=0.64\columnwidth]{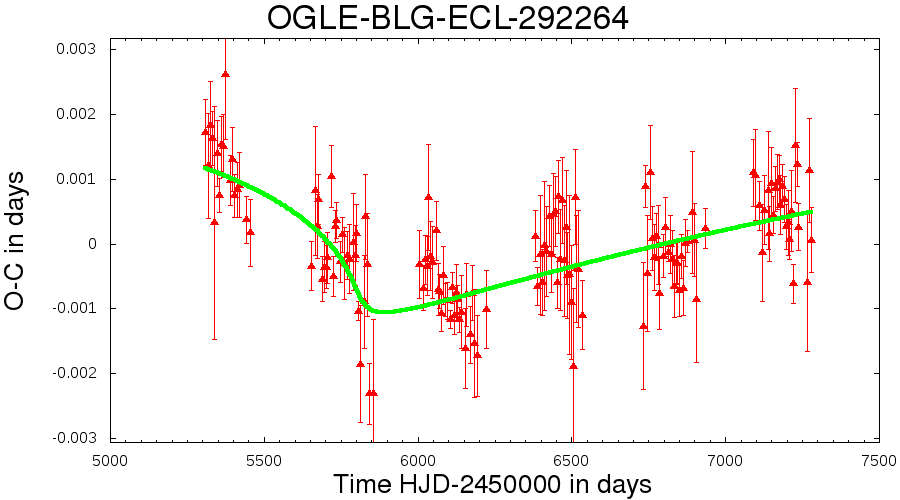}
\includegraphics[width=0.64\columnwidth]{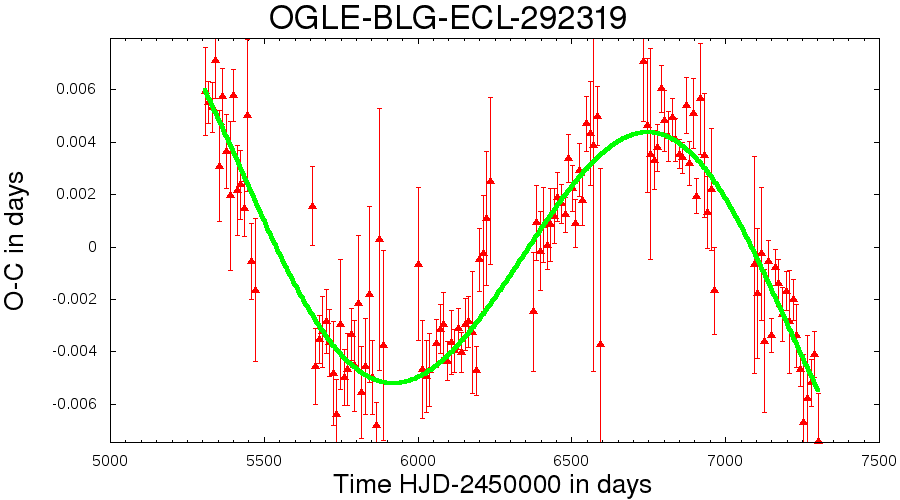}

\includegraphics[width=0.64\columnwidth]{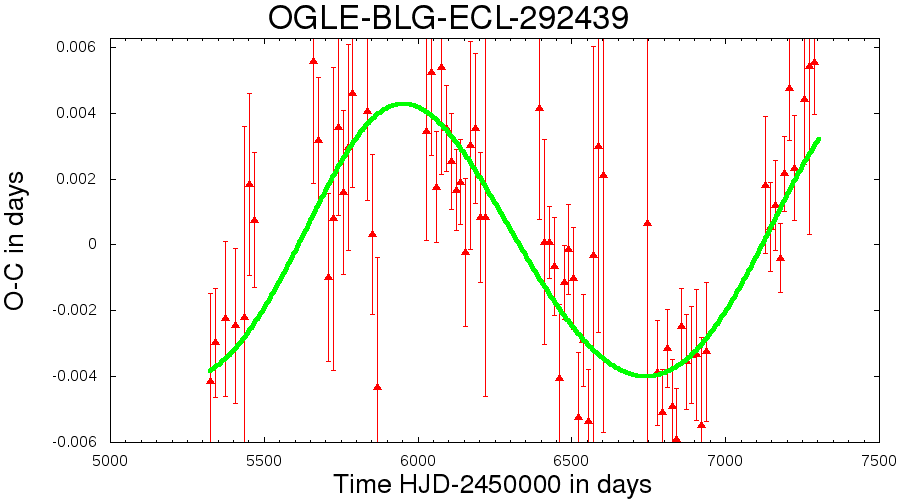}
\includegraphics[width=0.64\columnwidth]{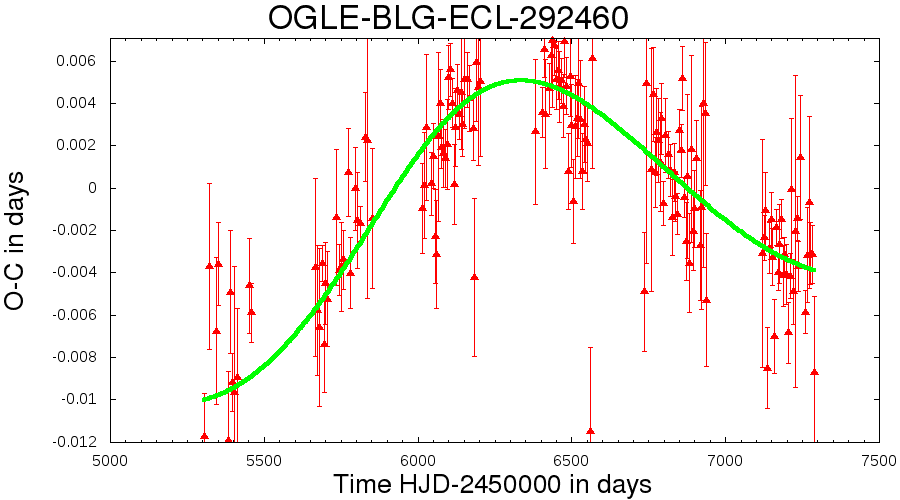}
\includegraphics[width=0.64\columnwidth]{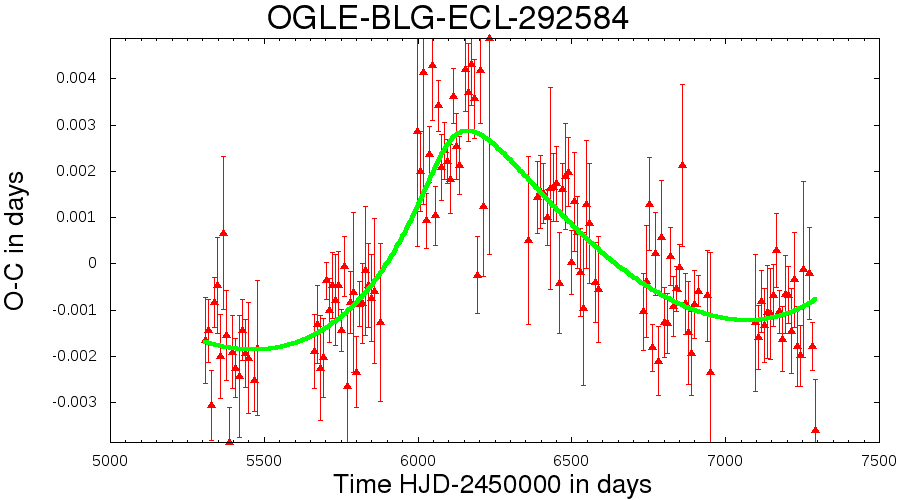}

\includegraphics[width=0.64\columnwidth]{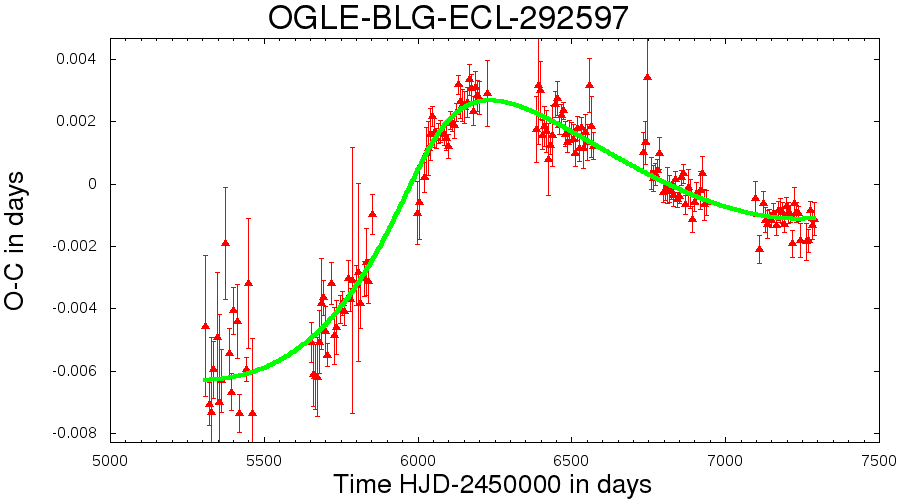}
\includegraphics[width=0.64\columnwidth]{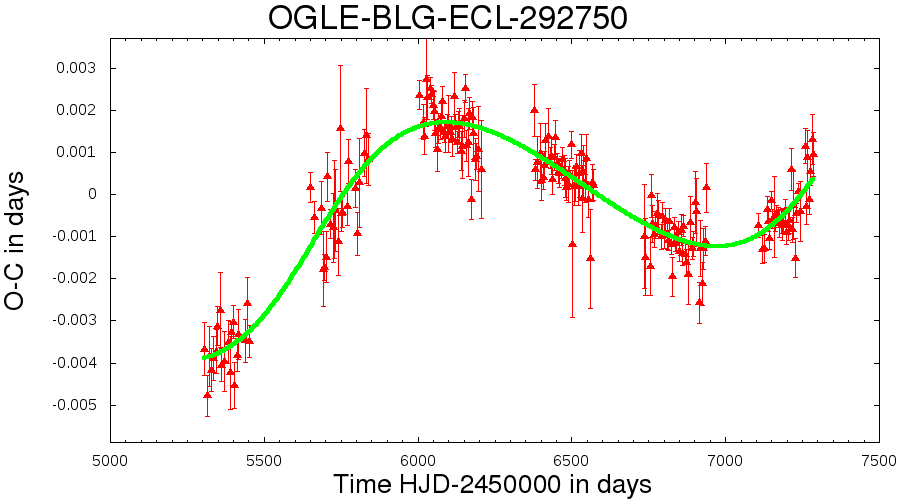}
\includegraphics[width=0.64\columnwidth]{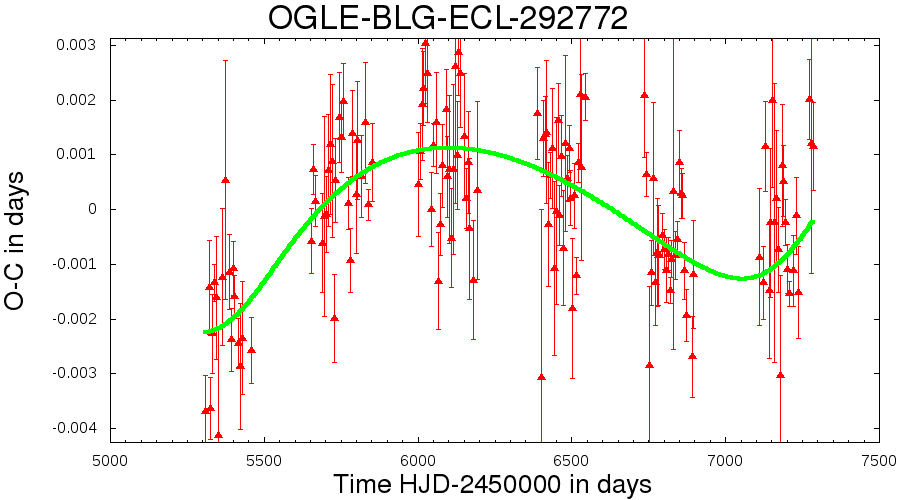}

\includegraphics[width=0.64\columnwidth]{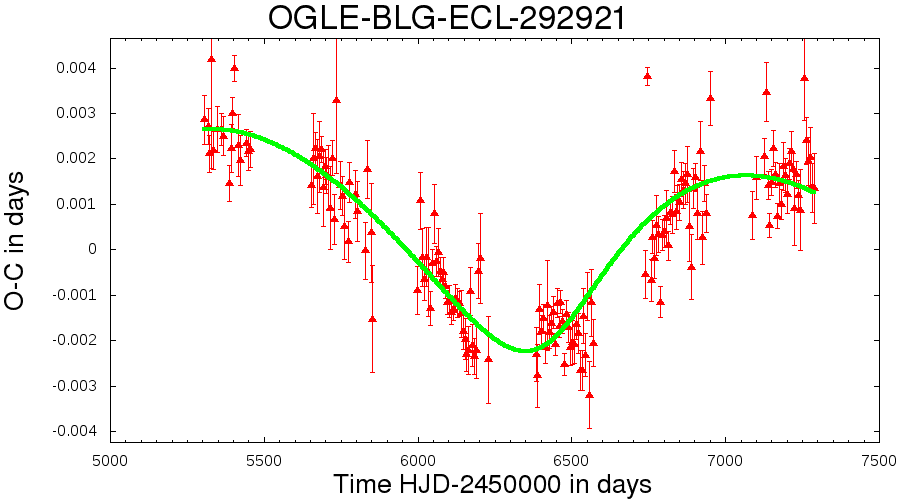}
\includegraphics[width=0.64\columnwidth]{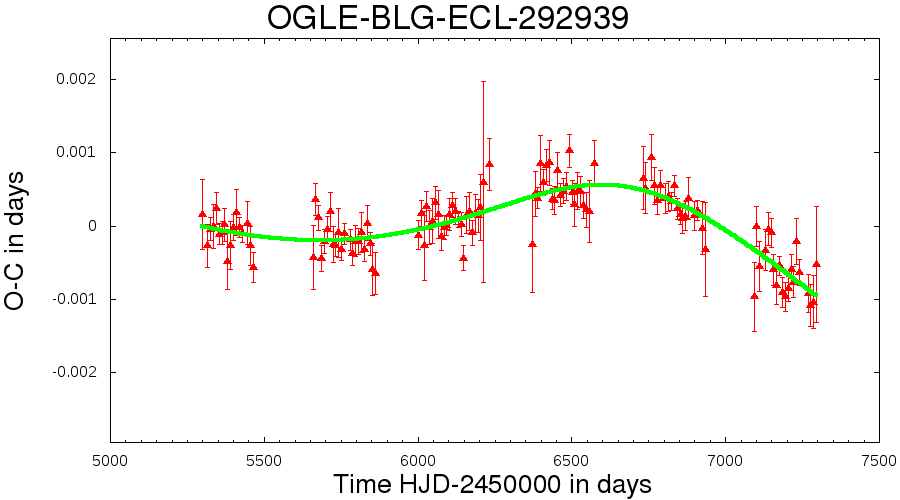}
\includegraphics[width=0.64\columnwidth]{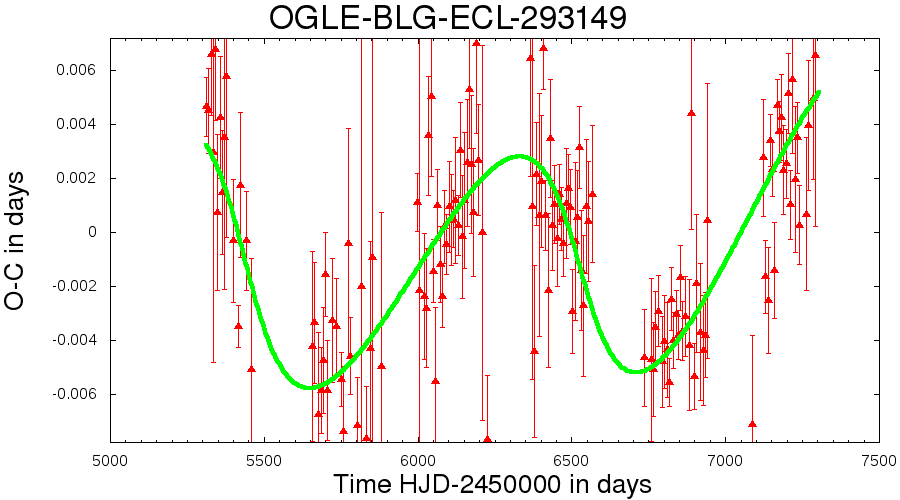}

\includegraphics[width=0.64\columnwidth]{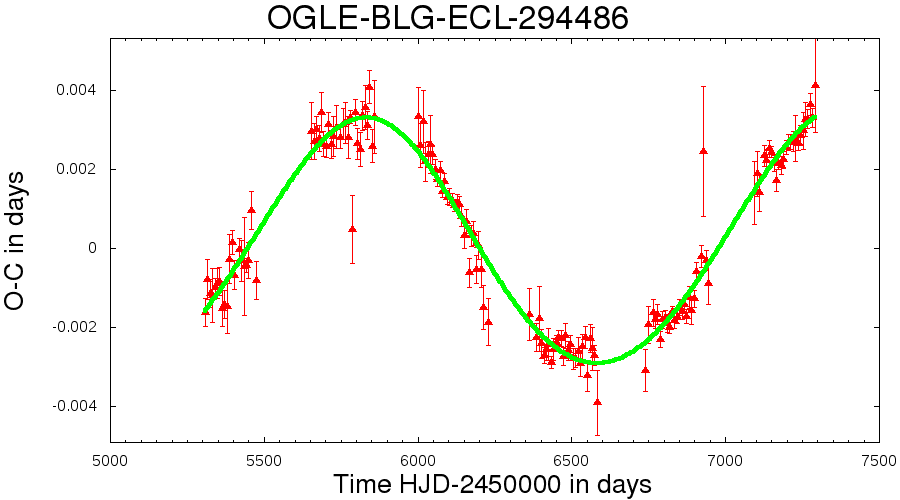}
\includegraphics[width=0.64\columnwidth]{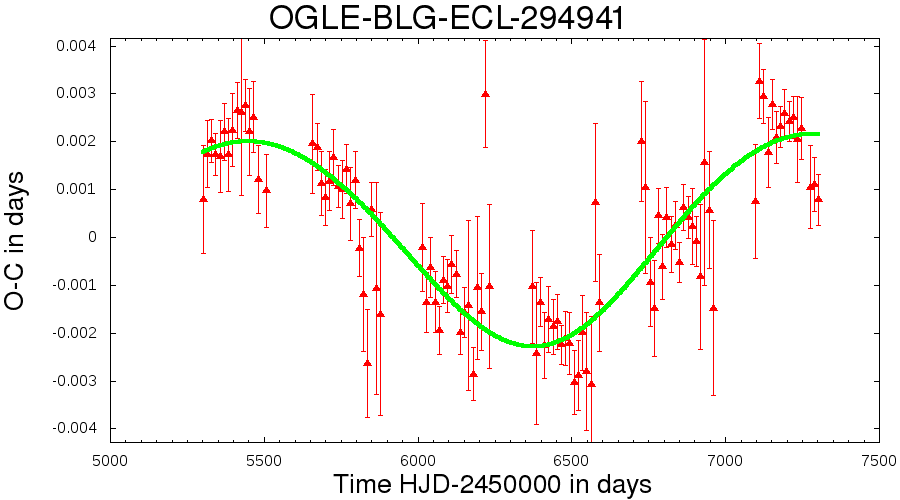}
\includegraphics[width=0.64\columnwidth]{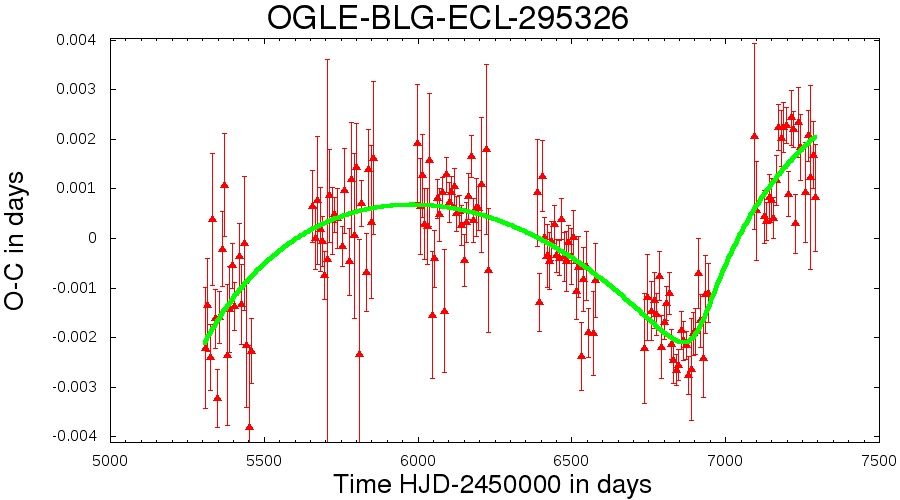}

\end{figure*}
\clearpage

\begin{figure*}
\includegraphics[width=0.64\columnwidth]{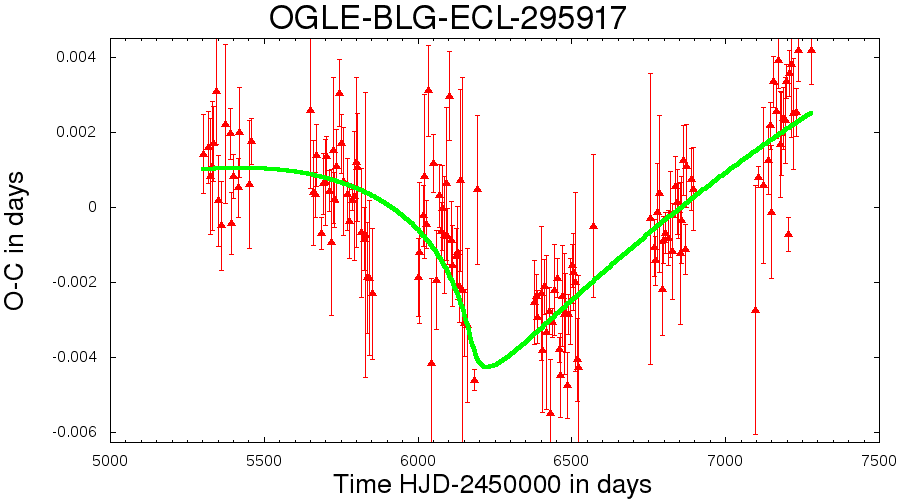}
\includegraphics[width=0.64\columnwidth]{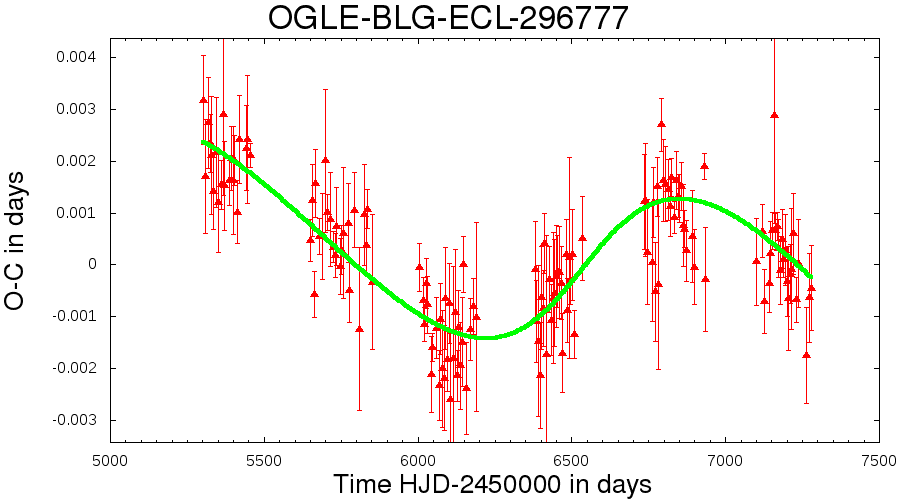}
\includegraphics[width=0.64\columnwidth]{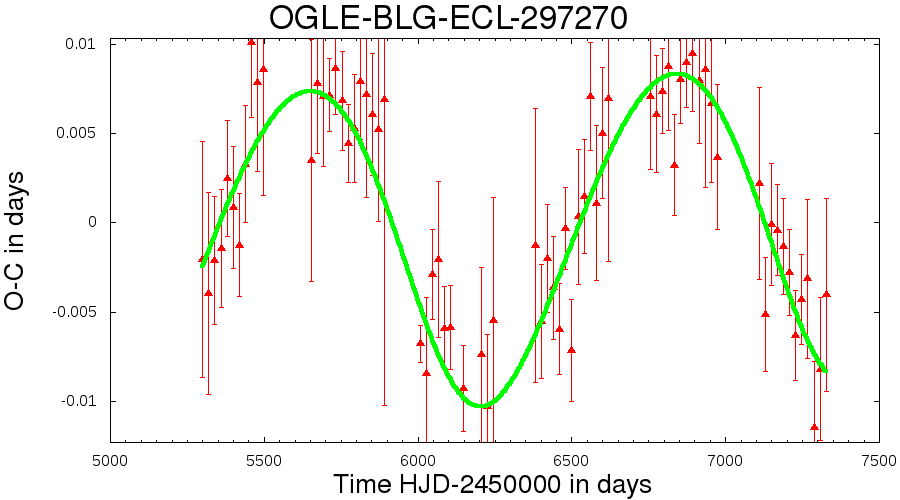}

\includegraphics[width=0.64\columnwidth]{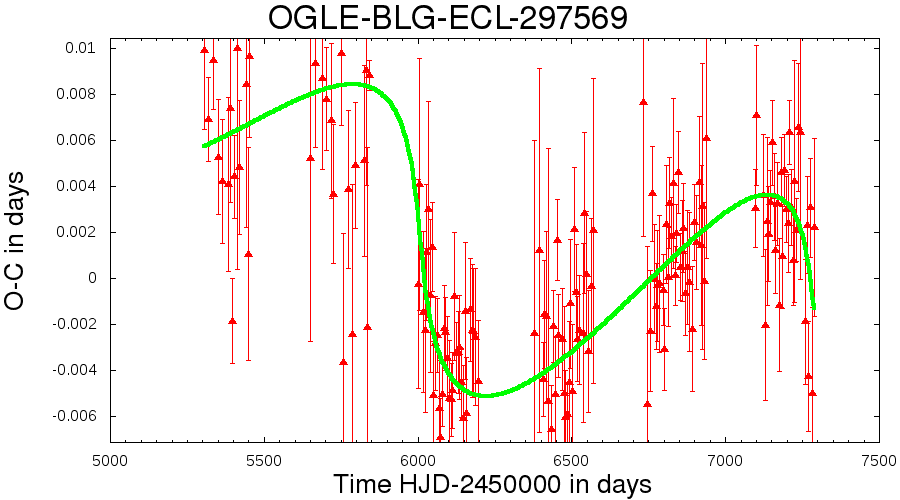}
\includegraphics[width=0.64\columnwidth]{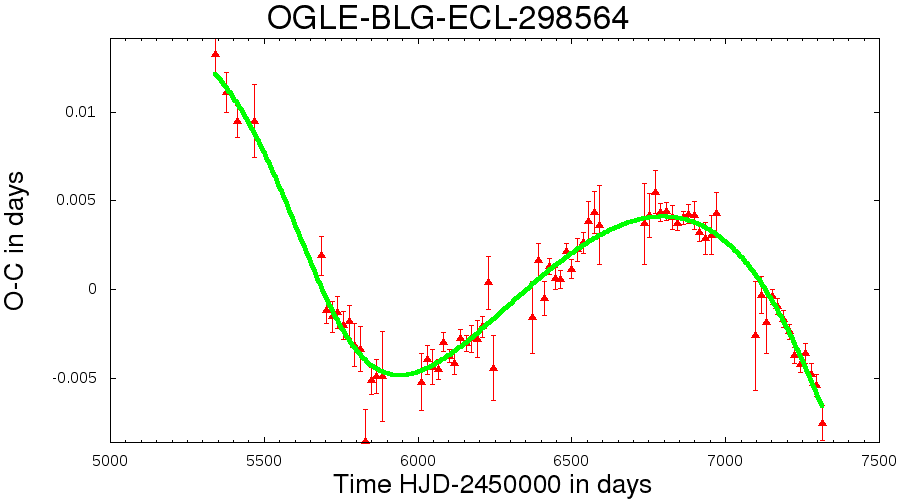}
\includegraphics[width=0.64\columnwidth]{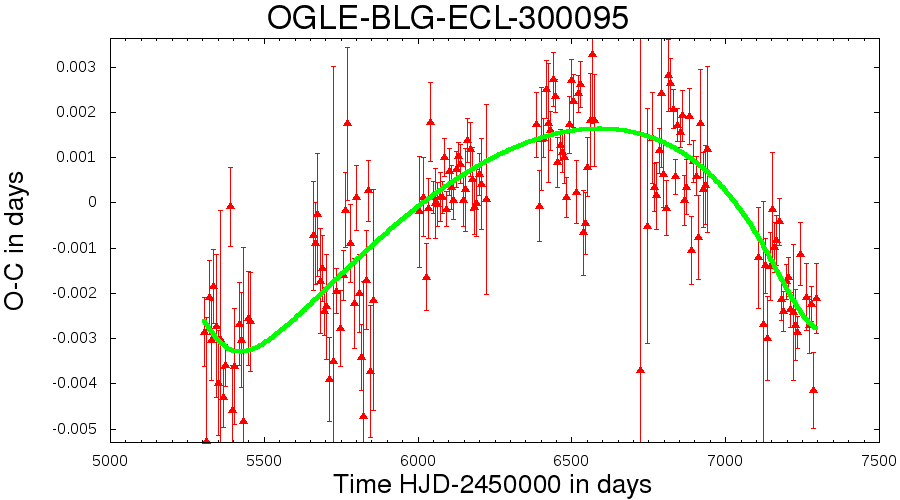}

\includegraphics[width=0.64\columnwidth]{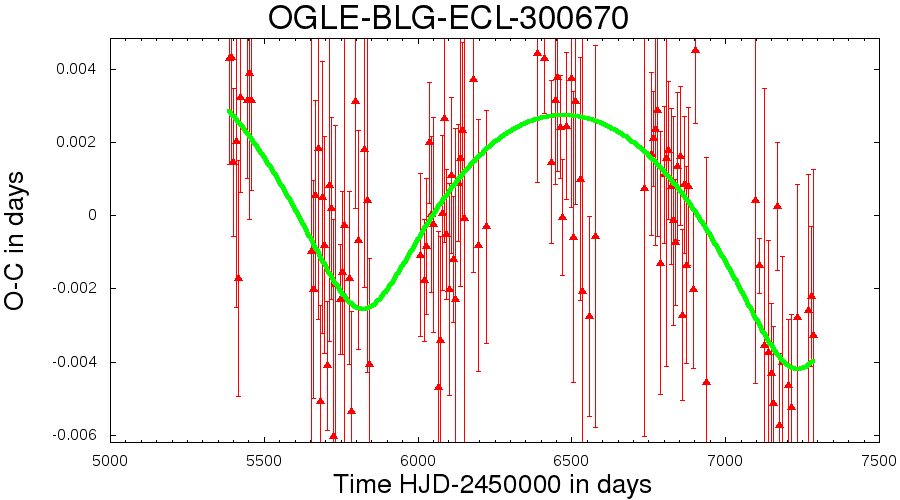}
\includegraphics[width=0.64\columnwidth]{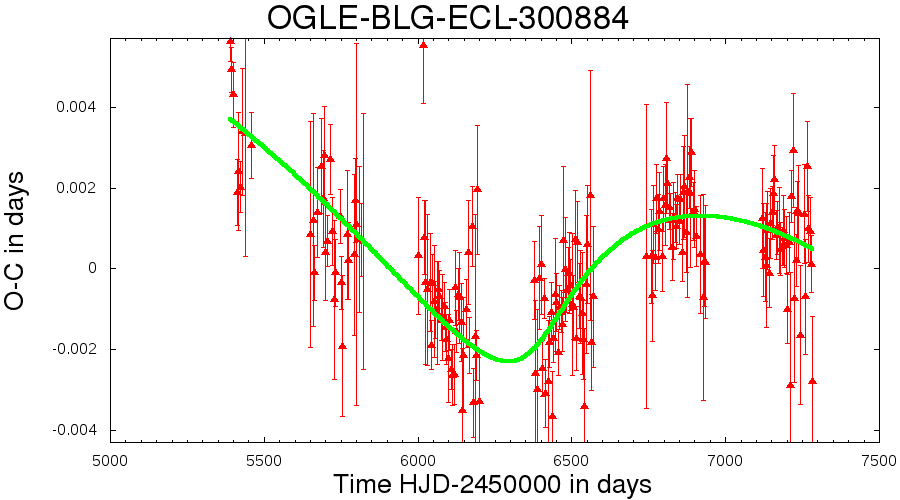}
\includegraphics[width=0.64\columnwidth]{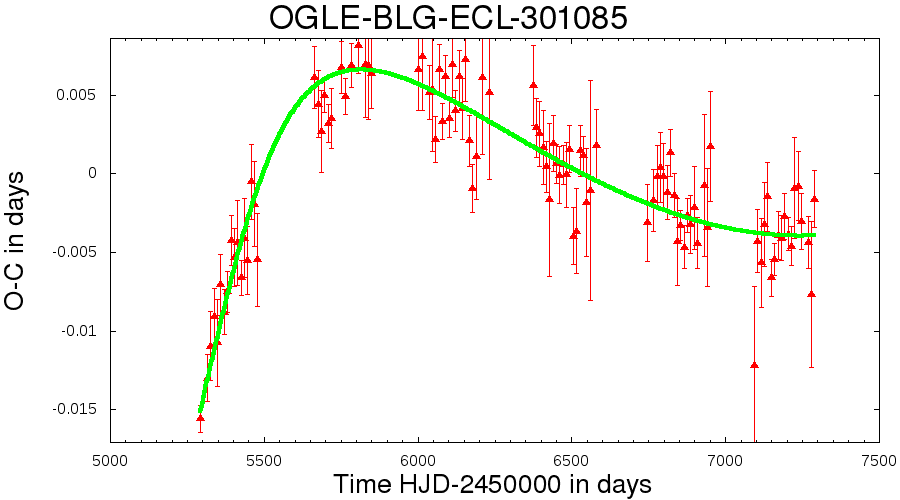}

\includegraphics[width=0.64\columnwidth]{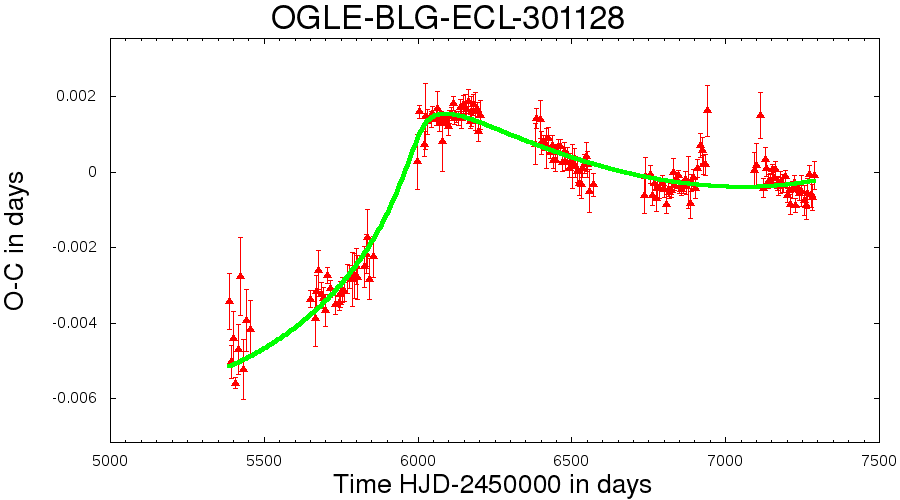}
\includegraphics[width=0.64\columnwidth]{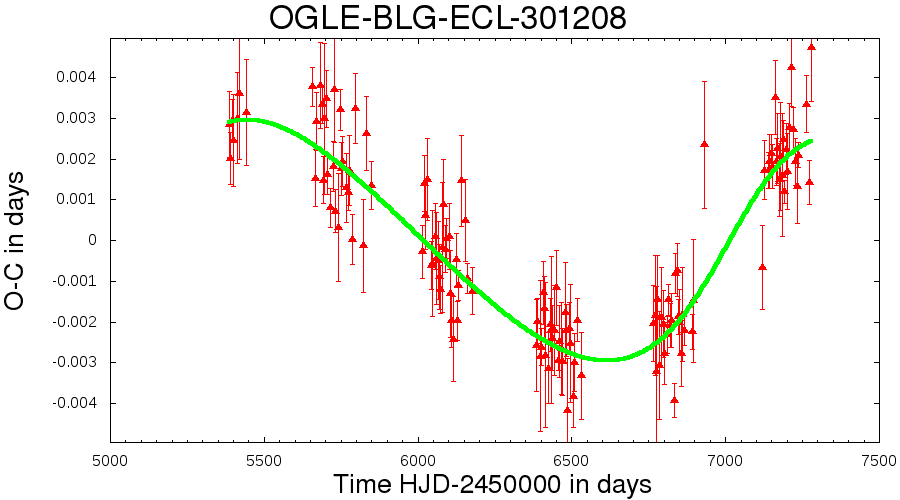}
\includegraphics[width=0.64\columnwidth]{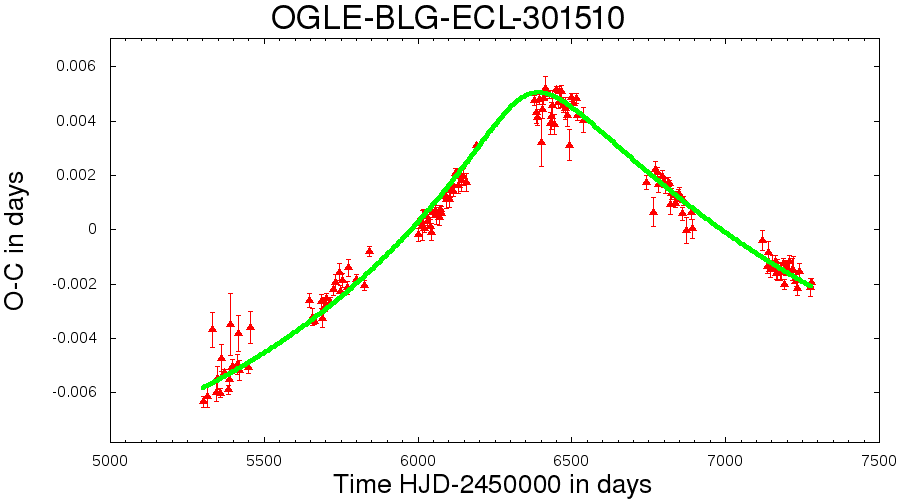}

\includegraphics[width=0.64\columnwidth]{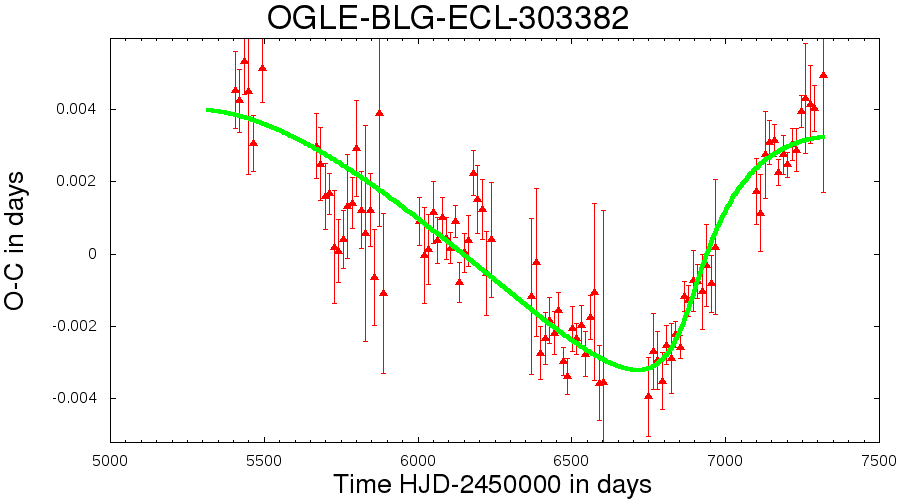}
\end{figure*}

\label{lastpage}
\end{document}